\documentclass[11pt]{article}
\usepackage{makeidx}
\usepackage{amsfonts}
\usepackage{amssymb}
\usepackage{graphicx}
\usepackage{amsmath}
\usepackage{geometry}
\usepackage{appendix}

\DeclareMathSizes{10.95}{10}{7}{6}

\begin{document}

\title{Financial Interactions and Collective States:\\
Part III. Description of Dynamic Collective States}
\author{Pierre Gosselin\thanks{%
Pierre Gosselin : Institut Fourier, UMR 5582 CNRS-UGA, Universit\'{e}
Grenoble Alpes, BP 74, 38402 St Martin d'H\`{e}res, France.\ E-Mail:
Pierre.Gosselin@univ-grenoble-alpes.fr} \and A\"{\i}leen Lotz\thanks{%
A\"{\i}leen Lotz: Cerca Trova, BP 114, 38001 Grenoble Cedex 1, France.\
E-mail: a.lotz@cercatrova.eu}}
\date{April 2026}
\maketitle

\begin{abstract}
In a series of papers, we analyzed the mechanisms of capital allocation and
accumulation among firms, investors, and banks within the framework of Field
Economics. At the macro level, we found that multiple collective states may
emerge, each organized in groups of agents characterized by distinct levels
of disposable capital, returns, and stakes across sectors. This multiplicity
of collective states reflects the inherent instability and default-prone
nature of financial markets. It also underscores the necessity of
transitions among these states.

These transitions were not fully accounted for in our previous description
of collective states, where we only considered marginal fluctuations around
equilibrium levels. To fill this gap, the present paper aims to characterize
the possible dynamic mechanisms underlying transitions between these
collective states. It reviews and extends our previous analysis by examining
their probability and stability. By introducing the notion of dynamic
sub-collective states, it explores potential patterns of transition---such
as morphing, merging, or splitting---among collective states. This study
thus provides the foundational tools and analytical framework that will be
employed to develop a field-theoretic description of dynamic collective
states.


\end{abstract}

\section{Introduction}

In previous work, we studied the mechanisms of capital allocation among a
large number of investors, firms, and banks in a sector space through the
notion of collective states. As a first approximation, a collective state is
decomposed into independent groups of sectors, each characterized by
specific levels of cross-sectoral stakes, disposable capital, and sectoral
returns. These variables jointly define the phase the is group in. We refer
to a group in a given phase as a sub-collective state. A collective state is
therefore defined as a collection of sub-collective states.

Since there exists an infinite number of possible decompositions of the
sector space into groups, and since each group may occupy several distinct
phases, the set of sub-collective states is infinite. As a consequence, the
number of collective states is itself infinite.

Both sub-collective states and the collective states they compose are
inherently prone to instability. Even small perturbations in the level of
disposable capital or returns within a given sector may propagate to other
sectors, triggering adjustments in their capital allocation and returns,
which in turn amplify the initial perturbation. Consequently, neither
sub-collective nor collective states can be regarded as independent or
isolated. Instead, they belong to a fully interconnected system of
interacting states, characterized by nonlinear propagation mechanisms and
feedback effects.

Within this system, states undergo medium- to long-term transformations,
including transitions, recomposition, and defaults, through which the system
evolves from one collective state to another.

Accounting for the multiplicity of sub-collective states, their
interactions, and the transitions between them requires a specific
formalism. Before introducing such a framework, however, it is necessary to
identify the properties it must satisfy. To this end, the present paper
consolidates our previous results, provides a systematic characterization of
sub-collective states and their associated default configurations, and
extends the analysis of their stability. We then broaden this description to
incorporate dynamical deviations around sub-collective states, which we
refer to as \emph{deviation states}. Within this broader perspective, a
collective state can be understood as the combination of a\ \emph{basis state%
}---describing a given partition of stakes among investors, banks, and
firms---and \emph{fiber states}, which encapsulate the multiple
configurations of capital and returns compatible with that partition, as
studied in Gosselin and Lotz (2025a,b). Two distinct types of fields must
therefore be introduced: fields associated with basis states and fields
associated with fiber states.

The formalism we aim to develop is a field-based model in which deviation
states constitute the fundamental variables. Within this framework, a given
group of interconnected investors may undergo multiple variations and
transitions, switching from one phase to another, potentially leading to
default. Sub-collective states may interact, merge, or reorganize, implying
that the sectoral composition of a group is itself endogenous. As a
consequence, sectors may appear or disappear within a group, leading to
transitions between collective states in which the decomposition into groups
is modified, as demonstrated in Gosselin and Lotz (2023d). Transitions
between collective states may be induced either by fiber or basis
transitions, and may occur through sequences of successive transformations
involving both types of fields, each type of transition potentially
triggering the other. This structure reveals that the system is inherently
unstable: the permanent occurrence of discontinuous transitions generates an
internal dynamics that continuously explores the infinite space of possible
collective states, thereby endogenously realizing it.

Based on these properties, we propose a preliminary formalism, which will be
fully developed in subsequent work. In this framework, each group is modeled
by a field that encodes both the set of the possible phases of the group and
the fluctuations around these phases. The resulting formalism therefore
involves an infinite number of fields, corresponding to all possible
decompositions of the sectoral space into groups of agents, together with
their associated dynamics and transitions.

This construction leads to a model that is substantially broader than our
original framework. It allows for the explicit treatment of interactions
between relatively independent groups, their consequences in terms of
transitions between sub-collective states, and, ultimately, transitions
between collective states. In this sense, the proposed formalism synthesizes
and extends the core features developed in our previous works\footnote{%
Gosselin and Lotz ().} (Gosselin and Lotz (2024)).

This work is organized as follows. After a brief literature review in
Section 2, Part I reviews and extends our previous results. In Section 3, we
introduce the notion of collective states, and Section 4 describes these
states in terms of their basis and fiber variables. Sections 5 and 6 study
the probabilities and stability ranges of these states. Section 7 presents a
compact notation for collective states that accounts for possible
interactions between groups. Part II extends the framework to a fully
dynamic model of collective states. Section 8 motivates the need for these
extensions, Section 9 derives the possible forms of dynamic collective
states, and Section 10 describes the expected patterns of transitions
between them. In Section 11, we summarize the characteristics required of
the field-theoretic formalism to describe these transitions. Finally,
Section 12 concludes the paper.

\section{Litterature Review}

Five major directions are related to our approach. The first direction
addresses heterogeneity among agents through distributions of agents modeled
by probability densities. In mean field games (MFG) and mean field control,
individual agents are negligible compared with the population but interact
through aggregate variables (see Bensoussan, Frehse, and Yam (2018) and
Lasry and Lions (2007, 2010a, 2010b)).\ This approach has been applied in
the dynamic programming framework developed by Gomes, Vilanova, and others
(see Gomes et al., 2015; Achdou et al., 2014). Heterogeneous Agent New
Keynesian (HANK) models incorporate similar heterogeneity into macroeconomic
structures.\ An equilibrium probability distribution is derived from a set
of optimizing heterogeneous agents\ in a new Keynesian context (see Kaplan
and Violante (2018), as well as Kaplan, Moll, and Violante (2018) for
quantitative implementations). Information-theoretic approaches build on
Sims' (2003, 2006) rational inattention theory to model agents optimizing
under informational constraints (see Yang (2018) and Matejka and McKay
(2015)). This information theoretic approach considers probabilistic states
around the equilibrium and replaces the Walrasian equilibrium with a
statistical equilibrium derived from an entropy maximisation program.\ In
these three types of models,\ probability distributions can be seen as
particular types of collective states postulated \textit{a priori} as
equilibria of the microeconomics setup.

Field economics, on the contrary, builds on the interactions between agents
at the microeconomic level. We do not postulate an equilibrium probability
distribution for each type of agent.\ Rather we consider any possible
probability density for the entire system of $N$ agents and their
interactions, and translate these probability densities in terms of fields.
Since the fields encompass all the possible densities of agents as their
realizations\footnote{%
In our formalism, the notion of fields refers to some abstract complex
functions defined on the state space and is similar to the
second-quantized-wave functions of quantum theory.}, the state-space in
field economics is thus much larger than those considered in the above
approaches. This allows us to study the agents' economic structural
relations and the emergence of the collective states induced by these
specific micro-relations, which in turn impact each agent's stochastic
dynamics at the microeconomic level. These emerging collective states are in
general multiple with their own characteristics, average quantities, and
distributions of agents, etc.

Interacting agents with heterogeneous behavioral rules have been dealt with
by multi-agent systems, particularly agent-based models (ABMs), with an
emphasis on non-equilibrium dynamics and bounded rationality (Gaffard and
Napoletano (2012), Delli Gatti et al. (2005)). Mandel, Landini, and
Gallegati (2010, 2012) further develop ABMs to capture innovation, sectoral
dynamics, and macroeconomic fluctuations. The field of economic networks,
notably Jackson (2010, 2014), focuses on the structural properties of agent
interactions within economic systems. Both approaches are highly numerical
and model-dependent.\ They also rely on microeconomic relations, such as 
\textit{ad hoc }reaction functions, that may be too simplistic. Field
economics, on the contrary, accounts for transitions between scales.
Macroeconomic patterns do not emerge solely from the dynamics of a large
population of agents: they are grounded in behaviours and interaction
structures. Describing these structures in terms of field theory allows for
the emergence of phases at the macro scale, and the study of their impact at
the individual level.

Econophysics applies statistical physics methods to socio-economic systems
(see Abergel et al. (2011a, 2011b) for reviews of the field's developments).
Lux (2008, 2016) explores stylized facts and agent-based dynamics, while
Kleinert (2009) uses path integral formulations to analyze financial
markets. Other relevant works include Bardoscia et al. (2017), Bouchaud and M%
\'{e}zard (2000), Chakraborti et al. (2011), Chakraborti et al. (2013).
However, Econophysics does not apply the full potential of field theory to
economic systems.\ Instead, it seeks to reveal empirical laws in economic
systems. Field economics, in contrast, keeps track of usual microeconomic
concepts, such as utility functions, expectations, and forward-looking
behaviors.\ It integrates these behaviours into the analytical treatment of
multi-agent systems by translating the main characteristics of optimizing
agents in terms of statistical systems. Closer to our approach, Bardoscia et
al. (2017) study a general equilibrium model for a large economy in the
context of statistical mechanics, and show that phase transitions may occur
in the system. Our issue is similar, but our use of field theory addresses a
larger class of dynamic models.

The second direction addresses interactions between Finance and the Real
Economy. The interactions between financial frictions and capital
accumulation have typically been modeled within DSGE frameworks, enriched
with credit constraints, incomplete markets, or firm-level heterogeneity.
Cochrane (2006) provides a review of asset pricing in macro models.
Bernanke, Gertler, and Gilchrist (1999) model the financial accelerator,
linking firm balance sheets to investment dynamics. Holmstrom and Tirole
(1997) study liquidity provision under moral hazard. More recent work
includes Campello et al. (2010) and Quadrini (2012). Other contributions
extend this framework to heterogeneous agents and endogenous risk (Grassetti
et al. (2022), Grosshans and Zeisberger (2018), B\"{o}hm, Kuehn, and
Schmedders (2008), Khan and Thomas (2013), Monacelli et al. (2011) and Moll
(2014)).

Field economics differs from DSGE models in that their models are micro
models that stand for the entire set of agents.\ This does not allow the
study of the diffusion and circulation of capital among agents. Field
economics, on the contrary, studies the entire system of agents.\ When
dealing with capital, it describes the different states that result from
this diffusion and circulation of capital. Besides, the relative
(un)stability of capital allocation can be assessed globally: it is the
relative distribution of returns or external conditions that determine
investors' allocation of capital to firms. We can study how a local interest
rate or a change in sectoral returns would impact the equilibrium of one
sector, and the whole system.

The third direction covers the literature on default and systemic risk, and
their contagion within financial networks. Early theoretical models were
developed by Allen and Gale (2000), Cifuentes, Ferrucci, and Shin (2005),
and Gai and Kapadia (2010). The network-based approach has been formalized
by Acemoglu, Ozdaglar, and Tahbaz-Salehi (2015) and extended by Bardoscia et
al. (2019) and Glasserman and Young (2015) using graph-theoretic
stress-testing techniques and feedback loop dynamics (see also Battiston et
al. (2012, 2020) for recursive losses in interconnected systems and Haldane
and May (2011) for an ecology and epidemiology approach to financial
instability). Empirically, contributions include Reinhart and Rogoff (2009),
Gennaioli, Martin, and Rossi (2012, 2018), Adrian and Brunnermeier (2016),
Langfield, Liu, and Ota (2020).

This paper adds to these studies by considering firms, investors and banks
in the analysis. Moreover, collective states are described both in terms of
global averages and sectoral quantities,\ allowing disparities in firms'
returns and borrowing conditions across agents to be explicitly accounted
for. Our analysis therefore moves back and forth between the micro and macro
levels to identify the conditions under which micro-level defaults propagate
to the macroeconomic level.

A fourth strand of the literature emphasizes the intermediary role of banks
and the regulatory frameworks---especially macroprudential
policies---designed to stabilize the financial system. It includes theories
of banking (Diamond and Dybvig (1983); Diamond and Rajan (2001); Freixas,
Parigi and Rochet (2000); Freixas and Rochet (2008)) and modern
macro-financial models (Gertler and Kiyotaki (2010); Brunnermeier and
Sannikov (2014); Adrian and Brunnermeier (2016)). Macroprudential tools are
explicitly aimed at reducing the likelihood that idiosyncratic shocks
trigger widespread defaults. Empirical studies and policy evaluations (Borio
and Drehmann (2009), Brunnermeier, Crockett, Goodhart, Persaud, and Shin
(2009); Acharya and Richardson (2012); Galati and Moessner (2013); Drehmann
and Juselius (2014); IMF (2014); Cerutti, Claessens and Laeven (2017))
indicate that well-designed macroprudential measures can lower the
probability of default, although their effectiveness depends on timing,
calibration, and cross-border coordination.

Our work differs from these approaches in that, in our framework,
instability is inherent to collective states, and the apparently more
capitalized, high-yield investors or banks are themselves sources of
systemic fragility. Since risk is an ad-hoc concept rather than an absolute
measure, restricting access to credit to low-return investors and firms in
the name of prudence may in fact strengthen high-return investors and banks,
thereby increasing structural instability.

The present work is specifically related to a fifth direction of research:
the field formalism for collective states developped in Gosselin and Lotz
(2023a,b,c,d). This litterature is specifically based on the assumption that
collective states can be decomposed in sub-collective states, that is the
combination of a group and the phase it is in. Using this assumption, an
effective field formalism is developped that associates one specific field
to each group of the decomposition along with all the possible phases this
group can take. The number of possible groups composing the collective
states of the system, being infinite, the number of fields in this formalism
will also be infinite. In this context, the interactions between fields
describe the transitions, modifications and creations of the groups The
closest modeling tool to the one described here, although completely
different in the objects studied would be string field theory, particularly
its description in terms of an infinite family of fields (see Erbin 2018):
string states can be seen as collections of large number of oscillators of
multiple frequencies, and each field in this infinite family of fields
correspond to one of these state and its potential variations and dynamics.
The analogy with this branch of theoretical physics is only formal but
stresses the fact that describing interactions between large number of
emerging states needs an enlargment of the formalism, by introducing a sort
of field of fields.

\part*{Collective States: Review and Extensions}

In the context of field economics, multiple collective states may emerge.
Each is composed of several groups of agents in a particular phase,
characterized by their level of sectoral disposable capital and returns and
cross-sectoral stakes. These collective states are not necessarily stable,
and can experience transitions towards other collective states. In this part
we gather our previous results and describe the characteristics of
sub-collective states, including their possible default states. We also
present and extend the study of the stability of sub-collective states.
Ultimately, we introduce a compact notation for these states that accounts
for their possible interactions.

\section{Collective States}

In view of our previous works on financial interactions, we present the main
characteristics of emerging collective states in systems of financial
interactions, detailed in the Appendices\footnote{%
The model, the action functionals, a summary of the main equations and the
principle of resolution is given in Appendix 1, while technical details are
given in Appendices 2 to 4.}. The system is composed of several fields, each
describing the different types of agents and their main characteristics:
disposable capital, sectors of location, and connections with other agents,
here stakes. Besides, we can associate to each type of agent an action
functional, which translates, in terms of fields, the agents' microeconomic
behaviour. The system of these action functionals is, what we call, the
field model. The field realizations that minimize these action functionals
are called the background fields of the system and the set of background
fields are called the collective states of the system.

\subsection{Fields}

There are two types of fields involved in the present model\footnote{%
See Gosselin and Lotz (2024, 2025a,b).}: the three fields that describe
capital distribution in each sector, among firms, investors, and banks,
respectively; and the two fields representing the possible connections
between agents of different sectors, i.e. the stakes, for investors and
banks respectively.

\subsubsection{Fields of capital}

The fields describing the capital distribution among agents within each
sector of the sector space are, respectively:

\paragraph{Investors}

\begin{equation*}
\hat{\Psi}\left( \hat{K},X\right)
\end{equation*}

\paragraph{Banks}

\begin{equation*}
\bar{\Psi}\left( \bar{K},X\right)
\end{equation*}

\paragraph{Firms}

\begin{equation*}
\Psi \left( K,X\right)
\end{equation*}%
The arguments of these fields are the level of disposable capital per agent
and sector, so that each realization of these fields $\hat{\Psi}$,$\bar{\Psi}
$ and $\Psi $ yields a distribution of disposable capital $\hat{K}$, $\bar{K}
$ and $K$ for investors, banks and firms respectively in each sector $X$.

The dynamical properties of the fields are summarized in their respective
action functionals, $\hat{S}\left( \hat{\Psi}\right) $, $\bar{S}\left( \bar{%
\Psi}\right) $, and $S\left( \Psi \right) $ that describe the accumulation
process of each type of agents, depending on their returns and their
connections to other sectors\footnote{%
The formula for the field action functionals are given in Appendix 1.}.

\subsubsection{Fields of stakes}

The endogenous connections between agents are their respective stakes. In
our setup firms are exlusively producers, and any investment activity of a
firm is deemed as that of an investor. As a result, there will be only two
distinct fields of stakes in our model, one for investors and one for banks%
\footnote{%
See Gosselin and Lotz (2025a,b).}.

\paragraph{Investors}

The field of stakes for investors is written as:%
\begin{equation*}
\Gamma \left( \hat{S}^{\left( T\right) },X^{\prime },X\right)
\end{equation*}%
where $X$ is the sector of origin of the stakes, and $X^{\prime }$ is the
sector invested in, and $\hat{S}^{\left( T\right) }$ is the vector of all
possible stakes of investors:%
\begin{equation*}
\hat{S}^{\left( T\right) }=\left( S_{1},\hat{S}_{1},S_{2},\hat{S}_{2}\right)
\end{equation*}%
In this vector, the shares and loans taken by an investor $X$ in firms $X$
are denoted $S_{1}$ and $S_{2}$ respectively, whereas the shares and loans
taken by an investor $X$ in an investor $X^{\prime }$ are denoted $\hat{S}%
_{1}$ and $\hat{S}_{2}$ respectively.

\paragraph{Banks:}

The field of stakes for banks is written as:%
\begin{equation*}
\bar{\Gamma}\left( \bar{S}^{\left( T\right) },X^{\prime },X\right)
\end{equation*}%
where $X$ is the sector of origin of the stakes, and $X^{\prime }$ is the
sector invested in, and $\bar{S}^{\left( T\right) }$ is the vector of all
possible stakes of banks:%
\begin{equation*}
\bar{S}^{\left( T\right) }=\left( \bar{S}_{E},S_{E}^{B},\hat{S}_{E}^{B},\bar{%
S}_{L},S_{L}^{B},\hat{S}_{L}^{B}\right)
\end{equation*}%
In this vector, the shares and loans taken by a bank $X$ in firms $X$ are
denoted $S_{E}^{B}$ and $S_{L}^{B}$ respectively, whereas the shares and
loans taken by a bank $X$ in an investor $X^{\prime }$ are denoted $\hat{S}%
_{E}^{B}$ and $\hat{S}_{L}^{B}$ respectively and the shares and loans taken
by a bank $X$ in a bank $X^{\prime }$ are denoted $\bar{S}_{E}$ and $\bar{S}%
_{L}$ respectively.

The fields of stakes $\Gamma $ and $\bar{\Gamma}$\ describe the possible
repartitions of investments' across sectors for investors and banks
respectively and the dynamical aspects of these fields are described by the
two action functionals $S\left( \Gamma \right) $ and $S\left( \bar{\Gamma}%
\right) $\footnote{%
See Appendix 1, formula (\ref{GCt}) and (\ref{GT}).}.

\subsection{Background Fields}

The minimizations of the action functionals $S\left( \Psi \right) $, $\hat{S}%
\left( \hat{\Psi}\right) $ and $\bar{S}\left( \bar{\Psi}\right) $ for
disposable capital and $S\left( \Gamma \right) $ and $S\left( \bar{\Gamma}%
\right) $ for stakes yield the background fields of the system, that are the
most probable configurations for each field $\Psi $, $\hat{\Psi}$ and $\bar{%
\Psi}$, $\Gamma $ and $\bar{\Gamma}$. These background fields are\textbf{\ }%
written $\hat{\Psi}_{0}\left( K,X\right) $, $\bar{\Psi}_{0}\left( K,X\right) 
$\ and $\Psi _{0}\left( K,X\right) $,\ and describe the most likely
distribution of capital for investors, banks and firms, whereas $\Gamma
_{0,X^{\prime },X}\left( \hat{S}^{\left( T\right) }\right) $\ and $\Gamma
_{0,X^{\prime },X}\left( \bar{S}^{\left( T\right) }\right) $\ represent the
most likely distribution of stakes of investors and banks across sectors.

The complete collection of possible set of background fields:\textbf{\ }%
\begin{equation*}
\left\{ \hat{\Psi}_{0}\left( K,X\right) \mathbf{,\ }\bar{\Psi}_{0}\left(
K,X\right) \mathbf{,\ }\Psi _{0}\left( K,X\right) \mathbf{,\ }\Gamma
_{0,X^{\prime },X}\left( \hat{S}^{\left( T\right) }\right) \mathbf{,\ }%
\Gamma _{0,X^{\prime },X}\left( \bar{S}^{\left( T\right) }\right) \right\}
\end{equation*}%
\textbf{\ }define the collective states of the system. However, these fields
are not independent. They are interconnected with each other. This reflects
that disposable capital depends on stakes, and that stakes between sectors
depend also on disposable capital and returns. We can show that these sets
of background fields depend on average sectoral stakes\footnote{%
See Appendix 1.} that solve the return equations of the system. These return
equations were derived in Gosselin and Lotz (2025a,b)\footnote{%
See Appendix 2 for a summary.}. Therefore, these sectoral stakes become the
fundamental variables of the system.

In each collective state, and for any quantity, such as stakes, returns and
disposable capital, both overall and sectoral averages can be obtained.

\subsubsection{Background fields of stakes}

The minimization of the action functionals $S\left( \Gamma \right) $ and $%
S\left( \bar{\Gamma}\right) $ yield the most probable realizations of the
fields of stakes, the background fields. These background fields describe
the distribution of stakes in each sector. They are gaussian distributions
centered around the sectoral averages $\hat{S}_{\eta }^{\left( T\right)
}\left( X^{\prime },X\right) $ and $\bar{S}_{\eta }^{\left( T\right) }\left(
X^{\prime },X\right) $, that characterize the collective states. Since there
exist several sectoral averages, there are several background fields and
each of these background fields is defined by a set of values $\left\{ \hat{S%
}_{\eta }^{\left( T\right) }\left( X^{\prime },X\right) ,\bar{S}_{\eta
}^{\left( T\right) }\left( X^{\prime },X\right) \right\} $ across sectors.

Thus, solving the model and finding the collective states amounts to finding
the possible sets $\left\{ \hat{S}_{\eta }^{\left( T\right) }\left(
X^{\prime },X\right) ,\bar{S}_{\eta }^{\left( T\right) }\left( X^{\prime
},X\right) \right\} $.

\paragraph{Investors}

The background field of stakes for investors is given by: 
\begin{equation}
\Gamma _{0,X^{\prime },X}\left( \hat{S}^{\left( T\right) }\right) =N\exp
\left( -\sum_{\eta }\frac{\left( \hat{S}_{\eta }^{\left( T\right) }-\hat{S}%
_{\eta }^{\left( T\right) }\left( X^{\prime },X\right) \right) ^{2}}{2\sigma
_{\hat{K}}^{2}}\right)  \label{Gmz}
\end{equation}%
where $N$ is a normalization factor that describes the overall number of
connections between investors.

\paragraph{Banks}

The background field of stakes for banks is given by:%
\begin{equation}
\Gamma _{0,X^{\prime },X}\left( \bar{S}^{\left( T\right) }\right) =\bar{N}%
\exp \left( -\sum_{\eta }\frac{\left( \bar{S}_{\eta }^{\left( T\right) }-%
\bar{S}_{\eta }^{\left( T\right) }\left( X^{\prime },X\right) \right) ^{2}}{%
2\sigma _{\hat{K}}^{2}}\right)  \label{Gmb}
\end{equation}%
where $\bar{N}$\ is a normalization factor that describes the overall number
of connections between banks and between banks and investors.

\subsubsection{Background fields of capital}

The background fields\ $\bar{\Psi}$ and $\hat{\Psi}$ describe the
distribution of capital in each sector, from which the average disposable
capital per sector and per type of agent in the collective state can be
derived. As for stakes, these values depend on the sectoral averages for
stakes, so that the collective state can, in last analysis, be described in
terms of stakes' sectoral averages.

The background fields $\Psi \left( K,X\right) $, $\hat{\Psi}\left(
K,X\right) $, and $\bar{\Psi}\left( K,X\right) $ are expressed in terms of
several types of variables. First, in terms of returns, for investors,
denoted $\hat{f}\left( X\right) $, and for banks, denoted $\bar{f}\left(
X\right) $. Second, in terms of the averages of these returns, denoted $%
\left\langle \hat{f}\right\rangle $ and $\left\langle \bar{f}\right\rangle $
respectively. Third, in terms of the sectoral stakes and their averages%
\footnote{%
See Gosselin and Lotz (2025a, b).}. Finally, in terms of the average fields
for investors and banks\footnote{%
Computed in Gosselin and Lotz (2025a, b).}, $\left\Vert \hat{\Psi}%
\right\Vert ^{2}$ and $\left\Vert \bar{\Psi}\right\Vert ^{2}$. In first
approximation, the formulas for the background fields can be written as
follows:

\paragraph{Firms}

The background field for firms is given by:

\begin{equation}
\left\vert \Psi \left( K,X\right) \right\vert ^{2}=\left\vert \Psi
_{0}\left( X\right) \right\vert ^{2}-\epsilon \left( \frac{\left(
f_{1}^{\left( e\right) }\left( X\right) K-\bar{C}\left( X\right) \right) ^{2}%
}{\sigma _{\hat{K}}^{2}}+\frac{f_{1}^{\left( e\right) }\left( X\right) }{2}%
\right)  \label{SLf}
\end{equation}%
where $f_{1}^{\left( e\right) }\left( X\right) $ is the firm's net return of
production once loans have been repaid, and $\bar{C}\left( X\right) $ is the
average cost per unit of capital.

\paragraph{Investors}

The background field for investors is given by:%
\begin{equation}
\left\vert \hat{\Psi}\left( \hat{K},X\right) \right\vert ^{2}=\left\Vert 
\hat{\Psi}_{0}\left( X\right) \right\Vert ^{2}-\hat{\mu}\left\{ \left( \frac{%
\hat{K}_{1}^{2}\hat{g}^{2}\left( X\right) }{2\sigma _{\hat{K}}^{2}}+\frac{%
\hat{g}\left( X\right) }{2}\right) -\left( \frac{\left\langle \hat{K}%
\right\rangle ^{2}\left\langle \hat{g}\right\rangle }{\sigma _{\hat{K}}^{2}}+%
\frac{1}{2}\right) \left\langle \hat{g}^{ef}\right\rangle \frac{\hat{K}}{%
\left\langle \hat{K}\right\rangle }\right\}  \label{sli}
\end{equation}%
where $\hat{g}\left( X\right) $ is defined as:%
\begin{equation*}
\hat{g}\left( X\right) =\frac{\hat{f}\left( X\right) }{\left\langle \bar{f}%
\right\rangle }+\frac{\left\langle \hat{S}\left( X^{\prime },X\right)
\right\rangle _{X^{\prime }}}{1-\left\langle \hat{S}\right\rangle }+\frac{%
\left( \left\langle \hat{S}_{1}^{B}\left( X,X^{\prime }\right) \right\rangle
_{X^{\prime }}+\left\langle \hat{S}_{2}^{B}\left( X,X^{\prime }\right)
\right\rangle _{X^{\prime }}+\frac{\left\langle \hat{S}_{1}^{B}\right\rangle
+\left\langle \hat{S}_{2}^{B}\right\rangle }{1-\left\langle \hat{S}%
\right\rangle }\right) \frac{\left\langle \bar{K}\right\rangle \left\Vert 
\bar{\Psi}\right\Vert ^{2}}{\left\langle \hat{K}\right\rangle \left\Vert 
\hat{\Psi}\right\Vert ^{2}}\left\langle \bar{f}\right\rangle }{\left(
1-\left\langle \bar{S}\right\rangle \right) \left\langle \bar{f}%
\right\rangle }
\end{equation*}%
The coefficients $\left\langle \hat{g}^{ef}\right\rangle $ is an average
coefficient that measures the impact on returns of any modification in the
field\footnote{%
In general, these coefficients are negligible.}.

\paragraph{Banks}

The background field for banks is given by:%
\begin{equation}
\left\vert \bar{\Psi}\left( \bar{K},X\right) \right\vert ^{2}=\left\vert 
\bar{\Psi}_{0}\left( X\right) \right\vert ^{2}-\hat{\mu}\left\{ \left( \frac{%
\bar{K}_{1}^{2}\bar{g}^{2}\left( X\right) }{\sigma _{\hat{K}}^{2}}+\frac{%
\bar{g}\left( X\right) }{2}\right) -\left( \frac{\left\langle \hat{K}%
\right\rangle ^{2}\left\langle \hat{g}\right\rangle }{\sigma _{\hat{K}}^{2}}+%
\frac{1}{2}\right) \left\langle \hat{g}^{Bef}\right\rangle \frac{\bar{K}}{%
\left\langle \hat{K}\right\rangle }\right\}  \label{slb}
\end{equation}%
where:%
\begin{equation*}
\bar{g}\left( X\right) =\left( \left( 1-\left\langle \bar{S}\right\rangle
\right) \bar{f}\left( X\right) +\left\langle \bar{S}\left( X^{\prime
},X\right) \right\rangle _{X^{\prime }}\left\langle \bar{f}\right\rangle
\right) \frac{\left\langle \bar{S}\left( X^{\prime },X\right) \right\rangle 
}{1-\left\langle \bar{S}\left( X^{\prime },X\right) \right\rangle }
\end{equation*}%
and $\left\langle \hat{g}^{Bef}\right\rangle $ measures the impact on
returns of any modification in the field\footnote{%
In general, these coefficients are negligible.}.

In the above expressions, the functions $\hat{g}\left( X\right) $ and $\bar{g%
}\left( X\right) $ are mixed returns composed of several factors: first,
investors' and banks' returns $\hat{f}\left( X\right) $ and $\bar{f}\left(
X\right) $ and their averages $\left\langle \hat{f}\right\rangle $ and $%
\left\langle \bar{f}\right\rangle $; second, the agents' stakes and their
averages\footnote{%
See Gosselin and Lotz (2025a, b).}; and third, the average fields\footnote{%
Computed in Gosselin and Lotz (2025a, b).} $\left\Vert \hat{\Psi}\right\Vert
^{2}$ and $\left\Vert \bar{\Psi}\right\Vert ^{2}$. Therefore, the functions $%
\hat{g}\left( X\right) $ and $\bar{g}\left( X\right) $ describe how
investors' and banks' sectoral returns $\hat{f}\left( X\right) $ and $\bar{f}%
\left( X\right) $ spread and combine through the stakes $\hat{S}_{\eta
}^{\left( T\right) }\left( X^{\prime },X\right) $, $\bar{S}_{\eta }^{\left(
T\right) }\left( X^{\prime },X\right) $ and their respective averages.

Note that the coefficients $\left\langle \hat{g}^{ef}\right\rangle $ and $%
\left\langle \hat{g}^{Bef}\right\rangle $ are average coefficients that
measure the impact on returns of any modification in the field\footnote{%
In general, these coefficients are negligible.}.

\subsection{Sub-Collective States}

\subsubsection{Collective States as Collections of Sub-Collective States}

Under uncertainty, agents connect to only a finite number of neighbouring
agents\footnote{%
See Gosselin and Lotz (2025a,b),}, so that agents are organized into several
and loosely-connected groups. The background fields (\ref{Gmz}), (\ref{Gmb}%
), (\ref{SLf}), (\ref{sli}), (\ref{slb}), and the collective states they
represent, must therefore be computed for any arbitrary group $G$ of agents.

For any group \footnote{%
See Appendix 1,}\textbf{\ }$G$, multiple solutions to the action
functionals' minimization equations exist, each defining a background field
and a corresponding phase. Therefore background fields are not unique.

Alternately, each phase of a group is defined by one set of background
fields (\ref{Gmz}), (\ref{Gmb}), (\ref{SLf}), (\ref{sli}), (\ref{slb}) among
multiple possibilities. We will define a sub-collective state as a group
within a specific phase.

The full collective states of the system will thus consist, first in a
decomposition of the sector space in groups, and, second, in a collection of
subcollective states associated to this decomposition.

\subsubsection{Recovering collective state from subcollective states}

Thus, each collective state is associated with an arbitrary decomposition of
the sector space into a collection $\left\{ G\right\} $ of several groups,
each in a phase defined by its background fields:%
\begin{equation*}
\Gamma _{0,G}\left( \hat{S},X^{\prime },X\right) \text{,}\bar{\Gamma}%
_{0,G}\left( \bar{S},X^{\prime },X\right) \text{,}\hat{\Psi}_{0,G}\left( 
\hat{K},X\right) \text{,}\bar{\Psi}_{0,G}\left( \bar{K},X\right) \text{,}%
\Psi _{0,G}\left( K,X\right)
\end{equation*}%
In this context, the background fields associated to a given decomposition
of the whole sector space in groups $\left\{ G\right\} $\ are the product of
the background fields for each group of the decomposition.

The collective state for the whole set of agents is thus defined by the
product of fields of stakes stakes for investors and banks respectively:%
\begin{equation}
\prod_{G}\Gamma _{0,G}\left( \hat{S},X^{\prime },X\right) \text{, }\prod_{G}%
\bar{\Gamma}_{0,G}\left( \bar{S},X^{\prime },X\right)  \label{Prn}
\end{equation}%
and by the product of the fields of disposable capital for firms, investors
and banks respectively:%
\begin{equation}
\prod_{G}\hat{\Psi}_{0,G}\left( \hat{K},X\right) \text{, }\prod_{G}\bar{\Psi}%
_{0,G}\left( \bar{K},X\right) \text{, }\prod_{G}\Psi _{0,G}\left( K,X\right)
\label{Prd}
\end{equation}

\section{Basis and Fiber Variables}

\subsection{Collective States as Collections of Basis and Fiber Variables}

The collective states described previously depend, for each group, on the
levels of average stakes, returns and disposable capital per sector. Thus,
we must determine, for each decomposition of the sector space in groups, the
levels of disposable capital and returns for each type of agents for each
sector for each group, and the distribution of stakes across all these
sectors. Taken together, these quantities will define one collective state
of the system.

Since returns and disposable capital per sector can be derived from the
levels of stakes between investors, a collective state is described by a set
of values of stakes for investors. We will call these variables the \textit{%
basis variables}, while the remaining variables, that are functions of these
basis variables in a given collective state, will be called \textit{fiber
variables}.

For a given collective state, the fiber variables are determined by the
distribution of the basis variables, here the sectoral stakes. However,
their expression is part of the complete description of the collective state.

\subsection{Basis background fields and variables}

\subsubsection{Basis background fields}

The organization of collective states in groups depends on the level of
stakes within the system. The basis background fields are thus the fields of
stakes for the groups composing the collective state:%
\begin{equation*}
\left\{ \Gamma _{0,G}\left( \hat{S},X^{\prime },X\right) ,\bar{\Gamma}%
_{0,G}\left( \bar{S},X^{\prime },X\right) \right\} _{G}
\end{equation*}

\subsubsection{Basis variables}

A collective state is characterized by its basis variables, here the
sectoral stakes of investors on the one hand, and on the other hand the
sectoral stakes of banks wether in other banks or investors:%
\begin{eqnarray}
&&\left\{ \hat{S}_{E}\left( X^{\prime },X\right) ,\hat{S}\left( X^{\prime
},X\right) ,S_{E}\left( X,X\right) ,S\left( X,X\right) ,\right.  \label{stf}
\\
&&\left. \hat{S}_{E}^{B}\left( X^{\prime },X\right) ,\hat{S}^{B}\left(
X^{\prime },X\right) ,S_{E}^{B}\left( X,X\right) ,S^{B}\left( X,X\right) ,%
\bar{S}_{E}\left( X^{\prime },X\right) ,\bar{S}^{B}\left( X^{\prime
},X\right) \right\} _{_{\left( X^{\prime },X\right) \in S}}  \notag
\end{eqnarray}%
These values close the formulas (\ref{Gmz}), (\ref{Gmb}), (\ref{sli}), (\ref%
{slb}), (\ref{SLf}) for the background fields.

These sectoral stakes are governed by the return equations of investors and
banks\footnote{%
See Appendix 1.}. These are non-local equations that depend on the
connections within each group. However, since under uncertainty, agents
connect to only a finite number of neighbouring agents, these agents are
organize into several and loosely-connected groups. Each collection of
groups describes a certain decomposition of the sector space $S$. Besides,
each of these groups $G$ can be in one among several possible phase and each
of these phases is defined by a distribution of stakes between agents in $G$:%
\begin{eqnarray}
&&\left\{ \hat{S}_{E}\left( X^{\prime },X\right) ,\hat{S}\left( X^{\prime
},X\right) ,S_{E}\left( X,X\right) ,S\left( X,X\right) ,\right.  \label{SG}
\\
&&\left. \hat{S}_{E}^{B}\left( X^{\prime },X\right) ,\hat{S}^{B}\left(
X^{\prime },X\right) ,S_{E}^{B}\left( X,X\right) ,S^{B}\left( X,X\right) ,%
\bar{S}_{E}\left( X^{\prime },X\right) ,\bar{S}^{B}\left( X^{\prime
},X\right) \right\} _{\left( X^{\prime },X\right) \in G}  \notag
\end{eqnarray}%
Thus, under the assumption of independent groups, a collective state is a
set of sub-collective states, that writes in terms of stakes:%
\begin{eqnarray}
&&\cup _{G}\left\{ \hat{S}_{E}\left( X^{\prime },X\right) ,\hat{S}\left(
X^{\prime },X\right) ,S_{E}\left( X,X\right) ,S\left( X,X\right) ,\right.
\label{SC} \\
&&\left. \hat{S}_{E}^{B}\left( X^{\prime },X\right) ,\hat{S}^{B}\left(
X^{\prime },X\right) ,S_{E}^{B}\left( X,X\right) ,S^{B}\left( X,X\right) ,%
\bar{S}_{E}\left( X^{\prime },X\right) ,\bar{S}^{B}\left( X^{\prime
},X\right) \right\} _{\left( X^{\prime },X\right) \in G}  \notag
\end{eqnarray}%
Since there is an infinite number of organization of the sector space into
collections of groups $\left\{ G\right\} $, and since for each organization,
multiple distinct phases are possible for each group, there is an infinite
number of possible collective states (\ref{SC}).

\subsection{Fiber background fields and variables}

Once the decomposition into groups and the values of sectoral stakes for
this decomposition fixed, the collective states are completely determined.
The remaining variables describing the state, the fiber variables, can then
be derived for each collective state. These fiber variables are the sectoral
background fields, i.e. the capital-average of the background fields of
capital the average disposable capital per sector, the returns for firms,
investors and banks per sectors. For the sake of simplicity, we only present
the two first of these variables, the other ones are given in Appendix 3.

\subsubsection{Fiber background fields}

The fiber background fields of the system complete the description of the
collective state. They consist in the collection of background fields for
each group of the collective state:%
\begin{equation*}
\left\{ \hat{\Psi}_{0,G}\left( \hat{K},X\right) \text{, }\bar{\Psi}%
_{0,G}\left( \bar{K},X\right) \text{, }\Psi _{0,G}\left( K,X\right) \right\}
_{G}
\end{equation*}

\subsubsection{Fiber variables}

\paragraph{First fiber variable: Sectoral background fields}

The first fiber variable considered are, the average over the disposable
capital $K$ of $\hat{\Psi}_{0,G}\left( \hat{K},X\right) $ and $\bar{\Psi}%
_{0,G}\left( \bar{K},X\right) $ respectively, for each sector. These average
sectoral fields are denoted $\hat{\Psi}\left( X\right) $, $\bar{\Psi}\left(
X\right) $ and defined by:

\begin{equation}
\left\vert \hat{\Psi}_{0}\left( X\right) \right\vert ^{2}\simeq \left( \frac{%
\left\langle \hat{g}\right\rangle }{\hat{g}\left( X\right) }\hat{I}%
_{X/\left\langle X^{\prime }\right\rangle }\right) ^{\frac{3}{2}}\left\Vert 
\hat{\Psi}\right\Vert ^{2}  \label{FPN}
\end{equation}%
and:%
\begin{equation}
\left\vert \bar{\Psi}_{0}\left( X\right) \right\vert ^{2}\simeq \left( \frac{%
\left\langle \bar{g}\right\rangle }{\bar{g}\left( X\right) }\bar{I}%
_{X/\left\langle X^{\prime }\right\rangle }\right) ^{\frac{3}{2}}\left\Vert 
\bar{\Psi}\right\Vert ^{2}  \label{FPD}
\end{equation}%
where the variables $\hat{I}_{X/\left\langle X^{\prime }\right\rangle }$ and
\ $\bar{I}_{X/\left\langle X^{\prime }\right\rangle }$ measure the relative
investment of investors and banks $X^{\prime }$ in investors and banks $X$
and are written as:%
\begin{equation}
I_{X/\left\langle X^{\prime }\right\rangle }=\frac{\frac{\left\langle \hat{S}%
\left( X,X^{\prime }\right) \right\rangle _{X^{\prime }}}{1-\left\langle 
\hat{S}\left( X,X^{\prime }\right) \right\rangle _{X^{\prime }}}}{\frac{%
\left\langle \hat{S}\left( X,X^{\prime }\right) \right\rangle }{%
1-\left\langle \hat{S}\left( X,X^{\prime }\right) \right\rangle }}
\label{IDef}
\end{equation}%
and:%
\begin{equation*}
\bar{I}_{X/\left\langle X^{\prime }\right\rangle }=\frac{\frac{\left\langle 
\bar{S}\left( X^{\prime },X\right) \right\rangle _{X}}{1-\left\langle \bar{S}%
\left( X^{\prime },X\right) \right\rangle _{X^{\prime }}}}{\frac{%
\left\langle \bar{S}\left( X^{\prime },X\right) \right\rangle }{%
1-\left\langle \bar{S}\left( X^{\prime },X\right) \right\rangle }}
\end{equation*}%
Thus, the sectoral background fields depend on the stakes, and their average
over the whole sector space.

\paragraph{Second Fiber Variable: Disposable capital}

The second fiber variable considered is the disposable capital per type of
agent per sector. It is the sum over the whole set of one type of agents in
a given sector of their disposable capital. It comprises the agents private
capital and the capital invested in them by the whole set of cross-sectoral
agents.

For a given distribution of stakes and returns, and defining the average
disposable capital\footnote{%
defined in Gosselin and Lotz (2025b).} over the whole sector space for banks
and investors by $\left\langle \bar{K}\right\rangle \left\Vert \bar{\Psi}%
\right\Vert ^{2}$ and $\left\langle \hat{K}\right\rangle \left\Vert \hat{\Psi%
}\right\Vert ^{2}$, respectively, the disposable capital can be written, for
banks:%
\begin{equation}
\bar{K}_{X}\left\vert \bar{\Psi}\left( X\right) \right\vert ^{2}\simeq
\left\langle \bar{K}\right\rangle \left\Vert \bar{\Psi}\right\Vert
^{2}\left( \frac{\left\langle \bar{g}\right\rangle }{\bar{g}\left( X\right) }%
\bar{I}_{X/\left\langle X^{\prime }\right\rangle }\right) ^{2}  \label{DC}
\end{equation}%
for investors:%
\begin{equation}
\hat{K}_{X}\left\vert \hat{\Psi}\left( X\right) \right\vert ^{2}\simeq
\left\langle \hat{K}\right\rangle \left\Vert \hat{\Psi}\right\Vert
^{2}\left( \frac{\left\langle \hat{g}\right\rangle }{\hat{g}\left( X\right) }%
I_{X/\left\langle X^{\prime }\right\rangle }\right) ^{2}  \label{DCV}
\end{equation}%
and for firms:%
\begin{equation}
K_{X}\left\vert \Psi \left( X\right) \right\vert ^{2}\simeq \left( 1-\left(
S\left( X,X\right) \frac{\hat{K}_{X}\left\vert \hat{\Psi}\left( X\right)
\right\vert ^{2}}{\left( K_{X}\left\Vert \Psi \left( X\right) \right\Vert
^{2}\right) _{0}}+S^{B}\left( X,X\right) \frac{\bar{K}_{X}\left\vert \bar{%
\Psi}\left( X\right) \right\vert ^{2}}{\left( K_{X}\left\Vert \Psi \left(
X\right) \right\Vert ^{2}\right) _{0}}\right) \right) \left( K_{X}\left\Vert
\Psi \left( X\right) \right\Vert ^{2}\right) _{0}
\end{equation}%
where $\left( K_{X}\left\Vert \Psi \left( X\right) \right\Vert ^{2}\right)
_{0}$ represents the disposable capital of firms $X$ in the absence of
investor participation\footnote{%
See Appendix 1.}, i.e. when\textbf{\ }$S\left( X,X\right) =0$.

These disposable capital depend on the level of stakes, so that the
description of the collective states relies, in last analysis, on the
sectoral stakes.

\paragraph{Third Fiber Variable: Returns}

The firms, investors and banks returns are the third fiber variables. In a
collective state, they are functions of averages stakes and some externla
parameters, interest rates per sector and type of agents $r\left( X\right) $%
, $\hat{r}\left( X\right) $, $\bar{r}\left( X\right) $ and firms
productivity per sector $f_{1}\left( X\right) $

\subparagraph{Firms returns}

The firms' excess return, $R_{exc}\left( X\right) $\ under decreasing
returns to scale, is given by:%
\begin{equation*}
R_{exc}\left( X\right) =\frac{1}{2}f_{a}-\frac{1}{2}f_{b}\left( \frac{C_{0}+%
\bar{r}}{f_{1}\left( X\right) }\right) ^{\frac{1}{r}}
\end{equation*}
where $f_{a}$ and $f_{b}$\ have been computed in Gosselin and Lotz (2025b), $%
C_{0}$ is a fixed cost, and $\bar{r}$ average interest rate faced by firms.

\subparagraph{Investors' returns}

Investors' returns per sector depend on three coefficients. First, the
investors' estimation of risks in investors stakes, denoted $z$ and defined
by\footnote{%
The link between $z$ and $z_{0}$, the value of $\left\langle \hat{S}%
_{E}\left( X^{\prime },X\right) \right\rangle $ under the constraint $%
\left\langle f\left( X\right) \right\rangle -\left\langle \bar{r}\left(
X\right) \right\rangle =0$, is given in Appendix 5.4.1 of Gosselin and Lotz
(2025). In first approximation, we can consider $z=z_{0}$.}:%
\begin{equation*}
z=\left\langle \hat{S}_{E}\left( X^{\prime },X\right) \right\rangle
\end{equation*}%
Second, the relative strength of banks' investments in investors, denoted%
\footnote{%
To the first approximation we identify $x=\left\langle \bar{S}_{E}\left(
X^{\prime },X\right) \right\rangle =\left\langle \bar{S}_{E}\right\rangle
_{0}$ with $\left\langle \bar{S}_{E}\right\rangle _{0}$ defined in section
12.1.3.. First order corrections to this approximation are defined in
Appendix 6.5.2.} $x$:%
\begin{equation*}
x=\left\langle \bar{S}_{E}\left( X^{\prime },X\right) \right\rangle
\end{equation*}%
Third, the relative average number of banks compared to investors, denoted $%
P $, defined by:%
\begin{equation*}
P=\frac{\left\Vert \bar{\Psi}_{0}\right\Vert ^{2}}{\left\Vert \hat{\Psi}%
_{0}\right\Vert ^{2}}
\end{equation*}%
The higher the ratio $P$, the higher the importance of the banking system
and its impact on intermediation. The higher the uncertainty of banks
regarding investors, the lower are $x$ and the impact of banks on investors.
Under high uncertainty, banks reduce both participations and loans.
Conversely, the higher the coefficient $z$, the higher is the intermediation
among investors, and thereby the higher the impact of banks on investors.

Depending on the value of $P$, several solutions exist for
investors'returns. We refer to Gosselin and Lotz (2025b) for a complete
description and focus on the main case.

When the relative average number of banks is below a certain threshold:%
\begin{equation*}
P<Th
\end{equation*}%
the investors' return per sector have two solutions, as in Gosselin and Lotz
(2025). These two solutions\ for investors' excess returns depend on firms'
excess returns per sector as well as the average returns in the system. They
are given by\footnote{%
See Appendix 7 in Gosselin and Lotz (2025) for the derivation.}:%
\begin{equation}
\hat{R}_{exc}^{H}\left( X\right) =\frac{\hat{R}_{exc}^{H,0}\left( X\right) }{%
2}+\sqrt{\left( \frac{\hat{R}_{exc}^{H,0}\left( X\right) }{2}\right) ^{2}-%
\frac{1-2z}{1-z}\frac{\left\langle \bar{r}\left( X\right) \right\rangle }{%
\left\vert a\left( z,P\right) \right\vert }\hat{R}_{exc}^{L,0}\left(
X\right) }  \label{SLna}
\end{equation}%
and:%
\begin{equation}
\hat{R}_{exc}^{L}\left( X\right) =\frac{\hat{R}_{exc}^{H,0}\left( X\right) }{%
2}-\sqrt{\left( \frac{\hat{R}_{exc}^{H,0}\left( X\right) }{2}\right) ^{2}-%
\frac{1-2z}{1-z}\frac{\left\langle \bar{r}\left( X\right) \right\rangle }{%
\left\vert a\left( z,P\right) \right\vert }\hat{R}_{exc}^{L,0}\left(
X\right) }  \label{Slpa}
\end{equation}%
with:%
\begin{equation}
\hat{R}_{exc}^{H,0}\left( X\right) =\frac{\frac{1-2z}{1-z}\left\langle \bar{r%
}\left( X\right) \right\rangle +z\left( b\left( z,L\right) \left\langle \hat{%
R}_{exc}\left( X^{\prime }\right) \right\rangle +\frac{1}{2}\frac{1-2z}{%
\left( z-1\right) ^{2}}R_{exc}\left( X\right) \right) }{2\left\vert a\left(
z,P\right) \right\vert }  \label{dernier1b}
\end{equation}%
and:%
\begin{equation}
\hat{R}_{exc}^{L,0}\left( X\right) =z\left\langle \hat{R}_{exc}\left(
X^{\prime }\right) \right\rangle +\frac{\left( 1-z\right) }{2}R_{exc}\left(
X\right)  \label{dernier2b}
\end{equation}%
Here, the functions $a\left( z,P\right) $\ and $b\left( z,L\right) $\ are
defined in Gosselin and Lotz (2025b).

\subparagraph{Banks' returns}

For banks, there are two solutions that have a form similar to those of
investors:%
\begin{equation*}
\bar{R}_{exc}\left( X\right) =\frac{\bar{R}_{exc}^{H,0}\left( X\right) }{2}%
\pm \sqrt{\left( \left( \frac{\bar{R}_{exc}^{H,0}\left( X\right) }{2}\right)
^{2}-\left( \kappa +2\right) \left( \kappa +1\right) \frac{1-x}{x}\bar{R}%
_{exc}^{L,0}\left( X\right) \right) ^{2}}
\end{equation*}%
with:%
\begin{equation*}
\bar{R}_{exc}^{H,0}\left( X\right) =\left( \kappa +2\right) \left( \kappa
+1\right) \left( \frac{1-x}{2x}\left\langle \bar{r}\left( X^{\prime }\right)
\right\rangle +\frac{\left( \left\langle f\left( X\right) \right\rangle
-\left\langle \bar{r}\left( X^{\prime }\right) \right\rangle \right) }{4}%
+\left( \left\langle \hat{f}\left( X^{\prime }\right) \right\rangle
-\left\langle \bar{r}\left( X^{\prime }\right) \right\rangle \right) x\right)
\end{equation*}%
and:%
\begin{eqnarray*}
&&\bar{R}_{exc}^{L,0}\left( X\right) =\left\langle \bar{S}_{E}\left(
X^{\prime },X\right) \right\rangle _{X^{\prime }}\left\langle \overline{DF}%
\left( X^{\prime }\right) \right\rangle \left( \left\langle \bar{f}\left(
X^{\prime }\right) \right\rangle -\bar{r}\right) \\
&&+\left\langle \hat{S}_{E}^{B}\left( X^{\prime },X\right) \right\rangle
_{X^{\prime }}\left\langle \widehat{DF}\left( X^{\prime }\right)
\right\rangle \left( \left\langle \hat{f}\left( X^{\prime }\right)
\right\rangle -\bar{r}\right) +S_{E}^{B}\left( X,X\right) \left( f_{1}\left(
X\right) _{dr}-\bar{r}\right)
\end{eqnarray*}

\subsection{Defaults in Collective States}

Defaults states do not emerge, as such, in collective states. They can be
specifically accounted for, as part of the equilibrium, or they the
resolution can reveal sub-collective states presenting negative returns in
some sectors. In both cases, one must go back to the resolution and assume
ex-ante one or several defaulting firms in some sectors. The default state
can then be built recursively from this initial impact, to take into account
the fact that, once the initial default has propagated, the returns in each
sector will be deviations from what would have been the returns in a
non-default scenario.

As such, default states are modifications of non-default states. Three
relations characterize the default states\footnote{%
Details are given in Appendix 1.}.

\subsubsection{Condition for the propagation of defaults}

\paragraph{From firms to investors and banks}

For defaults to propagate from firms to investors and banks, firms' excess
returns must be below a threshold $D_{Th}$, so that:

\begin{equation*}
R\left( X\right) <D_{Th}
\end{equation*}%
This threshold is a function of the stakes of the group\footnote{%
The formula for the threshold is given in Appendix 1.}, and the default
condition can be written in terms of stakes which will be useful later on to
account for defaults in transitions.

\paragraph{From investors and banks to other investors and banks}

Given an initial default among investors and banks in some sectors, the
condition for default to propagate to other investors and banks can be
written as:%
\begin{eqnarray}
&&\mathbf{H}\left( \left\langle \left( \hat{S}_{L}\left( X^{\prime
},X\right) \right) \right\rangle _{X^{\prime }}+S_{L}\left( X,X\right)
\right) \\
&&+\mathbf{G}\left( \left\langle \bar{S}_{L}\left( X^{\prime },X\right)
\right\rangle _{X^{\prime }}+\left\langle \hat{S}_{L}^{B}\left( X^{\prime
},X\right) \right\rangle _{X^{\prime }}+S_{L}^{B}\left( X,X\right) \right)
>2\left\langle \bar{f}\right\rangle  \notag
\end{eqnarray}%
where $\mathbf{H}$ and $\mathbf{G}$ are functions of stakes, uncertainty and
capital ratios under the non-default scenario\footnote{%
Their formula are in Appendix 1.}.

\subsubsection{Losses in a state of realized default}

Ultimately, the average loss incured by the remaining banks and investors
relatively to a non-default scenario can be obtained as a shift in agents'
returns. Denoting $d\hat{f}$, $d\bar{f}$ and $df$ are the average losses
incured by the remaining investors, banks, and firms, for each defaulting
sector, and $\mu $ is the fraction of investors impacted by the default%
\footnote{%
The detailed formulas are in Appendix 1.}, we have the following expressions.

\paragraph{Losses incured by investors}

\begin{equation}
\hat{f}\left( X\right) \rightarrow \hat{f}\left( X\right) -\mu d\hat{f}
\label{Sh1}
\end{equation}

\paragraph{Losses incured by banks}

\begin{equation}
\bar{f}\left( X\right) \rightarrow \bar{f}\left( X\right) -\mu d\bar{f}
\end{equation}

\paragraph{Losses incured by firms}

\begin{equation}
f\left( X\right) \rightarrow f\left( X\right) -\mu df
\end{equation}

These three relations are structural and characterize a group in a given
state defined by its sectoral stakes. They are relevant in this work, since
these relations will impact the possibility for a state to experience a
transition towards a default state.

\section{Probabilities}

\subsection{Probabilty of Sub-Collective States}

Minimizing the action functional of the system for any group $G_{a}$ yields
several solutions. Each are local maxima of probability and as such define a
sub-collective state. However, these local maxima and the collective state
they define are not equally probable.

Any sub-collective state defined by $\left\{ \Gamma _{0},\bar{\Gamma}%
_{0},\Psi _{0},\hat{\Psi}_{0},\bar{\Psi}_{0}\right\} $ has an associated
probability given by\footnote{%
The probability of the collective state is, up to a normalization factor:%
\begin{equation*}
\exp \left( -S\left( \Gamma _{0}\right) -S\left( \bar{\Gamma}_{0}\right) -%
\bar{S}\left( \bar{\Psi}_{0}\right) -\hat{S}\left( \hat{\Psi}_{0}\right)
\right)
\end{equation*}%
Note that, in the context of Gosselin and Lotz (2025a,b), the fields $\bar{%
\Psi}_{0}$ and $\hat{\Psi}_{0}$ were determined by the stakes, so that the
probability was proportional to:%
\begin{equation*}
\exp \left( -S\left( \Gamma _{0}\right) -S\left( \bar{\Gamma}_{0}\right)
\right)
\end{equation*}%
}:%
\begin{equation}
\exp \left( -S\left( \Gamma _{0}\right) -S\left( \bar{\Gamma}_{0}\right)
-S\left( \Psi _{0}\right) -S\left( \hat{\Psi}_{0}\right) -S\left( \bar{\Psi}%
_{0}\right) \right)  \label{Prb}
\end{equation}%
where the action functionals $S\left( \Gamma _{0}\right) $,.... involved in (%
\ref{Prb}) are estimated by:%
\begin{equation}
S\left( \Gamma _{0}\right) \simeq -2\sigma _{\hat{K}}^{2}4\int \left\vert
\Gamma _{0}\left( \hat{S}_{\eta }^{\left( T\right) },X^{\prime },X\right)
\right\vert ^{2}d\left( \hat{S}_{\eta }^{\left( T\right) },X^{\prime
},X\right)  \label{SGn}
\end{equation}%
\begin{equation}
S\left( \bar{\Gamma}_{0}\right) =-3\sigma _{\hat{K}}^{2}\int_{0}^{\dag
}\left\vert \Gamma _{0}\left( \bar{S}^{\left( T\right) },X^{\prime
},X\right) \right\vert ^{2}d\left( \bar{S}^{\left( T\right) },X^{\prime
},X\right)  \label{SGt}
\end{equation}%
\begin{equation}
\bar{S}\left( \bar{\Psi}_{0}\right) \simeq \int \left( \left( \frac{\bar{g}%
^{2}\left( X\right) }{2\sigma _{\bar{K}}^{2}}+\frac{\bar{g}\left( X\right) }{%
2\bar{K}}\right) \left\vert \bar{\Psi}\left( X\right) \right\vert
^{2}\right) dX
\end{equation}%
\begin{equation}
\hat{S}\left( \hat{\Psi}_{0}\right) =\int \left( \left( \frac{\hat{g}%
^{2}\left( X\right) }{2\sigma _{\hat{K}}^{2}}+\frac{\hat{g}\left( X\right) }{%
2\hat{K}}\right) \left\vert \hat{\Psi}\left( X\right) \right\vert
^{2}\right) dX
\end{equation}%
When we will consider the dynamic deviations of the fields around their
background fields, these action functionals will be expanded\ around the
background fields.

\subsection{Probability of\ Collective States}

Any collective state refers to a decomposition of the sector space in groups 
$\left\{ G\right\} $. The background field of this collective state is, in
first approximation, the product of the background fields (\ref{Prn}) and (%
\ref{Prd}) of these groups.

Since the background fields of these group are independent, the probability
of the collective state will be the product of the probability (\ref{Prb})
of these sub-collective states: 
\begin{equation}
\prod_{G}\exp \left( -S\left( \Gamma _{0,G}\right) -S\left( \bar{\Gamma}%
_{0,G}\right) -S\left( \Psi _{0,G}\right) -S\left( \hat{\Psi}_{0,G}\right)
-S\left( \bar{\Psi}_{0,G}\right) \right)
\end{equation}%
that is:%
\begin{equation}
\exp \left( -\sum\limits_{G}\left( S\left( \Gamma _{0,G}\right) +S\left( 
\bar{\Gamma}_{0,G}\right) +S\left( \Psi _{0,G}\right) +S\left( \hat{\Psi}%
_{0,G}\right) +S\left( \bar{\Psi}_{0,G}\right) \right) \right)  \label{smp}
\end{equation}%
where the index $G$ in (\ref{smp}) runs over the groups involved in the
collective state. Formula (\ref{smp}) reveals that the action functional for
the collective state is actually the sum:%
\begin{equation*}
\sum\limits_{G}\left( S\left( \Gamma _{0,G}\right) +S\left( \bar{\Gamma}%
_{0,G}\right) +S\left( \Psi _{0,G}\right) +S\left( \hat{\Psi}_{0,G}\right)
+S\left( \bar{\Psi}_{0,G}\right) \right)
\end{equation*}

\section{Fluctuations Around Collective States}

We showed\footnote{%
Our analysis of default states is given in Gosselin and Lotz (2025a,b). This
set-up is recalled here..} that the set of defaulting agents in a group
ultimately results from a cascade of defaults among firms, investors and
banks. This suggests that shocks could trigger transitions in sub-collective
states from one phase to an other and that some underlying dynamics exist
between these phases. The possibility of transitions between these phases
should be revealed by studying the instabilities in fluctuations of phases.

To study fluctuations around static collective states we must therefore
study the phase dynamics of its sub-collective states. To do so we must
transform their static return equations into dynamic ones. By introducing a
time parameter, we can analyse the deviations of any group around its actual
static phase, and in turn, the stability of this phase. This analysis will
include the range of stability of fluctuations.

\subsection{Fluctuations in Basis Variables}

\subsubsection{Stability: First-order dynamics}

The\ stability of any collective state can be studied through a linear
dynamic system of first-order fluctuations in the basis variables, here the
stakes, around the collective state static equilibrium.

Defining $S^{T}\left( X,\theta -1\right) $ the total stake of capital
invested in firms by:%
\begin{equation}
S^{T}\left( X,\theta -1\right) =S\left( X,\theta -1\right) +S_{E}^{B}\left(
X,\theta -1\right) +S_{L}^{B}\left( X,\theta -1\right)  \label{Cps}
\end{equation}%
the linear dynamic system\footnote{%
The derivation is given in Gosselin and Lotz (2025b)} will involve the set
of fluctuations variables: 
\begin{equation*}
\left( \delta \bar{S}\left( X^{\prime },X,\theta \right) ,\delta \hat{S}%
\left( X^{\prime },X,\theta \right) ,\delta S^{T}\left( X,\theta -1\right)
\right)
\end{equation*}%
and writes\footnote{%
See Appendix 1.}:

\begin{equation*}
\mathbf{\delta S}\left( X^{\prime },X,\theta \right) =M\left( X\right) 
\mathbf{\delta S}\left( X^{\prime },X,\theta -1\right) +\int dX^{\prime
}N\left( X,X^{\prime }\right) \mathbf{\delta S}\left( X^{\prime \prime
},X,\theta -1\right)
\end{equation*}%
where:%
\begin{equation*}
\mathbf{\delta S}\left( X^{\prime },X,\theta \right) =\left( 
\begin{array}{c}
\delta \bar{S}\left( X^{\prime },X,\theta \right) \\ 
\delta \hat{S}\left( X^{\prime },X,\theta \right) \\ 
\delta S^{T}\left( X,,\theta -1\right)%
\end{array}%
\right)
\end{equation*}%
where the matrix $M\left( X\right) $ captures the dynamics for average
stakes, and the matrix $N\left( X,X^{\prime }\right) $ measures the
propagation of fluctuations within the system\footnote{%
The various matrices are defined in Appendix 4.}. This system allows the
computation of the system's eigenvalues, and the analysis of stability of
the collective states\footnote{%
The results were presented in Gosselin and Lotz (2025a,b).}.

\subsubsection{Range of stability: second-order dynamics}

The first-order dynamics determined the conditions of stability for
collective states. However, even a stable collective state can be modified
by sufficiently large fluctuations. To study the range of stability of a
collective state, i.e. the magnitude of fluctuations that do not modify the
equilibrium, we must consider a second-order dynamical system.

\paragraph{Quadratic corrections}

To study the stability range of equilibria, we complete this system by
adding second-order terms in fluctuations, so that the system is written in
matricial form as:%
\begin{eqnarray*}
\mathbf{\delta S}\left( X^{\prime },X,\theta \right) &=&M\left( X\right) 
\mathbf{\delta S}\left( X^{\prime },X,\theta -1\right) +\int dX^{\prime
}N\left( X,X^{\prime }\right) \mathbf{\delta S}\left( X^{\prime },X,\theta
-1\right) \\
&&+\left( \mathbf{\delta S}\left( X^{\prime },X,\theta -1\right) \right) ^{t}%
\left[ 
\begin{array}{c}
\bar{M} \\ 
\hat{M} \\ 
M^{T}%
\end{array}%
\right] \mathbf{\delta S}\left( X^{\prime },X,\theta -1\right)
\end{eqnarray*}%
The non-linear squared contribution corrects the linear approximations and
allows to define approximatively the convergence zone of the collective
states.

Such a system can be studied at two levels. At the average level, its
eigenvalues determine the overall stability properties of a group. At the
sectoral level, the fluctuations of several sectors in a given group can be
considered and their local instability may induce modification of the system
globally.

\paragraph{Range of stability}

The range of stability of the collective states can be evaluated from the
non-linear part of the dynamics. A system is stable when its fluctuations
are below a certain range. If we consider the average of the squared
eigenvalues of the first-order system, denoted $\left\langle \lambda
^{2}\right\rangle $, and a vector of fluctuations of stakes, written as:%
\begin{equation*}
\mathbf{\delta S}=\left( 
\begin{array}{c}
\delta \bar{S}\left( X^{\prime },X\right) \\ 
\delta \hat{S}\left( X^{\prime },X\right) \\ 
\delta S^{T}\left( X\right)%
\end{array}%
\right)
\end{equation*}%
this system will remain stable as long as the magnitudes of these
fluctuations satisfy the condition: 
\begin{equation}
\left( 1-\left\langle \lambda ^{2}\right\rangle \right) \left( \mathbf{%
\delta S}\right) ^{2}>\left( \mathbf{\delta S}^{t}\bar{M}\mathbf{\delta S}%
\right) ^{2}+\left( \mathbf{\delta S}^{t}\hat{M}\mathbf{\delta S}\right)
^{2}+\left( \mathbf{\delta S}^{t}M^{T}\mathbf{\delta S}\right) ^{2}
\label{Dmn}
\end{equation}%
The detailed formulas for this condition (\ref{Dmn}) is given in Appendix 5.

\paragraph{Particular cases}

As a first approximation, we can consider two particular cases separately.
First, the stability range for fluctuations in banks' stakes, and second,
for fluctuations in investors stakes.

\subparagraph{Banks' stakes fluctuations}

Fluctuations in banks stakes will remain stable if their magnitudes satisfy
the following condition: 
\begin{eqnarray*}
&&\left( 1-\left\langle \lambda ^{2}\right\rangle \right) \left( \left(
\delta \bar{S}\left( X^{\prime },X\right) \right) ^{2}+\left( \delta
S^{T}\left( X\right) \right) ^{2}\right) \\
&>&\left( \left\{ S_{1}^{B}\left( X,X\right) \left( \frac{\bar{h}\left(
X\right) }{2}A\left( \delta S^{T}\left( X\right) \right) ^{2}+B\left( \delta 
\bar{S}\left( X^{\prime },X\right) \right) ^{2}+\delta S^{T}\left( X\right)
C\delta \bar{S}\left( X^{\prime },X\right) \right) \right\} ^{2}\right. \\
&&+\left. \left\{ S_{1}\left( X,X\right) \left( \frac{\hat{h}\left( X\right)
D\left( \delta S^{T}\left( X\right) \right) ^{2}}{2}+L\left( \delta \bar{S}%
\left( X^{\prime },X\right) \right) ^{2}\right) \right\} ^{2}\right) \\
&&\times \left\{ \left( \varepsilon _{1}\left( X\right) \right) ^{3}\left(
h_{1}^{B}\left( X\right) \left( \left\langle \bar{h}\left( X\right)
\right\rangle +\left\langle \hat{h}_{1}^{B}\left( X\right) \right\rangle
\right) +\frac{\hat{h}\left( X\right) }{2}S\left( X\right) \right) \right\}
^{2}
\end{eqnarray*}%
where the coefficients $A$, $B$, $C$, $D$, and $L$ are defined in Appendix 4
and $\varepsilon _{i}\left( X\right) =\frac{\partial f\left( X\right) }{%
\partial S^{T}\left( X,\theta -i\right) }$ representnt les\ elasticites des
retours en fonction des stakes passees. In first approximation:%
\begin{equation}
1-\left\langle \lambda ^{2}\right\rangle \simeq 1-\left( S^{T}\left(
X\right) \frac{\partial _{f\left( X\right) }K_{X}\left\vert \Psi \left(
X\right) \right\vert ^{2}}{K_{X}\left\vert \Psi \left( X\right) \right\vert
^{2}}\varepsilon _{2}\left( X\right) \right) ^{2}  \label{Br}
\end{equation}

This condition is structural: it provides an upper bound (\ref{Br}) under
which fluctuations remain stable. This bound depends on the total level of
stakes $S^{T}\left( X\right) $ invested in $X$ firms and on the elasticities 
$\varepsilon _{i}\left( X\right) $. It decreases with $S^{T}\left( X\right) $
and $\varepsilon _{2}\left( X\right) $, the larger the shares received by
firms in the sector, and the more reactive returns are to variations in
these shares, the smaller the stability domain of fluctuations.

\subparagraph{Investors' stakes fluctuations}

Fluctuations in investors stakes will remain stable if their magnitudes
satisfy the following condition:%
\begin{eqnarray*}
&&\left( 1-\left\langle \lambda ^{2}\right\rangle \right) \frac{\left(
\delta \bar{S}\left( X^{\prime },X\right) \right) ^{2}+\left( \delta
S^{T}\left( X\right) \right) ^{2}}{\left( \varepsilon _{1}\left( X\right)
\right) ^{2}} \\
&>&\left\{ \frac{8S_{1}\left( X,X\right) \frac{1-\left( \hat{S}_{1}\left(
X\right) \right) }{1-\left( \hat{S}\left( X\right) \right) }\left( \delta 
\hat{S}\left( X\right) \right) ^{2}}{\left( \hat{f}\left( X,\theta -1\right)
\right) ^{2}}-\frac{\hat{h}\left( X^{\prime },X\right) \left( \frac{\hat{h}%
\left( X\right) }{2}S\left( X\right) -\frac{1}{f\left( X\right) +C_{0}}%
\right) \left( \varepsilon _{2}\left( X\right) \delta S\left( X\right)
\right) ^{2}}{\hat{f}\left( X,\theta -1\right) }\right\} ^{2}
\end{eqnarray*}%
These two cases show that the range of stability increases with risk
perception. It increases with $\hat{h}\left( X^{\prime },X\right) $ and
decreases with $\varepsilon _{1}\left( X\right) $, $\frac{\partial _{f\left(
X\right) }K_{X}\left\vert \Psi \left( X\right) \right\vert ^{2}}{%
K_{X}\left\vert \Psi \left( X\right) \right\vert ^{2}}$, $\frac{\partial
f\left( X\right) }{\partial S^{T}\left( X,\theta -2\right) }$, and $%
\varepsilon _{2}\left( X\right) $, so that the higher the averages and the
elasticities of both returns and capital with respect to stakes, the lower
the range of stability.

\subsection{Fluctuations in Fiber Variables}

In all of the above, we only considered the fluctuations in basis variables,
i.e. the stakes. However, fluctuations in fiber variables can also impact
the whole system. Thus, we must now include fluctuations in fiber variables,
considered as independent variables. This will expand the previous dynamical
system to a $6$ dimensional system.

\subsubsection{Stability: first-order dynamics}

As for basis variables, the stability of static solutions is studied by a
first order dynamical system, in which fluctuations in fiber variables are
accounted for. This leads to the modified first-order dynamical system:%
\begin{equation*}
X\left( \theta \right) =MX\left( \theta -1\right)
\end{equation*}%
where:

\begin{equation*}
X\left( \theta \right) =\left( \delta \bar{S}\left( X,\theta \right) ,\delta 
\hat{S}\left( X,\theta \right) ,\delta S^{T}\left( X,,\theta -1\right) ,\bar{%
\Delta},\hat{\Delta},\Delta \right) ^{t}
\end{equation*}%
with:%
\begin{equation*}
\Delta _{\bar{K}_{X}}=\frac{\delta \bar{K}_{X}\left\vert \bar{\Psi}\left(
X\right) \right\vert ^{2}}{\bar{K}_{X}\left\vert \bar{\Psi}\left( X\right)
\right\vert ^{2}}
\end{equation*}%
\begin{equation*}
\Delta _{\hat{K}_{X}}=\frac{\delta \hat{K}_{X}\left\vert \hat{\Psi}\left(
X\right) \right\vert ^{2}}{\hat{K}_{X}\left\vert \hat{\Psi}\left( X\right)
\right\vert ^{2}}
\end{equation*}%
and:%
\begin{equation*}
\Delta _{K_{X}}=\frac{\delta K_{X}\left\vert \Psi \left( X\right)
\right\vert ^{2}}{K_{X}\left\vert \Psi \left( X\right) \right\vert ^{2}}
\end{equation*}%
the relative variation of disposable capital for banks, investors and firms
respectively. The matrix $M$ is defined in Appendix 5. This dynamical system
is in general stable\footnote{%
See Appendix 5.}.

\subsubsection{Range of stability: second-order dynamics}

\paragraph{Quadratic corrections}

As for the basis states, we consider the dynamics including quadratic
corrections for the whole system of basis and fiber variables:%
\begin{equation*}
X\left( \theta \right) =MX\left( \theta -1\right) +X^{t}\left( \theta
-1\right) \mathbf{M}X\left( \theta -1\right)
\end{equation*}%
with: 
\begin{equation}
\mathbf{M=}\left( \bar{M},\hat{M},M^{T},M_{\bar{K}},M_{\hat{K}},M_{K}\right)
^{t}  \label{df}
\end{equation}%
The matrices are given in Appendix 5.

\paragraph{Range of stability}

The estimation of the range of stability is similar to that for basis
states, expanded to account for fiber variables fluctuations:%
\begin{equation}
\left( 1-\left\langle \lambda ^{2}\right\rangle \right) \left( \mathbf{%
\delta }X\right) ^{2}>\sum_{i}\left( \mathbf{\delta }X^{t}\mathbf{M}_{i}%
\mathbf{\delta }X\right) ^{2}
\end{equation}%
where the matrices $\mathbf{M}_{i}$ are the components of $\mathbf{M}$
defined in (\ref{df}).

\paragraph{Particular cases}

As a first approximation, we will consider successively fluctuations in
banks' variables only, then fluctuations in investors' variables.

\subparagraph{Fluctuations in banks' fiber variables}

Considering banks only, the stability condition involving fiber fluctuations
can be approximated by:

\begin{eqnarray*}
&&\left( 1-\left\langle \lambda ^{2}\right\rangle \right) \\
&>&\frac{A\left( \delta \bar{S}\left( X\right) \delta S^{T}\left( X\right)
\right) ^{2}+\left( B_{1}\left( \delta S^{T}\left( X\right) \right)
^{2}+B_{2}\delta S^{T}\left( X\right) \left( \bar{\Delta}-\Delta \right)
\right) ^{2}+C\left( \left( \delta S^{T}\left( X\right) \right) ^{2}\right)
^{2}}{\left( \mathbf{\delta S}\right) ^{2}+\left( \mathbf{\delta K}%
_{X}\left\vert \mathbf{\Psi }\left( X\right) \right\vert ^{2}\right) ^{2}}
\end{eqnarray*}%
where the matrices $A$, $B_{1}$, $B_{2}$, and $C$ given in Appendix 5, and:%
\begin{equation*}
\left( \mathbf{\delta K}_{X}\left\vert \mathbf{\Psi }\left( X\right)
\right\vert ^{2}\right) ^{2}=\left( \delta \bar{K}_{X}\left\vert \bar{\Psi}%
\left( X\right) \right\vert ^{2}\right) ^{2}+\left( \delta \hat{K}%
_{X}\left\vert \hat{\Psi}\left( X\right) \right\vert ^{2}\right) ^{2}+\left(
\delta K_{X}\left\vert \Psi \left( X\right) \right\vert ^{2}\right) ^{2}
\end{equation*}%
is the total squared variation of disposable capital.

\subparagraph{Fluctuations in investors' fiber variables}

When investors are considered in isolation, the domain is approximated by:%
\begin{eqnarray*}
&&\left( 1-\left\langle \lambda ^{2}\right\rangle \right) \\
&>&\frac{A\left( \left( \delta S\left( X\right) \right) ^{2}\right)
^{2}+\left( B_{1}\delta S\left( X\right) +B_{2}\left( \hat{\Delta}-\Delta
\right) \right) ^{2}}{\left( \mathbf{\delta S}\right) ^{2}+\left( \mathbf{%
\delta K}_{X}\left\vert \mathbf{\Psi }\left( X\right) \right\vert
^{2}\right) ^{2}}
\end{eqnarray*}%
where the matrices $A$, $B_{1}$, $B_{2}$ given in Appendix 5 and the squared
variation:%
\begin{eqnarray*}
&&\left( \mathbf{\delta S}\right) ^{2}+\left( \mathbf{\delta K}%
_{X}\left\vert \mathbf{\Psi }\left( X\right) \right\vert ^{2}\right) ^{2} \\
&=&\left( \left( \delta \hat{S}\left( X^{\prime },X\right) \right)
^{2}+\left( \delta S^{T}\left( X\right) \right) ^{2}\right) +\left( \delta 
\hat{K}_{X}\left\vert \hat{\Psi}\left( X\right) \right\vert ^{2}\right)
^{2}+\left( \delta K_{X}\left\vert \Psi \left( X\right) \right\vert
^{2}\right) ^{2}
\end{eqnarray*}%
is reduced to investors and firms.

\section{Description of Collective States with Interacting Groups}

Introducing a distinction between basis and fiber variables fluctuations
amongs to doubling the variables of the system. We will thus introduce a
more compact notation to describe the collective states, notation that will
also allow us for interactions between groups.

\subsection{General Description}

Each collective state is described by a collection of background fields :(%
\ref{Gmz}), (\ref{Gmb}), (\ref{sli}), (\ref{slb}), (\ref{SLf}):%
\begin{equation*}
\left\{ \left\vert \Psi \left( K,X\right) \right\vert ^{2},\left\vert \bar{%
\Psi}\left( \bar{K},X\right) \right\vert ^{2},\left\vert \hat{\Psi}\left( 
\hat{K},X\right) \right\vert ^{2},\Gamma _{0,X^{\prime },X}\left( \hat{S}%
^{\left( T\right) }\right) ,\Gamma _{0,X^{\prime },X}\left( \bar{S}^{\left(
T\right) }\right) \right\} _{\left( X^{\prime },X\right) }
\end{equation*}%
with the set of average sectoral stakes of the form (\ref{stf}): 
\begin{equation}
\left\{ \bar{S}_{\eta }\left( X^{\prime },X\right) ,S_{\eta }^{B}\left(
X,X\right) ,\hat{S}_{\eta }^{B}\left( X^{\prime },X\right) ,\hat{S}_{\eta
}\left( X^{\prime },X\right) ,S_{\eta }\left( X,X\right) \right\}
\end{equation}%
where $\eta =E,L$ describes shares and loans, respectively.

The fiber variables, here returns and disposable capital, can be expressed
as functions of these basis variables. These levels of stakes of a
collective state are not arbitrary, they are the solutions of the various
minimization equations of the system, also called \emph{saddle point stakes}%
, and will be denoted, for later purpose, with a subscript to distinguish
them from any arbitrary levels of stakes taken by the fluctuations around
the saddle point stakes. Thus, a description of a collective state is given
ultimately by a collection of basis variables:%
\begin{eqnarray*}
&&\left\{ \underline{\bar{S}_{\eta }}\left( X^{\prime },X\right) ,\underline{%
S_{\eta }^{B}}\left( X,X\right) ,\underline{\hat{S}_{\eta }}^{B}\left(
X^{\prime },X\right) ,\underline{\hat{S}_{\eta }}\left( X^{\prime },X\right)
,\underline{S_{\eta }}\left( X,X\right) \right\} \\
&\equiv &\left[ \underline{\bar{S}_{\eta }},\underline{S_{\eta }^{B}},%
\underline{\hat{S}_{\eta }}^{B},\underline{\hat{S}_{\eta }},\underline{%
S_{\eta }}\right]
\end{eqnarray*}%
When there will be no ambiguity, we will denote this collection:%
\begin{equation*}
\left[ \underline{\mathbf{S}}\right]
\end{equation*}

In the sequel, to keep track that a collective state refers to some
realization of the fields, and not only on a set of stakes we will write a
collective state as:%
\begin{equation*}
\prod_{\left( X^{\prime },X\right) }\left\vert \underline{\bar{S}_{\eta }}%
\left( X^{\prime },X\right) ,\underline{S_{\eta }^{B}}\left( X,X\right) ,%
\underline{\hat{S}_{\eta }}^{B}\left( X^{\prime },X\right) ,\underline{\hat{S%
}_{\eta }}\left( X^{\prime },X\right) ,\underline{S_{\eta }}\left(
X,X\right) \right\rangle
\end{equation*}%
This notation will be useful to include interactions between groups.

This collective state can also be written alternately:%
\begin{eqnarray*}
&&\left\vert \left( \underline{\bar{S}_{\eta }}\left( X^{\prime },X\right) ,%
\underline{S_{\eta }^{B}}\left( X,X\right) ,\underline{\hat{S}_{\eta }}%
^{B}\left( X^{\prime },X\right) ,\underline{\hat{S}_{\eta }}\left( X^{\prime
},X\right) ,\underline{S_{\eta }}\left( X,X\right) \right) _{\left(
X^{\prime },X\right) }\right\rangle \\
&\equiv &\left\vert \left[ \underline{\mathbf{S}}\right] \right\rangle
\end{eqnarray*}%
This notation will be modified when needed to adjust the specificities of
the model\footnote{%
For later purpose, when only investors and firms will be considered, we will
write a collective state as:%
\begin{equation*}
\prod_{\left( X^{\prime },X\right) }\left\vert \underline{S_{\eta }}\left(
X,X\right) ,\underline{\hat{S}_{\eta }}\left( X^{\prime },X\right)
\right\rangle
\end{equation*}%
or alternately:%
\begin{equation*}
\left\vert \left( \underline{S_{\eta }}\left( X,X\right) ,\underline{\hat{S}%
_{\eta }}\left( X^{\prime },X\right) \right) _{\left( X^{\prime },X\right)
}\right\rangle
\end{equation*}%
Such a state describing the distribution of stakes for all sectors: 
\begin{equation*}
\left\{ \underline{S_{E}}\left( \hat{X},\hat{X}\right) ,\underline{\hat{S}%
_{E}}\left( \hat{X}^{\prime },\hat{X}\right) ,\underline{S_{L}}\left( \hat{X}%
,\hat{X}\right) ,\underline{\hat{S}_{L}}\left( \hat{X}^{\prime },\hat{X}%
\right) \right\}
\end{equation*}%
}.

\subsection{Decomposition in sub-collective states}

When a collective state is decomposed of several interacting group, the
notation has to be modified to account for their interactions. We develop
the notations by first describing a single sub-collective state associated
to one group, then considering multiple independent groups, and then the
full case of multiple interacting groups.

\subsubsection{Single Group}

For a sub-collective, we use the same notation, restricted to the given
group. A sub-collective state will therefore be denoted:%
\begin{equation*}
\left\vert \left( \underline{\bar{S}_{\eta }}\left( X^{\prime },X\right) ,%
\underline{S_{\eta }^{B}}\left( X,X\right) ,\underline{\hat{S}_{\eta }}%
^{B}\left( X^{\prime },X\right) ,\underline{\hat{S}_{\eta }}\left( X^{\prime
},X\right) ,\underline{S_{\eta }}\left( X,X\right) \right) _{\left(
X^{\prime }\in G,X\in G\right) }\right\rangle
\end{equation*}%
or, more compactly:%
\begin{equation}
\left\vert \left[ \underline{\mathbf{S}},G\right] \right\rangle  \label{abc}
\end{equation}

\subsubsection{Multiple Independent Groups}

Since collective states organize into groups that are independent, with
their particular connections, we must consider states that are decomposed
into products of several states:%
\begin{eqnarray}
&&\prod\limits_{G}\left\vert \left( \underline{\bar{S}_{\eta }}\left(
X^{\prime },X\right) ,\underline{S_{\eta }^{B}}\left( X,X\right) ,\underline{%
\hat{S}_{\eta }}^{B}\left( X^{\prime },X\right) ,\underline{\hat{S}_{\eta }}%
\left( X^{\prime },X\right) ,\underline{S_{\eta }}\left( X,X\right) \right)
_{\left( X^{\prime }\in G,X\in G\right) }\right\rangle  \notag \\
&\equiv &\prod\limits_{G}\left\vert \left[ \underline{\mathbf{S}},G\right]
\right\rangle  \label{Nc}
\end{eqnarray}%
corresponding to any arbitrary decomposition in groups $G$ of sectors, to
describe the collection of product of background fields \ref{Prn}) and (\ref%
{Prd}):%
\begin{equation*}
\left\{ \prod_{G}\Gamma _{0,G}\left( \hat{S},X^{\prime },X\right) ,\prod_{G}%
\bar{\Gamma}_{0,G}\left( \bar{S},X^{\prime },X\right) ,\prod_{G}\hat{\Psi}%
_{0,G}\left( \hat{K},X\right) ,\prod_{G}\bar{\Psi}_{0,G}\left( \bar{K}%
,X\right) ,\prod_{G}\Psi _{0,G}\left( K,X\right) \right\}
\end{equation*}

\subsubsection{Multiple Interacting Groups}

Until now, our notation has only described independent groups. However, in a
collective state, groups may be interconnected To account for these
connections, we must extend our notation. To do so, we denote the collection
of bilateral stakes between groups $i$ and $j$ by $\delta S_{G_{i},G_{j}}$
and we write the states including these\ bilateral stakes as:%
\begin{equation}
\overset{S_{G_{i},G_{j}}}{\overbrace{\prod\limits_{G}\prod_{\left( \left(
X^{\prime },X\right) ,\left( X^{\prime },X\right) \right) \in G}\left\vert 
\underline{\bar{S}_{\eta }}\left( X^{\prime },X\right) ,\underline{S_{\eta
}^{B}}\left( X,X\right) ,\underline{\hat{S}_{\eta }}^{B}\left( X^{\prime
},X\right) ,\underline{\hat{S}_{\eta }}\left( X^{\prime },X\right) ,%
\underline{S_{\eta }}\left( X,X\right) \right\rangle }}  \label{STts}
\end{equation}%
or, in a more compact form:%
\begin{equation}
\overset{S_{G_{i},G_{j}}}{\overbrace{\prod\limits_{G}\left\vert \left[ 
\underline{\mathbf{S}},G\right] \right\rangle }}  \label{Stv}
\end{equation}%
When groups are organized into independent clusters of several interacting
groups, to each cluster corresponds a \emph{cluster state }of the form (\ref%
{Stv}), and the full collective state is the product of these cluster
states, written as:%
\begin{equation}
\prod \overset{S_{G_{i},G_{j}}}{\overbrace{\prod\limits_{G}\left\vert \left[ 
\underline{\mathbf{S}},G\right] \right\rangle }}  \label{STtsb}
\end{equation}%
where the first product $\prod $ runs over the collection of independent
clusters.

\part*{Towards Fully Dynamic Collective States}

In the description above, collective states were considered static to
reflect that, as aggregates, their time scale is fundamentally larger than
the microeconomic time scale. However, the emergence of defaults and the
stability of collective states are inherently dynamic processes. Thus, our
framework should be extended to account for these dynamics.

To do so, we must introduce the concepts and assumptions needed to analyse
collective states as dynamic objects, along with their expected patterns of
transition, before summarizing the field theory to be developped.

\section{What dynamic collective states should account for}

To analyze the dynamics of collective states, we must consider them as
collections of interacting sub-collective states and develop the notion of
deviation states, which will be the basic necessary dynamic objects, and
show how these deviation states can be formalized within a field-theoretic
framework, in which each static sub-collective state is associated with its
distinct field.

\subsection{Dynamic sub-collective states}

\subsubsection{Sub-collective states}

Transitions between collective states arise from fluctuations within
sub-collective states. Even loose interactions between groups propagate
these fluctuations\footnote{%
See Gosselin and Lotz (2025a,b).}: for two groups with different stability
ranges, fluctuations originating in one group - although internally stable -
may push another group beyond its own stability range, thereby potentially
initiating a sequence of transitions affecting this group and others.

Thus, the transitions between collective states arise from their
modifications and from the propagation of transitions among groups. They
therefore must be studied at the level of sub-collective states. To simplify
our analysis, we will assume that groups are interacting but loosely
connected. For each sub-collective state, interactions induce fluctuations
around their phase. Besides, exogenous modifications in basis variables may
also induce shifts in equilibrium by the propagation of these modifications
to the whole sub-collective state.

\subsubsection{Interacting groups}

Among these independent groups, interactions materialize through the stakes
taken by investors of one group in investors of other groups, what we should
call inter-group investors' stakes. This implies assuming some distributions
of inter-group stakes. To do so, we must describe collective states as a set
of sub-collective states including some additional stakes between groups.\
This describes collective states as larger structures than the collection of
sub-collective states.

\subsection{Deviation states}

\emph{Deviation states }describe the possible fluctuations around collective
states, or sub-collective states when only a subset of sectors is
considered. As with static collective states, deviation states decompose
into \emph{basis} and \emph{fiber deviation states}, which may both
experience transitions.

\subsubsection{Basis deviation states}

A collective state is a set of groups, each described by their \emph{basis
variables} - in our model investors' and banks' stakes - gathered in a \emph{%
basis collective state}.\ This basis collective state determines the types
of fiber variables of the model, which are, in our model, returns, capital,
and sectoral background fields.

To introduce dynamic deviations around such a basis collective state, we
must introduce the notion of \emph{basis deviations variables}, that
encompasses all the basis variables of the system, $\mathbf{S}$, i.e. the
stakes, whose values differ from their equilibrium values, denoted $%
\underline{\mathbf{S}}$.\ We define a \emph{basis deviation state} for a
given group $G$ a sub-collective state where the variables have been
extended to account for $\mathbf{S\ }$\ and $\underline{\mathbf{S}}$.

\subsubsection{Fiber deviation states}

In a static framework, the basis collective states determine the values of
the fiber variables of the model. However, in a dynamical context, any
deviation in basis variables from their equilibrium values generates a
multiplicity of possible values for the fiber variables. To account for this
multiplicity and fluctuations in fiber variables themselves, a dynamic
framework should treat fiber variables as autonomous.

In a dynamic context, for each set of basis deviation variables of a group $%
G $,\ the fiber variables of the model can asume multiple values, each
describing a \emph{fiber equilibrium state}.

Moreover, as for basis variables, some \emph{fiber deviation states} arise
around these fiber equilibrium states. they will be described by a set of 
\emph{fiber deviation variables}.

The dynamic description of the fiber will thus both depend on the fiber
equilibrium states and the associated fiber deviation states. The dynamics
thus acts on both the basis and the fiber states, with changes and
transitions in each type, and transitions in each type being mutually
dependent.

\subsubsection{Fibered deviation states}

Each dynamic collective state should combine a basis deviation state and one
among its possible associated fiber deviation state. We will denote this
combination a \emph{fibered \ deviation state.} There are an infinite number
of fibered deviation states.

Since these two deviation states are not independent, and influence each
other, a fibered deviation state is not merely the sum of the deviation
ststaes that compose it, but also includes the impact of their interactions.

\subsection{Deviation fields}

We have considered deviation around static collective states.\ To model
their dynamics, we have to consider these states as being themselves the
realizations of some fields.\ Since we have introduced two deviation states,
basis and fiber, we will introduce a \emph{basis deviation field }and a 
\emph{fiber deviation field}.

\subsubsection{Basis deviation field}

Static collective states were defined by realizations of the fields $\bar{%
\Gamma}_{0}$ \ and $\Gamma _{0}$. Similarly, we introduce some basis
deviation fields around $\bar{\Gamma}_{0}$ \ and $\Gamma _{0}$. These fields
are characterized by their arguments, the basis deviation variables, here
the stakes, $\left( \bar{S}_{\eta },S_{\eta }^{B},\hat{S}_{\eta }^{B},\hat{S}%
_{\eta },S_{\eta }\right) $, and their action functionals, obtained by
computing the second order expansion of the initial static action
functionals $\bar{S}\left( \bar{\Gamma}\right) $ and $S\left( \Gamma \right) 
$ around $\bar{\Gamma}_{0}$ \ and $\Gamma _{0}$\footnote{%
See Appendix 1.}.

\subsubsection{Fiber deviation field}

However, merely introducing the basis deviation fields\ is not sufficient,
since they only describe dynamics in the basis states. Actually, for each
arbitrary distribution of stakes, an infinite number of possible
distributions of capital level and return per sector\footnote{%
See Gosselin and Lotz (2024),} may exist. We must therefore introduce some
deviation fields for the fiber variables, that will account for deviation
states around the multiple distributions of the fiber variables.

Our formalism will therefore be extended along two directions. First, a
field theory describing the basis fluctuations, and second, a field theory
for fiber fluctuations, i.e. fluctuations in capital and returns. This field
theory depends explicitely on the basis fluctuations\footnote{%
These fibers fluctuations are described by the field theory of Gosselin and
Lotz (2024).}. Two fields and their action functionals will model the
fluctuations in capital and returns of investors and banks.

Together, these two field theories provide a description of the deviation
states around a given collective state $\bar{\Gamma}_{0}$ \ and $\Gamma _{0}$%
. Since it depends explicitely on the collective state, we denote this field
theory $\left( \Gamma ,\Psi \right) \left[ \bar{\Gamma}_{0},\Gamma _{0}%
\right] $.

\subsubsection{Fibered deviation field}

The fibered deviation states should be seen as realizations of a field
denoted \emph{fibered deviation field}. However, fibered deviation states
being, combinations of basis and fiber deviation states and of their
interactions, any fibered deviation field can be seen as a vector of two
components, the basis and fiber deviation fields.

Studying fiber deviation states will therefore amount to considering these
two deviation fields and their interactions and, in the following, we will
directly consider their two components, the basis et fiber deviation fields.

\subsection{Dynamic patterns in basis, fiber and fibered states}

Various possible dynamics and transitions may exist for each group and, by
extension, for all groups within a collective state. We will describe here
the types of dynamical pattern that should be expected and modeled.

\subsubsection{Dynamic patterns in basis deviation states}

Fluctuations in the basis variables reflect changes in agents' optimization
which implies only marginal modifications in the equilibrium levels of the
basis variables of the model. These fluctuations being small, they are
necessarily present in any emerging transition.

However, these basis fluctuations, although marginal, can, through the
interplay of group interactions, induce more radical fluctuations in the
collective states and the system's equilibrium levels.\ Indeed, when
amplified by group interactions, fluctuations can alter the phase of a
sub-collective state and shift its equilibrium by driving a group beyond its
stability range. Such a shift may trigger a phase transition for the group,
which can in turn affect other groups and the collective state as a whole.

\subsubsection{Dynamic patterns in fiber deviation states}

Besides, even for relatively stable basis variables, fiber variables can
experience large shifts. When levels of basis variables become arbitrary,
fiber variables can assume multiple equilibria. Transitions can occur
between these multiple equilibria and be triggered by inter-sector agents'
interactions. These possibly sudden transitions may in turn impact the basis
variables. In this two-step mechanism, fluctuations in basis variables first
induce transitions in fiber variables, then in basis variables. And
conversely, shifts in fiber variables can induce fluctuations in basis
variables, indirectly triggering further fluctuations in fiber variables.
This yields the following sequence of events: fluctuations in the fiber
variables, trigger fluctuations in the basis variables, leading to a
transition in the fiber variables, then in the basis variables.

Now that we have defined the dynamical patterns of both basis and fiber
states, we can specify how transitions may emerge within static collective
states.

\subsubsection{Dynamic patterns in fibered deviation states}

The interactions between the dynamics of the fiber deviation state and the
dynamic of the basis deviation state is caracteristic of the dynamic of a
full deviation state, that is of the fibered deviation state.\ The study of
the dynamic collective states will therefore be the study of the fibered
deviation states.

\section{Modeling dynamic Collective States}

Transitions between two collective states $\left( \bar{\Gamma}_{0},\Gamma
_{0}\right) $ and $\left( \bar{\Gamma}_{0}^{\prime },\Gamma _{0}^{\prime
}\right) $ are dynamic processes that involve both field theories $\left(
\Gamma ,\Psi \right) \left[ \bar{\Gamma}_{0},\Gamma _{0}\right] $ and $%
\left( \Gamma ,\Psi \right) \left[ \bar{\Gamma}_{0}^{\prime },\Gamma
_{0}^{\prime }\right] $. To account for these dynamic processes, we must
review, define and denote the extended collective states that can emerge
from including the deviations of the system's variables around their
equilibrium values.

\subsection{Dynamics in Basis Deviation States}

As mentionned above, deviations in the basis variables of the system may
emerge.\ Including these variations into the initial collective state yields
what we will denote \emph{basis deviation states}.

\subsubsection{Basis Deviation States}

A collective state is a set of groups, each described by their basis
variables, gathered in a basis collective state.\ In the above, it was
written:%
\begin{equation}
\left\vert \left[ \underline{\mathbf{S}},G\right] \right\rangle =\left\vert %
\left[ \underline{\bar{S}_{\eta }},\underline{S_{\eta }^{B}},\underline{\hat{%
S}_{\eta }}^{B},\underline{\hat{S}_{\eta }},\underline{S_{\eta }},G\right]
\right\rangle
\end{equation}%
This basis collective state determines the types of fiber variables of the
model, which were in the above the returns $R\left( X\right) $, $\hat{R}%
\left( X\right) $, $\bar{R}\left( X\right) $; the capital $\hat{K}%
_{X}\left\vert \hat{\Psi}\left( X\right) \right\vert ^{2}$, $\bar{K}%
_{X}\left\vert \bar{\Psi}\left( X\right) \right\vert ^{2}$; and the average
sectoral\ background fields $\left\vert \hat{\Psi}\left( X\right)
\right\vert ^{2}$, $\left\vert \bar{\Psi}\left( X\right) \right\vert ^{2}$.

To introduce dynamic deviations around such a basis collective state, we
must introduce the \emph{basis deviations variables}, $\mathbf{S}$, that
encompasses all the basis variables of the system, i.e. the stakes:%
\begin{equation*}
\mathbf{S}\equiv \left( \bar{S}_{\eta },S_{\eta }^{B},\hat{S}_{\eta }^{B},%
\hat{S}_{\eta },S_{\eta }\right)
\end{equation*}%
The values of these basis deviation variables differ from their static
equilibrium values, $\underline{\mathbf{S}}$, so that a \emph{basis
deviation state }for a given group will include these variables and will be
written:%
\begin{equation*}
\left\vert \left[ \mathbf{S,}\underline{\mathbf{S}},G\right] \right\rangle
\end{equation*}

Basis deviation states will write differently depending on the composition
of the initial collective state. Therefore, their notation will depend on
the number of groups and the nature of their interactions.

\paragraph{Single group}

The basis deviation state associated to the group $G=\left\{ \left(
X^{\prime },X\right) \right\} $, with the stakes $S_{\eta }\left( X,X\right)
,\hat{S}_{\eta }\left( X^{\prime },X\right) $ as basis variables, deviating
from their equilibrium values $\underline{S_{\eta }}\left( X,X\right) ,%
\underline{\hat{S}_{\eta }}\left( X^{\prime },X\right) $ will be denoted%
\footnote{%
In systems with two types of agents, investors and firms, we denote the
sub-collective associated to a given group $G=\left\{ \left( X^{\prime
},X\right) \right\} $ by:%
\begin{equation*}
\prod_{\left( X^{\prime },X\right) \in G}\left\vert \underline{S_{\eta }}%
\left( X,X\right) ,\underline{\hat{S}_{\eta }}\left( X^{\prime },X\right)
\right\rangle
\end{equation*}%
Similarly, we denote the deviation state associated to the group $G=\left\{
\left( X^{\prime },X\right) \right\} $, with stakes $S_{\eta }\left(
X,X\right) ,\hat{S}_{\eta }\left( X^{\prime },X\right) $ deviating from
their equilibrium values $\underline{S_{\eta }}\left( X,X\right) ,\underline{%
\hat{S}_{\eta }}\left( X^{\prime },X\right) $ as:%
\begin{equation*}
\prod_{\left( \hat{X}^{\prime },\hat{X}\right) \in G}\left\vert S_{\eta
}\left( X,X\right) ,\hat{S}_{\eta }\left( X^{\prime },X\right) ,\underline{%
S_{\eta }}\left( X,X\right) ,\underline{\hat{S}_{\eta }}\left( X^{\prime
},X\right) ,\hat{K}\left[ X\right] \right\rangle
\end{equation*}%
}:

\begin{equation*}
\left\vert \left[ \bar{S}_{\eta }\left( X^{\prime },X\right) ,S_{\eta
}^{B}\left( X,X\right) ,\hat{S}_{\eta }^{B\left( i\right) }\left( X^{\prime
},X\right) ,\hat{S}_{\eta }^{\left( i\right) }\left( X^{\prime },X\right)
,S_{\eta }^{\left( i\right) }\left( X,X\right) ,\underline{\bar{S}_{\eta }},%
\underline{S_{\eta }}^{B},\underline{\hat{S}_{\eta }}^{B},\hat{S}_{\eta },%
\underline{S_{\eta }}\right] \right\rangle
\end{equation*}%
or, using the compact notation:%
\begin{equation*}
\left\vert \left[ \mathbf{S},\underline{\mathbf{S}},G\right] \right\rangle
\end{equation*}%
where the parameters $\underline{\mathbf{S}}$ represent the possible of
phases of the group, and $\mathbf{S}$ the values taken by the stakes when
then deviate from the phase. Some of these states are unstables, and can
trigger the instability of the whole collective state.

A compact expression for the stability condition of the sub-collective state
characterized by $\underline{\mathbf{S}}$\ can be written as:%
\begin{equation*}
C\left( \underline{\mathbf{S}},\mathbf{S}\right) <0
\end{equation*}%
To account for dynamic collective states, we will include $S$\ as an
additional variable describing these states.

\paragraph{Multiple independent groups}

For collective states composed of a collection $\left\{ G\right\} $ of
several independent groups, the deviation state will be denoted:%
\begin{equation*}
\prod\limits_{G}\left\vert \left[ \mathbf{S},\underline{\mathbf{S}},G\right]
\right\rangle
\end{equation*}%
that is, the product over the groups of the deviation states described by
the set $\left[ \mathbf{S},\underline{\mathbf{S}},G\right] $.

\paragraph{Multiple interacting groups}

When a collective state is composed of a collection $\left\{ G\right\} $ of
interacting groups, the interactions between groups must be explicitly
accounted for. However, these interactions do not necessarily span the
entire set of groups: heterogeneity in the intensity and structure of
interactions may exist and should be taken into account. We can therefore
suppose that groups organize into several independent clusters, each cluster
being composed of a collection of interacting groups.

In this context, the deviation state for a cluster of independant groups
would be represented by the notation:%
\begin{equation*}
\prod\limits_{G\in Cl}\left\vert \left[ \mathbf{S},\underline{\mathbf{S}},G%
\right] \right\rangle
\end{equation*}%
However, to account for intra-cluster interactions, we use the brace:%
\begin{equation*}
\overset{\mathbf{S}_{G\in Cl,G^{\prime }\in Cl}}{\overbrace{.............}}
\end{equation*}%
where $\mathbf{S}_{G,G^{\prime }}$ represents the cross-stakes between
groups within a given cluster, so that the deviation state of this cluster
of interacting groups is written: 
\begin{equation*}
\overset{\mathbf{S}_{G\in Cl,G^{\prime }\in Cl}}{\overbrace{%
\prod\limits_{G\in Cl}\left\vert \left[ \mathbf{S},\underline{\mathbf{S}},G%
\right] \right\rangle }}
\end{equation*}%
We will call this state a \emph{cluster deviation state}.

Building on this notation, the deviation state describing several
independent clusters will be the product of their respective cluster
deviations states:%
\begin{equation}
\prod_{Cl}\left( \overset{\mathbf{S}_{G\in Cl,G^{\prime }\in Cl}}{\overbrace{%
\prod\limits_{G\in Cl}\left\vert \left[ \mathbf{S},\underline{\mathbf{S}},G%
\right] \right\rangle }}\right)  \label{Bt}
\end{equation}%
where $\prod_{Cl}$ denotes the product over independent clusters. Note that,
when all the groups of a collective state interact, the previous notation (%
\ref{Bt}) reduces to:%
\begin{equation}
\overset{\mathbf{S}_{G,G^{\prime }}}{\overbrace{\prod\limits_{G}\left\vert %
\left[ \mathbf{S},\underline{\mathbf{S}},G\right] \right\rangle }}
\end{equation}%
In the expressions above, the interactions between groups within each
cluster are considered as relatively weak, which implies that the deviation
stakes are smaller than the interactions within groups, so that:%
\begin{eqnarray*}
\mathbf{S}_{G\in Cl,G^{\prime }\in Cl} &<&<\mathbf{S}_{G} \\
\mathbf{S}_{G\in Cl,G^{\prime }\in Cl} &<&<\mathbf{S}_{G^{\prime }}
\end{eqnarray*}

\subsubsection{Basis deviation fields}

Deviations states are defined with respect to a reference distribution of
basis variables - the stakes - $\left\{ \underline{\mathbf{S}}\right\} $ for
the set of groups $\left\{ G\right\} $. Said differently, for each group $G$%
, a distribution $\mathbf{S}$ is a deviation from the values $\underline{%
\mathbf{S}}$ taken by the stakes in a given phase of the group $G$.

In the same way that this distribution $\underline{\mathbf{S}}$ corresponds
to a given background field, the distributions of these deviations $\mathbf{S%
}$ should themselves be described by some fields, along with their
associated field action functionals. Besides, these fields should depend on
the static values $\underline{\mathbf{S}}$, but also explicitely on the
groups $G$, to specify the sectors on which the fields are defined. We must
therefore define, for each phase distribution $\underline{\mathbf{S}}$ of
the group $G$, the deviation field:%
\begin{equation*}
\Lambda \left( \mathbf{S},\underline{\mathbf{S}},G\right)
\end{equation*}%
The action functional for these deviation fields will be given by second
order expansion of the action functionals $S\left( \Gamma \right) $ and $%
S\left( \bar{\Gamma}\right) $ around the background fields representing a
static collective state. Depending on the number of groups and their
interactions, these action functionals can adopt various forms.

\paragraph{Single group}

For a single group, the action functionals for the deviation fields will be
obtained by first, computing the second derivatives evaluated at their
saddle point value:%
\begin{equation*}
\frac{\delta ^{2}}{\delta \Gamma \delta \Gamma ^{\dag }}S\left( \Gamma
\right)
\end{equation*}%
and:%
\begin{equation*}
\frac{\delta ^{2}}{\delta \bar{\Gamma}\delta \bar{\Gamma}^{\dag }}S\left( 
\bar{\Gamma}\right)
\end{equation*}%
then, by considering the second-order expansion around the background fields
of $S\left( \Gamma \right) $ and $S\left( \bar{\Gamma}\right) $\footnote{%
The formula are given by (\ref{GCt}) and (\ref{GT}) in Appendix 1.}.

This defines the action functional for the deviation field as:%
\begin{eqnarray}
S\left( \Lambda \left( \mathbf{S},\underline{\mathbf{S}},G\right) \right)
&=&\sum_{G}\int \Lambda ^{\dag }\left( \mathbf{S},\underline{\mathbf{S}}%
,G\right) \left[ \frac{\delta ^{2}}{\delta \Gamma \delta \Gamma ^{\dag }}%
S\left( \Gamma \right) \right] \Lambda \left( \mathbf{S},\underline{\mathbf{S%
}},G\right) d\left( \mathbf{S},X^{\prime },X\right)  \label{Bc} \\
&&+\sum_{G}\int \Lambda ^{\dag }\left( \mathbf{S},\underline{\mathbf{S}}%
,G\right) \left[ \frac{\delta ^{2}}{\delta \bar{\Gamma}\delta \bar{\Gamma}%
^{\dag }}S\left( \bar{\Gamma}\right) \right] \Lambda \left( \mathbf{S},%
\underline{\mathbf{S}},G\right) d\left( \mathbf{S},X^{\prime },X\right) 
\notag
\end{eqnarray}%
where the integrals in the action functional are restricted to $G$ and the
second order derivatives are given in Appendix 2.

For each group $G$ in any phase $\underline{\mathbf{S}}$, the field model
that we will develop will be defined by a field $\Lambda \left( \mathbf{S},%
\underline{\mathbf{S}}\right) $ and its action functional $S\left( \Lambda
\left( \mathbf{S},\underline{\mathbf{S}}\right) \right) $.

To sum up, for any given group $G$, the complete field description of its
basis states is given by the set of all deviation fields $\Lambda \left( 
\mathbf{S},\underline{\mathbf{S}}\right) $ with their action functionals,
and the corresponding field model will be denoted:

\begin{equation}
\left[ S\left( \Lambda \left( \mathbf{S},\underline{\mathbf{S}},G\right)
,G\right) \right] _{\underline{\mathbf{S}}}  \label{bs}
\end{equation}

\paragraph{Multiple independent groups}

For a collective state defined by a collection of independent groups $%
\left\{ G\right\} $ with fields $\Lambda \left( \mathbf{S},\underline{%
\mathbf{S}},G\right) $, the full description of its deviation states is
given by the collection, for the whole set of groups, of the fields $\Lambda
\left( \mathbf{S},\underline{\mathbf{S}}\right) $ and their action
functional. This action functional is obtained by summing the groups'
independent action functionals:%
\begin{equation}
\sum_{G}S\left( \left( \Lambda \left( \mathbf{S},\underline{\mathbf{S}}%
,G\right) ,G\right) \right)  \label{ctl}
\end{equation}

\paragraph{Multiple interacting groups}

To model the interactions between groups and develop a complete formalism
for dynamic collective states\footnote{%
These interactions will be investigated in the sequel of this work, while
developing the field formalism for deviations stakes.}, we must modify (\ref%
{ctl}) to account for interactions between groups and modifications in
groups. To do so, we add to (\ref{ctl}) interaction terms of the form $%
\delta S_{G,G^{\prime }}\left( \Lambda \left( \mathbf{S},\underline{\mathbf{S%
}},G\right) ,\Lambda \left( \mathbf{S},\underline{\mathbf{S}},G\right)
\right) $, which yields:%
\begin{equation}
\sum_{G}S\left( \Lambda \left( \mathbf{S},\underline{\mathbf{S}},G\right)
\right) +\sum_{G,G^{\prime }}\delta S_{G,G^{\prime }}\left( \Lambda \left( 
\mathbf{S},\underline{\mathbf{S}},G\right) ,\Lambda \left( \mathbf{S},%
\underline{\mathbf{S}},G\right) \right)  \label{ct}
\end{equation}%
The corresponding field model can be written as:%
\begin{equation}
\prod \left[ S\left( \Lambda \left( \mathbf{S},\underline{\mathbf{S}}%
,G\right) \right) ,\delta S_{G,G^{\prime }}\left( \Lambda \left( \mathbf{S},%
\underline{\mathbf{S}},G\right) ,\Lambda \left( \mathbf{S},\underline{%
\mathbf{S}},G\right) \right) \right]  \label{FTR}
\end{equation}%
These interaction terms $\delta S_{G,G^{\prime }}\left( \Lambda \left( 
\mathbf{S},\underline{\mathbf{S}},G\right) ,\Lambda \left( \mathbf{S},%
\underline{\mathbf{S}},G\right) \right) $ will be conditioned by the
transitions between sub-collective states that appear in the context of
multiple interaction groups\footnote{%
The precise form of such terms will be developed in our the next paper of
this series.}.

\subsection{Dynamics in Fiber Deviation States}

In a deviation state, the basis variable $\mathbf{S}$ differ from those
defining the collective states $\underline{\mathbf{S}}.$ Indeed, deviations
of basis variables from their equilibrium break the set of binding
optimization relations between both basis and fiber variables - stakes,
capital and returns - so that for any arbitrary values of $\mathbf{S}$,
there are multiple equilibrium values for the fiber variables $R$ and $K$%
\footnote{%
See Gosselin and Lotz (2024).}.

\subsubsection{Fiber Deviation States}

Recall that in a dynamical context, the deviations of basis variables from
their equilibrium values generate a multiplicity of possible values for the
fiber variables, so that a dynamic framework has to consider fiber variables
as autonomous.

In our model, the fiber variables for a given level of stakes in a group $G$
are the returns:

\begin{equation*}
\underline{\mathbf{R}}\mathbf{\equiv }\left( R\left( X\right) ,\hat{R}\left(
X\right) ,\bar{R}\left( X\right) \right) _{X\in G}
\end{equation*}%
the capital:%
\begin{equation*}
\underline{\mathbf{K}}\equiv \left( K_{X},\hat{K}_{X},\bar{K}_{X}\right)
_{X\in G}
\end{equation*}%
and the sectoral background fields:%
\begin{equation*}
\underline{\mathbf{\Psi }}\equiv \left( \left\vert \Psi \left( X\right)
\right\vert ^{2}\left\vert \hat{\Psi}\left( X\right) \right\vert
^{2},\left\vert \bar{\Psi}\left( X\right) \right\vert ^{2}\right) _{X\in G}
\end{equation*}%
so that, for any group $G$\ we must consider multiple equilibrium fiber
states of the form:%
\begin{equation*}
\left\vert \underline{\mathbf{R}},\underline{\mathbf{\Psi }},\underline{%
\mathbf{K}}\right\rangle _{\left[ \mathbf{S},G\right] }
\end{equation*}%
where $\left[ \mathbf{S},G\right] $ stands for the values of the basis
variables $\mathbf{S}$ in the group $G$, and determine the equilibrium
values of the various fiber variables of the model. However, as for basis
variables, some \emph{fiber deviations variables} $\left( \mathbf{R},\mathbf{%
\Psi },\mathbf{K}\right) $ arise around these equilibrium.

In the sequel, we will use the notation: 
\begin{equation*}
\mathbf{F=}\left( \mathbf{R},\mathbf{\Psi },\mathbf{K}\right)
\end{equation*}%
and:%
\begin{equation*}
\mathbf{\underline{\mathbf{F}}=}\left( \underline{\mathbf{R}},\underline{%
\mathbf{\Psi }},\underline{\mathbf{K}}\right)
\end{equation*}%
The fiber deviation state will therefore be written:%
\begin{equation*}
\left\vert \mathbf{F},\underline{\mathbf{F}}\right\rangle _{\left[ \mathbf{S}%
,G\right] }
\end{equation*}%
The dynamics thus acts on both the basis and the fiber states, with changes
and transitions in each type, and transitions in each type being mutually
dependent. Unlike for basis deviation states, fiber deviation states are
described solely by the set of values taken by the fiber equilibrium and
deviation variables associated to a given set of basis variables of a given
given group $G$.

\subsubsection{Fiber deviation field}

The deviation states of a single group are given by the equation (\ref{Expn}%
), in which the fiber states for any given level of stakes $\mathbf{S}$:%
\begin{equation*}
\left\vert \mathbf{F,}\underline{\mathbf{F}},G\right\rangle _{\left[ \mathbf{%
S}\right] }
\end{equation*}%
decomposes into two parts, the equilibrium part of the fiber variables and
the deviation part of the fiber variables. On doit d\'{e}finir un champs
pour chacune de ces variables.

\paragraph{Fiber equilibrium field}

First, the equilibrium fiber variables $\underline{\mathbf{F}}$ arise from a
fiber field theory:%
\begin{equation}
\left[ 
\begin{array}{c}
\hat{\Psi},\bar{\Psi} \\ 
\hat{S}\left( \hat{\Psi}\right) ,\bar{S}\left( \hat{\Psi},\bar{\Psi}\right)%
\end{array}%
\right] _{\left[ \mathbf{S},G\right] }  \label{fbr}
\end{equation}%
where the realizations of the fields $\hat{\Psi},\bar{\Psi}$ and $\Psi $ are
three among the saddle point solutions of the action functional given by%
\footnote{%
The expressions of the field action functionals $\hat{S}_{\left[ \mathbf{S}%
\right] }\left( \hat{\Psi}\right) $ for the investors' field, and $\bar{S}_{%
\left[ \mathbf{S}\right] }\left( \bar{\Psi},\hat{\Psi}\right) $ for banks'
fields are given in Appendix 1.}:%
\begin{equation*}
\hat{S}_{\left[ \mathbf{S}\right] }\left( \hat{\Psi}\left( K,X\right)
\right) +\bar{S}_{\left[ \mathbf{S}\right] }\left( \hat{\Psi}\left(
K,X\right) ,\bar{\Psi}\left( K,X\right) \right)
\end{equation*}%
where the action functionals\footnote{%
The expressions of the field action functionals $\hat{S}_{\left[ \mathbf{S}%
\right] }\left( \hat{\Psi}\right) $ for the investors' field, and $\bar{S}_{%
\left[ \mathbf{S}\right] }\left( \bar{\Psi},\hat{\Psi}\right) $ for banks'
fields are given in Appendix 1.} $\hat{S}_{\left[ \mathbf{S}\right] }$ and $%
\bar{S}_{\left[ \mathbf{S}\right] }$ associated to the fiber deviation
fields $\bar{\Psi}$ and $\hat{\Psi}$ depend on the level of the basis
variables $\mathbf{S}$.

\paragraph{Fiber deviation field}

Second, the variables $\mathbf{F}$ should themselves arise from a fiber
field theory. Given an equilibrium level an equilibrium level $\underline{%
\mathbf{F}}$ of fiber variables, The deviations should be described by a
second order expansion of the action functionals $\hat{S}\left( \hat{\Psi}%
\left( \hat{K},X\right) \right) $ and $\bar{S}\left( \hat{\Psi}\left( \hat{K}%
,X\right) ,\bar{\Psi}\left( \bar{K},X\right) \right) $ for a fiber field:%
\begin{equation*}
\mathbf{\Lambda }\left( \mathbf{F,}\underline{\mathbf{F}},G\right)
\end{equation*}%
The action for this field is thus given by:%
\begin{eqnarray}
&&S_{F}\left( \mathbf{\Lambda }\left( \mathbf{F,}\underline{\mathbf{F}}%
,G\right) \right)  \label{smf} \\
&=&\mathbf{\Lambda }^{\dag }\left( \mathbf{F,}\underline{\mathbf{F}}%
,G\right) \delta ^{2}S\left( \bar{\Psi},\hat{\Psi}\right) \mathbf{\Lambda }%
\left( \mathbf{F,}\underline{\mathbf{F}},G\right)  \notag
\end{eqnarray}%
where $\delta ^{2}S\left( \bar{\Psi},\hat{\Psi}\right) $ is the second order
derivative of the action functional for banks and investors fields:%
\begin{equation*}
\delta ^{2}S\left( \bar{\Psi},\hat{\Psi}\right) =\frac{\delta ^{2}\left( 
\hat{S}\left( \hat{\Psi}\right) +\bar{S}\left( \bar{\Psi}\right) \right) }{%
\delta \hat{\Psi}\left( \hat{K},X\right) \delta \hat{\Psi}^{\dag }\left( 
\hat{K},X\right) }+\frac{\delta ^{2}\bar{S}\left( \bar{\Psi}\right) \left( 
\hat{S}\left( \hat{\Psi}\right) +\bar{S}\left( \bar{\Psi}\right) \right) }{%
\delta \bar{\Psi}\left( \bar{K},X\right) \delta \bar{\Psi}^{\dag }\left( 
\bar{K},X\right) }
\end{equation*}%
This action functional corresponds to a fiber deviation field theory:%
\begin{equation}
\left[ 
\begin{array}{c}
\mathbf{\Lambda }\left( \mathbf{F,}\underline{\mathbf{F}},G\right) \\ 
\delta ^{2}S\left( \bar{\Psi},\hat{\Psi}\right)%
\end{array}%
\right] _{\left[ \mathbf{S},G\right] }  \label{fbs}
\end{equation}

\paragraph{Gathering fiber equilibrium and deviation fields}

The field-theoretical framework that jointly considers the equilibrium fiber
field theory (\ref{fbr}) and the fiber deviation field $\mathbf{\Lambda }%
\left( \mathbf{F,}\underline{\mathbf{F}},G\right) $\textbf{\ }is given by
the sum of the action functionals:%
\begin{equation}
S^{F}=\hat{S}_{\left[ \mathbf{S}\right] }\left( \hat{\Psi}\left( K,X\right)
\right) +\bar{S}_{\left[ \mathbf{S}\right] }\left( \hat{\Psi}\left(
K,X\right) ,\bar{\Psi}\left( K,X\right) \right) +\mathbf{\Lambda }^{\dag
}\left( \mathbf{F,}\underline{\mathbf{F}},G\right) \delta ^{2}S\left( \bar{%
\Psi},\hat{\Psi}\right) \mathbf{\Lambda }\left( \mathbf{F,}\underline{%
\mathbf{F}},G\right)  \label{Fct}
\end{equation}%
corresponding to the field theory gathering (\ref{fbr}) and (\ref{fbs}):%
\begin{equation}
\left[ 
\begin{array}{c}
\hat{\Psi},\bar{\Psi},\mathbf{\Lambda }\left( \mathbf{F,}\underline{\mathbf{F%
}},G\right) \\ 
\hat{S}\left( \hat{\Psi}\right) ,\bar{S}\left( \hat{\Psi},\bar{\Psi}\right)
,S_{F}\left( \mathbf{\Lambda }\right)%
\end{array}%
\right] _{\left[ \mathbf{S},G\right] }  \label{fbt}
\end{equation}

\subsection{Dynamics in Fibered Deviation States}

The full dynamics of the system should incorporate both basis and fiber
deviation states and the combination of these states leads to the notion of
fibered deviation states.

\subsubsection{Fibered Deviation States}

By combining basis and fiber states, a complete description of fibered
deviation states must therefore, as in the case of basis deviation fields,
distinguish between the number of groups and the nature of their
interactions.

\paragraph{Single group}

The deviation states for a single group should account for the level of
disposable capital, sectoral background fields and return, and be written as:%
\begin{equation}
\left[ 
\begin{array}{c}
\underset{\updownarrow }{\left\vert \mathbf{F,}\underline{\mathbf{F}}%
,G\right\rangle _{\left[ \mathbf{S}\right] }} \\ 
\left\vert \left[ \mathbf{S},\underline{\mathbf{S}},G\right] \right\rangle%
\end{array}%
\right]  \label{Expn}
\end{equation}%
where the notation $\left[ 
\begin{array}{c}
\updownarrow \\ 
\end{array}%
\right] $ indicates that any state of the system is composed of two
interacting deviation states, basis and fiber.

This notation can be read at two levels.

First, the level of the basis state, which describes a basis deviation state
for the stakes of group $G$:%
\begin{equation*}
\left\vert \left[ \mathbf{S},\underline{\mathbf{S}},G\right] \right\rangle
\end{equation*}%
Second, the level of the fiber deviation state, which describes one possible
fiber deviation state, here capital, sectoral background fields and returns%
\footnote{%
They are described in Gosselin and Lotz (2024).} for investors, banks and
firms, for the given stakes:%
\begin{equation*}
\left\vert \mathbf{F,}\underline{\mathbf{F}},G\right\rangle _{\left[ \mathbf{%
S}\right] }
\end{equation*}%
These possible fiber deviation states are described by the fields $\Psi
\left( K,X\right) ,\hat{\Psi}\left( \hat{K},X\right) $ and $\bar{\Psi}\left( 
\bar{K},X\right) $ minimizing the action functionals $\hat{S}\left( \hat{\Psi%
}\left( \hat{K},X\right) \right) $ and $\bar{S}\left( \hat{\Psi}\left( \hat{K%
},X\right) ,\bar{\Psi}\left( \bar{K},X\right) \right) $ with average fiber
variables $\underline{\mathbf{F}}$ and by deviations $\mathbf{F}$ from these
averages.

\paragraph{Multiple independent groups}

The description of the full dynamic collective state of multiple independent
groups is given by the product of independent states: 
\begin{equation}
\prod\limits_{G}\left[ 
\begin{array}{c}
\underset{\updownarrow }{\left\vert \mathbf{F,}\underline{\mathbf{F}}%
,G\right\rangle _{\left[ \mathbf{S}\right] }} \\ 
\left\vert \left[ \mathbf{S},\underline{\mathbf{S}},G\right] \right\rangle%
\end{array}%
\right]  \label{Expnp}
\end{equation}%
Here again, this notation can be read at two levels.

First, at the level of the basis deviation state.\ It describes the product
of all the basis deviation states:%
\begin{equation*}
\prod\limits_{G}\left\vert \left[ \mathbf{S},\underline{\mathbf{S}},G\right]
\right\rangle
\end{equation*}%
Second, at the level of the fiber deviation state.\ It describes the product
of all the fiber deviation states:%
\begin{equation*}
\prod\limits_{G}\left\vert \mathbf{F,}\underline{\mathbf{F}},G\right\rangle
_{\left[ \mathbf{S}\right] }
\end{equation*}%
for a given set of values of the basis variables: 
\begin{equation*}
\left( \mathbf{S},\underline{\mathbf{S}}\right)
\end{equation*}

\paragraph{Multiple interacting groups}

For collective states composed of multiple interacting groups, we complete
the above notation by accounting for the connections across a collection of
groups:%
\begin{equation*}
\left[ \prod_{G}\left[ \mathbf{S},G\right] ,\prod_{G,G^{\prime }}\mathbf{S}%
_{G,G^{\prime }}\right]
\end{equation*}%
where $\left[ \mathbf{S},G\right] $ denotes the stakes within groups, and $%
\mathbf{S}_{G,G^{\prime }}$ denotes the stakes between groups $G$ and $%
G^{\prime }$, or, in a more compact form:%
\begin{equation*}
\left[ \mathbf{S,}\left\{ G,G^{\prime }\right\} \right]
\end{equation*}%
where the collection $\left\{ G,G^{\prime }\right\} $ refers to the group
interactions.

\subparagraph{\textbf{Without clusters}}

When all groups interact with each others, a deviation state includes these
interactions and involves all interacting basis and fiber deviation states
as a composite deviation state. It will thus be written:%
\begin{equation}
\left[ 
\begin{array}{c}
\underset{\updownarrow }{\prod_{G}\left\vert \mathbf{F,}\underline{\mathbf{F}%
},G\right\rangle _{\left[ \mathbf{S,}\left\{ G,G^{\prime }\right\} \right] }}
\\ 
\overset{\mathbf{S}_{G,G^{\prime }}}{\overbrace{\prod_{G}\left\vert \left[ 
\mathbf{S},\underline{\mathbf{S}},G\right] \right\rangle }}%
\end{array}%
\right]  \label{Exp}
\end{equation}%
where the expression: 
\begin{equation}
\prod_{G}\left\vert \mathbf{F,}\underline{\mathbf{F}},G\right\rangle _{\left[
\mathbf{S,}\left\{ G,G^{\prime }\right\} \right] }  \label{Expnd}
\end{equation}%
is the fiber deviation state above the cluster of basis deviation states:%
\begin{equation*}
\overset{\mathbf{S}_{G,G^{\prime }}}{\overbrace{\prod_{G}\left\vert \left[ 
\mathbf{S},\underline{\mathbf{S}},G\right] \right\rangle }}
\end{equation*}%
that were described in (\ref{Bt}).

\subparagraph{Including clusters}

However, when groups are organized in independent clusters, the deviation
states are products of independent clusters of interacting deviation states,
and the notation becomes:%
\begin{equation}
\prod\limits_{Cl}\left[ 
\begin{array}{c}
\underset{\updownarrow }{\prod_{G\in Cl}\left\vert \mathbf{F,}\underline{%
\mathbf{F}},G\right\rangle _{\left[ \mathbf{S,}\left\{ G\in Cl,G^{\prime
}\in Cl\right\} \right] }} \\ 
\overset{\mathbf{S}_{G,G^{\prime }}}{\overbrace{\prod_{G\in Cl}\left\vert %
\left[ \mathbf{S},\underline{\mathbf{S}},G\right] \right\rangle }}%
\end{array}%
\right]  \label{fbc}
\end{equation}%
where the product $\prod\limits_{Cl}$\ is over the clusters composing the
collective state.

\subsubsection{Fibered deviation fields}

The fibered states (\ref{Exp}) and (\ref{fbc}) should be described by a
field theory gathering both the basis deviation fields and the fiber
equilibrium and deviation fields. We will start by describing the fibered
deviation fields for one group, then for multiple independent groups, and
multiple interacting groups.

\paragraph{Single group}

The complete description encompassing both basis and fiber deviation states
is obtained by accounting for all possible levels of the basis variable and
then combining the fiber and basis deviation fields.

The fibered deviation states for one group $G$ defined in (\ref{Expn})
should therefore emerge from a field-theoretical framework that jointly
considers the fiber field theory (\ref{fbt})\textbf{\ }and the basis field
theory (\ref{bs}) to which it is attached. We refer to this unified
framework as the \emph{fibered field theory}, which we denote as: 
\begin{equation}
\left[ 
\begin{array}{c}
\left[ 
\begin{array}{c}
\hat{\Psi},\bar{\Psi},\mathbf{\Lambda }\left( \mathbf{F,}\underline{\mathbf{F%
}},G\right) \\ 
\hat{S}\left( \hat{\Psi}\right) ,\bar{S}\left( \hat{\Psi},\bar{\Psi}\right)
,S_{F}\left( \mathbf{\Lambda }\right)%
\end{array}%
\right] _{\left[ \mathbf{S},G\right] } \\ 
\updownarrow \\ 
\left[ S\left( \Lambda \left( \mathbf{S},\underline{\mathbf{S}}\right)
,G\right) \right]%
\end{array}%
\right]  \label{Bpf}
\end{equation}%
This expression describes a field theory involving both basis and fiber
fields. In this expression, the symbol $\updownarrow $ indicates that the
relationship between the two theories is not unidirectional, and that the
fields $\hat{\Psi}\left( \hat{K},X\right) $, $\bar{\Psi}\left( \bar{K}%
,X\right) $ and $\mathbf{\Lambda }\left( \mathbf{F,}\underline{\mathbf{F}}%
,G\right) $\ are conditioned by the basis states and the values of sectoral
stakes.

Conversely, the fiber states also influence the basis states. Indeed, some
fiber transitions can trigger basis transitions. In the present model, the
action functional of this fibered field theory should therefore be the sum
of the action functional for basis deviation fields (\ref{Bc}) and the
action functional for fiber equilibrium and deviation fields (\ref{Fct}):%
\begin{equation}
S\left( \Lambda \left( \mathbf{S},\underline{\mathbf{S}},G\right) \right) +%
\hat{S}_{\left[ \mathbf{S}\right] }\left( \hat{\Psi}\left( K,X\right)
\right) +\bar{S}_{\left[ \mathbf{S}\right] }\left( \hat{\Psi}\left(
K,X\right) ,\bar{\Psi}\left( K,X\right) \right) +\mathbf{\Lambda }^{\dag
}\left( \mathbf{F,}\underline{\mathbf{F}},G\right) \delta ^{2}S\left( \bar{%
\Psi},\hat{\Psi}\right) \mathbf{\Lambda }\left( \mathbf{F,}\underline{%
\mathbf{F}},G\right)  \label{ql}
\end{equation}%
In expression (\ref{Bpf}), the fiber fields form a superstructure determined
by the landscape of connections, i.e. the stakes. Consequently, the full
field theory for collective states should account both for modifications of
the basis---as outlined in the previous section describing transitions
between groups---and for changes in the superstructure, which may switch
from one state to another and thereby modify the underlying basis.

Thus, collective states are, in fact, states of stakes with an associated
fiber, each fiber being itself a share-dependent field theory. This fibered
field theory, in turn, describes the possible states of an economy, with
agents interacting according to their dynamics, potential defaults, and
transitions arising from the diffusion of returns and disposable capital.

\paragraph{\textbf{M}ultiple independent groups}

For multiple independent groups, the previous argument generalizes directly,
and the states (\ref{Expnp}) are described an action action functional which
is the sum of the action functionals of each group: 
\begin{eqnarray}
&&\sum_{G}\left( S\left( \Lambda \left( \mathbf{S},\underline{\mathbf{S}}%
,G\right) \right) +\hat{S}_{\left[ \mathbf{S},G\right] }\left( \hat{\Psi}%
\left( K,X\right) \right) +\bar{S}_{\left[ \mathbf{S},G\right] }\left( \hat{%
\Psi}\left( K,X\right) ,\bar{\Psi}\left( K,X\right) \right) \right)
\label{Glb} \\
&&+\sum_{G}S_{F}\left( \left\{ \mathbf{\Lambda }\left( \mathbf{F,}\underline{%
\mathbf{F}},G\right) \right\} _{G}\right)  \notag
\end{eqnarray}%
This first part of the action functional corresponds to the sum of
individual groups actions (\ref{ql}), while the second part is obtained a a
sum of terms (\ref{smf}):%
\begin{eqnarray}
&&S_{F}\left( \left\{ \mathbf{\Lambda }\left( \mathbf{F,}\underline{\mathbf{F%
}},G\right) \right\} _{G}\right) \\
&=&\sum_{G}\mathbf{\Lambda }^{\dag }\left( \mathbf{F,}\underline{\mathbf{F}}%
,G\right) \delta ^{2}S\left( \bar{\Psi},\hat{\Psi}\right) \mathbf{\Lambda }%
\left( \mathbf{F,}\underline{\mathbf{F}},G\right)  \notag
\end{eqnarray}%
where $\left\{ \mathbf{\Lambda }\left( \mathbf{F,}\underline{\mathbf{F}}%
,G\right) \right\} _{G}$ is the collection of deviation fiber fields for the
whole set of groups.

The action functional (\ref{Glb}) corresponds to a fibered field theory
which is the product of several independent fibered field theories:%
\begin{equation}
\prod_{G}\left[ 
\begin{array}{c}
\left[ 
\begin{array}{c}
\hat{\Psi},\bar{\Psi},\mathbf{\Lambda }\left( \mathbf{F,}\underline{\mathbf{F%
}},G\right) \\ 
\hat{S}\left( \hat{\Psi}\right) ,\bar{S}\left( \hat{\Psi},\bar{\Psi}\right)
,S_{F}\left( \mathbf{\Lambda }\right)%
\end{array}%
\right] _{\left[ \mathbf{S},G\right] } \\ 
\updownarrow \\ 
\left[ S\left( \Lambda \left( \mathbf{S},\underline{\mathbf{S}}\right)
,G\right) \right]%
\end{array}%
\right]  \label{Ftr}
\end{equation}

\paragraph{\textbf{M}ultiple interacting groups}

Recall that the full deviation states for multiple interacting groups are
given by a combination of the form (\ref{Exp}). As before, the fiber states
in (\ref{Exp}) for investors and banks are defined over the distribution $%
\left[ \mathbf{S,}\left\{ G,G^{\prime }\right\} \right] $:%
\begin{equation*}
\left\vert \mathbf{F,}\underline{\mathbf{F}}\right\rangle _{\left[ \mathbf{S,%
}\left\{ G,G^{\prime }\right\} \right] }
\end{equation*}%
and should arise from a field theory similar to (\ref{Ftr}). However, as
with the states, since the stakes $\left[ \mathbf{S,}\left\{ G,G^{\prime
}\right\} \right] $ connect the various groups, we cannot consider the full
theory describing interacting groups as a product of independent field
theories. For each basis state:%
\begin{equation*}
\prod\limits_{Cl}\overset{\mathbf{S}_{G,G^{\prime }}}{\overbrace{\prod_{G\in
Cl}\left\vert \left[ \mathbf{S},\underline{\mathbf{S}},G\right]
\right\rangle }}
\end{equation*}%
we should consider the associated field theory as being described by a fully
interacting set of fiber fields:%
\begin{equation}
\prod\limits_{Cl}\left[ 
\begin{array}{c}
\prod_{G\in Cl}\left[ 
\begin{array}{c}
\hat{\Psi},\bar{\Psi},\mathbf{\Lambda }\left( \mathbf{F,}\underline{\mathbf{F%
}},G\right) \\ 
\hat{S}\left( \hat{\Psi}\right) ,\bar{S}\left( \hat{\Psi},\bar{\Psi}\right)
,S_{F}\left( \mathbf{\Lambda }\right)%
\end{array}%
\right] _{\left[ \mathbf{S,}\left\{ G,G^{\prime }\right\} \right] } \\ 
\updownarrow \\ 
\overset{\mathbf{S}_{G,G^{\prime }}}{\overbrace{\prod_{G\in Cl}\left\vert %
\left[ \mathbf{S},\underline{\mathbf{S}},G\right] \right\rangle }}%
\end{array}%
\right]  \label{Ps}
\end{equation}%
Such states should arise from a fibered field theory encompassing both basis
and fiber fields:%
\begin{equation}
\prod\limits_{Cl}\left[ 
\begin{array}{c}
\left[ \prod_{G\in Cl}\left[ 
\begin{array}{c}
\hat{\Psi},\bar{\Psi},\mathbf{\Lambda }\left( \mathbf{F,}\underline{\mathbf{F%
}},G\right) \\ 
S\left( \left\{ \hat{\Psi},\bar{\Psi},\mathbf{\Lambda }\right\} \right)%
\end{array}%
\right] _{\left[ \mathbf{S,}\left\{ G,G^{\prime }\right\} \right] }\right]
\\ 
\updownarrow \\ 
\left[ \prod_{G\in Cl}\left[ S\left( \left\{ \Lambda \left( \mathbf{S},%
\underline{\mathbf{S}},G\right) \right\} \right) \right] \right]%
\end{array}%
\right]  \label{Ts}
\end{equation}%
where both the basis theory:%
\begin{equation*}
\left[ \prod_{G\in Cl}\left[ S\left( \left\{ \Lambda \left( \mathbf{S},%
\underline{\mathbf{S}},G\right) \right\} \right) \right] \right]
\end{equation*}%
and fiber theory:%
\begin{equation*}
\left[ \prod_{G\in Cl}\left[ 
\begin{array}{c}
\hat{\Psi},\bar{\Psi},\mathbf{\Lambda }\left( \mathbf{F,}\underline{\mathbf{F%
}},G\right) \\ 
S\left( \left\{ \hat{\Psi},\bar{\Psi},\mathbf{\Lambda }\right\} \right)%
\end{array}%
\right] _{\left[ \mathbf{S,}\left\{ G,G^{\prime }\right\} \right] }\right]
\end{equation*}%
Condition each other, which justifies the symbol $\updownarrow $ in (\ref{Ts}%
)\ to indicate that the relationship between the two theories is not
unidirectional and that the fiber also influences the basis states. The use
of curly brackets\textbf{\ }$\left\{ {}\right\} $\ in the notations:%
\begin{equation*}
S\left( \left\{ \Lambda \left( \mathbf{S},\underline{\mathbf{S}},G\right)
\right\} \right)
\end{equation*}%
and:%
\begin{equation*}
S\left( \left\{ \hat{\Psi},\bar{\Psi},\mathbf{\Lambda }\right\} \right)
\end{equation*}

emphasizes the presence of intra-cluster interactions: the basis and fiber
action functionals must incorporate not only the action functionals $S\left(
\Lambda \left( \mathbf{S},\underline{\mathbf{S}},G\right) \right) $\ and $%
S\left( \hat{\Psi},\bar{\Psi},\mathbf{\Lambda }\right) $ of the individual
groups, but also the interactions between groups, which involve the entire
set of basis deviation and fiber fields.

Such a field theory should be represented by an action functional of the
form:%
\begin{eqnarray}
&&S^{B,F}=\sum_{G}S\left( \Lambda \left( \mathbf{S},\underline{\mathbf{S}}%
\right) ,G\right) +\sum_{G,G^{\prime }}S_{G,G^{\prime }}\left( \left(
\Lambda \left( \mathbf{S}^{\left( G/G^{\prime }\right) },\underline{\mathbf{S%
}}^{\left( G/G^{\prime }\right) }\right) \right) \right)  \label{Ftv} \\
&&+\sum_{G}\hat{S}_{\left[ \mathbf{S,}\left\{ G,G^{\prime }\right\} \right]
}\left( \hat{\Psi}\right) +\sum_{G}\bar{S}_{\left[ \mathbf{S,}\left\{
G,G^{\prime }\right\} \right] }\left( \hat{\Psi},\bar{\Psi}\right)  \notag \\
&&+\sum_{G,G^{\prime }}\delta S_{G,G^{\prime }}\left( \left( \hat{\Psi}%
^{\left( G\right) },\bar{\Psi}^{\left( G\right) }\right) ,\left( \hat{\Psi}%
^{\left( G^{\prime }\right) },\bar{\Psi}^{\left( G^{\prime }\right) }\right)
\right)  \notag \\
&&+\sum_{G}\mathbf{\Lambda }^{\dag }\left( \mathbf{F,}\underline{\mathbf{F}}%
,G\right) \delta ^{2}S\left( \bar{\Psi},\hat{\Psi}\right) \mathbf{\Lambda }%
\left( \mathbf{F,}\underline{\mathbf{F}},G\right)  \notag \\
&&+\sum_{G,G^{\prime }}\delta ^{2}S_{G,G^{\prime }}\left( \bar{\Psi},\hat{%
\Psi},\mathbf{\Lambda }\left( G\right) ,\mathbf{\Lambda }^{\prime }\left(
G\right) \right)  \notag
\end{eqnarray}%
where the fourth term:%
\begin{equation*}
\sum_{G,G^{\prime }}\delta S_{G,G^{\prime }}\left( \left( \hat{\Psi}^{\left(
G\right) },\bar{\Psi}^{\left( G\right) }\right) ,\left( \hat{\Psi}^{\left(
G^{\prime }\right) },\bar{\Psi}^{\left( G^{\prime }\right) }\right) \right)
\end{equation*}%
describes the interactions between the fibers of the various groups and the
last terms:%
\begin{equation*}
\sum_{G,G^{\prime }}\delta ^{2}S_{G,G^{\prime }}\left( \bar{\Psi},\hat{\Psi},%
\mathbf{\Lambda }\left( G\right) ,\mathbf{\Lambda }^{\prime }\left( G\right)
\right)
\end{equation*}%
the interactions between deviations around the equilibrium values of fiber
variables. The possible forms of these interaction terms will be considered
in subsequent work on the subject.

In formula (\ref{Ftv}), the expression:%
\begin{equation*}
\hat{S}_{\left[ \mathbf{S,}\left\{ G,G^{\prime }\right\} \right] }\left( 
\hat{\Psi}\left( \hat{K},X\right) \right)
\end{equation*}%
describes the action functional for the investors and:%
\begin{equation*}
\bar{S}_{\left[ \mathbf{S,}\left\{ G,G^{\prime }\right\} \right] }\left( 
\hat{\Psi}\left( \hat{K},X\right) ,\bar{\Psi}\left( \bar{K},X\right) \right)
\end{equation*}%
is the action functional for banks fields. They are given by formula (\ref%
{BFC}) and (\ref{BFD}) in Appendix 1. The stakes involved in these formula
are set to the values $\left[ \mathbf{S,}\left\{ G,G^{\prime }\right\} %
\right] $ and the second term in (\ref{Ftv}) describes the interactions
between basis states.

To conclude, note that:

1) The hypothetical field theory operates on a much larger space than the
various fields:%
\begin{equation*}
\prod_{G}\left[ S\left( \Lambda \left( \mathbf{S},\underline{\mathbf{S}}%
\right) ,G\right) \right]
\end{equation*}%
since such field theories are themselves variables and should be considered
as the arguments of a field of fields.

2) The total probability for the collective state (\ref{Exp}) should be:%
\begin{equation}
\exp \left( -S_{full}\left( \Lambda \left( \mathbf{S},\underline{\mathbf{S}}%
\right) ,\hat{\Psi},\bar{\Psi},\mathbf{\Lambda }\left( \mathbf{F,}\underline{%
\mathbf{F}},G\right) ,G\right) \right)
\end{equation}%
where $S_{full}$ is defined by (\ref{Ftv}) and includes the fiber variables.

\section{Transitions Between Collective States}

Since collective states are composed of interacting groups of agents in
specific phases, they can be considered dynamic in the long run. Even in the
short run, they may experience fluctuations that do not alter the intrinsic
characteristics of the groups but may change their phases at any time.

In this context, instability may arise from fluctuations or exogenous
modifications in the system. It may also occur when the strength of agents'
interactions within a group varies-for instance, when some agents strengthen
their connections at the expense of others.

Moreover, the number of groups within the system may itself vary. Any
initial structure of the system may either expand or contract. For instance,
emerging groups may merge with existing ones, thereby modifying the
stability of the system. These various patterns were not captured in our
initial description of fluctuations around static collective states and the
following paragraphs will fill this gap by exploring the various types of
transitions for basis and fibers that could be expected. This typology will
provide a framework to develop later a field theory for sub-collective
states, that is, for groups and the phases they occupy accounting for the
possible pattrns f transitions.

\subsection{Transitions for Basis States}

In collective states, transitions stem from interactions between agents-here
the stakes\footnote{%
Technically, instability and transitions arise from modification of
eigenvalues due to changes in connections.}. Within a given group, circular
over-investments across several sectors may induce instability and trigger
the entire group toward another, possibly default, state. This occurs when
investment returns in one sector significantly exceed the firms' returns,
reflecting the fact that investors favour cross-sector investment returns at
the expense of their own sector's capital expenditures. In this case, any
modification in cross- sector investment returns may lead to a sequence of
defaults across the entire group\footnote{%
See Gosselin and Lotz (2025a,b).}.

More generally, interactions betweeen groups may trigger transition
mechanisms involving several groups. These transitions may induce some
default states, but also the modifications in the composition of the groups,
including merging or disscociatin of groups.

We detail different types of transitions in the case of a single group,
multiple independent or interacting grps..

\subsubsection{Transitions for a single group}

Due to internal dynamics, or external perturbations, a sub-collective state
may experience a transition.\ In general, this transition does not affect
the group, since in this case it is unique, but modifies the phase it is in.
Thus, the transition only affects the phase of the sub-collective state. A
transition will only modify the associated group in the default mechanism,
in which the number of sectors diminishes.

\paragraph{Without modification of the group}

In general, for a given level of uncertainty and a given stability range, a
deviation state undergoes a transition of its phase $\underline{\mathbf{S}}$%
, without any change in the structure of the group $G$. Considering a
deviation state:%
\begin{equation*}
\left\vert \left[ \mathbf{S},\underline{\mathbf{S}},G\right] \right\rangle
\end{equation*}%
this deviation state becomes unstable when fluctuations $\mathbf{S}$ move
outside the stability range around $\underline{\mathbf{S}}$. Fluctuations
may be initiated by some exogenous modifications and induce a transition
between deviation states, so that:%
\begin{equation*}
\left\vert \left[ \mathbf{S},\underline{\mathbf{S}},G\right] \right\rangle
\rightarrow \left\vert \left[ \mathbf{S}^{\prime },\underline{\mathbf{S}}%
^{\prime },G\right] \right\rangle
\end{equation*}%
where the phase $\underline{\mathbf{S}}$ of the group $G$ shifts to a new
phase $\underline{\mathbf{S}}^{\prime }$.

\paragraph{With modification of the group}

For a single independent group, the only modification of the group arising
from internal fluctuations is one among several possible default states.
Such transition toward defaults can be writen as:%
\begin{equation*}
\left\vert \left[ \mathbf{S},\underline{\mathbf{S}},G\right] \right\rangle
\rightarrow \left\vert \left[ \mathbf{S}^{\prime },\underline{\mathbf{S}}%
^{DF},G/DF\right] \right\rangle
\end{equation*}%
where $G/DF$ is the set of remaining sectors after default.

\subsubsection{Transitions of multiple independent groups}

\paragraph{Without modification of groups}

When several independent groups compose a collective state, internal
transitions occuring in multiple groups may be considered, and the
transition can be written as:%
\begin{equation*}
\prod\limits_{G}\left\vert \left[ \mathbf{S},\underline{\mathbf{S}},G\right]
\right\rangle \rightarrow \prod\limits_{G}\left\vert \left[ \mathbf{S}%
^{\prime },\underline{\mathbf{S}}^{\prime },G\right] \right\rangle
\end{equation*}%
Each of these transitions arise independently from the other.

\paragraph{With modification of groups}

The general form of such transition is:

\begin{equation*}
\prod\limits_{G}\left\vert \left[ \mathbf{S},\underline{\mathbf{S}},G\right]
\right\rangle \rightarrow \prod\limits_{G^{\prime }}\left\vert \left[ 
\mathbf{S}^{\prime },\underline{\mathbf{S}}^{\prime },G^{\prime }\right]
\right\rangle
\end{equation*}%
where a product of states corresponding to a given collection of groups is
transformed into a product associated with another collection of groups. We
will specifically presetn here the case of mergers and dissociations.

\subparagraph{Mergers}

Groups may merge, which corresponds to the following transition in terms of
states:%
\begin{equation*}
\prod\limits_{G_{1}}\left\vert \left[ \mathbf{S}_{1},\underline{\mathbf{S}}%
_{1},G_{1}\right] \right\rangle \prod\limits_{G_{2}}\left\vert \left[ 
\mathbf{S}_{2},\underline{\mathbf{S}}_{2},G_{2}\right] \right\rangle
\rightarrow \prod\limits_{G}\left\vert \left[ \mathbf{S},\underline{\mathbf{S%
}},G\right] \right\rangle
\end{equation*}%
where $G$ denotes the aggregation of two or more groups. This binding
between groups can be described more precisely in terms of stakes.
Considering two merging groups:%
\begin{equation*}
\left\vert \left[ \mathbf{S}_{1},\underline{\mathbf{S}}_{1},G_{1}\right]
\right\rangle \left\vert \left[ \mathbf{S}_{2},\underline{\mathbf{S}}%
_{2},G_{2}\right] \right\rangle \rightarrow \left\vert \mathbf{S},\underline{%
\mathbf{S}},G_{1}\cup G_{2}\right\rangle
\end{equation*}%
the deviation stakes and the phase: 
\begin{equation*}
\left( \mathbf{S},\underline{\mathbf{S}}\right)
\end{equation*}%
are modifications in the deviations and phases for the two groups:%
\begin{equation*}
\left( \mathbf{S}_{1},\underline{\mathbf{S}}_{1}\right) ,\left( \mathbf{S}%
_{2},\underline{\mathbf{S}}_{2}\right)
\end{equation*}%
together with some inter-groups connections $\underline{S}_{12}$. Such
transitions may be induced either by direct interactions between the two
groups or by the temporary binding\ of a third group with $G_{1}$ and $G_{2}$%
.%
\begin{eqnarray*}
&&\left\vert \left[ \mathbf{S}_{1},\underline{\mathbf{S}}_{1},G_{1}\right]
\right\rangle \left\vert \left[ \mathbf{S}_{3},\underline{\mathbf{S}}%
_{3},G_{3}\right] \right\rangle \times \left\vert \left[ \mathbf{S}_{2},%
\underline{\mathbf{S}}_{2},G_{2}\right] \right\rangle \left\vert \left[ 
\mathbf{S}_{3},\underline{\mathbf{S}}_{3},G_{3}\right] \right\rangle \\
&\rightarrow &\left\vert \left[ \mathbf{S},\underline{\mathbf{S}},G_{1}\cup
G_{2}\cup G_{3}\right] \right\rangle \rightarrow \left( \left\vert \left[ 
\mathbf{S}_{1},\underline{\mathbf{S}}_{1},G_{1}\right] \right\rangle
\left\vert \left[ \mathbf{S}_{2},\underline{\mathbf{S}}_{2},G_{2}\right]
\right\rangle \right) \left\vert \left[ \mathbf{S}_{3},\underline{\mathbf{S}}%
_{3},G_{3}\right] \right\rangle
\end{eqnarray*}

\subparagraph{Dissociation}

Several groups may also split, which corresponds to the following
transitions:%
\begin{equation*}
\prod\limits_{G}\left\vert \left[ \mathbf{S},\underline{\mathbf{S}},G\right]
\right\rangle \rightarrow \prod\limits_{G_{1}}\left\vert \left[ \mathbf{S}%
_{1},\underline{\mathbf{S}}_{1},G_{1}\right] \right\rangle
\prod\limits_{G_{2}}\left\vert \left[ \mathbf{S}_{2},\underline{\mathbf{S}}%
_{2},G_{2}\right] \right\rangle
\end{equation*}%
This describes the reverse mechanism of merging states. For two groups:%
\begin{equation*}
\left\vert \mathbf{S},\underline{\mathbf{S}},G_{1}\cup G_{2}\right\rangle
\rightarrow \left\vert \left[ \mathbf{S}_{1},\underline{\mathbf{S}}_{1},G_{1}%
\right] \right\rangle \left\vert \left[ \mathbf{S}_{2},\underline{\mathbf{S}}%
_{2},G_{2}\right] \right\rangle
\end{equation*}%
this corresponds to inter-groups connections $\underline{S}_{12}$ that
gradually fade away.

\subparagraph{Defaults}

When transitions toward default state arise for some of these independent
groups, this can be writen as:%
\begin{equation*}
\prod\limits_{G}\left\vert \left[ \mathbf{S},\underline{\mathbf{S}},G\right]
\right\rangle \rightarrow \prod\limits_{G}\left\vert \left[ \mathbf{S}%
^{\prime },\underline{\mathbf{S}}^{\prime },G\right] \right\rangle
\prod\limits_{G}\left\vert \left[ \mathbf{S}^{\prime },\underline{\mathbf{S}}%
^{DF},G/DF\right] \right\rangle
\end{equation*}%
where the first product involves the non-defaulting groups, and the second
involves the defaulting groups.

\subparagraph{Uncertainty-induced}

In the formation of groups, the uncertainties $\zeta ^{2}$ and $\hat{h}%
_{1}^{\left( 0\right) }\left( X^{\prime },X_{m-1}\right) $\ about
investments play an important role, as they determine the stakes through the
coefficients $\hat{w}\left( X^{\prime },X\right) $. When $\zeta ^{2}$ or $%
\hat{w}_{1}^{\left( 0\right) }\left( X^{\prime },X_{m-1}\right) $ decrease
within a given group, the connections satisfying the saddle-point equations
are modified, which may trigger shifts in the system corresponding to a
reorganization of groups. In terms of transitions, this can be written as:%
\begin{equation*}
\prod\limits_{G_{1}}\left\vert \left[ \mathbf{S}_{1},\underline{\mathbf{S}}%
_{1},G_{1}\right] \right\rangle \rightarrow \prod\limits_{G_{2}}\left\vert %
\left[ \mathbf{S}_{2},\underline{\mathbf{S}}_{2},G_{2}\right] \right\rangle
\end{equation*}%
where the indices $1$ and $2$ refer to different organization in groups and
stakes. As a result, the forms of several groups are modified.

\subsubsection{Transitions of multiple interacting groups}

When groups interact in clusters, the fluctuations may propagate from one
group to another, leading to several simultaneous transitions. The
mechanisms described above still apply, but must be adapted to account for
interactions. In the following, we will focus on the general case involving
multiple clusters of states, in which the case with a single cluster is a
particular case.

\paragraph{Without modification of groups}

The principle is the same as before. The decomposition in clusters remains
globally unchanged, but inside each cluster, groups experience a change in
their level of stakes, $\mathbf{S}$, and in their phase $\underline{\mathbf{S%
}}$, so that the corresponding transition writes: 
\begin{equation}
\prod_{Cl}\left( \overset{\mathbf{S}_{G\in Cl,G^{\prime }\in Cl}}{\overbrace{%
\prod\limits_{G\in Cl}\left\vert \left[ \mathbf{S},\underline{\mathbf{S}},G%
\right] \right\rangle }}\right) \rightarrow \prod_{Cl}\left( \overset{%
\mathbf{S}_{G\in Cl,G^{\prime }\in Cl}^{\prime }}{\overbrace{%
\prod\limits_{G\in Cl}\left\vert \left[ \mathbf{S}^{\prime },\underline{%
\mathbf{S}}^{\prime },G\right] \right\rangle }}\right)
\end{equation}%
However, in the present setting, transitions within a cluster are not
independent and may depend on interactions between groups. Fluctuations of a
given group, although they do not modify the phase of that group, may induce
transitions in neighboring groups. These inter-group transition mechanisms
will be studied in greater detail in a future work.

\paragraph{With modification of groups}

Among interacting groups, the transitions associated with modifications in
groups are similar to those observed in the case of multiple independent
groups. However, we can distinguish between modifications that occur within
a cluster and those that occur between clusters.

\subparagraph{Intra-clusters modifications}

Here again, the groups of a cluster may merge or dissociate, due to
modification of uncertainty or interactions. The two mechanisms are
reciprocal, so that:%
\begin{equation*}
\left( \overset{\mathbf{S}_{G\in Cl,G^{\prime }\in Cl}}{\overbrace{%
\prod\limits_{G_{1}\in Cl}\left\vert \left[ \mathbf{S}_{1},\underline{%
\mathbf{S}}_{1},G_{1}\right] \right\rangle \prod\limits_{G_{2}\in
Cl}\left\vert \left[ \mathbf{S}_{2},\underline{\mathbf{S}}_{2},G_{2}\right]
\right\rangle }}\right) \rightleftharpoons \left( \overset{\mathbf{S}_{G\in
Cl,G^{\prime }\in Cl}}{\overbrace{\prod\limits_{G\in Cl}\left\vert \left[ 
\mathbf{S},\underline{\mathbf{S}},G\right] \right\rangle }}\right)
\end{equation*}%
where the double arrow $\rightleftharpoons $ accounts for the possibility of
transition in both directions.

\subparagraph{Inter-clusters modifications}

Uncertainty can induce reorganization of groups, such as clusters merging or
dissociating. This type of transitions are written:%
\begin{equation*}
\left( \overset{\mathbf{S}_{G\in Cl,G^{\prime }\in Cl}}{\overbrace{%
\prod\limits_{G_{1}\in Cl}\left\vert \left[ \mathbf{S}_{1},\underline{%
\mathbf{S}}_{1},G_{1}\right] \right\rangle }}\overset{\mathbf{S}_{G\in
Cl,G^{\prime }\in Cl}}{\overbrace{\prod\limits_{G_{2}\in Cl}\left\vert \left[
\mathbf{S}_{2},\underline{\mathbf{S}}_{2},G_{2}\right] \right\rangle }}%
\right) \rightleftharpoons \left( \overset{\mathbf{S}_{G\in Cl,G^{\prime
}\in Cl}}{\overbrace{\prod\limits_{G\in Cl}\left\vert \left[ \mathbf{S},%
\underline{\mathbf{S}},G\right] \right\rangle }}\right)
\end{equation*}

\subparagraph{Defaults}

Among the transitions induced by the interactions, some groups of a cluster
may be driven to a default state, and the general form of such type of
transition is:

\subparagraph{%
\protect\begin{equation}
\prod_{Cl}\left( \protect\overset{\mathbf{S}_{G\in Cl,G^{\prime }\in Cl}}{%
\protect\overbrace{\prod\limits_{G\in Cl}\left\vert \left[ \mathbf{S},%
\protect\underline{\mathbf{S}},G\right] \right\rangle }}\right) \rightarrow
\prod_{Cl}\left( \protect\overset{\mathbf{S}_{G\in Cl,G^{\prime }\in Cl}}{%
\protect\overbrace{\prod\limits_{G\in Cl}\left\vert \left[ \mathbf{S}%
^{\prime },\protect\underline{\mathbf{S}}^{\prime },G\right] \right\rangle
\prod\limits_{G\in Cl}\left\vert \left[ \mathbf{S}^{\prime },\protect%
\underline{\mathbf{S}}^{DF},G/DF\right] \right\rangle }}\right) 
\protect\end{equation}%
Uncertainty-induced}

We have seen that uncertainty may modify the composition of groups and thus,
in turn, the form of the clusters: groups may reorganize and change the
decomposition of the collective state into clusters. The mechanisms involved
are similar to the case of independent groups, and the corresponding
clusters' transition writes:

\begin{equation}
\prod_{Cl_{1}}\left( \overset{\mathbf{S}_{G_{1}\in Cl,_{1}G_{1}^{\prime }\in
Cl_{1}}}{\overbrace{\prod\limits_{G_{1}\in Cl_{1}}\left\vert \left[ \mathbf{S%
}_{1},\underline{\mathbf{S}}_{1},G_{1}\right] \right\rangle }}\right)
\rightarrow \prod_{Cl_{2}}\left( \overset{\mathbf{S}_{G_{2}\in
Cl,_{2}G_{2}^{\prime }\in Cl_{2}}}{\overbrace{\prod\limits_{G_{2}\in
Cl_{2}}\left\vert \left[ \mathbf{S}_{2},\underline{\mathbf{S}}_{2},G_{2}%
\right] \right\rangle }}\right)
\end{equation}

\subsubsection{Probabilities of transitions for basis states}

\paragraph{Principle}

The transitions studied above are quantified by the \emph{probabilities of
transition}, which compute the probability that a transition occurs from an
initial state:%
\begin{equation*}
\prod_{G_{1}}\left\vert \left[ \mathbf{S}_{1},\underline{\mathbf{S}}%
_{1},G_{1}\right] \right\rangle
\end{equation*}%
toward a final state:%
\begin{equation*}
\prod_{G_{2}}\left\vert \left[ \mathbf{S}_{2},\underline{\mathbf{S}}%
_{2},G_{2}\right] \right\rangle
\end{equation*}%
Formally, such a probability is written as:%
\begin{equation}
\prod_{G_{1}}\left\langle \left[ \mathbf{S}_{1},\underline{\mathbf{S}}%
_{1},G_{1}\right] \right\vert \exp \left( -S\right) \prod_{G_{2}}\left\vert %
\left[ \mathbf{S}_{2},\underline{\mathbf{S}}_{2},G_{2}\right] \right\rangle
\label{Tr}
\end{equation}%
where the action functional $S$ is given by (\ref{ct}):%
\begin{equation}
S=\sum_{G}S\left( \Lambda \left( \mathbf{S},\underline{\mathbf{S}},G\right)
\right) +\sum_{G,G^{\prime }}\delta S_{G,G^{\prime }}\left( \Lambda \left( 
\mathbf{S},\underline{\mathbf{S}},G\right) ,\Lambda \left( \mathbf{S},%
\underline{\mathbf{S}},G\right) \right)
\end{equation}%
For example, in the case of merging groups, the probability of transition
takes the form:%
\begin{equation*}
\prod_{G}\left\langle \left[ \mathbf{S},\underline{\mathbf{S}},G\right]
\right\vert \exp \left( -S\right) \prod_{G_{1}}\left\vert \left[ \mathbf{S}%
_{1},\underline{\mathbf{S}}_{1},G_{1}\right] \right\rangle
\prod_{G_{2}}\left\vert \left[ \mathbf{S}_{2},\underline{\mathbf{S}}%
_{2},G_{2}\right] \right\rangle
\end{equation*}%
Conversely, for splitting groups, the probability of transition becomes:%
\begin{equation*}
\prod_{G_{1}}\left\langle \left[ \mathbf{S}_{1},\underline{\mathbf{S}}%
_{1},G_{1}\right] \right\vert \prod_{G_{2}}\left\langle \left[ \mathbf{S}%
_{2},\underline{\mathbf{S}}_{2},G_{2}\right] \right\vert \exp \left(
-S\right) \prod_{G}\left\vert \left[ \mathbf{S},\underline{\mathbf{S}},G%
\right] \right\rangle
\end{equation*}

\paragraph{Derivation}

The derivation of such probability of transition was presented in Gosselin
and Lotz (2024). Formally, the probability (\ref{Tr}) is computed as:%
\begin{eqnarray}
&&\prod_{G_{1}}\left\langle \left[ \mathbf{S},\underline{\mathbf{S}},G_{1}%
\right] \right\vert \exp \left( -S\right) \prod_{G_{2}}\left\vert \left[ 
\mathbf{S}_{2},\underline{\mathbf{S}}_{2},G_{2}\right] \right\rangle
\label{Tch} \\
&=&\int_{\substack{ \left( \Lambda _{-\infty },\Lambda _{-\infty }^{\dag
}\right) =\left\vert \left[ \mathbf{S}_{1},\underline{\mathbf{S}}_{1},G_{1}%
\right] \right\rangle  \\ \left( \Lambda _{-\infty },\Lambda _{-\infty
}^{\dag }\right) =\left\vert \left[ \mathbf{S}_{2},\underline{\mathbf{S}}%
_{2},G_{2}\right] \right\rangle }}\exp \left( -\int S_{t}dt\right) D\Lambda
_{t}D\Lambda _{t}^{\dag }  \notag
\end{eqnarray}%
Here $\Lambda _{t}$ and $\Lambda _{t}^{\dag }$\ are extensions of the fields 
$\Lambda $ and $\Lambda ^{\dag }$\ including an explicit time variable. The
action functional $S_{t}$ is the functional $S$ evaluated at $\Lambda _{t}$
and $\Lambda _{t}^{\dag }$:%
\begin{equation*}
S_{t}=\sum_{G}S\left( \Lambda _{t}\left( \mathbf{S},\underline{\mathbf{S}}%
,G\right) \right) +\sum_{G,G^{\prime }}\delta S_{G,G^{\prime }}\left(
\Lambda _{t}\left( \mathbf{S},\underline{\mathbf{S}},G\right) ,\Lambda
_{t}\left( \mathbf{S},\underline{\mathbf{S}},G\right) \right)
\end{equation*}%
The measure $D\Lambda _{t}D\Lambda _{t}^{\dag }$ represents the functional
integration over all realizations of the fields $\Lambda $ and $\Lambda
^{\dag }$. The integral is performed under the boundary conditions:%
\begin{eqnarray*}
\left( \Lambda _{-\infty },\Lambda _{-\infty }^{\dag }\right) &=&\left\vert 
\left[ \mathbf{S}_{1},\underline{\mathbf{S}}_{1},G_{1}\right] \right\rangle
\\
\left( \Lambda _{-\infty },\Lambda _{-\infty }^{\dag }\right) &=&\left\vert 
\left[ \mathbf{S}_{2},\underline{\mathbf{S}}_{2},G_{2}\right] \right\rangle
\end{eqnarray*}%
that is, the fields realizations are constrained by background fields
corresponding to the states $\left\vert \left[ \mathbf{S}_{1},\underline{%
\mathbf{S}}_{1},G_{1}\right] \right\rangle $ and $\left\vert \left[ \mathbf{S%
}_{2},\underline{\mathbf{S}}_{2},G_{2}\right] \right\rangle $ as boundary
conditions. The techniques required to compute these probabilities will be
developed in the fourth part of this series.

\subsection{Transitions for Fiber States}

In a dynamic context, the fiber variables, eventhough depending they depend
on basis variables, are considered as independent variables. As a
consequence, fiber states $\left\vert \mathbf{F,}\underline{\mathbf{F}}%
,G\right\rangle $\ can themselves undergo two types of transitions. First,
they can experience specific-fiber transitions, that describe internal
modifications of the fiber, for a given basis state.\ Second, fiber states
can experience fiber transitions induced by basis transitions.

Since each fiber deviation state is attached to a basis state, and as such
depends on a given group, we will describe both type of fiber transitions
for a given basis deviation state - and thus a given group $G$ . Therefore,
we restrict our attention to the case of single-group transition.

\subsubsection{Types of transitions}

We first describe the fiber specific transitions, that are independent of
the basis, then the transitions induced by the basis.

\paragraph{Fiber-specifc transition}

Transitions in fiber states may be spontaneous or externally driven. Return
equations generally admit multiple solutions for a given distribution of
stakes for which transitions of the form:%
\begin{equation}
\left\vert \mathbf{F,}\underline{\mathbf{F}},G\right\rangle \rightarrow
\left\vert \mathbf{F,}\underline{\mathbf{F}},G\right\rangle ^{\prime }
\label{Ft}
\end{equation}%
may therefore occur without basis transitions.

Such transitions may also involve defaults for some groups. In this case,
the transition:%
\begin{equation}
\left\vert \mathbf{F,}\underline{\mathbf{F}},G\right\rangle \rightarrow
\left\vert \mathbf{F,}\underline{\mathbf{F}},G\right\rangle ^{DF}
\label{Ftd}
\end{equation}%
should induce a corresponding basis transition:%
\begin{equation*}
\left\vert \left[ \mathbf{S},\underline{\mathbf{S}},G\right] \right\rangle
\rightarrow \left\vert \left[ \mathbf{S}^{\prime },\underline{\mathbf{S}}%
^{DF},G/DF\right] \right\rangle
\end{equation*}%
Remark that transitions (\ref{Ft}) may in turn induce transitions in the
basis states:%
\begin{equation*}
\left\vert \left[ \mathbf{S},\underline{\mathbf{S}},G\right] \right\rangle
\rightarrow \left\vert \left[ \mathbf{S}^{\prime },\underline{\mathbf{S}}%
^{\prime },G\right] \right\rangle
\end{equation*}

\paragraph{Basis-induced transition}

Reciprocally, fiber states may undergo transitions induced by changes in
stakes:%
\begin{equation*}
\left\vert \left[ \mathbf{S},\underline{\mathbf{S}},G\right] \right\rangle
\rightarrow \left\vert \left[ \mathbf{S}^{\prime },\underline{\mathbf{S}},G%
\right] \right\rangle
\end{equation*}%
which describe fluctuations, or transitions in stakes:%
\begin{equation*}
\left\vert \left[ \mathbf{S},\underline{\mathbf{S}},G\right] \right\rangle
\rightarrow \left\vert \left[ \mathbf{S}^{\prime },\underline{\mathbf{S}}%
^{\prime },G\right] \right\rangle
\end{equation*}%
Both modifications may induce fiber transitions:%
\begin{equation}
\left\vert \mathbf{F,}\underline{\mathbf{F}},G\right\rangle \rightarrow
\left\vert \mathbf{F}^{\prime }\mathbf{,}\underline{\mathbf{F}}^{\prime
},G\right\rangle  \label{Fbt}
\end{equation}

\subsubsection{Probabilities of transition for fiber states}

\paragraph{Principle}

As for the basis states, we can associate a probability of transition to
each transition of a fiber state. For two fiber states:%
\begin{equation*}
\left\vert \mathbf{F,}\underline{\mathbf{F}},G\right\rangle
\end{equation*}%
and:%
\begin{equation*}
\left\vert \mathbf{F}^{\prime }\mathbf{,}\underline{\mathbf{F}}^{\prime
},G\right\rangle
\end{equation*}%
it is computed by evaluating:%
\begin{equation}
\left\langle \mathbf{F}^{\prime }\mathbf{,}\underline{\mathbf{F}}^{\prime
},G\right\vert \exp \left( -S^{F}\right) \left\vert \mathbf{F,}\underline{%
\mathbf{F}},G\right\rangle  \label{mpf}
\end{equation}%
where the action functional $S^{F}$ is given by (\ref{Fct}).

\paragraph{Derivation}

The derivation of the probability of transition between fiber states (\ref%
{mpf}) is similar to that of the probability of transition between basis
states (\ref{Tch}). Formally, it is computed as:%
\begin{eqnarray}
&&\left\langle \mathbf{F}^{\prime }\mathbf{,}\underline{\mathbf{F}}^{\prime
},G\right\vert \exp \left( -S^{F}\right) \left\vert \mathbf{F,}\underline{%
\mathbf{F}},G\right\rangle \\
&=&\int_{\substack{ \left( \Lambda _{-\infty },\Lambda _{-\infty }^{\dag
}\right) =\left\vert \mathbf{F,}\underline{\mathbf{F}},G\right\rangle  \\ %
\left( \Lambda _{-\infty },\Lambda _{-\infty }^{\dag }\right) =\left\vert 
\mathbf{F}^{\prime }\mathbf{,}\underline{\mathbf{F}}^{\prime
},G\right\rangle }}\exp \left( -\int S_{t}dt\right) D\mathbf{\Lambda }_{t}D%
\mathbf{\Lambda }_{t}^{\dag }D\hat{\Psi}_{t}D\hat{\Psi}_{t}^{\dag }D\bar{\Psi%
}_{t}D\bar{\Psi}_{t}^{\dag }  \notag
\end{eqnarray}%
Here $\mathbf{\Lambda }_{t}$, $\mathbf{\Lambda }_{t}^{\dag }$, $\hat{\Psi}%
_{t}$, $\hat{\Psi}_{t}^{\dag }$ and $\bar{\Psi}_{t}$, $\bar{\Psi}_{t}^{\dag
} $ are extensions of the fields $\mathbf{\Lambda }$, $\mathbf{\Lambda }%
^{\dag }$, $\hat{\Psi}$, $\hat{\Psi}^{\dag }$ and $\bar{\Psi}$, $\bar{\Psi}%
^{\dag }$\ including an explicit time variable. The action functional $%
S_{t}^{F}$ is the functional $S^{F}$ evaluated at $\left( \mathbf{\Lambda }%
_{t},\mathbf{\Lambda }_{t}^{\dag },\hat{\Psi}_{t},\hat{\Psi}_{t}^{\dag },%
\bar{\Psi}_{t},\bar{\Psi}_{t}^{\dag }\right) $:%
\begin{equation}
S_{t}^{F}=\hat{S}_{\left[ \mathbf{S}\right] }\left( \hat{\Psi}_{t}\left(
K,X\right) \right) +\bar{S}_{\left[ \mathbf{S}\right] }\left( \hat{\Psi}%
_{t}\left( K,X\right) ,\bar{\Psi}_{t}\left( K,X\right) \right) +\mathbf{%
\Lambda }_{t}^{\dag }\left( \mathbf{F,}\underline{\mathbf{F}},G\right)
\delta ^{2}S\left( \bar{\Psi},\hat{\Psi}\right) \mathbf{\Lambda }_{t}\left( 
\mathbf{F,}\underline{\mathbf{F}},G\right)
\end{equation}%
\begin{equation*}
S_{t}=\sum_{G}S\left( \Lambda _{t}\left( \mathbf{S},\underline{\mathbf{S}}%
,G\right) \right) +\sum_{G,G^{\prime }}\delta S_{G,G^{\prime }}\left(
\Lambda _{t}\left( \mathbf{S},\underline{\mathbf{S}},G\right) ,\Lambda
_{t}\left( \mathbf{S},\underline{\mathbf{S}},G\right) \right)
\end{equation*}%
The measure:%
\begin{equation*}
D\mathbf{\Lambda }_{t}D\mathbf{\Lambda }_{t}^{\dag }D\hat{\Psi}_{t}D\hat{\Psi%
}_{t}^{\dag }D\bar{\Psi}_{t}D\bar{\Psi}_{t}^{\dag }
\end{equation*}%
represents the functional integration over all realizations of the fields:%
\begin{equation*}
\left[ \mathbf{\Lambda }_{t},\mathbf{\Lambda }_{t}^{\dag }\right] ^{F}\equiv
\left( \mathbf{\Lambda }_{t},\mathbf{\Lambda }_{t}^{\dag },\hat{\Psi}_{t},%
\hat{\Psi}_{t}^{\dag },\bar{\Psi}_{t},\bar{\Psi}_{t}^{\dag }\right)
\end{equation*}%
. The integral is performed under the boundary conditions:%
\begin{eqnarray*}
\left[ \mathbf{\Lambda }_{-\infty },\mathbf{\Lambda }_{-\infty }^{\dag }%
\right] ^{F} &=&\left\vert \mathbf{F,}\underline{\mathbf{F}},G\right\rangle
\\
\left[ \mathbf{\Lambda }_{\infty },\mathbf{\Lambda }_{\infty }^{\dag }\right]
^{F} &=&\left\vert \mathbf{F}^{\prime }\mathbf{,}\underline{\mathbf{F}}%
^{\prime },G\right\rangle
\end{eqnarray*}

\subsection{Transitions for Fibered states}

Since fibered states describe the system in its globality, they encompass
the transitions both of the basis and the fiber states. However, as basis
transitions may induce fiber transitions, we can also consider that fiber
transitions induce basis transitions. These reciprocal interactions imply
possible sequences of retroaction and transitions between fibered states.

\subsubsection{Transitions for a single group}

Here, the two possible cases of basis transitions, without or with
modification of group can be considered. However we have also to account
that these transitions will induce fiber transitions. Moreover, a fiber
transition may itself impact the basis and induce a transition for a full
fibered state.

\paragraph{Without modification of the group}

In this case, the basis transition:%
\begin{equation*}
\left\vert \left[ \mathbf{S},\underline{\mathbf{S}},G\right] \right\rangle
\rightarrow \left\vert \left[ \mathbf{S}^{\prime },\underline{\mathbf{S}}%
^{\prime },G\right] \right\rangle
\end{equation*}%
induces a fiber transition (\ref{Fbt}), so that at the level of fibered
states, the transition writes:%
\begin{equation}
\left[ 
\begin{array}{c}
\underset{\updownarrow }{\left\vert \mathbf{F,}\underline{\mathbf{F}}\mathbf{%
,}G\right\rangle _{\left[ \mathbf{S}\right] }} \\ 
\left\vert \left[ \mathbf{S},\underline{\mathbf{S}},G\right] \right\rangle%
\end{array}%
\right] {\huge \rightarrow }\left[ 
\begin{array}{c}
\underset{\updownarrow }{\left\vert \mathbf{F}^{\prime }\mathbf{,}\underline{%
\mathbf{F}}^{\prime }\mathbf{,}G\right\rangle _{\left[ \mathbf{S}^{\prime }%
\right] }} \\ 
\left\vert \left[ \mathbf{S}^{\prime },\underline{\mathbf{S}}^{\prime },G%
\right] \right\rangle%
\end{array}%
\right]  \label{gr}
\end{equation}%
Note that this notation also encompasses the transtion initiated by the
fiber, through the mechanism:%
\begin{equation}
\left\vert \mathbf{F,}\underline{\mathbf{F}},G\right\rangle \rightarrow
\left\vert \mathbf{F}^{\prime }\mathbf{,}\underline{\mathbf{F}}^{\prime
},G\right\rangle
\end{equation}%
followed by basis transition:%
\begin{equation*}
\left\vert \left[ \mathbf{S},\underline{\mathbf{S}},G\right] \right\rangle
\rightarrow \left\vert \left[ \mathbf{S}^{\prime },\underline{\mathbf{S}}%
^{\prime },G\right] \right\rangle
\end{equation*}%
By including both possibilities, notation (\ref{gr}) underscores the fact
that transitions are actually characteristics of the whole state, seen as
the combination of basis and fiber states.

\paragraph{With modification of the group}

The transition towards default:%
\begin{equation*}
\left\vert \left[ \mathbf{S},\underline{\mathbf{S}},G\right] \right\rangle
\rightarrow \left\vert \left[ \mathbf{S}^{\prime },\underline{\mathbf{S}}%
^{DF},G/DF\right] \right\rangle
\end{equation*}%
may also impact the whole fibered state by:%
\begin{equation}
\left[ 
\begin{array}{c}
\underset{\updownarrow }{\left\vert \mathbf{F,}\underline{\mathbf{F}}%
,G\right\rangle _{\left[ \mathbf{S}\right] }} \\ 
\left\vert \left[ \mathbf{S},\underline{\mathbf{S}},G\right] \right\rangle%
\end{array}%
\right] {\huge \rightarrow }\left[ 
\begin{array}{c}
\underset{\updownarrow }{\left\vert \mathbf{F}^{\prime }\mathbf{,}\underline{%
\mathbf{F}}^{DF}\mathbf{,}G\right\rangle _{\left[ \mathbf{S}^{\prime }\right]
}} \\ 
\left\vert \left[ \mathbf{S}^{\prime },\underline{\mathbf{S}}^{DF},G/DF%
\right] \right\rangle%
\end{array}%
\right]  \label{grp}
\end{equation}%
As in the previous paragraph, this notation also encompasses the possibility
of a transition towards default starting with the fiber state.

\paragraph{Multiple independent groups}

When multiple groups are independent, the transition mechanisms described
for a single group, without and with modification, (\ref{gr}) and (\ref{grp}%
) respectively, apply to each group, so that the transition without groups
modifications writes:

\begin{equation}
\prod_{G}\left[ 
\begin{array}{c}
\underset{\updownarrow }{\left\vert \mathbf{F,}\underline{\mathbf{F}}\mathbf{%
,}G\right\rangle _{\left[ \mathbf{S}\right] }} \\ 
\left\vert \left[ \mathbf{S},\underline{\mathbf{S}},G\right] \right\rangle%
\end{array}%
\right] {\huge \rightarrow }\prod_{G}\left[ 
\begin{array}{c}
\underset{\updownarrow }{\left\vert \mathbf{F}^{\prime }\mathbf{,}\underline{%
\mathbf{F}}^{\prime }\mathbf{,}G\right\rangle _{\left[ \mathbf{S}^{\prime }%
\right] }} \\ 
\left\vert \left[ \mathbf{S}^{\prime },\underline{\mathbf{S}}^{\prime },G%
\right] \right\rangle%
\end{array}%
\right]  \label{grm}
\end{equation}%
and the transition with group modification writes:%
\begin{equation}
\prod_{G}\left[ 
\begin{array}{c}
\underset{\updownarrow }{\left\vert \mathbf{F,}\underline{\mathbf{F}}%
,G\right\rangle _{\left[ \mathbf{S}\right] }} \\ 
\left\vert \left[ \mathbf{S},\underline{\mathbf{S}},G\right] \right\rangle%
\end{array}%
\right] {\huge \rightarrow }\prod_{G}\left[ 
\begin{array}{c}
\underset{\updownarrow }{\left\vert \mathbf{F}^{\prime }\mathbf{,}\underline{%
\mathbf{F}}^{DF}\mathbf{,}G\right\rangle _{\left[ \mathbf{S}^{\prime }\right]
}^{\prime }} \\ 
\left\vert \left[ \mathbf{S}^{\prime },\underline{\mathbf{S}}^{DF},G/DF%
\right] \right\rangle%
\end{array}%
\right]  \label{grpm}
\end{equation}%
For multiple groups, the mechanisms of mergers and dissociations also apply,
and can be written in a compact form:%
\begin{equation}
\prod_{G}\left[ 
\begin{array}{c}
\underset{\updownarrow }{\left\vert \mathbf{F,}\underline{\mathbf{F}}%
,G\right\rangle _{\left[ \mathbf{S}\right] }} \\ 
\left\vert \left[ \mathbf{S},\underline{\mathbf{S}},G\right] \right\rangle%
\end{array}%
\right] {\huge \rightarrow }\prod_{G^{\prime }}\left[ 
\begin{array}{c}
\underset{\updownarrow }{\left\vert \mathbf{F}^{\prime }\mathbf{,}\underline{%
\mathbf{F}}^{\prime },G^{\prime }\right\rangle _{\left[ \mathbf{S}^{\prime }%
\right] }} \\ 
\left\vert \left[ \mathbf{S}^{\prime },\underline{\mathbf{S}}^{\prime
},G^{\prime }\right] \right\rangle%
\end{array}%
\right]  \label{mch}
\end{equation}%
where the products over the groups include the possible increase
(dissociation) or decrease (merger) in the number of groups.

\paragraph{Multiple interacting groups}

Among interacting groups, the various transition mechanisms can be gathered
in generalization of the transition of mutliple independent groups (\ref{mch}%
), in which clusters and transitions in fiber states are included:%
\begin{equation*}
\left[ 
\begin{array}{c}
\underset{\updownarrow }{\prod_{G}\left\vert \mathbf{F,}\underline{\mathbf{F}%
}\mathbf{,}G\right\rangle _{\left[ \mathbf{S,}\left\{ G,G^{\prime }\right\} %
\right] }} \\ 
\overset{\mathbf{S}_{G,G^{\prime }}}{\overbrace{\prod_{G}\left\vert \left[ 
\mathbf{S},\underline{\mathbf{S}},G\right] \right\rangle }}%
\end{array}%
\right] {\huge \rightarrow }\left[ 
\begin{array}{c}
\underset{\updownarrow }{\prod_{G}\left\vert \mathbf{F,}\underline{\mathbf{F}%
}\mathbf{,}G\right\rangle _{\left[ \mathbf{S}^{\prime }\mathbf{,}\left\{
G,G^{\prime }\right\} \right] }^{\prime }} \\ 
\overset{\mathbf{S}_{G^{\prime },G^{\prime \prime }}^{\prime }}{\overbrace{%
\prod_{G^{\prime }}\left\vert \left[ \mathbf{S}^{\prime },\underline{\mathbf{%
S}}^{\prime },G^{\prime }\right] \right\rangle }}%
\end{array}%
\right]
\end{equation*}%
This generalization encompasses transitions in fiber and basis states, so
that groups phases and fiber states can experience simultaneous transitions.
These mechanisms can be explored in detail by considering a sequence of
successive transitions.

\paragraph{Sequences of transitions}

Basis fluctuations or transitions may induce fiber transitions, but fiber
transitions may also induce basis transitions. These reciprocal interactions
may trigger the following sequence of transitions:%
\begin{equation*}
\begin{array}{cccc}
\left[ 
\begin{array}{c}
\underset{\updownarrow }{\prod_{G}\left\vert \mathbf{F,}\underline{\mathbf{F}%
}\mathbf{,}G\right\rangle _{\left[ \mathbf{S,}\left\{ G,G^{\prime }\right\} %
\right] }} \\ 
\overset{\mathbf{S}_{G,G^{\prime }}}{\overbrace{\prod_{G}\left\vert \left[ 
\mathbf{S},\underline{\mathbf{S}},G\right] \right\rangle }}%
\end{array}%
\right] & {\huge \rightarrow } & \left[ 
\begin{array}{c}
\underset{\updownarrow }{\prod_{G}\left\vert \mathbf{F,}\underline{\mathbf{F}%
}\mathbf{,}G\right\rangle _{\left[ \mathbf{S}^{\prime }\mathbf{,}\left\{
G,G^{\prime }\right\} \right] }^{\prime }} \\ 
\overset{\mathbf{S}_{G,G^{\prime }}}{\overbrace{\prod_{G}\left\vert \left[ 
\mathbf{S}^{\prime },\underline{\mathbf{S}},G\right] \right\rangle }}%
\end{array}%
\right] &  \\ 
&  & {\huge \downarrow } &  \\ 
&  & \left[ 
\begin{array}{c}
\underset{\updownarrow }{\prod_{G}\left\vert \mathbf{F,}\underline{\mathbf{F}%
}\mathbf{,}G\right\rangle _{\left[ \mathbf{S}^{\prime }\mathbf{,}\left\{
G,G^{\prime }\right\} \right] }^{\prime }} \\ 
\overset{\mathbf{S}_{G,G^{\prime }}}{\overbrace{\prod_{G}\left\vert \left[ 
\mathbf{S}^{\prime },\underline{\mathbf{S}}^{\prime },G\right] \right\rangle 
}}%
\end{array}%
\right] & {\huge \rightarrow ...}%
\end{array}%
\end{equation*}

In the above, a fluctuation in the basis state induces a first transition on
the fiber state.\ This transition in the fiber state induces in turn a
transition in the basis state, and so on... The changes in stakes induced by
these several successive transitions may drive the system into an
instability regime further triggering a default state, so that a transition:%
\begin{equation}
\left\vert \mathbf{F,}\underline{\mathbf{F}}\right\rangle ^{\prime
}\rightarrow \left\vert \mathbf{F,}\underline{\mathbf{F}}\right\rangle ^{DF}
\end{equation}%
may occur. This default will itself impact the basis state, which will in
turn enter a default state. This mecanism will emerge in the sequence of
transitions through the appearance of the following term:%
\begin{equation*}
\left[ 
\begin{array}{c}
\underset{\updownarrow }{\prod\limits_{G}\left\vert \mathbf{F,}\underline{%
\mathbf{F}}\right\rangle _{\left[ \mathbf{S}^{\prime }\mathbf{,}\left\{
G,G^{\prime }\right\} \right] }^{DF}} \\ 
\overset{\mathbf{S}_{G,G^{\prime }}}{\overbrace{\prod_{G}\left\vert \left[ 
\mathbf{S}^{\prime },\underline{\mathbf{S}}^{\prime },G\right] \right\rangle
^{DF}}}%
\end{array}%
\right]
\end{equation*}

\subsubsection{Probabilities of transitions for fibered states}

\paragraph{Principle}

The \emph{probabilities of transition} for fibered states are similar to
that of the basis or fiber states. Considering an initial state: 
\begin{equation*}
\left\vert In\right\rangle =\left[ 
\begin{array}{c}
\underset{\updownarrow }{\prod_{G}\left\vert \mathbf{F,}\underline{\mathbf{F}%
}\mathbf{,}G\right\rangle _{\left[ \mathbf{S,}\left\{ G,G^{\prime }\right\} %
\right] }} \\ 
\overset{\mathbf{S}_{G,G^{\prime }}}{\overbrace{\prod_{G}\left\vert \left[ 
\mathbf{S},\underline{\mathbf{S}},G\right] \right\rangle }}%
\end{array}%
\right]
\end{equation*}%
the transition towards a final state:%
\begin{equation*}
\left\vert Fn\right\rangle =\left[ 
\begin{array}{c}
\underset{\updownarrow }{\prod_{G}\left\vert \mathbf{F,}\underline{\mathbf{F}%
}\mathbf{,}G\right\rangle _{\left[ \mathbf{S}^{\prime }\mathbf{,}\left\{
G,G^{\prime }\right\} \right] }^{\prime }} \\ 
\overset{\mathbf{S}_{G,G^{\prime }}}{\overbrace{\prod_{G^{\prime
}}\left\vert \left[ \mathbf{S}^{\prime },\underline{\mathbf{S}}^{\prime
},G^{\prime }\right] \right\rangle }}%
\end{array}%
\right]
\end{equation*}%
is written formally as:%
\begin{equation}
\left\langle Fn\right\vert \exp \left( -S^{B,F}\right) \left\vert
In\right\rangle  \label{frmb}
\end{equation}%
with the action functional $S^{B,F}$ for the fibered fields is given by (\ref%
{Ftv}).

\paragraph{Derivation}

To derive the probabilities of transition of the fibered states (\ref{frmb}%
), we must extend the definition of each field by including a time variable
in its arguments.\ The probabilities are then computed by the following
integral:%
\begin{eqnarray}
&&\left\langle Fn\right\vert \exp \left( -S^{B,F}\right) \left\vert
In\right\rangle \\
&=&\int_{\substack{ \left( \Lambda _{-\infty },\Lambda _{-\infty }^{\dag
}\right) =\prod\limits_{G}\left\vert \left[ \mathbf{S},\underline{\mathbf{S}}%
,G\right] \right\rangle \left\vert \mathbf{F,}\underline{\mathbf{F}}%
,G\right\rangle  \\ \left( \Lambda _{-\infty },\Lambda _{-\infty }^{\dag
}\right) =\prod\limits_{G^{\prime }}\left\vert \left[ \mathbf{S}^{\prime },%
\underline{\mathbf{S}}^{\prime },G^{\prime }\right] \right\rangle \left\vert 
\mathbf{F}^{\prime }\mathbf{,}\underline{\mathbf{F}}^{\prime
},G\right\rangle }}\exp \left( -\int S^{B,F}dt\right) D\left( \Lambda
_{-t},\Lambda _{-t}^{\dag }\right)  \notag
\end{eqnarray}%
where:%
\begin{equation*}
\left( \Lambda _{-t},\Lambda _{-t}^{\dag }\right)
\end{equation*}%
represent the time extension of the whole set of fields involved during the
transition:%
\begin{equation*}
\left( \Lambda _{-t},\Lambda _{-t}^{\dag }\right) =\left\{ \left( \Lambda
_{t},\Lambda _{t}^{\dag },\mathbf{\Lambda }_{t},\mathbf{\Lambda }_{t}^{\dag
},\hat{\Psi}_{t},\hat{\Psi}_{t}^{\dag },\bar{\Psi}_{t},\bar{\Psi}_{t}^{\dag
}\right) \right\} _{\prod G,\prod G^{\prime }}
\end{equation*}

\paragraph{Remark}

For independent basis and fiber states, this probability would factor as:%
\begin{eqnarray*}
&&\overset{\mathbf{S}_{G,G^{\prime }}}{\overbrace{\prod_{G^{\prime
}}\left\langle \left[ \mathbf{S}^{\prime },\underline{\mathbf{S}}^{\prime
},G^{\prime }\right] \right\vert }\exp \left( -S_{B}\right) }\overset{%
\mathbf{S}_{G,G^{\prime }}}{\overbrace{\prod_{G}\left\vert \left[ \mathbf{S},%
\underline{\mathbf{S}},G\right] \right\rangle }} \\
&&\times \prod_{G}\left. {}\right. _{\left[ \prod_{G}\left[ \mathbf{S}%
^{\prime },G\right] ,\prod_{G,G^{\prime }}\mathbf{S}_{G,G^{\prime }}^{\prime
}\right] }\left\langle \mathbf{F}^{\prime }\mathbf{,}\underline{\mathbf{F}}%
^{\prime }\right\vert \exp \left( -S_{F}\right) \prod_{G}\left\vert \mathbf{%
F,}\underline{\mathbf{F}}\right\rangle _{\left[ \mathbf{S,}\left\{
G,G^{\prime }\right\} \right] }
\end{eqnarray*}%
where:%
\begin{equation*}
S_{B}=\sum_{G}S\left( \Lambda \left( \mathbf{S},\underline{\mathbf{S}}%
\right) ,G\right) +\sum_{G,G^{\prime }}S_{G,G^{\prime }}\left( \left(
\Lambda \left( \mathbf{S}^{\left( G/G^{\prime }\right) },\underline{\mathbf{S%
}}^{\left( G/G^{\prime }\right) }\right) \right) \right)
\end{equation*}%
and:%
\begin{eqnarray*}
S_{F} &=&\sum_{G}\hat{S}_{\left[ \mathbf{S,}\left\{ G,G^{\prime }\right\} %
\right] }\left( \hat{\Psi}\left( \hat{K},X\right) \right) +\sum_{G}\bar{S}_{%
\left[ \mathbf{S,}\left\{ G,G^{\prime }\right\} \right] }\left( \hat{\Psi}%
\left( \hat{K},X\right) ,\bar{\Psi}\left( \bar{K},X\right) \right) \\
&&+\sum_{G}\mathbf{\Lambda }^{\dag }\left( \mathbf{F,}\underline{\mathbf{F}}%
,G\right) \\
&&\times \left( \frac{\delta ^{2}\left( \hat{S}\left( \hat{\Psi}\right) +%
\bar{S}\left( \bar{\Psi}\right) \right) }{\delta \hat{\Psi}\left( \hat{K}%
,X\right) \delta \hat{\Psi}^{\dag }\left( \hat{K},X\right) }+\frac{\delta
^{2}\bar{S}\left( \bar{\Psi}\right) \left( \hat{S}\left( \hat{\Psi}\right) +%
\bar{S}\left( \bar{\Psi}\right) \right) }{\delta \bar{\Psi}\left( \bar{K}%
,X\right) \delta \bar{\Psi}^{\dag }\left( \bar{K},X\right) }\right) \mathbf{%
\Lambda }\left( \mathbf{F,}\underline{\mathbf{F}},G\right)
\end{eqnarray*}%
However the interaction terms:%
\begin{equation*}
\sum_{G,G^{\prime }}\delta S_{G,G^{\prime }}\left( \left( \hat{\Psi}^{\left(
G\right) }\left( \hat{K},X\right) ,\bar{\Psi}^{\left( G\right) }\left( \bar{K%
},X\right) \right) ,\left( \hat{\Psi}^{\left( G^{\prime }\right) }\left( 
\hat{K},X\right) ,\bar{\Psi}^{\left( G^{\prime }\right) }\left( \bar{K}%
,X\right) \right) \right)
\end{equation*}%
intertwin the two probabilities that are not independent.

We have described the various properties that a field theory for collective
states should encompass. The next section synthezises the results.

\section{Synthesis}

To describe a complete field-theoretic formalism, we reinterpreted
collective states as collections of interacting sub-collective states,
namely groups and their associated phases, and differenciated the variables
of these sub-collective states\ between basis and fiber variables. Viewed in
this light, the dynamics of collective states evolve along two complementary
and interacting directions: that of the basis and that of the fiber
collective states. For each, we provided specific notations, to account for
deviations around the static collective states.

The dynamics of the basis states and of their transitions is described by an
infinite set of basis deviation fields, each associated with a possible
equilibrium $\underline{\mathbf{S}}$ of the basis variables:%
\begin{equation*}
\Lambda \left( \mathbf{S},\underline{\mathbf{S}},G\right)
\end{equation*}%
and their realizations, the basis deviation states:%
\begin{equation*}
\left\vert \left[ \mathbf{S},\underline{\mathbf{S}},G\right] \right\rangle
\end{equation*}%
The dynamics of the fiber states and of their transitions is described by
the fiber equilibrium fields describing the with multiple equilibria of the
fiber state.\ These fields are denoted:%
\begin{equation*}
\hat{\Psi},\bar{\Psi}
\end{equation*}%
These fiber equilibrium fields are completed by the infinite set of fiber
deviation fields, one for each value of the fiber equilibrium $\underline{%
\mathbf{F}}$:%
\begin{equation*}
\mathbf{\Lambda }\left( \mathbf{F,}\underline{\mathbf{F}},G\right)
\end{equation*}%
The realizations of this set of fields are the fiber deviation states given
by:%
\begin{equation}
\left\vert \mathbf{F,}\underline{\mathbf{F}},G\right\rangle
\end{equation}%
Ultimately, the combinations of basis and fiber deviation states, the
so-called fibered deviation states, have their own dynamics, which emerge
from the interactions between basis and fiber deviation fields. These
fibered deviation states, denoted as:%
\begin{equation*}
\left[ 
\begin{array}{c}
\underset{\updownarrow }{\prod_{G}\left\vert \mathbf{F,}\underline{\mathbf{F}%
}\mathbf{,}G\right\rangle _{\left[ \mathbf{S,}\left\{ G,G^{\prime }\right\} %
\right] }} \\ 
\overset{\mathbf{S}_{G,G^{\prime }}}{\overbrace{\prod_{G}\left\vert \left[ 
\mathbf{S},\underline{\mathbf{S}},G\right] \right\rangle }}%
\end{array}%
\right]
\end{equation*}%
describe the system as a whole and are the realizations of the interactions
between basis deviation fields and fiber equilibrium and deviation fields.

The dynamics of these fibered deviation states, of their interactions and of
their transitions are formulated through three action functionals $S_{B}$, $%
S_{F}$ and $S^{B,F}$ for the basis, fiber, and fibered states, respectively.
The action functionals $S_{B}$\ and $S_{F}$\ describe the internal dynamics
of their respective deviations states, and the internal interactions among
their sub-collective states, whereas the action functional $S^{B,F}$ of the
fibered fields is the sum of the action functionals $S_{B}$\ and $S_{F}$, to
which some terms have been added to incorporate the interactions between the
basis and fiber fields.

Using this unified framework, we can determine the transition probabilities
between any initial and final fibered deviation state of the system, denoted 
$\left\vert In\right\rangle $ and $\left\vert Fn\right\rangle $
respectively, and given by:%
\begin{equation*}
\left\langle Fn\right\vert \exp \left( -S^{B,F}\right) \left\vert
In\right\rangle
\end{equation*}%
Since the action functional computes the transition probabilities between
initial and final fibered deviation states, we can also interpret its terms
to inspect the various transition mechanisms of the fibered deviation
states, as well as their possible sequences.\ Moreover, the action
functional of the fibered fields also allows to detect the most stable
sub-collective deviation states of a system, and analyze the internal
reorganization of a collective state.

However, this general structure of the various action functionals of the
system does not specify the precise forms that interactions between groups,
or between basis and fiber components, should take. It is the detailed study
of these interaction structures that will determine the mechanisms governing
state transitions and enable the explicit computation of transition
probabilities. Such an analysis constitutes the next stage of this research.

\section{Conclusion}

In this work, we have summarized the main properties of collective states in
systems with a large number of agents, firms, investors, and banks, and we
have detailed the analysis of the stability and instability of such states.
In order to account for the possibility of unstable fluctuations of these
collective states, inducing transitions between them, we were led to
consider the necessity of an extension of our formalism aimed at describing
collective states in their entirety, in a dynamical and interacting
framework. We have identified the essential elements of such an extension:
an infinity of deviation fields representing deviation states around
collective states; the need to distinguish these fields for basis states and
fiber states; a typology of the different possible types of transitions
between collective states; and the general form of the action functionals
for deviation fields inducing these transitions. On the basis of this
framework, the detailed development of the formalism and its applications
will be presented in the fourth part of this series.\pagebreak

\section*{References}

%\begin{thebibliography}{Gehrkelager 2000}

\begin{description}
\item Abergel, F., Chakraborti, A., Muni Toke, I., \& Patriarca, M. (2011a).
Econophysics review: I. Empirical facts. Quantitative Finance, 11(7),
991--1012.

\item Abergel, F., Chakraborti, A., Muni Toke, I., \& Patriarca, M. (2011b).
Econophysics review: II. Agent-based models. Quantitative Finance, 11(7),
1013--1041.

\item Acemoglu, D., Ozdaglar, A., \& Tahbaz-Salehi, A. (2015). Systemic risk
and stability in financial networks. American Economic Review, 105(2),
564--608.

\item Acharya, V. V., Engle, R. \& Richardson, M. (2012). Capital shortfall:
A new approach to ranking and regulating systemic risks. (Quantifying
capital shortfalls under stress).

\item Acharya, V. V., Pedersen, L. H., Philippon, T., \& Richardson, M.
(2017). Measuring systemic risk. The Review of Financial Studies, 30(1),
2--47.

\item Adrian, T., \& Brunnermeier, M. K. (2016). CoVaR. American Economic
Review, 106(7), 1705--1741.

\item Allen, F., \& Gale, D. (2000). Financial contagion. Journal of
Political Economy, 108(1), 1--33.

\item Bardoscia, M., Livan, G., \& Marsili, M. (2017). Statistical mechanics
of complex economies. Journal of Statistical Mechanics: Theory and
Experiment, 2017, 1--21.

\item Bardoscia, M., Battiston, S., Caccioli, F., \& Caldarelli, G. (2019).
Pathways towards instability in financial networks. Nature Communications,
10(1), 1--9.

\item Battiston, S., Puliga, M., Kaushik, R., Tasca, P., \& Caldarelli, G.
(2012). Debtrank: Too central to fail? Financial networks, the FED and
systemic risk. Scientific Reports, 2, 541.

\item Battiston, S., Caldarelli, G., May, R. M., Roukny, T., \& Stiglitz, J.
E. (2020). The price of complexity in financial networks. Proceedings of the
National Academy of Sciences, 117(52), 32779--32786.

\item Bernanke, B., Gertler, M., \& Gilchrist, S. (1999). The financial
accelerator in a quantitative business cycle framework. In J. B. Taylor \&
M. Woodford (Eds.), Handbook of Macroeconomics (Vol. 1, Part C, pp.
1341--1393). Elsevier.

\item Bensoussan, A., Frehse, J., \& Yam, P. (2018). Mean Field Games and
Mean Field Type Control Theory. Springer, New York.

\item B\"{o}hm, V., Kikuchi, T., \& Vachadze, G. (2008). Asset pricing and
productivity growth: The role of consumption scenarios. Computational
Economics, 32, 163--181.

\item Brunnermeier, M. K. \& Sannikov, Y. (2014). A Macroeconomic Model with
a Financial Sector. (Seminal macro-financial modelling showing financial
amplification mechanisms)

\item Brunnermeier, M. K., Crockett, A., Goodhart, C., Persaud, A. D. \&
Shin, H. S. (2009). The fundamental principles of financial regulation.
Geneva Report on the World Economy (report outlining regulatory principles
post-crisis).

\item Borio, C. \& Drehmann, M. (2009). Assessing the risk of banking crises
--- revisited. BIS Quarterly Review.

\item Caggese, A., \& Orive, A. P. (2015). The interaction between household
and firm dynamics and the amplification of financial shocks. Barcelona GSE
Working Paper Series, Working Paper n%
%TCIMACRO{\U{ba} }%
%BeginExpansion
${{}^o}$
%EndExpansion
866.

\item Campello, M., Graham, J., \& Harvey, C. R. (2010). The real effects of
financial constraints: Evidence from a financial crisis. Journal of
Financial Economics, 97(3), 470--487.

\item Cerutti, E., Claessens, S. \& Laeven, L. (2017). The use and
effectiveness of macroprudential policies: New evidence. IMF Working Paper /
Journal version.

\item Cifuentes, R., Ferrucci, G., \& Shin, H. S. (2005). Liquidity risk and
contagion. Journal of the European Economic Association, 3(2--3), 556--566.

\item Claessens, S., Ghosh, S. R. \& Mihet, R. (2013). Macro-prudential
policy: What instruments and how effective? IMF Economic Review / background
reports on macroprudential toolkits.

\item Cochrane, J. H. (Ed.). (2006). Financial markets and the real economy.
Edward Elgar, International Library of Critical Writings in Financial
Economics, Vol. 18.

\item Diamond, D. W. \& Dybvig, P. H. (1983). Bank runs, deposit insurance,
and liquidity. Journal of Political Economy.

\item Diamond, D. W. \& Rajan, R. G. (2001). Liquidity risk, liquidity
creation and financial fragility: A theory of banking. Journal of Political
Economy.

\item Drehmann, M. \& Juselius, M. (2014). Evaluating early warning
indicators of banking crises: Sizing up the credit-to-GDP gap. International
Journal of Forecasting / BIS discussion.

\item Elliott, M., Golub, B., \& Jackson, M. O. (2014). Financial networks
and contagion. American Economic Review, 104(10), 3115--3153.

\item Erbin H., String Field Theory: A Modern Introduction. Springer Lecture
Notes in Physics, 2021.

\item Freixas, X. \& Rochet, J.-C. (2008). Microeconomics of Banking.
(Second edition). MIT Press.

\item Freixas, X., Parigi, B. M. \& Rochet, J.-C. (2000). Systemic risk,
interbank relations, and liquidity provision by the central bank. Journal of
Money, Credit and Banking.

\item Gaffard, J.-L., \& Napoletano, M. (Eds.). (2012). Agent-based models
and economic policy. OFCE, Paris.

\item Gai, P., \& Kapadia, S. (2010). Contagion in financial networks.
Proceedings of the Royal Society A: Mathematical, Physical and Engineering
Sciences, 466(2120), 2401--2423.

\item Galati, G. \& Moessner, R. (2013). Macroprudential policy --- a
literature review. BIS Working Papers / BIS Quarterly Review articles.

\item Gennaioli, N., Shleifer, A., \& Vishny, R. W. (2012). Neglected risks,
financial innovation, and financial fragility. Journal of Financial
Economics, 104(3), 452--468.

\item Gertler, M. \& Kiyotaki, N. (2010). Financial intermediation and
credit policy in business cycle analysis. (Survey and modelling of the
financial accelerator and bank-mediated transmission).

\item Gomes, D. A., Nurbekyan, L., \& Pimentel, E. A. (2015). Economic
models and mean-field games theory. Publica\c{c}\~{o}es Matem\'{a}ticas do
IMPA, 30%
%TCIMACRO{\U{ba} }%
%BeginExpansion
${{}^o}$
%EndExpansion
Col\'{o}quio Brasileiro de Matem\'{a}tica, Rio de Janeiro.

\item Gosselin P., \& Lotz, A. (2025a) Financial Interactions and Collective
States: Part I. Investors and Firms. hal-05243522v3

\item Gosselin P., \& Lotz, A. (2025b) Financial Interactions and Collective
States: Part II. Banks, Investors and Firms. hal-05321420v2

\item Gosselin P., \& Lotz, A. (2024) Financial Interactions and Capital
Accumulation. hal-04573829.

\item Gosselin, P., \& Lotz, A. (2023a). A statistical field theory and
neural structures dynamics I: Action functionals, background states and
external perturbations. hal-04301171.

\item Gosselin, P., \& Lotz, A. (2023b). Statistical field theory and neural
structures dynamics II: Signals propagation, interferences, bound states.
hal-04301181v1.

\item Gosselin, P., \& Lotz, A. (2023c). A statistical field theory and
neural structures dynamics III: Effective action for connectivities,
interactions and emerging collective states. hal-04301185.

\item Gosselin, P., \& Lotz, A. (2023d). A statistical field theory and
neural structures dynamics IV: Field-theoretic formalism for interacting
collective states. hal-04301191.

\item Gosselin, P., Lotz, A., \& Wambst, M. (2017). A path integral approach
to interacting economic systems with multiple heterogeneous agents.
IF\_PREPUB, hal-01549586v2.

\item Gosselin, P., Lotz, A., \& Wambst, M. (2020). A path integral approach
to business cycle models with large number of agents. Journal of Economic
Interaction and Coordination, 15, 899--942.

\item Gosselin, P., Lotz, A., \& Wambst, M. (2021). A statistical field
approach to capital accumulation. Journal of Economic Interaction and
Coordination, 16, 817--908.

\item Grassetti, F., Mammana, C., \& Michetti, E. (2022). A dynamical model
for real economy and finance. Mathematical Finance and Economics.
https://doi.org/10.1007/s11579-021-00311-3

\item Greenwood, R., \& Hanson, S. G. (2013). Issuer quality and corporate
bond returns. The Review of Financial Studies, 26(6), 1483--1525.

\item Grosshans, D., \& Zeisberger, S. (2018). All's well that ends well? On
the importance of how returns are achieved. Journal of Banking \& Finance,
87, 397--410.

\item Haldane, A. G., \& May, R. M. (2011). Systemic risk in banking
ecosystems. Nature, 469(7330), 351--355.

\item Holmstrom, B., \& Tirole, J. (1997). Financial intermediation,
loanable funds, and the real sector. Quarterly Journal of Economics, 112(3),
663--691.

\item IMF (2014). Macroprudential policy: An organizing framework. IMF
Policy Paper (survey of policy objectives and instruments).

\item Jackson, M. O. (2010). Social and economic networks. Princeton
University Press, Princeton.

\item Jermann, U. J., \& Quadrini, V. (2012). Macroeconomic effects of
financial shocks. American Economic Review, 102(1), 238--271.

\item Khan, A., \& Thomas, J. K. (2013). Credit shocks and aggregate
fluctuations in an economy with production heterogeneity. Journal of
Political Economy, 121(6), 1055--1107.

\item Kaplan, G., \& Violante, L. (2018). Microeconomic heterogeneity and
macroeconomic shocks. Journal of Economic Perspectives, 32(3), 167--194.

\item Kleinert, H. (1989). Gauge fields in condensed matter (Vols. I--II).
World Scientific, Singapore.

\item Kleinert, H. (2009). Path integrals in quantum mechanics, statistics,
polymer physics, and financial markets (5th ed.). World Scientific,
Singapore.

\item Krugman, P. (1991). Increasing returns and economic geography. Journal
of Political Economy, 99(3), 483--499.

\item Langfield, S., Liu, Z., \& Ota, T. (2020). Mapping the network of
financial linkages: An industry-level analysis of the UK. Journal of Banking
\& Finance, 118, 105946.

\item Lasry, J.-M., Lions, P.-L., \& Gu\'{e}ant, O. (2010a). Application of
mean field games to growth theory. hal-00348376.

\item Lasry, J.-M., Lions, P.-L., \& Gu\'{e}ant, O. (2010b). Mean field
games and applications. In Paris-Princeton lectures on Mathematical Finance.
Springer. hal-01393103.

\item Lux, T. (2008). Applications of statistical physics in finance and
economics. Kiel Institute for the World Economy (IfW), Kiel.

\item Lux, T. (2016). Applications of statistical physics methods in
economics: Current state and perspectives. European Physical Journal Special
Topics, 225, 3255. https://doi.org/10.1140/epjst/e2016-60101-x

\item Mandel, A., Jaeger, C., F\"{u}rst, S., Lass, W., Lincke, D., Meissner,
F., Pablo-Marti, F., \& Wolf, S. (2010). Agent-based dynamics in
disaggregated growth models. Documents de travail du Centre d'Economie de la
Sorbonne, Paris.

\item Mandel, A. (2012). Agent-based dynamics in the general equilibrium
model. Complexity Economics, 1, 105--121.

\item Monacelli, T., Quadrini, V., \& Trigari, A. (2011). Financial markets
and unemployment. NBER Working Papers, 17389. National Bureau of Economic
Research.

\item Reinhart, C. M., \& Rogoff, K. S. (2009). This time is different:
Eight centuries of financial folly. Princeton University Press.

\item Sims, C. A. (2006). Rational inattention: Beyond the linear quadratic
case. American Economic Review, 96(2), 158--163.

\item Yang, J. (2018). Information theoretic approaches to economics.
Journal of Economic Surveys, 32(3), 940--960.
\end{description}

\pagebreak

\section*{Appendix 1}

We present a summary of the models developped in Gosselin and Lotz (2024,
2025,a,b). Some technical details are given in the subsequent Appendices.

\subsection*{A1.1 Fields and action functionals}

Let us first gather all the fields involved in the collective states in
Gosselin and Lotz (2025a,b)\footnote{%
Some details about the microfundations and their fields translations are
given in Appendix 2.}. They are the field of investors and banks, and the
field of their connections, the stakes. We have described the fields of
investors $\hat{\Psi}\left( K,X\right) $ and the field of banks $\bar{\Psi}%
\left( K,X\right) $ in Gosselin and Lotz (2024).\ Their argument are their
respective sector and level of capital and sector.

The dynamics of the level of capital is determined by the structure of
stakes between sectors, along with investors and banks' respective shares of
firms' returns. This dynamics is encompassed in the fields action functionals%
\footnote{%
A third field $\Psi \left( K,X\right) $ describing the firms was considered
in Gosselin and Lotz (2024). In Gosselin and Lotz (2025) and in this work
the relevant firms quantities were directly integrated in the equations for
investors and banks}. Gosselin and Lotz (2025a,b) endogeneized their
respective stakes by two fields, $\Gamma $ et $\bar{\Gamma}$. In the
following, we present a complete description of these fields and their
action functionals, for banks, investors and stakes.

\subsubsection*{A1.1.1 Banks}

The field $\bar{\Psi}\ $describes the banking system.\ It depends on the two
variables, the level of disposable capital $\bar{K}$ and the sector $X$. Its
action functional is given by:%
\begin{equation}
\bar{S}\left( \bar{\Psi}\right) =\int \left( -\bar{\Psi}^{\dag }\left( \bar{K%
},X\right) \nabla ^{2}\bar{\Psi}\left( \bar{K},X\right) +\left( \frac{\bar{g}%
^{2}\left( \bar{K},X\right) }{2\sigma _{\bar{K}}^{2}}+\frac{\bar{g}\left( 
\bar{K},X\right) }{2\bar{K}}\right) \left\vert \bar{\Psi}\left( \bar{K}%
,X\right) \right\vert ^{2}\right) d\left( \bar{K},X\right)  \label{BFC}
\end{equation}%
where:%
\begin{equation*}
\bar{g}\left( \hat{K},X\right) =\left( 1-\bar{M}\left\vert \bar{\Psi}\left( 
\bar{K},X\right) \right\vert ^{2}\right) ^{-1}\bar{f}\left( \bar{K},X\right)
\end{equation*}%
where the function $\bar{f}\left( \bar{K},X\right) $ is the return of a bank 
$X$ with disposable capital $\bar{K}$, and the function $\bar{g}\left( \hat{K%
},X\right) $ is an induced return that accounts for the propagation of the
return among agents, the diffusion being realized\footnote{%
See appendix 2.} through the matrix $M$.

The action (\ref{BFC}) describes the accumulation of capital $\bar{K}$ by
banks in a sector $X$. The squared field amplitude $\left\vert \bar{\Psi}%
\left( \bar{K},X\right) \right\vert ^{2}$\ measures the density of agents
holding capital $\bar{K}$ in sector $X$. By determining the fields $\bar{\Psi%
}\left( \bar{K},X\right) $\ that minimize (\ref{BFC}), one obtains the most
probable distributions of capital among banks across the different sectors,
as well as the average level of accumulated capital in each sector.

\subsubsection*{A1.1.2 Investors}

The field $\hat{\Psi}\ $describes the network of investors.\ It depends on
the level of disposable capital $\hat{K}$ and the sector $X$.\ Its action
functional is given by:%
\begin{equation}
\hat{S}\left( \hat{\Psi}\right) =\int \left( -\hat{\Psi}^{\dag }\left( \hat{K%
},X\right) \nabla ^{2}\hat{\Psi}\left( \hat{K},X\right) +\left( \frac{\hat{g}%
^{2}\left( \hat{K},X\right) }{2\sigma _{\hat{K}}^{2}}+\frac{\hat{g}\left( 
\hat{K},X\right) }{2\hat{K}}\right) \left\vert \hat{\Psi}\left( \hat{K}%
,X\right) \right\vert ^{2}\right) d\left( \hat{K},X\right)  \label{BFD}
\end{equation}%
where we define the induced return $\hat{g}$ as:%
\begin{equation*}
\hat{g}\left( \hat{K},X\right) =\left( 1-\hat{M}\left\vert \hat{\Psi}\left( 
\hat{K},X\right) \right\vert ^{2}\right) ^{-1}\hat{f}\left( \hat{K},X\right)
+\left( 1-\hat{M}\right) ^{-1}N\left( 1-\bar{M}\right) ^{-1}\bar{f}\left( 
\bar{K},X\right)
\end{equation*}%
This formula describes how the returns of investors, $\hat{f}\left( \hat{K}%
,X\right) $, and of banks $\bar{f}\left( \bar{K},X\right) $ are diffused to
other investors through the matrices $\bar{M}$, $\hat{M}$ and $N$\footnote{%
The field definition of these matrices $\bar{M}$, $\hat{M}$ and $N$ are
given in Appendix 2.}. Both formula (\ref{BFC}) and (\ref{BFD}) depend on
the network of connections between agents, that is, on their respective
stakes. The minimization of this action functional will be constrained by
the return equation for investors.

The action (\ref{BFD}) describes the accumulation of capital $\hat{K}$ by
investors in a sector $X$. The squared field amplitude $\left\vert \hat{\Psi}%
\left( \hat{K},X\right) \right\vert ^{2}$\ measures the density of agents
holding capital $\hat{K}$ in sector $X$. By determining the fields $\bar{\Psi%
}\left( \bar{K},X\right) $\ that minimize (\ref{BFD}), one obtains the most
probable distributions of capital among investors across the different
sectors, as well as the average level of accumulated capital in each sector.

\subsubsection*{A1.1.3 Firms}

Similarly the action functional for firms%
\begin{equation}
-\Psi ^{\dag }\left( K,X\right) \left( \nabla _{K_{p}}\left( \sigma
_{K}^{2}\nabla _{K_{p}}-f_{1}^{\prime }\left( K,X\right) K_{p}\right)
\right) \Psi \left( K,X\right) +\frac{1}{2\epsilon }\left( \left\vert \Psi
\left( K,X\right) \right\vert ^{2}-\left\vert \Psi _{0}\left( X\right)
\right\vert ^{2}\right) ^{2}  \label{Sft}
\end{equation}%
where:%
\begin{eqnarray*}
f_{1}^{\prime }\left( X\right) &=&\left( 1+\underline{k}_{2}\left( X\right)
\right) f_{1}\left( X\right) -\bar{r}\underline{k}_{2}\left( X\right) \\
&=&f_{1}\left( X\right) +\left( f_{1}\left( X\right) -\bar{r}\right) \int
k_{2}\left( X,X_{1}\right) \hat{K}_{1}\left\vert \hat{\Psi}\left( \hat{K}%
_{1},X_{1}\right) \right\vert ^{2}d\hat{K}_{1}dX_{1}
\end{eqnarray*}%
computes the firms returns, net from loans repaiements

\subsection*{A1.2 Stakes}

The connections between investors and banks have been endogeneized in
Gosselin and Lotz (2025a,b).\ They are described by two fields, one for the
stakes owned by investors, the other for the stakes owned by banks, whose
variables are the banks and investors' stakes respectively, along with the
positions in the sector space.

\subsubsection{A1.2.1 Investors}

\paragraph*{The field of investors' stakes}

The field of investors writes: 
\begin{equation*}
\Gamma \left( \hat{S}^{\left( T\right) },X^{\prime },X\right) \equiv \Gamma
\left( S_{1},\hat{S}_{1},S_{2},\hat{S}_{2},X^{\prime },X\right)
\end{equation*}%
where the four possible types of stakes are gathered in a vector $\hat{S}%
^{\left( T\right) }$:%
\begin{equation*}
\hat{S}^{\left( T\right) }=\left( S_{1},\hat{S}_{1},S_{2},\hat{S}_{2}\right)
\end{equation*}%
in which the arguments $S_{1}$ and $S_{2}$\ are the shares and loans taken
by investor $X$ in his sector firms, respectively, while $\hat{S}_{1}$ and $%
\hat{S}_{2}$ are those taken in an investor $X^{\prime }$. This field
describes the possible repatitions of investors investments' across sectors.

\paragraph*{The action functional}

The action functionals $S\left( \Gamma \right) $ for the field $\Gamma
\left( \hat{S}^{\left( T\right) },X^{\prime },X\right) $ is\footnote{%
Derived in Gosselin and Lotz (2025a,b).}:%
\begin{eqnarray}
S\left( \Gamma \right) &=&-\int \sigma _{\hat{K}}^{2}\Gamma ^{\dag }\left( 
\hat{S}^{\left( T\right) },X^{\prime },X\right) \nabla _{\hat{S}_{\eta
}^{\left( T\right) }}^{2}\Gamma \left( \hat{S}^{\left( T\right) },X^{\prime
},X\right) d\left( \hat{S}^{\left( T\right) },X^{\prime },X\right)
\label{GCt} \\
&&-\int \beta \left\vert \Gamma \left( \hat{S}^{\left( T\right) },X^{\prime
},X\right) \right\vert ^{2}d\left( \hat{S}^{\left( T\right) },X^{\prime
},X\right)  \notag \\
&&+\sum_{\eta }\int \left( \frac{\left( \hat{S}_{\eta }^{\left( T\right)
}\right) ^{2}}{2\hat{w}_{\eta }^{T}\left( X^{\prime },X\right) }-\hat{V}%
_{\eta }\hat{S}_{\eta }^{\left( T\right) }-\beta \right) \left\vert \Gamma
\left( \hat{S}^{\left( T\right) },X^{\prime },X\right) \right\vert
^{2}d\left( \hat{S}^{\left( T\right) },X^{\prime },X\right)  \notag \\
&&+\int \lambda \left( X\right) \left( \sum_{\eta }\int \hat{S}_{\eta
}^{\left( T\right) }\left\vert \Gamma \left( \hat{S}^{\left( T\right)
},X^{\prime },X\right) \right\vert ^{2}dX^{\prime }d\hat{S}^{\left( T\right)
}-1\right) \left\vert \Gamma \left( \hat{S}^{\left( T\right) },X^{\prime
},X\right) \right\vert ^{2}d\left( \hat{S}^{\left( T\right) },X^{\prime
},X\right)  \notag
\end{eqnarray}%
where the components of the vector $\left( \hat{w}_{\eta }^{T}\left(
X^{\prime },X\right) \right) $ are the field-translated (inverse)
uncertainty coefficients $\hat{w}_{\eta ij}$ and $w_{\eta ik}$:%
\begin{equation*}
\left( \hat{w}_{\eta }^{T}\left( X^{\prime },X\right) \right) =\left( \hat{w}%
_{E}\left( X^{\prime },X\right) ,\hat{w}_{L}\left( X^{\prime },X\right)
,w_{E}\left( X,X\right) ,w_{L}\left( X,X\right) \right)
\end{equation*}%
which reflect the uncertainty perceived by investors $X$ about their stakes
(either equity or loans) in investors $X^{\prime }$\ and in firms $X$. The
functions $\hat{V}_{\eta }\left( \hat{S}_{\eta }\right) $ involved in (\ref%
{GCt}) are defined in Appendix 2.\ 

The Lagrange multipliers $\lambda \left( X\right) $ implement the constraint:%
\begin{equation*}
\sum_{\eta }\int \hat{S}_{\eta }^{\left( T\right) }\left\vert \Gamma \left( 
\hat{S}^{\left( T\right) },X^{\prime },X\right) \right\vert ^{2}dX^{\prime }d%
\hat{S}^{\left( T\right) }=1
\end{equation*}%
where the integral accounts for the sum of stakes of an investor $X$
investing in various sectors $X^{\prime }$, and the factor $\left\vert
\Gamma \left( \hat{S}^{\left( T\right) },X^{\prime },X\right) \right\vert
^{2}$ weighs the stakes $\hat{S}_{\eta }^{\left( T\right) }$ by the number
of investors $X^{\prime }$ benefiting from these stakes.

We define the partial averages:%
\begin{eqnarray*}
S_{\eta }\left( X,X\right) &=&\int S_{\eta }\left\vert \Gamma \left( S_{E},%
\hat{S}_{E},S_{L},\hat{S}_{L},X^{\prime },X\right) \right\vert ^{2}d\left(
S_{E},\hat{S}_{E},S_{L},\hat{S}_{L},X^{\prime }\right) \\
\hat{S}_{\eta }\left( X^{\prime },X\right) &=&\int \hat{S}_{\eta }\left\vert
\Gamma \left( S_{E},\hat{S}_{E},S_{L},\hat{S}_{L},X^{\prime },X\right)
\right\vert ^{2}d\left( S_{E},\hat{S}_{E},S_{L},\hat{S}_{L}\right)
\end{eqnarray*}%
so that the constraint on allocation writes\footnote{%
Appendix 2 in Gosselin and Lotz (2025) derives the minimization equation of (%
\ref{GCt}) in terms of the sectoral averages:%
\begin{equation*}
\hat{S}_{\eta }^{\left( T\right) }\left( \hat{X}^{\prime },\hat{X}\right) =%
\frac{\int \hat{S}_{\eta }^{\left( T\right) }\left\vert \Gamma \left( \hat{S}%
^{\left( T\right) },\hat{X}^{\prime },\hat{X}\right) \right\vert ^{2}d\hat{S}%
^{\left( T\right) }}{\int \left\vert \Gamma \left( \hat{S}^{\left( T\right)
},\hat{X}^{\prime },\hat{X}\right) \right\vert ^{2}d\hat{S}^{\left( T\right)
}}
\end{equation*}%
}:%
\begin{equation}
\int \left( \hat{S}_{E}\left( X^{\prime },X\right) +\hat{S}_{L}\left(
X^{\prime },X\right) \right) dX^{\prime }+\int \left( S_{E}\left( X^{\prime
},X\right) +S_{L}\left( X^{\prime },X\right) \right) dX^{\prime }=1
\label{CSt}
\end{equation}

\subsubsection*{A1.2.2 Banks}

\paragraph*{The field of banks' stakes}

The field of banks is defined as:%
\begin{equation*}
\bar{\Gamma}\left( \bar{S}^{\left( T\right) },X^{\prime },X\right) =\bar{%
\Gamma}\left( \bar{S}_{E},S_{E}^{B},\hat{S}_{E}^{B},\bar{S}_{L},S_{L}^{B},%
\hat{S}_{L}^{B},X^{\prime },X\right)
\end{equation*}%
with:%
\begin{equation*}
\bar{S}^{\left( T\right) }=\left( \bar{S}_{E},S_{E}^{B},\hat{S}_{E}^{B},\bar{%
S}_{L},S_{L}^{B},\hat{S}_{L}^{B}\right)
\end{equation*}%
The arguments $S_{E}^{B}$ and $S_{L}^{B}$\ are the stakes taken by a bank $X$
in firms of the same sector, through shares and loans respectively, while $%
\hat{S}_{E}^{B}$ and $S_{L}^{B}$ are those taken in an investor $X^{\prime }$%
, and $\bar{S}_{E}$ and $\bar{S}_{L}$ are the stakes taken in a bank $%
X^{\prime }$. As for investors, this field describes the possible
repartitions of banks investments' across sectors.

\paragraph*{The action functional}

The action functional $S\left( \bar{\Gamma}\right) $ for the field $\bar{%
\Gamma}\left( \bar{S}^{\left( T\right) },X^{\prime },X\right) $ of
banks'stakes is defined as\footnote{%
See Appendix 1.}:

\begin{eqnarray}
S\left( \bar{\Gamma}\right) &=&-\sigma _{\hat{K}}^{2}\sum_{\eta }\int \bar{%
\Gamma}^{\dag }\left( \bar{S}^{\left( T\right) },X^{\prime },X\right) \nabla
_{\bar{S}_{\eta }^{\left( T\right) }}^{2}\bar{\Gamma}\left( \bar{S}^{\left(
T\right) },X^{\prime },X\right) d\left( \bar{S}^{\left( T\right) },X^{\prime
},X\right)  \label{GT} \\
&&+\int \left( \sum_{\eta }\left( \frac{\left( \bar{S}_{\eta }^{\left(
T\right) }\right) ^{2}}{2\bar{w}_{\eta }\left( X^{\prime },X\right) }-\bar{V}%
_{\eta }\bar{S}_{\eta }^{\left( T\right) }\right) -\beta \right) \left\vert 
\bar{\Gamma}\left( \bar{S}^{\left( T\right) },X^{\prime },X\right)
\right\vert ^{2}d\left( \bar{S}^{\left( T\right) },X^{\prime },X\right) 
\notag \\
&&+\int \lambda \left( X\right) \left( \sum_{\eta =E,L}\int \bar{S}_{\eta
}^{B}\left( X^{\prime },X\right) dX^{\prime }+\int \hat{S}_{E}^{B}\left(
X^{\prime },X\right) dX^{\prime }+S_{E}^{B}\left( X,X\right) -1\right)
\left\vert \bar{\Gamma}\left( \bar{S}^{\left( T\right) },X^{\prime
},X\right) \right\vert ^{2}  \notag \\
&&+\int \lambda ^{\prime }\left( X\right) \left( \int \hat{S}_{L}^{B}\left(
X^{\prime },X\right) dX^{\prime }+S_{L}^{B}\left( X,X\right) -\kappa \left(
1-\bar{S}_{\eta }^{B}\left( X\right) \right) \right) \left\vert \bar{\Gamma}%
\left( \bar{S}^{\left( T\right) },X^{\prime },X\right) \right\vert ^{2} 
\notag
\end{eqnarray}%
where we defined the various averages in the states defined by the fields:%
\begin{eqnarray*}
\bar{S}_{\eta }\left( X^{\prime },X\right) &=&\int \bar{S}_{\eta }\left\vert 
\bar{\Gamma}\right\vert ^{2}d\left( \bar{S}^{\left( T\right) },X^{\prime
}\right) \\
S_{\eta }^{B}\left( X,X\right) &=&\int S_{\eta }^{B}\left\vert \bar{\Gamma}%
\right\vert ^{2}d\left( \bar{S}^{\left( T\right) },X^{\prime }\right) \\
\hat{S}_{\eta }^{B}\left( X^{\prime },X\right) &=&\int \hat{S}_{\eta
}^{B}\left\vert \bar{\Gamma}\right\vert ^{2}d\left( \bar{S}^{\left( T\right)
},X^{\prime }\right)
\end{eqnarray*}%
and the coefficients:%
\begin{equation*}
\bar{w}_{E}^{T}\left( X^{\prime },X\right) =\bar{w}_{E}\left( X^{\prime
},X\right) \text{, }\bar{w}_{L}^{T}\left( X^{\prime },X\right) =\bar{w}%
_{L}\left( X^{\prime },X\right)
\end{equation*}%
\begin{equation*}
\hat{w}_{3}^{T}\left( X^{\prime },X\right) =\hat{w}_{E}\left( X^{\prime
},X\right) \text{, }\bar{w}_{4}^{T}\left( X^{\prime },X\right) =\hat{w}%
_{L}\left( X^{\prime },X\right)
\end{equation*}%
\begin{equation*}
\hat{w}_{5}^{T}\left( X^{\prime },X\right) =w_{E}\left( X,X\right) \text{, }%
\bar{w}_{6}^{T}\left( X^{\prime },X\right) =w_{L}\left( X,X\right)
\end{equation*}%
are the field translation of the inter- and intra-sectoral uncertainty
coefficients.

The Lagrange multipliers $\lambda \left( X\right) $ and $\lambda ^{\prime
}\left( X\right) $ implement the constraints on allocations:%
\begin{equation}
\sum_{\eta }\int \bar{S}_{\eta }^{B}\left( X^{\prime },X\right) dX^{\prime
}+\int \hat{S}_{E}^{B}\left( X^{\prime },X\right) dX^{\prime
}+S_{E}^{B}\left( X,X\right) =1  \label{CSr}
\end{equation}%
\begin{equation*}
\int \hat{S}_{L}^{B}\left( X^{\prime },X\right) dX^{\prime }+S_{L}^{B}\left(
X,X\right) =\kappa \left( 1-\bar{S}_{\eta }^{B}\left( X\right) \right)
\end{equation*}%
\begin{equation*}
\bar{S}^{\left( T\right) }\left( X^{\prime },X\right) \left\vert \Gamma
\left( X^{\prime },X\right) \right\vert ^{2}=\int \bar{S}^{\left( T\right)
}\left\vert \Gamma \left( \bar{S}^{\left( T\right) },X^{\prime },X\right)
\right\vert ^{2}
\end{equation*}%
The functions $\bar{V}_{\eta }\left( \bar{S}\right) $, involved in the
expression of $S\left( \bar{\Gamma}\right) $ are defined in Appendix 2.

\subsection*{Appendix 1.3 Minimization equations of action functionals}

The collective states of the system are defined by minimization equations of
each field of the system. We first consider the minimization equations for (%
\ref{BFC}) and (\ref{BFD}) leading to the field return equations for bank
investors and banks\footnote{%
See Appendix 2 for some details.}, and then the minimization equations for
fields of stakes, leading to find the stakes defining the collective state.

\subsubsection*{A.1.3.1 Firms, investors and banks}

\paragraph*{A.1.3.1.1 Field equations}

The minimization for investors and banks fields are obtained by minimizing:%
\begin{equation*}
\bar{S}\left( \bar{\Psi}\right) +\hat{S}\left( \hat{\Psi}\right)
\end{equation*}%
leading to the two equations:%
\begin{equation*}
\frac{\delta }{\delta \bar{\Psi}\left( \bar{K},X\right) }\left( \bar{S}%
\left( \bar{\Psi}\right) +\hat{S}\left( \hat{\Psi}\right) \right) =0
\end{equation*}%
for banks, and: 
\begin{equation*}
\frac{\delta }{\delta \hat{\Psi}\left( \hat{K},X\right) }\left( \bar{S}%
\left( \bar{\Psi}\right) +\hat{S}\left( \hat{\Psi}\right) \right) =0
\end{equation*}%
for investors. These equations write:%
\begin{align}
0& =\frac{\hat{K}_{1}^{2}\hat{g}^{2}\left( \hat{K}_{1},X_{1}\right) }{%
2\sigma _{\hat{K}}^{2}}+\frac{\hat{g}\left( \hat{K}_{1},X_{1}\right) }{2}
\label{SPn} \\
& +\int \left\vert \hat{\Psi}\left( \hat{K},X\right) \right\vert ^{2}\left( 
\frac{\hat{K}^{2}\hat{g}\left( \hat{K},X,\Psi ,\hat{\Psi}\right) }{\sigma _{%
\hat{K}}^{2}}+\frac{1}{2}\right) \frac{\delta \hat{g}\left( \hat{K},X\right) 
}{\delta \left\vert \hat{\Psi}\left( \hat{K}_{1},X_{1}\right) \right\vert
^{2}}+\frac{1}{\hat{\mu}}\left( \left\vert \hat{\Psi}\left( \hat{K},X\right)
\right\vert ^{2}-\left\vert \hat{\Psi}_{0}\left( X\right) \right\vert
^{2}\right)  \notag
\end{align}%
\begin{eqnarray}
0 &=&\left( \frac{\bar{K}_{1}^{2}\bar{g}^{2}\left( \bar{K}_{1},X_{1}\right) 
}{\sigma _{\hat{K}}^{2}}+\frac{\bar{g}\left( \bar{K}_{1},X_{1}\right) }{2}%
\right)  \label{SPt} \\
&&+\int \left\vert \hat{\Psi}\left( \hat{K},X\right) \right\vert ^{2}\left( 
\frac{\hat{K}^{2}\hat{g}\left( \hat{K},X\right) }{\sigma _{\hat{K}}^{2}}+%
\frac{1}{2}\right) \frac{\delta \hat{g}\left( \hat{K},X\right) }{\delta
\left\vert \bar{\Psi}\left( \bar{K}_{1},X_{1}\right) \right\vert ^{2}}+\frac{%
1}{\hat{\mu}}\left( \left\vert \bar{\Psi}\left( \bar{K}_{1},X_{1}\right)
\right\vert ^{2}-\left\vert \bar{\Psi}_{0}\left( X_{1}\right) \right\vert
^{2}\right)  \notag
\end{eqnarray}%
This yields the backgrounds fields $\bar{\Psi}$ and $\hat{\Psi}$ under the
constraint by the return equations:

These non-local equations link the distribution of capital of investors and
banks in one sector, measured by $\left\vert \hat{\Psi}\left( \hat{K}%
,X\right) \right\vert ^{2}$ and $\left\vert \bar{\Psi}\left( \bar{K}%
_{1},X_{1}\right) \right\vert ^{2}$ respectively, to the returns of provided
by the various sectors, through cross investment between banks and investors.

These equations have to be constrained by the return equations for investors
and banks. These equations translate that investors and banks returns are
constrained across sectors due to agents mutual investments (see Appendix 2
for details).

Similarly, the minimization equation equations, for (\ref{Sft}) reduce to:%
\begin{equation}
0=\left( \frac{f_{1}^{\prime 2}\left( K,X\right) }{\sigma _{\hat{K}}^{2}}+%
\frac{\frac{d}{dK}f_{1}^{\prime }\left( K,X\right) }{2}\right) +\frac{1}{%
\epsilon }\left( \left\vert \Psi \left( K,X\right) \right\vert
^{2}-\left\vert \Psi _{0}\left( X\right) \right\vert ^{2}\right)  \label{fbd}
\end{equation}

\paragraph*{A.1.3.1.2 The return equations}

The return equations for banks and investors constrain the previous
minimization equations by linking returns of investors and banks through
mutual participations and depend on the return of firms in the various
sectors\footnote{%
More details are given in Gosselin and Lotz (2025a,b). Appendix 2 presents
these returns equations as constraints.}.

Assuming\footnote{%
As in Gosselin and Lotz (2024).} investors and banks invest in neighbouring
firms, we can write:%
\begin{eqnarray*}
S_{E}\left( X^{\prime },X\right) &=&S_{E}\left( X,X\right) \delta \left(
X^{\prime }-X\right) \\
S_{L}\left( X^{\prime },X\right) &=&S_{L}\left( X,X\right) \delta \left(
X^{\prime }-X\right)
\end{eqnarray*}%
\begin{eqnarray*}
S_{E}^{B}\left( X^{\prime },X\right) &=&S_{E}^{B}\left( X,X\right) \delta
\left( X^{\prime }-X\right) \\
S_{L}^{B}\left( X^{\prime },X\right) &=&S_{L}^{B}\left( X,X\right) \delta
\left( X^{\prime }-X\right)
\end{eqnarray*}%
where $\delta \left( X^{\prime }-X\right) $\ is the Dirac function. It is
equal to $0$\ for $X^{\prime }\neq X$. Under these assumptions, the
constraint (\ref{CSt}) simplifies to:%
\begin{equation*}
\int \left( \hat{S}_{E}\left( X^{\prime },X\right) +\hat{S}_{L}\left(
X^{\prime },X\right) \right) dX^{\prime }+S_{E}\left( X,X\right)
+S_{L}\left( X,X\right) =1
\end{equation*}%
and the constraint (\ref{CSr}) simplifies as:%
\begin{equation*}
\int \left( \bar{S}_{E}\left( X^{\prime },X\right) +\bar{S}_{L}\left(
X^{\prime },X\right) \right) dX^{\prime }+\int \hat{S}_{E}^{B}\left(
X^{\prime },X\right) dX^{\prime }+S_{E}^{B}\left( X,X\right) =1
\end{equation*}%
We showed in Gosselin and Lotz (2025a,b) that the return equation for
investors writes\footnote{%
See Appendix 1 for details. The case of no default, or default are studied
in Appendices 1.2, 1.3.}:%
\begin{equation}
0=\int \left( \delta \left( X^{\prime }-X\right) -\hat{S}_{E}\left(
X^{\prime },X\right) \right) \widehat{DF}\left( X^{\prime }\right) \hat{R}%
_{exc}\left( X^{\prime }\right) dX^{\prime }-S_{E}\left( X,X\right)
R_{exc}\left( X\right)  \label{Rtn}
\end{equation}%
where $\widehat{DF}\left( X\right) $ is the investor's discount factor for
debt repayment and writes\footnote{%
In the following, we will also encounter the firm's discount factor for debt
repayment: 
\begin{equation*}
DF\left( X\right) =\frac{1-S\left( X\right) }{1-S_{E}\left( X\right) }
\end{equation*}%
}:%
\begin{equation*}
\widehat{DF}\left( X\right) =\frac{1-\hat{S}\left( X\right) }{1-\hat{S}%
_{E}\left( X\right) }
\end{equation*}%
and:%
\begin{equation*}
\hat{R}_{exc}\left( X^{\prime }\right) =f\left( X^{\prime }\right) -\bar{r}
\end{equation*}%
\begin{equation*}
R_{exc}\left( X\right) =f\left( X\right) -\bar{r}
\end{equation*}%
are the investors and firms excess returns respectively.

Similarly, the banks return equation without default for banks writes:%
\begin{eqnarray}
0 &=&\int \left( \delta \left( X^{\prime }-X\right) -\bar{S}_{E}\left(
X^{\prime },X\right) \right) \overline{DF}\left( X^{\prime }\right) \bar{R}%
_{exc}\left( X^{\prime }\right)  \label{RTb} \\
&&-\int \hat{S}_{E}^{B}\left( X^{\prime },X\right) \widehat{DF}\left(
X^{\prime }\right) \hat{R}_{exc}\left( X^{\prime }\right) -S_{E}^{B}\left(
X^{\prime },X\right) R_{exc}\left( X\right)  \notag
\end{eqnarray}%
where $\overline{DF}\left( X\right) $ is the bank's discount factor for debt
repayment and writes:%
\begin{equation*}
\overline{DF}\left( X\right) =\frac{1-\bar{S}\left( X\right) }{1-\bar{S}%
_{E}\left( X\right) }
\end{equation*}%
The excess returns $\bar{R}_{exc}\left( X^{\prime }\right) $ of banks is
given by:%
\begin{equation*}
\bar{R}_{exc}\left( X^{\prime }\right) =\bar{f}\left( X^{\prime }\right)
-\left( 1+\kappa \right) \bar{r}
\end{equation*}%
The full equations with default\footnote{%
See Appendix 1.} will be presented later in the text. Note that, in the
following, we will also encounter the firm's discount factor for debt
repayment: 
\begin{equation*}
DF\left( X\right) =\frac{1-\left( S\left( X\right) +S^{B}\left( X\right)
\right) }{1-S_{E}\left( X\right) -S_{E}^{B}\left( X\right) }
\end{equation*}

\subsubsection*{A.1.3.2 Stakes field equations}

The collective states correspond to the solution of the minimization
equations for the stakes' investors field $\Gamma \left( \hat{S}^{\left(
T\right) },X^{\prime },X\right) $ and $\bar{\Gamma}\left( \bar{S}^{\left(
T\right) },X^{\prime },X\right) $:%
\begin{eqnarray*}
\frac{\delta S\left( \Gamma \right) }{\delta \Gamma \left( \hat{S}^{\left(
T\right) },X^{\prime },X\right) } &=&0 \\
\frac{\delta S\left( \bar{\Gamma}\right) }{\delta \bar{\Gamma}\left( \bar{S}%
^{\left( T\right) },X^{\prime },X\right) } &=&0
\end{eqnarray*}%
which write:%
\begin{eqnarray}
0 &=&\left( \sum_{\eta }\left( -\sigma _{\hat{K}}^{2}\nabla _{\hat{S}_{\eta
}^{\left( T\right) }}^{2}+\frac{\left( \hat{S}_{\eta }^{\left( T\right)
}\right) ^{2}}{2\hat{w}_{\eta }}-\hat{V}_{\eta }\hat{S}_{\eta }^{\left(
T\right) }+\lambda \left( X\right) \left\Vert \Gamma \left( \hat{S}^{\left(
T\right) },X^{\prime },X\right) \right\Vert _{X}^{2}\hat{S}_{\eta }^{\left(
T\right) }\right) -\beta \right)  \label{MZn} \\
&&\Gamma \left( \hat{S}^{\left( T\right) },X^{\prime },X\right)  \notag
\end{eqnarray}%
for investors stakes fields, with $\lambda \left( X\right) $ a Lagrange
multiplier implementing the constraint (\ref{CSt}) on stakes, and:%
\begin{eqnarray}
0 &=&-\sigma _{\hat{K}}^{2}\sum_{\eta }\nabla _{\bar{S}_{\eta }^{\left(
T\right) }}^{2}\bar{\Gamma}\left( \bar{S}^{\left( T\right) },X^{\prime
},X\right)  \label{MZt} \\
&&+\left( \sum_{\eta }\left( \frac{\left( \bar{S}_{\eta }^{\left( T\right)
}\right) ^{2}}{2\bar{w}_{\eta }}-\bar{V}_{\eta }^{\left( T\right) }\bar{S}%
_{\eta }^{T}+\lambda _{\eta }\left( X\right) \bar{S}_{\eta }^{T}\right)
-\beta \right) \bar{\Gamma}\left( \bar{S}^{\left( T\right) },X^{\prime
},X\right)  \notag
\end{eqnarray}%
for banks field of stakes $\bar{\Gamma}\left( \bar{S}^{\left( T\right)
},X^{\prime },X\right) $, where:%
\begin{eqnarray*}
\lambda _{\eta }\left( X\right) &=&\lambda \left( X\right) \text{ for }\eta
=1,2,3,5 \\
\lambda _{\eta }\left( X\right) &=&\lambda ^{\prime }\left( X\right) \text{
for }\eta =5,6
\end{eqnarray*}%
are Lagrange multiplier implementing the constraints (\ref{CSt}) and (\ref%
{CSr}).

\subsubsection*{A1.3.3 Equations in terms of sectoral stakes}

We show in Gosselin and Lotz (2025a,b) that the minimization equations for
the field of stakes $\Gamma \left( \hat{S}^{\left( T\right) },X^{\prime
},X\right) $ and $\bar{\Gamma}\left( \bar{S}^{\left( T\right) },X^{\prime
},X\right) $ can be rewritten in terms of sectoral stakes.

Actually, in first approximation, the solutions of (\ref{MZn}) take the form:%
\begin{equation}
\Gamma _{0,X^{\prime },X}\left( \hat{S}^{\left( T\right) }\right) =N\exp
\left( -\sum_{\eta }\frac{\left( \hat{S}_{\eta }^{\left( T\right) }-\hat{S}%
_{\eta }^{\left( T\right) }\left( X^{\prime },X\right) \right) ^{2}}{2\sigma
_{\hat{K}}^{2}}\right)  \label{gmz}
\end{equation}%
where $N$\ is a normalization factor. Solving the minimization equations
thus reduces to finding the stakes invested by sector $X$\ into sector $%
X^{\prime }$, i.e. the sectoral averages $\hat{S}_{\eta }^{\left( T\right)
}\left( X^{\prime },X\right) $.\ They satisfy\footnote{%
See Appendix 2 in Gosselin and Lotz (2025).}:%
\begin{equation}
\hat{S}_{\eta }^{\left( T\right) }\left( X^{\prime },X\right) =\hat{w}_{\eta
}\left( X^{\prime },X\right) \left( \hat{V}_{\eta }\left( X^{\prime
},X\right) +\lambda \left( X\right) \right)  \label{mzn}
\end{equation}%
and depend on the Lagrange multipliers\footnote{%
See Appendix 2 in Gosselin and Lotz (2025).}.

The minimization equation for banks is similar to the investors equation,
and the solutions have the form\footnote{%
See Appendix 4.}:%
\begin{equation*}
\Gamma _{0,X^{\prime },X}\left( \bar{S}^{\left( T\right) }\right) \Gamma
\left( X^{\prime },X\right)
\end{equation*}%
where:%
\begin{equation}
\Gamma _{0,X^{\prime },X}\left( \bar{S}^{\left( T\right) }\right) =\bar{N}%
\exp \left( -\sum_{\eta }\frac{\left( \bar{S}_{\eta }^{\left( T\right) }-%
\bar{S}_{\eta }^{\left( T\right) }\left( X^{\prime },X\right) \right) ^{2}}{%
2\sigma _{\hat{K}}^{2}}\right)  \label{gmb}
\end{equation}%
where $\bar{N}$\ is a normalization factor. Solving the minimization
equations thus reduces to finding the stakes invested by sector $X$\ into
sector $X^{\prime }$, i.e.the sectoral averages $\bar{S}_{\eta }^{\left(
T\right) }\left( X^{\prime },X\right) $. We show in appendix 16 of Gosselin
and Lotz (2025b) that $\bar{S}_{\eta }^{\left( T\right) }\left( X^{\prime
},X\right) $ satisfies:%
\begin{equation}
\bar{S}_{\eta }^{\left( T\right) }\left( X^{\prime },X\right) =\bar{w}_{\eta
}\left( X^{\prime },X\right) \left( \bar{V}_{\eta }\left( X^{\prime
},X\right) +\lambda _{\eta }\left( X\right) \right)  \label{mzt}
\end{equation}%
and depend on the Lagrange multipliers $\lambda _{\eta }\left( X\right) $.

\subsubsection*{A1.3.4 The equations for uncertainty}

Uncertainty is measured by the coefficients $\hat{w}$, $\bar{w}$, which are
endogenous. These coefficients themselves depend on the stakes; we therefore
simply recall their functional form.

\paragraph*{A.1.3.4.1 Investors}

The equation that defines the weight of inverse uncertainty for
cross-investment between investors, derived in Gosselin and Lotz (2025a,b),
is:%
\begin{equation}
\hat{w}\left( X^{\prime },X\right) =\frac{2\left( 1-\left( \gamma
\left\langle \hat{S}_{E}\left( X\right) \right\rangle \right) ^{2}\right) 
\hat{w}_{E}^{\left( 0\right) }\left( X^{\prime },X\right) }{1+\hat{w}%
_{E}^{\left( 0\right) }\left( X^{\prime },X\right) \left( 1-\left( \gamma
\left\langle \hat{S}_{E}\left( X\right) \right\rangle \right) ^{2}\right)
+\Delta \left( \gamma \left\langle \hat{S}_{E}\left( X_{1},X^{\prime
}\right) \right\rangle _{X_{1}}\right) ^{2}}  \label{hb}
\end{equation}%
The coefficient $\hat{w}_{E}^{\left( 0\right) }\left( X^{\prime },X\right) $%
\ is a factor of local inverse uncertainty: it is a factor of confidence of
investors $X$\ in investors $X^{\prime }$. The higher this factor, the
higher the investments in sector $X^{\prime }$. As such, this coefficient
captures the characteristics of local uncertainty that induce deviations
from average behavior. The parameter $\gamma $ is the average uncertainty of
the distance-dependent investment paths and writes: 
\begin{equation*}
\gamma ^{2}\simeq \left( \hat{w}_{E}^{\left( 0\right) }\left( \left(
X^{\prime }\right) ^{\prime },X_{m-1}\right) ...\hat{w}_{E}^{\left( 0\right)
}\left( X_{1},X^{\prime }\right) \right) ^{-\frac{1}{m}}
\end{equation*}

The weight of inverse uncertainty for investments in firms is given by:%
\begin{equation}
w\left( X,X\right) =1-\left\langle w\left( X^{\prime },X\right)
\right\rangle _{X^{\prime }}  \label{NCH}
\end{equation}%
The contribution:\textbf{\ }%
\begin{equation*}
\Delta \left( \gamma \left\langle \hat{S}_{E}\left( X_{1},X^{\prime }\right)
\right\rangle _{X_{1}}\right) ^{2}=\left( \gamma \left\langle \hat{S}%
_{E}\left( X_{1},X^{\prime }\right) \right\rangle _{X_{1}}\right)
^{2}-\left( \gamma \left\langle \hat{S}_{E}\left( X\right) \right\rangle
\right) ^{2}
\end{equation*}%
in equation (\ref{hb}) represents the gap between the risk perception of
investor $X^{\prime }$\ and that of the market. It increases with the risk
perception of investor $X^{\prime }$, so that the higher this gap, the lower
the coefficient $\hat{w}\left( X^{\prime },X\right) $, and the lower the
investment of investor $X^{\prime }$ in investor $X$.

\paragraph{A.1.3.4.2 Banks}

Similarly the weights for inverse uncertainty of banks are:%
\begin{equation}
\left( \bar{w}\left( X^{\prime },X\right) \right) ^{-1}=1+\frac{1}{2}\left( 
\frac{\overline{IRG}\left( X^{\prime },X\right) }{\widehat{IR^{B}}\left(
X^{\prime },X\right) }+\frac{\overline{IRG}\left( X^{\prime },X\right) }{\xi
^{2}}\right)  \label{Cfn}
\end{equation}%
and those for banks investing investors are given by:%
\begin{equation}
\left( \hat{w}_{E}^{B}\left( X^{\prime },X\right) \right) ^{-1}=1+2\frac{%
\widehat{IR^{B}}\left( X^{\prime },X\right) }{\overline{IRG}\left( X^{\prime
},X\right) }+\frac{\widehat{IR^{B}}\left( X^{\prime },X\right) }{\xi ^{2}}
\label{Cft}
\end{equation}%
\begin{equation}
w_{E}^{B}\left( X^{\prime },X\right) =1-\bar{w}\left( X^{\prime },X\right) -%
\hat{w}_{E}^{B}\left( X^{\prime },X\right)  \label{chv}
\end{equation}%
where the various quantities are defined in Appendix 3.

Since the functions $\hat{w}$ and $w$\ both depend on the stakes, their
expression complete the minimization equations and yield $\hat{S}_{E}\left(
X^{\prime },X\right) $, $\hat{S}\left( X^{\prime },X\right) $, $S_{E}\left(
X,X\right) $, $S\left( X,X\right) $.

\subsection*{Appendix 1.4 Resolution and Collective states: no-default
scenario}

\subsubsection*{A1.4.1 Principle of resolution}

The collective states are defined by the solutions of (\ref{SPn}), (\ref{SPt}%
), (\ref{Rtn}), (\ref{RTb}), (\ref{mzn}), (\ref{mzt}).

The resolution is in made of two blocks. First, the collective states for
investors and banks are defined by the solutions of (\ref{SPn}), (\ref{SPt}%
). The resolution defines also what we will call fiber states. The fields $%
\hat{\Psi}\left( \hat{K},X\right) $ and $\bar{\Psi}\left( \bar{K},X\right) $
will describe the distribution of capital in each sector, but will be
dependent on the distribution of stakes and returns across the sector space.
The formula were derived in Gosselin and Lotz (2024).

Once the formulas for $\hat{\Psi}\left( \hat{K},X\right) $ and $\bar{\Psi}%
\left( \bar{K},X\right) $ are obtained, the second block of equations is
given by (\ref{Rtn}), (\ref{RTb}), (\ref{mzn}), (\ref{mzt}). This set of
equations, is solved in several steps.

The return equations in sector $X$, equations (\ref{Rtn}), (\ref{RTb}),
connect all investors in the sector space, and as such comprises three types
of variables: the inward aggregate stakes on investors and banks in sector $%
X $; the returns of investors $X$, banks $X$ and firms $X$, taken
independently; and the average returns of all investors of the sector space.

The resolution follows a four-step process:

1. We average the return equation over all sectors.\ This yields the
averages of all the variables of the equation: average stakes, levels of
capital and returns.

2. We then replace these results in the return equation, and further replace
the aggregate stakes in terms of returns, so that the return equation
depends only on returns.

3. Solving this equation yields the returns per sector and, in turn, the
aggregate stakes and the disposable capital per sector.

4. Using equations (\ref{mzn}), (\ref{mzt}), we reconstruct the remaining
quantities of the model.

\subsubsection*{A1.4.2 Form of the solution: groups and sub-collective states%
}

To solve the full system, we must determine the levels of capital and
returns of each type of agents for each sector of the sector space $S$,
along with the distribution of stakes across all these sectors. Taken
together, these quantities define a collective state of the system.

Since capital and returns can be derived from the level of stakes between
investors, a collective state is described by a set of values of stakes for
investors: 
\begin{equation*}
\left\{ \hat{S}_{E}\left( X^{\prime },X\right) ,\hat{S}\left( X^{\prime
},X\right) ,S_{E}\left( X,X\right) ,S\left( X,X\right) \right\} _{\left(
X^{\prime },X\right) \in S}
\end{equation*}%
and for banks:%
\begin{equation*}
\left\{ \hat{S}_{E}^{B}\left( X^{\prime },X\right) ,\hat{S}^{B}\left(
X^{\prime },X\right) ,S_{E}^{B}\left( X,X\right) ,S^{B}\left( X,X\right) ,%
\bar{S}_{E}\left( X^{\prime },X\right) ,\bar{S}^{B}\left( X^{\prime
},X\right) \right\} _{\left( X^{\prime },X\right) \in S}
\end{equation*}%
These quantities are governed by equations (\ref{Rtn}) and (\ref{RTb}).\
These are non-local equations that involve the network of connections within
each group.

We have seen that, under uncertainty, agents connect to only a finite number
of neighbors, so that agents are organized into several and
loosely-connected groups. Each of these groups $G$ can be in one among
several possible phase, each defined by a distribution of stakes between the
agents of $G$:%
\begin{eqnarray*}
&&\left\{ \hat{S}_{E}\left( X^{\prime },X\right) ,\hat{S}\left( X^{\prime
},X\right) ,S_{E}\left( X,X\right) ,S\left( X,X\right) ,\right. \\
&&\left. \hat{S}_{E}^{B}\left( X^{\prime },X\right) ,\hat{S}^{B}\left(
X^{\prime },X\right) ,S_{E}^{B}\left( X,X\right) ,S^{B}\left( X,X\right) ,%
\bar{S}_{E}\left( X^{\prime },X\right) ,\bar{S}^{B}\left( X^{\prime
},X\right) \right\} _{\left( X^{\prime },X\right) \in G}
\end{eqnarray*}%
We will call a \emph{sub-collective state} the combination of one group $G$
and one of its possible phases. Thus, collective states organize into
several distinct, weakly-interacting sub-collective states, so that, in
first approximation, a collective state can be seen as a set of
sub-collective states, that writes:%
\begin{eqnarray*}
\cup _{\alpha } &&\left\{ \hat{S}_{E}\left( X^{\prime },X\right) ,\hat{S}%
\left( X^{\prime },X\right) ,S_{E}\left( X,X\right) ,S\left( X,X\right)
,\right. \\
&&\left. \hat{S}_{E}^{B}\left( X^{\prime },X\right) ,\hat{S}^{B}\left(
X^{\prime },X\right) ,S_{E}^{B}\left( X,X\right) ,S^{B}\left( X,X\right) ,%
\bar{S}_{E}\left( X^{\prime },X\right) ,\bar{S}^{B}\left( X^{\prime
},X\right) \right\} _{\left( X^{\prime },X\right) \in G}
\end{eqnarray*}%
where the collection of groups $\left\{ G\right\} $ describe a certain
organization of the sector space $S$.

The organization of the sector space into groups $\left\{ G\right\} $ is not
unique: there are an infinite number of such organizations. Moreover, for
each organization, multiple distinct phases are possible for each group: the
uncertainty combined with the interconnectivity of sectors, leads to an
infinite number of possible solutions, among which, some correspond to
default states\footnote{%
See Gosselin and Lotz (2024).}. Thus, there are an infinite number of
possible collective states, and each collective state is one among an
infinite number of possible configurations of the full system, as composed
of multiple groups, each of them in a possible phase.

Any group associated with the phase it is in is called a \emph{%
sub-collective state}.\ For any group, there may be several sub-collective
states. The following results are to be considered for each group.

\subsubsection*{A1.4.3 Caracteristics of collective states}

We present the results without providing all explicit formulas, referring
the reader instead to the appendices or to Gosselin and Lotz (2025a,b). The
purpose is to show that a sub-collective state consists of a collection of
stakes associated with the elements of the group, which describe the basis
variables of the collective state, whereas the other quantities---returns,
capital, and so forth---are functions depending on this basis variables and
characterizing the fiber variables. The elements of the fiber are fixed by
the basis in the collective state under consideration, but they will become
dynamical variables again when we study fluctuations of the collective
states.

\paragraph*{A1.4.3.1 Averages}

For a given sub-collective state, the averages depend on firms' productivity
and on uncertainty. For a given group, the averages of the stakes are
denoted:%
\begin{eqnarray*}
&&\left\{ \left\langle \hat{S}_{E}\left( X^{\prime },X\right) \right\rangle
,\left\langle \hat{S}\left( X^{\prime },X\right) \right\rangle ,\left\langle
S_{E}\left( X,X\right) \right\rangle ,\left\langle S\left( X,X\right)
\right\rangle ,\right. \\
&&\left. \left\langle \hat{S}_{E}^{B}\left( X^{\prime },X\right)
\right\rangle ,\left\langle \hat{S}^{B}\left( X^{\prime },X\right)
\right\rangle ,\left\langle S_{E}^{B}\left( X,X\right) \right\rangle
,\left\langle S^{B}\left( X,X\right) \right\rangle ,\left\langle \bar{S}%
_{E}\left( X^{\prime },X\right) \right\rangle ,\left\langle \bar{S}%
^{B}\left( X^{\prime },X\right) \right\rangle \right\} _{\left( X^{\prime
},X\right) \in G}
\end{eqnarray*}%
A sub collective state, i.e. the distribution: 
\begin{eqnarray*}
&&\cup _{G}\left\{ \hat{S}_{E}\left( X^{\prime },X\right) ,\hat{S}\left(
X^{\prime },X\right) ,S_{E}\left( X,X\right) ,S\left( X,X\right) ,\right. \\
&&\left. \hat{S}_{E}^{B}\left( X^{\prime },X\right) ,\hat{S}^{B}\left(
X^{\prime },X\right) ,S_{E}^{B}\left( X,X\right) ,S^{B}\left( X,X\right) ,%
\bar{S}_{E}\left( X^{\prime },X\right) ,\bar{S}^{B}\left( X^{\prime
},X\right) \right\} _{\left( X^{\prime },X\right) \in G}
\end{eqnarray*}%
is organized around these averages.

On average, the levels of shares and loans depend on uncertainty and on the
firms' average returns within the group. These averages make it possible to
determine the average returns and the average levels of available capital $%
\left\langle \bar{K}\right\rangle \left\Vert \bar{\Psi}\right\Vert ^{2}$ and 
$\left\langle \hat{K}\right\rangle \left\Vert \hat{\Psi}\right\Vert ^{2}$
for the group, as given in Appendix 3. Once these averages are known, one
can determine the levels of returns and stakes by sector, that is, for any
sub-collective state.

\paragraph{A1.4.3.2 Investors and banks' returns and stakes}

Once the averages of stakes and returns of investors and banks have been
obtained, the values of returns and stakes can be derived. They are obtained
by solving the return equations and, once the sectoral returns are
determined, by substituting them into (\ref{mzn}) and (\ref{mzt}) . The
return equations generally yield two possible solutions per sector and per
agent type:%
\begin{equation*}
\hat{R}_{exc}^{L}\left( X\right) ,\hat{R}_{exc}^{H}\left( X\right)
\end{equation*}%
\begin{equation*}
\bar{R}_{exc}^{L}\left( X\right) ,\bar{R}_{exc}^{H}\left( X\right)
\end{equation*}%
This corresponds, for a given average, to a multiplicity of distribution of
stakes, high or low:

\begin{eqnarray}
\cup _{\alpha } &&\left\{ \hat{S}_{E}^{H/L}\left( X^{\prime },X\right) ,\hat{%
S}^{H/L}\left( X^{\prime },X\right) ,S_{E}^{H/L}\left( X,X\right)
,S^{H/L}\left( X,X\right) ,\right.  \label{ps} \\
&&\left. \hat{S}_{E}^{BH/L}\left( X^{\prime },X\right) ,\hat{S}^{BH/L}\left(
X^{\prime },X\right) ,S_{E}^{BH/L}\left( X,X\right) ,S^{B}H/L\left(
X,X\right) ,\bar{S}_{E}^{H/L}\left( X^{\prime },X\right) ,\bar{S}%
^{BH/L}\left( X^{\prime },X\right) \right\} _{\left( X^{\prime },X\right)
\in G_{\alpha }}  \notag
\end{eqnarray}%
for a given group.

Given that there are multiple types of possible returns per sector, there
exists an infinite number of collective states possible for a group, on
average. The values for the relevant variables are provided in Appendix 3.

Each collective state is specified by a possible collection (\ref{ps}) and
such collection forms the basis of the collective state; returns, disposable
capital per sector, and other such quantities, are functions of these data
and constitute the fiber of the collective state.

\subsubsection*{A1.4.4 Background fields and disposable capital for
investors and banks}

\paragraph*{\protect\bigskip A1.4.4.1 Background fields}

The backgrounds fields $\bar{\Psi}$ and $\hat{\Psi}$ are found in terms of
returns in Gosselin and Lotz (2024). The solutions of (\ref{SPn}) and (\ref%
{SPt}) for the background fields are:%
\begin{eqnarray*}
\left\vert \hat{\Psi}\left( \hat{K}_{1},X_{1}\right) \right\vert ^{2}
&=&\left\Vert \hat{\Psi}_{0}\left( X_{1}\right) \right\Vert ^{2}-\hat{\mu}%
\left\{ \left( \frac{\hat{K}_{1}^{2}\hat{g}^{2}\left( X_{1}\right) }{2\sigma
_{\hat{K}}^{2}}+\frac{\hat{g}\left( X_{1}\right) }{2}\right) \right. \\
&&\left. +\left( \frac{\left\langle \bar{K}\right\rangle ^{2}\left\langle 
\bar{g}\right\rangle }{\sigma _{\hat{K}}^{2}}+\frac{1}{2}\right) \Delta
\left( \hat{k}^{B}\left( X_{1},\left\langle X\right\rangle \right) A\right) 
\frac{\left\Vert \bar{\Psi}\right\Vert ^{2}\hat{K}_{1}}{\left\Vert \hat{\Psi}%
\right\Vert ^{2}\left\langle \hat{K}\right\rangle }-\left( \frac{%
\left\langle \hat{K}\right\rangle ^{2}\left\langle \hat{g}\right\rangle }{%
\sigma _{\hat{K}}^{2}}+\frac{1}{2}\right) \left\langle \hat{g}%
^{ef}\right\rangle \frac{\hat{K}_{1}}{\left\langle \hat{K}\right\rangle }%
\right\}
\end{eqnarray*}%
\begin{eqnarray*}
\left\vert \bar{\Psi}\left( \bar{K}_{1},X_{1}\right) \right\vert ^{2}
&=&\left\vert \bar{\Psi}_{0}\left( X_{1}\right) \right\vert ^{2}-\hat{\mu}%
\left\{ \left( \frac{\bar{K}_{1}^{2}\bar{g}^{2}\left( X_{1}\right) }{\sigma
_{\hat{K}}^{2}}+\frac{\bar{g}\left( X_{1}\right) }{2}\right) \right. \\
&&+\left( \frac{\left\langle \bar{K}\right\rangle ^{2}\left\langle \bar{g}%
\right\rangle }{\sigma _{\hat{K}}^{2}}+\frac{1}{2}\right) \left(
\left\langle \underline{\hat{k}}^{B}\right\rangle ^{ef}\Delta \left( \hat{k}%
^{B}\left( X_{1},\left\langle X\right\rangle \right) A\right) +\frac{\Delta 
\bar{k}_{2}\left( \left\langle X\right\rangle ,X\right) }{\left(
1-\left\langle \bar{k}_{1}\right\rangle \right) \left\Vert \bar{\Psi}%
\right\Vert ^{2}\left\langle \bar{K}\right\rangle }\left\langle \bar{g}%
\right\rangle \right) \frac{\left\Vert \bar{\Psi}\right\Vert ^{2}\bar{K}_{1}%
}{\left\Vert \hat{\Psi}\right\Vert ^{2}\left\langle \hat{K}\right\rangle } \\
&&\left. -\left( \frac{\left\langle \hat{K}\right\rangle ^{2}\left\langle 
\hat{g}\right\rangle }{\sigma _{\hat{K}}^{2}}+\frac{1}{2}\right)
\left\langle \hat{g}^{Bef}\right\rangle \frac{\bar{K}_{1}}{\left\langle \hat{%
K}\right\rangle }\right\}
\end{eqnarray*}%
where, for any quantity $Q$, the notation $\Delta \left( Q\left( X\right)
\right) $ stands for its deviation from the average: 
\begin{equation*}
\Delta \left( Q\left( X\right) \right) =\left( Q\left( X\right) \right)
-\left\langle \Delta \left( Q\left( X\right) \right) \right\rangle
\end{equation*}

The background field for firms is obtained similarly as solution (\ref{fbd}):%
\begin{equation}
\left\vert \Psi \left( K,X\right) \right\vert ^{2}=\left\vert \Psi
_{0}\left( X\right) \right\vert ^{2}-\epsilon \left( \frac{\left(
f_{1}^{\left( e\right) }\left( X\right) K-\bar{C}\left( X\right) \right) ^{2}%
}{\sigma _{\hat{K}}^{2}}+\frac{f_{1}^{\left( e\right) }\left( X\right) }{2}%
\right)
\end{equation}

More importantly for the description of the collective state are the average
over capital of the fields by sector. These are obtained for given levels of
stakes characteristic of a collective state. We obtain the following formula
for the average sectoral fields:%
\begin{equation}
\left\vert \hat{\Psi}_{0}\left( X\right) \right\vert ^{2}\simeq \frac{%
\left\Vert \hat{\Psi}\right\Vert ^{2}\left( I_{X/\left\langle X^{\prime
}\right\rangle }\right) ^{\frac{3}{2}}}{\left( 1-\bar{S}\right) ^{\frac{3}{2}%
}\left( \frac{\hat{f}\left( X\right) }{\left\langle \bar{f}\right\rangle }+%
\frac{\left\langle \hat{S}\left( X^{\prime },X\right) \right\rangle
_{X^{\prime }}}{1-\left\langle \hat{S}\right\rangle }+\frac{\left(
\left\langle \hat{S}_{1}^{B}\left( X,X^{\prime }\right) \right\rangle
_{X^{\prime }}+\left\langle \hat{S}_{2}^{B}\left( X,X^{\prime }\right)
\right\rangle _{X^{\prime }}+\frac{\left\langle \hat{S}_{1}^{B}\right\rangle
+\left\langle \hat{S}_{2}^{B}\right\rangle }{1-\left\langle \hat{S}%
\right\rangle }\right) \frac{\left\langle \bar{K}\right\rangle \left\Vert 
\bar{\Psi}\right\Vert ^{2}}{\left\langle \hat{K}\right\rangle \left\Vert 
\hat{\Psi}\right\Vert ^{2}}\left\langle \bar{f}\right\rangle }{\left(
1-\left\langle \bar{S}\right\rangle \right) \left\langle \bar{f}%
\right\rangle }\right) ^{\frac{3}{2}}}  \label{FPn}
\end{equation}%
\begin{equation}
\left\vert \bar{\Psi}_{0}\left( X\right) \right\vert ^{2}\simeq \left( \frac{%
\left\langle \bar{f}\right\rangle \frac{\left\langle \bar{S}\left( X^{\prime
},X\right) \right\rangle _{X}}{1-\bar{S}\left( X^{\prime },X\right) }}{%
\left( \left( 1-\left\langle \bar{S}\right\rangle \right) \bar{f}\left(
X\right) +\left\langle \bar{S}\left( X^{\prime },X\right) \right\rangle
_{X^{\prime }}\left\langle \bar{f}\right\rangle \right) \frac{\left\langle 
\bar{S}\left( X^{\prime },X\right) \right\rangle }{1-\left\langle \bar{S}%
\left( X^{\prime },X\right) \right\rangle }}\right) ^{\frac{3}{2}}\left\Vert 
\bar{\Psi}\right\Vert ^{2}  \label{FPd}
\end{equation}%
where:%
\begin{equation}
I_{X/\left\langle X^{\prime }\right\rangle }=\frac{\frac{\left\langle \hat{S}%
\left( X,X^{\prime }\right) \right\rangle _{X^{\prime }}}{1-\left\langle 
\hat{S}\left( X,X^{\prime }\right) \right\rangle _{X^{\prime }}}}{\frac{%
\left\langle \hat{S}\left( X,X^{\prime }\right) \right\rangle }{%
1-\left\langle \hat{S}\left( X,X^{\prime }\right) \right\rangle }}
\label{Idef1}
\end{equation}%
These solutions depends on the stakes and their average\footnote{%
See Gosselin and Lotz (2025a,b)} as well as the average fields\footnote{%
Computed in Gosselin and Lotz (2025a,b).} $\left\Vert \hat{\Psi}\right\Vert
^{2}$ and $\left\Vert \bar{\Psi}\right\Vert ^{2}$.

The results depend on the stakes and their average over the sector-space and
are replaced in the return equations.

\paragraph*{A1.4.4.2 Disposable capital}

For a given distribution of stakes and returns, and using the average
disposable capital $\left\langle \bar{K}\right\rangle \left\Vert \bar{\Psi}%
\right\Vert ^{2}$ and $\left\langle \hat{K}\right\rangle \left\Vert \hat{\Psi%
}\right\Vert ^{2}$ defined in Gosselin and Lotz (2025b), the disposable
capital per sector writes%
\begin{equation}
\bar{K}_{X}\left\vert \bar{\Psi}\left( X\right) \right\vert ^{2}\simeq
\left\langle \bar{K}\right\rangle \left\Vert \bar{\Psi}\right\Vert
^{2}\left( \frac{\left\langle \bar{f}\right\rangle \frac{\left\langle \bar{S}%
\left( X^{\prime },X\right) \right\rangle _{X}}{1-\bar{S}\left( X^{\prime
},X\right) }}{\left( \left( 1-\left\langle \bar{S}\right\rangle \right) \bar{%
f}\left( X\right) +\left\langle \bar{S}\left( X^{\prime },X\right)
\right\rangle _{X^{\prime }}\left\langle \bar{f}\right\rangle \right) \frac{%
\left\langle \bar{S}\left( X^{\prime },X\right) \right\rangle }{%
1-\left\langle \bar{S}\left( X^{\prime },X\right) \right\rangle }}\right)
^{2}  \label{Dc}
\end{equation}%
\begin{equation}
\hat{K}_{X}\left\vert \hat{\Psi}\left( X\right) \right\vert ^{2}\simeq \frac{%
\left\langle \hat{K}\right\rangle \left\Vert \hat{\Psi}\right\Vert
^{2}\left( \frac{\left\langle \hat{S}\left( X^{\prime },X\right)
\right\rangle _{X}}{1-\left\langle \hat{S}\left( X^{\prime },X\right)
\right\rangle _{X}}\frac{1-\left\langle \hat{S}\left( X^{\prime },X\right)
\right\rangle }{\left\langle \hat{S}\left( X^{\prime },X\right)
\right\rangle }\left\langle \bar{f}\right\rangle \right) ^{2}}{\left( 1-\bar{%
S}\right) ^{2}\left( \hat{f}\left( X\right) +\frac{\left\langle \hat{S}%
\left( X^{\prime },X\right) \right\rangle _{X^{\prime }}\left\langle \hat{f}%
\right\rangle }{1-\left\langle \hat{S}\right\rangle }+\frac{\left(
\left\langle \hat{S}_{1}^{B}\left( X,X^{\prime }\right) \right\rangle
_{X^{\prime }}+\left\langle \hat{S}_{2}^{B}\left( X,X^{\prime }\right)
\right\rangle _{X^{\prime }}+\frac{\left\langle \hat{S}_{1}^{B}\right\rangle
+\left\langle \hat{S}_{2}^{B}\right\rangle }{1-\left\langle \hat{S}%
\right\rangle }\right) \frac{\left\langle \bar{K}\right\rangle \left\Vert 
\bar{\Psi}\right\Vert ^{2}}{\left\langle \hat{K}\right\rangle \left\Vert 
\hat{\Psi}\right\Vert ^{2}}\left\langle \bar{f}\right\rangle }{%
1-\left\langle \bar{S}\right\rangle }\right) ^{2}}  \label{Dcv}
\end{equation}

Here too, results depend on the value of stakes, justifying the denomination
of fiber states.

\subsection*{A1.5 Resolution and collective states: default scenario}

Defaults states can be defined as modifications of non-default states.
Assuming one or several defaulting firms initially in one sector, the
default state is built recursively from this initial impact.\ Once the
initial default has propagated, the returns in each sector will be
deviations from what would have been the returns in a non-default scenario.

We will detail below the propagation mechanism and the condition for the
default to materialize and spread. The loss incured by investors and the
fraction of defaulting investors will then be computed.

\subsubsection*{A1.5.1 Principle of propagation}

The default states for investors have been described in Gosselin and Lotz
(2025). The principle of propagation is similar if we include banks. The
default states are found recursively. We first assume some realized initial
default states $S_{0}$, $\hat{S}_{0}$ and $\bar{S}_{0}$, when a set of
firms, investors and banks\ experience a complete loss of their private
capital. In terms of returns this corresponds to:%
\begin{eqnarray}
\hat{f}\left( X^{\prime }\right) &=&-1  \label{Cdf} \\
\bar{f}\left( X^{\prime }\right) &=&-1  \notag \\
f_{1}^{\prime }\left( X\right) &=&-1  \notag
\end{eqnarray}%
for investors, banks and firms in $\hat{S}_{0}$, $\bar{S}_{0}$ and $S_{0}$
respectively. The number $-1$ stands for a return of $-100$ percent, so that
the private capital for defaulting firms, banks and investors are $0$.

The possible default sets are then obtained from this initial set by the
limit of a sequence of equations. We assume that after $n$ iterations, the
sets of defaulting agents are $S_{n}$, $\hat{S}_{n}$ and $\bar{S}_{n}$. We
define the remaining sets of agents as $S/\hat{S}_{n}$, $S/\bar{S}_{n}$ and $%
S/S_{n}$.

Then these sets are included in the return equations for investors and banks:

\paragraph*{A1.5.1.1 Investors}

\begin{eqnarray}
0 &=&\int \left( \delta \left( X-X^{\prime }\right) -\hat{S}_{E}\left(
X^{\prime },X\right) \right) \widehat{DF}\left( X^{\prime }\right) \left( 
\hat{f}_{n+1}\left( X^{\prime }\right) -\bar{r}\right) dX^{\prime }
\label{dfn} \\
&&+\int_{\hat{S}_{n}}\frac{1-\left( \hat{S}\left( X^{\prime }\right) +\hat{S}%
_{E}^{B}\left( X^{\prime }\right) +\hat{S}_{L}^{B}\left( X^{\prime }\right)
\right) }{\hat{S}_{L}\left( X^{\prime }\right) }\hat{S}_{L}\left( X^{\prime
},X\right) dX^{\prime }  \notag \\
&&+\int_{S_{n}}\frac{1-\left( S\left( X^{\prime }\right) +S_{E}^{B}\left(
X^{\prime }\right) +S_{L}^{B}\left( X^{\prime }\right) \right) }{S_{L}\left(
X^{\prime }\right) }S_{L}\left( X^{\prime },X\right) dX^{\prime }-\int
S_{E}\left( X^{\prime },X\right) \left( \left( f\left( X^{\prime }\right) -%
\bar{r}\right) \right) dX^{\prime }  \notag
\end{eqnarray}

\paragraph*{A1.5.1.2 Banks}

\begin{eqnarray}
0 &=&\left( \delta \left( X-X^{\prime }\right) -\bar{S}_{E}\left( X^{\prime
},X\right) \right) \left( \bar{f}_{n+1}\left( X^{\prime }\right) -\bar{r}%
\right) \overline{DF}\left( X^{\prime }\right) -\widehat{DF}\left( X^{\prime
}\right) \hat{S}_{E}^{B}\left( X^{\prime },X\right) \left( \hat{f}\left(
X^{\prime }\right) -\bar{r}\right) \\
&&+\int_{\bar{S}_{n}}\bar{S}_{L}\left( X^{\prime },X\right) \frac{\left( 1-%
\bar{S}\left( X^{\prime }\right) \right) }{\bar{S}_{L}\left( X^{\prime
}\right) }+\int_{\hat{S}_{n}}\hat{S}_{L}^{B}\left( X^{\prime },X\right) 
\frac{1-\left( S\left( X^{\prime }\right) +S_{E}^{B}\left( X^{\prime
}\right) +S_{L}^{B}\left( X^{\prime }\right) \right) }{S_{L}\left( X^{\prime
}\right) +S_{L}^{B}\left( X^{\prime }\right) }  \notag \\
&&+\int_{S_{n}}S_{L}^{B}\left( X^{\prime },X\right) \frac{1-\left( \hat{S}%
\left( X^{\prime },X\right) +\hat{S}_{E}^{B}\left( X^{\prime }\right) +\hat{S%
}_{L}^{B}\left( X^{\prime }\right) \right) }{\hat{S}_{L}\left( X^{\prime
}\right) +\hat{S}_{L}^{B}\left( X^{\prime }\right) }-S_{E}^{B}\left(
X^{\prime },X\right) \left( f\left( X^{\prime }\right) -\bar{r}\right) 
\notag
\end{eqnarray}%
These equations define new sets of returns $\hat{f}_{n+1}\left( X^{\prime
}\right) $ and $\bar{f}_{n+1}\left( X^{\prime }\right) $ which in turn may
define new default sets $S_{n+1}$, $\hat{S}_{n+1}$ and $\bar{S}_{n+1}$. Once
the iteration stabilizes, we find a remaining sets of - non-defaulted -
agents $\left( \hat{S}_{\infty },\bar{S}_{\infty },S_{\infty }\right) $,
defined by the limit $\left( S/\hat{S}_{n},\bar{S}/\bar{S}%
_{n},S/S_{n}\right) \underset{n\rightarrow \infty }{\rightarrow }\left( \hat{%
S}_{\infty },\bar{S}_{\infty },S_{\infty }\right) $ with returns $\left\{ 
\hat{f}_{n}\left( X\right) \rightarrow \hat{f}\left( X\right) ,\bar{f}%
_{n}\left( X\right) \rightarrow \bar{f}\left( X\right) \right\} $ for which
the resulting disposable capitals are given by (\ref{Dcn}), (\ref{Dcf}) and (%
\ref{Dct}).

\subsubsection*{A1.5.2 Conditions for propagation}

Firms default when they meet the conditions for initial defaults. But, for
this default to propagate to the whole set of investors and banks, some
additional conditions must be met. First, the defaut must propagate to the
firm's immediate investors or banks. Then, it must propagate from these
investors to other investors and banks. The case of investors default was
considered in Gosselin and Lotz (2025a) and we focus on the banks default
only\footnote{%
See Appendix 11.3.}.

\paragraph*{A1.5.2.1 Propagation from firms to banks and investors}

Some firms may lack the disposable capital to face their costs, pushing them
into default. Whether these initial defaults may push into default their
intra-sectoral banks depends on the magnitude of loss of the defaulting
firms.

For the default to spread to the firm's banks, the level of loss a firm has
to experience, by lack of capital or increase in costs, must be below a
negative threshold\footnote{%
See Appendix 8 in Gosselin and Lotz (2025b).}:%
\begin{equation*}
R\left( X\right) <D_{Th}
\end{equation*}%
where the threshold $D_{Th}$ is:%
\begin{eqnarray*}
D_{Th} &=&-\frac{1}{S_{E}^{B}\left( X,X\right) }\left( 1+\bar{r}%
+\left\langle \bar{S}_{E}\left( X^{\prime },X\right) \right\rangle
_{X^{\prime }}\overline{DF}\left( X^{\prime }\right) \left( \left\langle 
\bar{f}\left( X^{\prime }\right) \right\rangle -\bar{r}\right) \right.  \\
&&+\left. \left\langle \hat{S}_{E}^{B}\left( X^{\prime },X\right)
\right\rangle _{X^{\prime }}\widehat{DF}\left( X^{\prime }\right) \left(
\left\langle \hat{f}\left( X^{\prime }\right) \right\rangle -\bar{r}\right)
\right) 
\end{eqnarray*}%
This threshold is negative, and the default zone includes all returns that
are below it.

\paragraph*{A1.5.2.2 Propagation from investors and banks to other sectors}

Given an initial default in investors and banks in certain sectors, the
condition for default to propagate to investors is the same as in part 1 and
the condition under which the default may spread to other banks is:%
\begin{eqnarray}
&&\mathbf{H}\left( \left\langle \left( \hat{S}_{L}\left( X^{\prime
},X\right) \right) \right\rangle _{X^{\prime }}+S_{L}\left( X,X\right)
\right) \\
&&+\mathbf{G}\left( \left\langle \bar{S}_{L}\left( X^{\prime },X\right)
\right\rangle _{X^{\prime }}+\left\langle \hat{S}_{L}^{B}\left( X^{\prime
},X\right) \right\rangle _{X^{\prime }}+S_{L}^{B}\left( X,X\right) \right)
>2\left\langle \bar{f}\right\rangle  \notag
\end{eqnarray}%
where $\mathbf{H}$ and $\mathbf{G}$ are functions of stakes, uncertainty and
capital ratios under the non-default scenario. They are given by:

\begin{equation*}
\mathbf{H=}\frac{1-S_{1}^{B}\left( X,X\right) \frac{A_{1}}{B}dC}{1-\bar{S}%
\left( X^{\prime }\right) }+H\left( 1-S_{1}\left( X,X\right) \frac{A_{1}}{B}%
dC\right)
\end{equation*}%
\begin{equation*}
\mathbf{G=}\left( \frac{1-S_{1}^{B}\left( X,X\right) \frac{A_{2}}{B}dC}{1-%
\bar{S}\left( X^{\prime }\right) }\right)
\end{equation*}%
and:%
\begin{equation*}
dC=\left( C-\frac{r\left( f_{1}\left( X\right) +\tau \Delta F_{\tau }\left( 
\bar{R}\left( K,X\right) \right) \right) }{\left( 1-S\left( X\right) \right)
^{1-r}}\right)
\end{equation*}%
The term:%
\begin{equation*}
H=\frac{S_{1}^{B}\left( X,X\right) }{1-\bar{S}\left( X^{\prime }\right) }%
\frac{\hat{S}_{1}^{B}\left( X^{\prime },X\right) }{1-\left\langle \hat{S}%
\left( X^{\prime }\right) \right\rangle }\frac{1-\left( \hat{S}\left(
X^{\prime }\right) +\hat{S}_{1}^{B}\left( X^{\prime }\right) +\hat{S}%
_{2}^{B}\left( X^{\prime }\right) \right) }{1-\left( \hat{S}_{1}\left(
X^{\prime }\right) +\hat{S}_{1}^{B}\left( X^{\prime }\right) \right) }
\end{equation*}%
increases with participations $\hat{S}_{1}\left( X^{\prime }\right) +\hat{S}%
_{1}^{B}\left( X^{\prime }\right) $ and $\bar{S}\left( X^{\prime }\right) $
and $\left\langle \hat{S}\left( X^{\prime }\right) \right\rangle $, while it
decreases with loans $\hat{S}_{2}\left( X^{\prime }\right) +$ $\hat{S}%
_{2}^{B}\left( X^{\prime }\right) $.

The coefficients $A_{1}$ and $A_{2}$: 
\begin{equation*}
A_{1}=\frac{\frac{h\left( X\right) \hat{h}\left( X\right) }{2}\frac{\hat{K}%
_{X}\left\vert \hat{\Psi}\left( X\right) \right\vert ^{2}}{K_{X}\left\vert
\Psi \left( X\right) \right\vert ^{2}}+h_{1}^{B}\left( X\right) \left\langle 
\hat{h}_{1}^{B}\left( X\right) \right\rangle \frac{\bar{K}_{X}\left\vert 
\bar{\Psi}\left( X\right) \right\vert ^{2}}{K_{X}\left\vert \Psi \left(
X\right) \right\vert ^{2}}}{1-\left\langle \hat{S}\left( X\right)
\right\rangle }\text{, }A_{2}=h_{1}^{B}\left( X\right) \left\langle \bar{h}%
\left( X\right) \right\rangle _{\bar{S}}\frac{\bar{K}_{X}\left\vert \bar{\Psi%
}\left( X\right) \right\vert ^{2}}{K_{X}\left\vert \Psi \left( X\right)
\right\vert ^{2}}
\end{equation*}%
increase with $\left\langle \hat{S}\left( X\right) \right\rangle $, with
capital ratios $\frac{\hat{K}_{X}\left\vert \hat{\Psi}\left( X\right)
\right\vert ^{2}}{K_{X}\left\vert \Psi \left( X\right) \right\vert ^{2}}$ $%
\frac{\bar{K}_{X}\left\vert \bar{\Psi}\left( X\right) \right\vert ^{2}}{%
K_{X}\left\vert \Psi \left( X\right) \right\vert ^{2}}$ and, decrease with
uncertainty since $h_{1}^{B}\left( X\right) \left\langle \hat{h}%
_{1}^{B}\left( X\right) \right\rangle $ and $h_{1}^{B}\left( X\right)
\left\langle \bar{h}\left( X\right) \right\rangle _{\bar{S}}$ decrease with
uncertainty $\gamma $.

\begin{eqnarray*}
B &=&\left( 1-2\frac{S\left( X\right) +S^{B}\left( X\right) }{1-\left(
S\left( X\right) +S^{B}\left( X\right) \right) }\right) \\
&&\times \left( 1-dC\frac{\left( \frac{h\left( X\right) \hat{h}\left(
X\right) }{2}\left( 1-\frac{\left\langle S_{1}\left( X,X\right)
\right\rangle }{1-\left\langle \hat{S}\left( X\right) \right\rangle }\right) 
\frac{\hat{K}_{X}\left\vert \hat{\Psi}\left( X\right) \right\vert ^{2}}{%
K_{X}\left\vert \Psi \left( X\right) \right\vert ^{2}}+h_{1}^{B}\left(
X\right) \left\langle \bar{h}\left( X\right) \right\rangle _{f\left(
X\right) }\frac{\bar{K}_{X}\left\vert \bar{\Psi}\left( X\right) \right\vert
^{2}}{K_{X}\left\vert \Psi \left( X\right) \right\vert ^{2}}\right) }{1-2%
\frac{S\left( X\right) +S^{B}\left( X\right) }{1-\left( S\left( X\right)
+S^{B}\left( X\right) \right) }}\right)
\end{eqnarray*}%
$B$ increases with uncertainty $\gamma $. Then using that 
\begin{equation*}
\frac{\left\langle S_{1}\left( X,X\right) \right\rangle }{1-\left\langle 
\hat{S}\left( X\right) \right\rangle }=\frac{1-\left\langle \hat{S}\left(
X\right) \right\rangle -\left\langle S_{2}\left( X,X\right) \right\rangle }{%
1-\left\langle \hat{S}\left( X\right) \right\rangle }=1-\frac{\left\langle
S_{2}\left( X,X\right) \right\rangle }{1-\left\langle \hat{S}\left( X\right)
\right\rangle }
\end{equation*}%
decreases with $\left\langle \hat{S}\left( X\right) \right\rangle $, we
obtain that $B$ decreases with $\left\langle \hat{S}\left( X\right)
\right\rangle $. $B$ also decreases with capital ratios $\frac{\hat{K}%
_{X}\left\vert \hat{\Psi}\left( X\right) \right\vert ^{2}}{K_{X}\left\vert
\Psi \left( X\right) \right\vert ^{2}}$ $\frac{\bar{K}_{X}\left\vert \bar{%
\Psi}\left( X\right) \right\vert ^{2}}{K_{X}\left\vert \Psi \left( X\right)
\right\vert ^{2}}$.

As a consequence, $\mathbf{G}$ increases as function of $\bar{S}\left(
X^{\prime }\right) $, $\left\langle \hat{S}\left( X\right) \right\rangle $,
with capital ratios $\frac{\hat{K}_{X}\left\vert \hat{\Psi}\left( X\right)
\right\vert ^{2}}{K_{X}\left\vert \Psi \left( X\right) \right\vert ^{2}}$ $%
\frac{\bar{K}_{X}\left\vert \bar{\Psi}\left( X\right) \right\vert ^{2}}{%
K_{X}\left\vert \Psi \left( X\right) \right\vert ^{2}}$ \ and decrease as a
function of $S_{1}^{B}\left( X,X\right) $ and uncertaint $\gamma $.

Moreover, $\mathbf{H}$ increases as function of $\bar{S}\left( X^{\prime
}\right) $, $\left\langle \hat{S}\left( X\right) \right\rangle $, $\hat{S}%
_{1}\left( X^{\prime }\right) $, $\hat{S}_{1}^{B}\left( X^{\prime }\right) $%
, with capital ratios $\frac{\hat{K}_{X}\left\vert \hat{\Psi}\left( X\right)
\right\vert ^{2}}{K_{X}\left\vert \Psi \left( X\right) \right\vert ^{2}}$, $%
\frac{\bar{K}_{X}\left\vert \bar{\Psi}\left( X\right) \right\vert ^{2}}{%
K_{X}\left\vert \Psi \left( X\right) \right\vert ^{2}}$ while it decreases
with loans $\hat{S}_{2}\left( X^{\prime }\right) +$ $\hat{S}_{2}^{B}\left(
X^{\prime }\right) $\ and decrease as a function of $S_{1}^{B}\left(
X,X\right) $, $S_{1}\left( X,X\right) $ and uncertainty $\gamma $.

\subsubsection*{A1.5.3 Description of the default state}

Once the default has materialized, the system is in a default state. This
state can be analyzed as a deviation from a no-default state. To do so, we
will compute, for the default state, both the average loss and the fraction
of investors affected.

When defaults occur, investors' returns are shifted by:

\begin{equation}
\hat{f}\left( X\right) \rightarrow \hat{f}\left( X\right) -\mu d\hat{f}
\label{Sh}
\end{equation}%
where $d\hat{f}$ is the loss incured by the remaining investors for each
investor defaulting, and $\mu $ is the fraction of investors impacted by the
default.

We find that:%
\begin{equation*}
d\hat{f}\left( X\right) =\frac{\left( \left\langle \hat{S}_{L}\left(
X^{\prime },X\right) \right\rangle +\left\langle S_{L}\left( X,X\right)
\right\rangle \right) \mu +\left\langle S_{E}\left( X^{\prime }\right)
\right\rangle \left\langle df\left( X^{\prime }\right) \right\rangle }{%
1-\left\langle \hat{S}\left( X^{\prime }\right) \right\rangle }
\end{equation*}%
and that the fraction of defaulting investors $\mu $ is:%
\begin{equation}
\mu =\frac{1-\left\langle \hat{S}\left( X^{\prime }\right) \right\rangle }{%
\left( \left\langle \hat{S}_{L}\left( X^{\prime },X\right) \right\rangle
+\left\langle S_{L}\left( X,X\right) \right\rangle \right) +\left\langle
S_{E}\left( X^{\prime }\right) \right\rangle \left\langle df\left( X^{\prime
}\right) \right\rangle -2\left\langle \hat{f}\right\rangle \left(
1-\left\langle \hat{S}\left( X^{\prime }\right) \right\rangle \right) }
\end{equation}%
where all parameter values are those of a non-default scenario.

\bigskip When defaults occur, banks' returns are shifted by:

\begin{equation}
\bar{f}\left( X\right) \rightarrow \bar{f}\left( X\right) -\mu d\bar{f}
\end{equation}%
where $d\bar{f}$ is the loss incured by the remaining investors for each
investor defaulting, and $\mu $ is the fraction of investors impacted by the
default.

We find that:%
\begin{eqnarray*}
&&\frac{\left\langle d\bar{f}\right\rangle }{\mu }=-\left( \frac{%
S_{E}^{B}\left( X,X\right) }{1-\bar{S}\left( X^{\prime }\right) }%
+S_{E}\left( X,X\right) \frac{S_{E}^{B}\left( X,X\right) }{1-\bar{S}\left(
X^{\prime }\right) }\frac{\hat{S}_{E}^{B}\left( X^{\prime },X\right) }{%
1-\left\langle \hat{S}\left( X^{\prime }\right) \right\rangle }\widehat{DF}%
\left( X^{\prime }\right) \right) \frac{A}{B}dC \\
&&+\left( \frac{1}{1-\bar{S}\left( X^{\prime }\right) }+\frac{%
S_{E}^{B}\left( X,X\right) }{1-\bar{S}\left( X^{\prime }\right) }\frac{\hat{S%
}_{E}^{B}\left( X^{\prime },X\right) }{1-\left\langle \hat{S}\left(
X^{\prime }\right) \right\rangle }\widehat{DF}\left( X^{\prime }\right)
\right) \left( \left\langle \left( \hat{S}_{L}\left( X^{\prime },X\right)
\right) \right\rangle _{X^{\prime }}+S_{L}\left( X,X\right) \right) \\
&&+\frac{1}{1-\bar{S}\left( X^{\prime }\right) }\left( \left\langle \bar{S}%
_{L}\left( X^{\prime },X\right) \right\rangle _{X^{\prime }}+\left\langle 
\hat{S}_{L}^{B}\left( X^{\prime },X\right) \right\rangle _{X^{\prime
}}+S_{L}^{B}\left( X,X\right) \right)
\end{eqnarray*}%
and the fraction of defaulting investors $\mu $ is given by:%
\begin{equation}
\mu =\frac{1}{\frac{\left\langle d\bar{f}\right\rangle }{\mu }-2\left\langle 
\bar{f}\right\rangle }
\end{equation}%
where all parameter values are those of a non-default scenario.%
\begin{equation*}
\left\langle d\bar{f}\right\rangle =\frac{1}{\frac{\left\langle d\bar{f}%
\right\rangle }{\mu }-2\left\langle \bar{f}\right\rangle }\frac{\left\langle
d\bar{f}\right\rangle }{\mu }
\end{equation*}

The average loss due to default for the entire group of firms can also be
derived:%
\begin{equation}
\left\langle dR\left( X\right) \right\rangle =-\frac{A}{B}\mu dC
\end{equation}%
with:%
\begin{equation*}
A=A_{1}\left( \left\langle \left( \hat{S}_{L}\left( X^{\prime },X\right)
\right) \right\rangle _{X^{\prime }}+S_{L}\left( X,X\right) \right)
+A_{2}\left( \left\langle \bar{S}_{L}\left( X^{\prime },X\right)
\right\rangle _{X^{\prime }}+\left\langle \hat{S}_{L}^{B}\left( X^{\prime
},X\right) \right\rangle _{X^{\prime }}+S_{L}^{B}\left( X,X\right) \right)
\end{equation*}%
\begin{equation*}
A_{1}=\frac{\frac{w\left( X\right) \hat{w}\left( X\right) }{2}\frac{\hat{K}%
_{X}\left\vert \hat{\Psi}\left( X\right) \right\vert ^{2}}{K_{X}\left\vert
\Psi \left( X\right) \right\vert ^{2}}+w_{E}^{B}\left( X\right) \left\langle 
\hat{w}_{E}^{B}\left( X\right) \right\rangle \frac{\bar{K}_{X}\left\vert 
\bar{\Psi}\left( X\right) \right\vert ^{2}}{K_{X}\left\vert \Psi \left(
X\right) \right\vert ^{2}}}{1-\left\langle \hat{S}\left( X\right)
\right\rangle }\text{, }A_{2}=w_{E}^{B}\left( X\right) \left\langle \bar{w}%
\left( X\right) \right\rangle _{\bar{S}}\frac{\bar{K}_{X}\left\vert \bar{\Psi%
}\left( X\right) \right\vert ^{2}}{K_{X}\left\vert \Psi \left( X\right)
\right\vert ^{2}}
\end{equation*}%
\begin{eqnarray*}
B &=&\left( 1-2\frac{S\left( X\right) +S^{B}\left( X\right) }{1-\left(
S\left( X\right) +S^{B}\left( X\right) \right) }\right) \\
&&\times \left( 1-dC\frac{\left( \frac{w\left( X\right) \hat{w}\left(
X\right) }{2}\left( 1-\frac{\left\langle S_{E}\left( X,X\right)
\right\rangle }{1-\left\langle \hat{S}\left( X\right) \right\rangle }\right) 
\frac{\hat{K}_{X}\left\vert \hat{\Psi}\left( X\right) \right\vert ^{2}}{%
K_{X}\left\vert \Psi \left( X\right) \right\vert ^{2}}+w_{E}^{B}\left(
X\right) \left\langle \bar{w}\left( X\right) \right\rangle _{f\left(
X\right) }\frac{\bar{K}_{X}\left\vert \bar{\Psi}\left( X\right) \right\vert
^{2}}{K_{X}\left\vert \Psi \left( X\right) \right\vert ^{2}}\right) }{1-2%
\frac{S\left( X\right) +S^{B}\left( X\right) }{1-\left( S\left( X\right)
+S^{B}\left( X\right) \right) }}\right)
\end{eqnarray*}

\bigskip

The average loss that results from the default for the entire group of firms
can also be derived:%
\begin{equation*}
\left\langle df\left( X^{\prime }\right) \right\rangle =-\frac{\left(
\left\langle \hat{S}_{L}\left( X^{\prime },X\right) \right\rangle
+\left\langle S_{L}\left( X,X\right) \right\rangle \right) \frac{N}{%
1-\left\langle \hat{S}\left( X\right) \right\rangle }dC}{1-N\left( 1-\frac{%
\left\langle S_{E}\left( X,X\right) \right\rangle }{1-\left\langle \hat{S}%
\left( X\right) \right\rangle }\right) dC}
\end{equation*}%
where the parameter $dC$ writes:%
\begin{equation}
dC=C-\frac{rf_{1}\left( X\right) }{\left( K_{X}\right) ^{r-1}}  \label{Dcr}
\end{equation}%
This parameter reflects the impact of defaults on the firms' cost
structure.\ It captures the relative weight of costs against marginal
returns: higher values of $dC$ are associated with reduced net capital and a
higher likelihood of default.

Similarly, the expressions above reveal that, the higher the investment
stake $S\left( X\right) $ in sector $X$, the smaller the value of $dC$.
Financially vulnerable states---characterized by a high $dC$---are more
prone to initiate a cascade of defaults. The resulting fraction of
defaulting investors will be given by equation.\bigskip

\subsection*{A1.6 Internal dynamics}

To reveal the internal dynamics of the system, we transform the static
return equations (\ref{Rtn}) and (\ref{RTb}) for each group into a dynamic
system. To do so, we perform a first-order perturbation of these return
equations.

\subsubsection*{A1.6.1 Dynamic return equation}

\paragraph*{A1.6.1.1 Investors}

To study the distribution of stakes within groups dynamically, we must
specify the temporal sequence of investments by replacing the static return
equations (\ref{Rtn}), (\ref{RTb}) with a dynamic formulation. For
investors, this can be expressed as:%
\begin{eqnarray}
0 &=&\widehat{DF}\left( X^{\prime },\theta -1\right) \left( \hat{f}\left(
X^{\prime },\theta \right) -\bar{r}\right)  \label{CPR} \\
&&-\left\langle \hat{S}_{E}\left( X^{\prime },X,\theta \right) \right\rangle
_{X^{\prime }}\left\langle \widehat{DF}\left( X,\theta -1\right)
\right\rangle \left( \left\langle \hat{f}\left( X^{\prime },\theta \right)
\right\rangle -\left\langle \hat{r}\left( X^{\prime }\right) \right\rangle
\right)  \notag \\
&&-S_{E}\left( X,X,\theta -1\right) \left( f\left( X,\theta -1\right)
-\left\langle \bar{r}\left( X\right) \right\rangle \right)  \notag
\end{eqnarray}%
where $\widehat{DF}\left( X,\theta -1\right) $ denotes the dynamic discount
factor of investor $X$ for debt repayment:

\begin{equation*}
\widehat{DF}\left( X,\theta -1\right) =\frac{1-\left\langle \hat{S}\left(
X^{\prime },\theta -1\right) \right\rangle +\left\langle \hat{S}%
_{E}^{B}\left( X^{\prime },\theta -1\right) \right\rangle +\left\langle \hat{%
S}_{L}^{B}\left( X^{\prime },\theta -1\right) \right\rangle }{1-\left\langle 
\hat{S}_{E}\left( X^{\prime },\theta -1\right) \right\rangle +\left\langle 
\hat{S}_{E}^{B}\left( X^{\prime },\theta -1\right) \right\rangle }
\end{equation*}%
and the firms' returns are assumed to exhibit decreasing returns to scale:%
\begin{equation*}
f\left( X,\theta -1\right) -\left\langle \bar{r}\left( X\right)
\right\rangle =\frac{f_{1}\left( X,\theta -1\right) +\tau \left(
\left\langle f_{1}\left( X\right) \right\rangle -\left\langle f_{1}\left(
X^{\prime }\right) \right\rangle \right) }{\left( K\left( X,,\theta
-1\right) \left\Vert \Psi \left( \left( X,,\theta -1\right) \right)
\right\Vert ^{2}\right) ^{r}}-\frac{C}{K\left( X,,\theta -1\right) }%
-C_{0}-\left\langle \bar{r}\left( X\right) \right\rangle
\end{equation*}

\paragraph*{A1.6.1.2 Banks}

The dynamic equation for the bank's return is given by:%
\begin{eqnarray}
0 &=&\overline{DF}\left( X,\theta -1\right) \left( \bar{f}\left( X,\theta
\right) -\left( 1+\kappa \right) \bar{r}\right)  \label{CPT} \\
&&-\left\langle \bar{S}_{E}\left( X^{\prime },X,\theta -1\right)
\right\rangle _{X^{\prime }}\left\langle \overline{DF}\left( X^{\prime
},\theta -1\right) \right\rangle \left( \left\langle \bar{f}\left( X^{\prime
},\theta \right) \right\rangle -\left( 1+\kappa \right) \bar{r}\right) 
\notag \\
&&-\int \left\langle \hat{S}_{E}^{B}\left( X^{\prime },X,\theta -1\right)
\right\rangle _{X^{\prime }}\left\langle \widehat{DF}\left( X^{\prime
},\theta -1\right) \right\rangle \left( \left\langle \hat{f}\left( X^{\prime
},\theta \right) \right\rangle -\left\langle \hat{r}\left( X^{\prime
}\right) \right\rangle \right) dX^{\prime }  \notag \\
&&-S_{E}^{B}\left( X,X,\theta -1\right) \left( f\left( X,\theta -1\right)
-\left\langle \bar{r}\left( X\right) \right\rangle \right)  \notag
\end{eqnarray}%
where $\overline{DF}\left( X,\theta -1\right) $ denotes the dynamic discount
factor of investor $X$ for debt repayment:%
\begin{equation*}
\overline{DF}\left( X,\theta -1\right) =\frac{1-\bar{S}\left( X,\theta
-1\right) }{1-\bar{S}_{E}\left( X,\theta -1\right) }
\end{equation*}%
and the returns they generate:%
\begin{equation*}
\hat{f}\left( X^{\prime },\theta \right) -\bar{r}
\end{equation*}%
The returns generated by firms are themselves determined by lagged
productivity:%
\begin{equation*}
f\left( X,\theta -1\right) -\bar{r}
\end{equation*}%
so that capital reallocation delays the effect on firms' returns.\footnote{%
In this study, we focus on the case of decreasing returns to scale.}%
\footnote{%
The formula for \ $f\left( \hat{X},\theta -1\right) -\bar{r}$ is provided in
Appendix 9 in Gosselin and Lotz (2025).}.

\section{Appendix 2 Some formulas for field models}

\subsection{A2.1 Field models for investors and banks}

\subsubsection*{A2.1.1 Micro model for investors and banks}

The micro-foundations of investors and banks dynamics have been developed in
Gosselin and Lotz (2024). We present the main elements leading to the field
model for $\hat{\Psi}$ and $\bar{\Psi}$.

The dynamics for $\hat{K}_{j}\left( t\right) $ is described by the equation:%
\begin{equation*}
\frac{d}{dt}\hat{K}_{j}\left( t\right) =\hat{R}_{j}+\sum_{l}\hat{M}_{jl}%
\frac{d}{dt}\hat{K}_{l}\left( t\right) -\sum_{l}\bar{N}_{jl}\frac{d}{dt}\bar{%
K}_{0l}\left( t\right)
\end{equation*}

The right-hand side of equation (\ref{DIP2}) indicates that the variation in
investor $j$'s disposable capital depends not only on its return $\hat{R}%
_{j} $, but also on the capital provided by other investors through the
matrix $\hat{M}$, as well as that provided by banks through the matrix $\bar{%
N}$. This equation can be reformulated as a dynamic process involving the
variation of all the capital at the investor's disposal:%
\begin{equation}
\sum_{l}\left( \delta _{jl}-\hat{M}\right) _{jl}\frac{d}{dt}\hat{K}%
_{l}\left( t\right) +\sum_{l}\bar{N}_{jl}\frac{d}{dt}\bar{K}_{0l}\left(
t\right) =\hat{R}_{j}  \label{DIP2}
\end{equation}%
where:%
\begin{equation}
\hat{M}_{jm}=\frac{\hat{k}_{jm}\hat{K}_{j}\left( t\right) }{1+\sum_{l}\hat{k}%
_{jl}\hat{K}_{l}\left( t\right) +\sum_{l}\hat{k}_{1jl}^{B}\bar{K}_{l0}\left(
t\right) +\kappa \sum_{l}\hat{k}_{2jl}^{B}\frac{\bar{K}_{l0}\left( t\right) 
}{1+\sum_{m}\overline{\bar{k}}_{lm}\bar{K}_{m0}\left( t\right) }}
\label{MATMIC}
\end{equation}%
and:%
\begin{equation}
\bar{N}_{jl}=\frac{\left( \hat{k}_{1jl}^{B}+\kappa \hat{k}_{2jl}^{B}\frac{1}{%
1+\sum_{m}\overline{\bar{k}}_{lm}\bar{K}_{m0}\left( t\right) }-\kappa \frac{%
\sum_{m}\hat{k}_{2jm}^{B}\bar{K}_{m0}\left( t\right) \overline{\bar{k}}_{ml}%
}{\left( 1+\sum_{n}\overline{\bar{k}}_{mn}\bar{K}_{n0}\left( t\right)
\right) ^{2}}\right) \hat{K}_{j}}{1+\sum_{l}\hat{k}_{jl}\hat{K}_{l}\left(
t\right) +\sum_{l}\hat{k}_{1jl}^{B}\bar{K}_{l0}\left( t\right) +\kappa
\sum_{l}\hat{k}_{2jl}^{B}\frac{\bar{K}_{j0}\left( t\right) }{1+\sum_{m}%
\overline{\bar{k}}_{lm}\bar{K}_{m0}\left( t\right) }}  \label{MATMIc}
\end{equation}

In the continuous approximation, the capital accumulation for banks writes: 
\begin{equation*}
\frac{d}{dt}\bar{K}_{0j}\left( t\right) =\bar{R}_{j}+\sum_{l}M_{jl}\frac{d}{%
dt}\bar{K}_{0l}\left( t\right)
\end{equation*}

Similarly to investors, the variation in available capital for bank $j$ is
the sum of its total return, denoted as $\bar{R}_{j}$, plus the variation in
participations and loans from banks investing in bank $j$, through the
matrix $\bar{M}$. As before, this equation can be reformulated as a dynamic
process involving the variation of all the capital at bank $j$'s disposal. 
\begin{equation}
\sum_{l}\left( \delta _{jl}-M_{jl}\frac{d}{dt}\right) \bar{K}_{0l}\left(
t\right) =\bar{R}_{j}^{\prime }  \label{DBP2}
\end{equation}%
with:%
\begin{equation}
\bar{M}_{jm}=\frac{\bar{k}_{jm}\bar{K}_{j0}\left( t\right) }{1+\sum_{\nu }%
\bar{k}_{jv}\hat{K}_{\nu }\left( t\right) }  \label{MATMic}
\end{equation}%
Here, the overall leverage effect $\bar{k}_{jm}$ provided by bank $m$ to
bank $j$ decomposes into the leverage effect associated with participations
and loans, denoted as $\bar{k}_{1jm}$ and $\bar{k}_{2jm}$ respectively:%
\begin{equation*}
\bar{k}_{1jm}+\bar{k}_{2jm}=\bar{k}_{jm}
\end{equation*}

\subsubsection*{A.2.1.2 Field translation}

\paragraph*{\protect\bigskip A.2.1.2.1 Field translation in term of
exogenous connections}

We showed previously that the dynamic equations (\ref{DIP2}), (\ref{DBP2})
can be translated into field action functional encompassing the dynamical
accumulation for each sector. Consider first the field $\bar{\Psi}$
describing the banking system.\ It depends on the two variables $\bar{K}$
and $X$, and its action functional is given by:%
\begin{equation}
-\bar{\Psi}^{\dag }\left( \bar{K},X\right) \nabla ^{2}\bar{\Psi}\left( \bar{K%
},X\right) +\left( \frac{\bar{g}^{2}\left( \bar{K},X\right) }{2\sigma _{\bar{%
K}}^{2}}+\frac{\bar{g}\left( \bar{K},X\right) }{2\bar{K}}\right) \left\vert 
\bar{\Psi}\left( \bar{K},X\right) \right\vert ^{2}
\end{equation}%
where:%
\begin{equation*}
\bar{g}\left( \hat{K},X\right) =\left( 1-\bar{M}\left\vert \bar{\Psi}\left( 
\bar{K},X\right) \right\vert ^{2}\right) ^{-1}\bar{f}\left( \bar{K},X\right)
\end{equation*}%
and $\bar{M}$ is the matrix translating its microeconomic equivalent (\ref%
{MATMIC}).

The field $\hat{\Psi}$ describing the investors depends on the two variables 
$\hat{K}$ and $X$, and its action functional is given by:%
\begin{equation}
-\hat{\Psi}^{\dag }\left( \hat{K},X\right) \nabla ^{2}\hat{\Psi}\left( \hat{K%
},X\right) +\left( \frac{\hat{g}^{2}\left( \hat{K},X\right) }{2\sigma _{\hat{%
K}}^{2}}+\frac{\hat{g}\left( \hat{K},X\right) }{2\hat{K}}\right) \left\vert 
\hat{\Psi}\left( \hat{K},X\right) \right\vert ^{2}  \label{NFC}
\end{equation}%
where we define:%
\begin{equation*}
\hat{g}\left( \hat{K},X\right) =\left( 1-\hat{M}\left\vert \hat{\Psi}\left( 
\hat{K},X\right) \right\vert ^{2}\right) ^{-1}\hat{f}\left( \hat{K},X\right)
+\left( 1-\hat{M}\right) ^{-1}\bar{N}\left( 1-\bar{M}\right) ^{-1}\bar{f}%
\left( \bar{K},X\right)
\end{equation*}%
The field definition of the matrices $\bar{M}$, $\hat{M}$ and $\bar{N}$ are
given here.

These matrices $\bar{M}\left( \left( \bar{K},X\right) ,\left( \bar{K}%
^{\prime },X^{\prime }\right) \right) $, $\hat{M}\left( \left( \hat{K}%
^{\prime },X^{\prime }\right) ,\left( \hat{K},X\right) \right) $ and $\bar{N}%
\left( \left( \hat{K},X\right) ,\left( \bar{K},X\right) \right) $, were
defined in GL 24 using exogenous set of connections. They were the direct
translation of (\ref{MATMIC}), (\ref{MATMIc}), (\ref{MATMic}):%
\begin{equation*}
\bar{M}\left( \left( \bar{K},X\right) ,\left( \bar{K}^{\prime },X^{\prime
}\right) \right) =\frac{\overline{\bar{k}}\left( X,X^{\prime }\right) \bar{K}%
_{0}}{1+\int \overline{\bar{k}}\left( X,X^{\prime }\right) \bar{K}%
_{0}^{\prime }\left\vert \bar{\Psi}\left( \bar{K}_{0}^{\prime },X^{\prime
}\right) \right\vert ^{2}}
\end{equation*}%
\begin{eqnarray*}
&&\hat{M}\left( \left( \hat{K}^{\prime },X^{\prime }\right) ,\left( \hat{K}%
,X\right) \right) \\
&=&\frac{\hat{k}\left( X,X^{\prime }\right) \hat{K}}{1+\int \hat{k}\left(
X,X^{\prime }\right) \left\vert \hat{\Psi}\left( \hat{K}^{\prime },X^{\prime
}\right) \right\vert ^{2}+\int \hat{k}_{1}^{B}\left( X,X^{\prime }\right) 
\bar{K}_{0}^{\prime }\left\vert \bar{\Psi}\left( \bar{K}_{0}^{\prime
},X^{\prime }\right) \right\vert ^{2}+\kappa \int \hat{k}_{2}^{B}\left(
X,X^{\prime }\right) \frac{\bar{K}_{0}^{\prime }\left\vert \bar{\Psi}\left( 
\bar{K}_{0}^{\prime },X^{\prime }\right) \right\vert ^{2}}{1+\int \overline{%
\bar{k}}\left( X^{\prime }\right) \bar{K}_{0}^{\prime \prime }\left\vert 
\bar{\Psi}\left( \bar{K}_{0}^{\prime \prime },X^{\prime \prime }\right)
\right\vert ^{2}}}
\end{eqnarray*}%
and ultimately:%
\begin{equation*}
\bar{N}=\frac{\left( \hat{k}_{1}^{B}\left( X,X^{\prime }\right) +\kappa 
\frac{\hat{k}_{2}^{B}\left( X,X^{\prime }\right) }{1+\int \overline{\bar{k}}%
\left( X^{\prime }\right) \bar{K}_{0}^{\prime \prime }\left\vert \bar{\Psi}%
\left( \bar{K}_{0}^{\prime \prime },X^{\prime \prime }\right) \right\vert
^{2}}-\kappa \int \frac{\hat{k}_{2}^{B}\left( X,X^{\prime \prime }\right) 
\bar{K}_{0}^{\prime \prime }\overline{\bar{k}}\left( X^{\prime \prime
},X^{\prime }\right) }{\left( 1+\int \overline{\bar{k}}\left( X^{\prime
\prime },\bar{Y}\right) \bar{K}_{0}^{Y}\left\vert \bar{\Psi}\left( \bar{K}%
_{0}^{Y},\bar{Y}\right) \right\vert ^{2}\right) ^{2}}\right) \hat{K}}{1+\int 
\hat{k}\left( X,X^{\prime }\right) \left\vert \hat{\Psi}\left( \hat{K}%
^{\prime },X^{\prime }\right) \right\vert ^{2}+\int \hat{k}_{1}^{B}\left(
X,X^{\prime }\right) \bar{K}_{0}^{\prime }\left\vert \bar{\Psi}\left( \bar{K}%
_{0}^{\prime },X^{\prime }\right) \right\vert ^{2}+\kappa \int \hat{k}%
_{2}^{B}\left( X,X^{\prime }\right) \frac{\bar{K}_{0}^{\prime }\left\vert 
\bar{\Psi}\left( \bar{K}_{0}^{\prime },X^{\prime }\right) \right\vert ^{2}}{%
1+\int \overline{\bar{k}}\left( X^{\prime }\right) \bar{K}_{0}^{\prime
\prime }\left\vert \bar{\Psi}\left( \bar{K}_{0}^{\prime \prime },X^{\prime
\prime }\right) \right\vert ^{2}}}
\end{equation*}

\paragraph*{A.2.1.2.1 Field translation in term of endogenous stakes}

Using the endogenous stakes defined in Gosselin and Lotz (2025b), we have:

the various matrices defined in the first part and arising in the return
equation can be rewritten in terms of stakes (see Gosselin and Lotz (2025b)):

\begin{eqnarray}
\hat{S}_{\eta }\left( X^{\prime },\hat{K}^{\prime },X\right) &=&\frac{\hat{K}%
^{\prime }\hat{k}_{\eta }\left( X^{\prime },X\right) \left\vert \hat{\Psi}%
\left( \hat{K}^{\prime },X^{\prime }\right) \right\vert ^{2}}{1+\underline{%
\hat{k}}\left( X^{\prime }\right) +\underline{\hat{k}}_{1}^{B}\left(
X^{\prime }\right) +\kappa \left[ \frac{\underline{\hat{k}}_{2}^{B}}{1+\bar{k%
}}\right] \left( X^{\prime }\right) }  \label{CF1} \\
\hat{S}_{\eta }\left( X^{\prime },X\right) &=&\int \frac{\hat{K}^{\prime }%
\hat{k}_{\eta }\left( X^{\prime },X\right) \left\vert \hat{\Psi}\left( \hat{K%
}^{\prime },X^{\prime }\right) \right\vert ^{2}}{1+\underline{\hat{k}}\left(
X^{\prime }\right) +\underline{\hat{k}}_{1}^{B}\left( X^{\prime }\right)
+\kappa \left[ \frac{\underline{\hat{k}}_{2}^{B}}{1+\bar{k}}\right] \left(
X^{\prime }\right) }d\hat{K}^{\prime }=\frac{\hat{K}_{X^{\prime }}\hat{k}%
_{\eta }\left( X^{\prime },X\right) \left\vert \hat{\Psi}\left( X^{\prime
}\right) \right\vert ^{2}}{1+\underline{\hat{k}}\left( X^{\prime }\right) +%
\underline{\hat{k}}_{1}^{B}\left( X^{\prime }\right) +\kappa \left[ \frac{%
\underline{\hat{k}}_{2}^{B}}{1+\bar{k}}\right] \left( X^{\prime }\right) } 
\notag
\end{eqnarray}%
and:%
\begin{eqnarray}
S_{\eta }\left( X^{\prime },K^{\prime },X\right) &=&\frac{k_{\eta }\left(
X^{\prime },X\right) K^{\prime }\left\vert \Psi \left( K^{\prime },X^{\prime
}\right) \right\vert ^{2}}{1+\underline{k}\left( X^{\prime }\right) +%
\underline{k}_{1}^{\left( B\right) }\left( X^{\prime }\right) +\kappa \left[ 
\frac{\underline{k}_{2}^{B}}{1+\bar{k}}\right] \left( X^{\prime }\right) }
\label{CF2} \\
S_{\eta }\left( X^{\prime },X\right) &=&\int \frac{k_{\eta }\left( X^{\prime
},X\right) K^{\prime }\left\vert \Psi \left( K^{\prime },X^{\prime }\right)
\right\vert ^{2}}{1+\underline{k}\left( X^{\prime }\right) +\underline{k}%
_{1}^{\left( B\right) }\left( X^{\prime }\right) +\kappa \left[ \frac{%
\underline{k}_{2}^{B}}{1+\bar{k}}\right] \left( X^{\prime }\right) }%
dK^{\prime }=\frac{k_{\eta }\left( X^{\prime },X\right) K_{X^{\prime
}}\left\vert \Psi \left( X^{\prime }\right) \right\vert ^{2}}{1+\underline{k}%
\left( X^{\prime }\right) +\underline{k}_{1}^{\left( B\right) }\left(
X^{\prime }\right) +\kappa \left[ \frac{\underline{k}_{2}^{B}}{1+\bar{k}}%
\right] \left( X^{\prime }\right) }  \notag
\end{eqnarray}%
for investors, and:%
\begin{eqnarray*}
\bar{S}_{\eta }\left( X^{\prime },\bar{K}^{\prime },X\right) &=&\frac{\bar{K}%
^{\prime }\bar{k}_{\eta }\left( X^{\prime },X\right) \left\vert \bar{\Psi}%
\left( \bar{K}^{\prime },X^{\prime }\right) \right\vert ^{2}}{1+\underline{%
\bar{k}}\left( X^{\prime }\right) } \\
\bar{S}_{\eta }\left( X^{\prime },X\right) &=&\int \frac{\bar{K}^{\prime }%
\bar{k}_{\eta }\left( X^{\prime },X\right) \left\vert \bar{\Psi}\left( \bar{K%
}^{\prime },X^{\prime }\right) \right\vert ^{2}}{1+\underline{\bar{k}}\left(
X^{\prime }\right) }d\bar{K}^{\prime }=\frac{\bar{K}_{X^{\prime }}\bar{k}%
_{\eta }\left( X^{\prime },X\right) \left\vert \bar{\Psi}\left( X^{\prime
}\right) \right\vert ^{2}}{1+\underline{\bar{k}}\left( X^{\prime }\right) }
\end{eqnarray*}%
\begin{eqnarray*}
\hat{S}_{1}^{B}\left( X^{\prime },\hat{K}^{\prime },X\right) &=&\frac{\hat{K}%
^{\prime }\underline{\hat{k}}_{1}^{B}\left( X^{\prime },X\right) \left\vert 
\hat{\Psi}\left( \hat{K}^{\prime },X^{\prime }\right) \right\vert ^{2}}{1+%
\underline{\hat{k}}\left( X^{\prime }\right) +\underline{\hat{k}}%
_{1}^{B}\left( X^{\prime }\right) +\kappa \left[ \frac{\underline{\hat{k}}%
_{2}^{B}}{1+\bar{k}}\right] \left( X^{\prime }\right) } \\
\hat{S}_{1}^{B}\left( X^{\prime },X\right) &=&\frac{\hat{K}_{X^{\prime }}%
\underline{\hat{k}}_{1}^{B}\left( X^{\prime },X\right) \left\vert \hat{\Psi}%
\left( X^{\prime }\right) \right\vert ^{2}}{1+\underline{\hat{k}}\left(
X^{\prime }\right) +\underline{\hat{k}}_{1}^{B}\left( X^{\prime }\right)
+\kappa \left[ \frac{\underline{\hat{k}}_{2}^{B}}{1+\bar{k}}\right] \left(
X^{\prime }\right) }
\end{eqnarray*}%
\begin{eqnarray*}
\hat{S}_{2}^{B}\left( X^{\prime },\hat{K}^{\prime },X\right) &=&\frac{\hat{K}%
^{\prime }\frac{\kappa \underline{\hat{k}}_{2}^{B}\left( X^{\prime
},X\right) }{1+\underline{\bar{k}}\left( X\right) }\left\vert \hat{\Psi}%
\left( \hat{K}^{\prime },X^{\prime }\right) \right\vert ^{2}}{1+\underline{%
\hat{k}}\left( X^{\prime }\right) +\underline{\hat{k}}_{1}^{B}\left(
X^{\prime }\right) +\kappa \left[ \frac{\underline{\hat{k}}_{2}^{B}}{1+\bar{k%
}}\right] \left( X^{\prime }\right) } \\
\hat{S}_{2}^{B}\left( X^{\prime },X\right) &=&\frac{\hat{K}_{X^{\prime }}%
\frac{\kappa \underline{\hat{k}}_{2}^{B}\left( X^{\prime },X\right) }{1+%
\underline{\bar{k}}\left( X\right) }\left\vert \hat{\Psi}\left( X^{\prime
}\right) \right\vert ^{2}}{1+\underline{\hat{k}}\left( X^{\prime }\right) +%
\underline{\hat{k}}_{1}^{B}\left( X^{\prime }\right) +\kappa \frac{%
\underline{\hat{k}}_{2}^{B}\left( X^{\prime }\right) }{1+\bar{k}}}
\end{eqnarray*}%
\begin{equation*}
\left[ \frac{\underline{k}_{2}^{B}}{1+\bar{k}}\right] \left( X^{\prime
}\right) =\int \frac{\underline{k}_{2}^{B}\left( X^{\prime },X\right) }{1+%
\bar{k}\left( X\right) }\bar{K}\left\vert \bar{\Psi}\left( \bar{K},X\right)
\right\vert ^{2}d\bar{K}dX
\end{equation*}%
\begin{eqnarray}
S_{1}^{B}\left( X^{\prime },K^{\prime },X\right) &=&\frac{K^{\prime }%
\underline{k}_{1}^{\left( B\right) }\left( X^{\prime },X\right) \left\vert
\Psi \left( K^{\prime },X^{\prime }\right) \right\vert ^{2}}{1+\underline{k}%
\left( X^{\prime }\right) +\underline{k}_{1}^{\left( B\right) }\left(
X^{\prime }\right) +\kappa \left[ \frac{\underline{k}_{2}^{\left( B\right) }%
}{1+\underline{\bar{k}}}\right] \left( X^{\prime }\right) }  \label{DFrApp}
\\
S_{2}^{B}\left( X^{\prime },K^{\prime },X\right) &=&\frac{\frac{\kappa 
\underline{k}_{2}^{\left( B\right) }\left( X^{\prime },X\right) }{1+%
\underline{\bar{k}}\left( X\right) }K^{\prime }\left\vert \Psi \left(
K^{\prime },X^{\prime }\right) \right\vert ^{2}}{1+\underline{k}\left(
X^{\prime }\right) +\underline{k}_{1}^{\left( B\right) }\left( X^{\prime
}\right) +\kappa \left[ \frac{\underline{k}_{2}^{\left( B\right) }}{1+%
\underline{\bar{k}}}\right] \left( X^{\prime }\right) }  \notag \\
S_{1}^{B}\left( X^{\prime },X\right) &=&\frac{K_{X^{\prime }}\underline{k}%
_{1}^{\left( B\right) }\left( X^{\prime },X\right) \left\vert \Psi \left(
K^{\prime },X^{\prime }\right) \right\vert ^{2}}{1+\underline{k}\left(
X^{\prime }\right) +\underline{k}_{1}^{\left( B\right) }\left( X^{\prime
}\right) +\kappa \left[ \frac{\underline{k}_{2}^{\left( B\right) }}{1+%
\underline{\bar{k}}}\right] \left( X^{\prime }\right) }\left\vert \Psi
\left( X^{\prime }\right) \right\vert ^{2}  \notag \\
S_{2}^{B}\left( X^{\prime },X\right) &=&\frac{\frac{\kappa \underline{k}%
_{2}^{\left( B\right) }\left( X^{\prime },X\right) }{1+\underline{\bar{k}}%
\left( X\right) }K_{X^{\prime }}\left\vert \Psi \left( X^{\prime }\right)
\right\vert ^{2}}{1+\underline{k}\left( X^{\prime }\right) +\underline{k}%
_{1}^{\left( B\right) }\left( X^{\prime }\right) +\kappa \left[ \frac{%
\underline{k}_{2}^{\left( B\right) }}{1+\underline{\bar{k}}}\right] \left(
X^{\prime }\right) }  \notag
\end{eqnarray}%
for banks. This leads directly to:

\begin{eqnarray*}
\bar{M}\left( \left( \bar{K},X\right) ,\left( \bar{K}^{\prime },X^{\prime
}\right) \right) &=&\frac{\overline{\bar{k}}\left( X,X^{\prime }\right) \bar{%
K}}{1+\int \overline{\bar{k}}\left( X,X^{\prime }\right) \bar{K}_{0}^{\prime
}\left\vert \bar{\Psi}\left( \bar{K}_{0}^{\prime },X^{\prime }\right)
\right\vert ^{2}} \\
&\rightarrow &\bar{S}\left( \left( \bar{K},X\right) ,\left( \bar{K}^{\prime
},X^{\prime }\right) \right)
\end{eqnarray*}%
\begin{eqnarray*}
&&\hat{M}\left( \left( \hat{K}^{\prime },X^{\prime }\right) ,\left( \hat{K}%
,X\right) \right) \\
&=&\frac{\hat{k}\left( X,X^{\prime }\right) \hat{K}}{1+\int \hat{k}\left(
X,X^{\prime }\right) \left\vert \hat{\Psi}\left( \hat{K}^{\prime },X^{\prime
}\right) \right\vert ^{2}+\int \hat{k}_{1}^{B}\left( X,X^{\prime }\right) 
\bar{K}_{0}^{\prime }\left\vert \bar{\Psi}\left( \bar{K}_{0}^{\prime
},X^{\prime }\right) \right\vert ^{2}+\int \hat{k}_{2}^{B}\left( X,X^{\prime
}\right) \frac{\bar{K}_{0}^{\prime }\left\vert \bar{\Psi}\left( \bar{K}%
_{0}^{\prime },X^{\prime }\right) \right\vert ^{2}}{1+\int \overline{\bar{k}}%
\left( X^{\prime }\right) \bar{K}_{0}^{\prime \prime }\left\vert \bar{\Psi}%
\left( \bar{K}_{0}^{\prime \prime },X^{\prime \prime }\right) \right\vert
^{2}}} \\
&\rightarrow &\hat{S}\left( \left( \hat{K}^{\prime },X^{\prime }\right)
,\left( \hat{K},X\right) \right)
\end{eqnarray*}%
and ultimately:%
\begin{eqnarray*}
\bar{N} &\rightarrow &\frac{\left( \hat{k}_{1}^{B}\left( X,X^{\prime
}\right) +\kappa \frac{\hat{k}_{2}^{B}\left( X,X^{\prime }\right) }{1+\int 
\overline{\bar{k}}\left( X^{\prime }\right) \bar{K}_{0}^{\prime \prime
}\left\vert \bar{\Psi}\left( \bar{K}_{0}^{\prime \prime },X^{\prime \prime
}\right) \right\vert ^{2}}-\kappa \int \frac{\hat{k}_{2}^{B}\left(
X,X^{\prime \prime }\right) \bar{K}_{0}^{\prime \prime }\overline{\bar{k}}%
\left( X^{\prime \prime },X^{\prime }\right) }{\left( 1+\int \overline{\bar{k%
}}\left( X^{\prime \prime },\bar{Y}\right) \bar{K}_{0}^{Y}\left\vert \bar{%
\Psi}\left( \bar{K}_{0}^{Y},\bar{Y}\right) \right\vert ^{2}\right) ^{2}}%
\right) \hat{K}}{1+\int \hat{k}\left( X,X^{\prime }\right) \left\vert \hat{%
\Psi}\left( \hat{K}^{\prime },X^{\prime }\right) \right\vert ^{2}+\int \hat{k%
}_{1}^{B}\left( X,X^{\prime }\right) \bar{K}_{0}^{\prime }\left\vert \bar{%
\Psi}\left( \bar{K}_{0}^{\prime },X^{\prime }\right) \right\vert ^{2}+\kappa
\int \hat{k}_{2}^{B}\left( X,X^{\prime }\right) \frac{\bar{K}_{0}^{\prime
}\left\vert \bar{\Psi}\left( \bar{K}_{0}^{\prime },X^{\prime }\right)
\right\vert ^{2}}{1+\int \overline{\bar{k}}\left( X^{\prime }\right) \bar{K}%
_{0}^{\prime \prime }\left\vert \bar{\Psi}\left( \bar{K}_{0}^{\prime \prime
},X^{\prime \prime }\right) \right\vert ^{2}}} \\
&\rightarrow &\hat{S}_{1}^{B}\left( \left( \hat{K},X\right) ,\left( \bar{K}%
^{\prime },X^{\prime }\right) \right) +\hat{S}_{2}^{B}\left( \left( \hat{K}%
,X\right) ,\left( \bar{K}^{\prime },X^{\prime }\right) \right)
\end{eqnarray*}

\subsubsection*{A2.1.3 Returns equations as constraints}

The return equations for investors and banks:%
\begin{equation}
0=\int \left( \delta \left( X^{\prime }-X\right) -\hat{S}_{E}\left(
X^{\prime },X\right) \right) \widehat{DF}\left( X^{\prime }\right) \hat{R}%
_{exc}\left( X^{\prime }\right) dX^{\prime }-S_{E}\left( X,X\right)
R_{exc}\left( X\right)
\end{equation}%
and:%
\begin{eqnarray}
0 &=&\int \left( \delta \left( X^{\prime }-X\right) -\bar{S}_{E}\left(
X^{\prime },X\right) \right) \overline{DF}\left( X^{\prime }\right) \bar{R}%
_{exc}\left( X^{\prime }\right) \\
&&-\int \hat{S}_{E}^{B}\left( X^{\prime },X\right) \widehat{DF}\left(
X^{\prime }\right) \hat{R}_{exc}\left( X^{\prime }\right) -S_{E}^{B}\left(
X^{\prime },X\right) R_{exc}\left( X\right)  \notag
\end{eqnarray}%
with:%
\begin{equation*}
\widehat{DF}\left( X\right) =\frac{1-\hat{S}\left( X\right) }{1-\hat{S}%
_{E}\left( X\right) }
\end{equation*}%
\begin{equation*}
\overline{DF}\left( X\right) =\frac{1-\bar{S}\left( X\right) }{1-\bar{S}%
_{E}\left( X\right) }
\end{equation*}%
and:%
\begin{equation*}
\hat{R}_{exc}\left( X^{\prime }\right) =f\left( X^{\prime }\right) -\bar{r}
\end{equation*}%
\begin{equation*}
R_{exc}\left( X\right) =f\left( X\right) -\bar{r}
\end{equation*}%
\begin{equation*}
\bar{R}_{exc}\left( X^{\prime }\right) =\bar{f}\left( X^{\prime }\right)
-\left( 1+\kappa \right) \bar{r}
\end{equation*}%
should be included in the action functional minimization equations through
Lagrange multipliers $\bar{\xi}\left( X\right) $ and $\hat{\xi}\left(
X\right) $ modifying these equations by terms of the form:%
\begin{eqnarray*}
&&\hat{\xi}\left( X\right) \int \left( \delta \left( X^{\prime }-X\right) -%
\hat{S}_{E}\left( X^{\prime },X\right) \right) \frac{\delta \widehat{DF}%
\left( X^{\prime }\right) }{\delta \hat{\Psi}\left( \hat{K},X\right) }\hat{R}%
_{exc}\left( X^{\prime }\right) dX^{\prime }-\frac{\delta S_{E}\left(
X,X\right) }{\delta \hat{\Psi}\left( K,X\right) }R_{exc}\left( X\right) \\
&&-\hat{\xi}\left( X\right) \hat{S}_{E}\left( X^{\prime },X\right) \widehat{%
DF}\left( X^{\prime }\right) \hat{R}_{exc}\left( X^{\prime }\right)
dX^{\prime }
\end{eqnarray*}%
and similar contributions for the derivatives with respects to $\bar{\Psi}%
\left( \bar{K},X\right) $.

We can neglect these contributions for two reasons. First, the stakes $%
S_{E}\left( X,X\right) $, $\hat{S}_{E}\left( X^{\prime },X\right) $ $\bar{S}%
_{E}\left( X^{\prime },X\right) $, $\hat{S}_{E}^{B}\left( X^{\prime
},X\right) $, $S_{E}^{B}\left( X^{\prime },X\right) $ as well as $\widehat{DF%
}\left( X\right) $ and $\overline{DF}$\ depend weakly on the fields $\hat{%
\Psi}\left( \hat{K},X\right) $ and $\bar{\Psi}\left( \bar{K},X\right) $,
given that they correspond to decision of investments with respect to other
agents level of return and uncertaint.

Second, these contributions are proportional to excess returns, so that they
corresponds to corrections relatively to the levels of returns. We can thus
neglect them in first approximation.

\subsubsection*{A2.1.4 Action functionals' second order derivatives}

The second order derivaives for investors and fields action functionals are
given by:%
\begin{eqnarray*}
\frac{\delta ^{2}}{\delta \Gamma \delta \Gamma ^{\dag }}S\left( \Gamma
\right) &\simeq &-\sum_{\eta }\left( \sigma _{\hat{K}}^{2}\nabla _{\hat{S}%
_{\eta }^{\left( T\right) }}^{2}-\left( \frac{\left( \hat{S}_{\eta }^{\left(
T\right) }\right) ^{2}}{2\hat{h}_{\eta }\left( X^{\prime },X\right) }-\hat{V}%
_{\eta }\hat{S}_{\eta }^{\left( T\right) }\right) -\beta \right) -\underline{%
\lambda \left( X\right) }\left( \sum \hat{S}_{\eta }^{\left( T\right)
}-1\right) \\
&&-\int \Gamma ^{\dag }\left( \hat{S}_{\eta }^{\left( T\right) }\right) %
\left[ \frac{\delta ^{2}}{\delta \Gamma \delta \Gamma ^{\dag }}\underline{%
\hat{V}_{\eta }}\hat{S}_{\eta }^{\left( T\right) }+\left( \frac{\delta }{%
\delta \Gamma }+\frac{\delta }{\delta \Gamma ^{\dag }}\right) \underline{%
\hat{V}_{\eta }}\hat{S}_{\eta }^{\left( T\right) }\left( \sum \hat{S}_{\eta
}^{\left( T\right) }-1\right) \right] \Gamma \left( \hat{S}_{\eta }^{\left(
T\right) }\right)
\end{eqnarray*}%
and:%
\begin{eqnarray*}
\frac{\delta ^{2}}{\delta \bar{\Gamma}\delta \bar{\Gamma}^{\dag }}S\left( 
\bar{\Gamma}\right) &=&-\sum_{\eta }\left( \sigma _{\hat{K}}^{2}\nabla _{%
\bar{S}_{\eta }^{\left( T\right) }}^{2}-\left( \frac{\left( \bar{S}_{\eta
}^{\left( T\right) }\right) ^{2}}{2\hat{h}_{\eta }\left( X^{\prime
},X\right) }-\bar{V}_{\eta }\bar{S}_{\eta }^{\left( T\right) }\right) -\beta
\right) -\underline{\lambda \left( X\right) }\left( \sum \bar{S}_{\eta
}^{\left( T\right) }-1\right) \\
&&-\bar{\Gamma}\left( \bar{S}_{\eta }^{\left( T\right) }\right) \left[ \frac{%
\delta ^{2}}{\delta \Gamma \delta \Gamma ^{\dag }}\bar{V}_{\eta }\bar{S}%
_{\eta }^{\left( T\right) }+\left( \frac{\delta }{\delta \bar{\Gamma}}+\frac{%
\delta }{\delta \bar{\Gamma}^{\dag }}\right) \bar{V}_{\eta }\bar{S}_{\eta
}^{\left( T\right) }\left( \sum \bar{S}_{\eta }^{\left( T\right) }-1\right) %
\right] \bar{\Gamma}^{\dag }\left( \bar{S}_{\eta }^{\left( T\right) }\right)
\end{eqnarray*}%
where $\beta $\ represents the inverse time scale of the system, i.e. the
inverse average persistence time of a collective state.

In the above expressions, the stakes $\hat{S}^{\left( T\right) }\left(
X^{\prime },X\right) $ and $\bar{S}^{\left( T\right) }\left( X^{\prime
},X\right) $ are respectively: 
\begin{equation*}
\hat{S}^{\left( T\right) }\left( X^{\prime },X\right) =\left[ \left\{ \hat{S}%
_{\eta }\left( X^{\prime },X\right) \right\} _{\eta =1,2},\left\{ S_{\eta
}\left( X,X\right) \right\} _{\eta =1,2}\right]
\end{equation*}%
and:%
\begin{equation*}
\bar{S}^{\left( T\right) }\left( X^{\prime },X\right) =\left[ \left\{ \bar{S}%
_{\eta }\left( X^{\prime },X\right) \right\} _{\eta =1,2},\left\{ \hat{S}%
_{\eta }\left( X^{\prime },X\right) \right\} _{\eta =1,2},\left\{ S_{\eta
}\left( X,X\right) \right\} _{\eta =1,2}\right]
\end{equation*}%
They are computed in the collective state defined by (\ref{gmz}) and (\ref%
{gmb}).

These formulas will be rewritten in a more compact form in the sequel of
this work, in order to study the dynamics of deviations states.

\subsection*{A2.2 Fields of stakes' action functionals and minimization
functions}

\subsubsection*{A2.2.1 Micro model for investors stakes}

We start with the sum over points of investors objective functions
integrated over time:%
\begin{equation*}
\sum_{j}\hat{S}_{1ij}\hat{f}_{j}+\sum_{j}\hat{S}_{2ij}\hat{r}_{j}-\frac{1}{2}%
\sum_{ij}\frac{\left( \hat{S}_{\eta ij}\right) ^{2}}{\hat{h}_{\eta
_{i}}\left( X_{j}\right) }+\sum_{k}S_{1ik}f_{k}+\sum_{k}S_{2ik}\bar{r}_{k}-%
\frac{1}{2}\sum_{ik}\frac{\left( S_{\eta ik}\right) ^{2}}{h_{\eta ik}\left(
X_{k}\right) }
\end{equation*}%
\begin{equation}
\int dt\left( \sum_{i,j,k}\hat{S}_{1ij}\hat{f}_{j}+\sum_{i,j,k}\hat{S}_{2ij}%
\hat{r}_{j}-\frac{1}{2}\sum_{i,j,k}\frac{\left( \hat{S}_{\eta ij}\right) ^{2}%
}{\hat{h}_{\eta _{ij}}\left( X_{j}\right) }+\sum_{i,k,j}S_{1ik}f_{k}+%
\sum_{i,k,j}S_{2ik}\bar{r}_{k}-\frac{1}{2}\sum_{ji,k}\frac{\left( S_{\eta
ik}\right) ^{2}}{h_{\eta ik}\left( X_{k}\right) }\right)  \label{Smt}
\end{equation}%
Then, we add an inertial term for all agents and a term corresponding to the
time scale in which collective states are considered:%
\begin{equation}
-\alpha \sum_{i,j}\left( \frac{d}{dt}\hat{S}_{\eta ij}\right) ^{2}-\alpha
\sum_{i,k}\left( \frac{d}{dt}S_{\eta ik}\right) ^{2}-\beta \int dt
\label{NTS}
\end{equation}

\subsubsection*{A2.2.2 Field translation and investors stakes action
functional}

Considering the field: 
\begin{equation*}
\Gamma \left( S_{1},\hat{S}_{1},S_{2},\hat{S}_{2},X^{\prime },X\right)
\end{equation*}%
written:%
\begin{equation*}
\Gamma \left( \hat{S}^{\left( T\right) },X^{\prime },X\right)
\end{equation*}%
with:%
\begin{equation*}
\hat{S}^{\left( T\right) }=\left( S_{1},\hat{S}_{1},S_{2},\hat{S}_{2}\right)
\end{equation*}%
the potential (\ref{Smt}) becomes in terms of field:%
\begin{equation*}
-\sum_{\eta }\int \Gamma ^{\dag }\left( \hat{S}^{\left( T\right) },X^{\prime
},X\right) \left( -\frac{\left( \hat{S}_{\eta }^{\left( T\right) }\right)
^{2}}{2\hat{h}_{\eta }^{T}\left( X^{\prime },X\right) }+\hat{V}_{\eta }\hat{S%
}_{\eta }^{\left( T\right) }\right) \Gamma \left( \hat{S}^{\left( T\right)
},X^{\prime },X\right) d\left( \hat{S}^{\left( T\right) },X^{\prime
},X\right)
\end{equation*}%
The coefficients $\hat{h}_{\eta }^{T}\left( X^{\prime },X\right) $
translates the uncertainty coefficients $\hat{h}_{\eta _{i}}\left(
X_{j}\right) $, $h_{\eta ik}\left( X_{k}\right) $ in inter, or intra,
sectoral functions, while $\hat{V}_{\eta }$ encompass the returns. We have: 
\begin{eqnarray*}
\hat{V}_{1} &=&\hat{f}\left( X^{\prime }\right) ,\hat{h}_{1}\left( X^{\prime
},X\right) =\hat{h}_{1}\left( X^{\prime },X\right) \\
\hat{V}_{2} &=&\hat{r}\left( X^{\prime }\right) ,\hat{h}_{2}\left( X^{\prime
},X\right) =\hat{h}_{2}\left( X^{\prime },X\right)
\end{eqnarray*}%
\begin{eqnarray*}
\hat{V}_{3} &=&f\left( X\right) ,\hat{h}_{3}\left( X^{\prime },X\right)
=h_{1}\left( X,X\right) \\
\hat{V}_{4} &=&r\left( X\right) ,\hat{h}_{4}\left( X^{\prime },X\right)
=h_{2}\left( X,X\right)
\end{eqnarray*}%
The contributions (\ref{NTS}) are translated by:%
\begin{equation*}
-\sigma _{\hat{K}}^{2}\sum_{\eta }\int \Gamma ^{\dag }\left( \hat{S}^{\left(
T\right) },X^{\prime },X\right) \nabla _{\hat{S}_{\eta }^{\left( T\right)
}}^{2}\Gamma \left( \hat{S}^{\left( T\right) },X^{\prime },X\right) d\left( 
\hat{S}^{\left( T\right) },X^{\prime },X\right) -\int \beta \left\vert
\Gamma \left( \hat{S}^{\left( T\right) },X^{\prime },X\right) \right\vert
^{2}d\left( \hat{S}^{\left( T\right) },X^{\prime },X\right)
\end{equation*}%
We also add the contraint:%
\begin{equation*}
\int \lambda \left( X\right) \left( \sum_{\eta }\int \hat{S}_{\eta }^{\left(
T\right) }\left\vert \Gamma \left( \hat{S}^{\left( T\right) },X^{\prime
},X\right) \right\vert ^{2}dX^{\prime }d\hat{S}^{\left( T\right) }-1\right)
\left\vert \Gamma \left( \hat{S}^{\left( T\right) },X^{\prime },X\right)
\right\vert ^{2}d\left( \hat{S}^{\left( T\right) },X^{\prime },X\right)
\end{equation*}%
The factor $\left\vert \Gamma \left( \hat{S}^{\left( T\right) },X^{\prime
},X\right) \right\vert ^{2}$ accounts for the number of agents in sector $%
X^{\prime }$ connected to $X$.

The full action functional is the sum of these contributions:%
\begin{eqnarray*}
&&-\sigma _{\hat{K}}^{2}\sum_{\eta }\int \Gamma ^{\dag }\left( \hat{S}%
^{\left( T\right) },X^{\prime },X\right) \nabla _{\hat{S}_{\eta }^{\left(
T\right) }}^{2}\Gamma \left( \hat{S}^{\left( T\right) },X^{\prime },X\right)
d\left( \hat{S}^{\left( T\right) },X^{\prime },X\right) -\int \beta
\left\vert \Gamma \left( \hat{S}^{\left( T\right) },X^{\prime },X\right)
\right\vert ^{2}d\left( \hat{S}^{\left( T\right) },X^{\prime },X\right) \\
&&+\sum_{\eta }\int \left( \frac{\left( \hat{S}_{\eta }^{\left( T\right)
}\right) ^{2}}{2\hat{h}_{\eta }}-\hat{V}_{\eta }\hat{S}_{\eta }^{\left(
T\right) }-\beta \right) \left\vert \Gamma \left( \hat{S}^{\left( T\right)
},X^{\prime },X\right) \right\vert ^{2}d\left( \hat{S}^{\left( T\right)
},X^{\prime },X\right) \\
&&\int \lambda \left( X\right) \left( \sum_{\eta }\int \hat{S}_{\eta
}^{\left( T\right) }\left\vert \Gamma \left( \hat{S}^{\left( T\right)
},X^{\prime },X\right) \right\vert ^{2}dX^{\prime }d\hat{S}^{\left( T\right)
}-1\right) \left\vert \Gamma \left( \hat{S}^{\left( T\right) },X^{\prime
},X\right) \right\vert ^{2}d\left( \hat{S}^{\left( T\right) },X^{\prime
},X\right)
\end{eqnarray*}

\subsubsection*{A2.2.3 Investors' stakes minimization equations and
background field}

The minimization equations is obtained by the derivative with respect to $%
\Gamma \left( \hat{S}^{\left( T\right) },X^{\prime },X\right) $: 
\begin{eqnarray}
0 &=&\left( \sum_{\eta }\left( -\sigma _{\hat{K}}^{2}\nabla _{\hat{S}_{\eta
}^{\left( T\right) }}^{2}+\frac{\left( \hat{S}_{\eta }^{\left( T\right)
}\right) ^{2}}{2\hat{h}_{\eta }}-\hat{V}_{\eta }\hat{S}_{\eta }^{\left(
T\right) }\right) -\beta \right) \Gamma \left( \hat{S}^{\left( T\right)
},X^{\prime },X\right)  \label{MNT} \\
&&+\lambda \left( X\right) \left( \sum_{\eta }\int \hat{S}_{\eta }^{\left(
T\right) }\left\vert \Gamma \left( \hat{S}^{\left( T\right) },X^{\prime
},X\right) \right\vert ^{2}dX^{\prime }d\hat{S}^{\left( T\right) }-1\right)
\Gamma \left( \hat{S}^{\left( T\right) },X^{\prime },X\right)  \notag \\
&&+\lambda \left( X\right) \Gamma \left( \hat{S}^{\left( T\right)
},X^{\prime },X\right) \hat{S}_{\eta }^{\left( T\right) }\left\Vert \Gamma
\left( \hat{S}^{\left( T\right) },X^{\prime },X\right) \right\Vert ^{2} 
\notag
\end{eqnarray}%
along with the derivative with respect to $\lambda \left( X\right) $ which
implies:%
\begin{equation}
\left( \sum_{\eta }\int \hat{S}_{\eta }^{\left( T\right) }\left\vert \Gamma
\left( \hat{S}^{\left( T\right) },X^{\prime },X\right) \right\vert
^{2}dX^{\prime }d\hat{S}^{\left( T\right) }-1\right) =0  \label{CNTs}
\end{equation}%
and (\ref{MNT}) simplifies as:%
\begin{eqnarray}
0 &=&\left( \sum_{\eta }\left( -\sigma _{\hat{K}}^{2}\nabla _{\hat{S}_{\eta
}^{\left( T\right) }}^{2}+\frac{\left( \hat{S}_{\eta }^{\left( T\right)
}\right) ^{2}}{2\hat{h}_{\eta }}-\hat{V}_{\eta }\hat{S}_{\eta }^{\left(
T\right) }+\lambda \left( X\right) \left\Vert \Gamma \left( \hat{S}^{\left(
T\right) },X^{\prime },X\right) \right\Vert _{X}^{2}\hat{S}_{\eta }^{\left(
T\right) }\right) -\beta \right) \\
&&\times \Gamma \left( \hat{S}^{\left( T\right) },X^{\prime },X\right) 
\notag
\end{eqnarray}%
where:%
\begin{equation*}
\left\Vert \Gamma \left( \hat{S}^{\left( T\right) },X^{\prime },X\right)
\right\Vert _{X}^{2}=\int \left\vert \Gamma \left( \hat{S}^{\left( T\right)
},X^{\prime },X\right) \right\vert ^{2}dX^{\prime }d\hat{S}^{\left( T\right)
}
\end{equation*}%
We can rescale $\lambda \left( X\right) $ and replace:%
\begin{equation*}
\lambda \left( X\right) \left\Vert \Gamma \left( \hat{S}^{\left( T\right)
},X^{\prime },X\right) \right\Vert _{X}^{2}\rightarrow \lambda \left(
X\right)
\end{equation*}%
Solutions of this equation have the form:%
\begin{equation*}
\Gamma _{0,X^{\prime },X}\left( \hat{S}^{\left( T\right) }\right) \Gamma
\left( X^{\prime },X\right)
\end{equation*}%
The function $\Gamma _{0,X}\left( \hat{S}^{\left( T\right) }\right) $ is a
parabolic cylinder function depending on $\beta $. To simplify, we can
adjust the parameter $\beta $ so that:%
\begin{equation*}
\Gamma _{0,X^{\prime },X}\left( \hat{S}^{\left( T\right) }\right) =N\exp
\left( -\sum_{\eta }\frac{\left( \hat{S}_{\eta }^{\left( T\right) }-%
\overline{\hat{S}_{\eta }^{\left( T\right) }}\left( X^{\prime },X\right)
\right) ^{2}}{2\sigma _{\hat{K}}^{2}}\right)
\end{equation*}%
with $N$ a normalization factor. The equation for $\overline{\hat{S}^{\left(
T\right) }}$ is the minimum of:%
\begin{equation*}
\frac{\left( \hat{S}_{\eta }^{\left( T\right) }\right) ^{2}}{2\hat{h}_{\eta
}\left( X^{\prime },X\right) }-\hat{V}_{\eta }\left( X^{\prime },X\right) 
\hat{S}_{\eta }^{\left( T\right) }-\beta +\lambda \left( X\right) \hat{S}%
_{\eta }^{\left( T\right) }
\end{equation*}%
so that:%
\begin{equation}
\overline{\hat{S}_{\eta }^{\left( T\right) }}\left( X^{\prime },X\right) =%
\hat{h}_{\eta }\left( X^{\prime },X\right) \left( \hat{V}_{\eta }\left(
X^{\prime },X\right) +\lambda \left( X\right) \right)
\end{equation}%
The average $\overline{\hat{S}^{\left( T\right) }}$ is also the average:%
\begin{equation}
\overline{\hat{S}_{\eta }^{\left( T\right) }}\left( X^{\prime },X\right) =%
\frac{\int \hat{S}_{\eta }^{\left( T\right) }\left\vert \Gamma _{0,X^{\prime
},X}\left( \hat{S}^{\left( T\right) }\right) \right\vert ^{2}d\hat{S}%
^{\left( T\right) }}{\int \left\vert \Gamma _{0,X^{\prime },X}\left( \hat{S}%
^{\left( T\right) }\right) \right\vert ^{2}d\hat{S}^{\left( T\right) }}=%
\frac{\int \hat{S}_{\eta }^{\left( T\right) }\left\vert \Gamma \left( \hat{S}%
^{\left( T\right) },X^{\prime },X\right) \right\vert ^{2}d\hat{S}^{\left(
T\right) }}{\int \left\vert \Gamma \left( \hat{S}^{\left( T\right)
},X^{\prime },X\right) \right\vert ^{2}d\hat{S}^{\left( T\right) }}
\end{equation}%
and (\ref{MNS}) writes:%
\begin{equation*}
\int \hat{S}_{\eta }^{\left( T\right) }\left\vert \Gamma _{0,X^{\prime
},X}\left( \hat{S}^{\left( T\right) }\right) \right\vert ^{2}d\hat{S}%
^{\left( T\right) }=\hat{h}_{\eta }\left( X^{\prime },X\right) \left( \hat{V}%
_{\eta }\left( X^{\prime },X\right) +\lambda \left( X\right) \right) \int
\left\vert \Gamma _{0,X^{\prime },X}\left( \hat{S}^{\left( T\right) }\right)
\right\vert ^{2}d\hat{S}^{\left( T\right) }
\end{equation*}%
Multiplying by $\Gamma \left( X^{\prime },X\right) $, we integrate with
respect to $X^{\prime }$ and sum over $\eta $ and we find:%
\begin{eqnarray*}
&&\sum_{\eta }\int \hat{S}_{\eta }^{\left( T\right) }\left\vert \Gamma
\left( \hat{S}^{\left( T\right) },X^{\prime },X\right) \right\vert
^{2}dX^{\prime }d\hat{S}^{\left( T\right) } \\
&=&\sum_{\eta }\int \hat{h}_{\eta }\left( X^{\prime },X\right) \hat{V}_{\eta
}\left( X^{\prime },X\right) \left\vert \Gamma \left( \hat{S}^{\left(
T\right) },X^{\prime },X\right) \right\vert ^{2}dX^{\prime }d\hat{S}^{\left(
T\right) }+\lambda \left( X\right) \int \hat{h}_{\eta }\left( X^{\prime
},X\right) \left\vert \Gamma \left( \hat{S}^{\left( T\right) },X^{\prime
},X\right) \right\vert ^{2}dX^{\prime }d\hat{S}^{\left( T\right) }
\end{eqnarray*}%
Then using the constraint (\ref{CNTs}) yields:%
\begin{equation*}
1=\sum_{\eta }\int \hat{h}_{\eta }\left( X^{\prime },X\right) \hat{V}_{\eta
}\left( X^{\prime },X\right) \left\vert \Gamma \left( \hat{S}^{\left(
T\right) },X^{\prime },X\right) \right\vert ^{2}dX^{\prime }d\hat{S}^{\left(
T\right) }+\lambda \left( X\right) \int \hat{h}_{\eta }\left( X^{\prime
},X\right) \left\vert \Gamma \left( \hat{S}^{\left( T\right) },X^{\prime
},X\right) \right\vert ^{2}dX^{\prime }d\hat{S}^{\left( T\right) }
\end{equation*}%
so that:%
\begin{equation*}
\lambda \left( X\right) =\frac{1-\sum_{\eta }\int \hat{h}_{\eta }\left(
X^{\prime },X\right) \hat{V}_{\eta }\left( X^{\prime },X\right) \left\vert
\Gamma \left( \hat{S}^{\left( T\right) },X^{\prime },X\right) \right\vert
^{2}dX^{\prime }d\hat{S}^{\left( T\right) }}{\sum_{\eta }\int \hat{h}_{\eta
}\left( X^{\prime },X\right) \left\vert \Gamma \left( \hat{S}^{\left(
T\right) },X^{\prime },X\right) \right\vert ^{2}dX^{\prime }d\hat{S}^{\left(
T\right) }}
\end{equation*}%
using the normalization over $\hat{S}^{\left( T\right) }$, this becomes:%
\begin{equation*}
\lambda \left( X\right) =\frac{1-\sum_{\eta }\int \hat{h}_{\eta }\left(
X^{\prime },X\right) \hat{V}_{\eta }\left( X^{\prime },X\right) \left\vert
\Gamma \left( X^{\prime },X\right) \right\vert ^{2}dX^{\prime }}{\sum_{\eta
}\int \hat{h}_{\eta }\left( X^{\prime },X\right) \left\vert \Gamma \left(
X^{\prime },X\right) \right\vert ^{2}dX^{\prime }}
\end{equation*}%
In the sequel we consider that the distribution $\left\vert \Gamma \left(
X^{\prime },X\right) \right\vert ^{2}$ varies slowly, so that:%
\begin{equation*}
\lambda \left( X\right) =\frac{1-\sum_{\eta }\int \hat{h}_{\eta }\left(
X^{\prime },X\right) \hat{V}_{\eta }\left( X^{\prime },X\right) dX^{\prime }%
}{\sum_{\eta }\int \hat{h}_{\eta }\left( X^{\prime },X\right) dX^{\prime }}
\end{equation*}%
Using the explicit form for $\hat{V}_{\eta }\left( X^{\prime },X\right) $
leads to the expanded form of $\lambda \left( X\right) $: 
\begin{eqnarray}
&&\lambda \left( X\right) \\
&\rightarrow &\frac{1-\int \hat{h}_{1}\left( X^{\prime },X\right) \hat{f}%
\left( X^{\prime }\right) -\int \hat{h}_{2}\left( X^{\prime },X\right) \hat{r%
}\left( X^{\prime }\right) -h_{1}\left( X,X\right) f\left( X\right)
-h_{2}\left( X,X\right) \bar{r}}{\hat{h}_{1}\left( X\right) +\hat{h}%
_{2}\left( X\right) +h_{1}\left( X\right) +h_{2}\left( X\right) }  \notag
\end{eqnarray}%
where we set:%
\begin{equation*}
\hat{h}_{\eta }\left( X\right) =\int \hat{h}_{\eta }\left( X^{\prime
},X\right) dX^{\prime }
\end{equation*}%
In the sequel we will simplify the notation, and $\overline{\hat{S}_{\eta
}^{\left( T\right) }}\left( X^{\prime },X\right) \rightarrow \hat{S}_{\eta
}^{\left( T\right) }\left( X^{\prime },X\right) $, so that the average
shares equations write:%
\begin{equation}
\hat{S}_{\eta }^{\left( T\right) }\left( X^{\prime },X\right) =\hat{h}_{\eta
}\left( X^{\prime },X\right) \left( \hat{V}_{\eta }\left( X^{\prime
},X\right) +\lambda \left( X\right) \right)  \label{qtm}
\end{equation}%
These minimizations equations for the shares will be studied in appendix 3.

Ultimately, remark that (\ref{VRD}) also writes:%
\begin{equation*}
\hat{S}_{\eta }\left( X^{\prime },X\right) =\int \hat{S}_{\eta }\left\vert
\Gamma \left( S_{1},\hat{S}_{1},S_{2},\hat{S}_{2},X^{\prime },X\right)
\right\vert ^{2}d\left( S_{1},\hat{S}_{1},S_{2},\hat{S}_{2}\right)
\end{equation*}%
for the shares in investors, and:%
\begin{equation*}
S_{\eta }\left( X\right) =\int S_{\eta }\left\vert \Gamma \left( S_{1},\hat{S%
}_{1},S_{2},\hat{S}_{2},X^{\prime },X\right) \right\vert ^{2}d\left( S_{1},%
\hat{S}_{1},S_{2},\hat{S}_{2},X^{\prime }\right)
\end{equation*}%
for the shares in firms.

The constraint on stakes writes:%
\begin{equation*}
\int \left( \hat{S}_{1}\left( X^{\prime },X\right) +\hat{S}_{2}\left(
X^{\prime },X\right) \right) dX^{\prime }+\int \left( S_{1}\left( X^{\prime
},X\right) +S_{2}\left( X^{\prime },X\right) \right) dX^{\prime }=1
\end{equation*}

\subsubsection*{A2.2.4 Micro model for banks' stakes}

Considering banks, we start with the full objective function for the system
of banks:%
\begin{eqnarray}
&&\int dt\sum_{i}\left( \sum_{j}\bar{S}_{1ij}^{B}\bar{f}_{j}+\sum_{j}\bar{S}%
_{2ij}^{B}\bar{r}_{j}-\frac{1}{2}\sum_{j}\frac{\left( \bar{S}_{\eta
ij}\right) ^{2}}{\bar{h}_{\eta _{i}}\left( X_{j}\right) }\right.  \label{MNF}
\\
&&+\left. \sum_{j}\hat{S}_{1ij}^{B}\hat{f}_{j}-\frac{1}{2}\sum_{j}\frac{%
\left( \hat{S}_{1ij}^{B}\right) ^{2}}{\hat{h}_{1i}\left( X_{j}\right) }%
+\sum_{k}S_{1ik}^{B}f_{k}-\frac{1}{2}\sum_{k}\frac{\left( S_{1ik}^{B}\right)
^{2}}{h_{\eta ik}\left( X_{k}\right) }\right)  \notag
\end{eqnarray}%
and for loans:%
\begin{equation}
\int dt\sum_{i}\left( \sum_{j}\hat{S}_{2ij}^{B}\hat{r}_{j}-\frac{1}{2}%
\sum_{j}\frac{\left( \hat{S}_{2ij}^{B}\right) ^{2}}{\hat{h}_{\eta
_{2i}}\left( X_{j}\right) }+\sum_{k}S_{2ik}^{B}\bar{r}_{k}-\frac{1}{2}%
\sum_{k}\frac{\left( S_{2ik}^{B}\right) ^{2}}{h_{\eta ik}\left( X_{k}\right) 
}\right)  \label{MNf}
\end{equation}

\subsubsection*{A2.2.5 Field translation and banks's stakes action functional%
}

Starting with banks shares fields:%
\begin{equation*}
\bar{\Gamma}\left( \bar{S}^{\left( T\right) },X^{\prime },X\right) =\bar{%
\Gamma}\left( \bar{S}_{1},S_{1},\hat{S}_{1},\bar{S}_{2},S_{2},\hat{S}%
_{2},X^{\prime },X\right)
\end{equation*}%
The functions (\ref{MNF}) and (\ref{MNf}) are translated by the action
functionals:%
\begin{equation}
\sum_{\eta }\int \left( \frac{\left( \bar{S}_{\eta }^{T}\right) ^{2}}{2\bar{h%
}_{\eta }^{T}\left( X^{\prime },X\right) }-\bar{V}_{\eta }\bar{S}_{\eta
}^{T}\right) \left\vert \bar{\Gamma}\left( \bar{S}^{\left( T\right)
},X^{\prime },X\right) \right\vert ^{2}d\left( \bar{S}^{\left( T\right)
},X^{\prime },X\right)  \label{CTF}
\end{equation}%
where:%
\begin{equation*}
\bar{V}_{1}\left( X^{\prime },X\right) =\bar{f}\left( X^{\prime }\right) 
\text{, }\bar{V}_{2}\left( X^{\prime },X\right) =\bar{r}\left( X^{\prime
}\right)
\end{equation*}%
\begin{equation*}
\bar{V}_{3}\left( X^{\prime },X\right) =\hat{f}\left( X^{\prime }\right) 
\text{, }\bar{V}_{4}\left( X^{\prime },X\right) =\hat{r}\left( X^{\prime
}\right)
\end{equation*}%
\begin{equation*}
\bar{V}_{5}\left( X^{\prime },X\right) =f\left( X\right) \text{, }\bar{V}%
_{6}\left( X^{\prime },X\right) =r\left( X\right)
\end{equation*}%
and:%
\begin{equation*}
\bar{h}_{1}^{T}\left( X^{\prime },X\right) =\bar{h}_{1}\left( X^{\prime
},X\right) \text{, }\bar{h}_{2}^{T}\left( X^{\prime },X\right) =\bar{h}%
_{2}\left( X^{\prime },X\right)
\end{equation*}%
\begin{equation*}
\hat{h}_{3}^{T}\left( X^{\prime },X\right) =\hat{h}_{1}\left( X^{\prime
},X\right) \text{, }\bar{h}_{4}^{T}\left( X^{\prime },X\right) =\hat{h}%
_{2}\left( X^{\prime },X\right)
\end{equation*}%
\begin{equation*}
\hat{h}_{5}^{T}\left( X^{\prime },X\right) =h_{1}\left( X,X\right) \text{, }%
\bar{h}_{6}^{T}\left( X^{\prime },X\right) =h_{2}\left( X,X\right)
\end{equation*}%
As for investors, we add to the functional (\ref{CTF}) a term accounting for
inertia and a contribution characteristic of the time scale considered for
the collective states: 
\begin{eqnarray*}
&&-\sigma _{\hat{K}}^{2}\sum_{\eta }\int \bar{\Gamma}^{\dag }\left( \bar{S}%
^{\left( T\right) },X^{\prime },X\right) \nabla _{\bar{S}_{\eta }^{\left(
T\right) }}^{2}\bar{\Gamma}\left( \bar{S}^{\left( T\right) },X^{\prime
},X\right) d\left( \bar{S}^{\left( T\right) },X^{\prime },X\right) \\
&&-\int \beta \left\vert \bar{\Gamma}\left( \bar{S}^{\left( T\right)
},X^{\prime },X\right) \right\vert ^{2}d\left( \bar{S}^{\left( T\right)
},X^{\prime },X\right)
\end{eqnarray*}%
Translating the constraints:%
\begin{equation*}
\bar{S}_{1ij}+\bar{S}_{2ij}+\hat{S}_{1ij}^{B}+S_{1ik}^{B}=1
\end{equation*}%
and:%
\begin{equation*}
\hat{S}_{2ij}^{B}+S_{2ik}^{B}=\kappa \left( 1-\sum_{j}\bar{S}%
_{2ij}^{B}\right)
\end{equation*}%
leads to the two following contributions:%
\begin{eqnarray*}
&&\int \lambda \left( X\right) \left\vert \bar{\Gamma}\left( \bar{S}^{\left(
T\right) },X^{\prime },X\right) \right\vert ^{2}d\left( \bar{S}^{\left(
T\right) },X^{\prime },X\right) \\
&&\times \left( \int \left\vert \bar{\Gamma}\left( \bar{S}^{\left( T\right)
},X^{\prime },X\right) \right\vert ^{2}\left( \bar{S}_{1}d\left( \bar{S}%
_{1},X^{\prime }\right) +\bar{S}_{2}d\left( \bar{S}_{2},X^{\prime }\right) +%
\hat{S}_{1}^{B}d\left( \hat{S}_{1}^{B},X^{\prime }\right) +S_{1}^{B}d\left(
S_{1}^{B},X^{\prime }\right) \right) -1\right) \\
&&+\int \lambda ^{\prime }\left( X\right) \left\vert \bar{\Gamma}\left( \bar{%
S}^{\left( T\right) },X^{\prime },X\right) \right\vert ^{2}d\left( \bar{S}%
^{\left( T\right) },X^{\prime },X\right) \\
&&\times \left( \int \left\vert \bar{\Gamma}\left( \bar{S}^{\left( T\right)
},X^{\prime },X\right) \right\vert ^{2}\left( \hat{S}_{2}^{B}d\left( \hat{S}%
_{1}^{B},X^{\prime }\right) +S_{2}^{B}d\left( S_{1}^{B},X^{\prime }\right)
\right) -\kappa \left( 1-\bar{S}_{2}^{B}\right) d\left( S_{2}^{B},X^{\prime
}\right) \right) \\
&&+\int \lambda \left( X\right) \left( \sum_{\eta }\int \bar{S}_{\eta
}\left( X^{\prime },X\right) dX^{\prime }+\int \hat{S}_{1}^{B}\left(
X^{\prime },X\right) dX^{\prime }+S_{1}^{B}\left( X,X\right) -1\right)
\left\vert \bar{\Gamma}\left( \bar{S}^{\left( T\right) },X^{\prime
},X\right) \right\vert ^{2} \\
&&+\int \lambda ^{\prime }\left( X\right) \left( \int \hat{S}_{2}^{B}\left(
X^{\prime },X\right) dX^{\prime }+S_{2}^{B}\left( X,X\right) -\kappa \left(
1-\bar{S}_{\eta }^{B}\left( X\right) \right) \right) \left\vert \bar{\Gamma}%
\left( \bar{S}^{\left( T\right) },X^{\prime },X\right) \right\vert ^{2}
\end{eqnarray*}%
we define the various averages in the states defined by the fields:%
\begin{eqnarray*}
\bar{S}_{\eta }^{B}\left( X^{\prime },X\right) &=&\int \bar{S}_{\eta
}\left\vert \bar{\Gamma}\right\vert ^{2}d\left( \bar{S}_{1},S_{1},\hat{S}%
_{1},\bar{S}_{2},S_{2},\hat{S}_{2},X^{\prime }\right) \\
S_{\eta }^{B}\left( X,X\right) &=&\int S_{\eta }^{B}\left\vert \bar{\Gamma}%
\right\vert ^{2}d\left( \bar{S}_{1},S_{1},\hat{S}_{1},\bar{S}_{2},S_{2},\hat{%
S}_{2},X^{\prime }\right) \\
\hat{S}_{\eta }^{B}\left( X^{\prime },X\right) &=&\int \hat{S}_{\eta
}^{B}\left\vert \bar{\Gamma}\right\vert ^{2}d\left( \bar{S}_{1},S_{1},\hat{S}%
_{1},\bar{S}_{2},S_{2},\hat{S}_{2},X^{\prime }\right)
\end{eqnarray*}%
We also define:%
\begin{equation*}
\bar{S}^{\left( T\right) }\left( X^{\prime },X\right) \left\vert \Gamma
\left( X^{\prime },X\right) \right\vert ^{2}=\int \bar{S}^{\left( T\right)
}\left\vert \Gamma \left( \bar{S},X^{\prime },X\right) \right\vert ^{2}
\end{equation*}%
\begin{eqnarray*}
\bar{S}_{\eta }^{B}\left( X\right) &=&\int \bar{S}_{\eta }^{B}\left(
X^{\prime },X\right) dX^{\prime } \\
\hat{S}_{\eta }^{B}\left( X\right) &=&\int \hat{S}_{\eta }^{B}\left(
X^{\prime },X\right) dX^{\prime }
\end{eqnarray*}%
and the constraints on allocation in this context writes:%
\begin{equation*}
\sum_{\eta =1}^{2}\int \bar{S}_{\eta }^{B}\left( X^{\prime },X\right)
dX^{\prime }+\int \hat{S}_{1}^{B}\left( X^{\prime },X\right) dX^{\prime
}+S_{1}^{B}\left( X,X\right) =1
\end{equation*}%
\begin{equation*}
\int \hat{S}_{2}^{B}\left( X^{\prime },X\right) dX^{\prime }+S_{2}^{B}\left(
X,X\right) =\kappa \left( 1-\bar{S}_{2}^{B}\left( X\right) \right)
\end{equation*}%
As a consequence, the contraint rewrite in terms of fields:

\begin{eqnarray*}
&&\int \lambda \left( X\right) \left( \sum_{\eta =1}^{2}\int \bar{S}_{\eta
}^{B}\left( X^{\prime },X\right) dX^{\prime }+\int \hat{S}_{1}^{B}\left(
X^{\prime },X\right) dX^{\prime }+S_{1}^{B}\left( X,X\right) -1\right)
\left\vert \bar{\Gamma}\left( \bar{S}^{\left( T\right) },X^{\prime
},X\right) \right\vert ^{2} \\
&&+\int \lambda ^{\prime }\left( X\right) \left( \int \hat{S}_{2}^{B}\left(
X^{\prime },X\right) dX^{\prime }+S_{2}^{B}\left( X,X\right) -\kappa \left(
1-\bar{S}_{\eta }^{B}\left( X\right) \right) \right) \left\vert \bar{\Gamma}%
\left( \bar{S}^{\left( T\right) },X^{\prime },X\right) \right\vert ^{2}
\end{eqnarray*}%
The action functional for shars is ths defined by:%
\begin{eqnarray}
S\left( \bar{\Gamma}\right) &=&-\sigma _{\hat{K}}^{2}\sum_{\eta }\int \bar{%
\Gamma}^{\dag }\left( \bar{S}^{\left( T\right) },X^{\prime },X\right) \nabla
_{\bar{S}_{\eta }^{\left( T\right) }}^{2}\bar{\Gamma}\left( \bar{S}^{\left(
T\right) },X^{\prime },X\right) d\left( \bar{S}^{\left( T\right) },X^{\prime
},X\right)  \label{CFB} \\
&&+\int \left( \sum_{\eta }\left( \frac{\left( \bar{S}_{\eta }^{\left(
T\right) }\right) ^{2}}{2\hat{h}_{\eta }\left( X^{\prime },X\right) }-\bar{V}%
_{\eta }\bar{S}_{\eta }^{\left( T\right) }\right) -\beta \right) \left\vert 
\bar{\Gamma}\left( \bar{S}^{\left( T\right) },X^{\prime },X\right)
\right\vert ^{2}d\left( \bar{S}^{\left( T\right) },X^{\prime },X\right) 
\notag \\
&&+\int \lambda \left( X\right) \left( \sum_{\eta =1}^{2}\int \bar{S}_{\eta
}^{B}\left( X^{\prime },X\right) dX^{\prime }+\int \hat{S}_{1}^{B}\left(
X^{\prime },X\right) dX^{\prime }+S_{1}^{B}\left( X,X\right) -1\right)
\left\vert \bar{\Gamma}\left( \bar{S}^{\left( T\right) },X^{\prime
},X\right) \right\vert ^{2}  \notag \\
&&+\int \lambda ^{\prime }\left( X\right) \left( \int \hat{S}_{2}^{B}\left(
X^{\prime },X\right) dX^{\prime }+S_{2}^{B}\left( X,X\right) -\kappa \left(
1-\bar{S}_{\eta }^{B}\left( X\right) \right) \right) \left\vert \bar{\Gamma}%
\left( \bar{S}^{\left( T\right) },X^{\prime },X\right) \right\vert ^{2} 
\notag
\end{eqnarray}

\begin{equation*}
\bar{S}^{\left( T\right) }=\left[ \bar{S}_{1},S_{1},\hat{S}_{1},\bar{S}%
_{2},S_{2},\hat{S}_{2}\right]
\end{equation*}

\subsubsection*{A2.2.6 Banks' stakes minimization equations and background
field}

The minimization equations for investors has been presented above (see (\ref%
{MNT})). The minimization equations for (\ref{CFB}) are similar. They are
obtained by using the derivatives with respect to $\Gamma \left( \hat{S}%
^{\left( T\right) },X^{\prime },X\right) $, $\lambda \left( X\right) $ and $%
\lambda ^{\prime }\left( X\right) $. These two derivatives implement the
constraints. We find:%
\begin{eqnarray*}
&&-\sigma _{\hat{K}}^{2}\sum_{\eta }\nabla _{\bar{S}_{\eta }^{\left(
T\right) }}^{2}\bar{\Gamma}\left( \bar{S}^{\left( T\right) },X^{\prime
},X\right) \\
&&+\left( \sum_{\eta }\left( \frac{\left( \bar{S}_{\eta }^{\left( T\right)
}\right) ^{2}}{2\bar{w}_{\eta }\left( X^{\prime },X\right) }-\bar{V}_{\eta
}^{\left( T\right) }\bar{S}_{\eta }^{T}+\lambda _{\eta }\left( X\right) \bar{%
S}_{\eta }^{T}\right) -\beta \right) \bar{\Gamma}\left( \bar{S}^{\left(
T\right) },X^{\prime },X\right)
\end{eqnarray*}%
with:%
\begin{eqnarray*}
\lambda _{\eta }\left( X\right) &=&\lambda \left( X\right) \text{ for }\eta
=1,2,3,5 \\
\lambda _{\eta }\left( X\right) &=&\lambda ^{\prime }\left( X\right) \text{
for }\eta =5,6
\end{eqnarray*}%
and where we rescaled:%
\begin{eqnarray*}
\lambda \left( X\right) \left\Vert \bar{\Gamma}\left( \bar{S}^{\left(
T\right) },X^{\prime },X\right) \right\Vert _{X}^{2} &\rightarrow &\lambda
\left( X\right) \\
\lambda ^{\prime }\left( X\right) \left\Vert \bar{\Gamma}\left( \bar{S}%
^{\left( T\right) },X^{\prime },X\right) \right\Vert _{X}^{2} &\rightarrow
&\lambda ^{\prime }\left( X\right)
\end{eqnarray*}%
with:%
\begin{equation*}
\left\Vert \bar{\Gamma}\left( \bar{S}^{\left( T\right) },X^{\prime
},X\right) \right\Vert _{X}^{2}=\int \left\vert \bar{\Gamma}\left( \bar{S}%
^{\left( T\right) },X^{\prime },X\right) \right\vert ^{2}d\left( X^{\prime
}\right) d\bar{S}^{\left( T\right) }
\end{equation*}%
As in Gosselin and Lotz (2025), we can adjust $\beta $, so that the solution
to the minimization equations have the form:%
\begin{equation*}
\Gamma _{0,X^{\prime },X}\left( \bar{S}^{\left( T\right) }\right) \Gamma
\left( X^{\prime },X\right)
\end{equation*}%
where:%
\begin{equation*}
\Gamma _{0,X^{\prime },X}\left( \bar{S}^{\left( T\right) }\right) =N\exp
\left( -\sum_{\eta }\frac{\left( \bar{S}_{\eta }^{\left( T\right) }-%
\overline{\bar{S}_{\eta }^{\left( T\right) }}\left( X^{\prime },X\right)
\right) ^{2}}{2\sigma _{\hat{K}}^{2}}\right)
\end{equation*}%
and the average $\overline{\bar{S}_{\eta }^{\left( T\right) }}$ is given by:%
\begin{equation}
\overline{\bar{S}_{\eta }^{\left( T\right) }}\left( X^{\prime },X\right) =%
\bar{w}_{\eta }\left( X^{\prime },X\right) \left( \bar{V}_{\eta }\left(
X^{\prime },X\right) +\lambda _{\eta }\left( X\right) \right)  \label{MNS}
\end{equation}%
The average $\overline{\hat{S}^{\left( T\right) }}$ s also the average:%
\begin{eqnarray}
\overline{\bar{S}_{\eta }^{\left( T\right) }}\left( X^{\prime },X\right) &=&%
\frac{\int \bar{S}_{\eta }^{\left( T\right) }\left\vert \Gamma _{0,X^{\prime
},X}\left( \bar{S}^{\left( T\right) }\right) \right\vert ^{2}d\bar{S}%
^{\left( T\right) }}{\int \left\vert \Gamma _{0,X^{\prime },X}\left( \bar{S}%
^{\left( T\right) }\right) \right\vert ^{2}d\bar{S}^{\left( T\right) }}
\label{VRD} \\
&=&\frac{\int \bar{S}_{\eta }^{\left( T\right) }\left\vert \Gamma \left( 
\bar{S}^{\left( T\right) },X^{\prime },X\right) \right\vert ^{2}d\bar{S}%
^{\left( T\right) }}{\int \left\vert \Gamma \left( \bar{S}^{\left( T\right)
},X^{\prime },X\right) \right\vert ^{2}d\bar{S}^{\left( T\right) }}  \notag
\end{eqnarray}

Multiplying by $\Gamma \left( X^{\prime },X\right) $, we integrate with
respect to $X^{\prime }$:%
\begin{eqnarray*}
&&\int \bar{S}_{\eta }^{\left( T\right) }\left\vert \bar{\Gamma}\left( \bar{S%
}^{\left( T\right) },X^{\prime },X\right) \right\vert ^{2}d\left( X^{\prime
}\right) d\bar{S}^{\left( T\right) } \\
&=&\int \bar{w}_{\eta }\left( X^{\prime },X\right) \bar{V}_{\eta }\left( 
\bar{S}^{\left( T\right) },X^{\prime },X\right) \left\vert \bar{\Gamma}%
\left( \bar{S}^{\left( T\right) },X^{\prime },X\right) \right\vert
^{2}d\left( X^{\prime }\right) d\bar{S}^{\left( T\right) } \\
&&+\lambda _{\eta }\left( X\right) \int \bar{w}_{\eta }\left( X^{\prime
},X\right) \left\vert \bar{\Gamma}\left( X^{\prime },X\right) \right\vert
^{2}d\left( X^{\prime }\right) d\bar{S}^{\left( T\right) }
\end{eqnarray*}%
Then the partial summation over $\eta $ and the constraints yield:%
\begin{eqnarray*}
1 &=&\sum_{\eta \in \left\{ 1,2,3,5,\right\} }\int \bar{w}_{\eta }\left(
X^{\prime },X\right) \bar{V}_{\eta }\left( X^{\prime },X\right) \left\vert 
\bar{\Gamma}\left( \bar{S}^{\left( T\right) },X^{\prime },X\right)
\right\vert ^{2}d\left( X^{\prime }\right) d\bar{S}^{\left( T\right) } \\
&&+\lambda \left( X\right) \sum_{\eta \in \left\{ 1,2,3,5,\right\} }\int 
\bar{w}_{\eta }\left( X^{\prime },X\right) \left\vert \bar{\Gamma}\left( 
\bar{S}^{\left( T\right) },X^{\prime },X\right) \right\vert ^{2}d\left(
X^{\prime }\right) d\bar{S}^{\left( T\right) }
\end{eqnarray*}%
\begin{eqnarray*}
&&\kappa \left( 1-\bar{S}_{L}^{B}\left( X\right) \right) \\
&=&\sum_{\eta \in \left\{ 4,6\right\} }\int \bar{w}_{\eta }\left( X^{\prime
},X\right) \bar{V}_{\eta }\left( X^{\prime },X\right) \left\vert \bar{\Gamma}%
\left( \bar{S}^{\left( T\right) },X^{\prime },X\right) \right\vert
^{2}d\left( X^{\prime }\right) d\bar{S}^{\left( T\right) } \\
&&+\lambda ^{\prime }\left( X\right) \sum_{\eta \in \left\{ 4,6\right\}
}\int \bar{w}_{\eta }\left( X^{\prime },X\right) \left\vert \bar{\Gamma}%
\left( \bar{S}^{\left( T\right) },X^{\prime },X\right) \right\vert
^{2}d\left( X^{\prime }\right) d\bar{S}^{\left( T\right) }
\end{eqnarray*}%
Using the normalization over $\hat{S}^{\left( T\right) }$, we find the
Lagrange multipliers:%
\begin{equation*}
\lambda \left( X\right) =\frac{1-\underset{\eta \in \left\{ 1,2,3,5,\right\} 
}{\sum }\int \bar{w}_{\eta }\left( X^{\prime },X\right) \bar{V}_{\eta
}\left( X^{\prime },X\right) \left\vert \bar{\Gamma}\left( X^{\prime
},X\right) \right\vert ^{2}d\left( X^{\prime }\right) }{\underset{\eta \in
\left\{ 1,2,3,5,\right\} }{\sum }\int \bar{w}_{\eta }\left( X^{\prime
},X\right) \bar{V}_{\eta }\left( X^{\prime },X\right) \left\vert \bar{\Gamma}%
\left( X^{\prime },X\right) \right\vert ^{2}d\left( X^{\prime }\right) }
\end{equation*}%
\begin{equation*}
\lambda ^{\prime }\left( X\right) =\frac{\kappa \left( 1-\bar{S}%
_{L}^{B}\left( X\right) \right) -\sum_{\eta \in \left\{ 4,6\right\} }\int 
\bar{w}_{\eta }\left( X^{\prime },X\right) \bar{V}_{\eta }\left( X^{\prime
},X\right) \left\vert \bar{\Gamma}\left( X^{\prime },X\right) \right\vert
^{2}d\left( X^{\prime }\right) d\bar{S}^{\left( T\right) }}{\sum_{\eta \in
\left\{ 4,6\right\} }\int \bar{w}_{\eta }\left( X^{\prime },X\right)
\left\vert \bar{\Gamma}\left( X^{\prime },X\right) \right\vert ^{2}d\left(
X^{\prime }\right) d\bar{S}^{\left( T\right) }}
\end{equation*}%
In the sequel we consider that the distribution $\left\vert \Gamma \left(
X^{\prime },X\right) \right\vert ^{2}$ varies slowly, so that, using the
explicit form for $\hat{V}_{\eta }\left( X^{\prime },X\right) $ leads to the
expanded form of $\lambda \left( X\right) $ and $\lambda ^{\prime }\left(
X\right) $:%
\begin{equation}
\lambda \left( X\right) =\frac{1-\int \bar{w}_{E}\left( X^{\prime },X\right) 
\bar{f}\left( X^{\prime }\right) -\int \bar{w}_{L}\left( X^{\prime
},X\right) \bar{r}\left( X^{\prime }\right) -\int \hat{w}_{E}^{B}\left(
X^{\prime },X\right) \hat{f}\left( X^{\prime }\right) -w_{E}^{B}\left(
X,X\right) f\left( X\right) }{\bar{w}_{E}\left( X\right) +\bar{w}_{E}\left(
X\right) +\hat{w}_{E}^{B}\left( X\right) +w_{E}^{B}\left( X\right) }
\label{LMN}
\end{equation}%
and:%
\begin{equation}
\lambda ^{\prime }\left( X\right) =\frac{\kappa \left( 1-\bar{S}%
_{L}^{B}\left( X\right) \right) -\int \hat{w}_{L}\left( X^{\prime },X\right) 
\hat{r}\left( X^{\prime }\right) -w_{L}\left( X,X\right) \bar{r}}{\hat{w}%
_{L}\left( X\right) +w_{L}\left( X\right) }
\end{equation}%
In the sequel, we will replace:%
\begin{equation*}
\overline{\bar{S}_{\eta }^{\left( T\right) }}\left( X^{\prime },X\right)
\rightarrow \bar{S}_{\eta }^{\left( T\right) }\left( X^{\prime },X\right)
\end{equation*}

\section*{Appendix 3 Solutions in terms of stakes}

\subsection*{A3.1 Formulas for shares between two sectors}

\subsubsection*{A3.1.1 Investors}

For investors' field, the saddle point equations are the same as in the
first part. \ Collective states satisfy the saddle point equations for $%
S\left( \Gamma \right) $. The expanded form of equation (\ref{qtm}) is: 
\begin{eqnarray*}
\hat{S}_{E}\left( X^{\prime },X\right) &=&\hat{w}_{E}\left( X^{\prime
},X\right) \left( \hat{f}\left( X^{\prime }\right) +\lambda \left( X\right)
\right) \\
\hat{S}_{L}\left( X^{\prime },X\right) &=&\hat{w}_{L}\left( X^{\prime
},X\right) \left( \bar{r}\left( X^{\prime },X\right) +\lambda \left(
X\right) \right)
\end{eqnarray*}%
for investors-investors shares, and:%
\begin{eqnarray*}
S_{E}\left( X,X\right) &=&w_{E}\left( X\right) \left( f\left( X\right)
+\lambda \left( X\right) \right) \\
S_{L}\left( X,X\right) &=&w_{E}\left( X\right) \left( \bar{r}\left(
X,X\right) +\lambda \left( X\right) \right)
\end{eqnarray*}%
for investors-firms shares. The Lagrange multiplier $\lambda \left( X\right) 
$ was given in (\ref{LMN}): 
\begin{eqnarray*}
&&\lambda \left( X\right) \\
&\rightarrow &\frac{1-\int \hat{w}_{E}\left( X^{\prime },X\right) \hat{f}%
\left( X^{\prime }\right) -\int \hat{w}_{L}\left( X^{\prime },X\right) \hat{r%
}\left( X^{\prime }\right) -w_{E}\left( X,X\right) f\left( X\right)
-w_{L}\left( X,X\right) \bar{r}}{\hat{w}_{E}\left( X\right) +\hat{w}%
_{L}\left( X\right) +w_{E}\left( X\right) +w_{L}\left( X\right) }
\end{eqnarray*}%
where we set:%
\begin{equation*}
\hat{w}_{\eta }\left( X\right) =\int \hat{w}_{\eta }\left( X^{\prime
},X\right)
\end{equation*}%
The solutions for shares are:%
\begin{eqnarray*}
&&\hat{S}_{E}\left( X^{\prime },X\right) \\
&=&\underline{\hat{S}}_{E}\left( X^{\prime },X\right) +\frac{\hat{w}%
_{E}\left( X^{\prime },X\right) }{\hat{w}_{E}\left( X\right) +\hat{w}%
_{L}\left( X\right) +w_{E}\left( X\right) +w_{L}\left( X\right) } \\
&&\times \left\{ \hat{w}_{E}\left( X\right) \left( \hat{f}\left( X^{\prime
}\right) -\left\langle \hat{f}\left( X^{\prime }\right) \right\rangle _{\hat{%
w}_{E}}\right) +\hat{w}_{L}\left( X\right) \left( \hat{f}\left( X^{\prime
}\right) -\left\langle \hat{r}\left( X^{\prime }\right) \right\rangle _{\hat{%
w}_{L}}\right) \right. \\
&&\left. +w_{E}\left( X\right) \left( \hat{f}\left( X^{\prime }\right)
-f\left( X\right) \right) +w_{L}\left( X\right) \left( \hat{f}\left(
X^{\prime }\right) -\bar{r}\right) \right\}
\end{eqnarray*}%
\begin{eqnarray*}
&&\hat{S}_{E}\left( X^{\prime },X\right) \\
&=&\underline{\hat{S}}_{E}\left( X^{\prime },X\right) +\frac{\hat{w}%
_{E}\left( X^{\prime },X\right) }{\hat{w}_{E}\left( X\right) +\hat{w}%
_{L}\left( X\right) +w_{E}\left( X\right) +w_{L}\left( X\right) } \\
&&\times \left\{ \hat{w}_{E}\left( X\right) \left( \hat{f}\left( X^{\prime
}\right) -\left\langle \hat{f}\left( X^{\prime }\right) \right\rangle _{\hat{%
w}_{E}}\right) +\hat{w}_{L}\left( X\right) \left( \hat{f}\left( X^{\prime
}\right) -\left\langle \hat{r}\left( X^{\prime }\right) \right\rangle _{\hat{%
w}_{L}}\right) \right. \\
&&\left. +w_{E}\left( X\right) \left( \hat{f}\left( X^{\prime }\right)
-f\left( X\right) \right) +w_{L}\left( X\right) \left( \hat{f}\left(
X^{\prime }\right) -\bar{r}\right) \right\}
\end{eqnarray*}%
\begin{eqnarray*}
\hat{S}_{L}\left( X^{\prime },X\right) &=&\underline{\hat{S}}_{L}\left(
X^{\prime },X\right) +\frac{\hat{w}_{L}\left( X^{\prime },X\right) }{\hat{w}%
_{E}\left( X\right) +\hat{w}_{L}\left( X\right) +w_{E}\left( X\right)
+w_{L}\left( X\right) } \\
&&\times \left\{ \hat{w}_{E}\left( X\right) \left( \hat{r}\left( X^{\prime
}\right) -\left\langle \hat{f}\left( X^{\prime }\right) \right\rangle _{\hat{%
w}_{E}}\right) +\hat{w}_{L}\left( X\right) \left( \hat{r}\left( X^{\prime
}\right) -\left\langle \hat{r}\left( X^{\prime }\right) \right\rangle _{\hat{%
w}_{L}}\right) \right. \\
&&\left. +w_{E}\left( X\right) \left( \hat{r}\left( X^{\prime }\right)
-f\left( X\right) \right) +w_{L}\left( X\right) \left( \hat{r}\left(
X^{\prime }\right) -\bar{r}\right) \right\}
\end{eqnarray*}%
with:%
\begin{equation*}
\left\langle F\left( X^{\prime }\right) \right\rangle _{\hat{w}_{\eta }}=%
\frac{\int F\left( X^{\prime }\right) \hat{w}_{\eta }\left( X^{\prime
},X\right) }{\hat{w}_{\eta }\left( X\right) }
\end{equation*}%
for any function $F\left( X^{\prime }\right) $ and:%
\begin{equation*}
\underline{\hat{S}}_{\eta }\left( X^{\prime },X\right) =\frac{\hat{w}_{\eta
}\left( X^{\prime },X\right) }{\hat{w}_{E}\left( X\right) +\hat{w}_{L}\left(
X\right) +w_{E}\left( X\right) +w_{L}\left( X\right) }
\end{equation*}%
The shares of investment in firms are given by:%
\begin{eqnarray*}
&&S_{E}\left( X,X\right) \\
&=&\underline{S}_{E}\left( X,X\right) +w_{E}\left( X\right) \\
&&\times \frac{\hat{w}_{E}\left( X\right) \left( f\left( X\right)
-\left\langle \hat{f}\left( X^{\prime }\right) \right\rangle _{\hat{w}%
_{E}}\right) +\hat{w}_{L}\left( X\right) \left( f\left( X\right)
-\left\langle \hat{r}\left( X^{\prime }\right) \right\rangle _{\hat{w}%
_{L}}\right) +w_{L}\left( X\right) \left( f\left( X\right) -\bar{r}\right) }{%
\hat{w}_{E}\left( X\right) +\hat{w}_{L}\left( X\right) +w_{E}\left( X\right)
+w_{L}\left( X\right) }
\end{eqnarray*}%
for participations, and:%
\begin{eqnarray*}
&&S_{L}\left( X,X\right) \\
&=&\underline{S}_{L}\left( X,X\right) +w_{L}\left( X\right) \\
&&\times \frac{\hat{w}_{E}\left( X\right) \left( \bar{r}\left( X\right)
-\left\langle \hat{f}\left( X^{\prime }\right) \right\rangle _{\hat{w}%
_{E}}\right) +\hat{w}_{L}\left( X\right) \left( \bar{r}\left( X\right)
-\left\langle \hat{r}\left( X^{\prime }\right) \right\rangle _{\hat{w}%
_{L}}\right) +w_{L}\left( X\right) \left( \bar{r}\left( X\right) -f\left(
X\right) \right) }{\hat{w}_{E}\left( X\right) +\hat{w}_{L}\left( X\right)
+w_{E}\left( X\right) +w_{L}\left( X\right) }
\end{eqnarray*}%
for loans, with:%
\begin{equation*}
\underline{S}_{\eta }\left( X,X\right) =\frac{w_{\eta }\left( X^{\prime
},X\right) }{\int \hat{w}_{E}\left( X^{\prime },X\right) +\int \hat{w}%
_{L}\left( X^{\prime },X\right) +w_{E}\left( X\right) +w_{L}\left( X\right) }
\end{equation*}%
The coefficients $\hat{w}_{\alpha }$, $w_{\alpha }$ are endogeneous since
they depend on the uncertainty on returns depending themselves on th $%
\underline{\hat{S}}_{\eta }$ $S_{\eta }$. Solving will be done by detailling
this uncertainties. Before doing so, we simplify slightly by imposing
several assmptions.

First, we normalize $\hat{w}_{E}\left( X\right) +\hat{w}_{L}\left( X\right)
+w_{E}\left( X\right) +w_{L}\left( X\right) =1$ and choose:%
\begin{equation*}
\hat{w}_{E}\left( X^{\prime },X\right) =\hat{w}_{L}\left( X^{\prime
},X\right) =\frac{1}{2}\hat{w}\left( X^{\prime },X\right)
\end{equation*}%
\begin{equation*}
w_{E}\left( X,X\right) =w_{L}\left( X,X\right) =\frac{w\left( X,X\right) }{2}
\end{equation*}%
\begin{equation*}
\hat{w}_{E}\left( X\right) =\hat{w}_{L}\left( X\right) =\frac{\hat{w}\left(
X\right) }{2}
\end{equation*}%
\begin{equation*}
w_{E}\left( X\right) =w_{L}\left( X\right) =\frac{w\left( X\right) }{2}
\end{equation*}%
The constraint on coefficients implies:%
\begin{equation*}
w\left( X\right) =1-\hat{w}\left( X\right)
\end{equation*}%
and the solutions writes ultimately:%
\begin{eqnarray*}
&&\hat{S}_{E}\left( X^{\prime },X\right) \\
&=&\frac{\underline{\hat{S}}\left( X^{\prime },X\right) }{2}+\frac{\hat{w}%
\left( X^{\prime },X\right) }{2}\left( \hat{w}\left( X\right) \left( \hat{f}%
\left( X^{\prime }\right) -\frac{\left\langle \hat{f}\left( X^{\prime
}\right) \right\rangle _{\hat{w}_{E}}+\left\langle \hat{r}\left( X^{\prime
}\right) \right\rangle _{\hat{w}_{L}}}{2}\right) +w\left( X\right) \left( 
\hat{f}\left( X^{\prime }\right) -\frac{f\left( X\right) +r\left( X\right) }{%
2}\right) \right) \\
&=&\frac{\underline{\hat{S}}\left( X^{\prime },X\right) }{2}+\frac{\hat{w}%
\left( X^{\prime },X\right) }{2}\left( \hat{f}\left( X^{\prime }\right) -%
\hat{w}\left( X\right) \frac{\left\langle \hat{f}\left( X^{\prime }\right)
\right\rangle _{\hat{w}_{E}}+\left\langle \hat{r}\left( X^{\prime }\right)
\right\rangle _{\hat{w}_{L}}}{2}-w\left( X\right) \frac{f\left( X\right)
+r\left( X\right) }{2}\right)
\end{eqnarray*}%
\begin{eqnarray*}
&&\hat{S}_{L}\left( X^{\prime },X\right) \\
&=&\frac{\underline{\hat{S}}\left( X^{\prime },X\right) }{2}+\frac{\hat{w}%
\left( X^{\prime },X\right) }{2}\left( \hat{w}\left( X\right) \left( \hat{r}%
\left( X^{\prime }\right) -\frac{\left\langle \hat{f}\left( X^{\prime
}\right) \right\rangle _{\hat{w}_{E}}+\left\langle \hat{r}\left( X^{\prime
}\right) \right\rangle _{\hat{w}_{L}}}{2}\right) +w\left( X\right) \left( 
\hat{r}\left( X^{\prime }\right) -\frac{f\left( X\right) +r\left( X\right) }{%
2}\right) \right) \\
&=&\frac{\underline{\hat{S}}\left( X^{\prime },X\right) }{2}+\frac{\hat{w}%
\left( X^{\prime },X\right) }{2}\left( \hat{r}\left( X^{\prime }\right) -%
\hat{w}\left( X\right) \frac{\left\langle \hat{f}\left( X^{\prime }\right)
\right\rangle _{\hat{w}_{E}}+\left\langle \hat{r}\left( X^{\prime }\right)
\right\rangle _{\hat{w}_{L}}}{2}-w\left( X\right) \frac{f\left( X\right)
+r\left( X\right) }{2}\right)
\end{eqnarray*}%
with:%
\begin{equation*}
\underline{\hat{S}}_{E}\left( X^{\prime },X\right) =\underline{\hat{S}}%
_{L}\left( X^{\prime },X\right) =\frac{\underline{\hat{S}}\left( X^{\prime
},X\right) }{2}=\frac{1}{2}\hat{w}\left( X^{\prime },X\right)
\end{equation*}%
\begin{eqnarray*}
&&\hat{S}\left( X^{\prime },X\right) \\
&=&\underline{\hat{S}}\left( X^{\prime },X\right) +\hat{w}\left( X^{\prime
},X\right) \left( \frac{\hat{f}\left( X^{\prime }\right) +\hat{r}\left(
X^{\prime }\right) }{2}-\hat{w}\left( X\right) \frac{\left\langle \hat{f}%
\left( X^{\prime }\right) \right\rangle _{\hat{w}_{E}}+\left\langle \hat{r}%
\left( X^{\prime }\right) \right\rangle _{\hat{w}_{L}}}{2}-w\left( X\right) 
\frac{f\left( X\right) +r\left( X\right) }{2}\right)
\end{eqnarray*}%
\begin{eqnarray}
&&S_{E}\left( X,X\right)  \label{SNpa} \\
&=&\frac{\underline{S}\left( X,X\right) }{2}+\frac{w\left( X\right) }{2}%
\left( \hat{w}\left( X\right) \left( f\left( X\right) -\frac{\left\langle 
\hat{f}\left( X^{\prime }\right) \right\rangle _{\hat{w}_{E}}+\left\langle 
\hat{r}\left( X^{\prime }\right) \right\rangle _{\hat{w}_{L}}}{2}\right) +%
\frac{w\left( X\right) }{2}\left( f\left( X\right) -\bar{r}\left( X\right)
\right) \right)  \notag \\
&=&\frac{\underline{S}\left( X,X\right) }{2}+\frac{w\left( X\right) }{2}%
\left( f\left( X\right) -\hat{w}\left( X\right) \frac{\left\langle \hat{f}%
\left( X^{\prime }\right) \right\rangle _{\hat{w}_{E}}+\left\langle \hat{r}%
\left( X^{\prime }\right) \right\rangle _{\hat{w}_{L}}}{2}-w\left( X\right) 
\frac{f\left( X\right) +\bar{r}\left( X\right) }{2}\right)
\end{eqnarray}%
\begin{eqnarray*}
&&S_{L}\left( X,X\right) \\
&=&\frac{\underline{S}\left( X,X\right) }{2}+\frac{w\left( X\right) }{2}%
\left( \hat{w}\left( X\right) \left( \bar{r}\left( X\right) -\frac{%
\left\langle \hat{f}\left( X^{\prime }\right) \right\rangle _{\hat{w}%
_{E}}+\left\langle \hat{r}\left( X^{\prime }\right) \right\rangle _{\hat{w}%
_{L}}}{2}\right) +\frac{w\left( X\right) }{2}\left( \bar{r}\left( X\right)
-f\left( X\right) \right) \right) \\
&=&\frac{\underline{S}\left( X,X\right) }{2}+\frac{w\left( X\right) }{2}%
\left( \bar{r}\left( X\right) -\hat{w}\left( X\right) \frac{\left\langle 
\hat{f}\left( X^{\prime }\right) \right\rangle _{\hat{w}_{E}}+\left\langle 
\hat{r}\left( X^{\prime }\right) \right\rangle _{\hat{w}_{L}}}{2}-w\left(
X\right) \frac{f\left( X\right) +\bar{r}\left( X\right) }{2}\right)
\end{eqnarray*}%
with:%
\begin{equation*}
\underline{S}_{E}\left( X,X\right) =\underline{S}_{L}\left( X,X\right) =%
\frac{\underline{S}\left( X,X\right) }{2}=\frac{w\left( X,X\right) }{2}
\end{equation*}%
\begin{equation}
S\left( X,X\right) =\underline{S}\left( X,X\right) +w\left( X\right) \left( 
\hat{w}\left( X\right) \left( \frac{f\left( X\right) +\bar{r}\left( X\right) 
}{2}-\frac{\left\langle \hat{f}\left( X^{\prime }\right) \right\rangle _{%
\hat{w}_{E}}+\left\langle \hat{r}\left( X^{\prime }\right) \right\rangle _{%
\hat{w}_{L}}}{2}\right) \right)  \label{STpa}
\end{equation}

\subsubsection*{A.3.1.2 Banks}

For banks, they write:

\begin{eqnarray*}
\bar{S}_{E}\left( X^{\prime },X\right) &=&\bar{w}_{E}\left( X^{\prime
},X\right) \left( \bar{f}\left( X^{\prime }\right) +\lambda \left( X\right)
\right) \\
\bar{S}_{L}\left( X^{\prime },X\right) &=&\bar{w}_{L}\left( X^{\prime
},X\right) \left( \bar{r}\left( X^{\prime }\right) +\lambda \left( X\right)
\right)
\end{eqnarray*}%
\begin{eqnarray*}
\hat{S}_{E}^{B}\left( X^{\prime },X\right) &=&\hat{w}_{E}^{B}\left(
X^{\prime },X\right) \left( \hat{f}\left( X^{\prime }\right) +\lambda \left(
X\right) \right) \\
\hat{S}_{L}^{B}\left( X^{\prime },X\right) &=&\hat{w}_{L}^{B}\left(
X^{\prime },X\right) \left( \hat{r}\left( X^{\prime }\right) +\lambda
^{\prime }\left( X\right) \right)
\end{eqnarray*}%
\begin{eqnarray*}
S_{E}^{B}\left( X,X\right) &=&w_{E}^{B}\left( X\right) \left( f\left(
X\right) +\lambda \left( X\right) \right) \\
S_{L}^{B}\left( X,X\right) &=&w_{L}^{B}\left( X\right) \left( r\left(
X\right) +\lambda ^{\prime }\left( X\right) \right)
\end{eqnarray*}%
and the multiplers satisfy:%
\begin{eqnarray*}
&&\lambda \left( X\right) \\
&=&\frac{1-\int \bar{w}_{E}\left( X^{\prime },X\right) \hat{f}\left(
X^{\prime }\right) -\int \bar{w}_{L}\left( X^{\prime },X\right) \hat{r}%
\left( X^{\prime }\right) -\int \hat{w}_{E}^{B}\left( X^{\prime },X\right) 
\hat{f}\left( X^{\prime }\right) -w_{E}^{B}\left( X,X\right) f\left(
X\right) }{\bar{w}_{E}\left( X\right) +\bar{w}_{L}\left( X\right) +\hat{w}%
_{E}^{B}\left( X\right) +w_{E}^{B}\left( X\right) }
\end{eqnarray*}%
and:%
\begin{eqnarray*}
&&\lambda ^{\prime }\left( X\right) \\
&=&\frac{\kappa \left( 1-\bar{S}\left( X\right) \right) -\int \hat{w}%
_{L}^{B}\left( X^{\prime },X\right) \hat{r}\left( X^{\prime }\right)
-w_{L}^{B}\left( X,X\right) \bar{r}}{\hat{w}_{L}^{B}\left( X\right)
+w_{L}^{B}\left( X\right) }
\end{eqnarray*}%
we define the average coefficients:%
\begin{equation*}
\bar{w}_{\eta }\left( X\right) =\int \bar{w}_{\eta }\left( X^{\prime
},X\right)
\end{equation*}%
and:%
\begin{equation*}
\hat{w}_{\eta }^{B}\left( X\right) =\int \hat{w}_{\eta }^{B}\left( X^{\prime
},X\right)
\end{equation*}%
so that $\bar{S}_{E}\left( X^{\prime },X\right) $ and $\bar{S}_{L}\left(
X^{\prime },X\right) $ are given by:

\begin{eqnarray*}
&&\bar{S}_{E}\left( X^{\prime },X\right) \\
&=&\bar{S}_{E}\left( X^{\prime },X\right) +\frac{\bar{w}_{E}\left( X^{\prime
},X\right) }{\bar{w}_{E}\left( X\right) +\bar{w}_{L}\left( X\right) +\hat{w}%
_{E}^{B}\left( X\right) +w_{E}^{B}\left( X\right) } \\
&&\times \left\{ \bar{w}_{E}\left( X\right) \left( \bar{f}\left( X^{\prime
}\right) -\left\langle \bar{f}\left( X^{\prime }\right) \right\rangle _{\bar{%
w}_{E}}\right) +\bar{w}_{L}\left( X\right) \left( \bar{f}\left( X^{\prime
}\right) -\left\langle \bar{r}\left( X^{\prime }\right) \right\rangle _{\bar{%
w}_{L}}\right) \right. \\
&&\left. +\hat{w}_{E}^{B}\left( X\right) \left( \bar{f}\left( X^{\prime
}\right) -\left\langle \hat{f}\left( X^{\prime }\right) \right\rangle _{\hat{%
w}_{E}}\right) +w_{E}^{B}\left( X\right) \left( \bar{f}\left( X^{\prime
}\right) -f\left( X\right) \right) \right\}
\end{eqnarray*}%
\begin{eqnarray*}
&&\bar{S}_{L}\left( X^{\prime },X\right) \\
&=&\bar{S}_{L}\left( X^{\prime },X\right) +\frac{\bar{w}_{L}\left( X^{\prime
},X\right) }{\bar{w}_{E}\left( X\right) +\bar{w}_{L}\left( X\right) +\hat{w}%
_{E}^{B}\left( X\right) +w_{E}^{B}\left( X\right) } \\
&&\times \left\{ \bar{w}_{E}\left( X\right) \left( \bar{r}\left( X^{\prime
}\right) -\left\langle \bar{f}\left( X^{\prime }\right) \right\rangle _{\bar{%
w}_{E}}\right) +\bar{w}_{L}\left( X\right) \left( \bar{r}\left( X^{\prime
}\right) -\left\langle \bar{r}\left( X^{\prime }\right) \right\rangle _{\bar{%
w}_{L}}\right) \right. \\
&&\left. +\hat{w}_{E}^{B}\left( X\right) \left( \bar{r}\left( X^{\prime
}\right) -\left\langle \hat{f}\left( X^{\prime }\right) \right\rangle _{\hat{%
w}_{E}^{B}}\right) +w_{E}^{B}\left( X\right) \left( \bar{r}\left( X^{\prime
}\right) -f\left( X\right) \right) \right\}
\end{eqnarray*}%
Defining the weighted averages of a function by:%
\begin{equation*}
\left\langle F\left( X^{\prime }\right) \right\rangle _{\bar{w}_{\eta }}=%
\frac{\int F\left( X^{\prime }\right) \bar{w}_{\eta }\left( X^{\prime
},X\right) }{\bar{w}_{\eta }\left( X\right) }
\end{equation*}%
\begin{equation*}
\left\langle F\left( X^{\prime }\right) \right\rangle _{\hat{w}_{\eta }}=%
\frac{\int F\left( X^{\prime }\right) \hat{w}_{\eta }^{B}\left( X^{\prime
},X\right) }{\hat{w}_{\eta }^{B}\left( X\right) }
\end{equation*}%
for any functions $F\left( X^{\prime }\right) $, $F\left( X^{\prime }\right) 
$ and:%
\begin{equation*}
\underline{\bar{S}}_{\eta }\left( X^{\prime },X\right) =\frac{\bar{w}_{\eta
}\left( X^{\prime },X\right) }{\bar{w}_{E}\left( X\right) +\bar{w}_{L}\left(
X\right) +\hat{w}_{E}^{B}\left( X\right) +w_{E}^{B}\left( X\right) }
\end{equation*}%
the shares $\hat{S}_{E}^{B}\left( X^{\prime },X\right) $ and $\hat{S}%
_{L}^{B}\left( X^{\prime },X\right) $ are given by:

\begin{eqnarray*}
&&\hat{S}_{E}^{B}\left( X^{\prime },X\right) \\
&=&\underline{\hat{S}}_{E}^{B}\left( X^{\prime },X\right) +\frac{\hat{w}%
_{E}^{B}\left( X^{\prime },X\right) }{\bar{w}_{E}\left( X\right) +\bar{w}%
_{L}\left( X\right) +\hat{w}_{E}^{B}\left( X\right) +w_{E}^{B}\left(
X\right) } \\
&&\times \left\{ \bar{w}_{E}\left( X\right) \left( \hat{f}\left( X^{\prime
}\right) -\left\langle \bar{f}\left( X^{\prime }\right) \right\rangle _{\bar{%
w}_{E}}\right) +\bar{w}_{L}\left( X\right) \left( \hat{f}\left( X^{\prime
}\right) -\left\langle \bar{r}\left( X^{\prime }\right) \right\rangle _{\bar{%
w}_{L}}\right) \right. \\
&&\left. +\hat{w}_{E}^{B}\left( X\right) \left( \hat{f}\left( X^{\prime
}\right) -\left\langle \hat{f}\left( X^{\prime }\right) \right\rangle _{\hat{%
w}_{E}}\right) +w_{E}^{B}\left( X\right) \left( \hat{f}\left( X^{\prime
}\right) -f\left( X\right) \right) \right\}
\end{eqnarray*}%
and:%
\begin{eqnarray*}
&&\frac{\hat{S}_{L}^{B}\left( X^{\prime },X\right) }{\kappa \left( 1-\bar{S}%
\left( X\right) \right) }=\underline{\hat{S}}_{L}^{B}\left( X^{\prime
},X\right) \\
&&+\frac{\hat{w}_{L}^{B}\left( X^{\prime },X\right) }{\hat{w}_{L}^{B}\left(
X\right) +w_{L}^{B}\left( X\right) }\left[ \hat{w}_{L}^{B}\left( X\right)
\left( \hat{r}\left( X^{\prime }\right) -\left\langle \hat{r}\left(
X^{\prime }\right) \right\rangle _{\hat{w}_{E}}\right) +w_{L}^{B}\left(
X\right) \left( \hat{r}\left( X^{\prime }\right) -\left\langle r\left(
X^{\prime }\right) \right\rangle \right) \right] \\
&\simeq &\underline{\hat{S}}_{L}^{B}\left( X^{\prime },X\right) +\frac{\hat{w%
}_{L}^{B}\left( X^{\prime },X\right) }{\hat{w}_{L}^{B}\left( X\right)
+w_{L}^{B}\left( X\right) }w_{L}^{B}\left( X\right) \left( \hat{r}\left(
X^{\prime }\right) -\left\langle r\left( X^{\prime }\right) \right\rangle
\right)
\end{eqnarray*}%
where:%
\begin{equation*}
\underline{\hat{S}}_{E}\left( X^{\prime },X\right) =\frac{\hat{w}%
_{E}^{B}\left( X^{\prime },X\right) }{\bar{w}_{E}\left( X\right) +\bar{w}%
_{L}\left( X\right) +\hat{w}_{E}^{B}\left( X\right) +w_{E}^{B}\left(
X\right) }
\end{equation*}%
\begin{equation*}
\underline{\hat{S}}_{L}^{B}\left( X^{\prime },X\right) =\frac{\hat{w}%
_{L}^{B}\left( X^{\prime },X\right) }{\hat{w}_{L}^{B}\left( X\right)
+w_{L}^{B}\left( X\right) }
\end{equation*}%
The invested shares in firms are given by:%
\begin{eqnarray*}
&&S_{E}^{B}\left( X,X\right) \\
&=&\underline{S}_{E}^{B}\left( X,X\right) +\frac{w_{E}^{B}\left( X\right) }{%
\bar{w}_{E}\left( X\right) +\bar{w}_{L}\left( X\right) +\hat{w}%
_{E}^{B}\left( X\right) +w_{E}^{B}\left( X\right) } \\
&&\times \left\{ \bar{w}_{E}\left( X\right) \left( f\left( X\right)
-\left\langle \bar{f}\left( X^{\prime }\right) \right\rangle _{\bar{w}%
_{E}}\right) +\bar{w}_{L}\left( X\right) \left( f\left( X\right)
-\left\langle \bar{r}\left( X^{\prime }\right) \right\rangle _{\bar{w}%
_{L}}\right) \right. \\
&&\left. +\hat{w}_{E}^{B}\left( X\right) \left( f\left( X\right)
-\left\langle \hat{f}\left( X^{\prime }\right) \right\rangle _{\hat{w}%
_{E}}\right) \right\}
\end{eqnarray*}%
\begin{eqnarray*}
&&\frac{S_{L}^{B}\left( X,X\right) }{\kappa \left( 1-\bar{S}\left( X\right)
\right) } \\
&=&\underline{S}_{L}^{B}\left( X,X\right) +\frac{w_{L}^{B}\left( X\right) }{%
\hat{w}_{L}^{B}\left( X\right) +w_{L}^{B}\left( X\right) }\left( \hat{w}%
_{L}^{B}\left( X\right) \left( r\left( X\right) -\left\langle \hat{r}\left(
X^{\prime }\right) \right\rangle _{\hat{w}_{L}}\right) +w_{L}^{B}\left(
X\right) \left( r\left( X\right) -\left\langle r\left( X\right)
\right\rangle \right) \right) \\
&\simeq &\underline{S}_{L}^{B}\left( X,X\right) +\frac{w_{L}^{B}\left(
X\right) }{\hat{w}_{L}^{B}\left( X\right) +w_{L}^{B}\left( X\right) }\hat{w}%
_{L}^{B}\left( X\right) \left( r\left( X\right) -\left\langle \hat{r}\left(
X^{\prime }\right) \right\rangle _{\hat{w}_{L}}\right)
\end{eqnarray*}%
with:%
\begin{equation*}
\underline{S}_{E}^{B}\left( X,X\right) =\frac{w_{E}^{B}\left( X,X\right) }{%
\bar{w}_{E}\left( X\right) +\bar{w}_{L}\left( X\right) +\hat{w}%
_{E}^{B}\left( X\right) +w_{E}^{B}\left( X\right) }
\end{equation*}%
\begin{equation*}
\underline{S}_{L}^{B}\left( X,X\right) =\frac{w_{L}^{B}\left( X,X\right) }{%
\hat{w}_{L}^{B}\left( X\right) +w_{L}^{B}\left( X\right) }
\end{equation*}%
The coefficients $\hat{w}_{\alpha }$, $w_{\alpha }$ are endogeneous since
they depend on the uncertainty on returns depending themselves on the $%
\underline{\hat{S}}_{\eta }$ $S_{\eta }$. Solving will be done by detailling
this uncertainties. Before doing so, we simplify slightly by imposing
several assumptions.

First, we normalize: 
\begin{equation*}
\bar{w}_{E}\left( X\right) +\bar{w}_{L}\left( X\right) +\hat{w}%
_{E}^{B}\left( X\right) +w_{E}^{B}\left( X\right) =1
\end{equation*}%
\begin{equation*}
\hat{w}_{L}^{B}\left( X\right) +w_{L}^{B}\left( X\right) =1
\end{equation*}%
and choose for the sake of simplicity:%
\begin{equation*}
\bar{w}_{E}\left( X^{\prime },X\right) =\bar{w}_{L}\left( X^{\prime
},X\right) =\frac{1}{2}\bar{w}\left( X^{\prime },X\right)
\end{equation*}%
which describes that the perceived uncertainty for participations and loans
are equal. This implies that:%
\begin{equation*}
\underline{\bar{S}}_{E}\left( X^{\prime },X\right) =\underline{\bar{S}}%
_{L}\left( X^{\prime },X\right) =\frac{\underline{\bar{S}}\left( X^{\prime
},X\right) }{2}=\frac{1}{2}\bar{w}\left( X^{\prime },X\right)
\end{equation*}%
Then defining: 
\begin{eqnarray*}
\bar{S}_{\eta }\left( X^{\prime }\right) &=&\int \bar{S}_{\eta }\left(
X^{\prime },X\right) \frac{\bar{K}_{X}\left\vert \bar{\Psi}\left( X\right)
\right\vert ^{2}}{\bar{K}_{X^{\prime }}\left\vert \bar{\Psi}\left( X^{\prime
}\right) \right\vert ^{2}}dX \\
&\simeq &\int \bar{S}_{\eta }\left( X^{\prime },X\right) dX\frac{%
\left\langle \bar{K}\right\rangle \left\Vert \bar{\Psi}\right\Vert ^{2}}{%
\bar{K}_{X^{\prime }}\left\vert \bar{\Psi}\left( X^{\prime }\right)
\right\vert ^{2}}
\end{eqnarray*}%
\begin{equation*}
\bar{w}\left( X^{\prime }\right) =\int \bar{w}\left( X^{\prime },X\right) dX
\end{equation*}%
\begin{eqnarray*}
\hat{w}\left( X^{\prime }\right) &=&\int \hat{w}\left( X^{\prime },X\right)
dX \\
\hat{w} &=&\int \hat{w}\left( X\right) dX
\end{eqnarray*}%
the stakes are functions of the returns and risk coefficients. Solving the
return equations will yield the formula for stakes.

\subsection*{A3.2 Banks investments risks}

The weights of inverse uncertainty between banks for both loans and stakes%
\footnote{%
See section 5.2 for details.} are given by:%
\begin{equation}
\left( \bar{w}\left( X^{\prime },X\right) \right) ^{-1}=1+\frac{1}{2}\left( 
\frac{\overline{IRG}\left( X^{\prime },X\right) }{\widehat{IR^{B}}\left(
X^{\prime },X\right) }+\frac{\overline{IRG}\left( X^{\prime },X\right) }{\xi
^{2}}\right)  \label{cfn}
\end{equation}%
and those for banks investing investors are given by:%
\begin{equation}
\left( \hat{w}_{E}^{B}\left( X^{\prime },X\right) \right) ^{-1}=1+2\frac{%
\widehat{IR^{B}}\left( X^{\prime },X\right) }{\overline{IRG}\left( X^{\prime
},X\right) }+\frac{\widehat{IR^{B}}\left( X^{\prime },X\right) }{\xi ^{2}}
\label{cft}
\end{equation}%
where:%
\begin{equation}
w_{E}^{B}\left( X^{\prime },X\right) =1-\bar{w}\left( X^{\prime },X\right) -%
\hat{w}_{E}^{B}\left( X^{\prime },X\right)  \label{cfv}
\end{equation}%
are the risk of investing in investors $X^{\prime }$ and banks $X^{\prime }$%
, respectively. These risks are depend on the level of investments these
agents will themselves realize. These level are weighted by the level of
uncertainty $\gamma $ and $\bar{\gamma}$.

The factor $\overline{IRG}\left( X^{\prime }\right) $ : 
\begin{equation*}
\overline{IRG}\left( X^{\prime }\right) =\left( \frac{\bar{\zeta}^{2}%
\overline{IR}\left( X^{\prime }\right) }{\left\langle \hat{w}_{E}^{\left(
0\right) B}\left( \left( X^{\prime }\right) ^{\prime },X^{\prime }\right)
\right\rangle _{\left( X^{\prime }\right) ^{\prime }}}+\frac{\xi ^{2}}{\zeta
^{2}}\frac{\overline{IR}\left( X^{\prime }\right) }{\widehat{IR^{B}}\left(
X^{\prime }\right) }\right)
\end{equation*}%
stands for the global risk of investing. It combines the relative risk to
invest in a bank $X^{\prime }$ rather than in an investor $X^{\prime }$, $%
\frac{\overline{IR}\left( X^{\prime }\right) }{\widehat{IR^{B}}\left(
X^{\prime }\right) }$, and the general propensity for banks to invest in $%
X^{\prime }$ compared to any other investment $\frac{\bar{\zeta}^{2}\zeta
^{2}\overline{IR}\left( X^{\prime }\right) }{\left\langle \hat{w}%
_{E}^{\left( 0\right) B}\left( \left( X^{\prime }\right) ^{\prime
},X^{\prime }\right) \right\rangle _{\left( X^{\prime }\right) ^{\prime }}}$.

In the above, the coefficient $\bar{\gamma}$ is the average uncertainty
associated with distance-dependent investment paths for banks: 
\begin{equation*}
\bar{\gamma}^{2}\simeq \left( \frac{1}{\bar{w}_{1}^{\left( 0\right) }\left(
\left( X^{\prime }\right) ^{\prime },X_{m-1}^{\prime }\right) ...\bar{w}%
_{1}^{\left( 0\right) }\left( X_{1}^{\prime },X^{\prime }\right) }\right) ^{%
\frac{1}{m}}
\end{equation*}%
the coefficients $\bar{w}_{1}^{\left( 0\right) }\left( X_{1}^{\prime
},X^{\prime }\right) $\ captures the local component of uncertainty
perceived in neighboring banks, whereas $\zeta ^{2}$\ and $\xi ^{2}$\
capture the global uncertainty in investors and firms returns
respectively.Since they are considered constant across agents, they
characterize the two types of agents' uncertainty

\subsection*{A3.3 Averages}

\subsubsection*{A3.3.1 Average investors stakes}

We denote $z_{0}$ the value of $\left\langle \hat{S}_{E}\left( X^{\prime
},X\right) \right\rangle $ and $x$, the value of \ $\left\langle \bar{S}%
_{E}\left( X^{\prime },X\right) \right\rangle $\ \ for $\left\langle
R_{exc}\left( X\right) \right\rangle =0$. Their values are functions of the
level of uncertainty $\gamma $.

The average shares taken in, and loans granted by, investors in other
investors are, respectively:%
\begin{eqnarray}
\left\langle \hat{S}_{E}\left( X^{\prime },X\right) \right\rangle
&=&\left\langle \hat{S}_{E}\left( X^{\prime }\right) \right\rangle
=z_{0}\left( 1+\frac{1}{2}\frac{z_{0}^{2}}{D}\left\langle R_{exc}\left(
X\right) \right\rangle \right) \\
&&  \notag \\
\left\langle \hat{S}_{L}\left( X^{\prime },X\right) \right\rangle
&=&z_{0}\left( 1-\frac{1}{2}\left( 1-\frac{z_{0}^{2}}{D}\right) \left\langle
R_{exc}\left( X\right) \right\rangle \right)  \notag
\end{eqnarray}%
where:%
\begin{equation*}
D=1-5z_{0}+8z_{0}^{2}
\end{equation*}%
and the total average stake is:%
\begin{eqnarray}
\left\langle \hat{S}\left( X^{\prime },X\right) \right\rangle
&=&\left\langle \hat{S}_{E}\left( X^{\prime },X\right) \right\rangle
+\left\langle \hat{S}_{L}\left( X^{\prime },X\right) \right\rangle \\
&=&2z_{0}\left( 1-\frac{1}{4}\frac{\left( 1-3z_{0}\right) \left(
1-2z_{0}\right) }{D}\left\langle R_{exc}\left( X\right) \right\rangle \right)
\notag
\end{eqnarray}%
The stakes in firms are given by: 
\begin{eqnarray}
\left\langle S_{E}\left( X,X\right) \right\rangle &=&\frac{1-2z_{0}}{2}%
\left( 1+\left( z_{0}\varepsilon \left( z_{0}\right) +\left( \frac{3}{4}%
-z_{0}\right) \right) \left\langle R_{exc}\left( X\right) \right\rangle
\right) \\
&&  \notag \\
\left\langle S_{L}\left( X,X\right) \right\rangle &=&\frac{1-2z_{0}}{2}%
\left( 1+\left( z_{0}-\varepsilon \left( z_{0}\right) \left( \frac{1}{2}%
-z_{0}\right) -\frac{3}{8}\right) \left\langle R_{exc}\left( X\right)
\right\rangle \right)  \notag \\
&&  \notag \\
\left\langle S\left( X,X\right) \right\rangle &=&\left\langle S_{E}\left(
X,X\right) \right\rangle +\left\langle S_{L}\left( X,X\right) \right\rangle 
\notag \\
&=&\left( 1-2z_{0}\right) \left( 1+\left( \frac{1}{2}+\varepsilon \left(
z_{0}\right) \right) z_{0}\left\langle R_{exc}\left( X\right) \right\rangle
\right)  \notag
\end{eqnarray}%
where:%
\begin{equation*}
\varepsilon \left( z_{0}\right) =\frac{z_{0}^{2}\left( 1-4z_{0}\right) }{%
\left( 1-5z_{0}+8z_{0}^{2}\right) }
\end{equation*}%
Under excess return, firms seeking fresh capital will turn to equity
creation rather than debt. Shares in firms $\left\langle S_{E}\left(
X,X\right) \right\rangle $ increase, whereas loans $\left\langle S_{L}\left(
X,X\right) \right\rangle $ decline. Nonetheless, total investment in firms $%
\left\langle S\left( X,X\right) \right\rangle $ increase. Besides, a
relative increase in uncertainty will increase direct investment in firms.

\subsubsection*{A.3.3.2 Average banks stakes}

For banks, the cross-shares in banks are given by:%
\begin{equation}
\left\langle \bar{S}_{E}\left( X^{\prime },X\right) \right\rangle =\left(
1+\left( \frac{1-\left\langle \bar{S}_{E}\right\rangle _{0}}{\left(
1-2\left\langle \bar{S}_{E}\right\rangle \right) }\left\langle S_{E}^{\left(
e\right) }\right\rangle _{0}-\left\langle \bar{S}_{E}^{B}\right\rangle _{0}-%
\frac{\left\langle \hat{S}_{E}^{B}\right\rangle _{0}}{2}\right) \left\langle
R_{exc}\left( X\right) \right\rangle \right) \left\langle \bar{S}%
_{E}\right\rangle _{0}  \label{FRP}
\end{equation}%
and cross-loans by:%
\begin{equation}
\left\langle \bar{S}_{L}\left( X^{\prime },X\right) \right\rangle =\left(
1+\left( \frac{\left\langle \bar{S}_{E}\right\rangle _{0}}{\left(
1-2\left\langle \bar{S}_{E}\right\rangle \right) }\left\langle S_{E}^{\left(
e\right) }\right\rangle _{0}-\left\langle \bar{S}_{E}^{B}\right\rangle _{0}-%
\frac{\left\langle \hat{S}_{E}^{B}\right\rangle _{0}}{2}\right) \left\langle
R_{exc}\left( X\right) \right\rangle \right) \left\langle \bar{S}%
_{E}\right\rangle _{0}  \label{FRDm}
\end{equation}%
\begin{eqnarray}
\left\langle \bar{S}\left( X^{\prime },X\right) \right\rangle
&=&2\left\langle \bar{S}_{E}\left( X^{\prime },X\right) \right\rangle -\frac{%
\left\langle \bar{S}_{E}\left( X^{\prime },X\right) \right\rangle }{\left(
1-2\left\langle \bar{S}_{E}\left( X^{\prime },X\right) \right\rangle \right) 
}\left\langle S_{E}^{\left( e\right) }\left( X,X\right) \right\rangle
_{0}\left\langle R_{exc}\left( X\right) \right\rangle  \label{FRS} \\
&\simeq &2x-2x\left( \frac{\left\langle \hat{S}_{E}^{B}\right\rangle _{0}}{2}%
+\frac{\left\langle \bar{S}_{E}^{B}\right\rangle _{0}}{2\left(
1-\left\langle \bar{S}_{E}\right\rangle _{0}\right) }\right) \left\langle
R_{exc}\left( X\right) \right\rangle  \notag
\end{eqnarray}%
These results differ from the investors case. When firms' returns increase,
banks\ gain to directly invest in firms or investors.\ This is explicit in
the formula for banks shares in firms and investors:%
\begin{eqnarray}
\left\langle S_{E}^{B}\right\rangle &\simeq &1-6x+\frac{7}{4}x^{2}\left(
1+5x\right) \left\langle R_{exc}\left( X\right) \right\rangle \\
&&-\frac{1}{2}\left( 1-x\right) ^{2}x\left\langle \hat{R}_{exc}\left(
X\right) \right\rangle -\frac{1}{4}\left( 1-x\right) x^{2}\left\langle \bar{R%
}_{exc}\left( X\right) \right\rangle  \notag
\end{eqnarray}%
and similarly:%
\begin{eqnarray*}
\left\langle \hat{S}_{E}^{B}\left( X^{\prime },X\right) \right\rangle
&=&1-\left\langle S_{E}^{B}\right\rangle -\left\langle \bar{S}\left(
X^{\prime },X\right) \right\rangle \\
&\simeq &4x+2x\left( \frac{4-11x}{2}\right) \left\langle R_{exc}\left(
X\right) \right\rangle \\
&&+\frac{1}{2}\left( 1-x\right) ^{2}x\left\langle \hat{R}_{exc}\left(
X\right) \right\rangle +\frac{1}{4}\left( 1-x\right) x^{2}\left\langle \bar{R%
}_{exc}\left( X\right) \right\rangle
\end{eqnarray*}%
As expected, banks loans to investors and firms depend positively on the
firms' returns $\left\langle f\left( X\right) \right\rangle $ and on the
investors' shares\ $z_{0}$. They write: 
\begin{eqnarray}
&&\left\langle \hat{S}_{L}^{B}\left( X^{\prime }\right) \right\rangle
\label{LNP} \\
&=&\kappa \left( 1-2z_{0}\right) 2z_{0}\left( 1+z_{0}\left( 1-\frac{\left(
1-4z_{0}\right) \left( 1-2z_{0}\right) }{2\left( 1-5z_{0}+8z_{0}^{2}\right) }%
\right) \left\langle R_{exc}\left( X\right) \right\rangle \right.  \notag \\
&&\left. +\left( 1-2z_{0}\right) \left( \left\langle \hat{r}\left( X^{\prime
}\right) \right\rangle -\left\langle r\left( X\right) \right\rangle \right)
\right)  \notag
\end{eqnarray}%
\begin{eqnarray}
&&\left\langle S_{L}^{B}\left( X,X\right) \right\rangle  \label{LNDM} \\
&=&\kappa \left( 1-2z_{0}\right) ^{2}  \notag \\
&&\times \left( 1+z_{0}\left( 1+\frac{\left( 1-4z_{0}\right) z_{0}}{\left(
1-5z_{0}+8z_{0}^{2}\right) }\right) \left\langle R_{exc}\left( X\right)
\right\rangle +2z_{0}\left( \left\langle r\left( X\right) \right\rangle
-\left\langle \hat{r}\left( X^{\prime }\right) \right\rangle _{\hat{w}%
_{2}}\right) \right)  \notag
\end{eqnarray}%
The increase of returns calls for more capital, both directly and
indirectly, so that loans increase in proportion of the multiplier $\kappa $%
. The increase in loans is nonetheless bounded by the assumption of
decreasing returns.

\subsubsection*{A3.3.3 Average disposable capital for investors and
associated agregate stakes}

Disposable capital for investors is:%
\begin{equation}
\left\langle \hat{K}\right\rangle \left\Vert \hat{\Psi}\right\Vert
^{2}\simeq \frac{\hat{\mu}V\sigma _{\hat{K}}^{2}}{2}\left( \frac{\frac{%
\left\Vert \hat{\Psi}_{0}\right\Vert ^{2}}{\hat{\mu}}\left( 1-\hat{S}\right)
+\tau \left\langle \hat{f}\right\rangle \hat{S}}{\left\langle \hat{f}%
\right\rangle \left( 1-\hat{S}\right) }\right) ^{2}
\end{equation}%
where $\left\langle \hat{f}\right\rangle $ is the average return for
investors:%
\begin{equation}
\left\langle \hat{f}\right\rangle \simeq \left\langle \hat{r}\left(
X^{\prime }\right) \right\rangle +\frac{f_{a}-f_{b}\left( \frac{C_{0}+\bar{r}%
}{f_{1}\left( X\right) }\right) ^{\frac{2}{r}}}{2}  \label{Rpl}
\end{equation}%
and where $f_{a}$ and $f_{b}$, $\tau $ are some coefficients\footnote{%
These coefficients are defined in Apppendix 5.4 in Gosselin and Lotz (2025).
\par
.}, and the parameter $\left\Vert \hat{\Psi}_{0}\right\Vert ^{2}$ represents
the average number of investors per sectors.

In the above,the parameter $\hat{\mu}$\ represents the fluctuations in the
number of agents within each sector. When $\hat{\mu}$\ is high, the number
of agents may fluctuate widely.\ In this case, the average is no longer a
defining term for the level of disposable capital.

The associated average aggregate stakes are:%
\begin{eqnarray*}
\left\langle \hat{S}_{E}\left( X^{\prime }\right) \right\rangle
&=&\left\langle \hat{S}_{E}\left( X^{\prime },X\right) \right\rangle \\
\left\langle \hat{S}_{L}\left( X^{\prime }\right) \right\rangle
&=&\left\langle \hat{S}_{L}\left( X^{\prime },X\right) \right\rangle
\end{eqnarray*}%
\begin{eqnarray*}
\left\langle S_{E}\left( X\right) \right\rangle &=&\left\langle \hat{S}%
_{E}\left( X,X\right) \right\rangle \frac{\left\langle \hat{K}\right\rangle
\left\Vert \hat{\Psi}\right\Vert ^{2}}{\left\langle K\right\rangle
\left\Vert \Psi \right\Vert ^{2}} \\
\left\langle S_{L}\left( X\right) \right\rangle &=&\left\langle S_{L}\left(
X,X\right) \right\rangle \frac{\left\langle \hat{K}\right\rangle \left\Vert 
\hat{\Psi}\right\Vert ^{2}}{\left\langle K\right\rangle \left\Vert \Psi
\right\Vert ^{2}}
\end{eqnarray*}

\subsubsection*{A3.3.4 Disposable capital for banks and associated aggregate
stakes}

The disposable capital for banks $\left\langle \bar{K}\right\rangle
\left\Vert \bar{\Psi}\right\Vert ^{2}$ is given by:%
\begin{equation*}
\left\langle \bar{K}\right\rangle \left\Vert \bar{\Psi}\right\Vert ^{2}=9%
\frac{\sigma _{\hat{K}}^{2}V\left\Vert \bar{\Psi}_{0}\right\Vert ^{4}\left(
1-\bar{S}\right) ^{2}}{2\hat{\mu}\left\langle \bar{f}\right\rangle ^{2}}
\end{equation*}%
where the bank returns are given in first approximation by their returns of
loans:%
\begin{equation*}
\left\langle \bar{f}\right\rangle \simeq \left( 1+\kappa \right) \bar{r}
\end{equation*}

\subsection*{A3.4 Sectoral returns and disposable capital for investors,
banks and firms.}

Having established the average solutions, we now refine the analysis to
identify sector-specific equilibria. In contrast with the previous study,
the incorporation of banks alters the distribution of investors' returns.
The principle of a dual solution---whereby high and low return states
coexist within the same sectors---remains valid for both banks and investors.

For investors, however, such a configuration arises only under specific
conditions. When these conditions are not met, the system converges toward a
unique solution characterized by returns clustered around the average across
all investors. In what follows, we examine separately the return structures
of investors and of banks.

\subsubsection*{A.3.4.1 Solutions of returns equations}

\paragraph*{A.3.4.1.1 Investors returns}

Investors' return per sector depend on three coefficients: the investors'
estimation of risks in investors stakes, denoted $z$ and defined by\footnote{%
The link between $z$ and $z_{0}$, the value of $\left\langle \hat{S}%
_{E}\left( X^{\prime },X\right) \right\rangle $ under the constraint $%
\left\langle f\left( X\right) \right\rangle -\left\langle \bar{r}\left(
X\right) \right\rangle =0$, is given in Appendix 5.4.1 of Gosselin and Lotz
(2025). In first approximation, we can consider $z=z_{0}$.}:%
\begin{equation*}
z=\left\langle \hat{S}_{E}\left( X^{\prime },X\right) \right\rangle
\end{equation*}%
the relative strength of banks' investments in investors, denoted\footnote{%
To the first approximation we identify $x=\left\langle \bar{S}_{E}\left(
X^{\prime },X\right) \right\rangle =\left\langle \bar{S}_{E}\right\rangle
_{0}$ with $\left\langle \bar{S}_{E}\right\rangle _{0}$ defined in section
12.1.3.. First order corrections to this approximation are defined in
Appendix 6.5.2.} $x$:%
\begin{equation*}
x=\left\langle \bar{S}_{E}\left( X^{\prime },X\right) \right\rangle
\end{equation*}%
and the relative average number of banks over that of investors, denoted $P$%
, defined by:%
\begin{equation*}
P=\frac{\left\Vert \bar{\Psi}_{0}\right\Vert ^{2}}{\left\Vert \hat{\Psi}%
_{0}\right\Vert ^{2}}
\end{equation*}%
The higher the ratio $P$, the higher the importance of banking system and
its impact on intermediation. The higher the uncertainty of banks about
investors, the lower is $x$,\ and the lower is the impact of banks on
investors. Under high uncertainty, banks reduce participations and loans.
The higher the coefficient $z$, the higher will be the intermediation among
investors, and consequently, the higher the impact of banks on investors.

Two types of solutions arise depending on these three parameters $P$ and $x$
and $z$.

\paragraph*{First case: a unique solution}

We show that when $P$ is above some threshold\footnote{%
See Appendix 9.}:%
\begin{equation*}
P>Th
\end{equation*}%
with:%
\begin{equation*}
Th=\frac{2\left( 2z-1\right) ^{2}\left( A-1\right) }{\left( 1-2x\right) }+%
\frac{\sqrt{\left( 2\left( 2z-1\right) ^{2}\left( A-1\right) \right)
^{2}-\left( 1-2x\right) \left( \left( A-1\right) \left( 4-4z-\frac{1}{z}%
\right) \right) }}{\left( 1-2x\right) }
\end{equation*}%
and:%
\begin{equation*}
A=\frac{2-24z+112z^{2}-224z^{3}+160z^{4}}{1-9z+26z^{2}-12z^{3}+8z^{4}}
\end{equation*}%
investors' return equations per sector (\ref{Rtn}) have only a unique
solution, which is close to the average solution:%
\begin{equation}
\hat{R}_{exc}\left( X\right) \simeq z\left( \left\langle \hat{R}_{exc}\left(
X^{\prime }\right) \right\rangle \right) +\left( 1-z\right) R_{exc}\left(
X\right)  \label{UniqInv}
\end{equation}

\paragraph*{Second case: two solutions}

On the contrary, when the relative average number of banks is below the
threshold:%
\begin{equation*}
P<Th
\end{equation*}%
investors' return equations per sector (\ref{Rtn}) have two solutions, as in
Gosselin and Lotz (2025). These two solutions\ for investors' excess returns
depend on firms' excess returns per sector and average returns in the
system. They write\footnote{%
See Appendix 7 in Gosselin and Lotz (2025) for the derivation.}:%
\begin{equation}
\hat{R}_{exc}^{H}\left( X\right) =\frac{\hat{R}_{exc}^{H,0}\left( X\right) }{%
2}+\sqrt{\left( \frac{\hat{R}_{exc}^{H,0}\left( X\right) }{2}\right) ^{2}-%
\frac{1-2z}{1-z}\frac{\left\langle \bar{r}\left( X\right) \right\rangle }{%
\left\vert a\left( z,P\right) \right\vert }\hat{R}_{exc}^{L,0}\left(
X\right) }  \label{SLn}
\end{equation}%
and:%
\begin{equation}
\hat{R}_{exc}^{L}\left( X\right) =\frac{\hat{R}_{exc}^{H,0}\left( X\right) }{%
2}-\sqrt{\left( \frac{\hat{R}_{exc}^{H,0}\left( X\right) }{2}\right) ^{2}-%
\frac{1-2z}{1-z}\frac{\left\langle \bar{r}\left( X\right) \right\rangle }{%
\left\vert a\left( z,P\right) \right\vert }\hat{R}_{exc}^{L,0}\left(
X\right) }  \label{Slp}
\end{equation}%
with:%
\begin{equation}
\hat{R}_{exc}^{H,0}\left( X\right) =\frac{\frac{1-2z}{1-z}\left\langle \bar{r%
}\left( X\right) \right\rangle +z\left( b\left( z,L\right) \left\langle \hat{%
R}_{exc}\left( X^{\prime }\right) \right\rangle +\frac{1}{2}\frac{1-2z}{%
\left( z-1\right) ^{2}}R_{exc}\left( X\right) \right) }{2\left\vert a\left(
z,P\right) \right\vert }  \label{dernier1}
\end{equation}%
and:%
\begin{equation}
\hat{R}_{exc}^{L,0}\left( X\right) =z\left\langle \hat{R}_{exc}\left(
X^{\prime }\right) \right\rangle +\frac{\left( 1-z\right) }{2}R_{exc}\left(
X\right)  \label{dernier2}
\end{equation}%
where the functions $a\left( z,P\right) $\ and $b\left( z,L\right) $\ are
defined in Appendix 10.5. Note that the excess return of firms $%
R_{exc}\left( X\right) $\ is given by $\left( f_{1}\left( X\right)
-\left\langle \hat{r}\left( X^{\prime }\right) \right\rangle \right) $\
under constant returns to scale, and by $\frac{1}{2}f_{a}-\frac{1}{2}%
f_{b}\left( \frac{C_{0}+\bar{r}}{f_{1}\left( X\right) }\right) ^{\frac{1}{r}%
} $\ under decreasing returns to scale\footnote{%
See Appendix 8.2.}. The interpretations will be similar in both cases, but
with dampened amplitudes for decreasing returns.

Obviously, the two solutions (\ref{SLn}) and (\ref{Slp}) differ whether we
add or substract the second term:%
\begin{equation}
\sqrt{\left( \frac{\hat{R}_{exc}^{H,0}\left( X\right) }{2}\right) ^{2}-\frac{%
1-2z}{1-z}\frac{\left\langle \bar{r}\left( X\right) \right\rangle }{%
\left\vert a\left( z,P\right) \right\vert }\hat{R}_{exc}^{L,0}\left(
X\right) }  \label{dernier3}
\end{equation}%
When it is added, it is the high-return solution (\ref{SLn}), in which
returns are relatively higher than otherwise. When it is substracted, it
yields the low-return solution (\ref{Slp}).

When excess returns are relatively small compared to the interest rate,
these two solutions can be expressed as:%
\begin{equation}
\hat{R}_{exc}^{H}\left( X\right) =\hat{R}_{exc}^{H,0}\left( X\right) -\hat{R}%
_{exc}^{L,0}\left( X\right)  \label{HRet}
\end{equation}%
and:%
\begin{equation}
\hat{R}_{exc}^{L}\left( X\right) =\hat{R}_{exc}^{L,0}\left( X\right) \left(
1-\frac{2\left\vert a\left( z,P\right) \right\vert \left( 1-z\right) }{%
\left( 1-2z\right) }\left( \frac{\left\langle \hat{R}_{exc}\left( X\right)
\right\rangle }{\left\langle \bar{r}\right\rangle }+\frac{R_{exc}\left(
X\right) }{\left\langle \bar{r}\right\rangle }\right) \right)  \label{LRet}
\end{equation}%
Note that the high-return solution only occurs under some conditions. For
instance, when uncertainty is high, returns $\hat{R}_{exc}^{H}\left(
X\right) $\ would have to be very high to compensate for high uncertainty,
which rules out this solution practically. The high-return solution
therefore only occurs when the risk perception of investors is low.

\paragraph*{A.3.4.1.2 Banks returns}

For banks, there are two solutions that have a similar form to those of
investors:%
\begin{equation*}
\bar{R}_{exc}\left( X\right) =\frac{\bar{R}_{exc}^{H,0}\left( X\right) }{2}%
\pm \sqrt{\left( \left( \frac{\bar{R}_{exc}^{H,0}\left( X\right) }{2}\right)
^{2}-\left( \kappa +2\right) \left( \kappa +1\right) \frac{1-x}{x}\bar{R}%
_{exc}^{L,0}\left( X\right) \right) ^{2}}
\end{equation*}%
with:%
\begin{equation*}
\bar{R}_{exc}^{H,0}\left( X\right) =\left( \kappa +2\right) \left( \kappa
+1\right) \left( \frac{1-x}{2x}\left\langle \bar{r}\left( X^{\prime }\right)
\right\rangle +\frac{\left( \left\langle f\left( X\right) \right\rangle
-\left\langle \bar{r}\left( X^{\prime }\right) \right\rangle \right) }{4}%
+\left( \left\langle \hat{f}\left( X^{\prime }\right) \right\rangle
-\left\langle \bar{r}\left( X^{\prime }\right) \right\rangle \right) x\right)
\end{equation*}%
and:%
\begin{eqnarray*}
&&\bar{R}_{exc}^{L,0}\left( X\right) =\left\langle \bar{S}_{E}\left(
X^{\prime },X\right) \right\rangle _{X^{\prime }}\left\langle \overline{DF}%
\left( X^{\prime }\right) \right\rangle \left( \left\langle \bar{f}\left(
X^{\prime }\right) \right\rangle -\bar{r}\right) \\
&&+\left\langle \hat{S}_{E}^{B}\left( X^{\prime },X\right) \right\rangle
_{X^{\prime }}\left\langle \widehat{DF}\left( X^{\prime }\right)
\right\rangle \left( \left\langle \hat{f}\left( X^{\prime }\right)
\right\rangle -\bar{r}\right) +S_{E}^{B}\left( X,X\right) \left( f_{1}\left(
X\right) _{dr}-\bar{r}\right)
\end{eqnarray*}%
Since $\frac{2x}{\left( \kappa +2\right) \left( \kappa +1\right) }<<1$, the
low-return solution is close to the average:%
\begin{eqnarray*}
&&\bar{R}_{exc}^{L}\left( X\right) \\
&\simeq &\hat{R}_{exc}^{L,0}\left( X\right) \left( 1-\frac{2x}{\left(
1-x\right) \left( \kappa +2\right) \left( \kappa +1\right) }\bar{R}%
_{exc}^{H,0}\left( X\right) \right) \\
&\simeq &\bar{R}_{exc}^{L,0}\left( X\right)
\end{eqnarray*}%
and this rewrites in terms of banks uncertainty $x$ as:%
\begin{equation*}
\bar{R}_{exc}\left( X\right) \simeq x\left( \left\langle \bar{f}\left(
X^{\prime }\right) \right\rangle -\left( \kappa +1\right) \bar{r}\right)
+4x\left( \left\langle \hat{f}\left( X^{\prime }\right) \right\rangle -\bar{r%
}\right) +\left( 1-6x\right) \left( f\left( X\right) -\bar{r}\right)
\end{equation*}%
As an investor, that is loans excluded, banks' returns combine their
investments in other banks, in investors and in firms. The coefficients
depend on uncertainty. For high uncertainty, banks mainly invest in firms,
and diversify under low uncertainty. In this case the shares in investors
are larger than in other banks, since the intermediation through investors
is more direct than a path bank-investors-firms.

The high-return solution writes:%
\begin{equation*}
\bar{R}_{exc}^{H}\left( X\right) \simeq \bar{R}_{exc}^{H,0}\left( X\right)
\end{equation*}%
The returns are written in terms of sectoral returns and their averages. The
model is thus specified by the values of these quantities.

\paragraph*{A.3.4.1.3 Firms returns}

Under decreasing returns to scale, firms' returns are:%
\begin{equation}
f\left( X\right) =\frac{f_{1}\left( X\right) }{\left( K_{X}\left\vert \hat{%
\Psi}\left( X\right) \right\vert ^{2}\right) ^{r}}-\frac{C}{K_{X}}-C_{0}
\label{Frt}
\end{equation}%
These returns depend on the sectoral disposable capital that will be
obtained below.

\subsubsection*{A3.4.2 Disposable capital per sector}

Under decreasing returns to scale, firms' returns are given by (\ref{Frt}).
For each solution (\ref{SLn}), the disposable capital for banks is given by:%
\begin{equation}
\bar{K}_{X}\left\vert \bar{\Psi}\left( X\right) \right\vert ^{2}\simeq
\left\langle \bar{K}\right\rangle \left\Vert \bar{\Psi}\right\Vert
^{2}\left( \frac{\left\langle \bar{f}\right\rangle }{\left( \left(
1-\left\langle \bar{S}\right\rangle \right) \bar{f}\left( X\right)
+\left\langle \bar{S}\left( X^{\prime },X\right) \right\rangle _{X^{\prime
}}\left\langle \bar{f}\right\rangle \right) }\bar{I}_{X/\left\langle
X^{\prime }\right\rangle }\right) ^{2}  \label{Dcf}
\end{equation}%
where:%
\begin{equation*}
\bar{I}_{X/\left\langle X^{\prime }\right\rangle }=\frac{\frac{\left\langle 
\bar{S}\left( X^{\prime },X\right) \right\rangle _{X}}{1-\left\langle \bar{S}%
\left( X^{\prime },X\right) \right\rangle _{X^{\prime }}}}{\frac{%
\left\langle \bar{S}\left( X^{\prime },X\right) \right\rangle }{%
1-\left\langle \bar{S}\left( X^{\prime },X\right) \right\rangle }}
\end{equation*}%
which measures the level of investment in investor $X$\ by other sectors, $%
\frac{\left\langle \bar{S}\left( X^{\prime },X\right) \right\rangle _{X}}{%
1-\left\langle \bar{S}\left( X^{\prime },X\right) \right\rangle _{X^{\prime
}}}$, with respect to the level of investment in the rest of the market, $%
\frac{\left\langle \bar{S}\left( X^{\prime },X\right) \right\rangle }{%
1-\left\langle \bar{S}\left( X^{\prime },X\right) \right\rangle }$. It is
also weighted by the ratio:%
\begin{equation*}
\frac{\left\langle \bar{f}\right\rangle }{\left( \left( 1-\left\langle \bar{S%
}\right\rangle \right) \bar{f}\left( X\right) +\left\langle \bar{S}\left(
X^{\prime },X\right) \right\rangle _{X^{\prime }}\left\langle \bar{f}%
\right\rangle \right) }
\end{equation*}%
that measures a saturation effect: a high level of returns limits the level
of capital that produces this level of return. It is not solely the return $%
\bar{f}\left( X\right) $ that matters, but its weighted combination with
average returns $\left\langle \bar{S}\left( X^{\prime },X\right)
\right\rangle _{X^{\prime }}\left\langle \bar{f}\right\rangle $\ which
translates the share of capital reinvested in the market by banks $X$.

The capital levels per sector for investors are:%
\begin{equation}
\hat{K}_{X}\left\vert \hat{\Psi}\left( X\right) \right\vert ^{2}\simeq
\left\langle \hat{K}\right\rangle \left\Vert \hat{\Psi}\right\Vert
^{2}\left( \frac{\left\langle \hat{g}\right\rangle }{\hat{g}\left( X\right) }%
I_{X/\left\langle X^{\prime }\right\rangle }\right) ^{2}  \label{Dcn}
\end{equation}%
where $\left\langle \hat{g}\right\rangle $ and $\hat{g}\left( X\right) $\
are given by:%
\begin{equation*}
\left\langle \hat{g}\right\rangle =\left\langle \hat{f}\right\rangle +\frac{%
\left( \hat{S}_{E}^{B}+\hat{S}_{L}^{B}\right) }{1-\bar{S}}\frac{\left\langle 
\bar{K}\right\rangle \left\Vert \bar{\Psi}\right\Vert ^{2}}{\left\langle 
\hat{K}\right\rangle \left\Vert \hat{\Psi}\right\Vert ^{2}}\left\langle \bar{%
f}\right\rangle
\end{equation*}%
\begin{equation*}
\hat{g}\left( X\right) =\left( \hat{f}\left( X\right) +\frac{\left(
\left\langle \hat{S}_{E}^{B}\left( X,X^{\prime }\right) \right\rangle
_{X^{\prime }}+\left\langle \hat{S}_{L}^{B}\left( X,X^{\prime }\right)
\right\rangle _{X^{\prime }}\right) \frac{\left\langle \bar{K}\right\rangle
\left\Vert \bar{\Psi}\right\Vert ^{2}}{\left\langle \hat{K}\right\rangle
\left\Vert \hat{\Psi}\right\Vert ^{2}}\left\langle \bar{f}\right\rangle }{%
1-\left\langle \bar{S}\right\rangle }\right) +\left\langle \hat{S}\left(
X^{\prime },X\right) \right\rangle _{X^{\prime }}\left\langle \hat{g}%
\right\rangle
\end{equation*}%
The average investors return are defined by\ (\ref{Rpl}) and investors $X$\
returns $\hat{f}\left( X\right) $ are given by:%
\begin{equation*}
\hat{f}\left( X\right) \simeq \hat{r}\left( X\right) +\frac{1}{2}f_{a}-\frac{%
1}{2}f_{b}\left( \frac{C_{0}+\hat{r}\left( X\right) }{f_{1}\left( X\right) }%
\right) ^{\frac{1}{r}}
\end{equation*}%
Disposable capital (\ref{Dcn}) is proportional to the ratio:%
\begin{equation}
\hat{I}_{X/\left\langle X^{\prime }\right\rangle }=\frac{\frac{\left\langle 
\hat{S}\left( X,X^{\prime }\right) \right\rangle _{X^{\prime }}}{%
1-\left\langle \hat{S}\left( X,X^{\prime }\right) \right\rangle _{X^{\prime
}}}}{\frac{\left\langle \hat{S}\left( X,X^{\prime }\right) \right\rangle }{%
1-\left\langle \hat{S}\left( X,X^{\prime }\right) \right\rangle }}
\label{Idef2}
\end{equation}%
which measures the level of investment in investor $X$\ by other sectors, $%
\frac{\left\langle \hat{S}\left( X,X^{\prime }\right) \right\rangle
_{X^{\prime }}}{1-\left\langle \hat{S}\left( X,X^{\prime }\right)
\right\rangle _{X^{\prime }}}$, with respect to the level of investment in
the rest of the market, $\frac{\left\langle \hat{S}\left( X,X^{\prime
}\right) \right\rangle }{1-\left\langle \hat{S}\left( X,X^{\prime }\right)
\right\rangle }$.

It is also weighted by the ratio $\frac{\left\langle \hat{g}\right\rangle }{%
\hat{g}\left( X\right) }$ that measures a saturation effect: a high level of
returns limits the level of capital that produces this level of return. It
is not solely the return $\hat{f}\left( X\right) $ that matters, but its
weighted combination with average returns $\left\langle \hat{S}\left(
X^{\prime },X\right) \right\rangle _{X^{\prime }}\left\langle \hat{f}%
\right\rangle $\ which translates the share of capital reinvested in the
market by investors $X$.

The disposable capital for firms is given by:%
\begin{equation}
K_{X}\left\vert \Psi \left( X\right) \right\vert ^{2}\simeq \left( 1-\left(
S\left( X,X\right) \frac{\hat{K}_{X}\left\vert \hat{\Psi}\left( X\right)
\right\vert ^{2}}{\left( K_{X}\left\Vert \Psi \left( X\right) \right\Vert
^{2}\right) _{0}}+S^{B}\left( X,X\right) \frac{\bar{K}_{X}\left\vert \bar{%
\Psi}\left( X\right) \right\vert ^{2}}{\left( K_{X}\left\Vert \Psi \left(
X\right) \right\Vert ^{2}\right) _{0}}\right) \right) \left( K_{X}\left\Vert
\Psi \left( X\right) \right\Vert ^{2}\right) _{0}  \label{Dct}
\end{equation}%
where:%
\begin{equation*}
\left( K_{X}\left\Vert \Psi \left( X\right) \right\Vert ^{2}\right)
_{0}=\left( \left( \frac{2\epsilon }{3\sigma _{\hat{K}}^{2}}\right) ^{\frac{r%
}{2}}\frac{f_{1}\left( X\right) }{C_{0}+\bar{r}}\right) ^{\frac{2}{r}}
\end{equation*}%
is the firms $X$ disposable capital in the absence of investor
participation, i.e. when\textbf{\ }$S\left( X,X\right) =0$.

Under both solutions, firms' capital increases with firms' productivity $%
f_{1}\left( X\right) $, and investors' $X$ disposable capital increase with
firms' productivity $f_{1}\left( X\right) $. This refines our results for
the average capital found in step 1, since\ this increase in firms'
productivity in any given sector $X$ increases investments in investors $X$
from other sectors.\textbf{\ }However, investors' disposable capital is
higher under the high-return solution than under the low-return solution,
since investors attract more capital so\ that, under this solution the ratio 
$I_{X/\left\langle X^{\prime }\right\rangle }$\textbf{\ } is particularly
high.

\subsubsection*{A3.4.3 Capital ratio per sector}

\begin{equation}
\frac{\left\langle \bar{K}\right\rangle \left\Vert \bar{\Psi}\right\Vert ^{2}%
}{\bar{K}_{X}\left\vert \bar{\Psi}\left( X\right) \right\vert ^{2}}\simeq
\left( \frac{\left( \left( 1-\left\langle \bar{S}\right\rangle \right) \bar{f%
}\left( X\right) +\left\langle \bar{S}\left( X^{\prime },X\right)
\right\rangle _{X^{\prime }}\left\langle \bar{f}\right\rangle \right) }{%
\left\langle \bar{f}\right\rangle }\bar{I}_{X/\left\langle X^{\prime
}\right\rangle }\right) ^{2}  \label{CR1}
\end{equation}%
and using (\ref{Dcn}), the ratio bank's average-investors' sector is: 
\begin{equation}
\frac{\left\langle \bar{K}\right\rangle \left\Vert \bar{\Psi}\right\Vert ^{2}%
}{\hat{K}_{X}\left\vert \hat{\Psi}\left( X\right) \right\vert ^{2}}\simeq
\left( \frac{\hat{g}\left( X\right) }{\left\langle \bar{g}\right\rangle }%
\right) ^{2}\frac{\left\Vert \bar{\Psi}_{0}\right\Vert ^{4}}{\left\Vert \hat{%
\Psi}_{0}\right\Vert ^{4}\left( 1+\frac{\left\Vert \bar{\Psi}_{0}\right\Vert
^{2}}{\left\Vert \hat{\Psi}_{0}\right\Vert ^{2}}\left\langle \hat{S}%
_{L}^{B}\right\rangle \right) }  \label{CR2}
\end{equation}%
The ratio banks disposable capital to firms disposabl capital:%
\begin{equation}
\frac{\bar{K}_{X}\left\vert \bar{\Psi}\left( X\right) \right\vert ^{2}}{%
K_{X}\left\vert \Psi \left( X\right) \right\vert ^{2}}\simeq \frac{18\sigma
_{\hat{K}}^{2}V\left\Vert \bar{\Psi}_{0}\left( X\right) \right\Vert ^{4}}{%
\left( \bar{f}\left( X\right) +\frac{\left\langle \bar{S}\left( X^{\prime
},X\right) \right\rangle _{X^{\prime }}}{\left( 1-\left\langle \bar{S}%
\right\rangle \right) }\left\langle \bar{f}\right\rangle \right) ^{2}\hat{\mu%
}\left( \left( \frac{2\epsilon }{3\sigma _{\hat{K}}^{2}}\right) ^{\frac{r}{2}%
}\frac{f_{1}\left( X\right) }{C_{0}+\frac{S_{L}\left( X\right) }{%
1-S_{E}\left( X\right) }\bar{r}}\right) ^{\frac{2}{r}}}  \label{CR3}
\end{equation}%
and the ratio investors disposable capital to firms disposable capital:%
\begin{equation*}
\frac{\hat{K}_{X}\left\vert \hat{\Psi}\left( X\right) \right\vert ^{2}}{%
K_{X}\left\vert \Psi \left( X\right) \right\vert ^{2}}\simeq \frac{18\sigma
_{\hat{K}}^{2}V\left\Vert \hat{\Psi}_{0}\left( X\right) \right\Vert ^{4}}{%
\hat{\mu}F_{1}^{2}\left( \left( \frac{2\epsilon }{3\sigma _{\hat{K}}^{2}}%
\right) ^{\frac{r}{2}}\frac{f_{1}\left( X\right) }{C_{0}+\frac{S_{L}\left(
X\right) }{1-S_{E}\left( X\right) }\bar{r}}\right) ^{\frac{2}{r}}}
\end{equation*}%
where $F_{1}$ is defined in Appendix 9.

\subsection*{A3.5 Formula for stakes}

Once the sectoral returns, the sectoral levels of disposable capital and the
uncertainty coefficients (\ref{cfn}), (\ref{cft}), (\ref{cfv}) are obtained,
the cross-sectoral stakes are derived using the stakes equations given in
Appendix 3.1. The formulas for these cross-sectoral stakes involve several
type of relative returns for banks and investors. The detailed results are
not necessary in the present context since we consider the stakes values as
some given parameters from which the remaining fiber variables are defined.
The precise formula are given in Gosselin and Lotz (2025b).

\section*{Appendix 4 Basis Stability}

\subsection*{A4.1 Quadratic order dynamics for banks}

We consider the dynamic fluctuations:%
\begin{eqnarray*}
\left( 
\begin{array}{c}
\delta \bar{S}\left( X^{\prime },X,\theta \right) \\ 
\delta \hat{S}\left( X^{\prime },X,\theta \right) \\ 
\delta S^{T}\left( X,,\theta -1\right)%
\end{array}%
\right) &=&\left[ 
\begin{array}{ccc}
a & 0 & \frac{\bar{h}\left( X^{\prime },X\right) }{2}c \\ 
\frac{\hat{h}\left( X^{\prime },X\right) }{\bar{h}\left( X^{\prime
},X\right) }d & -1 & \frac{\hat{h}\left( X^{\prime },X\right) }{2}f \\ 
\frac{g}{\frac{\bar{h}\left( X^{\prime },X\right) }{2}} & 0 & i%
\end{array}%
\right] \left( 
\begin{array}{c}
\delta \bar{S}\left( X^{\prime },X,\theta -1\right) \\ 
\delta \hat{S}\left( X^{\prime },X,\theta -1\right) \\ 
\delta S^{T}\left( X,,\theta -2\right)%
\end{array}%
\right) \\
&&+\int dX^{\prime \prime }\left[ 
\begin{array}{ccc}
v\left( X,X^{\prime }\right) & w\left( X,X^{\prime }\right) & 0 \\ 
0 & t\left( X,X^{\prime }\right) & 0 \\ 
0 & 0 & 0%
\end{array}%
\right] \left( 
\begin{array}{c}
\delta \bar{S}\left( X^{\prime \prime },X,\theta -1\right) \\ 
\delta \hat{S}\left( X^{\prime \prime },X,\theta -1\right) \\ 
\delta S^{T}\left( X,,\theta -2\right)%
\end{array}%
\right) \\
&&+\left( 
\begin{array}{c}
\delta \bar{S}\left( X^{\prime },X,\theta \right) \\ 
\delta \hat{S}\left( X^{\prime },X,\theta \right) \\ 
\delta S^{T}\left( X,,\theta -1\right)%
\end{array}%
\right) ^{t}\left[ 
\begin{array}{c}
\bar{M} \\ 
\hat{M} \\ 
M^{T}%
\end{array}%
\right] \left( 
\begin{array}{c}
\delta \bar{S}\left( X^{\prime },X,\theta \right) \\ 
\delta \hat{S}\left( X^{\prime },X,\theta \right) \\ 
\delta S^{T}\left( X,,\theta -1\right)%
\end{array}%
\right)
\end{eqnarray*}%
with:%
\begin{eqnarray*}
a &=&\left\{ \left( \frac{1}{2}H\left( X\right) \bar{S}\left( X\right) -%
\frac{\bar{S}\left( X^{\prime },\theta -1\right) \partial _{\bar{f}\left(
X\right) }\left\langle \bar{K}_{X}\right\rangle \left\vert \bar{\Psi}\left(
X\right) \right\vert ^{2}}{\left\langle \bar{K}_{X^{\prime }}\right\rangle
\left\vert \bar{\Psi}\left( X^{\prime }\right) \right\vert ^{2}}\right)
\right. \frac{\left( \bar{f}\left( X\right) -\bar{r}\right) }{1-\left( \bar{S%
}_{1}\left( X^{\prime },\theta -1\right) \right) } \\
&&-\left( H\left( X\right) \bar{S}_{1}\left( X\right) -\frac{\bar{S}%
_{1}\left( X^{\prime },\theta -1\right) \partial _{\bar{f}\left( X\right)
}\left\langle \bar{K}_{X}\right\rangle \left\vert \bar{\Psi}\left( X\right)
\right\vert ^{2}}{\left\langle \bar{K}_{X^{\prime }}\right\rangle \left\vert 
\bar{\Psi}\left( X^{\prime }\right) \right\vert ^{2}}\right) \frac{\left(
1-\left( \bar{S}\left( X,\theta -1\right) \right) \right) \left( \bar{f}%
\left( X\right) -\bar{r}\right) }{\left( 1-\left( \bar{S}_{1}\left(
X^{\prime },\theta -1\right) \right) \right) ^{2}} \\
&&\left. +\left( S_{1}^{B}\left( X,\theta -1\right) +S_{2}^{B}\left(
X,\theta -1\right) \right) \frac{\partial _{\bar{f}\left( X\right) }\bar{K}%
_{X}\left\vert \bar{\Psi}\left( X\right) \right\vert ^{2}}{\bar{K}%
_{X}\left\vert \bar{\Psi}\left( X\right) \right\vert ^{2}}S_{1}^{B}\left(
X,X\right) \frac{\partial f\left( X\right) }{\partial S^{T}\left( X,\theta
-1\right) }\right\}
\end{eqnarray*}%
\begin{eqnarray*}
c &=&\left\{ h_{1}^{B}\left( X\right) \left( \left\langle \bar{h}\left(
X\right) \right\rangle +\left\langle \hat{h}_{1}^{B}\left( X\right)
\right\rangle \right) \left( f\left( X\right) -\bar{r}\right) \frac{\partial
f\left( X\right) }{\partial S^{T}\left( X,\theta -2\right) }-\frac{%
\left\langle \bar{h}\left( X^{\prime },X\right) \right\rangle }{2}%
h_{1}^{B}\left( X\right) \frac{1-\left\langle \bar{S}\left( X^{\prime
}\right) \right\rangle }{1-\left\langle \bar{S}_{1}\left( X^{\prime }\right)
\right\rangle }\left( \left\langle \bar{f}\left( X^{\prime }\right)
\right\rangle -\bar{r}\right) \right. \\
&&-\left\langle \hat{h}_{1}^{B}\left( X^{\prime },X\right) \right\rangle
\left\langle h_{1}^{B}\left( X\right) \right\rangle \frac{1-\left\langle 
\hat{S}\left( X^{\prime }\right) \right\rangle +\left\langle \hat{S}%
_{1}^{B}\left( X^{\prime }\right) \right\rangle +\left\langle \hat{S}%
_{2}^{B}\left( X^{\prime }\right) \right\rangle }{1-\left\langle \hat{S}%
_{1}\left( X^{\prime }\right) \right\rangle +\left\langle \hat{S}%
_{1}^{B}\left( X^{\prime }\right) \right\rangle }\left( \left\langle \hat{f}%
\left( X^{\prime }\right) \right\rangle -\bar{r}\right) \\
&&+S_{1}^{B}\left( X,X\right) \frac{\partial f\left( X\right) }{\partial
S^{T}\left( X,\theta -1\right) }\left\{ h_{1}^{B}\left( X\right) \left(
\left\langle \bar{h}\left( X\right) \right\rangle +\left\langle \hat{h}%
_{1}^{B}\left( X\right) \right\rangle \right) \right. \\
&&-\left( S_{1}^{B}\left( X,\theta -1\right) +S_{2}^{B}\left( X,\theta
-1\right) \right) \frac{\partial _{f\left( X\right) }K_{X}\left\vert \Psi
\left( X\right) \right\vert ^{2}}{K_{X}\left\vert \Psi \left( X\right)
\right\vert ^{2}} \\
&&\left. \left. +\left( \frac{\frac{\hat{h}\left( X\right) }{2}S\left(
X\right) }{1+\left( \hat{h}\left( X\right) \left( \frac{f\left( X\right) +%
\bar{r}\left( X\right) }{2}-\frac{\left\langle \hat{f}\left( X^{\prime
}\right) \right\rangle _{\hat{h}_{1}}+\left\langle \hat{r}\left( X^{\prime
}\right) \right\rangle _{\hat{h}_{2}}}{2}\right) \right) }+\frac{\partial
_{f\left( X\right) }\left( \frac{\hat{K}_{X}\left\vert \hat{\Psi}\left(
X\right) \right\vert ^{2}}{K_{X}\left\vert \Psi \left( X\right) \right\vert
^{2}}\right) S\left( X\right) }{\frac{\hat{K}_{X}\left\vert \hat{\Psi}\left(
X\right) \right\vert ^{2}}{K_{X}\left\vert \Psi \left( X\right) \right\vert
^{2}}}\right) \right\} \right\} \frac{\partial f\left( X\right) }{\partial
S^{T}\left( X,\theta -2\right) }
\end{eqnarray*}

\begin{equation*}
d=S_{1}\left( X,X,\theta -1\right) \left( S_{1}^{B}\left( X,\theta -1\right)
+S_{2}^{B}\left( X,\theta -1\right) \right) \frac{\partial f\left( X\right) 
}{\partial S^{T}\left( X,\theta -1\right) }\frac{\partial _{\bar{f}\left(
X\right) }\bar{K}_{X}\left\vert \bar{\Psi}\left( X\right) \right\vert ^{2}}{%
\bar{K}_{X}\left\vert \bar{\Psi}\left( X\right) \right\vert ^{2}}<<1
\end{equation*}%
\bigskip 
\begin{eqnarray*}
&&f=\left\{ -\frac{\left( 1-\left\langle \hat{S}\left( X^{\prime }\right)
\right\rangle \right) \left( \left\langle \hat{f}\left( X^{\prime }\right)
\right\rangle -\bar{r}\right) }{1-\left\langle \hat{S}_{1}\left( X^{\prime
}\right) \right\rangle }\frac{\left\langle \hat{h}\left( X^{\prime
},X\right) \right\rangle \left\langle h\left( X\right) \right\rangle }{4}%
\right. \\
&&+\frac{\frac{h\left( X\right) }{4}\left( 1+\hat{h}\left( X\right) \right)
S_{1}\left( X\right) }{1+\left( \hat{h}\left( X\right) \left( f\left(
X\right) -\frac{\left\langle \hat{f}\left( X^{\prime }\right) \right\rangle
_{\hat{h}_{1}}+\left\langle \hat{r}\left( X^{\prime }\right) \right\rangle _{%
\hat{h}_{2}}}{2}\right) +\frac{h\left( X\right) }{2}\left( f\left( X\right) -%
\bar{r}\left( X\right) \right) \right) }\left( f\left( X\right) -\bar{r}%
\right) \\
&&+S_{1}\left( X,X,\theta -1\right) \frac{\partial f\left( X\right) }{%
\partial S^{T}\left( X,\theta -1\right) } \\
&&\times \left( h_{1}^{B}\left( X\right) \left( \left\langle \bar{h}\left(
X\right) \right\rangle +\left\langle \hat{h}_{1}^{B}\left( X\right)
\right\rangle \right) -\left( S_{1}^{B}\left( X,\theta -1\right)
+S_{2}^{B}\left( X,\theta -1\right) \right) \frac{\partial _{f\left(
X\right) }K_{X}\left\vert \Psi \left( X\right) \right\vert ^{2}}{%
K_{X}\left\vert \Psi \left( X\right) \right\vert ^{2}}\right. \\
&&\left. \left. +\frac{\frac{\hat{h}\left( X\right) }{2}S\left( X\right) }{%
1+\left( \hat{h}\left( X\right) \left( \frac{f\left( X\right) +\bar{r}\left(
X\right) }{2}-\frac{\left\langle \hat{f}\left( X^{\prime }\right)
\right\rangle _{\hat{h}_{1}}+\left\langle \hat{r}\left( X^{\prime }\right)
\right\rangle _{\hat{h}_{2}}}{2}\right) \right) }+\frac{\partial _{f\left(
X\right) }\left( \frac{\hat{K}_{X}\left\vert \hat{\Psi}\left( X\right)
\right\vert ^{2}}{K_{X}\left\vert \Psi \left( X\right) \right\vert ^{2}}%
\right) S\left( X\right) }{\frac{\hat{K}_{X}\left\vert \hat{\Psi}\left(
X\right) \right\vert ^{2}}{K_{X}\left\vert \Psi \left( X\right) \right\vert
^{2}}}\right) \right\} \\
&&\times \frac{\partial f\left( X\right) }{\partial S^{T}\left( X,\theta
-2\right) }
\end{eqnarray*}%
\begin{eqnarray*}
g &=&\left( S_{1}^{B}\left( X,\theta -1\right) +S_{2}^{B}\left( X,\theta
-1\right) \right) \frac{\partial _{\bar{f}\left( X\right) }\bar{K}%
_{X}\left\vert \bar{\Psi}\left( X\right) \right\vert ^{2}}{\bar{K}%
_{X}\left\vert \bar{\Psi}\left( X\right) \right\vert ^{2}} \\
i &=&\left\{ h_{1}^{B}\left( X\right) \left( \left\langle \bar{h}\left(
X\right) \right\rangle +\left\langle \hat{h}_{1}^{B}\left( X\right)
\right\rangle \right) -\left( S_{1}^{B}\left( X,\theta -1\right)
+S_{2}^{B}\left( X,\theta -1\right) \right) \frac{\partial _{f\left(
X\right) }K_{X}\left\vert \Psi \left( X\right) \right\vert ^{2}}{%
K_{X}\left\vert \Psi \left( X\right) \right\vert ^{2}}\right. \\
&&\left. +\left( \frac{\frac{\hat{h}\left( X\right) }{2}S\left( X\right) }{%
1+\left( \hat{h}\left( X\right) \left( \frac{f\left( X\right) +\bar{r}\left(
X\right) }{2}-\frac{\left\langle \hat{f}\left( X^{\prime }\right)
\right\rangle _{\hat{h}_{1}}+\left\langle \hat{r}\left( X^{\prime }\right)
\right\rangle _{\hat{h}_{2}}}{2}\right) \right) }+\frac{\partial _{f\left(
X\right) }\left( \frac{\hat{K}_{X}\left\vert \hat{\Psi}\left( X\right)
\right\vert ^{2}}{K_{X}\left\vert \Psi \left( X\right) \right\vert ^{2}}%
\right) S\left( X\right) }{\frac{\hat{K}_{X}\left\vert \hat{\Psi}\left(
X\right) \right\vert ^{2}}{K_{X}\left\vert \Psi \left( X\right) \right\vert
^{2}}}\right) \right\} \frac{\partial f\left( X\right) }{\partial
S^{T}\left( X,\theta -2\right) }
\end{eqnarray*}%
and the non local interactions:%
\begin{eqnarray*}
v\left( X,X^{\prime }\right) &=&\frac{\bar{h}\left( X^{\prime \prime
},X^{\prime }\right) }{\bar{h}\left( X^{\prime },X\right) }\bar{S}_{1}\left(
X^{\prime },X\right) \frac{1-\left\langle \bar{S}\left( X^{\prime }\right)
\right\rangle }{1-\left\langle \bar{S}_{1}\left( X^{\prime }\right)
\right\rangle } \\
w\left( X,X^{\prime }\right) &=&\frac{\bar{h}\left( X^{\prime \prime
},X^{\prime }\right) }{\hat{h}\left( X^{\prime },X\right) }\hat{S}%
_{1}^{B}\left( X^{\prime },X\right) \frac{1-\left\langle \hat{S}\left(
X^{\prime }\right) \right\rangle +\left\langle \hat{S}_{1}^{B}\left(
X^{\prime }\right) \right\rangle +\left\langle \hat{S}_{2}^{B}\left(
X^{\prime }\right) \right\rangle }{1-\left\langle \hat{S}_{1}\left(
X^{\prime }\right) \right\rangle +\left\langle \hat{S}_{1}^{B}\left(
X^{\prime }\right) \right\rangle }
\end{eqnarray*}

\begin{equation*}
t\left( X,X^{\prime }\right) =\frac{\hat{h}\left( X^{\prime \prime
},X^{\prime }\right) }{\hat{h}\left( X^{\prime },X\right) }\hat{S}_{1}\left(
X^{\prime },X\right) \frac{1-\left\langle \hat{S}\left( X^{\prime }\right)
\right\rangle }{1-\left\langle \hat{S}_{1}\left( X^{\prime }\right)
\right\rangle }
\end{equation*}%
The quadratic corrections are given by the matrices:

\begin{eqnarray*}
\bar{M} &=&\left[ 
\begin{array}{ccc}
\frac{\delta }{\delta \bar{f}\left( X,\theta -1\right) }a & 0 & \frac{\bar{h}%
\left( X^{\prime },X\right) }{2}\frac{\delta }{\delta \bar{f}\left( X,\theta
-1\right) }c \\ 
\frac{\delta }{\delta \hat{f}\left( X,\theta -1\right) }a & 0 & \frac{\bar{h}%
\left( X^{\prime },X\right) }{2}\frac{\delta }{\delta \hat{f}\left( X,\theta
-1\right) }c \\ 
\frac{\delta }{\delta S^{T}\left( X,,\theta -2\right) }a & 0 & \frac{\bar{h}%
\left( X^{\prime },X\right) }{2}\frac{\delta }{\delta S^{T}\left( X,,\theta
-2\right) }c%
\end{array}%
\right] ,\hat{M}=\left[ 
\begin{array}{ccc}
\frac{\hat{h}\left( X^{\prime },X\right) }{\bar{h}\left( X^{\prime
},X\right) }\frac{\delta }{\delta \bar{f}\left( X,\theta -1\right) }d & 0 & 
\frac{\hat{h}\left( X^{\prime },X\right) }{2}\frac{\delta }{\delta \bar{f}%
\left( X,\theta -1\right) }f \\ 
\frac{\hat{h}\left( X^{\prime },X\right) }{\bar{h}\left( X^{\prime
},X\right) }\frac{\delta }{\delta \hat{f}\left( X,\theta -1\right) }d & 0 & 
\frac{\hat{h}\left( X^{\prime },X\right) }{2}\frac{\delta }{\delta \hat{f}%
\left( X,\theta -1\right) }f \\ 
\frac{\hat{h}\left( X^{\prime },X\right) }{\bar{h}\left( X^{\prime
},X\right) }\frac{\delta }{\delta S^{T}\left( X,,\theta -2\right) }d & 0 & 
\frac{\hat{h}\left( X^{\prime },X\right) }{2}\frac{\delta }{\delta
S^{T}\left( X,,\theta -2\right) }f%
\end{array}%
\right] \\
M^{T} &=&\left[ 
\begin{array}{ccc}
\frac{2}{\bar{h}\left( X^{\prime },X\right) }\frac{\delta }{\delta \bar{f}%
\left( X,\theta -1\right) }g & 0 & \frac{\delta }{\delta \bar{f}\left(
X,\theta -1\right) }i \\ 
\frac{2}{\bar{h}\left( X^{\prime },X\right) }\frac{\delta }{\delta \hat{f}%
\left( X,\theta -1\right) }g & 0 & \frac{\delta }{\delta \hat{f}\left(
X,\theta -1\right) }i \\ 
\frac{2}{\bar{h}\left( X^{\prime },X\right) }\frac{\delta }{\delta
S^{T}\left( X,,\theta -2\right) }g & 0 & \frac{\delta }{\delta S^{T}\left(
X,,\theta -2\right) }i%
\end{array}%
\right]
\end{eqnarray*}%
We assume $\left( \bar{f}\left( X\right) -\bar{r}\right) <<1$ and $\left( 
\hat{f}\left( X\right) -\bar{r}\right) <<1$ due to decreasing returns and
large amounts of loans. We will use:%
\begin{eqnarray*}
&&\delta S_{1}\left( X,X,\theta -1\right) \\
&=&\left( \frac{\frac{\hat{h}\left( X\right) }{2}S_{1}\left( X,X\right) }{%
1+\left( \hat{h}\left( X\right) \left( \frac{f\left( X\right) +\bar{r}\left(
X\right) }{2}-\frac{\left\langle \hat{f}\left( X^{\prime }\right)
\right\rangle _{\hat{h}_{1}}+\left\langle \hat{r}\left( X^{\prime }\right)
\right\rangle _{\hat{h}_{2}}}{2}\right) \right) }\right) \frac{\partial
f\left( X\right) }{\partial S^{T}\left( X,\theta -2\right) }\delta
S^{T}\left( X,\theta -2\right)
\end{eqnarray*}%
writtn as:%
\begin{eqnarray*}
&&\delta S_{1}\left( X,\theta -1\right) \\
&=&\left( \frac{\frac{\hat{h}\left( X\right) }{2}S_{1}\left( X\right) }{%
1+\left( \hat{h}\left( X\right) \left( \frac{f\left( X\right) +\bar{r}\left(
X\right) }{2}-\frac{\left\langle \hat{f}\left( X^{\prime }\right)
\right\rangle _{\hat{h}_{1}}+\left\langle \hat{r}\left( X^{\prime }\right)
\right\rangle _{\hat{h}_{2}}}{2}\right) \right) }+\frac{\partial _{f\left(
X\right) }\left( \frac{\hat{K}_{X}\left\vert \hat{\Psi}\left( X\right)
\right\vert ^{2}}{K_{X}\left\vert \Psi \left( X\right) \right\vert ^{2}}%
\right) S_{1}\left( X\right) }{\frac{\hat{K}_{X}\left\vert \hat{\Psi}\left(
X\right) \right\vert ^{2}}{K_{X}\left\vert \Psi \left( X\right) \right\vert
^{2}}}\right) \\
&&\times \frac{\partial f\left( X\right) }{\partial S^{T}\left( X,\theta
-2\right) }\delta S^{T}\left( X,\theta -2\right) \\
&&+\frac{\partial _{\hat{f}\left( X\right) }\left( \frac{\hat{K}%
_{X}\left\vert \hat{\Psi}\left( X\right) \right\vert ^{2}}{K_{X}\left\vert
\Psi \left( X\right) \right\vert ^{2}}\right) S_{1}\left( X\right) }{\frac{%
\hat{K}_{X}\left\vert \hat{\Psi}\left( X\right) \right\vert ^{2}}{%
K_{X}\left\vert \Psi \left( X\right) \right\vert ^{2}}}\delta \hat{f}\left(
X,\theta -1\right)
\end{eqnarray*}%
\begin{eqnarray*}
&&\delta S\left( X,\theta -1\right) \\
&=&\left( \frac{\frac{\hat{h}\left( X\right) }{2}S\left( X\right) }{1+\left( 
\hat{h}\left( X\right) \left( \frac{f\left( X\right) +\bar{r}\left( X\right) 
}{2}-\frac{\left\langle \hat{f}\left( X^{\prime }\right) \right\rangle _{%
\hat{h}_{1}}+\left\langle \hat{r}\left( X^{\prime }\right) \right\rangle _{%
\hat{h}_{2}}}{2}\right) \right) }+\frac{\partial _{f\left( X\right) }\left( 
\frac{\hat{K}_{X}\left\vert \hat{\Psi}\left( X\right) \right\vert ^{2}}{%
K_{X}\left\vert \Psi \left( X\right) \right\vert ^{2}}\right) S\left(
X\right) }{\frac{\hat{K}_{X}\left\vert \hat{\Psi}\left( X\right) \right\vert
^{2}}{K_{X}\left\vert \Psi \left( X\right) \right\vert ^{2}}}\right) \\
&&\times \frac{\partial f\left( X\right) }{\partial S^{T}\left( X,\theta
-2\right) }\delta S^{T}\left( X,\theta -2\right) \\
&&+\frac{\partial _{\hat{f}\left( X\right) }\left( \frac{\hat{K}%
_{X}\left\vert \hat{\Psi}\left( X\right) \right\vert ^{2}}{K_{X}\left\vert
\Psi \left( X\right) \right\vert ^{2}}\right) S\left( X\right) }{\frac{\hat{K%
}_{X}\left\vert \hat{\Psi}\left( X\right) \right\vert ^{2}}{K_{X}\left\vert
\Psi \left( X\right) \right\vert ^{2}}}\delta \hat{f}\left( X,\theta
-1\right)
\end{eqnarray*}%
\bigskip

\begin{eqnarray*}
&&\delta S_{1}^{B}\left( X,X,\theta -1\right) \\
&=&h_{1}^{B}\left( X\right) \left( \left\langle \bar{h}\left( X\right)
\right\rangle +\left\langle \hat{h}_{1}^{B}\left( X\right) \right\rangle
\right) \frac{\partial f\left( X\right) }{\partial S^{T}\left( X,\theta
-2\right) }\delta S^{T}\left( X,\theta -2\right)
\end{eqnarray*}%
\begin{eqnarray*}
&&\delta S_{1}^{B}\left( X,\theta -1\right) \\
&=&\left( h_{1}^{B}\left( X\right) \left( \left\langle \bar{h}\left(
X\right) \right\rangle +\left\langle \hat{h}_{1}^{B}\left( X\right)
\right\rangle \right) -S_{1}^{B}\left( X,\theta -1\right) \frac{\partial
_{f\left( X\right) }K_{X}\left\vert \Psi \left( X\right) \right\vert ^{2}}{%
K_{X}\left\vert \Psi \left( X\right) \right\vert ^{2}}\right) \frac{\partial
f\left( X\right) }{\partial S^{T}\left( X,\theta -2\right) }\delta
S^{T}\left( X,\theta -2\right) \\
&&+S_{1}^{B}\left( X,\theta -1\right) \frac{\partial _{\bar{f}\left(
X\right) }\bar{K}_{X}\left\vert \bar{\Psi}\left( X\right) \right\vert ^{2}}{%
\bar{K}_{X}\left\vert \bar{\Psi}\left( X\right) \right\vert ^{2}}\delta \bar{%
f}\left( X,\theta -1\right)
\end{eqnarray*}%
\begin{eqnarray*}
\delta S_{2}^{B}\left( X,\theta -1\right) &=&S_{2}^{B}\left( X,\theta
-1\right) \frac{\partial _{\bar{f}\left( X\right) }\bar{K}_{X}\left\vert 
\bar{\Psi}\left( X\right) \right\vert ^{2}}{\bar{K}_{X}\left\vert \bar{\Psi}%
\left( X\right) \right\vert ^{2}}\delta \bar{f}\left( X,\theta -1\right) \\
&&-S_{2}^{B}\left( X,\theta -1\right) \frac{\partial _{f\left( X\right)
}K_{X}\left\vert \Psi \left( X\right) \right\vert ^{2}}{K_{X}\left\vert \Psi
\left( X\right) \right\vert ^{2}}\frac{\partial f\left( X\right) }{\partial
S^{T}\left( X,\theta -2\right) }\delta S^{T}\left( X,\theta -2\right)
\end{eqnarray*}%
Moreover:%
\begin{equation*}
\frac{\partial _{\bar{f}\left( X\right) }\bar{K}_{X}\left\vert \bar{\Psi}%
\left( X\right) \right\vert ^{2}}{\bar{K}_{X}\left\vert \bar{\Psi}\left(
X\right) \right\vert ^{2}}\simeq -\frac{2}{\bar{f}\left( X\right) \left(
1-\left\langle \bar{S}\right\rangle \right) +\left\langle \bar{S}\left(
X^{\prime },X\right) \right\rangle _{X^{\prime }}\left\langle \bar{f}%
\right\rangle }
\end{equation*}%
and:%
\begin{equation*}
\partial _{\bar{f}\left( X\right) }\frac{\partial _{\bar{f}\left( X\right) }%
\bar{K}_{X}\left\vert \bar{\Psi}\left( X\right) \right\vert ^{2}}{\bar{K}%
_{X}\left\vert \bar{\Psi}\left( X\right) \right\vert ^{2}}<<1
\end{equation*}%
for $\kappa >>1$. Ultimately we will need the formula for $\frac{\partial
_{f\left( X\right) }K_{X}\left\vert \Psi \left( X\right) \right\vert ^{2}}{%
K_{X}\left\vert \Psi \left( X\right) \right\vert ^{2}}$ and: $\frac{\partial
_{f\left( X\right) }\left( \frac{\hat{K}_{X}\left\vert \hat{\Psi}\left(
X\right) \right\vert ^{2}}{K_{X}\left\vert \Psi \left( X\right) \right\vert
^{2}}\right) S_{1}\left( X\right) }{\frac{\hat{K}_{X}\left\vert \hat{\Psi}%
\left( X\right) \right\vert ^{2}}{K_{X}\left\vert \Psi \left( X\right)
\right\vert ^{2}}}$. We have:%
\begin{equation*}
K_{X}\left\vert \Psi \left( X\right) \right\vert ^{2}\simeq \left( \left( 
\frac{2\epsilon }{3\sigma _{\hat{K}}^{2}}\right) ^{\frac{r}{2}}\frac{%
f_{1}\left( X\right) }{C_{0}+\frac{S_{2}\left( X\right) }{1-S_{1}\left(
X\right) }\bar{r}}\right) ^{\frac{2}{r}}
\end{equation*}%
which implies:%
\begin{eqnarray*}
\frac{\partial _{f\left( X\right) }K_{X}\left\vert \Psi \left( X\right)
\right\vert ^{2}}{K_{X}\left\vert \Psi \left( X\right) \right\vert ^{2}}
&=&\left( \partial _{f\left( X\right) }f_{1}\left( X\right) \right) \frac{%
\partial _{f_{1}\left( X\right) }K_{X}\left\vert \Psi \left( X\right)
\right\vert ^{2}}{K_{X}\left\vert \Psi \left( X\right) \right\vert ^{2}} \\
&=&\frac{2}{r}\left( \partial _{f\left( X\right) }f_{1}\left( X\right)
\right)
\end{eqnarray*}%
For decreasing returns:%
\begin{equation*}
f_{1}\left( X\right) =K_{X}^{r}f^{dr}\left( X\right) +\frac{C}{K_{X}^{1-r}}%
+C_{0}K_{X}^{r}
\end{equation*}%
\begin{equation*}
\partial _{f\left( X\right) }f_{1}\left( X\right) =\frac{K_{X}^{r}}{%
f_{1}\left( X\right) }=\frac{1}{f^{dr}\left( X\right) +\frac{C}{K_{X}}+C_{0}}%
\simeq \frac{1}{\left( f^{dr}\left( X\right) +C_{0}\right) }
\end{equation*}%
and:%
\begin{equation*}
\frac{\partial _{f\left( X\right) }K_{X}\left\vert \Psi \left( X\right)
\right\vert ^{2}}{K_{X}\left\vert \Psi \left( X\right) \right\vert ^{2}}%
\simeq \frac{2}{r}\frac{1}{f\left( X\right) +C_{0}}
\end{equation*}%
Moreover, we use:%
\begin{equation*}
\frac{\hat{K}_{X}\left\vert \hat{\Psi}\left( X\right) \right\vert ^{2}}{%
K_{X}\left\vert \Psi \left( X\right) \right\vert ^{2}}\simeq \frac{18\sigma
_{\hat{K}}^{2}V\left\Vert \bar{\Psi}_{0}\left( X\right) \right\Vert ^{4}}{%
\hat{\mu}F_{1}^{2}\left( \left( \frac{2\epsilon }{3\sigma _{\hat{K}}^{2}}%
\right) ^{\frac{r}{2}}\frac{f_{1}\left( X\right) }{C_{0}+\frac{S_{2}\left(
X\right) }{1-S_{1}\left( X\right) }\bar{r}}\right) ^{\frac{2}{r}}}
\end{equation*}%
and thus:%
\begin{equation*}
\frac{\partial _{f\left( X\right) }\left( \frac{\hat{K}_{X}\left\vert \hat{%
\Psi}\left( X\right) \right\vert ^{2}}{K_{X}\left\vert \Psi \left( X\right)
\right\vert ^{2}}\right) }{\frac{\hat{K}_{X}\left\vert \hat{\Psi}\left(
X\right) \right\vert ^{2}}{K_{X}\left\vert \Psi \left( X\right) \right\vert
^{2}}}\simeq -\frac{2}{r}\frac{1}{f\left( X\right) +C_{0}}
\end{equation*}%
\begin{equation*}
\frac{\delta }{\delta S^{T}\left( X,,\theta -2\right) }\frac{\partial
_{f\left( X\right) }\left( \frac{\hat{K}_{X}\left\vert \hat{\Psi}\left(
X\right) \right\vert ^{2}}{K_{X}\left\vert \Psi \left( X\right) \right\vert
^{2}}\right) }{\frac{\hat{K}_{X}\left\vert \hat{\Psi}\left( X\right)
\right\vert ^{2}}{K_{X}\left\vert \Psi \left( X\right) \right\vert ^{2}}}%
\simeq -\frac{1}{f\left( X\right) +C_{0}}\frac{\partial _{f\left( X\right)
}\left( \frac{\hat{K}_{X}\left\vert \hat{\Psi}\left( X\right) \right\vert
^{2}}{K_{X}\left\vert \Psi \left( X\right) \right\vert ^{2}}\right) }{\frac{%
\hat{K}_{X}\left\vert \hat{\Psi}\left( X\right) \right\vert ^{2}}{%
K_{X}\left\vert \Psi \left( X\right) \right\vert ^{2}}}\frac{\delta f\left(
X\right) }{\delta S^{T}\left( X,,\theta -2\right) }
\end{equation*}%
\begin{equation*}
\frac{\delta }{\delta S^{T}\left( X,,\theta -2\right) }\frac{\partial
_{f\left( X\right) }K_{X}\left\vert \Psi \left( X\right) \right\vert ^{2}}{%
K_{X}\left\vert \Psi \left( X\right) \right\vert ^{2}}\simeq -\frac{1}{%
f\left( X\right) +C_{0}}\frac{\partial _{f\left( X\right) }K_{X}\left\vert
\Psi \left( X\right) \right\vert ^{2}}{K_{X}\left\vert \Psi \left( X\right)
\right\vert ^{2}}\frac{\delta f\left( X\right) }{\delta S^{T}\left(
X,,\theta -2\right) }
\end{equation*}%
\begin{eqnarray*}
&&\delta \left( \left( S_{1}^{B}\left( X,\theta -1\right) +S_{2}^{B}\left(
X,\theta -1\right) \right) \frac{\partial _{f\left( X\right)
}K_{X}\left\vert \Psi \left( X\right) \right\vert ^{2}}{K_{X}\left\vert \Psi
\left( X\right) \right\vert ^{2}}\right) \\
&=&\left( h_{1}^{B}\left( X\right) \left( \left\langle \bar{h}\left(
X\right) \right\rangle +\left\langle \hat{h}_{1}^{B}\left( X\right)
\right\rangle \right) -\left( S_{2}^{B}\left( X,\theta -1\right)
+S_{1}^{B}\left( X,\theta -1\right) \right) \frac{\partial _{f\left(
X\right) }K_{X}\left\vert \Psi \left( X\right) \right\vert ^{2}}{%
K_{X}\left\vert \Psi \left( X\right) \right\vert ^{2}}\left( 1+\frac{\frac{1%
}{f\left( X\right) +C_{0}}}{\frac{\partial _{f\left( X\right)
}K_{X}\left\vert \Psi \left( X\right) \right\vert ^{2}}{K_{X}\left\vert \Psi
\left( X\right) \right\vert ^{2}}}\right) \right) \\
&&\times \left( \frac{\partial _{f\left( X\right) }K_{X}\left\vert \Psi
\left( X\right) \right\vert ^{2}}{K_{X}\left\vert \Psi \left( X\right)
\right\vert ^{2}}\right) \frac{\partial f\left( X\right) }{\partial
S^{T}\left( X,\theta -2\right) }\delta S^{T}\left( X,\theta -2\right) \\
&&+\left( S_{1}^{B}\left( X,\theta -1\right) +S_{2}^{B}\left( X,\theta
-1\right) \right) \frac{\partial _{f\left( X\right) }K_{X}\left\vert \Psi
\left( X\right) \right\vert ^{2}}{K_{X}\left\vert \Psi \left( X\right)
\right\vert ^{2}}\frac{\partial _{\bar{f}\left( X\right) }\bar{K}%
_{X}\left\vert \bar{\Psi}\left( X\right) \right\vert ^{2}}{\bar{K}%
_{X}\left\vert \bar{\Psi}\left( X\right) \right\vert ^{2}}\delta \bar{f}%
\left( X,\theta -1\right)
\end{eqnarray*}%
\begin{eqnarray*}
&&\frac{\partial }{\partial S^{T}\left( X,\theta -2\right) }\left(
S_{1}^{B}\left( X,X\right) \left\{ h_{1}^{B}\left( X\right) \left(
\left\langle \bar{h}\left( X\right) \right\rangle +\left\langle \hat{h}%
_{1}^{B}\left( X\right) \right\rangle \right) \right. \right. \\
&&\left. \left. -\left( S_{1}^{B}\left( X,\theta -1\right) +S_{2}^{B}\left(
X,\theta -1\right) \right) \frac{\partial _{f\left( X\right)
}K_{X}\left\vert \Psi \left( X\right) \right\vert ^{2}}{K_{X}\left\vert \Psi
\left( X\right) \right\vert ^{2}}\right\} \right) \\
&=&h_{1}^{B}\left( X\right) \left( \left\langle \bar{h}\left( X\right)
\right\rangle +\left\langle \hat{h}_{1}^{B}\left( X\right) \right\rangle
\right) \\
&&\times \frac{\partial f\left( X\right) }{\partial S^{T}\left( X,\theta
-1\right) }\left( h_{1}^{B}\left( X\right) \left( \left\langle \bar{h}\left(
X\right) \right\rangle +\left\langle \hat{h}_{1}^{B}\left( X\right)
\right\rangle \right) -\left( S_{1}^{B}\left( X,\theta -1\right)
+S_{2}^{B}\left( X,\theta -1\right) \right) \frac{\partial _{f\left(
X\right) }K_{X}\left\vert \Psi \left( X\right) \right\vert ^{2}}{%
K_{X}\left\vert \Psi \left( X\right) \right\vert ^{2}}\right) \\
&&-S_{1}^{B}\left( X,X\right) \frac{\partial f\left( X\right) }{\partial
S^{T}\left( X,\theta -1\right) }\left( h_{1}^{B}\left( X\right) \left(
\left\langle \bar{h}\left( X\right) \right\rangle +\left\langle \hat{h}%
_{1}^{B}\left( X\right) \right\rangle \right) \right. \\
&&\left. -\left( S_{1}^{B}\left( X,\theta -1\right) +S_{2}^{B}\left(
X,\theta -1\right) \right) \left( 1+\frac{1}{f\left( X\right) +C_{0}}\right) 
\frac{\partial _{f\left( X\right) }K_{X}\left\vert \Psi \left( X\right)
\right\vert ^{2}}{K_{X}\left\vert \Psi \left( X\right) \right\vert ^{2}}%
\right)
\end{eqnarray*}%
\begin{eqnarray*}
&&\frac{\delta }{\delta S^{T}\left( X,,\theta -2\right) }\left(
S_{1}^{B}\left( X,X\right) \left( S_{1}^{B}\left( X,\theta -1\right)
+S_{2}^{B}\left( X,\theta -1\right) \right) \right) \\
&=&S_{1}^{B}\left( X,X\right) \left( h_{1}^{B}\left( X\right) \left(
\left\langle \bar{h}\left( X\right) \right\rangle +\left\langle \hat{h}%
_{1}^{B}\left( X\right) \right\rangle \right) -\left( S_{1}^{B}\left(
X,\theta -1\right) +S_{2}^{B}\left( X,\theta -1\right) \right) \frac{%
\partial _{f\left( X\right) }K_{X}\left\vert \Psi \left( X\right)
\right\vert ^{2}}{K_{X}\left\vert \Psi \left( X\right) \right\vert ^{2}}%
\right) \\
&&\times \frac{\partial f\left( X\right) }{\partial S^{T}\left( X,\theta
-2\right) }+\left( S_{1}^{B}\left( X,\theta -1\right) +S_{2}^{B}\left(
X,\theta -1\right) \right) h_{1}^{B}\left( X\right) \left( \left\langle \bar{%
h}\left( X\right) \right\rangle +\left\langle \hat{h}_{1}^{B}\left( X\right)
\right\rangle \right) \frac{\partial f\left( X\right) }{\partial S^{T}\left(
X,\theta -2\right) }
\end{eqnarray*}%
The derivatives are obtained as:%
\begin{eqnarray*}
&&\frac{\delta }{\delta \bar{f}\left( X,\theta -1\right) }a \\
&\simeq &\left\{ \left( \frac{1}{2}H\left( X\right) \bar{S}\left( X\right) -%
\frac{\bar{S}\left( X^{\prime },\theta -1\right) \partial _{\bar{f}\left(
X\right) }\left\langle \bar{K}_{X}\right\rangle \left\vert \bar{\Psi}\left(
X\right) \right\vert ^{2}}{\left\langle \bar{K}_{X^{\prime }}\right\rangle
\left\vert \bar{\Psi}\left( X^{\prime }\right) \right\vert ^{2}}\right)
\right. \frac{1}{1-\left( \bar{S}_{1}\left( X^{\prime },\theta -1\right)
\right) } \\
&&-\left( H\left( X\right) \bar{S}_{1}\left( X\right) -\frac{\bar{S}%
_{1}\left( X^{\prime },\theta -1\right) \partial _{\bar{f}\left( X\right)
}\left\langle \bar{K}_{X}\right\rangle \left\vert \bar{\Psi}\left( X\right)
\right\vert ^{2}}{\left\langle \bar{K}_{X^{\prime }}\right\rangle \left\vert 
\bar{\Psi}\left( X^{\prime }\right) \right\vert ^{2}}\right) \frac{\left(
1-\left( \bar{S}\left( X,\theta -1\right) \right) \right) }{\left( 1-\left( 
\bar{S}_{1}\left( X^{\prime },\theta -1\right) \right) \right) ^{2}} \\
&&+\left( S_{1}^{B}\left( X,\theta -1\right) +S_{2}^{B}\left( X,\theta
-1\right) \right) \\
&&\left. \times \left( \frac{\partial _{\bar{f}\left( X\right) }\bar{K}%
_{X}\left\vert \bar{\Psi}\left( X\right) \right\vert ^{2}}{\bar{K}%
_{X}\left\vert \bar{\Psi}\left( X\right) \right\vert ^{2}}S_{1}^{B}\left(
X,X\right) +h_{1}^{B}\left( X\right) \left( \left\langle \bar{h}\left(
X\right) \right\rangle +\left\langle \hat{h}_{1}^{B}\left( X\right)
\right\rangle \right) \right) \frac{\partial _{\bar{f}\left( X\right) }\bar{K%
}_{X}\left\vert \bar{\Psi}\left( X\right) \right\vert ^{2}}{\bar{K}%
_{X}\left\vert \bar{\Psi}\left( X\right) \right\vert ^{2}}\frac{\partial
f\left( X\right) }{\partial S^{T}\left( X,\theta -1\right) }\right\}
\end{eqnarray*}%
\begin{equation*}
\frac{\delta }{\delta \hat{f}\left( X,\theta -1\right) }a\simeq 0
\end{equation*}%
\begin{eqnarray*}
&&\frac{\delta }{\delta S^{T}\left( X,\theta -2\right) }\left(
S_{1}^{B}\left( X,X\right) \left( S_{1}^{B}\left( X,\theta -1\right)
+S_{2}^{B}\left( X,\theta -1\right) \right) \right) \\
&=&S_{1}^{B}\left( X,X\right) \left( h_{1}^{B}\left( X\right) \left(
\left\langle \bar{h}\left( X\right) \right\rangle +\left\langle \hat{h}%
_{1}^{B}\left( X\right) \right\rangle \right) -\left( S_{1}^{B}\left(
X,\theta -1\right) +S_{2}^{B}\left( X,\theta -1\right) \right) \frac{%
\partial _{f\left( X\right) }K_{X}\left\vert \Psi \left( X\right)
\right\vert ^{2}}{K_{X}\left\vert \Psi \left( X\right) \right\vert ^{2}}%
\right) \\
&&\times \frac{\partial f\left( X\right) }{\partial S^{T}\left( X,\theta
-2\right) }+\left( S_{1}^{B}\left( X,\theta -1\right) +S_{2}^{B}\left(
X,\theta -1\right) \right) h_{1}^{B}\left( X\right) \left( \left\langle \bar{%
h}\left( X\right) \right\rangle +\left\langle \hat{h}_{1}^{B}\left( X\right)
\right\rangle \right) \frac{\partial f\left( X\right) }{\partial S^{T}\left(
X,\theta -2\right) }
\end{eqnarray*}%
\begin{eqnarray*}
&&\frac{\delta }{\delta S^{T}\left( X,\theta -2\right) }a \\
&\simeq &\left( S_{1}^{B}\left( X,X\right) \left( h_{1}^{B}\left( X\right)
\left( \left\langle \bar{h}\left( X\right) \right\rangle +\left\langle \hat{h%
}_{1}^{B}\left( X\right) \right\rangle \right) -\left( S_{1}^{B}\left(
X,\theta -1\right) +S_{2}^{B}\left( X,\theta -1\right) \right) \frac{%
\partial _{f\left( X\right) }K_{X}\left\vert \Psi \left( X\right)
\right\vert ^{2}}{K_{X}\left\vert \Psi \left( X\right) \right\vert ^{2}}%
\right) \right. \\
&&\left. +\left( S_{1}^{B}\left( X,\theta -1\right) +S_{2}^{B}\left(
X,\theta -1\right) \right) h_{1}^{B}\left( X\right) \left( \left\langle \bar{%
h}\left( X\right) \right\rangle +\left\langle \hat{h}_{1}^{B}\left( X\right)
\right\rangle \right) \right) \left( \frac{\partial f\left( X\right) }{%
\partial S^{T}\left( X,\theta -2\right) }\right) ^{2}\frac{\partial
_{f\left( X\right) }K_{X}\left\vert \Psi \left( X\right) \right\vert ^{2}}{%
K_{X}\left\vert \Psi \left( X\right) \right\vert ^{2}}
\end{eqnarray*}

\begin{eqnarray*}
\frac{\delta }{\delta \bar{f}\left( X,\theta -1\right) }c &=&\left\{
h_{1}^{B}\left( X\right) \left( \left\langle \bar{h}\left( X\right)
\right\rangle +\left\langle \hat{h}_{1}^{B}\left( X\right) \right\rangle
\right) \frac{\partial f\left( X\right) }{\partial S^{T}\left( X,\theta
-2\right) }-\frac{\left\langle \bar{h}\left( X^{\prime },X\right)
\right\rangle }{2}h_{1}^{B}\left( X\right) \frac{1-\left\langle \bar{S}%
\left( X^{\prime }\right) \right\rangle }{1-\left\langle \bar{S}_{1}\left(
X^{\prime }\right) \right\rangle }\right\} \\
&&\times \frac{\partial f\left( X\right) }{\partial S^{T}\left( X,\theta
-2\right) }
\end{eqnarray*}%
\begin{eqnarray*}
\frac{\delta }{\delta \hat{f}\left( X,\theta -1\right) }c &\simeq
&-\left\langle \hat{h}_{1}^{B}\left( X^{\prime },X\right) \right\rangle
\left\langle h_{1}^{B}\left( X\right) \right\rangle \frac{1-\left\langle 
\hat{S}\left( X^{\prime }\right) \right\rangle +\left\langle \hat{S}%
_{1}^{B}\left( X^{\prime }\right) \right\rangle +\left\langle \hat{S}%
_{2}^{B}\left( X^{\prime }\right) \right\rangle }{1-\left\langle \hat{S}%
_{1}\left( X^{\prime }\right) \right\rangle +\left\langle \hat{S}%
_{1}^{B}\left( X^{\prime }\right) \right\rangle } \\
&&+ZS_{1}^{B}\left( X,X\right) \frac{\partial _{\hat{f}\left( X\right)
}\left( \frac{\hat{K}_{X}\left\vert \hat{\Psi}\left( X\right) \right\vert
^{2}}{K_{X}\left\vert \Psi \left( X\right) \right\vert ^{2}}\right) S\left(
X\right) }{\frac{\hat{K}_{X}\left\vert \hat{\Psi}\left( X\right) \right\vert
^{2}}{K_{X}\left\vert \Psi \left( X\right) \right\vert ^{2}}}\left( \frac{%
\partial f\left( X\right) }{\partial S^{T}\left( X,\theta -2\right) }\right)
^{2}
\end{eqnarray*}%
with:%
\begin{equation*}
Z=\left( \frac{\frac{\hat{h}\left( X\right) }{2}}{1+\left( \hat{h}\left(
X\right) \left( \frac{f\left( X\right) +\bar{r}\left( X\right) }{2}-\frac{%
\left\langle \hat{f}\left( X^{\prime }\right) \right\rangle _{\hat{h}%
_{1}}+\left\langle \hat{r}\left( X^{\prime }\right) \right\rangle _{\hat{h}%
_{2}}}{2}\right) \right) }+\frac{\partial _{f\left( X\right) }\left( \frac{%
\hat{K}_{X}\left\vert \hat{\Psi}\left( X\right) \right\vert ^{2}}{%
K_{X}\left\vert \Psi \left( X\right) \right\vert ^{2}}\right) }{\frac{\hat{K}%
_{X}\left\vert \hat{\Psi}\left( X\right) \right\vert ^{2}}{K_{X}\left\vert
\Psi \left( X\right) \right\vert ^{2}}}\right)
\end{equation*}

\begin{eqnarray*}
&&\frac{\delta }{\delta S^{T}\left( X,,\theta -2\right) }c \\
&\simeq &\left( \frac{\partial f\left( X\right) }{\partial S^{T}\left(
X,\theta -1\right) }\right) ^{3} \\
&&\times \left\{ h_{1}^{B}\left( X\right) \left( \left\langle \bar{h}\left(
X\right) \right\rangle +\left\langle \hat{h}_{1}^{B}\left( X\right)
\right\rangle \right) \left( h_{1}^{B}\left( X\right) \left( \left\langle 
\bar{h}\left( X\right) \right\rangle +\left\langle \hat{h}_{1}^{B}\left(
X\right) \right\rangle \right) \right. \right. \\
&&\left. -\left( S_{1}^{B}\left( X,\theta -1\right) +S_{2}^{B}\left(
X,\theta -1\right) \right) \frac{\partial _{f\left( X\right)
}K_{X}\left\vert \Psi \left( X\right) \right\vert ^{2}}{K_{X}\left\vert \Psi
\left( X\right) \right\vert ^{2}}+Z\right) \\
&&+S_{1}^{B}\left( X,X\right) \left( -\frac{\partial _{f\left( X\right)
}K_{X}\left\vert \Psi \left( X\right) \right\vert ^{2}}{K_{X}\left\vert \Psi
\left( X\right) \right\vert ^{2}}\left( h_{1}^{B}\left( X\right) \left(
\left\langle \bar{h}\left( X\right) \right\rangle +\left\langle \hat{h}%
_{1}^{B}\left( X\right) \right\rangle \right) \right. \right. \\
&&\left. -\left( S_{1}^{B}\left( X,\theta -1\right) +S_{2}^{B}\left(
X,\theta -1\right) \right) \left( 1+\frac{\frac{1}{f\left( X\right) +C_{0}}}{%
\frac{\partial _{f\left( X\right) }K_{X}\left\vert \Psi \left( X\right)
\right\vert ^{2}}{K_{X}\left\vert \Psi \left( X\right) \right\vert ^{2}}}%
\right) \frac{\partial _{f\left( X\right) }K_{X}\left\vert \Psi \left(
X\right) \right\vert ^{2}}{K_{X}\left\vert \Psi \left( X\right) \right\vert
^{2}}\right) \\
&&+\left. \left. Z^{2}S\left( X\right) -\frac{S\left( X\right) \partial
_{f\left( X\right) }\left( \frac{\hat{K}_{X}\left\vert \hat{\Psi}\left(
X\right) \right\vert ^{2}}{K_{X}\left\vert \Psi \left( X\right) \right\vert
^{2}}\right) }{\left( f\left( X\right) +C_{0}\right) \frac{\hat{K}%
_{X}\left\vert \hat{\Psi}\left( X\right) \right\vert ^{2}}{K_{X}\left\vert
\Psi \left( X\right) \right\vert ^{2}}}\right) \right\} \\
&&+c\frac{\frac{\partial ^{2}f\left( X\right) }{\partial \left( S^{T}\left(
X,\theta -2\right) \right) ^{2}}}{\frac{\partial f\left( X\right) }{\partial
S^{T}\left( X,\theta -2\right) }}+S_{1}^{B}\left( X,X\right) \frac{\partial
f\left( X\right) }{\partial S^{T}\left( X,\theta -1\right) }\left\{
h_{1}^{B}\left( X\right) \left( \left\langle \bar{h}\left( X\right)
\right\rangle +\left\langle \hat{h}_{1}^{B}\left( X\right) \right\rangle
\right) \right. \\
&&\left. -\left( S_{1}^{B}\left( X,\theta -1\right) +S_{2}^{B}\left(
X,\theta -1\right) \right) \frac{\partial _{f\left( X\right)
}K_{X}\left\vert \Psi \left( X\right) \right\vert ^{2}}{K_{X}\left\vert \Psi
\left( X\right) \right\vert ^{2}}+Z\right\} \frac{\partial ^{2}f\left(
X\right) }{\partial \left( S^{T}\left( X,\theta -2\right) \right) ^{2}}
\end{eqnarray*}%
with:%
\begin{eqnarray*}
&&c\frac{\frac{\partial ^{2}f\left( X\right) }{\partial \left( S^{T}\left(
X,\theta -2\right) \right) ^{2}}}{\frac{\partial f\left( X\right) }{\partial
S^{T}\left( X,\theta -2\right) }} \\
&\simeq &S_{1}^{B}\left( X,X\right) \frac{\partial f\left( X\right) }{%
\partial S^{T}\left( X,\theta -1\right) }\left\{ h_{1}^{B}\left( X\right)
\left( \left\langle \bar{h}\left( X\right) \right\rangle +\left\langle \hat{h%
}_{1}^{B}\left( X\right) \right\rangle \right) \right. \\
&&\left. -\left( S_{1}^{B}\left( X,\theta -1\right) +S_{2}^{B}\left(
X,\theta -1\right) \right) \frac{\partial _{f\left( X\right)
}K_{X}\left\vert \Psi \left( X\right) \right\vert ^{2}}{K_{X}\left\vert \Psi
\left( X\right) \right\vert ^{2}}+Z\right\} \frac{\partial ^{2}f\left(
X\right) }{\partial \left( S^{T}\left( X,\theta -2\right) \right) ^{2}}
\end{eqnarray*}%
\begin{eqnarray*}
&&\frac{\delta }{\delta \bar{f}\left( X,\theta -1\right) }d \\
&\simeq &\frac{\partial _{\bar{f}\left( X\right) }\bar{K}_{X}\left\vert \bar{%
\Psi}\left( X\right) \right\vert ^{2}}{\bar{K}_{X}\left\vert \bar{\Psi}%
\left( X\right) \right\vert ^{2}}d \\
&=&S_{1}\left( X,X,\theta -1\right) \left( S_{1}^{B}\left( X,\theta
-1\right) +S_{2}^{B}\left( X,\theta -1\right) \right) \frac{\partial f\left(
X\right) }{\partial S^{T}\left( X,\theta -1\right) }\left( \frac{\partial _{%
\bar{f}\left( X\right) }\bar{K}_{X}\left\vert \bar{\Psi}\left( X\right)
\right\vert ^{2}}{\bar{K}_{X}\left\vert \bar{\Psi}\left( X\right)
\right\vert ^{2}}\right) ^{2}<<1
\end{eqnarray*}%
so that we set:%
\begin{equation*}
\frac{\delta }{\delta \hat{f}\left( X,\theta -1\right) }d\simeq 0
\end{equation*}%
\begin{eqnarray*}
&&\frac{\delta }{\delta S^{T}\left( X,,\theta -2\right) }d \\
&\simeq &\left( \frac{\frac{\hat{h}\left( X\right) }{2}}{1+\left( \hat{h}%
\left( X\right) \left( \frac{f\left( X\right) +\bar{r}\left( X\right) }{2}-%
\frac{\left\langle \hat{f}\left( X^{\prime }\right) \right\rangle _{\hat{h}%
_{1}}+\left\langle \hat{r}\left( X^{\prime }\right) \right\rangle _{\hat{h}%
_{2}}}{2}\right) \right) }\right) \frac{\partial f\left( X\right) }{\partial
S^{T}\left( X,\theta -2\right) }d \\
&&+\left( h_{1}^{B}\left( X\right) \left( \left\langle \bar{h}\left(
X\right) \right\rangle +\left\langle \hat{h}_{1}^{B}\left( X\right)
\right\rangle \right) -\left( S_{1}^{B}\left( X,\theta -1\right)
+S_{2}^{B}\left( X,\theta -1\right) \right) \frac{\left( 1-\frac{1}{f\left(
X\right) +C_{0}}\right) \partial _{f\left( X\right) }K_{X}\left\vert \Psi
\left( X\right) \right\vert ^{2}}{K_{X}\left\vert \Psi \left( X\right)
\right\vert ^{2}}\right) \\
&&\times \left( \frac{\partial f\left( X\right) }{\partial S^{T}\left(
X,\theta -2\right) }\right) ^{2}+\frac{\frac{\partial ^{2}f\left( X\right) }{%
\partial \left( S^{T}\left( X,\theta -2\right) \right) ^{2}}}{\frac{\partial
f\left( X\right) }{\partial S^{T}\left( X,\theta -2\right) }}d \\
&\rightarrow &\left( \frac{\frac{\hat{h}\left( X\right) }{2}}{1+\left( \hat{h%
}\left( X\right) \left( \frac{f\left( X\right) +\bar{r}\left( X\right) }{2}-%
\frac{\left\langle \hat{f}\left( X^{\prime }\right) \right\rangle _{\hat{h}%
_{1}}+\left\langle \hat{r}\left( X^{\prime }\right) \right\rangle _{\hat{h}%
_{2}}}{2}\right) \right) }+\frac{\left( 1-\frac{1}{f\left( X\right) +C_{0}}%
\right) h_{1}^{B}\left( X\right) \left( \left\langle \bar{h}\left( X\right)
\right\rangle +\left\langle \hat{h}_{1}^{B}\left( X\right) \right\rangle
\right) }{S_{1}^{B}\left( X,\theta -1\right) +S_{2}^{B}\left( X,\theta
-1\right) }\right) \\
&&\times \frac{\partial f\left( X\right) }{\partial S^{T}\left( X,\theta
-2\right) }d+\frac{\frac{\partial ^{2}f\left( X\right) }{\partial \left(
S^{T}\left( X,\theta -2\right) \right) ^{2}}}{\frac{\partial f\left(
X\right) }{\partial S^{T}\left( X,\theta -2\right) }}d
\end{eqnarray*}%
\bigskip 
\begin{eqnarray*}
\frac{\delta }{\delta \bar{f}\left( X,\theta -1\right) }f &\simeq
&-S_{1}\left( X,X,\theta -1\right) \left( S_{1}^{B}\left( X,\theta -1\right)
+S_{2}^{B}\left( X,\theta -1\right) \right) \\
&&\times \left[ \frac{\partial f\left( X\right) }{\partial S^{T}\left(
X,\theta -1\right) }\right] ^{2}\frac{\partial _{\bar{f}\left( X\right) }%
\bar{K}_{X}\left\vert \bar{\Psi}\left( X\right) \right\vert ^{2}}{\bar{K}%
_{X}\left\vert \bar{\Psi}\left( X\right) \right\vert ^{2}}\frac{\partial
_{f\left( X\right) }K_{X}\left\vert \Psi \left( X\right) \right\vert ^{2}}{%
K_{X}\left\vert \Psi \left( X\right) \right\vert ^{2}}
\end{eqnarray*}%
\begin{eqnarray*}
&&\frac{\delta }{\delta \hat{f}\left( X,\theta -1\right) }f \\
&\simeq &-\frac{\left( 1-\left\langle \hat{S}\left( X^{\prime }\right)
\right\rangle \right) }{1-\left\langle \hat{S}_{1}\left( X^{\prime }\right)
\right\rangle }\frac{\left\langle \hat{h}\left( X^{\prime },X\right)
\right\rangle \left\langle h\left( X\right) \right\rangle }{4}\frac{\partial
f\left( X\right) }{\partial S^{T}\left( X,\theta -1\right) } \\
&&+S_{1}\left( X,X,\theta -1\right) \left[ \frac{\partial f\left( X\right) }{%
\partial S^{T}\left( X,\theta -1\right) }\right] ^{2} \\
&&\times \left( h_{1}^{B}\left( X\right) \left( \left\langle \bar{h}\left(
X\right) \right\rangle +\left\langle \hat{h}_{1}^{B}\left( X\right)
\right\rangle \right) -\left( S_{1}^{B}\left( X,\theta -1\right)
+S_{2}^{B}\left( X,\theta -1\right) \right) \frac{\partial _{f\left(
X\right) }K_{X}\left\vert \Psi \left( X\right) \right\vert ^{2}}{%
K_{X}\left\vert \Psi \left( X\right) \right\vert ^{2}}+2Z\right) \\
&&\times \frac{\partial _{\hat{f}\left( X\right) }\left( \frac{\hat{K}%
_{X}\left\vert \hat{\Psi}\left( X\right) \right\vert ^{2}}{K_{X}\left\vert
\Psi \left( X\right) \right\vert ^{2}}\right) }{\frac{\hat{K}_{X}\left\vert 
\hat{\Psi}\left( X\right) \right\vert ^{2}}{K_{X}\left\vert \Psi \left(
X\right) \right\vert ^{2}}}
\end{eqnarray*}%
\begin{eqnarray*}
&&\frac{\delta }{\delta S^{T}\left( X,,\theta -2\right) }f \\
&\simeq &\frac{\frac{h\left( X\right) }{4}\left( 1+\hat{h}\left( X\right)
\right) S_{1}\left( X\right) }{1+\left( \hat{h}\left( X\right) \left(
f\left( X\right) -\frac{\left\langle \hat{f}\left( X^{\prime }\right)
\right\rangle _{\hat{h}_{1}}+\left\langle \hat{r}\left( X^{\prime }\right)
\right\rangle _{\hat{h}_{2}}}{2}\right) +\frac{h\left( X\right) }{2}\left(
f\left( X\right) -\bar{r}\left( X\right) \right) \right) }\left( \frac{%
\partial f\left( X\right) }{\partial S^{T}\left( X,\theta -1\right) }\right)
^{2}\frac{\partial _{f\left( X\right) }K_{X}\left\vert \Psi \left( X\right)
\right\vert ^{2}}{K_{X}\left\vert \Psi \left( X\right) \right\vert ^{2}} \\
&&+S_{1}\left( X,X,\theta -1\right) \left( \frac{\partial f\left( X\right) }{%
\partial S^{T}\left( X,\theta -1\right) }\right) ^{3}\frac{\frac{\hat{h}%
\left( X\right) }{2}}{1+\left( \hat{h}\left( X\right) \left( \frac{f\left(
X\right) +\bar{r}\left( X\right) }{2}-\frac{\left\langle \hat{f}\left(
X^{\prime }\right) \right\rangle _{\hat{h}_{1}}+\left\langle \hat{r}\left(
X^{\prime }\right) \right\rangle _{\hat{h}_{2}}}{2}\right) \right) } \\
&&\times \left( \left( h_{1}^{B}\left( X\right) \left( \left\langle \bar{h}%
\left( X\right) \right\rangle +\left\langle \hat{h}_{1}^{B}\left( X\right)
\right\rangle \right) -\left( S_{1}^{B}\left( X,\theta -1\right)
+S_{2}^{B}\left( X,\theta -1\right) \right) \frac{\partial _{f\left(
X\right) }K_{X}\left\vert \Psi \left( X\right) \right\vert ^{2}}{%
K_{X}\left\vert \Psi \left( X\right) \right\vert ^{2}}+ZS\left( X\right)
\right) \right. \\
&&+S_{1}\left( X,X,\theta -1\right) \left( \frac{\partial f\left( X\right) }{%
\partial S^{T}\left( X,\theta -1\right) }\right) ^{3} \\
&&\times \left( -\frac{\partial _{f\left( X\right) }K_{X}\left\vert \Psi
\left( X\right) \right\vert ^{2}}{K_{X}\left\vert \Psi \left( X\right)
\right\vert ^{2}}\left( h_{1}^{B}\left( X\right) \left( \left\langle \bar{h}%
\left( X\right) \right\rangle +\left\langle \hat{h}_{1}^{B}\left( X\right)
\right\rangle \right) \right. \right. \\
&&\left. -\left( S_{2}^{B}\left( X\right) +S_{1}^{B}\left( X\right) \right) 
\frac{\partial _{f\left( X\right) }K_{X}\left\vert \Psi \left( X\right)
\right\vert ^{2}}{K_{X}\left\vert \Psi \left( X\right) \right\vert ^{2}}%
\left( 1+\frac{\frac{1}{f\left( X\right) +C_{0}}}{\frac{\partial _{f\left(
X\right) }K_{X}\left\vert \Psi \left( X\right) \right\vert ^{2}}{%
K_{X}\left\vert \Psi \left( X\right) \right\vert ^{2}}}\right) \right) \\
&&\left. \left. +Z^{2}S\left( X\right) -\frac{S\left( X\right) \partial
_{f\left( X\right) }\left( \frac{\hat{K}_{X}\left\vert \hat{\Psi}\left(
X\right) \right\vert ^{2}}{K_{X}\left\vert \Psi \left( X\right) \right\vert
^{2}}\right) }{\left( f\left( X\right) +C_{0}\right) \frac{\hat{K}%
_{X}\left\vert \hat{\Psi}\left( X\right) \right\vert ^{2}}{K_{X}\left\vert
\Psi \left( X\right) \right\vert ^{2}}}\right) \right\} +2S_{1}\left(
X,X,\theta -1\right) \frac{\partial f\left( X\right) }{\partial S^{T}\left(
X,\theta -2\right) }\frac{\partial ^{2}f\left( X\right) }{\partial \left(
S^{T}\left( X,\theta -2\right) \right) ^{2}} \\
&&\times \left( h_{1}^{B}\left( X\right) \left( \left\langle \bar{h}\left(
X\right) \right\rangle +\left\langle \hat{h}_{1}^{B}\left( X\right)
\right\rangle \right) -\left( S_{1}^{B}\left( X,\theta -1\right)
+S_{2}^{B}\left( X,\theta -1\right) \right) \frac{\partial _{f\left(
X\right) }K_{X}\left\vert \Psi \left( X\right) \right\vert ^{2}}{%
K_{X}\left\vert \Psi \left( X\right) \right\vert ^{2}}+Z\right)
\end{eqnarray*}%
\begin{equation*}
\frac{\delta }{\delta \bar{f}\left( X,\theta -1\right) }g\simeq g\frac{%
\partial _{\bar{f}\left( X\right) }\bar{K}_{X}\left\vert \bar{\Psi}\left(
X\right) \right\vert ^{2}}{\bar{K}_{X}\left\vert \bar{\Psi}\left( X\right)
\right\vert ^{2}}
\end{equation*}%
and we consider:%
\begin{equation*}
\frac{\delta }{\delta \hat{f}\left( X,\theta -1\right) }g\simeq 0
\end{equation*}%
\begin{eqnarray*}
&&\frac{\delta }{\delta S^{T}\left( X,,\theta -2\right) }g \\
&\simeq &\left( h_{1}^{B}\left( X\right) \left( \left\langle \bar{h}\left(
X\right) \right\rangle +\left\langle \hat{h}_{1}^{B}\left( X\right)
\right\rangle \right) -\left( S_{1}^{B}\left( X,\theta -1\right)
+S_{2}^{B}\left( X,\theta -1\right) \right) \left( 1+\frac{1}{f\left(
X\right) +C_{0}}\right) \frac{\partial _{f\left( X\right) }K_{X}\left\vert
\Psi \left( X\right) \right\vert ^{2}}{K_{X}\left\vert \Psi \left( X\right)
\right\vert ^{2}}\right) \\
&&\times \frac{\partial _{\bar{f}\left( X\right) }\bar{K}_{X}\left\vert \bar{%
\Psi}\left( X\right) \right\vert ^{2}}{\bar{K}_{X}\left\vert \bar{\Psi}%
\left( X\right) \right\vert ^{2}}
\end{eqnarray*}%
\begin{equation*}
\frac{\delta }{\delta \bar{f}\left( X,\theta -1\right) }i\simeq -g\frac{%
\partial _{\bar{f}\left( X\right) }\bar{K}_{X}\left\vert \bar{\Psi}\left(
X\right) \right\vert ^{2}}{\bar{K}_{X}\left\vert \bar{\Psi}\left( X\right)
\right\vert ^{2}}\frac{\partial f\left( X\right) }{\partial S^{T}\left(
X,\theta -2\right) }
\end{equation*}%
\begin{eqnarray*}
&&\frac{\delta }{\delta \hat{f}\left( X,\theta -1\right) }i \\
&\simeq &Z\frac{\partial _{\hat{f}\left( X\right) }\left( \frac{\hat{K}%
_{X}\left\vert \hat{\Psi}\left( X\right) \right\vert ^{2}}{K_{X}\left\vert
\Psi \left( X\right) \right\vert ^{2}}\right) S\left( X\right) }{\frac{\hat{K%
}_{X}\left\vert \hat{\Psi}\left( X\right) \right\vert ^{2}}{K_{X}\left\vert
\Psi \left( X\right) \right\vert ^{2}}}\frac{\partial f\left( X\right) }{%
\partial S^{T}\left( X,\theta -2\right) }
\end{eqnarray*}%
\begin{eqnarray*}
&&\frac{\delta }{\delta S^{T}\left( X,,\theta -2\right) }i \\
&\simeq &-\left( h_{1}^{B}\left( X\right) \left( \left\langle \bar{h}\left(
X\right) \right\rangle +\left\langle \hat{h}_{1}^{B}\left( X\right)
\right\rangle \right) -\left( S_{1}^{B}\left( X,\theta -1\right)
+S_{2}^{B}\left( X,\theta -1\right) \right) \frac{\partial _{f\left(
X\right) }K_{X}\left\vert \Psi \left( X\right) \right\vert ^{2}}{%
K_{X}\left\vert \Psi \left( X\right) \right\vert ^{2}}\left( 1+\frac{1}{%
f\left( X\right) +C_{0}}\right) \right) \\
&&\times \frac{\partial _{\bar{f}\left( X\right) }\bar{K}_{X}\left\vert \bar{%
\Psi}\left( X\right) \right\vert ^{2}}{\bar{K}_{X}\left\vert \bar{\Psi}%
\left( X\right) \right\vert ^{2}}\frac{\partial f\left( X\right) }{\partial
S^{T}\left( X,\theta -2\right) } \\
&&+S\left( X\right) \left( Z^{2}S\left( X\right) -\frac{S\left( X\right)
\partial _{f\left( X\right) }\left( \frac{\hat{K}_{X}\left\vert \hat{\Psi}%
\left( X\right) \right\vert ^{2}}{K_{X}\left\vert \Psi \left( X\right)
\right\vert ^{2}}\right) }{\left( f\left( X\right) +C_{0}\right) \frac{\hat{K%
}_{X}\left\vert \hat{\Psi}\left( X\right) \right\vert ^{2}}{K_{X}\left\vert
\Psi \left( X\right) \right\vert ^{2}}}\right) \left[ \frac{\partial f\left(
X\right) }{\partial S^{T}\left( X,\theta -2\right) }\right] ^{2}+i\frac{%
\frac{\partial ^{2}f\left( X\right) }{\partial \left( S^{T}\left( X,\theta
-2\right) \right) ^{2}}}{\frac{\partial f\left( X\right) }{\partial
S^{T}\left( X,\theta -2\right) }}
\end{eqnarray*}

\subsection*{A4.2 \protect\bigskip Formula for the convergence domain}

We set:%
\begin{equation*}
X=\left( 
\begin{array}{c}
\delta \bar{f}\left( X,\theta \right) \\ 
\delta \hat{f}\left( X,\theta \right) \\ 
\delta S^{T}\left( X,,\theta -1\right)%
\end{array}%
\right) ,M=\left[ 
\begin{array}{ccc}
a & 0 & c \\ 
d & -1 & f \\ 
g & 0 & i%
\end{array}%
\right] ,\mathbf{M}=\left[ 
\begin{array}{c}
\bar{M} \\ 
\hat{M} \\ 
M^{T}%
\end{array}%
\right]
\end{equation*}%
In average, neglecting the non-local interactions, we write the system as:%
\begin{equation*}
X\left( \theta \right) =MX\left( \theta -1\right) +X^{t}\mathbf{M}X
\end{equation*}%
and consider the squared distance to the fixed point:%
\begin{equation*}
\left( X\left( \theta \right) \right) ^{t}X\left( \theta \right)
\end{equation*}%
The time variation of this distance is given by the dyamics:%
\begin{eqnarray*}
&&\left( X\left( \theta -1\right) \right) ^{t}X\left( \theta -1\right)
-\left( X\left( \theta \right) \right) ^{t}X\left( \theta \right) \\
&=&\left( X\left( \theta -1\right) \right) ^{t}\left( 1-M^{t}M\right) \left(
X\left( \theta -1\right) \right) \\
&&-\left( X\left( \theta -1\right) \right) ^{t}M^{t}\left( X^{t}\mathbf{M}%
X\right) -\left( X^{t}\mathbf{M}X\right) ^{t}MX\left( \theta -1\right)
-\left( X^{t}\mathbf{M}X\right) ^{t}\left( X^{t}\mathbf{M}X\right)
\end{eqnarray*}%
To the first approximation the range of the stability zone is given by the
equation:%
\begin{eqnarray*}
0 &<&\left( X\left( \theta -1\right) \right) ^{t}\left( 1-M^{t}M\right)
\left( X\left( \theta -1\right) \right) \\
&&-\left( X\left( \theta -1\right) \right) ^{t}M^{t}\left( X^{t}\mathbf{M}%
X\right) -\left( X^{t}\mathbf{M}X\right) ^{t}MX\left( \theta -1\right)
-\left( X^{t}\mathbf{M}X\right) ^{t}\left( X^{t}\mathbf{M}X\right)
\end{eqnarray*}%
To obtain some order of magnitude, we replace $X\left( \theta -1\right) $ by 
$X$ and:%
\begin{equation*}
M^{t}M\rightarrow \left\langle \lambda ^{2}\right\rangle
\end{equation*}%
the average of squared eigenvalues, and:%
\begin{equation*}
M\rightarrow \left\langle \lambda \right\rangle
\end{equation*}%
the average of eigenvalues. We obtain:%
\begin{eqnarray*}
&&x1\left( xax\right) +x2\left( xa2x\right) \\
&&x1\left( xax\right) +x2\left( xa2x\right)
\end{eqnarray*}%
\begin{equation*}
0<\left( 1-\left\langle \lambda ^{2}\right\rangle \right)
X^{t}X-\left\langle \lambda \right\rangle \left( X^{t}\left( X^{t}\mathbf{M}%
X\right) +\left( X^{t}\mathbf{M}X\right) ^{t}X\right) -\left( X^{t}\mathbf{M}%
X\right) ^{t}\left( X^{t}\mathbf{M}X\right)
\end{equation*}%
and this lead to the approximation:%
\begin{equation*}
\left( 1-\left\langle \lambda ^{2}\right\rangle \right) X^{t}X-\left( X^{t}%
\mathbf{M}X\right) ^{t}\left( X^{t}\mathbf{M}X\right) >0
\end{equation*}%
and the fluctuations are stable in the range:%
\begin{equation*}
\frac{\left( X^{t}\mathbf{M}X\right) ^{t}\left( X^{t}\mathbf{M}X\right) }{%
X^{t}X}<1-\left\langle \lambda ^{2}\right\rangle
\end{equation*}%
given that:%
\begin{equation*}
\left( X^{t}\mathbf{M}X\right) =X^{t}\left[ 
\begin{array}{c}
X^{t}\bar{M}X \\ 
X^{t}\hat{M}X \\ 
X^{t}M^{T}X%
\end{array}%
\right]
\end{equation*}%
\begin{equation*}
\frac{2\left\langle \lambda \right\rangle \mathbf{\delta S}^{t}\left[ \delta
S^{T}\left( X\right) M^{T}+\delta \hat{S}\left( X^{\prime },X\right) \hat{M}%
+\delta \bar{S}\left( X^{\prime },X\right) \bar{M}\right] \mathbf{\delta S}%
+\left( \mathbf{\delta S}^{t}\bar{M}\mathbf{\delta S}\right) ^{2}+\left( 
\mathbf{\delta S}^{t}\hat{M}\mathbf{\delta S}\right) ^{2}+\left( \mathbf{%
\delta S}^{t}M^{T}\mathbf{\delta S}\right) ^{2}}{\left( \delta \bar{S}\left(
X^{\prime },X\right) \right) ^{2}+\left( \delta \hat{S}\left( X^{\prime
},X\right) \right) ^{2}+\left( \delta S^{T}\left( X\right) \right) ^{2}}%
<1-\left\langle \lambda ^{2}\right\rangle
\end{equation*}%
where:%
\begin{equation*}
\mathbf{\delta S}=\left( 
\begin{array}{c}
\delta \bar{S}\left( X^{\prime },X\right) \\ 
\delta \hat{S}\left( X^{\prime },X\right) \\ 
\delta S^{T}\left( X\right)%
\end{array}%
\right)
\end{equation*}%
For relatively large fluctuations, this becomes:%
\begin{equation}
\frac{\left( \mathbf{\delta S}^{t}\bar{M}\mathbf{\delta S}\right)
^{2}+\left( \mathbf{\delta S}^{t}\hat{M}\mathbf{\delta S}\right) ^{2}+\left( 
\mathbf{\delta S}^{t}M^{T}\mathbf{\delta S}\right) ^{2}}{\left( \delta \bar{S%
}\left( X^{\prime },X\right) \right) ^{2}+\left( \delta \hat{S}\left(
X^{\prime },X\right) \right) ^{2}+\left( \delta S^{T}\left( X\right) \right)
^{2}}<1-\left\langle \lambda ^{2}\right\rangle  \label{cndt}
\end{equation}%
Given that we have neglected the role of investors we can set $\delta \hat{S}%
\left( X^{\prime },X\right) =0$ and this writes: 
\begin{equation}
\frac{\left( \mathbf{\delta S}^{t}\bar{M}\mathbf{\delta S}\right)
^{2}+\left( \mathbf{\delta S}^{t}\hat{M}\mathbf{\delta S}\right) ^{2}+\left( 
\mathbf{\delta S}^{t}M^{T}\mathbf{\delta S}\right) ^{2}}{\left( \delta \bar{S%
}\left( X^{\prime },X\right) \right) ^{2}+\left( \delta S^{T}\left( X\right)
\right) ^{2}}<1-\left\langle \lambda ^{2}\right\rangle
\end{equation}

\subsection*{A4.3 Estimation of the convergence domain}

In the system with banks, we can consider that $S^{T}\left( X,\theta
-1\right) \rightarrow 1$. Under decreasing return to scale, this corresponds
to consider that $\left\vert \frac{\partial f\left( X\right) }{\partial
S^{T}\left( X,\theta -1\right) }\right\vert >>1$. We can thus estimate the
corrections by keeping the higher order terms for each derivatives. We
disregard terms involving $\delta \hat{f}$ since we neglected impact of
investors fluctuations, that will be considered independently. This amount
to consider that:

\begin{eqnarray*}
\bar{M} &=&\left[ 
\begin{array}{ccc}
\frac{\delta }{\delta \bar{f}\left( X,\theta -1\right) }a & 0 & \frac{\bar{h}%
\left( X^{\prime },X\right) }{2}\frac{\delta }{\delta \bar{f}\left( X,\theta
-1\right) }c \\ 
0 & 0 & 0 \\ 
\frac{\delta }{\delta S^{T}\left( X,,\theta -2\right) }a & 0 & \frac{\bar{h}%
\left( X^{\prime },X\right) }{2}\frac{\delta }{\delta S^{T}\left( X,,\theta
-2\right) }c3%
\end{array}%
\right] ,\hat{M}=\left[ 
\begin{array}{ccc}
\frac{\hat{h}\left( X^{\prime },X\right) }{\bar{h}\left( X^{\prime
},X\right) }\frac{\delta }{\delta \bar{f}\left( X,\theta -1\right) }d & 0 & 
\frac{\bar{h}\left( X^{\prime },X\right) }{2}\frac{\delta }{\delta \bar{f}%
\left( X,\theta -1\right) }f \\ 
0 & 0 & 0 \\ 
\frac{\hat{h}\left( X^{\prime },X\right) }{\bar{h}\left( X^{\prime
},X\right) }\frac{\delta }{\delta S^{T}\left( X,,\theta -2\right) }d & 0 & 
\frac{\hat{h}\left( X^{\prime },X\right) }{2}\frac{\delta }{\delta
S^{T}\left( X,,\theta -2\right) }f%
\end{array}%
\right] \\
M^{T} &=&0
\end{eqnarray*}%
and the coefficients are approximated below.

\bigskip We assume $S_{1}^{B}\left( X,\theta -1\right) +S_{2}^{B}\left(
X,\theta -1\right) >>1$ due to bank loans:%
\begin{eqnarray*}
&&\frac{\delta }{\delta S^{T}\left( X,,\theta -2\right) }c \\
&\simeq &\left( \frac{\partial f\left( X\right) }{\partial S^{T}\left(
X,\theta -1\right) }\right) ^{3} \\
&&\times \left\{ h_{1}^{B}\left( X\right) \left( \left\langle \bar{h}\left(
X\right) \right\rangle +\left\langle \hat{h}_{1}^{B}\left( X\right)
\right\rangle \right) \right. \\
&&\times \left( h_{1}^{B}\left( X\right) \left( \left\langle \bar{h}\left(
X\right) \right\rangle +\left\langle \hat{h}_{1}^{B}\left( X\right)
\right\rangle \right) -\left( \left( S_{1}^{B}\left( X\right)
+S_{2}^{B}\left( X\right) +S\left( X\right) \right) \frac{\partial _{f\left(
X\right) }K_{X}\left\vert \Psi \left( X\right) \right\vert ^{2}}{%
K_{X}\left\vert \Psi \left( X\right) \right\vert ^{2}}-\frac{\hat{h}\left(
X\right) }{2}S\left( X\right) \right) \right) \\
&&+S_{1}^{B}\left( X,X\right) \left( \frac{\partial _{f\left( X\right)
}K_{X}\left\vert \Psi \left( X\right) \right\vert ^{2}}{K_{X}\left\vert \Psi
\left( X\right) \right\vert ^{2}}\right. \\
&&\times \left( \left( \left( S_{1}^{B}\left( X\right) +S_{2}^{B}\left(
X\right) \right) \left( 1+\frac{\frac{1}{f\left( X\right) +C_{0}}}{\frac{%
\partial _{f\left( X\right) }K_{X}\left\vert \Psi \left( X\right)
\right\vert ^{2}}{K_{X}\left\vert \Psi \left( X\right) \right\vert ^{2}}}%
\right) +\frac{\frac{1}{f\left( X\right) +C_{0}}}{\frac{\partial _{f\left(
X\right) }K_{X}\left\vert \Psi \left( X\right) \right\vert ^{2}}{%
K_{X}\left\vert \Psi \left( X\right) \right\vert ^{2}}}S\left( X\right)
\right) \frac{\partial _{f\left( X\right) }K_{X}\left\vert \Psi \left(
X\right) \right\vert ^{2}}{K_{X}\left\vert \Psi \left( X\right) \right\vert
^{2}}-h_{1}^{B}\left( X\right) \left( \left\langle \bar{h}\left( X\right)
\right\rangle +\left\langle \hat{h}_{1}^{B}\left( X\right) \right\rangle
\right) \right) \\
&&+\left. \left. \left( \frac{\hat{h}\left( X\right) }{2}-\frac{\partial
_{f\left( X\right) }K_{X}\left\vert \Psi \left( X\right) \right\vert ^{2}}{%
K_{X}\left\vert \Psi \left( X\right) \right\vert ^{2}}\right) ^{2}S\left(
X\right) \right) \right\} \\
&&+c\frac{\frac{\partial ^{2}f\left( X\right) }{\partial \left( S^{T}\left(
X,\theta -2\right) \right) ^{2}}}{\frac{\partial f\left( X\right) }{\partial
S^{T}\left( X,\theta -2\right) }}+S_{1}^{B}\left( X,X\right) \frac{\partial
f\left( X\right) }{\partial S^{T}\left( X,\theta -1\right) }\left\{
h_{1}^{B}\left( X\right) \left( \left\langle \bar{h}\left( X\right)
\right\rangle +\left\langle \hat{h}_{1}^{B}\left( X\right) \right\rangle
\right) \right. \\
&&\left. -\left( S_{1}^{B}\left( X,\theta -1\right) +S_{2}^{B}\left(
X,\theta -1\right) +S\left( X\right) \right) \frac{\partial _{f\left(
X\right) }K_{X}\left\vert \Psi \left( X\right) \right\vert ^{2}}{%
K_{X}\left\vert \Psi \left( X\right) \right\vert ^{2}}+\frac{\hat{h}\left(
X\right) }{2}S\left( X\right) \right\} \frac{\partial ^{2}f\left( X\right) }{%
\partial \left( S^{T}\left( X,\theta -2\right) \right) ^{2}}
\end{eqnarray*}

\bigskip 
\begin{eqnarray*}
&&\frac{\delta }{\delta S^{T}\left( X,,\theta -2\right) }f \\
&\simeq &S_{1}\left( X,X,\theta -1\right) \left( \frac{\partial f\left(
X\right) }{\partial S^{T}\left( X,\theta -1\right) }\right) ^{3}\frac{\hat{h}%
\left( X\right) }{2} \\
&&\times \left( \left( h_{1}^{B}\left( X\right) \left( \left\langle \bar{h}%
\left( X\right) \right\rangle +\left\langle \hat{h}_{1}^{B}\left( X\right)
\right\rangle \right) -\left( S_{1}^{B}\left( X,\theta -1\right)
+S_{2}^{B}\left( X,\theta -1\right) \right) \frac{\partial _{f\left(
X\right) }K_{X}\left\vert \Psi \left( X\right) \right\vert ^{2}}{%
K_{X}\left\vert \Psi \left( X\right) \right\vert ^{2}}+ZS\left( X\right)
\right) \right. \\
&&+S_{1}\left( X,X,\theta -1\right) \left( \frac{\partial f\left( X\right) }{%
\partial S^{T}\left( X,\theta -1\right) }\right) ^{3} \\
&&\times \left( -\frac{\partial _{f\left( X\right) }K_{X}\left\vert \Psi
\left( X\right) \right\vert ^{2}}{K_{X}\left\vert \Psi \left( X\right)
\right\vert ^{2}}\left( h_{1}^{B}\left( X\right) \left( \left\langle \bar{h}%
\left( X\right) \right\rangle +\left\langle \hat{h}_{1}^{B}\left( X\right)
\right\rangle \right) -\left( S_{2}^{B}\left( X,\theta -1\right)
+S_{1}^{B}\left( X,\theta -1\right) \right) \frac{\partial _{f\left(
X\right) }K_{X}\left\vert \Psi \left( X\right) \right\vert ^{2}}{%
K_{X}\left\vert \Psi \left( X\right) \right\vert ^{2}}\left( 1+\frac{\frac{1%
}{f\left( X\right) +C_{0}}}{\frac{\partial _{f\left( X\right)
}K_{X}\left\vert \Psi \left( X\right) \right\vert ^{2}}{K_{X}\left\vert \Psi
\left( X\right) \right\vert ^{2}}}\right) \right) \right. \\
&&\left. \left. +Z^{2}S\left( X\right) -\frac{S\left( X\right) \partial
_{f\left( X\right) }\left( \frac{\hat{K}_{X}\left\vert \hat{\Psi}\left(
X\right) \right\vert ^{2}}{K_{X}\left\vert \Psi \left( X\right) \right\vert
^{2}}\right) }{\left( f\left( X\right) +C_{0}\right) \frac{\hat{K}%
_{X}\left\vert \hat{\Psi}\left( X\right) \right\vert ^{2}}{K_{X}\left\vert
\Psi \left( X\right) \right\vert ^{2}}}\right) \right\} +2S_{1}\left(
X,X,\theta -1\right) \frac{\partial f\left( X\right) }{\partial S^{T}\left(
X,\theta -2\right) }\frac{\partial ^{2}f\left( X\right) }{\partial \left(
S^{T}\left( X,\theta -2\right) \right) ^{2}} \\
&&\times \left( h_{1}^{B}\left( X\right) \left( \left\langle \bar{h}\left(
X\right) \right\rangle +\left\langle \hat{h}_{1}^{B}\left( X\right)
\right\rangle \right) -\left( S_{1}^{B}\left( X,\theta -1\right)
+S_{2}^{B}\left( X,\theta -1\right) \right) \frac{\partial _{f\left(
X\right) }K_{X}\left\vert \Psi \left( X\right) \right\vert ^{2}}{%
K_{X}\left\vert \Psi \left( X\right) \right\vert ^{2}}+Z\right)
\end{eqnarray*}

and given that:%
\begin{equation*}
\frac{\partial _{\bar{f}\left( X\right) }\bar{K}_{X}\left\vert \bar{\Psi}%
\left( X\right) \right\vert ^{2}}{\bar{K}_{X}\left\vert \bar{\Psi}\left(
X\right) \right\vert ^{2}}<<1
\end{equation*}%
\begin{equation*}
\frac{\delta }{\delta \bar{f}\left( X,\theta -1\right) }c=\left( \frac{%
\partial f\left( X\right) }{\partial S^{T}\left( X,\theta -2\right) }\right)
^{2}h_{1}^{B}\left( X\right) \left( \left\langle \bar{h}\left( X\right)
\right\rangle +\left\langle \hat{h}_{1}^{B}\left( X\right) \right\rangle
\right)
\end{equation*}%
\begin{eqnarray*}
\frac{\delta }{\delta \bar{f}\left( X,\theta -1\right) }f &\simeq
&-S_{1}\left( X,X,\theta -1\right) \left( S_{1}^{B}\left( X,\theta -1\right)
+S_{2}^{B}\left( X,\theta -1\right) \right) \\
&&\times \left[ \frac{\partial f\left( X\right) }{\partial S^{T}\left(
X,\theta -1\right) }\right] ^{2}\frac{\partial _{\bar{f}\left( X\right) }%
\bar{K}_{X}\left\vert \bar{\Psi}\left( X\right) \right\vert ^{2}}{\bar{K}%
_{X}\left\vert \bar{\Psi}\left( X\right) \right\vert ^{2}}\frac{\partial
_{f\left( X\right) }K_{X}\left\vert \Psi \left( X\right) \right\vert ^{2}}{%
K_{X}\left\vert \Psi \left( X\right) \right\vert ^{2}}<<1
\end{eqnarray*}%
\begin{eqnarray*}
&&\frac{\delta }{\delta S^{T}\left( X,\theta -2\right) }a \\
&\simeq &\left( S_{1}^{B}\left( X,X\right) \left( h_{1}^{B}\left( X\right)
\left( \left\langle \bar{h}\left( X\right) \right\rangle +\left\langle \hat{h%
}_{1}^{B}\left( X\right) \right\rangle \right) -\left( S_{1}^{B}\left(
X,\theta -1\right) +S_{2}^{B}\left( X,\theta -1\right) \right) \frac{%
\partial _{f\left( X\right) }K_{X}\left\vert \Psi \left( X\right)
\right\vert ^{2}}{K_{X}\left\vert \Psi \left( X\right) \right\vert ^{2}}%
\right) \right. \\
&&\left. +\left( S_{1}^{B}\left( X,\theta -1\right) +S_{2}^{B}\left(
X,\theta -1\right) \right) h_{1}^{B}\left( X\right) \left( \left\langle \bar{%
h}\left( X\right) \right\rangle +\left\langle \hat{h}_{1}^{B}\left( X\right)
\right\rangle \right) \right) \left( \frac{\partial f\left( X\right) }{%
\partial S^{T}\left( X,\theta -2\right) }\right) ^{2}\frac{\partial
_{f\left( X\right) }K_{X}\left\vert \Psi \left( X\right) \right\vert ^{2}}{%
K_{X}\left\vert \Psi \left( X\right) \right\vert ^{2}}
\end{eqnarray*}%
\begin{eqnarray*}
&&\frac{\delta }{\delta \bar{f}\left( X,\theta -1\right) }a \\
&\simeq &\left( S_{1}^{B}\left( X,\theta -1\right) +S_{2}^{B}\left( X,\theta
-1\right) \right) \\
&&\times \left( \frac{\partial _{\bar{f}\left( X\right) }\bar{K}%
_{X}\left\vert \bar{\Psi}\left( X\right) \right\vert ^{2}}{\bar{K}%
_{X}\left\vert \bar{\Psi}\left( X\right) \right\vert ^{2}}S_{1}^{B}\left(
X,X\right) +h_{1}^{B}\left( X\right) \left( \left\langle \bar{h}\left(
X\right) \right\rangle +\left\langle \hat{h}_{1}^{B}\left( X\right)
\right\rangle \right) \right) \frac{\partial _{\bar{f}\left( X\right) }\bar{K%
}_{X}\left\vert \bar{\Psi}\left( X\right) \right\vert ^{2}}{\bar{K}%
_{X}\left\vert \bar{\Psi}\left( X\right) \right\vert ^{2}}\frac{\partial
f\left( X\right) }{\partial S^{T}\left( X,\theta -1\right) }
\end{eqnarray*}%
\begin{eqnarray*}
&&\frac{\delta }{\delta S^{T}\left( X,,\theta -2\right) }d \\
&\simeq &\left( \left( \frac{\hat{h}\left( X\right) }{2}+\frac{\left( 1-%
\frac{1}{f\left( X\right) +C_{0}}\right) h_{1}^{B}\left( X\right) \left(
\left\langle \bar{h}\left( X\right) \right\rangle +\left\langle \hat{h}%
_{1}^{B}\left( X\right) \right\rangle \right) }{S_{1}^{B}\left( X,\theta
-1\right) +S_{2}^{B}\left( X,\theta -1\right) }\right) \frac{\partial
f\left( X\right) }{\partial S^{T}\left( X,\theta -2\right) }+\frac{\frac{%
\partial ^{2}f\left( X\right) }{\partial \left( S^{T}\left( X,\theta
-2\right) \right) ^{2}}}{\frac{\partial f\left( X\right) }{\partial
S^{T}\left( X,\theta -2\right) }}\right) d \\
&=&\left( \left( \frac{\hat{h}\left( X\right) }{2}+\frac{\left( 1-\frac{1}{%
f\left( X\right) +C_{0}}\right) h_{1}^{B}\left( X\right) \left( \left\langle 
\bar{h}\left( X\right) \right\rangle +\left\langle \hat{h}_{1}^{B}\left(
X\right) \right\rangle \right) }{S_{1}^{B}\left( X,\theta -1\right)
+S_{2}^{B}\left( X,\theta -1\right) }\right) \frac{\partial f\left( X\right) 
}{\partial S^{T}\left( X,\theta -2\right) }+\frac{\frac{\partial ^{2}f\left(
X\right) }{\partial \left( S^{T}\left( X,\theta -2\right) \right) ^{2}}}{%
\frac{\partial f\left( X\right) }{\partial S^{T}\left( X,\theta -2\right) }}%
\right) \\
&&\times S_{1}\left( X,X,\theta -1\right) \left( S_{1}^{B}\left( X,\theta
-1\right) +S_{2}^{B}\left( X,\theta -1\right) \right) \frac{\partial f\left(
X\right) }{\partial S^{T}\left( X,\theta -1\right) }\frac{\partial _{\bar{f}%
\left( X\right) }\bar{K}_{X}\left\vert \bar{\Psi}\left( X\right) \right\vert
^{2}}{\bar{K}_{X}\left\vert \bar{\Psi}\left( X\right) \right\vert ^{2}} \\
&<&<1
\end{eqnarray*}%
\begin{eqnarray*}
&&\frac{\delta }{\delta \bar{f}\left( X,\theta -1\right) }d \\
&\simeq &S_{1}\left( X,X,\theta -1\right) \left( S_{1}^{B}\left( X,\theta
-1\right) +S_{2}^{B}\left( X,\theta -1\right) \right) \frac{\partial f\left(
X\right) }{\partial S^{T}\left( X,\theta -1\right) }\left( \frac{\partial _{%
\bar{f}\left( X\right) }\bar{K}_{X}\left\vert \bar{\Psi}\left( X\right)
\right\vert ^{2}}{\bar{K}_{X}\left\vert \bar{\Psi}\left( X\right)
\right\vert ^{2}}\right) ^{2}<<1
\end{eqnarray*}

\subsubsection*{A4.3.1 Case 1: $S^{T}\left( X\right) \frac{1}{f\left(
X\right) +C_{0}}>h_{1}^{B}\left( X\right) \left( \left\langle \bar{h}\left(
X\right) \right\rangle +\left\langle \hat{h}_{1}^{B}\left( X\right)
\right\rangle \right) $}

This case corresponds to a marginal return above a minimal value $\frac{1}{%
f_{0}\left( X\right) +C_{0}}$. This minimal value can be lower than $1$
since due to loans, $S^{T}\left( X\right) \rightarrow 1$ and $%
h_{1}^{B}\left( X\right) \left( \left\langle \bar{h}\left( X\right)
\right\rangle +\left\langle \hat{h}_{1}^{B}\left( X\right) \right\rangle
\right) <\frac{1}{4}$ (with uncertainty, this is in general lower than $%
\frac{1}{8}$).

\textbf{Derivation of }$\frac{\delta }{\delta S^{T}\left( X,,\theta
-2\right) }c$

When $S^{T}\left( X\right) \frac{1}{f\left( X\right) +C_{0}}>1$:\bigskip 
\begin{equation*}
S^{T}\left( X\right) =\left( S_{1}^{B}\left( X,\theta -1\right)
+S_{2}^{B}\left( X,\theta -1\right) +S\left( X\right) \right)
\end{equation*}

\begin{eqnarray*}
&&\frac{\delta }{\delta S^{T}\left( X,,\theta -2\right) }c \\
&\simeq &\left( \frac{\partial f\left( X\right) }{\partial S^{T}\left(
X,\theta -1\right) }\right) ^{3}\frac{\partial _{f\left( X\right)
}K_{X}\left\vert \Psi \left( X\right) \right\vert ^{2}}{K_{X}\left\vert \Psi
\left( X\right) \right\vert ^{2}} \\
&&\times \left\{ -h_{1}^{B}\left( X\right) \left( \left\langle \bar{h}\left(
X\right) \right\rangle +\left\langle \hat{h}_{1}^{B}\left( X\right)
\right\rangle \right) S^{T}\left( X\right) \right. \\
&&+S_{1}^{B}\left( X,X\right) \left( \left( \left( S_{1}^{B}\left( X,\theta
-1\right) +S_{2}^{B}\left( X,\theta -1\right) \right) \left( 1+\frac{\frac{1%
}{f\left( X\right) +C_{0}}}{\frac{\partial _{f\left( X\right)
}K_{X}\left\vert \Psi \left( X\right) \right\vert ^{2}}{K_{X}\left\vert \Psi
\left( X\right) \right\vert ^{2}}}\right) +\frac{\frac{1}{f\left( X\right)
+C_{0}}}{\frac{\partial _{f\left( X\right) }K_{X}\left\vert \Psi \left(
X\right) \right\vert ^{2}}{K_{X}\left\vert \Psi \left( X\right) \right\vert
^{2}}}S\left( X\right) \right) \right. \\
&&\times \frac{\partial _{f\left( X\right) }K_{X}\left\vert \Psi \left(
X\right) \right\vert ^{2}}{K_{X}\left\vert \Psi \left( X\right) \right\vert
^{2}}+\left. \left. \left( \frac{\partial _{f\left( X\right)
}K_{X}\left\vert \Psi \left( X\right) \right\vert ^{2}}{K_{X}\left\vert \Psi
\left( X\right) \right\vert ^{2}}\right) S\left( X\right) \right) \right\} \\
&&+c\frac{\frac{\partial ^{2}f\left( X\right) }{\partial \left( S^{T}\left(
X,\theta -2\right) \right) ^{2}}}{\frac{\partial f\left( X\right) }{\partial
S^{T}\left( X,\theta -2\right) }}+S_{1}^{B}\left( X,X\right) \frac{\partial
f\left( X\right) }{\partial S^{T}\left( X,\theta -1\right) }\left\{
h_{1}^{B}\left( X\right) \left( \left\langle \bar{h}\left( X\right)
\right\rangle +\left\langle \hat{h}_{1}^{B}\left( X\right) \right\rangle
\right) \right. \\
&&\left. -S^{T}\left( X\right) \frac{\partial _{f\left( X\right)
}K_{X}\left\vert \Psi \left( X\right) \right\vert ^{2}}{K_{X}\left\vert \Psi
\left( X\right) \right\vert ^{2}}+\frac{\hat{h}\left( X\right) }{2}S\left(
X\right) \right\} \frac{\partial ^{2}f\left( X\right) }{\partial \left(
S^{T}\left( X,\theta -2\right) \right) ^{2}}
\end{eqnarray*}%
and this rewrites:%
\begin{eqnarray*}
&&\frac{\delta }{\delta S^{T}\left( X,,\theta -2\right) }c \\
&\simeq &\left( \frac{\partial f\left( X\right) }{\partial S^{T}\left(
X,\theta -1\right) }\right) ^{3}\frac{\partial _{f\left( X\right)
}K_{X}\left\vert \Psi \left( X\right) \right\vert ^{2}}{K_{X}\left\vert \Psi
\left( X\right) \right\vert ^{2}} \\
&&\times \left\{ -h_{1}^{B}\left( X\right) \left( \left\langle \bar{h}\left(
X\right) \right\rangle +\left\langle \hat{h}_{1}^{B}\left( X\right)
\right\rangle \right) S^{T}\left( X\right) \right. \\
&&+\left. S_{1}^{B}\left( X,X\right) \left( S^{T}\left( X\right) \left( 1+%
\frac{\frac{1}{f\left( X\right) +C_{0}}}{\frac{\partial _{f\left( X\right)
}K_{X}\left\vert \Psi \left( X\right) \right\vert ^{2}}{K_{X}\left\vert \Psi
\left( X\right) \right\vert ^{2}}}\right) \frac{\partial _{f\left( X\right)
}K_{X}\left\vert \Psi \left( X\right) \right\vert ^{2}}{K_{X}\left\vert \Psi
\left( X\right) \right\vert ^{2}}\right. \right\} \\
&&+c\frac{\frac{\partial ^{2}f\left( X\right) }{\partial \left( S^{T}\left(
X,\theta -2\right) \right) ^{2}}}{\frac{\partial f\left( X\right) }{\partial
S^{T}\left( X,\theta -2\right) }}-S_{1}^{B}\left( X,X\right) \frac{\partial
f\left( X\right) }{\partial S^{T}\left( X,\theta -1\right) }S^{T}\left(
X\right) \frac{\partial _{f\left( X\right) }K_{X}\left\vert \Psi \left(
X\right) \right\vert ^{2}}{K_{X}\left\vert \Psi \left( X\right) \right\vert
^{2}}\frac{\partial ^{2}f\left( X\right) }{\partial \left( S^{T}\left(
X,\theta -2\right) \right) ^{2}}
\end{eqnarray*}%
Given that:%
\begin{eqnarray*}
&&c\frac{\frac{\partial ^{2}f\left( X\right) }{\partial \left( S^{T}\left(
X,\theta -2\right) \right) ^{2}}}{\frac{\partial f\left( X\right) }{\partial
S^{T}\left( X,\theta -2\right) }} \\
&\simeq &-S_{1}^{B}\left( X,X\right) \frac{\partial f\left( X\right) }{%
\partial S^{T}\left( X,\theta -1\right) }\left( S_{1}^{B}\left( X,\theta
-1\right) +S_{2}^{B}\left( X,\theta -1\right) +S\left( X\right) \right) 
\frac{\partial _{f\left( X\right) }K_{X}\left\vert \Psi \left( X\right)
\right\vert ^{2}}{K_{X}\left\vert \Psi \left( X\right) \right\vert ^{2}} \\
&&\times \frac{\partial ^{2}f\left( X\right) }{\partial \left( S^{T}\left(
X,\theta -2\right) \right) ^{2}} \\
&=&-S_{1}^{B}\left( X,X\right) S^{T}\left( X\right) \frac{\partial _{f\left(
X\right) }K_{X}\left\vert \Psi \left( X\right) \right\vert ^{2}}{%
K_{X}\left\vert \Psi \left( X\right) \right\vert ^{2}}\frac{\partial f\left(
X\right) }{\partial S^{T}\left( X,\theta -1\right) }\frac{\partial
^{2}f\left( X\right) }{\partial \left( S^{T}\left( X,\theta -2\right)
\right) ^{2}}
\end{eqnarray*}%
We obtain:%
\begin{eqnarray*}
&&\frac{\delta }{\delta S^{T}\left( X,,\theta -2\right) }c \\
&\simeq &\left( \frac{\partial f\left( X\right) }{\partial S^{T}\left(
X,\theta -1\right) }\right) ^{3}\frac{\partial _{f\left( X\right)
}K_{X}\left\vert \Psi \left( X\right) \right\vert ^{2}}{K_{X}\left\vert \Psi
\left( X\right) \right\vert ^{2}} \\
&&\times \left\{ -h_{1}^{B}\left( X\right) \left( \left\langle \bar{h}\left(
X\right) \right\rangle +\left\langle \hat{h}_{1}^{B}\left( X\right)
\right\rangle \right) S^{T}\left( X\right) \right. \\
&&+\left. S_{1}^{B}\left( X,X\right) \left( S^{T}\left( X\right) \left( 1+%
\frac{\frac{1}{f\left( X\right) +C_{0}}}{\frac{\partial _{f\left( X\right)
}K_{X}\left\vert \Psi \left( X\right) \right\vert ^{2}}{K_{X}\left\vert \Psi
\left( X\right) \right\vert ^{2}}}\right) \frac{\partial _{f\left( X\right)
}K_{X}\left\vert \Psi \left( X\right) \right\vert ^{2}}{K_{X}\left\vert \Psi
\left( X\right) \right\vert ^{2}}\right. \right\} \\
&&-2S_{1}^{B}\left( X,X\right) \frac{\partial f\left( X\right) }{\partial
S^{T}\left( X,\theta -1\right) }S^{T}\left( X\right) \frac{\partial
_{f\left( X\right) }K_{X}\left\vert \Psi \left( X\right) \right\vert ^{2}}{%
K_{X}\left\vert \Psi \left( X\right) \right\vert ^{2}}\frac{\partial
^{2}f\left( X\right) }{\partial \left( S^{T}\left( X,\theta -2\right)
\right) ^{2}}
\end{eqnarray*}%
and in an expanded form:%
\begin{eqnarray*}
&&\frac{\delta }{\delta S^{T}\left( X,,\theta -2\right) }c \\
&\simeq &\left( \frac{\partial f\left( X\right) }{\partial S^{T}\left(
X,\theta -1\right) }\right) ^{3}\left( \frac{\partial _{f\left( X\right)
}K_{X}\left\vert \Psi \left( X\right) \right\vert ^{2}}{K_{X}\left\vert \Psi
\left( X\right) \right\vert ^{2}}\right) ^{2}S_{1}^{B}\left( X,X\right)
S^{T}\left( X\right) \\
&&\times \left\{ \left( 1+\frac{\frac{1}{f\left( X\right) +C_{0}}}{\frac{%
\partial _{f\left( X\right) }K_{X}\left\vert \Psi \left( X\right)
\right\vert ^{2}}{K_{X}\left\vert \Psi \left( X\right) \right\vert ^{2}}}%
\right) -\frac{\frac{h_{1}^{B}\left( X\right) \left( \left\langle \bar{h}%
\left( X\right) \right\rangle +\left\langle \hat{h}_{1}^{B}\left( X\right)
\right\rangle \right) }{S_{1}^{B}\left( X,X\right) }+2\frac{\frac{\partial
^{2}f\left( X\right) }{\partial \left( S^{T}\left( X,\theta -2\right)
\right) ^{2}}}{\left( \frac{\partial f\left( X\right) }{\partial S^{T}\left(
X,\theta -2\right) }\right) ^{2}}}{\frac{\partial _{f\left( X\right)
}K_{X}\left\vert \Psi \left( X\right) \right\vert ^{2}}{K_{X}\left\vert \Psi
\left( X\right) \right\vert ^{2}}}\right\}
\end{eqnarray*}%
\textbf{Derivation of }$\frac{\delta }{\delta S^{T}\left( X,,\theta
-2\right) }f$

\begin{eqnarray*}
&&\frac{\delta }{\delta S^{T}\left( X,,\theta -2\right) }f \\
&\simeq &S_{1}\left( X,X,\theta -1\right) \left( \frac{\partial f\left(
X\right) }{\partial S^{T}\left( X,\theta -1\right) }\right) ^{3}\frac{\hat{h}%
\left( X\right) }{2}\left( -S^{T}\left( X\right) \frac{\partial _{f\left(
X\right) }K_{X}\left\vert \Psi \left( X\right) \right\vert ^{2}}{%
K_{X}\left\vert \Psi \left( X\right) \right\vert ^{2}}\right) \\
&&+S_{1}\left( X,X,\theta -1\right) \left( \frac{\partial f\left( X\right) }{%
\partial S^{T}\left( X,\theta -1\right) }\right) ^{3} \\
&&\times \left. \left( +\frac{\partial _{f\left( X\right) }K_{X}\left\vert
\Psi \left( X\right) \right\vert ^{2}}{K_{X}\left\vert \Psi \left( X\right)
\right\vert ^{2}}\left( \left( S^{T}\left( X\right) \right) \frac{\partial
_{f\left( X\right) }K_{X}\left\vert \Psi \left( X\right) \right\vert ^{2}}{%
K_{X}\left\vert \Psi \left( X\right) \right\vert ^{2}}\left( 1+\frac{\frac{1%
}{f\left( X\right) +C_{0}}}{\frac{\partial _{f\left( X\right)
}K_{X}\left\vert \Psi \left( X\right) \right\vert ^{2}}{K_{X}\left\vert \Psi
\left( X\right) \right\vert ^{2}}}\right) \right) \right. \right\} \\
&&+2S_{1}\left( X,X,\theta -1\right) \frac{\partial f\left( X\right) }{%
\partial S^{T}\left( X,\theta -2\right) }\frac{\partial ^{2}f\left( X\right) 
}{\partial \left( S^{T}\left( X,\theta -2\right) \right) ^{2}} \\
&&\times \left( h_{1}^{B}\left( X\right) \left( \left\langle \bar{h}\left(
X\right) \right\rangle +\left\langle \hat{h}_{1}^{B}\left( X\right)
\right\rangle \right) -\left( S_{1}^{B}\left( X,\theta -1\right)
+S_{2}^{B}\left( X,\theta -1\right) \right) \frac{\partial _{f\left(
X\right) }K_{X}\left\vert \Psi \left( X\right) \right\vert ^{2}}{%
K_{X}\left\vert \Psi \left( X\right) \right\vert ^{2}}+Z\right)
\end{eqnarray*}%
that writes:%
\begin{eqnarray*}
&&\frac{\delta }{\delta S^{T}\left( X,,\theta -2\right) }f \\
&\simeq &S_{1}\left( X,X,\theta -1\right) \left( \frac{\partial f\left(
X\right) }{\partial S^{T}\left( X,\theta -1\right) }\right) ^{3}\frac{\hat{h}%
\left( X\right) }{2}\left( -S^{T}\left( X\right) \frac{\partial _{f\left(
X\right) }K_{X}\left\vert \Psi \left( X\right) \right\vert ^{2}}{%
K_{X}\left\vert \Psi \left( X\right) \right\vert ^{2}}\right) \\
&&+S_{1}\left( X,X,\theta -1\right) \left( \frac{\partial f\left( X\right) }{%
\partial S^{T}\left( X,\theta -1\right) }\right) ^{3} \\
&&\times \left. \left( \frac{\partial _{f\left( X\right) }K_{X}\left\vert
\Psi \left( X\right) \right\vert ^{2}}{K_{X}\left\vert \Psi \left( X\right)
\right\vert ^{2}}\left( S^{T}\left( X\right) \frac{\partial _{f\left(
X\right) }K_{X}\left\vert \Psi \left( X\right) \right\vert ^{2}}{%
K_{X}\left\vert \Psi \left( X\right) \right\vert ^{2}}\left( 1+\frac{\frac{1%
}{f\left( X\right) +C_{0}}}{\frac{\partial _{f\left( X\right)
}K_{X}\left\vert \Psi \left( X\right) \right\vert ^{2}}{K_{X}\left\vert \Psi
\left( X\right) \right\vert ^{2}}}\right) \right) \right. \right\} \\
&&+2S_{1}\left( X,X,\theta -1\right) \frac{\partial f\left( X\right) }{%
\partial S^{T}\left( X,\theta -2\right) }\frac{\partial ^{2}f\left( X\right) 
}{\partial \left( S^{T}\left( X,\theta -2\right) \right) ^{2}}\left(
-S^{T}\left( X\right) \frac{\partial _{f\left( X\right) }K_{X}\left\vert
\Psi \left( X\right) \right\vert ^{2}}{K_{X}\left\vert \Psi \left( X\right)
\right\vert ^{2}}\right)
\end{eqnarray*}%
Gathering the various terms yields:%
\begin{eqnarray*}
&&\frac{\delta }{\delta S^{T}\left( X,,\theta -2\right) }f \\
&\simeq &S_{1}\left( X,X\right) S^{T}\left( X\right) \left( \frac{\partial
f\left( X\right) }{\partial S^{T}\left( X\right) }\right) ^{3}\left( \frac{%
\partial _{f\left( X\right) }K_{X}\left\vert \Psi \left( X\right)
\right\vert ^{2}}{K_{X}\left\vert \Psi \left( X\right) \right\vert ^{2}}%
\right) ^{2} \\
&&\times \left\{ \left( 1+\frac{\frac{1}{f\left( X\right) +C_{0}}}{\frac{%
\partial _{f\left( X\right) }K_{X}\left\vert \Psi \left( X\right)
\right\vert ^{2}}{K_{X}\left\vert \Psi \left( X\right) \right\vert ^{2}}}%
\right) -\frac{\frac{\hat{h}\left( X\right) }{2}+2\frac{\frac{\partial
^{2}f\left( X\right) }{\partial \left( S^{T}\left( X,\theta -2\right)
\right) ^{2}}}{\left( \frac{\partial f\left( X\right) }{\partial S^{T}\left(
X,\theta -1\right) }\right) ^{2}}}{\frac{\partial _{f\left( X\right)
}K_{X}\left\vert \Psi \left( X\right) \right\vert ^{2}}{K_{X}\left\vert \Psi
\left( X\right) \right\vert ^{2}}}\right\}
\end{eqnarray*}

\textbf{Estimation of }$\frac{\delta }{\delta \bar{f}\left( X,\theta
-1\right) }c$ \textbf{and }$\frac{\delta }{\delta \bar{f}\left( X,\theta
-1\right) }d$

\begin{equation*}
\frac{\delta }{\delta \bar{f}\left( X,\theta -1\right) }c=\left( \frac{%
\partial f\left( X\right) }{\partial S^{T}\left( X,\theta -2\right) }\right)
^{2}h_{1}^{B}\left( X\right) \left( \left\langle \bar{h}\left( X\right)
\right\rangle +\left\langle \hat{h}_{1}^{B}\left( X\right) \right\rangle
\right)
\end{equation*}%
\begin{eqnarray*}
&&\frac{\delta }{\delta \bar{f}\left( X,\theta -1\right) }d \\
&\simeq &S_{1}\left( X,X,\theta -1\right) \left( S_{1}^{B}\left( X,\theta
-1\right) +S_{2}^{B}\left( X,\theta -1\right) \right) \frac{\partial f\left(
X\right) }{\partial S^{T}\left( X,\theta -1\right) }\left( \frac{\partial _{%
\bar{f}\left( X\right) }\bar{K}_{X}\left\vert \bar{\Psi}\left( X\right)
\right\vert ^{2}}{\bar{K}_{X}\left\vert \bar{\Psi}\left( X\right)
\right\vert ^{2}}\right) ^{2}<<1
\end{eqnarray*}

\textbf{Estimation of }$\frac{\delta }{\delta S^{T}\left( X,\theta -2\right) 
}a$ \textbf{and} $\frac{\delta }{\delta \bar{f}\left( X,\theta -1\right) }a$ 
\begin{eqnarray*}
&&\frac{\delta }{\delta S^{T}\left( X,\theta -2\right) }a \\
&\simeq &-S_{1}^{B}\left( X,X\right) \left( S_{1}^{B}\left( X,\theta
-1\right) +S_{2}^{B}\left( X,\theta -1\right) \right) \times \left( \frac{%
\partial _{f\left( X\right) }K_{X}\left\vert \Psi \left( X\right)
\right\vert ^{2}}{K_{X}\left\vert \Psi \left( X\right) \right\vert ^{2}}%
\right) ^{2}\left( \frac{\partial f\left( X\right) }{\partial S^{T}\left(
X,\theta -2\right) }\right) ^{2}
\end{eqnarray*}%
\begin{eqnarray*}
&&\frac{\delta }{\delta \bar{f}\left( X,\theta -1\right) }a \\
&\simeq &S_{1}^{B}\left( X,X\right) \left( S_{1}^{B}\left( X,\theta
-1\right) +S_{2}^{B}\left( X,\theta -1\right) \right) \left( \frac{\partial
_{\bar{f}\left( X\right) }\bar{K}_{X}\left\vert \bar{\Psi}\left( X\right)
\right\vert ^{2}}{\bar{K}_{X}\left\vert \bar{\Psi}\left( X\right)
\right\vert ^{2}}\right) ^{2}\frac{\partial f\left( X\right) }{\partial
S^{T}\left( X,\theta -1\right) }
\end{eqnarray*}

\textbf{Estimation of the stability domain}

Condition (\ref{cndt}) becomes:%
\begin{eqnarray*}
&&\left( 1-\left\langle \lambda ^{2}\right\rangle \right) \left( \left(
\delta \bar{S}\left( X^{\prime },X\right) \right) ^{2}+\left( \delta
S^{T}\left( X\right) \right) ^{2}\right) \\
&>&\left\{ \frac{\frac{\left\langle \bar{h}\left( X,X\right) \right\rangle }{%
2}\delta c}{\delta S^{T}\left( X,,\theta -2\right) }\left( \delta
S^{T}\left( X\right) \right) ^{2}+\delta S^{T}\left( X\right) \left( \frac{%
\delta a}{\delta S^{T}\left( X,,\theta -2\right) }+\frac{\frac{\bar{h}\left(
X^{\prime },X\right) }{2}\delta c}{\delta \bar{f}\left( X,\theta -1\right) }%
\right) \delta \bar{S}\left( X\right) +\frac{\delta a}{\delta \bar{f}\left(
X,\theta -1\right) }\left( \delta S^{T}\left( X\right) \right) ^{2}\right\}
^{2} \\
&&+\left\{ \frac{\left\langle \hat{h}\left( X^{\prime },X\right)
\right\rangle }{2}\left( \frac{\delta f}{\delta S^{T}\left( X,,\theta
-2\right) }\right) \left( \delta S^{T}\left( X\right) \right) ^{2}+\frac{%
\hat{h}\left( X^{\prime },X\right) }{\bar{h}\left( X^{\prime },X\right) }%
\frac{\delta d}{\delta \bar{f}\left( X,\theta -1\right) }\left( \delta \bar{S%
}\left( X\right) \right) ^{2}\right\} ^{2}
\end{eqnarray*}%
and this writes:%
\begin{eqnarray*}
1-\left\langle \lambda ^{2}\right\rangle &>&\left( \left\{ S_{1}^{B}\left(
X,X\right) \left( A\left( \delta S^{T}\left( X\right) \right) ^{2}+B\left(
\delta \bar{S}\left( X^{\prime },X\right) \right) ^{2}+\delta S^{T}\left(
X\right) C\delta \bar{S}\left( X^{\prime },X\right) \right) \right\}
^{2}\right. \\
&&\left. +\left\{ S_{1}\left( X,X\right) \left( D\left( \delta S^{T}\left(
X\right) \right) ^{2}+L\left( \delta \bar{S}\left( X^{\prime },X\right)
\right) ^{2}\right) \right\} ^{2}\right) \\
&&\times \frac{\left\{ \left( \frac{\partial f\left( X\right) }{\partial
S^{T}\left( X,\theta -1\right) }\right) ^{3}\left( \frac{\partial _{f\left(
X\right) }K_{X}\left\vert \Psi \left( X\right) \right\vert ^{2}}{%
K_{X}\left\vert \Psi \left( X\right) \right\vert ^{2}}\right)
^{2}S^{T}\left( X\right) \right\} ^{2}}{\left( \delta \bar{S}\left(
X^{\prime },X\right) \right) ^{2}+\left( \delta S^{T}\left( X\right) \right)
^{2}}
\end{eqnarray*}%
with:%
\begin{eqnarray*}
A &=&\frac{\bar{h}\left( X\right) }{2}\left[ \left( 1+\frac{\frac{1}{f\left(
X\right) +C_{0}}}{\frac{\partial _{f\left( X\right) }K_{X}\left\vert \Psi
\left( X\right) \right\vert ^{2}}{K_{X}\left\vert \Psi \left( X\right)
\right\vert ^{2}}}\right) -\frac{\frac{h_{1}^{B}\left( X\right) \left(
\left\langle \bar{h}\left( X\right) \right\rangle +\left\langle \hat{h}%
_{1}^{B}\left( X\right) \right\rangle \right) }{S_{1}^{B}\left( X,X\right) }%
+2\frac{\frac{\partial ^{2}f\left( X\right) }{\partial \left( S^{T}\left(
X,\theta -2\right) \right) ^{2}}}{\left( \frac{\partial f\left( X\right) }{%
\partial S^{T}\left( X,\theta -2\right) }\right) ^{2}}}{\frac{\partial
_{f\left( X\right) }K_{X}\left\vert \Psi \left( X\right) \right\vert ^{2}}{%
K_{X}\left\vert \Psi \left( X\right) \right\vert ^{2}}}\right] \\
B &=&\frac{\left( S_{1}^{B}\left( X,\theta -1\right) +S_{2}^{B}\left(
X,\theta -1\right) \right) }{\frac{\partial f\left( X\right) }{\partial
S^{T}\left( X,\theta -2\right) }S^{T}\left( X\right) } \\
C &=&\frac{\frac{\bar{h}\left( X^{\prime },X\right) }{2}h_{1}^{B}\left(
X\right) \left( \left\langle \bar{h}\left( X\right) \right\rangle
+\left\langle \hat{h}_{1}^{B}\left( X\right) \right\rangle \right) }{\frac{%
\partial f\left( X\right) }{\partial S^{T}\left( X,\theta -2\right) }\left( 
\frac{\partial _{f\left( X\right) }K_{X}\left\vert \Psi \left( X\right)
\right\vert ^{2}}{K_{X}\left\vert \Psi \left( X\right) \right\vert ^{2}}%
\right) ^{2}S^{T}\left( X\right) S_{1}^{B}\left( X,X\right) }-\frac{\left(
S_{1}^{B}\left( X,\theta -1\right) +S_{2}^{B}\left( X,\theta -1\right)
\right) }{\frac{\partial f\left( X\right) }{\partial S^{T}\left( X,\theta
-2\right) }S^{T}\left( X\right) } \\
D &=&\frac{\hat{h}\left( X\right) }{2}\left[ \left( 1+\frac{\frac{1}{f\left(
X\right) +C_{0}}}{\frac{\partial _{f\left( X\right) }K_{X}\left\vert \Psi
\left( X\right) \right\vert ^{2}}{K_{X}\left\vert \Psi \left( X\right)
\right\vert ^{2}}}\right) -\frac{\frac{\hat{h}\left( X\right) }{2}+2\frac{%
\frac{\partial ^{2}f\left( X\right) }{\partial \left( S^{T}\left( X,\theta
-2\right) \right) ^{2}}}{\left( \frac{\partial f\left( X\right) }{\partial
S^{T}\left( X,\theta -1\right) }\right) ^{2}}}{\frac{\partial _{f\left(
X\right) }K_{X}\left\vert \Psi \left( X\right) \right\vert ^{2}}{%
K_{X}\left\vert \Psi \left( X\right) \right\vert ^{2}}}\right] \\
L &=&\frac{\hat{h}\left( X^{\prime },X\right) }{\bar{h}\left( X^{\prime
},X\right) }\frac{\left( S_{1}^{B}\left( X,\theta -1\right) +S_{2}^{B}\left(
X,\theta -1\right) \right) \left( \frac{\partial _{\bar{f}\left( X\right) }%
\bar{K}_{X}\left\vert \bar{\Psi}\left( X\right) \right\vert ^{2}}{\bar{K}%
_{X}\left\vert \bar{\Psi}\left( X\right) \right\vert ^{2}}\right) ^{2}}{%
S^{T}\left( X\right) \left( \frac{\partial f\left( X\right) }{\partial
S^{T}\left( X,\theta -2\right) }\right) ^{2}\left( \frac{\partial _{f\left(
X\right) }K_{X}\left\vert \Psi \left( X\right) \right\vert ^{2}}{%
K_{X}\left\vert \Psi \left( X\right) \right\vert ^{2}}\right) ^{2}}
\end{eqnarray*}%
\begin{equation*}
\lambda =\frac{1}{2}a+\frac{1}{2}i\pm \frac{1}{2}\sqrt{\left( a-i\right)
^{2}+4cg}
\end{equation*}%
\begin{equation*}
a\simeq \left( S_{1}^{B}\left( X,\theta -1\right) +S_{2}^{B}\left( X,\theta
-1\right) \right) \frac{\partial _{\bar{f}\left( X\right) }\bar{K}%
_{X}\left\vert \bar{\Psi}\left( X\right) \right\vert ^{2}}{\bar{K}%
_{X}\left\vert \bar{\Psi}\left( X\right) \right\vert ^{2}}S_{1}^{B}\left(
X,X\right) \frac{\partial f\left( X\right) }{\partial S^{T}\left( X,\theta
-1\right) }
\end{equation*}%
\begin{eqnarray*}
c &=&S_{1}^{B}\left( X,X\right) \frac{\partial f\left( X\right) }{\partial
S^{T}\left( X,\theta -1\right) }\left\{ h_{1}^{B}\left( X\right) \left(
\left\langle \bar{h}\left( X\right) \right\rangle +\left\langle \hat{h}%
_{1}^{B}\left( X\right) \right\rangle \right) \right. -S^{T}\left( X\right) 
\frac{\partial _{f\left( X\right) }K_{X}\left\vert \Psi \left( X\right)
\right\vert ^{2}}{K_{X}\left\vert \Psi \left( X\right) \right\vert ^{2}} \\
&&\left. \left. +\left( \frac{\hat{h}\left( X\right) }{2}S\left( X\right)
\right) \right\} \right\} \frac{\partial f\left( X\right) }{\partial
S^{T}\left( X,\theta -2\right) }
\end{eqnarray*}%
\begin{equation*}
i=\left( h_{1}^{B}\left( X\right) \left( \left\langle \bar{h}\left( X\right)
\right\rangle +\left\langle \hat{h}_{1}^{B}\left( X\right) \right\rangle
\right) -S^{T}\left( X\right) \frac{\partial _{f\left( X\right)
}K_{X}\left\vert \Psi \left( X\right) \right\vert ^{2}}{K_{X}\left\vert \Psi
\left( X\right) \right\vert ^{2}}+\left( \frac{\hat{h}\left( X\right) }{2}%
S\left( X\right) \right) \right) \frac{\partial f\left( X\right) }{\partial
S^{T}\left( X,\theta -2\right) }
\end{equation*}%
\begin{equation*}
\frac{\partial _{\bar{f}\left( X\right) }\bar{K}_{X}\left\vert \bar{\Psi}%
\left( X\right) \right\vert ^{2}}{\bar{K}_{X}\left\vert \bar{\Psi}\left(
X\right) \right\vert ^{2}}<<1
\end{equation*}%
which implies:%
\begin{equation*}
g=\left( S_{1}^{B}\left( X,\theta -1\right) +S_{2}^{B}\left( X,\theta
-1\right) \right) \frac{\partial _{\bar{f}\left( X\right) }\bar{K}%
_{X}\left\vert \bar{\Psi}\left( X\right) \right\vert ^{2}}{\bar{K}%
_{X}\left\vert \bar{\Psi}\left( X\right) \right\vert ^{2}}<<1
\end{equation*}%
\begin{equation*}
a<<i
\end{equation*}%
and:%
\begin{equation*}
\lambda \simeq \frac{1}{2}i\pm \frac{1}{2}i
\end{equation*}%
\begin{equation*}
\left\langle \lambda ^{2}\right\rangle \simeq \frac{i^{2}}{2}
\end{equation*}%
and:%
\begin{equation*}
1-\left\langle \lambda ^{2}\right\rangle \simeq 1-\left( S^{T}\left(
X\right) \frac{\partial _{f\left( X\right) }K_{X}\left\vert \Psi \left(
X\right) \right\vert ^{2}}{K_{X}\left\vert \Psi \left( X\right) \right\vert
^{2}}\frac{\partial f\left( X\right) }{\partial S^{T}\left( X,\theta
-2\right) }\right) ^{2}
\end{equation*}

\subsubsection*{A4.3.2 Case 2: $S^{TT}\left( X\right) \frac{1}{f\left(
X\right) +C_{0}}<h_{1}^{B}\left( X\right) \left( \left\langle \bar{h}\left(
X\right) \right\rangle +\left\langle \hat{h}_{1}^{B}\left( X\right)
\right\rangle \right) $}

\textbf{Derivation of }$\frac{\delta }{\delta S^{T}\left( X,,\theta
-2\right) }c$

When $\frac{1}{f\left( X\right) +C_{0}}<1$:\bigskip 
\begin{eqnarray*}
&&\frac{\delta }{\delta S^{T}\left( X,,\theta -2\right) }c \\
&\simeq &\left( \frac{\partial f\left( X\right) }{\partial S^{T}\left(
X,\theta -1\right) }\right) ^{3}h_{1}^{B}\left( X\right) \left( \left\langle 
\bar{h}\left( X\right) \right\rangle +\left\langle \hat{h}_{1}^{B}\left(
X\right) \right\rangle \right) \\
&&\times \left\{ h_{1}^{B}\left( X\right) \left( \left\langle \bar{h}\left(
X\right) \right\rangle +\left\langle \hat{h}_{1}^{B}\left( X\right)
\right\rangle \right) +\frac{\hat{h}\left( X\right) }{2}S\left( X\right)
\right. \\
&&\left. -S_{1}^{B}\left( X,X\right) \left( \frac{\partial _{f\left(
X\right) }K_{X}\left\vert \Psi \left( X\right) \right\vert ^{2}}{%
K_{X}\left\vert \Psi \left( X\right) \right\vert ^{2}}h_{1}^{B}\left(
X\right) \left( \left\langle \bar{h}\left( X\right) \right\rangle
+\left\langle \hat{h}_{1}^{B}\left( X\right) \right\rangle \right) -\left( 
\frac{\hat{h}\left( X\right) }{2}\right) ^{2}S\left( X\right) \right)
\right\} \\
&&+c\frac{\frac{\partial ^{2}f\left( X\right) }{\partial \left( S^{T}\left(
X,\theta -2\right) \right) ^{2}}}{\frac{\partial f\left( X\right) }{\partial
S^{T}\left( X,\theta -2\right) }}+S_{1}^{B}\left( X,X\right) \frac{\partial
f\left( X\right) }{\partial S^{T}\left( X,\theta -1\right) }\left\{
h_{1}^{B}\left( X\right) \left( \left\langle \bar{h}\left( X\right)
\right\rangle +\left\langle \hat{h}_{1}^{B}\left( X\right) \right\rangle
\right) \right. \\
&&\left. +\frac{\hat{h}\left( X\right) }{2}S\left( X\right) \right\} \frac{%
\partial ^{2}f\left( X\right) }{\partial \left( S^{T}\left( X,\theta
-2\right) \right) ^{2}}
\end{eqnarray*}%
We use that:%
\begin{eqnarray*}
&&c\frac{\frac{\partial ^{2}f\left( X\right) }{\partial \left( S^{T}\left(
X,\theta -2\right) \right) ^{2}}}{\frac{\partial f\left( X\right) }{\partial
S^{T}\left( X,\theta -2\right) }} \\
&\simeq &S_{1}^{B}\left( X,X\right) \frac{\partial f\left( X\right) }{%
\partial S^{T}\left( X,\theta -1\right) }\left( h_{1}^{B}\left( X\right)
\left( \left\langle \bar{h}\left( X\right) \right\rangle +\left\langle \hat{h%
}_{1}^{B}\left( X\right) \right\rangle \right) +\frac{\hat{h}\left( X\right) 
}{2}S\left( X\right) \right) \frac{\partial ^{2}f\left( X\right) }{\partial
\left( S^{T}\left( X,\theta -2\right) \right) ^{2}}
\end{eqnarray*}%
and we obtain:%
\begin{eqnarray*}
&&\frac{\delta }{\delta S^{T}\left( X,,\theta -2\right) }c \\
&\simeq &\left( \frac{\partial f\left( X\right) }{\partial S^{T}\left(
X,\theta -1\right) }\right) ^{3}h_{1}^{B}\left( X\right) \left( \left\langle 
\bar{h}\left( X\right) \right\rangle +\left\langle \hat{h}_{1}^{B}\left(
X\right) \right\rangle \right) \\
&&\times \left\{ h_{1}^{B}\left( X\right) \left( \left\langle \bar{h}\left(
X\right) \right\rangle +\left\langle \hat{h}_{1}^{B}\left( X\right)
\right\rangle \right) +\frac{\hat{h}\left( X\right) }{2}S\left( X\right)
\right. \\
&&\left. -S_{1}^{B}\left( X,X\right) \left( \frac{\partial _{f\left(
X\right) }K_{X}\left\vert \Psi \left( X\right) \right\vert ^{2}}{%
K_{X}\left\vert \Psi \left( X\right) \right\vert ^{2}}h_{1}^{B}\left(
X\right) \left( \left\langle \bar{h}\left( X\right) \right\rangle
+\left\langle \hat{h}_{1}^{B}\left( X\right) \right\rangle \right) -\left( 
\frac{\hat{h}\left( X\right) }{2}\right) ^{2}S\left( X\right) \right)
\right\} \\
&&+2S_{1}^{B}\left( X,X\right) \frac{\partial f\left( X\right) }{\partial
S^{T}\left( X,\theta -1\right) }\left( h_{1}^{B}\left( X\right) \left(
\left\langle \bar{h}\left( X\right) \right\rangle +\left\langle \hat{h}%
_{1}^{B}\left( X\right) \right\rangle \right) +\frac{\hat{h}\left( X\right) 
}{2}S\left( X\right) \right) \frac{\partial ^{2}f\left( X\right) }{\partial
\left( S^{T}\left( X,\theta -2\right) \right) ^{2}}
\end{eqnarray*}%
Ultimately this yields:%
\begin{eqnarray*}
&&\frac{\delta }{\delta S^{T}\left( X,,\theta -2\right) }c \\
&\simeq &\left( \frac{\partial f\left( X\right) }{\partial S^{T}\left(
X,\theta -1\right) }\right) ^{3}\left( h_{1}^{B}\left( X\right) \left(
\left\langle \bar{h}\left( X\right) \right\rangle +\left\langle \hat{h}%
_{1}^{B}\left( X\right) \right\rangle \right) +\frac{\hat{h}\left( X\right) 
}{2}S\left( X\right) \right) \\
&&\times \left\{ h_{1}^{B}\left( X\right) \left( \left\langle \bar{h}\left(
X\right) \right\rangle +\left\langle \hat{h}_{1}^{B}\left( X\right)
\right\rangle \right) +S_{1}^{B}\left( X,X\right) \frac{2\frac{\partial
^{2}f\left( X\right) }{\partial \left( S^{T}\left( X,\theta -2\right)
\right) ^{2}}}{\left( \frac{\partial f\left( X\right) }{\partial S^{T}\left(
X,\theta -1\right) }\right) ^{2}}-\frac{\left( \frac{\hat{h}\left( X\right) 
}{2}\right) ^{2}S\left( X\right) }{h_{1}^{B}\left( X\right) \left(
\left\langle \bar{h}\left( X\right) \right\rangle +\left\langle \hat{h}%
_{1}^{B}\left( X\right) \right\rangle \right) +\frac{\hat{h}\left( X\right) 
}{2}S\left( X\right) }\right\} \\
&\simeq &\left( \frac{\partial f\left( X\right) }{\partial S^{T}\left(
X,\theta -1\right) }\right) ^{3}\left( h_{1}^{B}\left( X\right) \left(
\left\langle \bar{h}\left( X\right) \right\rangle +\left\langle \hat{h}%
_{1}^{B}\left( X\right) \right\rangle \right) \right) h_{1}^{B}\left(
X\right) \left( \left\langle \bar{h}\left( X\right) \right\rangle
+\left\langle \hat{h}_{1}^{B}\left( X\right) \right\rangle \right)
\end{eqnarray*}%
where we used that $S\left( X\right) <<1$ given the bank loans.

\bigskip \textbf{Derivation of }$\frac{\delta }{\delta S^{T}\left( X,,\theta
-2\right) }f$

\begin{eqnarray*}
&&\frac{\delta }{\delta S^{T}\left( X,,\theta -2\right) }f \\
&\simeq &S_{1}\left( X,X\right) \left( \frac{\partial f\left( X\right) }{%
\partial S^{T}\left( X,\theta -1\right) }\right) ^{3}\frac{\hat{h}\left(
X\right) }{2}\left( h_{1}^{B}\left( X\right) \left( \left\langle \bar{h}%
\left( X\right) \right\rangle +\left\langle \hat{h}_{1}^{B}\left( X\right)
\right\rangle \right) +\frac{\hat{h}\left( X\right) }{2}S\left( X\right)
\right) \\
&&+S_{1}\left( X,X\right) \left( \frac{\partial f\left( X\right) }{\partial
S^{T}\left( X,\theta -1\right) }\right) ^{3}\left( -\frac{\partial _{f\left(
X\right) }K_{X}\left\vert \Psi \left( X\right) \right\vert ^{2}}{%
K_{X}\left\vert \Psi \left( X\right) \right\vert ^{2}}h_{1}^{B}\left(
X\right) \left( \left\langle \bar{h}\left( X\right) \right\rangle
+\left\langle \hat{h}_{1}^{B}\left( X\right) \right\rangle \right) +\left( 
\frac{\hat{h}\left( X\right) }{2}\right) ^{2}S\left( X\right) \right) \\
&&+2S_{1}\left( X,X\right) \frac{\partial f\left( X\right) }{\partial
S^{T}\left( X,\theta -2\right) }\frac{\partial ^{2}f\left( X\right) }{%
\partial \left( S^{T}\left( X,\theta -2\right) \right) ^{2}}\left(
h_{1}^{B}\left( X\right) \left( \left\langle \bar{h}\left( X\right)
\right\rangle +\left\langle \hat{h}_{1}^{B}\left( X\right) \right\rangle
\right) +\frac{\hat{h}\left( X\right) }{2}S\left( X\right) \right)
\end{eqnarray*}%
that is factored as:%
\begin{eqnarray*}
&&\frac{\delta }{\delta S^{T}\left( X,,\theta -2\right) }f \\
&\simeq &S_{1}\left( X,X\right) \left( \frac{\partial f\left( X\right) }{%
\partial S^{T}\left( X,\theta -1\right) }\right) ^{3}\left( h_{1}^{B}\left(
X\right) \left( \left\langle \bar{h}\left( X\right) \right\rangle
+\left\langle \hat{h}_{1}^{B}\left( X\right) \right\rangle \right) +\frac{%
\hat{h}\left( X\right) }{2}S\left( X\right) \right) \\
&&\times \left( \frac{\hat{h}\left( X\right) }{2}+\frac{\left( \frac{\hat{h}%
\left( X\right) }{2}\right) ^{2}S\left( X\right) -\frac{\partial _{f\left(
X\right) }K_{X}\left\vert \Psi \left( X\right) \right\vert ^{2}}{%
K_{X}\left\vert \Psi \left( X\right) \right\vert ^{2}}h_{1}^{B}\left(
X\right) \left( \left\langle \bar{h}\left( X\right) \right\rangle
+\left\langle \hat{h}_{1}^{B}\left( X\right) \right\rangle \right) }{%
h_{1}^{B}\left( X\right) \left( \left\langle \bar{h}\left( X\right)
\right\rangle +\left\langle \hat{h}_{1}^{B}\left( X\right) \right\rangle
\right) +\frac{\hat{h}\left( X\right) }{2}S\left( X\right) }+2\frac{\frac{%
\partial ^{2}f\left( X\right) }{\partial \left( S^{T}\left( X,\theta
-2\right) \right) ^{2}}}{\left( \frac{\partial f\left( X\right) }{\partial
S^{T}\left( X,\theta -1\right) }\right) ^{2}}\right) \\
&\simeq &S_{1}\left( X,X\right) \left( \frac{\partial f\left( X\right) }{%
\partial S^{T}\left( X,\theta -1\right) }\right) ^{3}\left( h_{1}^{B}\left(
X\right) \left( \left\langle \bar{h}\left( X\right) \right\rangle
+\left\langle \hat{h}_{1}^{B}\left( X\right) \right\rangle \right) \right) \\
&&\times \left( \frac{\hat{h}\left( X\right) }{2}-\frac{\partial _{f\left(
X\right) }K_{X}\left\vert \Psi \left( X\right) \right\vert ^{2}}{%
K_{X}\left\vert \Psi \left( X\right) \right\vert ^{2}}+2\frac{\frac{\partial
^{2}f\left( X\right) }{\partial \left( S^{T}\left( X,\theta -2\right)
\right) ^{2}}}{\left( \frac{\partial f\left( X\right) }{\partial S^{T}\left(
X,\theta -1\right) }\right) ^{2}}\right)
\end{eqnarray*}

\bigskip \textbf{Estimation of }$\frac{\delta }{\delta \bar{f}\left(
X,\theta -1\right) }c$ \textbf{and }$\frac{\delta }{\delta \bar{f}\left(
X,\theta -1\right) }d$

\begin{equation*}
\frac{\delta }{\delta \bar{f}\left( X,\theta -1\right) }c=\left( \frac{%
\partial f\left( X\right) }{\partial S^{T}\left( X,\theta -2\right) }\right)
^{2}h_{1}^{B}\left( X\right) \left( \left\langle \bar{h}\left( X\right)
\right\rangle +\left\langle \hat{h}_{1}^{B}\left( X\right) \right\rangle
\right)
\end{equation*}%
\begin{eqnarray*}
&&\frac{\delta }{\delta \bar{f}\left( X,\theta -1\right) }d \\
&\simeq &S_{1}\left( X,X,\theta -1\right) \left( S_{1}^{B}\left( X,\theta
-1\right) +S_{2}^{B}\left( X,\theta -1\right) \right) \frac{\partial f\left(
X\right) }{\partial S^{T}\left( X,\theta -1\right) }\left( \frac{\partial _{%
\bar{f}\left( X\right) }\bar{K}_{X}\left\vert \bar{\Psi}\left( X\right)
\right\vert ^{2}}{\bar{K}_{X}\left\vert \bar{\Psi}\left( X\right)
\right\vert ^{2}}\right) ^{2}<<1
\end{eqnarray*}

\textbf{Estimation of }$\frac{\delta }{\delta S^{T}\left( X,\theta -2\right) 
}a$ \textbf{and} $\frac{\delta }{\delta \bar{f}\left( X,\theta -1\right) }a$ 
\begin{eqnarray*}
&&\frac{\delta }{\delta S^{T}\left( X,\theta -2\right) }a \\
&\simeq &h_{1}^{B}\left( X\right) \left( \left\langle \bar{h}\left( X\right)
\right\rangle +\left\langle \hat{h}_{1}^{B}\left( X\right) \right\rangle
\right) \\
&&\times \left( S_{1}^{B}\left( X,X\right) +\left( S_{1}^{B}\left( X,\theta
-1\right) +S_{2}^{B}\left( X,\theta -1\right) \right) \left( \frac{\partial
f\left( X\right) }{\partial S^{T}\left( X,\theta -2\right) }\right) ^{2}%
\frac{\partial _{f\left( X\right) }K_{X}\left\vert \Psi \left( X\right)
\right\vert ^{2}}{K_{X}\left\vert \Psi \left( X\right) \right\vert ^{2}}%
\right) \\
&\rightarrow &h_{1}^{B}\left( X\right) \left( \left\langle \bar{h}\left(
X\right) \right\rangle +\left\langle \hat{h}_{1}^{B}\left( X\right)
\right\rangle \right) \left( \left( S_{1}^{B}\left( X,\theta -1\right)
+S_{2}^{B}\left( X,\theta -1\right) \right) \left( \frac{\partial f\left(
X\right) }{\partial S^{T}\left( X,\theta -2\right) }\right) ^{2}\frac{%
\partial _{f\left( X\right) }K_{X}\left\vert \Psi \left( X\right)
\right\vert ^{2}}{K_{X}\left\vert \Psi \left( X\right) \right\vert ^{2}}%
\right)
\end{eqnarray*}%
\begin{eqnarray*}
&&\frac{\delta }{\delta \bar{f}\left( X,\theta -1\right) }a \\
&\simeq &\left( S_{1}^{B}\left( X,\theta -1\right) +S_{2}^{B}\left( X,\theta
-1\right) \right) \\
&&\times \left( h_{1}^{B}\left( X\right) \left( \left\langle \bar{h}\left(
X\right) \right\rangle +\left\langle \hat{h}_{1}^{B}\left( X\right)
\right\rangle \right) \right) \frac{\partial _{\bar{f}\left( X\right) }\bar{K%
}_{X}\left\vert \bar{\Psi}\left( X\right) \right\vert ^{2}}{\bar{K}%
_{X}\left\vert \bar{\Psi}\left( X\right) \right\vert ^{2}}\frac{\partial
f\left( X\right) }{\partial S^{T}\left( X,\theta -1\right) }
\end{eqnarray*}

\textbf{Estimation of the stability domain}

Condition (\ref{cndt}) becomes:

\begin{eqnarray*}
&&\left( 1-\left\langle \lambda ^{2}\right\rangle \right) \left( \left(
\delta \bar{S}\left( X^{\prime },X\right) \right) ^{2}+\left( \delta
S^{T}\left( X\right) \right) ^{2}\right) \\
&>&\left\{ \frac{\left( \delta S^{T}\left( X\right) \right) ^{2}\frac{%
\left\langle \bar{h}\left( X,X\right) \right\rangle }{2}\delta c}{\delta
S^{T}\left( X,,\theta -2\right) }+\delta S^{T}\left( X\right) \left( \frac{%
\delta a}{\delta S^{T}\left( X,,\theta -2\right) }+\frac{\frac{\bar{h}\left(
X^{\prime },X\right) }{2}\delta c}{\delta \bar{f}\left( X,\theta -1\right) }%
\right) \delta \bar{S}\left( X\right) +\frac{\left( \delta S^{T}\left(
X\right) \right) ^{2}\delta a}{\delta \bar{f}\left( X,\theta -1\right) }%
\right\} ^{2} \\
&&+\left\{ \frac{\left\langle \hat{h}\left( X^{\prime },X\right)
\right\rangle }{2}\frac{\delta f}{\delta S^{T}\left( X,,\theta -2\right) }%
\left( \delta S^{T}\left( X\right) \right) ^{2}+\frac{\hat{h}\left(
X^{\prime },X\right) }{\bar{h}\left( X^{\prime },X\right) }\frac{\delta d}{%
\delta \bar{f}\left( X,\theta -1\right) }\left( \delta \bar{S}\left(
X\right) \right) ^{2}\right\} ^{2}
\end{eqnarray*}%
which can be developed as:%
\begin{eqnarray*}
1-\left\langle \lambda ^{2}\right\rangle &>&\left( \left\{ S_{1}^{B}\left(
X,X\right) \left( \frac{\bar{h}\left( X\right) }{2}A\left( \delta
S^{T}\left( X\right) \right) ^{2}+B\left( \delta \bar{S}\left( X^{\prime
},X\right) \right) ^{2}+\delta S^{T}\left( X\right) C\delta \bar{S}\left(
X^{\prime },X\right) \right) \right\} ^{2}\right. \\
&&+\left\{ S_{1}\left( X,X\right) \left( \left( \frac{\hat{h}\left( X\right) 
}{2}D\left( \delta S^{T}\left( X\right) \right) ^{2}\right) +L\left( \delta 
\bar{S}\left( X^{\prime },X\right) \right) ^{2}\right) \right\} ^{2} \\
&&\times \frac{\left\{ \left( \frac{\partial f\left( X\right) }{\partial
S^{T}\left( X,\theta -1\right) }\right) ^{3}\left( h_{1}^{B}\left( X\right)
\left( \left\langle \bar{h}\left( X\right) \right\rangle +\left\langle \hat{h%
}_{1}^{B}\left( X\right) \right\rangle \right) +\frac{\hat{h}\left( X\right) 
}{2}S\left( X\right) \right) \right\} ^{2}}{\left( \delta \bar{S}\left(
X^{\prime },X\right) \right) ^{2}+\left( \delta S^{T}\left( X\right) \right)
^{2}}
\end{eqnarray*}%
where the coefficients are defined by:%
\begin{eqnarray*}
A &=&h_{1}^{B}\left( X\right) \left( \left\langle \bar{h}\left( X\right)
\right\rangle +\left\langle \hat{h}_{1}^{B}\left( X\right) \right\rangle
\right) +S_{1}^{B}\left( X,X\right) \frac{2\frac{\partial ^{2}f\left(
X\right) }{\partial \left( S^{T}\left( X,\theta -2\right) \right) ^{2}}}{%
\left( \frac{\partial f\left( X\right) }{\partial S^{T}\left( X,\theta
-1\right) }\right) ^{2}}-\frac{\left( \frac{\hat{h}\left( X\right) }{2}%
\right) ^{2}S\left( X\right) }{h_{1}^{B}\left( X\right) \left( \left\langle 
\bar{h}\left( X\right) \right\rangle +\left\langle \hat{h}_{1}^{B}\left(
X\right) \right\rangle \right) +\frac{\hat{h}\left( X\right) }{2}S\left(
X\right) } \\
&\simeq &h_{1}^{B}\left( X\right) \left( \left\langle \bar{h}\left( X\right)
\right\rangle +\left\langle \hat{h}_{1}^{B}\left( X\right) \right\rangle
\right) +S_{1}^{B}\left( X,X\right) \frac{2\frac{\partial ^{2}f\left(
X\right) }{\partial \left( S^{T}\left( X,\theta -2\right) \right) ^{2}}}{%
\left( \frac{\partial f\left( X\right) }{\partial S^{T}\left( X,\theta
-1\right) }\right) ^{2}} \\
B &=&\frac{\left( S_{1}^{B}\left( X,\theta -1\right) +S_{2}^{B}\left(
X,\theta -1\right) \right) \left( h_{1}^{B}\left( X\right) \left(
\left\langle \bar{h}\left( X\right) \right\rangle +\left\langle \hat{h}%
_{1}^{B}\left( X\right) \right\rangle \right) \right) \frac{\partial _{\bar{f%
}\left( X\right) }\bar{K}_{X}\left\vert \bar{\Psi}\left( X\right)
\right\vert ^{2}}{\bar{K}_{X}\left\vert \bar{\Psi}\left( X\right)
\right\vert ^{2}}}{\left( \frac{\partial f\left( X\right) }{\partial
S^{T}\left( X,\theta -1\right) }\right) ^{2}\left( h_{1}^{B}\left( X\right)
\left( \left\langle \bar{h}\left( X\right) \right\rangle +\left\langle \hat{h%
}_{1}^{B}\left( X\right) \right\rangle \right) +\frac{\hat{h}\left( X\right) 
}{2}S\left( X\right) \right) } \\
&\simeq &\frac{\left( S_{1}^{B}\left( X,\theta -1\right) +S_{2}^{B}\left(
X,\theta -1\right) \right) \frac{\partial _{\bar{f}\left( X\right) }\bar{K}%
_{X}\left\vert \bar{\Psi}\left( X\right) \right\vert ^{2}}{\bar{K}%
_{X}\left\vert \bar{\Psi}\left( X\right) \right\vert ^{2}}}{\left( \frac{%
\partial f\left( X\right) }{\partial S^{T}\left( X,\theta -1\right) }\right)
^{2}} \\
C &=&\frac{h_{1}^{B}\left( X\right) \left( \left\langle \bar{h}\left(
X\right) \right\rangle +\left\langle \hat{h}_{1}^{B}\left( X\right)
\right\rangle \right) \left( \frac{\bar{h}\left( X^{\prime },X\right) }{2}%
+\left( S_{1}^{B}\left( X,\theta -1\right) +S_{2}^{B}\left( X,\theta
-1\right) \right) \frac{\frac{\partial _{\bar{f}\left( X\right) }\bar{K}%
_{X}\left\vert \bar{\Psi}\left( X\right) \right\vert ^{2}}{\bar{K}%
_{X}\left\vert \bar{\Psi}\left( X\right) \right\vert ^{2}}}{\frac{\partial
f\left( X\right) }{\partial S^{T}\left( X,\theta -1\right) }}\right) }{\frac{%
\partial f\left( X\right) }{\partial S^{T}\left( X,\theta -2\right) }%
h_{1}^{B}\left( X\right) \left( \left\langle \bar{h}\left( X\right)
\right\rangle +\left\langle \hat{h}_{1}^{B}\left( X\right) \right\rangle
\right) +\frac{\hat{h}\left( X\right) }{2}S\left( X\right) } \\
&\simeq &\frac{\bar{h}\left( X^{\prime },X\right) }{2\frac{\partial f\left(
X\right) }{\partial S^{T}\left( X,\theta -2\right) }}+\left( S_{1}^{B}\left(
X,\theta -1\right) +S_{2}^{B}\left( X,\theta -1\right) \right) \frac{\frac{%
\partial _{\bar{f}\left( X\right) }\bar{K}_{X}\left\vert \bar{\Psi}\left(
X\right) \right\vert ^{2}}{\bar{K}_{X}\left\vert \bar{\Psi}\left( X\right)
\right\vert ^{2}}}{\left( \frac{\partial f\left( X\right) }{\partial
S^{T}\left( X,\theta -1\right) }\right) ^{2}} \\
D &=&S_{1}\left( X,X\right) \left( \frac{\hat{h}\left( X\right) }{2}+\frac{%
\left( \frac{\hat{h}\left( X\right) }{2}\right) ^{2}S\left( X\right) -\frac{%
\partial _{f\left( X\right) }K_{X}\left\vert \Psi \left( X\right)
\right\vert ^{2}}{K_{X}\left\vert \Psi \left( X\right) \right\vert ^{2}}%
h_{1}^{B}\left( X\right) \left( \left\langle \bar{h}\left( X\right)
\right\rangle +\left\langle \hat{h}_{1}^{B}\left( X\right) \right\rangle
\right) }{h_{1}^{B}\left( X\right) \left( \left\langle \bar{h}\left(
X\right) \right\rangle +\left\langle \hat{h}_{1}^{B}\left( X\right)
\right\rangle \right) +\frac{\hat{h}\left( X\right) }{2}S\left( X\right) }+2%
\frac{\frac{\partial ^{2}f\left( X\right) }{\partial \left( S^{T}\left(
X,\theta -2\right) \right) ^{2}}}{\left( \frac{\partial f\left( X\right) }{%
\partial S^{T}\left( X,\theta -1\right) }\right) ^{2}}\right) \\
&\simeq &S_{1}\left( X,X\right) \frac{\hat{h}\left( X\right) }{2}-\frac{%
\partial _{f\left( X\right) }K_{X}\left\vert \Psi \left( X\right)
\right\vert ^{2}}{K_{X}\left\vert \Psi \left( X\right) \right\vert ^{2}}+2%
\frac{\frac{\partial ^{2}f\left( X\right) }{\partial \left( S^{T}\left(
X,\theta -2\right) \right) ^{2}}}{\left( \frac{\partial f\left( X\right) }{%
\partial S^{T}\left( X,\theta -1\right) }\right) ^{2}} \\
L &=&\frac{\hat{h}\left( X^{\prime },X\right) }{\bar{h}\left( X^{\prime
},X\right) }\frac{\left( S_{1}^{B}\left( X,\theta -1\right) +S_{2}^{B}\left(
X,\theta -1\right) \right) \left( \frac{\partial _{\bar{f}\left( X\right) }%
\bar{K}_{X}\left\vert \bar{\Psi}\left( X\right) \right\vert ^{2}}{\bar{K}%
_{X}\left\vert \bar{\Psi}\left( X\right) \right\vert ^{2}}\right) ^{2}}{%
\left( \frac{\partial f\left( X\right) }{\partial S^{T}\left( X,\theta
-2\right) }\right) ^{2}\left( h_{1}^{B}\left( X\right) \left( \left\langle 
\bar{h}\left( X\right) \right\rangle +\left\langle \hat{h}_{1}^{B}\left(
X\right) \right\rangle \right) +\frac{\hat{h}\left( X\right) }{2}S\left(
X\right) \right) } \\
&\simeq &\frac{\hat{h}\left( X^{\prime },X\right) }{\bar{h}\left( X^{\prime
},X\right) }\frac{\left( S_{1}^{B}\left( X,\theta -1\right) +S_{2}^{B}\left(
X,\theta -1\right) \right) \left( \frac{\partial _{\bar{f}\left( X\right) }%
\bar{K}_{X}\left\vert \bar{\Psi}\left( X\right) \right\vert ^{2}}{\bar{K}%
_{X}\left\vert \bar{\Psi}\left( X\right) \right\vert ^{2}}\right) ^{2}}{%
\left( \frac{\partial f\left( X\right) }{\partial S^{T}\left( X,\theta
-2\right) }\right) ^{2}\left( h_{1}^{B}\left( X\right) \left( \left\langle 
\bar{h}\left( X\right) \right\rangle +\left\langle \hat{h}_{1}^{B}\left(
X\right) \right\rangle \right) \right) }\frac{\hat{h}\left( X^{\prime
},X\right) }{\bar{h}\left( X^{\prime },X\right) }
\end{eqnarray*}%
Replacing the various quantities by their equilibrium static values yields:%
\begin{eqnarray*}
A &\simeq &h_{1}^{B}\left( X\right) \left( \left\langle \bar{h}\left(
X\right) \right\rangle +\left\langle \hat{h}_{1}^{B}\left( X\right)
\right\rangle \right) +S_{1}^{B}\left( X,X\right) \frac{2\frac{\partial
^{2}f\left( X\right) }{\partial \left( S^{T}\left( X,\theta -2\right)
\right) ^{2}}}{\left( \frac{\partial f\left( X\right) }{\partial S^{T}\left(
X,\theta -1\right) }\right) ^{2}} \\
B &\simeq &\frac{\left( S_{1}^{B}\left( X,\theta -1\right) +S_{2}^{B}\left(
X,\theta -1\right) \right) \frac{\partial _{\bar{f}\left( X\right) }\bar{K}%
_{X}\left\vert \bar{\Psi}\left( X\right) \right\vert ^{2}}{\bar{K}%
_{X}\left\vert \bar{\Psi}\left( X\right) \right\vert ^{2}}}{\left( \frac{%
\partial f\left( X\right) }{\partial S^{T}\left( X,\theta -1\right) }\right)
^{2}}
\end{eqnarray*}%
and:%
\begin{eqnarray*}
C &\simeq &\frac{\bar{h}\left( X^{\prime },X\right) }{2\frac{\partial
f\left( X\right) }{\partial S^{T}\left( X,\theta -2\right) }}+\left(
S_{1}^{B}\left( X\right) +S_{2}^{B}\left( X\right) \right) \frac{\frac{%
\partial _{\bar{f}\left( X\right) }\bar{K}_{X}\left\vert \bar{\Psi}\left(
X\right) \right\vert ^{2}}{\bar{K}_{X}\left\vert \bar{\Psi}\left( X\right)
\right\vert ^{2}}}{\left( \frac{\partial f\left( X\right) }{\partial
S^{T}\left( X,\theta -1\right) }\right) ^{2}} \\
D &\simeq &S_{1}\left( X,X\right) \frac{\hat{h}\left( X\right) }{2}-\frac{%
\partial _{f\left( X\right) }K_{X}\left\vert \Psi \left( X\right)
\right\vert ^{2}}{K_{X}\left\vert \Psi \left( X\right) \right\vert ^{2}}+2%
\frac{\frac{\partial ^{2}f\left( X\right) }{\partial \left( S^{T}\left(
X,\theta -2\right) \right) ^{2}}}{\left( \frac{\partial f\left( X\right) }{%
\partial S^{T}\left( X,\theta -1\right) }\right) ^{2}} \\
L &\simeq &\frac{\hat{h}\left( X^{\prime },X\right) }{\bar{h}\left(
X^{\prime },X\right) }\frac{\left( S_{1}^{B}\left( X\right) +S_{2}^{B}\left(
X\right) \right) \left( \frac{\partial _{\bar{f}\left( X\right) }\bar{K}%
_{X}\left\vert \bar{\Psi}\left( X\right) \right\vert ^{2}}{\bar{K}%
_{X}\left\vert \bar{\Psi}\left( X\right) \right\vert ^{2}}\right) ^{2}}{%
\left( \frac{\partial f\left( X\right) }{\partial S^{T}\left( X,\theta
-2\right) }\right) ^{2}\left( h_{1}^{B}\left( X\right) \left( \left\langle 
\bar{h}\left( X\right) \right\rangle +\left\langle \hat{h}_{1}^{B}\left(
X\right) \right\rangle \right) \right) }
\end{eqnarray*}

\subsection*{A 4.4 \ Quadratic order dynamics for investors}

\begin{eqnarray*}
\left( 
\begin{array}{c}
\delta \hat{S}\left( X^{\prime },X,\theta \right) \\ 
\delta S\left( X,,\theta -1\right)%
\end{array}%
\right) &=&\left[ 
\begin{array}{cc}
\alpha & \frac{\hat{h}\left( X^{\prime },X\right) }{2}c \\ 
\frac{h}{\frac{\hat{h}\left( X^{\prime },X\right) }{2}} & \beta%
\end{array}%
\right] \left( 
\begin{array}{c}
\delta \hat{S}\left( X^{\prime },X,\theta -1\right) \\ 
\delta S\left( X,,\theta -2\right)%
\end{array}%
\right) \\
&&+\left( 
\begin{array}{c}
\delta \hat{S}\left( X^{\prime },X,\theta \right) \\ 
\delta S\left( X,,\theta -1\right)%
\end{array}%
\right) ^{t}\left[ 
\begin{array}{c}
\hat{M} \\ 
M%
\end{array}%
\right] \left( 
\begin{array}{c}
\delta \hat{S}\left( X^{\prime },X,\theta \right) \\ 
\delta S\left( X,\theta -1\right)%
\end{array}%
\right)
\end{eqnarray*}%
\begin{equation*}
\hat{M}=\left[ 
\begin{array}{cc}
\frac{\delta }{\delta \hat{f}\left( X,\theta -1\right) }\alpha & \frac{\hat{h%
}\left( X^{\prime },X\right) }{2}\frac{\delta }{\delta \hat{f}\left(
X,\theta -1\right) }c \\ 
\frac{\delta }{\delta S\left( X,,\theta -1\right) }\alpha & \frac{\hat{h}%
\left( X^{\prime },X\right) }{2}\frac{\delta }{\delta S\left( X,,\theta
-1\right) }c%
\end{array}%
\right] ,M=\left[ 
\begin{array}{cc}
\frac{1}{\frac{\hat{h}\left( X^{\prime },X\right) }{2}}\frac{\delta }{\delta 
\hat{f}\left( X,\theta -1\right) }h & \frac{\delta }{\delta \hat{f}\left(
X,\theta -1\right) }\beta \\ 
\frac{1}{\frac{\hat{h}\left( X^{\prime },X\right) }{2}}\frac{\delta }{\delta
S\left( X,,\theta -1\right) }h & \frac{\delta }{\delta S\left( X,,\theta
-1\right) }\beta%
\end{array}%
\right]
\end{equation*}%
The condition for stability becomes:%
\begin{eqnarray*}
&&1-\left\langle \lambda ^{2}\right\rangle \\
&>&\left\{ \frac{\delta \alpha }{\delta \hat{f}\left( X\right) }\left(
\delta \hat{S}\left( X\right) \right) ^{2}+\delta \hat{S}\left( X\right)
\left( \frac{\frac{\hat{h}\left( X^{\prime },X\right) }{2}\delta c}{\delta 
\hat{f}\left( X\right) }+\frac{\delta \alpha }{\delta S\left( X\right) }%
\right) \delta S\left( X\right) +\frac{\frac{\hat{h}\left( X^{\prime
},X\right) }{2}\delta c}{\delta S\left( X,\theta -1\right) }\left( \delta
S\left( X\right) \right) ^{2}\right\} ^{2} \\
&&+\left\{ \frac{\delta h}{\frac{\hat{h}\left( X^{\prime },X\right) }{2}%
\delta \hat{f}\left( X\right) }\left( \delta \hat{S}\left( X\right) \right)
^{2}\right. \\
&&\left. +\delta \hat{S}\left( X\right) \left( \frac{\delta h}{\frac{\hat{h}%
\left( X^{\prime },X\right) }{2}\delta S\left( X,,\theta -1\right) }+\frac{%
\delta \beta }{\delta \hat{f}\left( X\right) }\right) \delta S\left(
X\right) +\frac{\delta \beta }{\delta S\left( X,,\theta -1\right) }\left(
\delta S\left( X\right) \right) ^{2}\right\} ^{2}
\end{eqnarray*}

where the coefficients are given in Gosselin and Lotz (2025a):%
\begin{eqnarray}
\alpha &=&\left( \frac{\hat{S}\left( X\right) }{\left( 1-\left( \hat{S}%
\left( X\right) \right) \right) }\left( \frac{1}{2\left( 1+\frac{\Delta \hat{%
f}\left( X^{\prime }\right) +\Delta \hat{r}\left( X^{\prime }\right) }{2}%
\right) }-\frac{\frac{\partial \hat{K}_{X^{\prime }}\left\vert \hat{\Psi}%
\left( X^{\prime }\right) \right\vert ^{2}}{\partial \hat{f}\left( X,\theta
-1\right) }}{\hat{K}_{X^{\prime }}\left\vert \hat{\Psi}\left( X^{\prime
}\right) \right\vert ^{2}}\right) \right.  \label{LP1} \\
&&\left. -\frac{\hat{S}_{1}\left( X\right) }{\left( 1-\left( \hat{S}%
_{1}\left( X^{\prime }\right) \right) \right) }\left( \frac{1}{1+\Delta \hat{%
f}\left( X\right) }-\frac{\frac{\partial \hat{K}_{X^{\prime }}\left\vert 
\hat{\Psi}\left( X^{\prime }\right) \right\vert ^{2}}{\partial \hat{f}\left(
X,\theta -1\right) }}{\hat{K}_{X^{\prime }}\left\vert \hat{\Psi}\left(
X^{\prime }\right) \right\vert ^{2}}\right) \right) \left( \hat{f}\left(
X\right) -\bar{r}\right)  \notag \\
&&+\frac{1-\left( \hat{S}_{1}\left( X\right) \right) }{1-\left( \hat{S}%
\left( X\right) \right) }S_{1}\left( X,X,\theta -1\right) \frac{\partial
f\left( X\right) }{\partial S\left( X,\theta -1\right) }\frac{\partial _{%
\hat{f}\left( X\right) }\left( \frac{\hat{K}_{X}\left\vert \hat{\Psi}\left(
X\right) \right\vert ^{2}}{K_{X}\left\vert \Psi \left( X\right) \right\vert
^{2}}\right) S\left( X\right) }{\frac{\hat{K}_{X}\left\vert \hat{\Psi}\left(
X\right) \right\vert ^{2}}{K_{X}\left\vert \Psi \left( X\right) \right\vert
^{2}}}  \notag
\end{eqnarray}%
\begin{eqnarray*}
\beta &=&\left( \frac{\frac{\hat{h}\left( X\right) }{2}S\left( X\right) }{%
1+\left( \hat{h}\left( X\right) \left( \frac{f\left( X\right) +\bar{r}\left(
X\right) }{2}-\frac{\left\langle \hat{f}\left( X^{\prime }\right)
\right\rangle _{\hat{h}_{1}}+\left\langle \hat{r}\left( X^{\prime }\right)
\right\rangle _{\hat{h}_{2}}}{2}\right) \right) }+\frac{\partial _{f\left(
X\right) }\left( \frac{\hat{K}_{X}\left\vert \hat{\Psi}\left( X\right)
\right\vert ^{2}}{K_{X}\left\vert \Psi \left( X\right) \right\vert ^{2}}%
\right) S\left( X\right) }{\frac{\hat{K}_{X}\left\vert \hat{\Psi}\left(
X\right) \right\vert ^{2}}{K_{X}\left\vert \Psi \left( X\right) \right\vert
^{2}}}\right) \frac{\partial f\left( X\right) }{\partial S\left( X,\theta
-2\right) } \\
&\simeq &\left( \frac{\hat{h}\left( X\right) }{2}S\left( X\right) +\frac{%
\partial _{f\left( X\right) }\left( \frac{\hat{K}_{X}\left\vert \hat{\Psi}%
\left( X\right) \right\vert ^{2}}{K_{X}\left\vert \Psi \left( X\right)
\right\vert ^{2}}\right) S\left( X\right) }{\frac{\hat{K}_{X}\left\vert \hat{%
\Psi}\left( X\right) \right\vert ^{2}}{K_{X}\left\vert \Psi \left( X\right)
\right\vert ^{2}}}\right) \frac{\partial f\left( X\right) }{\partial S\left(
X,\theta -2\right) }
\end{eqnarray*}%
\begin{eqnarray*}
&&c=\left\{ -\frac{\left( 1-\left\langle \hat{S}\left( X^{\prime }\right)
\right\rangle \right) \left( \left\langle \hat{f}\left( X^{\prime }\right)
\right\rangle -\bar{r}\right) }{1-\left\langle \hat{S}_{1}\left( X^{\prime
}\right) \right\rangle }\frac{\left\langle \hat{h}\left( X^{\prime
},X\right) \right\rangle \left\langle h\left( X\right) \right\rangle }{4}%
\right. \\
&&+\frac{\frac{h\left( X\right) }{4}\left( 1+\hat{h}\left( X\right) \right)
S_{1}\left( X\right) \left( f\left( X\right) -\bar{r}\right) }{1+\left( \hat{%
h}\left( X\right) \left( f\left( X\right) -\frac{\left\langle \hat{f}\left(
X^{\prime }\right) \right\rangle _{\hat{h}_{1}}+\left\langle \hat{r}\left(
X^{\prime }\right) \right\rangle _{\hat{h}_{2}}}{2}\right) +\frac{h\left(
X\right) }{2}\left( f\left( X\right) -\bar{r}\left( X\right) \right) \right) 
} \\
&&\left. +S_{1}\left( X\right) \left( \frac{\frac{\hat{h}\left( X\right) }{2}%
S\left( X\right) }{1+\left( \hat{h}\left( X\right) \left( \frac{f\left(
X\right) +\bar{r}\left( X\right) }{2}-\frac{\left\langle \hat{f}\left(
X^{\prime }\right) \right\rangle _{\hat{h}_{1}}+\left\langle \hat{r}\left(
X^{\prime }\right) \right\rangle _{\hat{h}_{2}}}{2}\right) \right) }+\frac{%
\partial _{f\left( X\right) }\left( \frac{\hat{K}_{X}\left\vert \hat{\Psi}%
\left( X\right) \right\vert ^{2}}{K_{X}\left\vert \Psi \left( X\right)
\right\vert ^{2}}\right) S\left( X\right) }{\frac{\hat{K}_{X}\left\vert \hat{%
\Psi}\left( X\right) \right\vert ^{2}}{K_{X}\left\vert \Psi \left( X\right)
\right\vert ^{2}}}\right) \frac{\partial f\left( X\right) }{\partial S\left(
X,\theta -2\right) }\right\} \\
&&\times \frac{\partial f\left( X\right) }{\partial S\left( X,\theta
-1\right) }
\end{eqnarray*}%
\begin{equation*}
h=\frac{\partial _{\hat{f}\left( X\right) }\left( \frac{\hat{K}%
_{X}\left\vert \hat{\Psi}\left( X\right) \right\vert ^{2}}{K_{X}\left\vert
\Psi \left( X\right) \right\vert ^{2}}\right) S\left( X\right) }{\frac{\hat{K%
}_{X}\left\vert \hat{\Psi}\left( X\right) \right\vert ^{2}}{K_{X}\left\vert
\Psi \left( X\right) \right\vert ^{2}}}
\end{equation*}

\subsubsection*{A4.4.1 Estimation of coefficients derivatives}

The variations of these coefficients are obtained by using the following
formula derived in (2025a).

\begin{eqnarray*}
&&\delta \hat{S}_{1}\left( X^{\prime },X,\theta -1\right) \\
&\rightarrow &\frac{\hat{h}\left( X^{\prime },X\right) }{2}\left( 1+\left(
\delta \hat{f}\left( X^{\prime }\right) -\frac{\left\langle h\left( X\right)
\right\rangle }{2}\delta f\left( X\right) \right) \right)
\end{eqnarray*}%
\begin{equation*}
\delta \hat{S}_{1}\left( X,\theta -1\right) \simeq \left( 1-\frac{\frac{%
\partial \hat{K}_{X^{\prime }}\left\vert \hat{\Psi}\left( X^{\prime }\right)
\right\vert ^{2}}{\partial \hat{f}\left( X,\theta -1\right) }}{\hat{K}%
_{X^{\prime }}\left\vert \hat{\Psi}\left( X^{\prime }\right) \right\vert ^{2}%
}\right) \hat{S}_{1}\left( X\right) \delta \hat{f}\left( X,\theta -1\right)
\end{equation*}%
\begin{equation*}
\delta \hat{S}\left( X,\theta -1\right) \simeq \left( \frac{1}{2}-\frac{%
\frac{\partial \hat{K}_{X^{\prime }}\left\vert \hat{\Psi}\left( X^{\prime
}\right) \right\vert ^{2}}{\partial \hat{f}\left( X,\theta -1\right) }}{\hat{%
K}_{X^{\prime }}\left\vert \hat{\Psi}\left( X^{\prime }\right) \right\vert
^{2}}\right) \hat{S}\left( X\right) \delta \hat{f}\left( X,\theta -1\right)
\end{equation*}%
\begin{eqnarray*}
&&\delta S_{1}\left( X,X,\theta -1\right) \\
&\rightarrow &\frac{h\left( X\right) }{2}\left( \hat{h}\left( X\right) +%
\frac{h\left( X\right) }{2}\right) S_{1}\left( X,X\right) \delta f\left(
X,\theta -1\right) \\
&&+\frac{\partial _{\hat{f}\left( X\right) }\left( \frac{\hat{K}%
_{X}\left\vert \hat{\Psi}\left( X\right) \right\vert ^{2}}{K_{X}\left\vert
\Psi \left( X\right) \right\vert ^{2}}\right) S_{1}\left( X\right) }{\frac{%
\hat{K}_{X}\left\vert \hat{\Psi}\left( X\right) \right\vert ^{2}}{%
K_{X}\left\vert \Psi \left( X\right) \right\vert ^{2}}}\delta \hat{f}\left(
X,\theta -1\right) +\frac{\partial _{f\left( X\right) }\left( \frac{\hat{K}%
_{X}\left\vert \hat{\Psi}\left( X\right) \right\vert ^{2}}{K_{X}\left\vert
\Psi \left( X\right) \right\vert ^{2}}\right) S_{1}\left( X\right) }{\frac{%
\hat{K}_{X}\left\vert \hat{\Psi}\left( X\right) \right\vert ^{2}}{%
K_{X}\left\vert \Psi \left( X\right) \right\vert ^{2}}}\frac{\partial
f\left( X\right) }{\partial S\left( X,\theta -2\right) }\delta S\left(
X,\theta -2\right) \\
&\rightarrow &\frac{\partial _{\hat{f}\left( X\right) }\left( \frac{\hat{K}%
_{X}\left\vert \hat{\Psi}\left( X\right) \right\vert ^{2}}{K_{X}\left\vert
\Psi \left( X\right) \right\vert ^{2}}\right) S_{1}\left( X\right) }{\frac{%
\hat{K}_{X}\left\vert \hat{\Psi}\left( X\right) \right\vert ^{2}}{%
K_{X}\left\vert \Psi \left( X\right) \right\vert ^{2}}}\delta \hat{f}\left(
X,\theta -1\right) \\
&&+\left( \frac{h\left( X\right) }{2}\left( \hat{h}\left( X\right) +\frac{%
h\left( X\right) }{2}\right) S_{1}\left( X,X\right) +\frac{\partial
_{f\left( X\right) }\left( \frac{\hat{K}_{X}\left\vert \hat{\Psi}\left(
X\right) \right\vert ^{2}}{K_{X}\left\vert \Psi \left( X\right) \right\vert
^{2}}\right) S_{1}\left( X\right) }{\frac{\hat{K}_{X}\left\vert \hat{\Psi}%
\left( X\right) \right\vert ^{2}}{K_{X}\left\vert \Psi \left( X\right)
\right\vert ^{2}}}\right) \frac{\partial f\left( X\right) }{\partial S\left(
X,\theta -2\right) }\delta S\left( X,\theta -2\right)
\end{eqnarray*}%
\begin{eqnarray*}
\delta S\left( X,\theta -1\right) &\simeq &\frac{\hat{h}\left( X\right) }{2}%
\delta f\left( X,\theta -1\right) S\left( X\right) +\frac{\delta \frac{\hat{K%
}_{X}\left\vert \hat{\Psi}\left( X\right) \right\vert ^{2}}{K_{X}\left\vert
\Psi \left( X\right) \right\vert ^{2}}}{\frac{\hat{K}_{X}\left\vert \hat{\Psi%
}\left( X\right) \right\vert ^{2}}{K_{X}\left\vert \Psi \left( X\right)
\right\vert ^{2}}}S\left( X\right) \\
&\simeq &\frac{\hat{h}\left( X\right) }{2}\delta f\left( X\right) S\left(
X\right) -2S\left( X\right) \frac{\delta \hat{f}\left( X,\theta -1\right) }{%
\hat{f}\left( X,\theta -1\right) }-\frac{\delta f\left( X\right) }{f\left(
X\right) +C_{0}}S\left( X\right) \\
&=&\left( \frac{\hat{h}\left( X\right) }{2}-\frac{1}{f\left( X\right) +C_{0}}%
\right) S\left( X\right) \frac{\partial f\left( X\right) }{\partial S\left(
X,\theta -2\right) }\delta S\left( X,\theta -2\right) -2S\left( X\right) 
\frac{\delta \hat{f}\left( X,\theta -1\right) }{\hat{f}\left( X,\theta
-1\right) }
\end{eqnarray*}

We use in first approximation that:%
\begin{eqnarray*}
\delta \frac{1-\left( \hat{S}_{1}\left( X\right) \right) }{1-\left( \hat{S}%
\left( X\right) \right) } &\simeq &\delta \frac{1-\frac{1}{2}\hat{S}\left(
X\right) }{1-\hat{S}\left( X\right) } \\
&=&\frac{1}{2\left( 1-\hat{S}\left( X\right) \right) ^{2}}\delta \hat{S}%
\left( X\right) \\
&\simeq &\frac{1}{2\left( 1-\hat{S}\left( X\right) \right) ^{2}}\left( \frac{%
1}{2}-\frac{\frac{\partial \hat{K}_{X^{\prime }}\left\vert \hat{\Psi}\left(
X^{\prime }\right) \right\vert ^{2}}{\partial \hat{f}\left( X,\theta
-1\right) }}{\hat{K}_{X^{\prime }}\left\vert \hat{\Psi}\left( X^{\prime
}\right) \right\vert ^{2}}\right) \hat{S}\left( X\right) \delta \hat{f}%
\left( X,\theta -1\right)
\end{eqnarray*}%
\begin{equation*}
\frac{\partial _{\hat{f}\left( X\right) }\left( \frac{\hat{K}_{X}\left\vert 
\hat{\Psi}\left( X\right) \right\vert ^{2}}{K_{X}\left\vert \Psi \left(
X\right) \right\vert ^{2}}\right) S\left( X\right) }{\frac{\hat{K}%
_{X}\left\vert \hat{\Psi}\left( X\right) \right\vert ^{2}}{K_{X}\left\vert
\Psi \left( X\right) \right\vert ^{2}}}\simeq -\frac{2S\left( X\right) }{%
\hat{f}\left( X,\theta -1\right) }
\end{equation*}%
\begin{equation*}
\delta \frac{\partial _{\hat{f}\left( X\right) }\left( \frac{\hat{K}%
_{X}\left\vert \hat{\Psi}\left( X\right) \right\vert ^{2}}{K_{X}\left\vert
\Psi \left( X\right) \right\vert ^{2}}\right) S\left( X\right) }{\frac{\hat{K%
}_{X}\left\vert \hat{\Psi}\left( X\right) \right\vert ^{2}}{K_{X}\left\vert
\Psi \left( X\right) \right\vert ^{2}}}\simeq -2\frac{\frac{\hat{h}\left(
X\right) }{2}\delta f\left( X,\theta -1\right) S\left( X\right) -2S\left(
X\right) \frac{\delta \hat{f}\left( X,\theta -1\right) }{\hat{f}\left(
X,\theta -1\right) }-\frac{\frac{\partial f\left( X\right) }{\partial
S\left( X,\theta -2\right) }}{f\left( X\right) +C_{0}}\delta S\left(
X,\theta -2\right) }{\hat{f}\left( X,\theta -1\right) }
\end{equation*}%
\begin{eqnarray}
\delta \alpha &\simeq &\left( \frac{\hat{S}\left( X\right) }{\left( 1-\left( 
\hat{S}\left( X\right) \right) \right) }\left( \frac{1}{2\left( 1+\frac{%
\Delta \hat{f}\left( X^{\prime }\right) +\Delta \hat{r}\left( X^{\prime
}\right) }{2}\right) }-\frac{\frac{\partial \hat{K}_{X^{\prime }}\left\vert 
\hat{\Psi}\left( X^{\prime }\right) \right\vert ^{2}}{\partial \hat{f}\left(
X,\theta -1\right) }}{\hat{K}_{X^{\prime }}\left\vert \hat{\Psi}\left(
X^{\prime }\right) \right\vert ^{2}}\right) \right. \\
&&\left. -\frac{\hat{S}_{1}\left( X\right) }{\left( 1-\left( \hat{S}%
_{1}\left( X^{\prime }\right) \right) \right) }\left( \frac{1}{1+\Delta \hat{%
f}\left( X\right) }-\frac{\frac{\partial \hat{K}_{X^{\prime }}\left\vert 
\hat{\Psi}\left( X^{\prime }\right) \right\vert ^{2}}{\partial \hat{f}\left(
X,\theta -1\right) }}{\hat{K}_{X^{\prime }}\left\vert \hat{\Psi}\left(
X^{\prime }\right) \right\vert ^{2}}\right) \right) \delta \hat{f}\left(
X\right)  \notag \\
&&+\delta \left[ \frac{1-\left( \hat{S}_{1}\left( X\right) \right) }{%
1-\left( \hat{S}\left( X\right) \right) }S_{1}\left( X,X,\theta -1\right) 
\frac{\partial f\left( X\right) }{\partial S\left( X,\theta -1\right) }\frac{%
\partial _{\hat{f}\left( X\right) }\left( \frac{\hat{K}_{X}\left\vert \hat{%
\Psi}\left( X\right) \right\vert ^{2}}{K_{X}\left\vert \Psi \left( X\right)
\right\vert ^{2}}\right) S\left( X\right) }{\frac{\hat{K}_{X}\left\vert \hat{%
\Psi}\left( X\right) \right\vert ^{2}}{K_{X}\left\vert \Psi \left( X\right)
\right\vert ^{2}}}\right]  \notag
\end{eqnarray}%
In all expressions, due to presence of banks, $\hat{f}\left( X\right) $
stands for:%
\begin{equation*}
\hat{f}\left( X\right) +\frac{\left\langle \hat{S}\left( X^{\prime
},X\right) \right\rangle _{X^{\prime }}\left\langle \hat{f}\right\rangle }{%
1-\left\langle \hat{S}\right\rangle }+\frac{\left( \left\langle \hat{S}%
_{1}^{B}\left( X,X\right) \right\rangle _{X}+\left\langle \hat{S}%
_{2}^{B}\left( X,X\right) \right\rangle _{X}+\left\langle \hat{S}\left(
X^{\prime },X\right) \right\rangle _{X^{\prime }}\frac{\left\langle \hat{S}%
_{1}^{B}\right\rangle +\left\langle \hat{S}_{2}^{B}\right\rangle }{%
1-\left\langle \hat{S}\right\rangle }\right) \frac{\left\langle \bar{K}%
\right\rangle \left\Vert \bar{\Psi}\right\Vert ^{2}}{\left\langle \hat{K}%
\right\rangle \left\Vert \hat{\Psi}\right\Vert ^{2}}\left\langle \bar{f}%
\right\rangle }{1-\left\langle \bar{S}\right\rangle }
\end{equation*}%
We first derive:%
\begin{eqnarray*}
&&\delta \left[ \frac{1-\left( \hat{S}_{1}\left( X\right) \right) }{1-\left( 
\hat{S}\left( X\right) \right) }S_{1}\left( X,X,\theta -1\right) \frac{%
\partial f\left( X\right) }{\partial S\left( X,\theta -1\right) }\frac{%
\partial _{\hat{f}\left( X\right) }\left( \frac{\hat{K}_{X}\left\vert \hat{%
\Psi}\left( X\right) \right\vert ^{2}}{K_{X}\left\vert \Psi \left( X\right)
\right\vert ^{2}}\right) S\left( X\right) }{\frac{\hat{K}_{X}\left\vert \hat{%
\Psi}\left( X\right) \right\vert ^{2}}{K_{X}\left\vert \Psi \left( X\right)
\right\vert ^{2}}}\right] \\
&\simeq &\left[ \frac{\hat{S}\left( X\right) }{2\left( 1-\hat{S}\left(
X\right) \right) ^{2}}\left( \frac{1}{2}-\frac{\frac{\partial \hat{K}%
_{X^{\prime }}\left\vert \hat{\Psi}\left( X^{\prime }\right) \right\vert ^{2}%
}{\partial \hat{f}\left( X,\theta -1\right) }}{\hat{K}_{X^{\prime
}}\left\vert \hat{\Psi}\left( X^{\prime }\right) \right\vert ^{2}}\right)
S_{1}\left( X,X,\theta -1\right) \frac{\partial f\left( X\right) }{\partial
S\left( X,\theta -1\right) }\frac{\partial _{\hat{f}\left( X\right) }\left( 
\frac{\hat{K}_{X}\left\vert \hat{\Psi}\left( X\right) \right\vert ^{2}}{%
K_{X}\left\vert \Psi \left( X\right) \right\vert ^{2}}\right) S\left(
X\right) }{\frac{\hat{K}_{X}\left\vert \hat{\Psi}\left( X\right) \right\vert
^{2}}{K_{X}\left\vert \Psi \left( X\right) \right\vert ^{2}}}\right] \delta 
\hat{f}\left( X,\theta -1\right) \\
&&+\left[ \frac{1-\left( \hat{S}_{1}\left( X\right) \right) }{1-\left( \hat{S%
}\left( X\right) \right) }S_{1}\left( X,X,\theta -1\right) \frac{\partial
f\left( X\right) }{\partial S\left( X,\theta -1\right) }\frac{\partial _{%
\hat{f}\left( X\right) }\left( \frac{\hat{K}_{X}\left\vert \hat{\Psi}\left(
X\right) \right\vert ^{2}}{K_{X}\left\vert \Psi \left( X\right) \right\vert
^{2}}\right) S\left( X\right) }{\frac{\hat{K}_{X}\left\vert \hat{\Psi}\left(
X\right) \right\vert ^{2}}{K_{X}\left\vert \Psi \left( X\right) \right\vert
^{2}}}\right] \\
&&\times \left( \frac{\partial _{\hat{f}\left( X\right) }\left( \frac{\hat{K}%
_{X}\left\vert \hat{\Psi}\left( X\right) \right\vert ^{2}}{K_{X}\left\vert
\Psi \left( X\right) \right\vert ^{2}}\right) }{\frac{\hat{K}_{X}\left\vert 
\hat{\Psi}\left( X\right) \right\vert ^{2}}{K_{X}\left\vert \Psi \left(
X\right) \right\vert ^{2}}}\delta \hat{f}\left( X,\theta -1\right) +\left( 
\frac{h\left( X\right) }{2}\left( \hat{h}\left( X\right) +\frac{h\left(
X\right) }{2}\right) +\frac{\partial _{f\left( X\right) }\left( \frac{\hat{K}%
_{X}\left\vert \hat{\Psi}\left( X\right) \right\vert ^{2}}{K_{X}\left\vert
\Psi \left( X\right) \right\vert ^{2}}\right) S_{1}\left( X\right) }{\frac{%
\hat{K}_{X}\left\vert \hat{\Psi}\left( X\right) \right\vert ^{2}}{%
K_{X}\left\vert \Psi \left( X\right) \right\vert ^{2}}}\right) \right) \\
&&\times \frac{\partial f\left( X\right) }{\partial S\left( X,\theta
-2\right) }\delta S\left( X,\theta -2\right) \\
&&-2\frac{\left( \frac{\hat{h}\left( X\right) }{2}S\left( X\right) -\frac{1}{%
f\left( X\right) +C_{0}}\right) \frac{\partial f\left( X\right) }{\partial
S\left( X,\theta -2\right) }\delta S\left( X,\theta -2\right) -2S\left(
X\right) \frac{\delta \hat{f}\left( X,\theta -1\right) }{\hat{f}\left(
X,\theta -1\right) }}{\hat{f}\left( X,\theta -1\right) } \\
&&\times \left[ \frac{1-\left( \hat{S}_{1}\left( X\right) \right) }{1-\left( 
\hat{S}\left( X\right) \right) }S_{1}\left( X,X,\theta -1\right) \frac{%
\partial f\left( X\right) }{\partial S\left( X,\theta -1\right) }\right] \\
&&+\left[ \frac{1-\left( \hat{S}_{1}\left( X\right) \right) }{1-\left( \hat{S%
}\left( X\right) \right) }S_{1}\left( X,X,\theta -1\right) \frac{\partial _{%
\hat{f}\left( X\right) }\left( \frac{\hat{K}_{X}\left\vert \hat{\Psi}\left(
X\right) \right\vert ^{2}}{K_{X}\left\vert \Psi \left( X\right) \right\vert
^{2}}\right) S\left( X\right) }{\frac{\hat{K}_{X}\left\vert \hat{\Psi}\left(
X\right) \right\vert ^{2}}{K_{X}\left\vert \Psi \left( X\right) \right\vert
^{2}}}\right] \frac{\partial ^{2}f\left( X\right) }{\partial \left( S\left(
X,\theta -2\right) \right) ^{2}}\delta S\left( X,\theta -2\right)
\end{eqnarray*}%
and gather the terms to obtain $\delta \alpha $: 
\begin{equation}
\delta \alpha \simeq T\delta \hat{f}\left( X\right) +\left( V\frac{\partial
f\left( X\right) }{\partial S\left( X,\theta -2\right) }+W\frac{\partial
^{2}f\left( X\right) }{\partial \left( S\left( X,\theta -2\right) \right)
^{2}}\right) \delta S\left( X,\theta -2\right)
\end{equation}%
\begin{eqnarray*}
T &=&\left( \frac{\hat{S}\left( X\right) }{\left( 1-\left( \hat{S}\left(
X\right) \right) \right) }-\frac{\hat{S}_{1}\left( X\right) }{\left(
1-\left( \hat{S}_{1}\left( X^{\prime }\right) \right) \right) }\right) \frac{%
2}{\hat{f}\left( X,\theta -1\right) } \\
&&+\left[ \frac{\hat{S}\left( X\right) }{\left( 1-\hat{S}\left( X\right)
\right) ^{2}\hat{f}\left( X,\theta -1\right) }S_{1}\left( X,X,\theta
-1\right) \frac{\partial f\left( X\right) }{\partial S\left( X,\theta
-1\right) }\frac{\partial _{\hat{f}\left( X\right) }\left( \frac{\hat{K}%
_{X}\left\vert \hat{\Psi}\left( X\right) \right\vert ^{2}}{K_{X}\left\vert
\Psi \left( X\right) \right\vert ^{2}}\right) S\left( X\right) }{\frac{\hat{K%
}_{X}\left\vert \hat{\Psi}\left( X\right) \right\vert ^{2}}{K_{X}\left\vert
\Psi \left( X\right) \right\vert ^{2}}}\right] \\
&&+\left[ \frac{1-\left( \hat{S}_{1}\left( X\right) \right) }{1-\left( \hat{S%
}\left( X\right) \right) }S_{1}\left( X,X,\theta -1\right) \frac{\partial
f\left( X\right) }{\partial S\left( X,\theta -1\right) }\frac{\partial _{%
\hat{f}\left( X\right) }\left( \frac{\hat{K}_{X}\left\vert \hat{\Psi}\left(
X\right) \right\vert ^{2}}{K_{X}\left\vert \Psi \left( X\right) \right\vert
^{2}}\right) S\left( X\right) }{\frac{\hat{K}_{X}\left\vert \hat{\Psi}\left(
X\right) \right\vert ^{2}}{K_{X}\left\vert \Psi \left( X\right) \right\vert
^{2}}}\right] \\
&&\times \left( \frac{\partial _{\hat{f}\left( X\right) }\left( \frac{\hat{K}%
_{X}\left\vert \hat{\Psi}\left( X\right) \right\vert ^{2}}{K_{X}\left\vert
\Psi \left( X\right) \right\vert ^{2}}\right) }{\frac{\hat{K}_{X}\left\vert 
\hat{\Psi}\left( X\right) \right\vert ^{2}}{K_{X}\left\vert \Psi \left(
X\right) \right\vert ^{2}}}+4\frac{1}{\left( \hat{f}\left( X,\theta
-1\right) \right) ^{2}\frac{\partial _{\hat{f}\left( X\right) }\left( \frac{%
\hat{K}_{X}\left\vert \hat{\Psi}\left( X\right) \right\vert ^{2}}{%
K_{X}\left\vert \Psi \left( X\right) \right\vert ^{2}}\right) }{\frac{\hat{K}%
_{X}\left\vert \hat{\Psi}\left( X\right) \right\vert ^{2}}{K_{X}\left\vert
\Psi \left( X\right) \right\vert ^{2}}}}\right)
\end{eqnarray*}%
\begin{eqnarray}
&&V=\left[ \frac{1-\left( \hat{S}_{1}\left( X\right) \right) }{1-\left( \hat{%
S}\left( X\right) \right) }S_{1}\left( X,X,\theta -1\right) \frac{\partial
f\left( X\right) }{\partial S\left( X,\theta -1\right) }\frac{\partial _{%
\hat{f}\left( X\right) }\left( \frac{\hat{K}_{X}\left\vert \hat{\Psi}\left(
X\right) \right\vert ^{2}}{K_{X}\left\vert \Psi \left( X\right) \right\vert
^{2}}\right) S\left( X\right) }{\frac{\hat{K}_{X}\left\vert \hat{\Psi}\left(
X\right) \right\vert ^{2}}{K_{X}\left\vert \Psi \left( X\right) \right\vert
^{2}}}\right]  \notag \\
&&\times \left( \frac{h\left( X\right) }{2}\left( \hat{h}\left( X\right) +%
\frac{h\left( X\right) }{2}\right) +\frac{\partial _{f\left( X\right)
}\left( \frac{\hat{K}_{X}\left\vert \hat{\Psi}\left( X\right) \right\vert
^{2}}{K_{X}\left\vert \Psi \left( X\right) \right\vert ^{2}}\right)
S_{1}\left( X\right) }{\frac{\hat{K}_{X}\left\vert \hat{\Psi}\left( X\right)
\right\vert ^{2}}{K_{X}\left\vert \Psi \left( X\right) \right\vert ^{2}}}-2%
\frac{\left( \frac{\hat{h}\left( X\right) }{2}S\left( X\right) -\frac{1}{%
f\left( X\right) +C_{0}}\right) }{\hat{f}\left( X,\theta -1\right) \frac{%
\partial _{\hat{f}\left( X\right) }\left( \frac{\hat{K}_{X}\left\vert \hat{%
\Psi}\left( X\right) \right\vert ^{2}}{K_{X}\left\vert \Psi \left( X\right)
\right\vert ^{2}}\right) S\left( X\right) }{\frac{\hat{K}_{X}\left\vert \hat{%
\Psi}\left( X\right) \right\vert ^{2}}{K_{X}\left\vert \Psi \left( X\right)
\right\vert ^{2}}}}\right)  \notag
\end{eqnarray}%
\begin{equation*}
W=\left[ \frac{1-\left( \hat{S}_{1}\left( X\right) \right) }{1-\left( \hat{S}%
\left( X\right) \right) }S_{1}\left( X,X,\theta -1\right) \frac{\partial _{%
\hat{f}\left( X\right) }\left( \frac{\hat{K}_{X}\left\vert \hat{\Psi}\left(
X\right) \right\vert ^{2}}{K_{X}\left\vert \Psi \left( X\right) \right\vert
^{2}}\right) S\left( X\right) }{\frac{\hat{K}_{X}\left\vert \hat{\Psi}\left(
X\right) \right\vert ^{2}}{K_{X}\left\vert \Psi \left( X\right) \right\vert
^{2}}}\right]
\end{equation*}%
\begin{eqnarray*}
\delta \beta &\simeq &\left( 4\frac{S\left( X\right) }{\left( \hat{f}\left(
X,\theta -1\right) \right) ^{2}}-\frac{\hat{h}\left( X\right) }{2}\frac{2}{%
\hat{f}\left( X,\theta -1\right) }S\left( X\right) \frac{\partial f\left(
X\right) }{\partial S\left( X,\theta -2\right) }\right) \delta \hat{f}\left(
X,\theta -1\right) \\
&&+\left( \frac{\hat{h}\left( X\right) }{2}S\left( X\right) -\frac{1}{%
f\left( X\right) +C_{0}}\right) \left( \frac{\hat{h}\left( X\right) }{2}-%
\frac{2}{\hat{f}\left( X,\theta -1\right) }\right) \left( \frac{\partial
f\left( X\right) }{\partial S\left( X,\theta -2\right) }\right) ^{2}\delta
S\left( X,\theta -2\right) +\beta \frac{\frac{\partial ^{2}f\left( X\right) 
}{\partial \left( S\left( X,\theta -2\right) \right) ^{2}}}{\frac{\partial
f\left( X\right) }{\partial S\left( X,\theta -2\right) }}
\end{eqnarray*}%
\begin{eqnarray*}
\delta c &\simeq &\left\{ \left( -\frac{\left( 1-\left\langle \hat{S}\left(
X^{\prime }\right) \right\rangle \right) }{1-\left\langle \hat{S}_{1}\left(
X^{\prime }\right) \right\rangle }\frac{\left\langle \hat{h}\left( X^{\prime
},X\right) \right\rangle \left\langle h\left( X\right) \right\rangle }{4}-%
\frac{2S_{1}\left( X\right) \beta }{\hat{f}\left( X,\theta -1\right) }%
\right) \delta \left\langle \hat{f}\left( X^{\prime }\right) \right\rangle
\right. \\
&&+\left( \frac{h\left( X\right) }{4}\left( 1+\hat{h}\left( X\right) \right)
S_{1}\left( X\right) +S_{1}\left( X\right) \beta \left( \frac{\hat{h}\left(
X\right) }{2}-\frac{1}{f\left( X\right) +C_{0}}\right) \right) \frac{%
\partial f\left( X\right) }{\partial S\left( X,\theta -2\right) }\delta
S\left( X,\theta -2\right) \\
&&\left. +S_{1}\left( X\right) \delta \beta \right\} \frac{\partial f\left(
X\right) }{\partial S\left( X,\theta -2\right) }+c\frac{\frac{\partial
^{2}f\left( X\right) }{\partial \left( S\left( X,\theta -2\right) \right)
^{2}}}{\frac{\partial f\left( X\right) }{\partial S\left( X,\theta -2\right) 
}} \\
&\simeq &S_{1}\left( X\right) \left( 4\frac{S\left( X\right) }{\left( \hat{f}%
\left( X,\theta -1\right) \right) ^{2}}-\frac{\hat{h}\left( X\right) }{2}%
\frac{2}{\hat{f}\left( X,\theta -1\right) }S\left( X\right) \frac{\partial
f\left( X\right) }{\partial S\left( X,\theta -2\right) }\right. \\
&&\left. -\left( \frac{\left( 1-\left\langle \hat{S}\left( X^{\prime
}\right) \right\rangle \right) }{1-\left\langle \hat{S}_{1}\left( X^{\prime
}\right) \right\rangle }\frac{\left\langle \hat{h}\left( X^{\prime
},X\right) \right\rangle \left\langle h\left( X\right) \right\rangle }{4}+%
\frac{2S_{1}\left( X\right) \beta }{\hat{f}\left( X,\theta -1\right) }%
\right) \right) \frac{\partial f\left( X\right) }{\partial S\left( X,\theta
-2\right) }\delta \left\langle \hat{f}\left( X^{\prime }\right) \right\rangle
\\
&&+\left( \left( \frac{\hat{h}\left( X\right) }{2}S\left( X\right) -\frac{1}{%
f\left( X\right) +C_{0}}\right) \left( \frac{\hat{h}\left( X\right) }{2}-%
\frac{2}{\hat{f}\left( X,\theta -1\right) }\right) \left( \frac{\partial
f\left( X\right) }{\partial S\left( X,\theta -2\right) }\right) ^{2}\right.
\\
&&\left. +\left( \frac{h\left( X\right) }{4}\left( 1+\hat{h}\left( X\right)
\right) S_{1}\left( X\right) +S_{1}\left( X\right) \beta \left( \frac{\hat{h}%
\left( X\right) }{2}-\frac{1}{f\left( X\right) +C_{0}}\right) \right) \frac{%
\partial f\left( X\right) }{\partial S\left( X,\theta -2\right) }\right) \\
&&\times \frac{\partial f\left( X\right) }{\partial S\left( X,\theta
-2\right) }\delta S\left( X,\theta -2\right) +c\frac{\frac{\partial
^{2}f\left( X\right) }{\partial \left( S\left( X,\theta -2\right) \right)
^{2}}}{\frac{\partial f\left( X\right) }{\partial S\left( X,\theta -2\right) 
}}\delta S\left( X,\theta -2\right)
\end{eqnarray*}%
\begin{equation*}
\delta h=4\frac{S\left( X\right) }{\left( \hat{f}\left( X,\theta -1\right)
\right) ^{2}}\delta \hat{f}\left( X,\theta -1\right) -2\frac{\left( \frac{%
\hat{h}\left( X\right) }{2}S\left( X\right) -\frac{1}{f\left( X\right) +C_{0}%
}\right) }{\hat{f}\left( X,\theta -1\right) }\frac{\partial f\left( X\right) 
}{\partial S\left( X,\theta -2\right) }\delta S\left( X,\theta -2\right)
\end{equation*}

\subsubsection*{A4.4.2 Dominant contributions}

The dominant contributions in $\frac{\partial f\left( X\right) }{\partial
S\left( X,\theta -1\right) }$ and $\frac{1}{\left( \hat{f}\left( X,\theta
-1\right) \right) }<<1$:%
\begin{eqnarray*}
T &\simeq &\left[ \frac{1-\left( \hat{S}_{1}\left( X\right) \right) }{%
1-\left( \hat{S}\left( X\right) \right) }S_{1}\left( X,X\right) \frac{%
\partial f\left( X\right) }{\partial S\left( X,\theta -1\right) }\frac{%
\partial _{\hat{f}\left( X\right) }\left( \frac{\hat{K}_{X}\left\vert \hat{%
\Psi}\left( X\right) \right\vert ^{2}}{K_{X}\left\vert \Psi \left( X\right)
\right\vert ^{2}}\right) S\left( X\right) }{\frac{\hat{K}_{X}\left\vert \hat{%
\Psi}\left( X\right) \right\vert ^{2}}{K_{X}\left\vert \Psi \left( X\right)
\right\vert ^{2}}}\right] \\
&&\times \left( \frac{\partial _{\hat{f}\left( X\right) }\left( \frac{\hat{K}%
_{X}\left\vert \hat{\Psi}\left( X\right) \right\vert ^{2}}{K_{X}\left\vert
\Psi \left( X\right) \right\vert ^{2}}\right) }{\frac{\hat{K}_{X}\left\vert 
\hat{\Psi}\left( X\right) \right\vert ^{2}}{K_{X}\left\vert \Psi \left(
X\right) \right\vert ^{2}}}+4\frac{1}{\left( \hat{f}\left( X,\theta
-1\right) \right) ^{2}\frac{\partial _{\hat{f}\left( X\right) }\left( \frac{%
\hat{K}_{X}\left\vert \hat{\Psi}\left( X\right) \right\vert ^{2}}{%
K_{X}\left\vert \Psi \left( X\right) \right\vert ^{2}}\right) }{\frac{\hat{K}%
_{X}\left\vert \hat{\Psi}\left( X\right) \right\vert ^{2}}{K_{X}\left\vert
\Psi \left( X\right) \right\vert ^{2}}}}\right) \\
&\simeq &\frac{1-\left( \hat{S}_{1}\left( X\right) \right) }{1-\left( \hat{S}%
\left( X\right) \right) }S_{1}\left( X,X\right) \frac{\partial f\left(
X\right) }{\partial S\left( X,\theta -1\right) }\frac{8}{\left( \hat{f}%
\left( X,\theta -1\right) \right) ^{2}}
\end{eqnarray*}%
\begin{eqnarray}
\delta \alpha &\simeq &S_{1}\left( X,X\right) \frac{1-\left( \hat{S}%
_{1}\left( X\right) \right) }{1-\left( \hat{S}\left( X\right) \right) }\frac{%
8}{\left( \hat{f}\left( X,\theta -1\right) \right) ^{2}}\frac{\partial
f\left( X\right) }{\partial S\left( X,\theta -1\right) }\delta \hat{f}\left(
X\right) \\
&&+\left[ \frac{1-\left( \hat{S}_{1}\left( X\right) \right) }{1-\left( \hat{S%
}\left( X\right) \right) }S_{1}\left( X,X\right) \frac{\partial _{\hat{f}%
\left( X\right) }\left( \frac{\hat{K}_{X}\left\vert \hat{\Psi}\left(
X\right) \right\vert ^{2}}{K_{X}\left\vert \Psi \left( X\right) \right\vert
^{2}}\right) S\left( X\right) }{\frac{\hat{K}_{X}\left\vert \hat{\Psi}\left(
X\right) \right\vert ^{2}}{K_{X}\left\vert \Psi \left( X\right) \right\vert
^{2}}}\right]  \notag \\
&&\times \left( \frac{h\left( X\right) }{2}\left( \hat{h}\left( X\right) +%
\frac{h\left( X\right) }{2}\right) +\frac{\partial _{f\left( X\right)
}\left( \frac{\hat{K}_{X}\left\vert \hat{\Psi}\left( X\right) \right\vert
^{2}}{K_{X}\left\vert \Psi \left( X\right) \right\vert ^{2}}\right)
S_{1}\left( X\right) }{\frac{\hat{K}_{X}\left\vert \hat{\Psi}\left( X\right)
\right\vert ^{2}}{K_{X}\left\vert \Psi \left( X\right) \right\vert ^{2}}}-2%
\frac{\left( \frac{\hat{h}\left( X\right) }{2}S\left( X\right) -\frac{1}{%
f\left( X\right) +C_{0}}\right) }{\hat{f}\left( X,\theta -1\right) \frac{%
\partial _{\hat{f}\left( X\right) }\left( \frac{\hat{K}_{X}\left\vert \hat{%
\Psi}\left( X\right) \right\vert ^{2}}{K_{X}\left\vert \Psi \left( X\right)
\right\vert ^{2}}\right) S\left( X\right) }{\frac{\hat{K}_{X}\left\vert \hat{%
\Psi}\left( X\right) \right\vert ^{2}}{K_{X}\left\vert \Psi \left( X\right)
\right\vert ^{2}}}}\right)  \notag \\
&&\times \left( \left( \frac{\partial f\left( X\right) }{\partial S\left(
X,\theta -2\right) }\right) ^{2}+\frac{\partial ^{2}f\left( X\right) }{%
\partial \left( S\left( X,\theta -2\right) \right) ^{2}}\right) \delta
S\left( X,\theta -2\right)  \notag
\end{eqnarray}%
\begin{eqnarray*}
\delta \beta &\simeq &S\left( X\right) \left( \frac{4}{\left( \hat{f}\left(
X,\theta -1\right) \right) ^{2}}-\frac{\hat{h}\left( X\right) }{\hat{f}%
\left( X,\theta -1\right) }\frac{\partial f\left( X\right) }{\partial
S\left( X,\theta -2\right) }\right) \delta \hat{f}\left( X,\theta -1\right)
\\
&&-\frac{2}{\hat{f}\left( X,\theta -1\right) }\left( \frac{\hat{h}\left(
X\right) }{2}S\left( X\right) -\frac{1}{f\left( X\right) +C_{0}}\right)
\left( \frac{\partial f\left( X\right) }{\partial S\left( X,\theta -2\right) 
}\right) ^{2}\delta S\left( X\right) +\beta \frac{\frac{\partial ^{2}f\left(
X\right) }{\partial \left( S\left( X,\theta -2\right) \right) ^{2}}}{\frac{%
\partial f\left( X\right) }{\partial S\left( X,\theta -2\right) }}\delta
S\left( X\right)
\end{eqnarray*}%
\begin{eqnarray*}
\delta c &\simeq &-\frac{2S_{1}\left( X\right) \beta }{\hat{f}\left(
X,\theta -1\right) }\delta \left\langle \hat{f}\left( X^{\prime }\right)
\right\rangle \\
&&+\left( \frac{\hat{h}\left( X\right) }{2}S\left( X\right) -\frac{1}{%
f\left( X\right) +C_{0}}\right) \left( \frac{\hat{h}\left( X\right) }{2}-%
\frac{2}{\hat{f}\left( X,\theta -1\right) }\right) \left( \frac{\partial
f\left( X\right) }{\partial S\left( X,\theta -2\right) }\right) ^{3}\delta
S\left( X\right) +c\frac{\frac{\partial ^{2}f\left( X\right) }{\partial
\left( S\left( X,\theta -2\right) \right) ^{2}}}{\frac{\partial f\left(
X\right) }{\partial S\left( X,\theta -2\right) }}\delta S\left( X\right) \\
&&-\left( \hat{h}\left( X\right) +2\left( \frac{\hat{h}\left( X\right) }{2}+%
\frac{\partial _{f\left( X\right) }\left( \frac{\hat{K}_{X}\left\vert \hat{%
\Psi}\left( X\right) \right\vert ^{2}}{K_{X}\left\vert \Psi \left( X\right)
\right\vert ^{2}}\right) }{\frac{\hat{K}_{X}\left\vert \hat{\Psi}\left(
X\right) \right\vert ^{2}}{K_{X}\left\vert \Psi \left( X\right) \right\vert
^{2}}}\right) \right) \frac{S_{1}\left( X\right) S\left( X\right) }{\hat{f}%
\left( X,\theta -1\right) }\left( \frac{\partial f\left( X\right) }{\partial
S\left( X,\theta -2\right) }\right) ^{2}\delta \left\langle \hat{f}\left(
X^{\prime }\right) \right\rangle
\end{eqnarray*}%
In general, due to bank loans: 
\begin{equation*}
S_{1}\left( X,X\right) S\left( X\right) <<1
\end{equation*}%
so that:%
\begin{eqnarray*}
\delta c &\simeq &-\frac{2S_{1}\left( X\right) \beta }{\hat{f}\left(
X,\theta -1\right) }\delta \left\langle \hat{f}\left( X^{\prime }\right)
\right\rangle \\
&&+\left( \frac{\hat{h}\left( X\right) }{2}S\left( X\right) -\frac{1}{%
f\left( X\right) +C_{0}}\right) \left( \frac{\hat{h}\left( X\right) }{2}-%
\frac{2}{\hat{f}\left( X,\theta -1\right) }\right) \left( \frac{\partial
f\left( X\right) }{\partial S\left( X,\theta -2\right) }\right) ^{3}\delta
S\left( X\right) +c\frac{\frac{\partial ^{2}f\left( X\right) }{\partial
\left( S\left( X,\theta -2\right) \right) ^{2}}}{\frac{\partial f\left(
X\right) }{\partial S\left( X,\theta -2\right) }}\delta S\left( X\right) \\
&\rightarrow &-\frac{2S_{1}\left( X\right) \beta }{\hat{f}\left( X,\theta
-1\right) }\delta \left\langle \hat{f}\left( X^{\prime }\right)
\right\rangle -\frac{2}{\hat{f}\left( X,\theta -1\right) }\left( \frac{\hat{h%
}\left( X\right) }{2}S\left( X\right) -\frac{1}{f\left( X\right) +C_{0}}%
\right) \left( \frac{\partial f\left( X\right) }{\partial S\left( X,\theta
-2\right) }\right) ^{3}\delta S\left( X\right)
\end{eqnarray*}%
\begin{equation*}
\delta h\simeq 4\frac{S\left( X\right) }{\left( \hat{f}\left( X,\theta
-1\right) \right) ^{2}}\delta \hat{f}\left( X,\theta -1\right) -2\frac{%
\left( \frac{\hat{h}\left( X\right) }{2}S\left( X\right) -\frac{1}{f\left(
X\right) +C_{0}}\right) }{\hat{f}\left( X,\theta -1\right) }\frac{\partial
f\left( X\right) }{\partial S\left( X,\theta -2\right) }\delta S\left(
X,\theta -2\right)
\end{equation*}%
\begin{eqnarray*}
\delta \beta &\simeq &\left( 4\frac{S\left( X\right) }{\left( \hat{f}\left(
X,\theta -1\right) \right) ^{2}}-\frac{\hat{h}\left( X\right) }{2}\frac{2}{%
\hat{f}\left( X,\theta -1\right) }S\left( X\right) \frac{\partial f\left(
X\right) }{\partial S\left( X,\theta -2\right) }\right) \delta \hat{f}\left(
X,\theta -1\right) \\
&&-\left( \frac{\hat{h}\left( X\right) }{2}S\left( X\right) -\frac{1}{%
f\left( X\right) +C_{0}}\right) \frac{2}{\hat{f}\left( X,\theta -1\right) }%
\left( \frac{\partial f\left( X\right) }{\partial S\left( X,\theta -2\right) 
}\right) ^{2}\delta S\left( X,\theta -2\right)
\end{eqnarray*}

\subsubsection*{A4.4.3 Stability condition}

As for banks, we can derive a stabilit condtion:%
\begin{eqnarray*}
&&\left( 1-\left\langle \lambda ^{2}\right\rangle \right) \left( \left(
\delta S\left( X\right) \right) ^{2}+\left( \delta \hat{S}\left( X\right)
\right) ^{2}\right) \\
&>&\left\{ \frac{\delta \alpha }{\delta \hat{f}\left( X\right) }\left(
\delta \hat{S}\left( X\right) \right) ^{2}+\delta \hat{S}\left( X\right)
\left( \frac{\delta \alpha }{\delta S\left( X\right) }+\frac{\hat{h}\left(
X^{\prime },X\right) }{2}\frac{\delta c}{\delta \hat{f}\left( X\right) }%
\right) \delta S\left( X\right) +\frac{\hat{h}\left( X^{\prime },X\right) }{2%
}\frac{\delta c}{\delta S\left( X\right) }\left( \delta S\left( X\right)
\right) ^{2}\right\} ^{2} \\
&&+\left\{ \frac{1}{\frac{\hat{h}\left( X^{\prime },X\right) }{2}}\frac{%
\delta h}{\delta \hat{f}\left( X,\theta -1\right) }\left( \delta \hat{S}%
\left( X\right) \right) ^{2}\right. \\
&&\left. +\delta \hat{S}\left( X\right) \left( \frac{1}{\frac{\hat{h}\left(
X^{\prime },X\right) }{2}}\frac{\delta }{\delta S\left( X\right) }h+\frac{%
\delta \beta }{\delta \hat{f}\left( X,\theta -1\right) }\right) \delta
S\left( X\right) +\frac{\delta \beta }{\delta S\left( X,,\theta -1\right) }%
\left( \delta S\left( X\right) \right) ^{2}\right\} ^{2}
\end{eqnarray*}%
For $\hat{f}\left( X,\theta -1\right) <<1$, $\left\vert \frac{\partial
f\left( X\right) }{\partial S\left( X,\theta -2\right) }\right\vert >1$ we
use that:%
\begin{equation*}
\frac{\delta \alpha }{\delta S\left( X\right) }+\frac{\hat{h}\left(
X^{\prime },X\right) }{2}\frac{\delta c}{\delta \hat{f}\left( X\right) }%
\simeq \frac{\delta \alpha }{\delta S\left( X\right) }
\end{equation*}%
is of order:%
\begin{equation*}
\frac{\left( \frac{\partial f\left( X\right) }{\partial S\left( X,\theta
-2\right) }\right) ^{2}+\frac{\partial ^{2}f\left( X\right) }{\partial
\left( S\left( X,\theta -2\right) \right) ^{2}}}{\hat{f}\left( X,\theta
-1\right) }
\end{equation*}%
is negligible with respect to:%
\begin{equation*}
\sqrt{\left\vert \frac{\delta \alpha }{\delta \hat{f}\left( X\right) }\frac{%
\delta c}{\delta S\left( X\right) }\right\vert }
\end{equation*}%
which is of order:%
\begin{equation*}
\sqrt{\frac{1}{\left( \hat{f}\left( X,\theta -1\right) \right) ^{3}}\left( 
\frac{\partial f\left( X\right) }{\partial S\left( X,\theta -2\right) }%
\right) ^{4}}
\end{equation*}%
Moreover:%
\begin{eqnarray*}
&&\left\vert \frac{\delta }{\delta S\left( X\right) }h\right\vert \\
&\rightarrow &2\frac{\left\vert \frac{\hat{h}\left( X\right) }{2}S\left(
X\right) -\frac{1}{f\left( X\right) +C_{0}}\right\vert }{\hat{f}\left(
X,\theta -1\right) }\frac{\partial f\left( X\right) }{\partial S\left(
X,\theta -2\right) }
\end{eqnarray*}%
\begin{equation*}
\left\vert \frac{\delta \beta }{\delta \hat{f}\left( X,\theta -1\right) }%
\right\vert \simeq S\left( X\right) \left\vert 4\frac{1}{\left( \hat{f}%
\left( X,\theta -1\right) \right) ^{2}}-\frac{\hat{h}\left( X\right) }{2}%
\frac{2}{\hat{f}\left( X,\theta -1\right) }\frac{\partial f\left( X\right) }{%
\partial S\left( X,\theta -2\right) }\right\vert
\end{equation*}%
and:%
\begin{eqnarray*}
&&\left\vert \frac{\delta h}{\delta \hat{f}\left( X,\theta -1\right) }\frac{%
\delta \beta }{\delta S\left( X,,\theta -1\right) }\right\vert \\
&\rightarrow &4\frac{S\left( X\right) }{\left( \hat{f}\left( X,\theta
-1\right) \right) ^{3}}\left( \left\vert \frac{\hat{h}\left( X\right) }{2}%
S\left( X\right) -\frac{1}{f\left( X\right) +C_{0}}\right\vert \right)
\left( \frac{\partial f\left( X\right) }{\partial S\left( X,\theta -2\right) 
}\right) ^{2}
\end{eqnarray*}

For: 
\begin{equation*}
\hat{f}\left( X,\theta -1\right) <<1,\left\vert \frac{\partial f\left(
X\right) }{\partial S\left( X,\theta -2\right) }\right\vert >1,S\left(
X\right) <<1,\frac{1}{f\left( X\right) +C_{0}}<<1
\end{equation*}%
we have in most cases:%
\begin{equation*}
\frac{\sqrt{S\left( X\right) }}{\hat{f}\left( X,\theta -1\right) }<<1
\end{equation*}%
\begin{eqnarray*}
\left\vert \frac{\delta }{\delta S\left( X\right) }h\right\vert &<&\sqrt{%
\left\vert \frac{\delta h}{\delta \hat{f}\left( X,\theta -1\right) }\frac{%
\delta \beta }{\delta S\left( X,,\theta -1\right) }\right\vert } \\
\left\vert \frac{\delta \beta }{\delta \hat{f}\left( X,\theta -1\right) }%
\right\vert &<&\sqrt{\left\vert \frac{\delta h}{\delta \hat{f}\left(
X,\theta -1\right) }\frac{\delta \beta }{\delta S\left( X,,\theta -1\right) }%
\right\vert }
\end{eqnarray*}%
As a consequence, the condition approximates:%
\begin{eqnarray*}
&&\left( 1-\left\langle \lambda ^{2}\right\rangle \right) \left( \left(
\delta S\left( X\right) \right) ^{2}+\left( \delta \hat{S}\left( X\right)
\right) ^{2}\right) \\
&>&\left\{ \frac{\delta \alpha }{\delta \hat{f}\left( X\right) }\left(
\delta \hat{S}\left( X\right) \right) ^{2}+\frac{\hat{h}\left( X^{\prime
},X\right) }{2}\frac{\delta c}{\delta S\left( X\right) }\left( \delta
S\left( X\right) \right) ^{2}\right\} ^{2} \\
&&+\left\{ \frac{1}{\frac{\hat{h}\left( X^{\prime },X\right) }{2}}\frac{%
\delta h}{\delta \hat{f}\left( X,\theta -1\right) }\left( \delta \hat{S}%
\left( X\right) \right) ^{2}+\frac{\delta \beta }{\delta S\left( X,,\theta
-1\right) }\left( \delta S\left( X\right) \right) ^{2}\right\} ^{2}
\end{eqnarray*}%
that is:%
\begin{eqnarray*}
&&\left( 1-\left\langle \lambda ^{2}\right\rangle \right) \left( \left(
\delta S\left( X\right) \right) ^{2}+\left( \delta \hat{S}\left( X\right)
\right) ^{2}\right) \\
&>&\left\{ \frac{8S_{1}\left( X,X\right) \frac{1-\left( \hat{S}_{1}\left(
X\right) \right) }{1-\left( \hat{S}\left( X\right) \right) }}{\left( \hat{f}%
\left( X,\theta -1\right) \right) ^{2}}\frac{\partial f\left( X\right) }{%
\partial S\left( X,\theta -1\right) }\left( \delta \hat{S}\left( X\right)
\right) ^{2}\right. \\
&&\left. -\frac{\hat{h}\left( X^{\prime },X\right) }{2}\frac{2\left( \frac{%
\hat{h}\left( X\right) }{2}S\left( X\right) -\frac{1}{f\left( X\right) +C_{0}%
}\right) \left( \frac{\partial f\left( X\right) }{\partial S\left( X,\theta
-2\right) }\right) ^{3}\left( \delta S\left( X\right) \right) ^{2}}{\hat{f}%
\left( X,\theta -1\right) }\right\} ^{2} \\
&&+\left\{ \frac{1}{\frac{\hat{h}\left( X^{\prime },X\right) }{2}}4\frac{%
S\left( X\right) \left( \delta \hat{S}\left( X\right) \right) ^{2}}{\left( 
\hat{f}\left( X,\theta -1\right) \right) ^{2}}-\frac{2\left( \frac{\hat{h}%
\left( X\right) }{2}S\left( X\right) -\frac{1}{f\left( X\right) +C_{0}}%
\right) }{\hat{f}\left( X,\theta -1\right) }\left( \frac{\partial f\left(
X\right) }{\partial S\left( X,\theta -2\right) }\right) ^{2}\left( \delta
S\left( X\right) \right) ^{2}\right\} ^{2}
\end{eqnarray*}%
Due to bank loans $S_{1}\left( X,X\right) <<1$ and for $\left\vert \frac{%
\partial f\left( X\right) }{\partial S\left( X,\theta -2\right) }\right\vert
>1$ this reduces to:%
\begin{eqnarray*}
&&1-\left\langle \lambda ^{2}\right\rangle \\
&>&\frac{\left\{ \frac{8S_{1}\left( X,X\right) \frac{1-\left( \hat{S}%
_{1}\left( X\right) \right) }{1-\left( \hat{S}\left( X\right) \right) }}{%
\left( \hat{f}\left( X,\theta -1\right) \right) ^{2}}\frac{\partial f\left(
X\right) }{\partial S\left( X,\theta -1\right) }\left( \delta \hat{S}\left(
X\right) \right) ^{2}-\frac{\hat{h}\left( X^{\prime },X\right) }{2}\frac{%
2\left( \frac{\hat{h}\left( X\right) }{2}S\left( X\right) -\frac{1}{f\left(
X\right) +C_{0}}\right) }{\hat{f}\left( X,\theta -1\right) }\left( \frac{%
\partial f\left( X\right) }{\partial S\left( X,\theta -2\right) }\right)
^{3}\left( \delta S\left( X\right) \right) ^{2}\right\} ^{2}}{\left( \delta
S\left( X\right) \right) ^{2}+\left( \delta \hat{S}\left( X\right) \right)
^{2}}
\end{eqnarray*}

\bigskip The estimation of $\left\langle \lambda ^{2}\right\rangle $ comes
from:%
\begin{equation}
\alpha =\frac{1-\left( \hat{S}_{1}\left( X\right) \right) }{1-\left( \hat{S}%
\left( X\right) \right) }S_{1}\left( X,X\right) \frac{\partial f\left(
X\right) }{\partial S\left( X\right) }\frac{\partial _{\hat{f}\left(
X\right) }\left( \frac{\hat{K}_{X}\left\vert \hat{\Psi}\left( X\right)
\right\vert ^{2}}{K_{X}\left\vert \Psi \left( X\right) \right\vert ^{2}}%
\right) S\left( X\right) }{\frac{\hat{K}_{X}\left\vert \hat{\Psi}\left(
X\right) \right\vert ^{2}}{K_{X}\left\vert \Psi \left( X\right) \right\vert
^{2}}}
\end{equation}%
\begin{eqnarray*}
\beta &\simeq &\left( \frac{\hat{h}\left( X\right) }{2}S\left( X\right) +%
\frac{\partial _{f\left( X\right) }\left( \frac{\hat{K}_{X}\left\vert \hat{%
\Psi}\left( X\right) \right\vert ^{2}}{K_{X}\left\vert \Psi \left( X\right)
\right\vert ^{2}}\right) S\left( X\right) }{\frac{\hat{K}_{X}\left\vert \hat{%
\Psi}\left( X\right) \right\vert ^{2}}{K_{X}\left\vert \Psi \left( X\right)
\right\vert ^{2}}}\right) \frac{\partial f\left( X\right) }{\partial S\left(
X\right) } \\
&\rightarrow &\left( \frac{\hat{h}\left( X\right) }{2}S\left( X\right) +%
\frac{1}{f\left( X\right) +C_{0}}\right) \frac{\partial f\left( X\right) }{%
\partial S\left( X\right) }
\end{eqnarray*}%
\begin{equation*}
c\simeq S_{1}\left( X\right) \left( \frac{\hat{h}\left( X\right) }{2}S\left(
X\right) +\frac{\partial _{f\left( X\right) }\left( \frac{\hat{K}%
_{X}\left\vert \hat{\Psi}\left( X\right) \right\vert ^{2}}{K_{X}\left\vert
\Psi \left( X\right) \right\vert ^{2}}\right) S\left( X\right) }{\frac{\hat{K%
}_{X}\left\vert \hat{\Psi}\left( X\right) \right\vert ^{2}}{K_{X}\left\vert
\Psi \left( X\right) \right\vert ^{2}}}\right) \left( \frac{\partial f\left(
X\right) }{\partial S\left( X,\theta -2\right) }\right) ^{2}
\end{equation*}%
\begin{eqnarray*}
h &=&\frac{\partial _{\hat{f}\left( X\right) }\left( \frac{\hat{K}%
_{X}\left\vert \hat{\Psi}\left( X\right) \right\vert ^{2}}{K_{X}\left\vert
\Psi \left( X\right) \right\vert ^{2}}\right) S\left( X\right) }{\frac{\hat{K%
}_{X}\left\vert \hat{\Psi}\left( X\right) \right\vert ^{2}}{K_{X}\left\vert
\Psi \left( X\right) \right\vert ^{2}}} \\
&\rightarrow &-2\frac{S\left( X\right) }{\hat{f}\left( X\right) }
\end{eqnarray*}

\begin{eqnarray*}
\left\langle \lambda \right\rangle &=&\frac{1}{2}\alpha +\frac{1}{2}\beta \\
&=&\frac{1}{2}\left( \left( \frac{\hat{h}\left( X\right) }{2}+\frac{1}{%
f\left( X\right) +C_{0}}\right) -\frac{1-\left( \hat{S}_{1}\left( X\right)
\right) }{1-\left( \hat{S}\left( X\right) \right) }S_{1}\left( X,X\right) 
\frac{2}{\hat{f}\left( X\right) }\right) S\left( X\right) \frac{\partial
f\left( X\right) }{\partial S\left( X,\theta -2\right) } \\
&\rightarrow &-\frac{2S_{1}\left( X,X\right) S\left( X\right) \left(
1-\left( \hat{S}_{1}\left( X\right) \right) \right) }{\hat{f}\left( X\right)
\left( 1-\left( \hat{S}\left( X\right) \right) \right) }\frac{\partial
f\left( X\right) }{\partial S\left( X,\theta -2\right) }>0
\end{eqnarray*}%
and the condition writes:%
\begin{eqnarray*}
&&1-\left( \frac{2S_{1}\left( X,X\right) S\left( X\right) \left( 1-\left( 
\hat{S}_{1}\left( X\right) \right) \right) }{\hat{f}\left( X\right) \left(
1-\left( \hat{S}\left( X\right) \right) \right) }\frac{\partial f\left(
X\right) }{\partial S\left( X,\theta -2\right) }\right) ^{2} \\
&>&\frac{\left\{ \frac{8S_{1}\left( X,X\right) \frac{1-\left( \hat{S}%
_{1}\left( X\right) \right) }{1-\left( \hat{S}\left( X\right) \right) }}{%
\left( \hat{f}\left( X,\theta -1\right) \right) ^{2}}\left( \delta \hat{S}%
\left( X\right) \right) ^{2}-\frac{\hat{h}\left( X^{\prime },X\right) }{2}%
\frac{2\left( \frac{\hat{h}\left( X\right) }{2}S\left( X\right) -\frac{1}{%
f\left( X\right) +C_{0}}\right) }{\hat{f}\left( X,\theta -1\right) }\left( 
\frac{\partial f\left( X\right) }{\partial S\left( X,\theta -2\right) }%
\right) ^{2}\left( \delta S\left( X\right) \right) ^{2}\right\} ^{2}\left( 
\frac{\partial f\left( X\right) }{\partial S\left( X,\theta -1\right) }%
\right) ^{2}}{\left( \delta S\left( X\right) \right) ^{2}+\left( \delta \hat{%
S}\left( X\right) \right) ^{2}}
\end{eqnarray*}

\section*{Appendix 5 Fiber stability}

\subsection*{A5.1 Banks}

\subsubsection*{\protect\bigskip A5.1.1 Fiber fluctuations}

The effect of fluctuations in investors stakes as been studied in Gosselin
and Lotz 25 a and we focus on banks only. To evaluate the effect of
introducing banks, we consider:%
\begin{eqnarray*}
S_{L}^{B}\left( X,\theta -1\right) &>&>S_{E}^{B}\left( X,\theta -1\right) \\
S_{L}^{B}\left( X,\theta -1\right) &>&>S\left( X\right) \\
&&\hat{S}\left( X\right)
\end{eqnarray*}%
\begin{eqnarray*}
a &>&>b \\
d &>&>e \\
h &>&>
\end{eqnarray*}%
so that in first approximation we replace the three dimensional system:%
\begin{equation*}
\left( 
\begin{array}{c}
\delta \bar{f}\left( X,\theta \right) \\ 
\delta \hat{f}\left( X,\theta \right) \\ 
\delta S^{T}\left( X,,\theta -1\right)%
\end{array}%
\right) =\left[ 
\begin{array}{ccc}
a & 0 & c \\ 
d & -1 & f \\ 
g & 0 & i%
\end{array}%
\right] \left( 
\begin{array}{c}
\delta \bar{f}\left( X,\theta -1\right) \\ 
\delta \hat{f}\left( X,\theta -1\right) \\ 
\delta S^{T}\left( X,,\theta -2\right)%
\end{array}%
\right)
\end{equation*}%
by:%
\begin{equation*}
X\left( \theta \right) =MX\left( \theta -1\right)
\end{equation*}%
with:

\begin{equation*}
X\left( \theta \right) =\left( 
\begin{array}{cccccc}
\delta \bar{f}\left( X,\theta \right) & \delta \hat{f}\left( X,\theta \right)
& \delta S^{T}\left( X,,\theta -1\right) & \frac{\delta \bar{K}%
_{X}\left\vert \bar{\Psi}\left( X\right) \right\vert ^{2}}{\bar{K}%
_{X}\left\vert \bar{\Psi}\left( X\right) \right\vert ^{2}} & \frac{\delta 
\hat{K}_{X}\left\vert \hat{\Psi}\left( X\right) \right\vert ^{2}}{\hat{K}%
_{X}\left\vert \hat{\Psi}\left( X\right) \right\vert ^{2}} & \frac{\delta
K_{X}\left\vert \Psi \left( X\right) \right\vert ^{2}}{K_{X}\left\vert \Psi
\left( X\right) \right\vert ^{2}}%
\end{array}%
\right) ^{t}
\end{equation*}%
The matrix $M$ is found by writing the various equatios. Their are similar
to the ones obtained above, but take into account that the disposable
capitals for banks, investors and firms are now independent.

We write:%
\begin{eqnarray*}
\frac{1-\bar{S}\left( X\right) }{1-\bar{S}_{E}\left( X\right) }\delta \bar{f}%
\left( X,\theta \right) &=&a\delta \bar{f}\left( X,\theta -1\right) +a_{0}%
\frac{\delta \frac{\bar{K}_{X}\left\vert \bar{\Psi}\left( X\right)
\right\vert ^{2}}{K_{X}\left\vert \Psi \left( X\right) \right\vert ^{2}}}{%
\frac{\bar{K}_{X}\left\vert \bar{\Psi}\left( X\right) \right\vert ^{2}}{%
K_{X}\left\vert \Psi \left( X\right) \right\vert ^{2}}}+a_{1}\frac{\delta 
\bar{K}_{X}\left\vert \bar{\Psi}\left( X\right) \right\vert ^{2}}{\bar{K}%
_{X}\left\vert \bar{\Psi}\left( X\right) \right\vert ^{2}} \\
&&+b\frac{\delta \left( \frac{\hat{K}_{X}\left\vert \hat{\Psi}\left(
X\right) \right\vert ^{2}}{K_{X}\left\vert \Psi \left( X\right) \right\vert
^{2}}\right) }{\frac{\hat{K}_{X}\left\vert \hat{\Psi}\left( X\right)
\right\vert ^{2}}{K_{X}\left\vert \Psi \left( X\right) \right\vert ^{2}}}%
+c\delta S^{T}\left( X,\theta -2\right)
\end{eqnarray*}%
\begin{eqnarray*}
a &=&\frac{1}{2}H\left( X\right) \bar{S}\left( X\right) \frac{\left( \bar{f}%
\left( X\right) -\bar{r}\right) }{1-\left( \bar{S}_{E}\left( X^{\prime
},\theta -1\right) \right) } \\
&&-\left( H\left( X\right) \bar{S}_{E}\left( X\right) \right) \frac{\left(
1-\left( \bar{S}\left( X,\theta -1\right) \right) \right) \left( \bar{f}%
\left( X\right) -\bar{r}\right) }{\left( 1-\left( \bar{S}_{E}\left(
X^{\prime },\theta -1\right) \right) \right) ^{2}}
\end{eqnarray*}%
\begin{equation*}
a_{0}=\left( S_{E}^{B}\left( X,\theta -1\right) +S_{L}^{B}\left( X,\theta
-1\right) \right) S_{E}^{B}\left( X,X\right)
\end{equation*}%
\begin{eqnarray*}
a_{1} &=&-\bar{S}\left( X^{\prime },\theta -1\right) \frac{\left( \bar{f}%
\left( X\right) -\bar{r}\right) }{1-\left( \bar{S}_{E}\left( X^{\prime
},\theta -1\right) \right) } \\
&&+\bar{S}_{E}\left( X^{\prime },\theta -1\right) \frac{\left( 1-\left( \bar{%
S}\left( X,\theta -1\right) \right) \right) \left( \bar{f}\left( X\right) -%
\bar{r}\right) }{\left( 1-\left( \bar{S}_{E}\left( X^{\prime },\theta
-1\right) \right) \right) ^{2}}
\end{eqnarray*}%
\begin{equation*}
b=S_{E}^{B}\left( X,X\right) S\left( X\right)
\end{equation*}%
\begin{eqnarray*}
c &=&\left\{ w_{1}^{B}\left( X\right) \left( \left\langle \bar{w}\left(
X\right) \right\rangle +\left\langle \hat{w}_{1}^{B}\left( X\right)
\right\rangle \right) \left( f\left( X\right) -\bar{r}\right) \frac{\partial
f\left( X\right) }{\partial S^{T}\left( X,\theta -2\right) }-\frac{%
\left\langle \bar{w}\left( X^{\prime },X\right) \right\rangle }{2}%
w_{1}^{B}\left( X\right) \frac{1-\left\langle \bar{S}\left( X^{\prime
}\right) \right\rangle }{1-\left\langle \bar{S}_{E}\left( X^{\prime }\right)
\right\rangle }\left( \left\langle \bar{f}\left( X^{\prime }\right)
\right\rangle -\bar{r}\right) \right. \\
&&-\left\langle \hat{w}_{1}^{B}\left( X^{\prime },X\right) \right\rangle
\left\langle w_{1}^{B}\left( X\right) \right\rangle \frac{1-\left\langle 
\hat{S}\left( X^{\prime }\right) \right\rangle +\left\langle \hat{S}%
_{E}^{B}\left( X^{\prime }\right) \right\rangle +\left\langle \hat{S}%
_{L}^{B}\left( X^{\prime }\right) \right\rangle }{1-\left\langle \hat{S}%
_{E}\left( X^{\prime }\right) \right\rangle +\left\langle \hat{S}%
_{E}^{B}\left( X^{\prime }\right) \right\rangle }\left( \left\langle \hat{f}%
\left( X^{\prime }\right) \right\rangle -\bar{r}\right) \\
&&+\left. \frac{S_{E}^{B}\left( X,X\right) \partial f\left( X\right) }{%
\partial S^{T}\left( X,\theta -1\right) }\left\{ w_{1}^{B}\left( X\right)
\left( \left\langle \bar{w}\left( X\right) \right\rangle +\left\langle \hat{w%
}_{1}^{B}\left( X\right) \right\rangle \right) +\frac{\frac{\hat{w}\left(
X\right) }{2}S\left( X\right) }{1+\left( \hat{w}\left( X\right) \left( \frac{%
f\left( X\right) +\bar{r}\left( X\right) }{2}-\frac{\left\langle \hat{f}%
\left( X^{\prime }\right) \right\rangle _{\hat{w}_{1}}+\left\langle \hat{r}%
\left( X^{\prime }\right) \right\rangle _{\hat{w}_{2}}}{2}\right) \right) }%
\right\} \right\} \\
&&\times \frac{\partial f\left( X\right) }{\partial S^{T}\left( X,\theta
-2\right) }
\end{eqnarray*}%
\begin{equation*}
\frac{1-\left( \hat{S}\left( X\right) \right) }{1-\left( \hat{S}_{E}\left(
X\right) \right) }\delta \hat{f}\left( X,\theta \right) =d\frac{\delta \frac{%
\bar{K}_{X}\left\vert \bar{\Psi}\left( X\right) \right\vert ^{2}}{%
K_{X}\left\vert \Psi \left( X\right) \right\vert ^{2}}}{\frac{\bar{K}%
_{X}\left\vert \bar{\Psi}\left( X\right) \right\vert ^{2}}{K_{X}\left\vert
\Psi \left( X\right) \right\vert ^{2}}}+e\delta \hat{f}\left( X,\theta
-1\right) +e_{0}\frac{\delta \hat{K}_{X}\left\vert \hat{\Psi}\left( X\right)
\right\vert ^{2}}{\hat{K}_{X}\left\vert \hat{\Psi}\left( X\right)
\right\vert ^{2}}+f\delta S^{T}\left( X,\theta -2\right)
\end{equation*}

\begin{equation*}
d=S_{E}\left( X,X,\theta -1\right) \left( S_{E}^{B}\left( X,\theta -1\right)
+S_{L}^{B}\left( X,\theta -1\right) \right)
\end{equation*}%
\begin{eqnarray*}
e &=&\left( \frac{\hat{S}\left( X\right) }{2\left( 1-\left( \hat{S}%
_{E}\left( X^{\prime }\right) \right) \right) }\frac{1}{1+\frac{\Delta \hat{f%
}\left( X^{\prime }\right) +\Delta \hat{r}\left( X^{\prime }\right) }{2}}%
\right. \\
&&\left. -\frac{\left( 1-\left( \hat{S}\left( X\right) \right) \right) \hat{S%
}_{E}\left( X\right) }{\left( 1-\left( \hat{S}_{E}\left( X^{\prime }\right)
\right) \right) ^{2}}\left( \frac{1}{1+\Delta \hat{f}\left( X\right) }%
\right) \right) \left( \hat{f}\left( X\right) -\bar{r}\right)
\end{eqnarray*}%
\begin{equation*}
e_{0}=\left( \frac{\left( 1-\left( \hat{S}\left( X\right) \right) \right) 
\hat{S}_{E}\left( X\right) }{\left( 1-\left( \hat{S}_{E}\left( X^{\prime
}\right) \right) \right) ^{2}}-\frac{\hat{S}\left( X\right) }{2\left(
1-\left( \hat{S}_{E}\left( X^{\prime }\right) \right) \right) }\right)
\left( \hat{f}\left( X\right) -\bar{r}\right) +S_{E}\left( X,X,\theta
-1\right)
\end{equation*}%
\bigskip 
\begin{eqnarray*}
&&f=\left\{ -\frac{\left( 1-\left\langle \hat{S}\left( X^{\prime }\right)
\right\rangle \right) \left( \left\langle \hat{f}\left( X^{\prime }\right)
\right\rangle -\bar{r}\right) }{1-\left\langle \hat{S}_{E}\left( X^{\prime
}\right) \right\rangle }\frac{\left\langle \hat{w}\left( X^{\prime
},X\right) \right\rangle \left\langle w\left( X\right) \right\rangle }{4}%
\right. \\
&&+\frac{\frac{w\left( X\right) }{4}\left( 1+\hat{w}\left( X\right) \right)
S_{E}\left( X\right) }{1+\left( \hat{w}\left( X\right) \left( f\left(
X\right) -\frac{\left\langle \hat{f}\left( X^{\prime }\right) \right\rangle
_{\hat{w}_{1}}+\left\langle \hat{r}\left( X^{\prime }\right) \right\rangle _{%
\hat{w}_{2}}}{2}\right) +\frac{w\left( X\right) }{2}\left( f\left( X\right) -%
\bar{r}\left( X\right) \right) \right) }\left( f\left( X\right) -\bar{r}%
\right) \\
&&+S_{E}\left( X,X,\theta -1\right) \frac{\partial f\left( X\right) }{%
\partial S^{T}\left( X,\theta -1\right) } \\
&&\left. \times \left( w_{1}^{B}\left( X\right) \left( \left\langle \bar{w}%
\left( X\right) \right\rangle +\left\langle \hat{w}_{1}^{B}\left( X\right)
\right\rangle \right) +\frac{\frac{\hat{w}\left( X\right) }{2}S\left(
X\right) }{1+\left( \hat{w}\left( X\right) \left( \frac{f\left( X\right) +%
\bar{r}\left( X\right) }{2}-\frac{\left\langle \hat{f}\left( X^{\prime
}\right) \right\rangle _{\hat{w}_{1}}+\left\langle \hat{r}\left( X^{\prime
}\right) \right\rangle _{\hat{w}_{2}}}{2}\right) \right) }\right) \right\} \\
&&\times \frac{\partial f\left( X\right) }{\partial S^{T}\left( X,\theta
-2\right) }
\end{eqnarray*}

\begin{equation*}
\delta S^{T}\left( X,,\theta -1\right) =g\frac{\delta \frac{\bar{K}%
_{X}\left\vert \bar{\Psi}\left( X\right) \right\vert ^{2}}{K_{X}\left\vert
\Psi \left( X\right) \right\vert ^{2}}}{\frac{\bar{K}_{X}\left\vert \bar{\Psi%
}\left( X\right) \right\vert ^{2}}{K_{X}\left\vert \Psi \left( X\right)
\right\vert ^{2}}}+h\frac{\delta \left( \frac{\hat{K}_{X}\left\vert \hat{\Psi%
}\left( X\right) \right\vert ^{2}}{K_{X}\left\vert \Psi \left( X\right)
\right\vert ^{2}}\right) }{\frac{\hat{K}_{X}\left\vert \hat{\Psi}\left(
X\right) \right\vert ^{2}}{K_{X}\left\vert \Psi \left( X\right) \right\vert
^{2}}}+i\delta S^{T}\left( X,\theta -2\right)
\end{equation*}%
\begin{eqnarray*}
g &=&\left( S_{E}^{B}\left( X,\theta -1\right) +S_{L}^{B}\left( X,\theta
-1\right) \right) \\
h &=&S\left( X\right) \\
i &=&\left\{ w_{1}^{B}\left( X\right) \left( \left\langle \bar{w}\left(
X\right) \right\rangle +\left\langle \hat{w}_{1}^{B}\left( X\right)
\right\rangle \right) \right. \\
&&\left. +\left( \frac{\frac{\hat{w}\left( X\right) }{2}S\left( X\right) }{%
1+\left( \hat{w}\left( X\right) \left( \frac{f\left( X\right) +\bar{r}\left(
X\right) }{2}-\frac{\left\langle \hat{f}\left( X^{\prime }\right)
\right\rangle _{\hat{w}_{1}}+\left\langle \hat{r}\left( X^{\prime }\right)
\right\rangle _{\hat{w}_{2}}}{2}\right) \right) }\right) \right\} \frac{%
\partial f\left( X\right) }{\partial S^{T}\left( X,\theta -2\right) }
\end{eqnarray*}

As before, we focus on banks only and consider the case of investors
independently. To evaluate the effect of introducing banks, we consider:%
\begin{eqnarray*}
S_{L}^{B}\left( X,\theta -1\right) &>&>S_{E}^{B}\left( X,\theta -1\right) \\
S_{L}^{B}\left( X,\theta -1\right) &>&>S\left( X\right) \\
&&\hat{S}\left( X\right)
\end{eqnarray*}%
\begin{eqnarray*}
a &>&>b \\
d &>&>e \\
h &>&>
\end{eqnarray*}%
so that in first approximation the system is:%
\begin{equation*}
X\left( \theta \right) =MX\left( \theta -1\right)
\end{equation*}%
with:%
\begin{equation*}
M=\left[ 
\begin{array}{cccccc}
a & 0 & c & a_{0}+a_{1} & b & -a_{0}-b \\ 
0 & -1 & f & d & e_{0} & -d \\ 
0 & 0 & i & g & h & -g-h \\ 
\frac{\partial _{\bar{f}\left( X\right) }\bar{K}_{X}\left\vert \bar{\Psi}%
\left( X\right) \right\vert ^{2}}{\bar{K}_{X}\left\vert \bar{\Psi}\left(
X\right) \right\vert ^{2}} & 0 & 0 & 0 & 0 & 0 \\ 
\frac{\partial _{\bar{f}\left( X\right) }\hat{K}_{X}\left\vert \hat{\Psi}%
\left( X\right) \right\vert ^{2}}{\hat{K}_{X}\left\vert \hat{\Psi}\left(
X\right) \right\vert ^{2}} & \frac{\partial _{\hat{f}\left( X\right) }\hat{K}%
_{X}\left\vert \hat{\Psi}\left( X\right) \right\vert ^{2}}{\hat{K}%
_{X}\left\vert \hat{\Psi}\left( X\right) \right\vert ^{2}} & 0 & 0 & 0 & 0
\\ 
0 & 0 & \frac{\partial _{f\left( X\right) }K_{X}\left\vert \Psi \left(
X\right) \right\vert ^{2}}{K_{X}\left\vert \Psi \left( X\right) \right\vert
^{2}}\frac{\partial f\left( X\right) }{\partial S^{T}\left( X,\theta
-2\right) } & 0 & 0 & 0%
\end{array}%
\right]
\end{equation*}

\subsubsection{A5.1.2 Eigenvalues}

As befr we assm that $\left\vert \frac{\partial f\left( X\right) }{\partial
S^{T}\left( X,\theta -1\right) }\right\vert >>1$. We set $z=\frac{\partial
f\left( X\right) }{\partial S^{T}\left( X,\theta -1\right) }$ and rescale
the matrix tht writes in first approximation:%
\begin{equation*}
M=\left( 
\begin{array}{cccccc}
0 & 0 & cz^{2} & a_{0} & b & -a-b \\ 
0 & -1 & fz^{2} & d & e_{0} & -d \\ 
0 & 0 & iz & g & h & -g-h \\ 
r & 0 & 0 & 0 & 0 & 0 \\ 
0 & s & 0 & 0 & 0 & 0 \\ 
0 & 0 & tz & 0 & 0 & 0%
\end{array}%
\right)
\end{equation*}%
with:%
\begin{equation*}
r=\frac{\partial _{\bar{f}\left( X\right) }\bar{K}_{X}\left\vert \bar{\Psi}%
\left( X\right) \right\vert ^{2}}{\bar{K}_{X}\left\vert \bar{\Psi}\left(
X\right) \right\vert ^{2}}\simeq -\frac{2}{\left( 1-\left\langle \bar{S}%
\right\rangle \right) \bar{f}\left( X\right) +\left\langle \bar{S}\left(
X^{\prime },X\right) \right\rangle _{X^{\prime }}\left\langle \bar{f}%
\right\rangle }<0
\end{equation*}%
\begin{equation*}
s=\frac{\partial _{\hat{f}\left( X\right) }\hat{K}_{X}\left\vert \hat{\Psi}%
\left( X\right) \right\vert ^{2}}{\hat{K}_{X}\left\vert \hat{\Psi}\left(
X\right) \right\vert ^{2}}\simeq -\frac{2}{\hat{f}\left( X\right) +\frac{%
\left( \left\langle \hat{S}_{E}^{B}\left( X,X^{\prime }\right) \right\rangle
_{X^{\prime }}+\left\langle \hat{S}_{L}^{B}\left( X,X^{\prime }\right)
\right\rangle _{X^{\prime }}\right) \frac{\left\langle \bar{K}\right\rangle
\left\Vert \bar{\Psi}\right\Vert ^{2}}{\left\langle \hat{K}\right\rangle
\left\Vert \hat{\Psi}\right\Vert ^{2}}\left\langle \bar{f}\right\rangle }{%
1-\left\langle \bar{S}\right\rangle }}<0
\end{equation*}%
\begin{equation*}
\left\vert \frac{\partial _{\bar{f}\left( X\right) }\hat{K}_{X}\left\vert 
\hat{\Psi}\left( X\right) \right\vert ^{2}}{\hat{K}_{X}\left\vert \hat{\Psi}%
\left( X\right) \right\vert ^{2}}\right\vert \simeq \frac{2\frac{\left(
\left\langle \hat{S}_{E}^{B}\left( X,X^{\prime }\right) \right\rangle
_{X^{\prime }}+\left\langle \hat{S}_{L}^{B}\left( X,X^{\prime }\right)
\right\rangle _{X^{\prime }}\right) \frac{\left\langle \bar{K}\right\rangle
\left\Vert \bar{\Psi}\right\Vert ^{2}}{\left\langle \hat{K}\right\rangle
\left\Vert \hat{\Psi}\right\Vert ^{2}}\left\langle \bar{f}\right\rangle }{%
1-\left\langle \bar{S}\right\rangle }}{\hat{f}\left( X\right) +\frac{\left(
\left\langle \hat{S}_{E}^{B}\left( X,X^{\prime }\right) \right\rangle
_{X^{\prime }}+\left\langle \hat{S}_{L}^{B}\left( X,X^{\prime }\right)
\right\rangle _{X^{\prime }}\right) \frac{\left\langle \bar{K}\right\rangle
\left\Vert \bar{\Psi}\right\Vert ^{2}}{\left\langle \hat{K}\right\rangle
\left\Vert \hat{\Psi}\right\Vert ^{2}}\left\langle \bar{f}\right\rangle }{%
1-\left\langle \bar{S}\right\rangle }}<<s
\end{equation*}%
\begin{equation*}
t=\frac{\partial _{f\left( X\right) }K_{X}\left\vert \Psi \left( X\right)
\right\vert ^{2}}{K_{X}\left\vert \Psi \left( X\right) \right\vert ^{2}}%
\simeq \frac{2}{r}\frac{1}{f\left( X\right) +C_{0}}>0
\end{equation*}%
At the lowest order the eigenvalue satisfy:%
\begin{equation*}
X\left( X^{2}\left( cgr+fhs\right) +Xcgr-cgrse-afhrs+bfgrs+cdhrs\right) =0
\end{equation*}%
\begin{equation*}
X^{2}+X\frac{cgr}{\left( cgr+fhs\right) }-\frac{rs}{\left( cgr+fhs\right) }%
\left( cge+afh-bfg-cdh\right)
\end{equation*}%
where:%
\begin{equation*}
\frac{cgr}{\left( cgr+fhs\right) }>0
\end{equation*}%
and:%
\begin{eqnarray*}
&&\left( cge+afh-bfg-cdh\right) \\
&=&c\left( ge-dh\right) +f\left( ah-bg\right) \\
&=&S_{E}\left( X,X,\theta -1\right) \left( S_{E}^{B}\left( X,\theta
-1\right) +S_{L}^{B}\left( X,\theta -1\right) \right) \left( 1-S\left(
X\right) \right) \\
&&+\left( S_{E}^{B}\left( X,\theta -1\right) +S_{L}^{B}\left( X,\theta
-1\right) \right) S_{E}^{B}\left( X,X\right) S\left( X\right)
-S_{E}^{B}\left( X,X\right) S\left( X\right) \left( S_{E}^{B}\left( X,\theta
-1\right) +S_{L}^{B}\left( X,\theta -1\right) \right) \\
&=&S_{E}\left( X,X,\theta -1\right) \left( S_{E}^{B}\left( X,\theta
-1\right) +S_{L}^{B}\left( X,\theta -1\right) \right) \left( 1-S\left(
X\right) \right) >0
\end{eqnarray*}%
then:%
\begin{equation*}
-\frac{rs}{\left( cgr+fhs\right) }\left( cge+afh-bfg-cdh\right) >0
\end{equation*}%
Eigenvalues are negative.

To obtain the eigenvalues close to $0$, \ we set $X=\frac{\epsilon }{z}$ and
the lowest order equation becomes:%
\begin{equation*}
\epsilon ^{2}+\left( z-iz^{2}\right) \epsilon -\left( sz^{2}e+arz^{2}\right)
=0
\end{equation*}%
Given that:%
\begin{equation*}
-\left( sz^{2}e+arz^{2}\right) >0
\end{equation*}%
and $i<0$: 
\begin{equation*}
\left( z-iz^{2}\right) >0
\end{equation*}%
Thus the eigenvalues are negative in general. This implies stability in main
cases.

\subsubsection*{A5.1.3 Range of stability}

\paragraph*{\protect\bigskip A5.1.3.1 Estimation of the coefficients}

As before we study the range os stability by considering the second order
system:%
\begin{equation*}
X\left( \theta \right) =MX\left( \theta -1\right) +X^{t}\left( \theta
-1\right) \mathbf{M}X\left( \theta -1\right)
\end{equation*}%
with:%
\begin{equation*}
\mathbf{M=}\left( 
\begin{array}{cccccc}
\bar{M} & \hat{M} & M^{T} & M_{\bar{K}} & M_{\hat{K}} & M_{K}%
\end{array}%
\right)
\end{equation*}%
where the matrices are defined below.

We consider that:

\begin{equation*}
S_{1}^{B}\left( X,\theta -1\right) +S_{2}^{B}\left( X,\theta -1\right) >>1
\end{equation*}%
due to bank loans. We consider also 
\begin{equation*}
\frac{\partial _{\bar{f}\left( X\right) }\bar{K}_{X}\left\vert \bar{\Psi}%
\left( X\right) \right\vert ^{2}}{\bar{K}_{X}\left\vert \bar{\Psi}\left(
X\right) \right\vert ^{2}}<<1
\end{equation*}%
and:%
\begin{equation*}
\frac{\partial _{f\left( X\right) }K_{X}\left\vert \Psi \left( X\right)
\right\vert ^{2}}{K_{X}\left\vert \Psi \left( X\right) \right\vert ^{2}}%
\simeq \frac{2}{r}\frac{1}{f\left( X\right) +C_{0}}
\end{equation*}%
Moreover, we use:%
\begin{equation*}
\frac{\hat{K}_{X}\left\vert \hat{\Psi}\left( X\right) \right\vert ^{2}}{%
K_{X}\left\vert \Psi \left( X\right) \right\vert ^{2}}\simeq \frac{18\sigma
_{\hat{K}}^{2}V\left\Vert \bar{\Psi}_{0}\left( X\right) \right\Vert ^{4}}{%
\hat{\mu}F_{1}^{2}\left( \left( \frac{2\epsilon }{3\sigma _{\hat{K}}^{2}}%
\right) ^{\frac{r}{2}}\frac{f_{1}\left( X\right) }{C_{0}+\frac{S_{2}\left(
X\right) }{1-S_{1}\left( X\right) }\bar{r}}\right) ^{\frac{2}{r}}}
\end{equation*}%
and thus:%
\begin{equation*}
\frac{\partial _{f\left( X\right) }\left( \frac{\hat{K}_{X}\left\vert \hat{%
\Psi}\left( X\right) \right\vert ^{2}}{K_{X}\left\vert \Psi \left( X\right)
\right\vert ^{2}}\right) }{\frac{\hat{K}_{X}\left\vert \hat{\Psi}\left(
X\right) \right\vert ^{2}}{K_{X}\left\vert \Psi \left( X\right) \right\vert
^{2}}}\simeq -\frac{2}{r}\frac{1}{f\left( X\right) +C_{0}}
\end{equation*}%
\begin{equation*}
\frac{\delta }{\delta S^{T}\left( X,,\theta -2\right) }\frac{\partial
_{f\left( X\right) }\left( \frac{\hat{K}_{X}\left\vert \hat{\Psi}\left(
X\right) \right\vert ^{2}}{K_{X}\left\vert \Psi \left( X\right) \right\vert
^{2}}\right) }{\frac{\hat{K}_{X}\left\vert \hat{\Psi}\left( X\right)
\right\vert ^{2}}{K_{X}\left\vert \Psi \left( X\right) \right\vert ^{2}}}%
\simeq \frac{2}{r}\frac{1}{\left( f\left( X\right) +C_{0}\right) ^{2}}\frac{%
\delta f\left( X\right) }{\delta S^{T}\left( X,,\theta -2\right) }
\end{equation*}%
\begin{equation*}
\frac{\delta }{\delta S^{T}\left( X,,\theta -2\right) }\frac{\partial
_{f\left( X\right) }K_{X}\left\vert \Psi \left( X\right) \right\vert ^{2}}{%
K_{X}\left\vert \Psi \left( X\right) \right\vert ^{2}}\simeq -\frac{2}{r}%
\frac{1}{\left( f\left( X\right) +C_{0}\right) ^{2}}\frac{\delta f\left(
X\right) }{\delta S^{T}\left( X,,\theta -2\right) }
\end{equation*}%
\begin{eqnarray*}
&&\frac{\partial }{\partial S^{T}\left( X,\theta -2\right) }\left(
S_{1}^{B}\left( X,X\right) h_{1}^{B}\left( X\right) \left( \left\langle \bar{%
h}\left( X\right) \right\rangle +\left\langle \hat{h}_{1}^{B}\left( X\right)
\right\rangle \right) \right) \\
&=&h_{1}^{B}\left( X\right) \left( \left\langle \bar{h}\left( X\right)
\right\rangle +\left\langle \hat{h}_{1}^{B}\left( X\right) \right\rangle
\right) \\
&&\times \frac{\partial f\left( X\right) }{\partial S^{T}\left( X,\theta
-1\right) }\left( h_{1}^{B}\left( X\right) \left( \left\langle \bar{h}\left(
X\right) \right\rangle +\left\langle \hat{h}_{1}^{B}\left( X\right)
\right\rangle \right) \right) \\
&&-S_{1}^{B}\left( X,X\right) \frac{\partial f\left( X\right) }{\partial
S^{T}\left( X,\theta -1\right) }\left( h_{1}^{B}\left( X\right) \left(
\left\langle \bar{h}\left( X\right) \right\rangle +\left\langle \hat{h}%
_{1}^{B}\left( X\right) \right\rangle \right) \right.
\end{eqnarray*}%
\begin{eqnarray*}
&&\frac{\delta }{\delta S^{T}\left( X,,\theta -2\right) }\left(
S_{1}^{B}\left( X,X\right) \left( S_{1}^{B}\left( X,\theta -1\right)
+S_{2}^{B}\left( X,\theta -1\right) \right) \right) \\
&=&S_{1}^{B}\left( X,X\right) \left( h_{1}^{B}\left( X\right) \left(
\left\langle \bar{h}\left( X\right) \right\rangle +\left\langle \hat{h}%
_{1}^{B}\left( X\right) \right\rangle \right) \right) \\
&&\times \frac{\partial f\left( X\right) }{\partial S^{T}\left( X,\theta
-2\right) }+\left( S_{1}^{B}\left( X,\theta -1\right) +S_{2}^{B}\left(
X,\theta -1\right) \right) h_{1}^{B}\left( X\right) \left( \left\langle \bar{%
h}\left( X\right) \right\rangle +\left\langle \hat{h}_{1}^{B}\left( X\right)
\right\rangle \right) \frac{\partial f\left( X\right) }{\partial S^{T}\left(
X,\theta -2\right) }
\end{eqnarray*}%
The derivatives are obtained as:%
\begin{eqnarray*}
&&\frac{\delta }{\delta \bar{f}\left( X,\theta -1\right) }a \\
&\simeq &\left( \frac{1}{2}H\left( X\right) \bar{S}\left( X\right) \right) 
\frac{1}{1-\left( \bar{S}_{1}\left( X^{\prime },\theta -1\right) \right) }%
-\left( H\left( X\right) \bar{S}_{1}\left( X\right) \right) \frac{\left(
1-\left( \bar{S}\left( X,\theta -1\right) \right) \right) }{\left( 1-\left( 
\bar{S}_{1}\left( X^{\prime },\theta -1\right) \right) \right) ^{2}}
\end{eqnarray*}%
\begin{equation*}
\frac{\delta }{\delta \bar{f}\left( X,\theta -1\right) }a_{0}=\left( \left(
2S_{1}^{B}\left( X,\theta -1\right) +S_{2}^{B}\left( X,\theta -1\right)
\right) h_{1}^{B}\left( X\right) \left( \left\langle \bar{h}\left( X\right)
\right\rangle +\left\langle \hat{h}_{1}^{B}\left( X\right) \right\rangle
\right) \right) \frac{\partial f\left( X\right) }{\partial S^{T}\left(
X,\theta -2\right) }
\end{equation*}%
\begin{equation*}
\frac{\delta }{\delta \bar{f}\left( X,\theta -1\right) }a_{1}=-\bar{S}\left(
X^{\prime },\theta -1\right) \frac{1}{1-\left( \bar{S}_{E}\left( X^{\prime
},\theta -1\right) \right) }+\bar{S}_{E}\left( X^{\prime },\theta -1\right) 
\frac{1-\left( \bar{S}\left( X,\theta -1\right) \right) }{\left( 1-\left( 
\bar{S}_{E}\left( X^{\prime },\theta -1\right) \right) \right) ^{2}}
\end{equation*}%
\begin{equation*}
\frac{\delta }{\delta \bar{f}\left( X,\theta -1\right) }b=0
\end{equation*}%
\begin{equation*}
\frac{\delta }{\delta \hat{f}\left( X,\theta -1\right) }a\simeq \frac{\delta 
}{\delta \hat{f}\left( X,\theta -1\right) }a_{0}\simeq \frac{\delta }{\delta 
\hat{f}\left( X,\theta -1\right) }a_{1}\simeq \frac{\delta }{\delta \hat{f}%
\left( X,\theta -1\right) }b\simeq 0
\end{equation*}%
\begin{equation*}
\frac{\delta }{\delta S^{T}\left( X,\theta -2\right) }a\simeq \frac{\delta }{%
\delta S^{T}\left( X,\theta -2\right) }a_{0}\simeq \frac{\delta }{\delta
S^{T}\left( X,\theta -2\right) }a_{1}\simeq \frac{\delta }{\delta
S^{T}\left( X,\theta -2\right) }b\simeq 0
\end{equation*}

\begin{eqnarray*}
\frac{\delta }{\delta \bar{f}\left( X,\theta -1\right) }c &=&\left\{
h_{1}^{B}\left( X\right) \left( \left\langle \bar{h}\left( X\right)
\right\rangle +\left\langle \hat{h}_{1}^{B}\left( X\right) \right\rangle
\right) \frac{\partial f\left( X\right) }{\partial S^{T}\left( X,\theta
-2\right) }-\frac{\left\langle \bar{h}\left( X^{\prime },X\right)
\right\rangle }{2}h_{1}^{B}\left( X\right) \frac{1-\left\langle \bar{S}%
\left( X^{\prime }\right) \right\rangle }{1-\left\langle \bar{S}_{1}\left(
X^{\prime }\right) \right\rangle }\right\} \\
&&\times \frac{\partial f\left( X\right) }{\partial S^{T}\left( X,\theta
-2\right) }
\end{eqnarray*}%
\begin{equation*}
\frac{\delta }{\delta \hat{f}\left( X,\theta -1\right) }c\simeq
-\left\langle \hat{h}_{1}^{B}\left( X^{\prime },X\right) \right\rangle
\left\langle h_{1}^{B}\left( X\right) \right\rangle \frac{1-\left\langle 
\hat{S}\left( X^{\prime }\right) \right\rangle +\left\langle \hat{S}%
_{1}^{B}\left( X^{\prime }\right) \right\rangle +\left\langle \hat{S}%
_{2}^{B}\left( X^{\prime }\right) \right\rangle }{1-\left\langle \hat{S}%
_{1}\left( X^{\prime }\right) \right\rangle +\left\langle \hat{S}%
_{1}^{B}\left( X^{\prime }\right) \right\rangle }
\end{equation*}

\begin{eqnarray*}
&&\frac{\delta }{\delta S^{T}\left( X,,\theta -2\right) }c \\
&\simeq &\left( \frac{\partial f\left( X\right) }{\partial S^{T}\left(
X,\theta -1\right) }\right) ^{3} \\
&&\times \left\{ h_{1}^{B}\left( X\right) \left( \left\langle \bar{h}\left(
X\right) \right\rangle +\left\langle \hat{h}_{1}^{B}\left( X\right)
\right\rangle \right) \left( h_{1}^{B}\left( X\right) \left( \left\langle 
\bar{h}\left( X\right) \right\rangle +\left\langle \hat{h}_{1}^{B}\left(
X\right) \right\rangle \right) +Z\right) +Z^{2}S\left( X\right) \right\} \\
&&+c\frac{\frac{\partial ^{2}f\left( X\right) }{\partial \left( S^{T}\left(
X,\theta -2\right) \right) ^{2}}}{\frac{\partial f\left( X\right) }{\partial
S^{T}\left( X,\theta -2\right) }}+S_{1}^{B}\left( X,X\right) \frac{\partial
f\left( X\right) }{\partial S^{T}\left( X,\theta -1\right) }\left\{
h_{1}^{B}\left( X\right) \left( \left\langle \bar{h}\left( X\right)
\right\rangle +\left\langle \hat{h}_{1}^{B}\left( X\right) \right\rangle
\right) +Z\right\} \frac{\partial ^{2}f\left( X\right) }{\partial \left(
S^{T}\left( X,\theta -2\right) \right) ^{2}}
\end{eqnarray*}%
with:%
\begin{equation*}
Z=\frac{\frac{\hat{h}\left( X\right) }{2}}{1+\left( \hat{h}\left( X\right)
\left( \frac{f\left( X\right) +\bar{r}\left( X\right) }{2}-\frac{%
\left\langle \hat{f}\left( X^{\prime }\right) \right\rangle _{\hat{h}%
_{1}}+\left\langle \hat{r}\left( X^{\prime }\right) \right\rangle _{\hat{h}%
_{2}}}{2}\right) \right) }
\end{equation*}%
and:%
\begin{eqnarray*}
&&c\frac{\frac{\partial ^{2}f\left( X\right) }{\partial \left( S^{T}\left(
X,\theta -2\right) \right) ^{2}}}{\frac{\partial f\left( X\right) }{\partial
S^{T}\left( X,\theta -2\right) }} \\
&\simeq &S_{1}^{B}\left( X,X\right) \frac{\partial f\left( X\right) }{%
\partial S^{T}\left( X,\theta -1\right) }\left\{ h_{1}^{B}\left( X\right)
\left( \left\langle \bar{h}\left( X\right) \right\rangle +\left\langle \hat{h%
}_{1}^{B}\left( X\right) \right\rangle \right) +Z\right\} \frac{\partial
^{2}f\left( X\right) }{\partial \left( S^{T}\left( X,\theta -2\right)
\right) ^{2}}
\end{eqnarray*}%
\begin{equation*}
d=S_{E}\left( X,X,\theta -1\right) \left( S_{E}^{B}\left( X,\theta -1\right)
+S_{L}^{B}\left( X,\theta -1\right) \right)
\end{equation*}%
\begin{equation*}
\frac{\delta }{\delta \bar{f}\left( X,\theta -1\right) }d\simeq 2S_{E}\left(
X,X,\theta -1\right) h_{1}^{B}\left( X\right)
\end{equation*}%
\begin{equation*}
e_{0}=\left( \frac{\left( 1-\left( \hat{S}\left( X\right) \right) \right) 
\hat{S}_{E}\left( X\right) }{\left( 1-\left( \hat{S}_{E}\left( X^{\prime
}\right) \right) \right) ^{2}}-\frac{\hat{S}\left( X\right) }{2\left(
1-\left( \hat{S}_{E}\left( X^{\prime }\right) \right) \right) }\right)
\left( \hat{f}\left( X\right) -\bar{r}\right) +S_{E}\left( X,X,\theta
-1\right)
\end{equation*}%
\begin{equation*}
\frac{\delta }{\delta \bar{f}\left( X,\theta -1\right) }e\simeq 0
\end{equation*}%
\begin{equation*}
\frac{\delta }{\delta \bar{f}\left( X,\theta -1\right) }e_{0}\simeq 0
\end{equation*}%
\begin{equation*}
\frac{\delta }{\delta \hat{f}\left( X,\theta -1\right) }d\simeq 0
\end{equation*}%
\begin{equation*}
\frac{\delta }{\delta \hat{f}\left( X,\theta -1\right) }e\simeq \frac{\hat{S}%
\left( X\right) }{2\left( 1-\left( \hat{S}_{E}\left( X^{\prime }\right)
\right) \right) }\frac{1}{1+\frac{\Delta \hat{f}\left( X^{\prime }\right)
+\Delta \hat{r}\left( X^{\prime }\right) }{2}}-\frac{\left( 1-\left( \hat{S}%
\left( X\right) \right) \right) \hat{S}_{E}\left( X\right) }{\left( 1-\left( 
\hat{S}_{E}\left( X^{\prime }\right) \right) \right) ^{2}}\left( \frac{1}{%
1+\Delta \hat{f}\left( X\right) }\right)
\end{equation*}%
\begin{equation*}
\frac{\delta }{\delta \hat{f}\left( X,\theta -1\right) }e_{0}\simeq \frac{%
\left( 1-\left( \hat{S}\left( X\right) \right) \right) \hat{S}_{E}\left(
X\right) }{\left( 1-\left( \hat{S}_{E}\left( X^{\prime }\right) \right)
\right) ^{2}}-\frac{\hat{S}\left( X\right) }{2\left( 1-\left( \hat{S}%
_{E}\left( X^{\prime }\right) \right) \right) }
\end{equation*}

\begin{eqnarray*}
&&\frac{\delta }{\delta S^{T}\left( X,,\theta -2\right) }d \\
&\simeq &\left( \frac{\frac{\hat{h}\left( X\right) }{2}}{1+\left( \hat{h}%
\left( X\right) \left( \frac{f\left( X\right) +\bar{r}\left( X\right) }{2}-%
\frac{\left\langle \hat{f}\left( X^{\prime }\right) \right\rangle _{\hat{h}%
_{1}}+\left\langle \hat{r}\left( X^{\prime }\right) \right\rangle _{\hat{h}%
_{2}}}{2}\right) \right) }\right) \frac{\partial f\left( X\right) }{\partial
S^{T}\left( X,\theta -2\right) }d \\
&&+\left( h_{1}^{B}\left( X\right) \left( \left\langle \bar{h}\left(
X\right) \right\rangle +\left\langle \hat{h}_{1}^{B}\left( X\right)
\right\rangle \right) \right) \left( \frac{\partial f\left( X\right) }{%
\partial S^{T}\left( X,\theta -2\right) }\right) ^{2}+\frac{\frac{\partial
^{2}f\left( X\right) }{\partial \left( S^{T}\left( X,\theta -2\right)
\right) ^{2}}}{\frac{\partial f\left( X\right) }{\partial S^{T}\left(
X,\theta -2\right) }}d \\
&\rightarrow &\left( \frac{\frac{\hat{h}\left( X\right) }{2}d}{1+\left( \hat{%
h}\left( X\right) \left( \frac{f\left( X\right) +\bar{r}\left( X\right) }{2}-%
\frac{\left\langle \hat{f}\left( X^{\prime }\right) \right\rangle _{\hat{h}%
_{1}}+\left\langle \hat{r}\left( X^{\prime }\right) \right\rangle _{\hat{h}%
_{2}}}{2}\right) \right) }+\frac{\frac{\partial f\left( X\right) }{\partial
S^{T}\left( X,\theta -2\right) }h_{1}^{B}\left( X\right) \left( \left\langle 
\bar{h}\left( X\right) \right\rangle +\left\langle \hat{h}_{1}^{B}\left(
X\right) \right\rangle \right) }{S_{1}^{B}\left( X,\theta -1\right)
+S_{2}^{B}\left( X,\theta -1\right) }\right) \\
&&\times \frac{\partial f\left( X\right) }{\partial S^{T}\left( X,\theta
-2\right) }+\frac{\frac{\partial ^{2}f\left( X\right) }{\partial \left(
S^{T}\left( X,\theta -2\right) \right) ^{2}}}{\frac{\partial f\left(
X\right) }{\partial S^{T}\left( X,\theta -2\right) }}d
\end{eqnarray*}%
\begin{equation*}
\frac{\delta }{\delta S^{T}\left( X,,\theta -2\right) }e\simeq 0
\end{equation*}%
\begin{eqnarray*}
\frac{\delta }{\delta S^{T}\left( X,,\theta -2\right) }e_{0} &\simeq &\frac{%
\delta }{\delta S^{T}\left( X,,\theta -2\right) }S_{E}\left( X,X,\theta
-1\right) \\
&\simeq &\frac{\delta }{\delta f\left( X\right) }S_{E}\left( X,X,\theta
-1\right) \frac{\partial f\left( X\right) }{\partial S^{T}\left( X,\theta
-2\right) } \\
&\simeq &h_{E}\left( X\right) \frac{\partial f\left( X\right) }{\partial
S^{T}\left( X,\theta -2\right) }
\end{eqnarray*}%
\bigskip \bigskip \bigskip 
\begin{eqnarray*}
\frac{\delta }{\delta \bar{f}\left( X,\theta -1\right) }f &\simeq
&-S_{1}\left( X,X,\theta -1\right) \left( S_{1}^{B}\left( X,\theta -1\right)
+S_{2}^{B}\left( X,\theta -1\right) \right) \\
&&\times \left[ \frac{\partial f\left( X\right) }{\partial S^{T}\left(
X,\theta -1\right) }\right] ^{2}\frac{\partial _{\bar{f}\left( X\right) }%
\bar{K}_{X}\left\vert \bar{\Psi}\left( X\right) \right\vert ^{2}}{\bar{K}%
_{X}\left\vert \bar{\Psi}\left( X\right) \right\vert ^{2}}\frac{\partial
_{f\left( X\right) }K_{X}\left\vert \Psi \left( X\right) \right\vert ^{2}}{%
K_{X}\left\vert \Psi \left( X\right) \right\vert ^{2}} \\
&\simeq &0
\end{eqnarray*}%
\begin{eqnarray*}
&&\frac{\delta }{\delta \hat{f}\left( X,\theta -1\right) }f \\
&\simeq &-\frac{\left( 1-\left\langle \hat{S}\left( X^{\prime }\right)
\right\rangle \right) }{1-\left\langle \hat{S}_{1}\left( X^{\prime }\right)
\right\rangle }\frac{\left\langle \hat{h}\left( X^{\prime },X\right)
\right\rangle \left\langle h\left( X\right) \right\rangle }{4}\frac{\partial
f\left( X\right) }{\partial S^{T}\left( X,\theta -1\right) }
\end{eqnarray*}%
\begin{equation*}
\frac{\delta }{\delta \bar{f}\left( X,\theta -1\right) }a_{0}=\left( \left(
2S_{1}^{B}\left( X,\theta -1\right) +S_{2}^{B}\left( X,\theta -1\right)
\right) h_{1}^{B}\left( X\right) \left( \left\langle \bar{h}\left( X\right)
\right\rangle +\left\langle \hat{h}_{1}^{B}\left( X\right) \right\rangle
\right) \right) \frac{\partial f\left( X\right) }{\partial S^{T}\left(
X,\theta -2\right) }
\end{equation*}

\begin{eqnarray*}
\frac{\delta }{\delta \bar{f}\left( X,\theta -1\right) }c &=&\left\{
h_{1}^{B}\left( X\right) \left( \left\langle \bar{h}\left( X\right)
\right\rangle +\left\langle \hat{h}_{1}^{B}\left( X\right) \right\rangle
\right) \frac{\partial f\left( X\right) }{\partial S^{T}\left( X,\theta
-2\right) }-\frac{\left\langle \bar{h}\left( X^{\prime },X\right)
\right\rangle }{2}h_{1}^{B}\left( X\right) \frac{1-\left\langle \bar{S}%
\left( X^{\prime }\right) \right\rangle }{1-\left\langle \bar{S}_{1}\left(
X^{\prime }\right) \right\rangle }\right\} \\
&&\times \frac{\partial f\left( X\right) }{\partial S^{T}\left( X,\theta
-2\right) }
\end{eqnarray*}%
\begin{eqnarray*}
&&\frac{\delta }{\delta S^{T}\left( X,,\theta -2\right) }c \\
&\simeq &\left( \frac{\partial f\left( X\right) }{\partial S^{T}\left(
X,\theta -1\right) }\right) ^{3} \\
&&\times \left\{ h_{1}^{B}\left( X\right) \left( \left\langle \bar{h}\left(
X\right) \right\rangle +\left\langle \hat{h}_{1}^{B}\left( X\right)
\right\rangle \right) \left( h_{1}^{B}\left( X\right) \left( \left\langle 
\bar{h}\left( X\right) \right\rangle +\left\langle \hat{h}_{1}^{B}\left(
X\right) \right\rangle \right) +Z\right) +Z^{2}S\left( X\right) \right\} \\
&&+c\frac{\frac{\partial ^{2}f\left( X\right) }{\partial \left( S^{T}\left(
X,\theta -2\right) \right) ^{2}}}{\frac{\partial f\left( X\right) }{\partial
S^{T}\left( X,\theta -2\right) }}+S_{1}^{B}\left( X,X\right) \frac{\partial
f\left( X\right) }{\partial S^{T}\left( X,\theta -1\right) }\left\{
h_{1}^{B}\left( X\right) \left( \left\langle \bar{h}\left( X\right)
\right\rangle +\left\langle \hat{h}_{1}^{B}\left( X\right) \right\rangle
\right) +Z\right\} \frac{\partial ^{2}f\left( X\right) }{\partial \left(
S^{T}\left( X,\theta -2\right) \right) ^{2}}
\end{eqnarray*}%
\begin{eqnarray*}
&&\frac{\delta }{\delta S^{T}\left( X,,\theta -2\right) }d \\
&\simeq &\left( \frac{\frac{\hat{h}\left( X\right) }{2}}{1+\left( \hat{h}%
\left( X\right) \left( \frac{f\left( X\right) +\bar{r}\left( X\right) }{2}-%
\frac{\left\langle \hat{f}\left( X^{\prime }\right) \right\rangle _{\hat{h}%
_{1}}+\left\langle \hat{r}\left( X^{\prime }\right) \right\rangle _{\hat{h}%
_{2}}}{2}\right) \right) }\right) \frac{\partial f\left( X\right) }{\partial
S^{T}\left( X,\theta -2\right) }d \\
&&+\left( h_{1}^{B}\left( X\right) \left( \left\langle \bar{h}\left(
X\right) \right\rangle +\left\langle \hat{h}_{1}^{B}\left( X\right)
\right\rangle \right) \right) \left( \frac{\partial f\left( X\right) }{%
\partial S^{T}\left( X,\theta -2\right) }\right) ^{2}+\frac{\frac{\partial
^{2}f\left( X\right) }{\partial \left( S^{T}\left( X,\theta -2\right)
\right) ^{2}}}{\frac{\partial f\left( X\right) }{\partial S^{T}\left(
X,\theta -2\right) }}d \\
&\rightarrow &\left( \frac{\frac{\hat{h}\left( X\right) }{2}d}{1+\left( \hat{%
h}\left( X\right) \left( \frac{f\left( X\right) +\bar{r}\left( X\right) }{2}-%
\frac{\left\langle \hat{f}\left( X^{\prime }\right) \right\rangle _{\hat{h}%
_{1}}+\left\langle \hat{r}\left( X^{\prime }\right) \right\rangle _{\hat{h}%
_{2}}}{2}\right) \right) }+\frac{\frac{\partial f\left( X\right) }{\partial
S^{T}\left( X,\theta -2\right) }h_{1}^{B}\left( X\right) \left( \left\langle 
\bar{h}\left( X\right) \right\rangle +\left\langle \hat{h}_{1}^{B}\left(
X\right) \right\rangle \right) }{S_{1}^{B}\left( X,\theta -1\right)
+S_{2}^{B}\left( X,\theta -1\right) }\right) \\
&&\times \frac{\partial f\left( X\right) }{\partial S^{T}\left( X,\theta
-2\right) }+\frac{\frac{\partial ^{2}f\left( X\right) }{\partial \left(
S^{T}\left( X,\theta -2\right) \right) ^{2}}}{\frac{\partial f\left(
X\right) }{\partial S^{T}\left( X,\theta -2\right) }}d
\end{eqnarray*}%
\begin{equation*}
\frac{\delta }{\delta S^{T}\left( X,,\theta -2\right) }e\simeq 0
\end{equation*}%
\begin{eqnarray*}
\frac{\delta }{\delta S^{T}\left( X,,\theta -2\right) }e_{0} &\simeq &\frac{%
\delta }{\delta S^{T}\left( X,,\theta -2\right) }S_{E}\left( X,X,\theta
-1\right) \\
&\simeq &\frac{\delta }{\delta f\left( X\right) }S_{E}\left( X,X,\theta
-1\right) \frac{\partial f\left( X\right) }{\partial S^{T}\left( X,\theta
-2\right) } \\
&\simeq &h_{E}\left( X\right) \frac{\partial f\left( X\right) }{\partial
S^{T}\left( X,\theta -2\right) }
\end{eqnarray*}%
\bigskip \bigskip \bigskip 
\begin{eqnarray*}
&&\frac{\delta }{\delta S^{T}\left( X,,\theta -2\right) }f \\
&\simeq &S_{1}\left( X,X,\theta -1\right) \left( \frac{\partial f\left(
X\right) }{\partial S^{T}\left( X,\theta -1\right) }\right) ^{3}\frac{\frac{%
\hat{h}\left( X\right) }{2}}{1+\left( \hat{h}\left( X\right) \left( \frac{%
f\left( X\right) +\bar{r}\left( X\right) }{2}-\frac{\left\langle \hat{f}%
\left( X^{\prime }\right) \right\rangle _{\hat{h}_{1}}+\left\langle \hat{r}%
\left( X^{\prime }\right) \right\rangle _{\hat{h}_{2}}}{2}\right) \right) }
\\
&&\times \left( h_{1}^{B}\left( X\right) \left( \left\langle \bar{h}\left(
X\right) \right\rangle +\left\langle \hat{h}_{1}^{B}\left( X\right)
\right\rangle \right) +ZS\left( X\right) \right) \\
&&+2S_{1}\left( X,X,\theta -1\right) \frac{\partial f\left( X\right) }{%
\partial S^{T}\left( X,\theta -2\right) }\frac{\partial ^{2}f\left( X\right) 
}{\partial \left( S^{T}\left( X,\theta -2\right) \right) ^{2}}\left(
h_{1}^{B}\left( X\right) \left( \left\langle \bar{h}\left( X\right)
\right\rangle +\left\langle \hat{h}_{1}^{B}\left( X\right) \right\rangle
\right) +Z\right)
\end{eqnarray*}%
\bigskip

\begin{equation*}
\frac{\delta }{\delta \hat{f}\left( X,\theta -1\right) }d\simeq 0
\end{equation*}%
\begin{equation*}
\frac{\delta }{\delta \bar{f}\left( X,\theta -1\right) }h\simeq 0
\end{equation*}%
\begin{equation*}
\frac{\delta }{\delta \hat{f}\left( X,\theta -1\right) }h\simeq 0
\end{equation*}%
\begin{equation*}
\frac{\delta }{\delta S^{T}\left( X,,\theta -2\right) }h\simeq \frac{%
\partial f\left( X,\theta -1\right) }{\partial S^{T}\left( X,,\theta
-2\right) }h\left( X\right)
\end{equation*}%
\bigskip 
\begin{equation*}
\frac{\delta }{\delta \bar{f}\left( X,\theta -1\right) }g\simeq 0
\end{equation*}%
\begin{equation*}
\frac{\delta }{\delta \hat{f}\left( X,\theta -1\right) }g\simeq 0
\end{equation*}%
\begin{equation*}
\frac{\delta }{\delta S^{T}\left( X,,\theta -2\right) }g\simeq 0
\end{equation*}%
\begin{equation*}
\frac{\delta }{\delta \bar{f}\left( X,\theta -1\right) }i\simeq 0
\end{equation*}%
\begin{equation*}
\frac{\delta }{\delta \hat{f}\left( X,\theta -1\right) }i\simeq 0
\end{equation*}%
\begin{equation*}
\frac{\delta }{\delta S^{T}\left( X,,\theta -2\right) }i\simeq S\left(
X\right) \left( Z^{2}S\left( X\right) \right) \left[ \frac{\partial f\left(
X\right) }{\partial S^{T}\left( X,\theta -2\right) }\right] ^{2}+i\frac{%
\frac{\partial ^{2}f\left( X\right) }{\partial \left( S^{T}\left( X,\theta
-2\right) \right) ^{2}}}{\frac{\partial f\left( X\right) }{\partial
S^{T}\left( X,\theta -2\right) }}
\end{equation*}%
Given that the derivatives of coefficients with respect to $\bar{K}%
_{X}\left\vert \bar{\Psi}\left( X\right) \right\vert ^{2}$, $\hat{K}%
_{X}\left\vert \hat{\Psi}\left( X\right) \right\vert ^{2}$ and $%
K_{X}\left\vert \Psi \left( X\right) \right\vert ^{2}$ are approximatively
the coefficients divided by $\bar{K}_{X}\left\vert \bar{\Psi}\left( X\right)
\right\vert ^{2}$, $\hat{K}_{X}\left\vert \hat{\Psi}\left( X\right)
\right\vert ^{2}$ and $K_{X}\left\vert \Psi \left( X\right) \right\vert ^{2}$
we can neglect the derivatives 
\begin{equation*}
\frac{\partial _{\bar{f}\left( X\right) }\bar{K}_{X}\left\vert \bar{\Psi}%
\left( X\right) \right\vert ^{2}}{\bar{K}_{X}\left\vert \bar{\Psi}\left(
X\right) \right\vert ^{2}}\simeq -\frac{2}{\left( 1-\left\langle \bar{S}%
\right\rangle \right) \bar{f}\left( X\right) +\left\langle \bar{S}\left(
X^{\prime },X\right) \right\rangle _{X^{\prime }}\left\langle \bar{f}%
\right\rangle }
\end{equation*}%
\begin{eqnarray*}
\frac{\partial _{\hat{f}\left( X\right) }\hat{K}_{X}\left\vert \hat{\Psi}%
\left( X\right) \right\vert ^{2}}{\hat{K}_{X}\left\vert \hat{\Psi}\left(
X\right) \right\vert ^{2}} &\simeq &-\frac{2}{\hat{f}\left( X\right) +\frac{%
\left( \left\langle \hat{S}_{E}^{B}\left( X,X^{\prime }\right) \right\rangle
_{X^{\prime }}+\left\langle \hat{S}_{L}^{B}\left( X,X^{\prime }\right)
\right\rangle _{X^{\prime }}\right) \frac{\left\langle \bar{K}\right\rangle
\left\Vert \bar{\Psi}\right\Vert ^{2}}{\left\langle \hat{K}\right\rangle
\left\Vert \hat{\Psi}\right\Vert ^{2}}\left\langle \bar{f}\right\rangle }{%
1-\left\langle \bar{S}\right\rangle }} \\
\frac{\partial _{\bar{f}\left( X\right) }\hat{K}_{X}\left\vert \hat{\Psi}%
\left( X\right) \right\vert ^{2}}{\hat{K}_{X}\left\vert \hat{\Psi}\left(
X\right) \right\vert ^{2}} &\simeq &-\frac{2\frac{\left( \left\langle \hat{S}%
_{E}^{B}\left( X,X^{\prime }\right) \right\rangle _{X^{\prime
}}+\left\langle \hat{S}_{L}^{B}\left( X,X^{\prime }\right) \right\rangle
_{X^{\prime }}\right) \frac{\left\langle \bar{K}\right\rangle \left\Vert 
\bar{\Psi}\right\Vert ^{2}}{\left\langle \hat{K}\right\rangle \left\Vert 
\hat{\Psi}\right\Vert ^{2}}\left\langle \bar{f}\right\rangle }{%
1-\left\langle \bar{S}\right\rangle }}{\hat{f}\left( X\right) +\frac{\left(
\left\langle \hat{S}_{E}^{B}\left( X,X^{\prime }\right) \right\rangle
_{X^{\prime }}+\left\langle \hat{S}_{L}^{B}\left( X,X^{\prime }\right)
\right\rangle _{X^{\prime }}\right) \frac{\left\langle \bar{K}\right\rangle
\left\Vert \bar{\Psi}\right\Vert ^{2}}{\left\langle \hat{K}\right\rangle
\left\Vert \hat{\Psi}\right\Vert ^{2}}\left\langle \bar{f}\right\rangle }{%
1-\left\langle \bar{S}\right\rangle }}<<1
\end{eqnarray*}%
\begin{equation*}
\frac{\partial _{f\left( X\right) }K_{X}\left\vert \Psi \left( X\right)
\right\vert ^{2}}{K_{X}\left\vert \Psi \left( X\right) \right\vert ^{2}}%
\simeq \frac{2}{r}\frac{1}{f\left( X\right) +C_{0}}
\end{equation*}%
As a consequence, the block involving disposable capital can be wrtten as:%
\begin{eqnarray*}
&&\left( 
\begin{array}{ccc}
\frac{\partial _{\bar{f}\left( X\right) }\bar{K}_{X}\left\vert \bar{\Psi}%
\left( X\right) \right\vert ^{2}}{\bar{K}_{X}\left\vert \bar{\Psi}\left(
X\right) \right\vert ^{2}} & 0 & 0 \\ 
\frac{\partial _{\bar{f}\left( X\right) }\hat{K}_{X}\left\vert \hat{\Psi}%
\left( X\right) \right\vert ^{2}}{\hat{K}_{X}\left\vert \hat{\Psi}\left(
X\right) \right\vert ^{2}} & \frac{\partial _{\hat{f}\left( X\right) }\hat{K}%
_{X}\left\vert \hat{\Psi}\left( X\right) \right\vert ^{2}}{\hat{K}%
_{X}\left\vert \hat{\Psi}\left( X\right) \right\vert ^{2}} & 0 \\ 
0 & 0 & \frac{\partial _{f\left( X\right) }K_{X}\left\vert \Psi \left(
X\right) \right\vert ^{2}}{K_{X}\left\vert \Psi \left( X\right) \right\vert
^{2}}\frac{\partial f\left( X\right) }{\partial S^{T}\left( X,\theta
-2\right) }%
\end{array}%
\right) \\
&\simeq &\left( 
\begin{array}{ccc}
-\frac{2}{\left( 1-\left\langle \bar{S}\right\rangle \right) \bar{f}\left(
X\right) +\left\langle \bar{S}\left( X^{\prime },X\right) \right\rangle
_{X^{\prime }}\left\langle \bar{f}\right\rangle } & 0 & 0 \\ 
0 & -\frac{2}{\hat{f}\left( X\right) +\frac{\left( \left\langle \hat{S}%
_{E}^{B}\left( X,X^{\prime }\right) \right\rangle _{X^{\prime
}}+\left\langle \hat{S}_{L}^{B}\left( X,X^{\prime }\right) \right\rangle
_{X^{\prime }}\right) \frac{\left\langle \bar{K}\right\rangle \left\Vert 
\bar{\Psi}\right\Vert ^{2}}{\left\langle \hat{K}\right\rangle \left\Vert 
\hat{\Psi}\right\Vert ^{2}}\left\langle \bar{f}\right\rangle }{%
1-\left\langle \bar{S}\right\rangle }} & 0 \\ 
0 & 0 & -\frac{2}{r}\frac{1}{f\left( X\right) +C_{0}}\frac{\partial f\left(
X\right) }{\partial S^{T}\left( X,\theta -2\right) }%
\end{array}%
\right)
\end{eqnarray*}

\paragraph*{A5.1.3.2 Estimation of the quadratic matrices}

The matrices involved in the quadratic term are:

\begin{eqnarray*}
\bar{M} &=&\left[ 
\begin{array}{c}
\begin{array}{cccccc}
\frac{\delta }{\delta \bar{f}\left( X,\theta -1\right) }a & 0 & \frac{\bar{h}%
\left( X^{\prime },X\right) }{2}\frac{\delta }{\delta \bar{f}\left( X,\theta
-1\right) }c & \frac{\delta \left( a_{0}+a_{1}\right) }{\delta \bar{f}\left(
X,\theta -1\right) } & 0 & -\frac{\delta a_{0}}{\delta \bar{f}\left(
X,\theta -1\right) } \\ 
0 & 0 & \frac{\delta }{\delta \hat{f}\left( X,\theta -1\right) }c & 0 & 0 & 0
\\ 
0 & 0 & \frac{\delta }{\delta S^{T}\left( X,,\theta -2\right) }c & 0 & 0 & 0%
\end{array}
\\ 
0\left[ 3\times 6\right]%
\end{array}%
\right] \\
,\hat{M} &=&\left( 
\begin{array}{c}
\begin{array}{cccccc}
0 & 0 & 0 & \frac{\delta }{\delta \bar{f}\left( X,\theta -1\right) }d & 0 & -%
\frac{\delta }{\delta \bar{f}\left( X,\theta -1\right) }d \\ 
0 & 0 & \frac{\delta }{\delta \hat{f}\left( X,\theta -1\right) }f & 0 & 
\frac{\delta }{\delta \hat{f}\left( X,\theta -1\right) }e_{0} & 0 \\ 
0 & 0 & \frac{\delta }{\delta S^{T}\left( X,,\theta -2\right) }f & \frac{%
\delta }{\delta S^{T}\left( X,,\theta -2\right) }d & \frac{\delta }{\delta
S^{T}\left( X,,\theta -2\right) }e_{0} & -\frac{\delta }{\delta S^{T}\left(
X,,\theta -2\right) }d%
\end{array}
\\ 
0\left[ 3\times 6\right]%
\end{array}%
\right) \\
M_{3,3}^{T} &=&\frac{\delta }{\delta S^{T}\left( X,,\theta -2\right) }i\text{%
, }M_{3,5}^{T}=\frac{\partial }{\partial S^{T}\left( X,,\theta -2\right) }h%
\text{ ,}M_{i,j}^{T}=0\text{ otherwise} \\
M_{\bar{K},1,1} &=&\partial _{\bar{f}\left( X\right) }\left( \frac{\partial
_{\bar{f}\left( X\right) }\bar{K}_{X}\left\vert \bar{\Psi}\left( X\right)
\right\vert ^{2}}{\bar{K}_{X}\left\vert \bar{\Psi}\left( X\right)
\right\vert ^{2}}\right) \text{, }M_{\bar{K},i,j}=0\text{ otherwise} \\
M_{\hat{K},2,2} &=&\partial _{\hat{f}\left( X\right) }\left( \frac{\partial
_{\hat{f}\left( X\right) }\hat{K}_{X}\left\vert \hat{\Psi}\left( X\right)
\right\vert ^{2}}{\hat{K}_{X}\left\vert \hat{\Psi}\left( X\right)
\right\vert ^{2}}\right) \text{, }M_{\hat{K},i,j}=0\text{ otherwise}
\end{eqnarray*}%
with:%
\begin{equation*}
\partial _{\bar{f}\left( X\right) }\left( \frac{\partial _{\bar{f}\left(
X\right) }\bar{K}_{X}\left\vert \bar{\Psi}\left( X\right) \right\vert ^{2}}{%
\bar{K}_{X}\left\vert \bar{\Psi}\left( X\right) \right\vert ^{2}}\right)
\simeq \frac{6}{\left( \bar{f}\left( X\right) \right) ^{4}}
\end{equation*}%
\begin{equation*}
\partial _{\hat{f}\left( X\right) }\left( \frac{\partial _{\hat{f}\left(
X\right) }\hat{K}_{X}\left\vert \hat{\Psi}\left( X\right) \right\vert ^{2}}{%
\hat{K}_{X}\left\vert \hat{\Psi}\left( X\right) \right\vert ^{2}}\right)
\simeq \frac{6}{\left( \hat{f}\left( X\right) \right) ^{4}}
\end{equation*}%
\begin{eqnarray*}
M_{K,3,3} &=&\partial _{f\left( X\right) }\left( \frac{\partial _{f\left(
X\right) }K_{X}\left\vert \Psi \left( X\right) \right\vert ^{2}}{%
K_{X}\left\vert \Psi \left( X\right) \right\vert ^{2}}\right) \left( \frac{%
\partial f\left( X\right) }{\partial S^{T}\left( X,\theta -2\right) }\right)
^{2} \\
&&+\frac{\partial _{f\left( X\right) }K_{X}\left\vert \Psi \left( X\right)
\right\vert ^{2}}{K_{X}\left\vert \Psi \left( X\right) \right\vert ^{2}}%
\frac{\partial ^{2}f\left( X\right) }{\partial \left( S^{T}\left( X,\theta
-2\right) \right) ^{2}} \\
&\simeq &-\frac{2}{r}\frac{1}{\left( f\left( X\right) +C_{0}\right) ^{2}}%
\left( \frac{\partial f\left( X\right) }{\partial S^{T}\left( X,\theta
-2\right) }\right) ^{3}
\end{eqnarray*}%
and $M_{K,i,j}=0$ otherwise.

\subsubsection{A5.1.4 Stability condition}

Fo relatively large fluctuations, the stability condition becomes:%
\begin{equation}
\frac{\left( \mathbf{\delta X}^{t}\bar{M}\mathbf{\delta X}\right)
^{2}+\left( \mathbf{\delta X}^{t}\hat{M}\mathbf{\delta X}\right) ^{2}+\left( 
\mathbf{\delta X}^{t}M^{T}\mathbf{\delta X}\right) ^{2}+\left( \mathbf{%
\delta X}^{t}M_{\bar{K}}\mathbf{\delta X}\right) ^{2}+\left( \mathbf{\delta X%
}^{t}M_{\hat{K}}\mathbf{\delta X}\right) ^{2}+\left( \mathbf{\delta X}%
^{t}M_{K}\mathbf{\delta X}\right) ^{2}}{\left( \delta \bar{S}\left( X\right)
\right) ^{2}+\left( \delta \hat{S}\left( X\right) \right) ^{2}+\left( \delta
S^{T}\left( X\right) \right) ^{2}+\left( \delta \bar{K}_{X}\left\vert \bar{%
\Psi}\left( X\right) \right\vert ^{2}\right) ^{2}+\left( \delta \hat{K}%
_{X}\left\vert \hat{\Psi}\left( X\right) \right\vert ^{2}\right) ^{2}+\left(
\delta K_{X}\left\vert \Psi \left( X\right) \right\vert ^{2}\right) ^{2}}%
<1-\left\langle \lambda ^{2}\right\rangle
\end{equation}%
\begin{eqnarray*}
&&\left( 1-\left\langle \lambda ^{2}\right\rangle \right) \left( \left(
\delta \bar{S}\left( X\right) \right) ^{2}+\left( \delta \hat{S}\left(
X\right) \right) ^{2}+\left( \delta S^{T}\left( X\right) \right) ^{2}+\left(
\delta \bar{K}_{X}\left\vert \bar{\Psi}\left( X\right) \right\vert
^{2}\right) ^{2}+\left( \delta \hat{K}_{X}\left\vert \hat{\Psi}\left(
X\right) \right\vert ^{2}\right) ^{2}+\left( \delta K_{X}\left\vert \Psi
\left( X\right) \right\vert ^{2}\right) ^{2}\right) \\
&>&\left\{ \frac{\left( \delta \bar{S}\left( X\right) \right) ^{2}\delta a}{%
\delta \bar{f}\left( X,\theta -1\right) }+\delta \bar{S}\left( X\right)
\left( \frac{\bar{h}\left( X^{\prime },X\right) }{2}\frac{\delta S^{T}\left(
X\right) \delta c}{\delta \bar{f}\left( X,\theta -1\right) }+\frac{\delta 
\bar{K}_{X}\left\vert \bar{\Psi}\left( X\right) \right\vert ^{2}\delta a_{1}%
}{\delta \bar{f}\left( X,\theta -1\right) }+\frac{\delta a_{0}}{\delta \bar{f%
}\left( X,\theta -1\right) }\frac{\delta \frac{\bar{K}_{X}\left\vert \bar{%
\Psi}\left( X\right) \right\vert ^{2}}{K_{X}\left\vert \Psi \left( X\right)
\right\vert ^{2}}}{\frac{\bar{K}_{X}\left\vert \bar{\Psi}\left( X\right)
\right\vert ^{2}}{K_{X}\left\vert \Psi \left( X\right) \right\vert ^{2}}}%
\right) \right. \\
&&+\left. \left( \delta \bar{S}\left( X\right) \frac{\delta }{\delta \hat{f}%
\left( X,\theta -1\right) }c+\delta \hat{S}\left( X\right) \frac{\delta }{%
\delta S^{T}\left( X,,\theta -2\right) }c\right) \delta S^{T}\left( X\right)
\right\} ^{2} \\
&&+\left\{ \delta \bar{S}\left( X\right) \frac{\delta d}{\delta \bar{f}%
\left( X,\theta -1\right) }\frac{\delta \frac{\bar{K}_{X}\left\vert \bar{\Psi%
}\left( X\right) \right\vert ^{2}}{K_{X}\left\vert \Psi \left( X\right)
\right\vert ^{2}}}{\frac{\bar{K}_{X}\left\vert \bar{\Psi}\left( X\right)
\right\vert ^{2}}{K_{X}\left\vert \Psi \left( X\right) \right\vert ^{2}}}%
+\delta \hat{S}\left( X\right) \left( \frac{\delta f}{\delta \hat{f}\left(
X,\theta -1\right) }\delta \bar{S}\left( X\right) +\frac{\delta e_{0}}{%
\delta \hat{f}\left( X,\theta -1\right) }\delta \hat{K}_{X}\left\vert \hat{%
\Psi}\left( X\right) \right\vert ^{2}\right) \right. \\
&&+\left. \delta S^{T}\left( X\right) \left( \frac{\delta f}{\delta
S^{T}\left( X,,\theta -2\right) }\delta S^{T}\left( X\right) +\frac{\delta
e_{0}}{\delta S^{T}\left( X,,\theta -2\right) }\delta \hat{K}_{X}\left\vert 
\hat{\Psi}\left( X\right) \right\vert ^{2}+\frac{\delta d}{\delta
S^{T}\left( X,,\theta -2\right) }\frac{\delta \frac{\bar{K}_{X}\left\vert 
\bar{\Psi}\left( X\right) \right\vert ^{2}}{K_{X}\left\vert \Psi \left(
X\right) \right\vert ^{2}}}{\frac{\bar{K}_{X}\left\vert \bar{\Psi}\left(
X\right) \right\vert ^{2}}{K_{X}\left\vert \Psi \left( X\right) \right\vert
^{2}}}\right) \right\} ^{2} \\
&&+\left\{ \delta S^{T}\left( X\right) \left( \frac{\delta d}{\delta
S^{T}\left( X,,\theta -2\right) }\frac{\delta \frac{\bar{K}_{X}\left\vert 
\bar{\Psi}\left( X\right) \right\vert ^{2}}{K_{X}\left\vert \Psi \left(
X\right) \right\vert ^{2}}}{\frac{\bar{K}_{X}\left\vert \bar{\Psi}\left(
X\right) \right\vert ^{2}}{K_{X}\left\vert \Psi \left( X\right) \right\vert
^{2}}}+\frac{\delta f}{\delta S^{T}\left( X,,\theta -2\right) }\delta
S^{T}\left( X\right) +\frac{\delta e_{0}}{\delta S^{T}\left( X,,\theta
-2\right) }\delta \hat{K}_{X}\left\vert \hat{\Psi}\left( X\right)
\right\vert ^{2}\right) \right\} ^{2} \\
&&+\left\{ \frac{\delta }{\delta S^{T}\left( X,,\theta -2\right) }i\left(
\delta S^{T}\left( X\right) \right) ^{2}+\frac{\partial h}{\partial
S^{T}\left( X,,\theta -2\right) }\delta S^{T}\left( X\right) \delta \hat{K}%
_{X}\left\vert \hat{\Psi}\left( X\right) \right\vert ^{2}\right\} ^{2} \\
&&+\left\{ \partial _{\bar{f}\left( X\right) }\left( \frac{\partial _{\bar{f}%
\left( X\right) }\bar{K}_{X}\left\vert \bar{\Psi}\left( X\right) \right\vert
^{2}}{\bar{K}_{X}\left\vert \bar{\Psi}\left( X\right) \right\vert ^{2}}%
\right) \left( \delta \bar{S}\left( X\right) \right) ^{2}\right\}
^{2}+\left\{ \partial _{\hat{f}\left( X\right) }\left( \frac{\partial _{\hat{%
f}\left( X\right) }\hat{K}_{X}\left\vert \hat{\Psi}\left( X\right)
\right\vert ^{2}}{\hat{K}_{X}\left\vert \hat{\Psi}\left( X\right)
\right\vert ^{2}}\right) \left( \delta \hat{S}\left( X\right) \right)
^{2}\right\} ^{2} \\
&&+\left\{ \left( \partial _{f\left( X\right) }\left( \frac{\partial
_{f\left( X\right) }K_{X}\left\vert \Psi \left( X\right) \right\vert ^{2}}{%
K_{X}\left\vert \Psi \left( X\right) \right\vert ^{2}}\right) \left( \frac{%
\partial f\left( X\right) }{\partial S^{T}\left( X,\theta -2\right) }\right)
^{2}+\frac{\partial _{f\left( X\right) }K_{X}\left\vert \Psi \left( X\right)
\right\vert ^{2}}{K_{X}\left\vert \Psi \left( X\right) \right\vert ^{2}}%
\frac{\partial ^{2}f\left( X\right) }{\partial \left( S^{T}\left( X,\theta
-2\right) \right) ^{2}}\right) \left( \delta S^{T}\left( X\right) \right)
^{2}\right\} ^{2}
\end{eqnarray*}%
We assume as in Appendix 4 that $\frac{\partial f\left( X\right) }{\partial
S^{T}\left( X,\theta -2\right) }>>1$.

The coefficients to consider in first approximation are:

\begin{equation*}
\frac{\delta }{\delta \bar{f}\left( X,\theta -1\right) }a_{0}=\left( \left(
2S_{1}^{B}\left( X,\theta -1\right) +S_{2}^{B}\left( X,\theta -1\right)
\right) h_{1}^{B}\left( X\right) \left( \left\langle \bar{h}\left( X\right)
\right\rangle +\left\langle \hat{h}_{1}^{B}\left( X\right) \right\rangle
\right) \right) \frac{\partial f\left( X\right) }{\partial S^{T}\left(
X,\theta -2\right) }
\end{equation*}

\begin{equation*}
\frac{\delta }{\delta \bar{f}\left( X,\theta -1\right) }c\simeq
h_{1}^{B}\left( X\right) \left( \left\langle \bar{h}\left( X\right)
\right\rangle +\left\langle \hat{h}_{1}^{B}\left( X\right) \right\rangle
\right) \left( \frac{\partial f\left( X\right) }{\partial S^{T}\left(
X,\theta -2\right) }\right) ^{2}
\end{equation*}%
\begin{equation*}
\frac{\delta }{\delta S^{T}\left( X,,\theta -2\right) }e_{0}\simeq
h_{E}\left( X\right) \frac{\partial f\left( X\right) }{\partial S^{T}\left(
X,\theta -2\right) }
\end{equation*}%
\begin{eqnarray*}
&&\frac{\delta }{\delta S^{T}\left( X,,\theta -2\right) }c \\
&\simeq &\left( \frac{\partial f\left( X\right) }{\partial S^{T}\left(
X,\theta -1\right) }\right) ^{3} \\
&&\times \left\{ h_{1}^{B}\left( X\right) \left( \left\langle \bar{h}\left(
X\right) \right\rangle +\left\langle \hat{h}_{1}^{B}\left( X\right)
\right\rangle \right) \left( h_{1}^{B}\left( X\right) \left( \left\langle 
\bar{h}\left( X\right) \right\rangle +\left\langle \hat{h}_{1}^{B}\left(
X\right) \right\rangle \right) +Z\right) +Z^{2}S\left( X\right) \right\}
\end{eqnarray*}%
\begin{eqnarray*}
&&\frac{\delta }{\delta S^{T}\left( X,,\theta -2\right) }d \\
&\simeq &\frac{h_{1}^{B}\left( X\right) \left( \left\langle \bar{h}\left(
X\right) \right\rangle +\left\langle \hat{h}_{1}^{B}\left( X\right)
\right\rangle \right) }{S_{1}^{B}\left( X,\theta -1\right) +S_{2}^{B}\left(
X,\theta -1\right) }\left( \frac{\partial f\left( X\right) }{\partial
S^{T}\left( X,\theta -2\right) }\right) ^{2}
\end{eqnarray*}%
\bigskip 
\begin{eqnarray*}
&&\frac{\delta }{\delta S^{T}\left( X,,\theta -2\right) }f \\
&\simeq &S_{1}\left( X,X,\theta -1\right) \left( \frac{\partial f\left(
X\right) }{\partial S^{T}\left( X,\theta -1\right) }\right) ^{3}\frac{\frac{%
\hat{h}\left( X\right) }{2}\left( h_{1}^{B}\left( X\right) \left(
\left\langle \bar{h}\left( X\right) \right\rangle +\left\langle \hat{h}%
_{1}^{B}\left( X\right) \right\rangle \right) +ZS\left( X\right) \right) }{%
1+\left( \hat{h}\left( X\right) \left( \frac{f\left( X\right) +\bar{r}\left(
X\right) }{2}-\frac{\left\langle \hat{f}\left( X^{\prime }\right)
\right\rangle _{\hat{h}_{1}}+\left\langle \hat{r}\left( X^{\prime }\right)
\right\rangle _{\hat{h}_{2}}}{2}\right) \right) }
\end{eqnarray*}%
\begin{equation*}
\frac{\delta }{\delta S^{T}\left( X,,\theta -2\right) }h\simeq \frac{%
\partial f\left( X,\theta -1\right) }{\partial S^{T}\left( X,,\theta
-2\right) }h\left( X\right)
\end{equation*}%
\bigskip 
\begin{equation*}
\frac{\delta }{\delta S^{T}\left( X,,\theta -2\right) }i\simeq S\left(
X\right) \left( Z^{2}S\left( X\right) \right) \left[ \frac{\partial f\left(
X\right) }{\partial S^{T}\left( X,\theta -2\right) }\right] ^{2}+i\frac{%
\frac{\partial ^{2}f\left( X\right) }{\partial \left( S^{T}\left( X,\theta
-2\right) \right) ^{2}}}{\frac{\partial f\left( X\right) }{\partial
S^{T}\left( X,\theta -2\right) }}
\end{equation*}%
The dominant contributions involving the fluctuations in $\bar{K}%
_{X}\left\vert \bar{\Psi}\left( X\right) \right\vert ^{2}$, $\hat{K}%
_{X}\left\vert \hat{\Psi}\left( X\right) \right\vert ^{2}$ \bigskip\ and $%
K_{X}\left\vert \Psi \left( X\right) \right\vert ^{2}$ are:%
\begin{eqnarray*}
&&\left\{ \frac{\left( \delta \bar{S}\left( X\right) \right) ^{2}\delta a}{%
\delta \bar{f}\left( X,\theta -1\right) }+\delta \bar{S}\left( X\right)
\left( \frac{\bar{h}\left( X^{\prime },X\right) }{2}\frac{\delta S^{T}\left(
X\right) \delta c}{\delta \bar{f}\left( X,\theta -1\right) }+\frac{\delta 
\bar{K}_{X}\left\vert \bar{\Psi}\left( X\right) \right\vert ^{2}\delta a_{1}%
}{\delta \bar{f}\left( X,\theta -1\right) }+\frac{\delta a_{0}}{\delta \bar{f%
}\left( X,\theta -1\right) }\frac{\delta \frac{\bar{K}_{X}\left\vert \bar{%
\Psi}\left( X\right) \right\vert ^{2}}{K_{X}\left\vert \Psi \left( X\right)
\right\vert ^{2}}}{\frac{\bar{K}_{X}\left\vert \bar{\Psi}\left( X\right)
\right\vert ^{2}}{K_{X}\left\vert \Psi \left( X\right) \right\vert ^{2}}}%
\right) \right. \\
&&+\left. \left( \delta \bar{S}\left( X\right) \frac{\delta }{\delta \hat{f}%
\left( X,\theta -1\right) }c+\delta \hat{S}\left( X\right) \frac{\delta }{%
\delta S^{T}\left( X,,\theta -2\right) }c\right) \delta S^{T}\left( X\right)
\right\} ^{2} \\
&\rightarrow &\left\{ \delta \bar{S}\left( X\right) h_{1}^{B}\left( X\right)
\left( \left\langle \bar{h}\left( X\right) \right\rangle +\left\langle \hat{h%
}_{1}^{B}\left( X\right) \right\rangle \right) \left( \left\{ \left(
h_{1}^{B}\left( X\right) \left( \left\langle \bar{h}\left( X\right)
\right\rangle +\left\langle \hat{h}_{1}^{B}\left( X\right) \right\rangle
\right) +Z\right) +Z^{2}S\left( X\right) \right\} \right. \right. \\
&&\left. \left. \times \left( \frac{\partial f\left( X\right) }{\partial
S^{T}\left( X,\theta -1\right) }\right) ^{3}\delta S^{T}\left( X\right)
+\left( 2S_{1}^{B}\left( X,\theta -1\right) +S_{2}^{B}\left( X,\theta
-1\right) \right) \frac{\partial f\left( X\right) }{\partial S^{T}\left(
X,\theta -2\right) }\frac{\delta \frac{\bar{K}_{X}\left\vert \bar{\Psi}%
\left( X\right) \right\vert ^{2}}{K_{X}\left\vert \Psi \left( X\right)
\right\vert ^{2}}}{\frac{\bar{K}_{X}\left\vert \bar{\Psi}\left( X\right)
\right\vert ^{2}}{K_{X}\left\vert \Psi \left( X\right) \right\vert ^{2}}}%
\right) \right\} ^{2} \\
&\rightarrow &\left( h_{1}^{B}\left( X\right) \left( \left\langle \bar{h}%
\left( X\right) \right\rangle +\left\langle \hat{h}_{1}^{B}\left( X\right)
\right\rangle \right) \left( \left( h_{1}^{B}\left( X\right) \left(
\left\langle \bar{h}\left( X\right) \right\rangle +\left\langle \hat{h}%
_{1}^{B}\left( X\right) \right\rangle \right) +Z\right) +Z^{2}S\left(
X\right) \right) \right) ^{2} \\
&&\times \left( \delta \bar{S}\left( X\right) \left( \frac{\partial f\left(
X\right) }{\partial S^{T}\left( X,\theta -1\right) }\right) ^{3}\delta
S^{T}\left( X\right) \right) ^{2}
\end{eqnarray*}%
\begin{eqnarray*}
&&\left\{ \delta \bar{S}\left( X\right) \frac{\delta d}{\delta \bar{f}\left(
X,\theta -1\right) }\frac{\delta \frac{\bar{K}_{X}\left\vert \bar{\Psi}%
\left( X\right) \right\vert ^{2}}{K_{X}\left\vert \Psi \left( X\right)
\right\vert ^{2}}}{\frac{\bar{K}_{X}\left\vert \bar{\Psi}\left( X\right)
\right\vert ^{2}}{K_{X}\left\vert \Psi \left( X\right) \right\vert ^{2}}}%
+\delta \hat{S}\left( X\right) \left( \frac{\delta f}{\delta \hat{f}\left(
X,\theta -1\right) }\delta \bar{S}\left( X\right) +\frac{\delta e_{0}}{%
\delta \hat{f}\left( X,\theta -1\right) }\delta \hat{K}_{X}\left\vert \hat{%
\Psi}\left( X\right) \right\vert ^{2}\right) \right. \\
&&+\left. \delta S^{T}\left( X\right) \left( \frac{\delta f}{\delta
S^{T}\left( X,,\theta -2\right) }\delta S^{T}\left( X\right) +\frac{\delta
e_{0}}{\delta S^{T}\left( X,,\theta -2\right) }\delta \hat{K}_{X}\left\vert 
\hat{\Psi}\left( X\right) \right\vert ^{2}+\frac{\delta d}{\delta
S^{T}\left( X,,\theta -2\right) }\frac{\delta \frac{\bar{K}_{X}\left\vert 
\bar{\Psi}\left( X\right) \right\vert ^{2}}{K_{X}\left\vert \Psi \left(
X\right) \right\vert ^{2}}}{\frac{\bar{K}_{X}\left\vert \bar{\Psi}\left(
X\right) \right\vert ^{2}}{K_{X}\left\vert \Psi \left( X\right) \right\vert
^{2}}}\right) \right\} ^{2} \\
&\rightarrow &\left\{ \delta S^{T}\left( X\right) \left( S_{1}\left(
X,X,\theta -1\right) \left( \frac{\partial f\left( X\right) }{\partial
S^{T}\left( X,\theta -1\right) }\right) ^{3}\frac{\frac{\hat{h}\left(
X\right) }{2}\left( h_{1}^{B}\left( X\right) \left( \left\langle \bar{h}%
\left( X\right) \right\rangle +\left\langle \hat{h}_{1}^{B}\left( X\right)
\right\rangle \right) +ZS\left( X\right) \right) }{1+\left( \hat{h}\left(
X\right) \left( \frac{f\left( X\right) +\bar{r}\left( X\right) }{2}-\frac{%
\left\langle \hat{f}\left( X^{\prime }\right) \right\rangle _{\hat{h}%
_{1}}+\left\langle \hat{r}\left( X^{\prime }\right) \right\rangle _{\hat{h}%
_{2}}}{2}\right) \right) }\delta S^{T}\left( X\right) \right. \right. \\
&&\left. \left. +\frac{h_{1}^{B}\left( X\right) \left( \left\langle \bar{h}%
\left( X\right) \right\rangle +\left\langle \hat{h}_{1}^{B}\left( X\right)
\right\rangle \right) }{S_{1}^{B}\left( X,\theta -1\right) +S_{2}^{B}\left(
X,\theta -1\right) }\left( \frac{\partial f\left( X\right) }{\partial
S^{T}\left( X,\theta -2\right) }\right) ^{2}\frac{\delta \frac{\bar{K}%
_{X}\left\vert \bar{\Psi}\left( X\right) \right\vert ^{2}}{K_{X}\left\vert
\Psi \left( X\right) \right\vert ^{2}}}{\frac{\bar{K}_{X}\left\vert \bar{\Psi%
}\left( X\right) \right\vert ^{2}}{K_{X}\left\vert \Psi \left( X\right)
\right\vert ^{2}}}\right) \right\} ^{2}
\end{eqnarray*}%
\begin{equation*}
\rightarrow \left( \delta S^{T}\left( X\right) S_{1}\left( X,X,\theta
-1\right) \left( \frac{\partial f\left( X\right) }{\partial S^{T}\left(
X,\theta -1\right) }\right) ^{3}\frac{\frac{\hat{h}\left( X\right) }{2}%
\left( h_{1}^{B}\left( X\right) \left( \left\langle \bar{h}\left( X\right)
\right\rangle +\left\langle \hat{h}_{1}^{B}\left( X\right) \right\rangle
\right) +ZS\left( X\right) \right) }{1+\left( \hat{h}\left( X\right) \left( 
\frac{f\left( X\right) +\bar{r}\left( X\right) }{2}-\frac{\left\langle \hat{f%
}\left( X^{\prime }\right) \right\rangle _{\hat{h}_{1}}+\left\langle \hat{r}%
\left( X^{\prime }\right) \right\rangle _{\hat{h}_{2}}}{2}\right) \right) }%
\delta S^{T}\left( X\right) \right) ^{2}
\end{equation*}%
\begin{eqnarray*}
&&\left\{ \frac{\delta }{\delta S^{T}\left( X,,\theta -2\right) }i\left(
\delta S^{T}\left( X\right) \right) ^{2}+\frac{\partial h}{\partial
S^{T}\left( X,,\theta -2\right) }\delta S^{T}\left( X\right) \delta \hat{K}%
_{X}\left\vert \hat{\Psi}\left( X\right) \right\vert ^{2}\right\} ^{2} \\
&\rightarrow &\left\{ S\left( X\right) \left( Z^{2}S\left( X\right) \right) 
\left[ \frac{\partial f\left( X\right) }{\partial S^{T}\left( X,\theta
-2\right) }\right] ^{2}\left( \delta S^{T}\left( X\right) \right)
^{2}\right\} ^{2}
\end{eqnarray*}%
The four last terms can be replaced by:%
\begin{equation*}
\left( \delta S^{T}\left( X\right) \frac{\frac{2}{r}}{\left( f\left(
X\right) +C_{0}\right) ^{2}}\left( \frac{\partial f\left( X\right) }{%
\partial S^{T}\left( X,\theta -2\right) }\right) ^{3}\delta S^{T}\left(
X\right) \right) ^{2}
\end{equation*}%
and we find the condition:%
\begin{eqnarray*}
&&\left( 1-\left\langle \lambda ^{2}\right\rangle \right) \left( \left(
\delta \bar{S}\left( X\right) \right) ^{2}+\left( \delta \hat{S}\left(
X\right) \right) ^{2}+\left( \delta S^{T}\left( X\right) \right) ^{2}+\left(
\delta \bar{K}_{X}\left\vert \bar{\Psi}\left( X\right) \right\vert
^{2}\right) ^{2}+\left( \delta \hat{K}_{X}\left\vert \hat{\Psi}\left(
X\right) \right\vert ^{2}\right) ^{2}+\left( \delta K_{X}\left\vert \Psi
\left( X\right) \right\vert ^{2}\right) ^{2}\right) \\
&>&A\left( \delta \bar{S}\left( X\right) \delta S^{T}\left( X\right) \right)
^{2}+\left( B_{1}\left( \delta S^{T}\left( X\right) \right) ^{2}+B_{2}\delta
S^{T}\left( X\right) \frac{\delta \frac{\bar{K}_{X}\left\vert \bar{\Psi}%
\left( X\right) \right\vert ^{2}}{K_{X}\left\vert \Psi \left( X\right)
\right\vert ^{2}}}{\frac{\bar{K}_{X}\left\vert \bar{\Psi}\left( X\right)
\right\vert ^{2}}{K_{X}\left\vert \Psi \left( X\right) \right\vert ^{2}}}%
\right) ^{2}+C\left( \left( \delta S^{T}\left( X\right) \right) ^{2}\right)
^{2}
\end{eqnarray*}%
wth:%
\begin{eqnarray*}
A &=&\left( h_{1}^{B}\left( X\right) \left( \left\langle \bar{h}\left(
X\right) \right\rangle +\left\langle \hat{h}_{1}^{B}\left( X\right)
\right\rangle \right) \left( \left( h_{1}^{B}\left( X\right) \left(
\left\langle \bar{h}\left( X\right) \right\rangle +\left\langle \hat{h}%
_{1}^{B}\left( X\right) \right\rangle \right) +Z\right) +Z^{2}S\left(
X\right) \right) \right) ^{2} \\
&&\times \left( \frac{\partial f\left( X\right) }{\partial S^{T}\left(
X,\theta -1\right) }\right) ^{6}
\end{eqnarray*}%
\begin{equation*}
B_{1}=S_{1}\left( X,X,\theta -1\right) \left( \frac{\partial f\left(
X\right) }{\partial S^{T}\left( X,\theta -1\right) }\right) ^{3}\frac{\frac{%
\hat{h}\left( X\right) }{2}\left( h_{1}^{B}\left( X\right) \left(
\left\langle \bar{h}\left( X\right) \right\rangle +\left\langle \hat{h}%
_{1}^{B}\left( X\right) \right\rangle \right) +ZS\left( X\right) \right) }{%
1+\left( \hat{h}\left( X\right) \left( \frac{f\left( X\right) +\bar{r}\left(
X\right) }{2}-\frac{\left\langle \hat{f}\left( X^{\prime }\right)
\right\rangle _{\hat{h}_{1}}+\left\langle \hat{r}\left( X^{\prime }\right)
\right\rangle _{\hat{h}_{2}}}{2}\right) \right) }
\end{equation*}%
\begin{equation*}
B_{2}=\frac{h_{1}^{B}\left( X\right) \left( \left\langle \bar{h}\left(
X\right) \right\rangle +\left\langle \hat{h}_{1}^{B}\left( X\right)
\right\rangle \right) }{S_{1}^{B}\left( X,\theta -1\right) +S_{2}^{B}\left(
X,\theta -1\right) }\left( \frac{\partial f\left( X\right) }{\partial
S^{T}\left( X,\theta -2\right) }\right) ^{2}\frac{\delta \frac{\bar{K}%
_{X}\left\vert \bar{\Psi}\left( X\right) \right\vert ^{2}}{K_{X}\left\vert
\Psi \left( X\right) \right\vert ^{2}}}{\frac{\bar{K}_{X}\left\vert \bar{\Psi%
}\left( X\right) \right\vert ^{2}}{K_{X}\left\vert \Psi \left( X\right)
\right\vert ^{2}}}
\end{equation*}%
\begin{equation*}
C=\left( \left( Z^{2}S\left( X\right) \right) \left[ \frac{\partial f\left(
X\right) }{\partial S^{T}\left( X,\theta -2\right) }\right] ^{2}\right)
^{2}+\left( \delta S^{T}\left( X\right) \frac{\frac{2}{r}}{\left( f\left(
X\right) +C_{0}\right) ^{2}}\left( \frac{\partial f\left( X\right) }{%
\partial S^{T}\left( X,\theta -2\right) }\right) ^{3}\delta S^{T}\left(
X\right) \right) ^{2}
\end{equation*}

\subsection*{A5.2 Investors}

The second order equation is:%
\begin{eqnarray*}
\left( 
\begin{array}{c}
\delta \hat{S}\left( X^{\prime },X,\theta \right) \\ 
\delta S\left( X,,\theta -1\right)%
\end{array}%
\right) &=&\left[ 
\begin{array}{cc}
\alpha & \frac{\hat{h}\left( X^{\prime },X\right) }{2}c \\ 
\frac{h}{\frac{\hat{h}\left( X^{\prime },X\right) }{2}} & \beta%
\end{array}%
\right] \left( 
\begin{array}{c}
\delta \hat{S}\left( X^{\prime },X,\theta -1\right) \\ 
\delta S\left( X,,\theta -2\right)%
\end{array}%
\right) \\
&&+\left( 
\begin{array}{c}
\delta \hat{S}\left( X^{\prime },X,\theta \right) \\ 
\delta S\left( X,,\theta -1\right)%
\end{array}%
\right) ^{t}\left[ 
\begin{array}{c}
\hat{M} \\ 
M%
\end{array}%
\right] \left( 
\begin{array}{c}
\delta \hat{S}\left( X^{\prime },X,\theta \right) \\ 
\delta S\left( X,\theta -1\right)%
\end{array}%
\right)
\end{eqnarray*}%
replaced by:%
\begin{eqnarray*}
&&\left( 
\begin{array}{c}
\delta \hat{S}\left( X^{\prime },X,\theta \right) \\ 
\delta S\left( X,,\theta -1\right) \\ 
\frac{\delta \hat{K}_{X}\left\vert \hat{\Psi}\left( X\right) \right\vert ^{2}%
}{\hat{K}_{X}\left\vert \hat{\Psi}\left( X\right) \right\vert ^{2}} \\ 
\frac{\delta K_{X}\left\vert \Psi \left( X\right) \right\vert ^{2}}{%
K_{X}\left\vert \Psi \left( X\right) \right\vert ^{2}}%
\end{array}%
\right) \\
&=&\left[ 
\begin{array}{cccc}
\alpha & \frac{\hat{h}\left( X^{\prime },X\right) }{2}c & \alpha _{0}+c_{0}
& \alpha _{1}-c_{0} \\ 
\frac{h}{\frac{\hat{h}\left( X^{\prime },X\right) }{2}} & \beta & \beta
_{0}+h & -\beta _{0}-h \\ 
\frac{\partial _{\hat{f}\left( X,\theta -1\right) }\hat{K}_{X}\left\vert 
\hat{\Psi}\left( X\right) \right\vert ^{2}}{\hat{K}_{X}\left\vert \hat{\Psi}%
\left( X\right) \right\vert ^{2}} & 0 & 0 & 0 \\ 
0 & \frac{\partial _{f\left( X\right) }K_{X}\left\vert \Psi \left( X\right)
\right\vert ^{2}}{K_{X}\left\vert \Psi \left( X\right) \right\vert ^{2}}%
\frac{\partial f\left( X\right) }{\partial S\left( X,,\theta -2\right) } & 0
& 0%
\end{array}%
\right] \left( 
\begin{array}{c}
\delta \hat{S}\left( X^{\prime },X,\theta -1\right) \\ 
\delta S\left( X,,\theta -2\right) \\ 
\frac{\delta \hat{K}_{X}\left\vert \hat{\Psi}\left( X\right) \right\vert ^{2}%
}{\hat{K}_{X}\left\vert \hat{\Psi}\left( X\right) \right\vert ^{2}} \\ 
\frac{\delta K_{X}\left\vert \Psi \left( X\right) \right\vert ^{2}}{%
K_{X}\left\vert \Psi \left( X\right) \right\vert ^{2}}%
\end{array}%
\right) \\
&&+\left( 
\begin{array}{c}
\delta \hat{S}\left( X^{\prime },X,\theta \right) \\ 
\delta S\left( X,,\theta -1\right) \\ 
\frac{\delta \hat{K}_{X}\left\vert \hat{\Psi}\left( X\right) \right\vert ^{2}%
}{\hat{K}_{X}\left\vert \hat{\Psi}\left( X\right) \right\vert ^{2}} \\ 
\frac{\delta K_{X}\left\vert \Psi \left( X\right) \right\vert ^{2}}{%
K_{X}\left\vert \Psi \left( X\right) \right\vert ^{2}}%
\end{array}%
\right) ^{t}\left[ 
\begin{array}{c}
\hat{M} \\ 
M \\ 
M_{\hat{K}} \\ 
M_{K}%
\end{array}%
\right] \left( 
\begin{array}{c}
\delta \hat{S}\left( X^{\prime },X,\theta -1\right) \\ 
\delta S\left( X,,\theta -2\right) \\ 
\frac{\delta \hat{K}_{X}\left\vert \hat{\Psi}\left( X\right) \right\vert ^{2}%
}{\hat{K}_{X}\left\vert \hat{\Psi}\left( X\right) \right\vert ^{2}} \\ 
\frac{\delta K_{X}\left\vert \Psi \left( X\right) \right\vert ^{2}}{%
K_{X}\left\vert \Psi \left( X\right) \right\vert ^{2}}%
\end{array}%
\right)
\end{eqnarray*}%
The coefficients and th matrices $\hat{M}$, $M$, $M_{\hat{K}}$, $M_{K}$ are
defined below.

\subsubsection*{A5.2.1 Coefficients}

The coefficients are given by:%
\begin{eqnarray}
\alpha &=&\left( \frac{\hat{S}\left( X\right) }{\left( 1-\left( \hat{S}%
\left( X\right) \right) \right) }\frac{1}{2\left( 1+\frac{\Delta \hat{f}%
\left( X^{\prime }\right) +\Delta \hat{r}\left( X^{\prime }\right) }{2}%
\right) }\right.  \label{LP} \\
&&\left. -\frac{\hat{S}_{1}\left( X\right) }{\left( 1-\left( \hat{S}%
_{1}\left( X^{\prime }\right) \right) \right) }\frac{1}{1+\Delta \hat{f}%
\left( X\right) }\right) \left( \hat{f}\left( X\right) -\bar{r}\right) 
\notag
\end{eqnarray}%
\begin{eqnarray}
\alpha _{0} &=&\left( \frac{\hat{S}_{1}\left( X\right) }{\left( 1-\left( 
\hat{S}_{1}\left( X^{\prime }\right) \right) \right) }-\frac{\hat{S}\left(
X\right) }{\left( 1-\left( \hat{S}\left( X\right) \right) \right) }\right)
\left( \hat{f}\left( X\right) -\bar{r}\right) \\
&&+\frac{1-\left( \hat{S}_{1}\left( X\right) \right) }{1-\left( \hat{S}%
\left( X\right) \right) }S_{1}\left( X,X,\theta -1\right) S\left( X\right) 
\notag
\end{eqnarray}%
\begin{equation*}
\alpha _{1}=-\frac{1-\left( \hat{S}_{1}\left( X\right) \right) }{1-\left( 
\hat{S}\left( X\right) \right) }S_{1}\left( X,X,\theta -1\right) S\left(
X\right)
\end{equation*}%
\begin{eqnarray*}
\beta &=&\left( \frac{\frac{\hat{h}\left( X\right) }{2}S\left( X\right) }{%
1+\left( \hat{h}\left( X\right) \left( \frac{f\left( X\right) +\bar{r}\left(
X\right) }{2}-\frac{\left\langle \hat{f}\left( X^{\prime }\right)
\right\rangle _{\hat{h}_{1}}+\left\langle \hat{r}\left( X^{\prime }\right)
\right\rangle _{\hat{h}_{2}}}{2}\right) \right) }\right) \frac{\partial
f\left( X\right) }{\partial S\left( X,\theta -2\right) } \\
&\simeq &\frac{\hat{h}\left( X\right) }{2}S\left( X\right) \frac{\partial
f\left( X\right) }{\partial S\left( X,\theta -2\right) }
\end{eqnarray*}%
\begin{equation*}
\beta _{0}\simeq S\left( X\right)
\end{equation*}%
\begin{eqnarray*}
&&c=\left\{ -\frac{\left( 1-\left\langle \hat{S}\left( X^{\prime }\right)
\right\rangle \right) \left( \left\langle \hat{f}\left( X^{\prime }\right)
\right\rangle -\bar{r}\right) }{1-\left\langle \hat{S}_{1}\left( X^{\prime
}\right) \right\rangle }\frac{\left\langle \hat{h}\left( X^{\prime
},X\right) \right\rangle \left\langle h\left( X\right) \right\rangle }{4}%
\right. \\
&&+\frac{\frac{h\left( X\right) }{4}\left( 1+\hat{h}\left( X\right) \right)
S_{1}\left( X\right) \left( f\left( X\right) -\bar{r}\right) }{1+\left( \hat{%
h}\left( X\right) \left( f\left( X\right) -\frac{\left\langle \hat{f}\left(
X^{\prime }\right) \right\rangle _{\hat{h}_{1}}+\left\langle \hat{r}\left(
X^{\prime }\right) \right\rangle _{\hat{h}_{2}}}{2}\right) +\frac{h\left(
X\right) }{2}\left( f\left( X\right) -\bar{r}\left( X\right) \right) \right) 
} \\
&&\left. +S_{1}\left( X\right) \left( \frac{\frac{\hat{h}\left( X\right) }{2}%
S\left( X\right) }{1+\left( \hat{h}\left( X\right) \left( \frac{f\left(
X\right) +\bar{r}\left( X\right) }{2}-\frac{\left\langle \hat{f}\left(
X^{\prime }\right) \right\rangle _{\hat{h}_{1}}+\left\langle \hat{r}\left(
X^{\prime }\right) \right\rangle _{\hat{h}_{2}}}{2}\right) \right) }\right) 
\frac{\partial f\left( X\right) }{\partial S\left( X,\theta -2\right) }%
\right\} \\
&&\times \frac{\partial f\left( X\right) }{\partial S\left( X,\theta
-1\right) }
\end{eqnarray*}%
\begin{equation*}
c_{0}=S_{1}\left( X\right) S\left( X\right) \frac{\partial f\left( X\right) 
}{\partial S\left( X,\theta -1\right) }
\end{equation*}%
\begin{equation*}
h=S\left( X\right)
\end{equation*}

\subsubsection{A5.2.2 Eigenvalues}

The eigenvalues equation:%
\begin{equation*}
\det \left( \left( 
\begin{array}{cccc}
0 & cz^{2} & vz & -vz \\ 
0 & \beta & w & -w \\ 
s & 0 & 0 & 0 \\ 
0 & tz & 0 & 0%
\end{array}%
\right) -X\left( 
\begin{array}{cccc}
1 & 0 & 0 & 0 \\ 
0 & 1 & 0 & 0 \\ 
0 & 0 & 1 & 0 \\ 
0 & 0 & 0 & 1%
\end{array}%
\right) \right) =0
\end{equation*}
\begin{equation*}
X^{4}-X^{3}\beta -X^{2}svz+X^{2}twz+Xsvz\beta -Xcswz^{2}\allowbreak =0
\end{equation*}%
$\allowbreak $

that is:%
\begin{equation*}
X^{3}-cswz^{2}=0
\end{equation*}%
with a solution:%
\begin{equation*}
X<0
\end{equation*}

\subsubsection*{A5.2.3 Estimation of coefficients derivatives and matrices}

The variations of these coefficients are obtained by using the following
formula derived in Gosselin and Lotz (2025a).

\begin{eqnarray*}
&&\delta \hat{S}_{1}\left( X^{\prime },X,\theta -1\right) \\
&\rightarrow &\frac{\hat{h}\left( X^{\prime },X\right) }{2}\left( 1+\left(
\delta \hat{f}\left( X^{\prime }\right) -\frac{\left\langle h\left( X\right)
\right\rangle }{2}\delta f\left( X\right) \right) \right)
\end{eqnarray*}%
\begin{equation*}
\delta \hat{S}_{1}\left( X,\theta -1\right) \simeq \left( \delta \hat{f}%
\left( X,\theta -1\right) -\frac{\delta \hat{K}_{X^{\prime }}\left\vert \hat{%
\Psi}\left( X^{\prime }\right) \right\vert ^{2}}{\hat{K}_{X^{\prime
}}\left\vert \hat{\Psi}\left( X^{\prime }\right) \right\vert ^{2}}\right) 
\hat{S}_{1}\left( X\right)
\end{equation*}%
\begin{equation*}
\delta \hat{S}\left( X,\theta -1\right) \simeq \left( \frac{1}{2}\delta \hat{%
f}\left( X,\theta -1\right) -\frac{\delta \hat{K}_{X^{\prime }}\left\vert 
\hat{\Psi}\left( X^{\prime }\right) \right\vert ^{2}}{\hat{K}_{X^{\prime
}}\left\vert \hat{\Psi}\left( X^{\prime }\right) \right\vert ^{2}}\right) 
\hat{S}\left( X\right)
\end{equation*}%
\begin{eqnarray*}
&&\delta S_{1}\left( X,X,\theta -1\right) \\
&\rightarrow &\frac{h\left( X\right) }{2}\left( \hat{h}\left( X\right) +%
\frac{h\left( X\right) }{2}\right) S_{1}\left( X,X\right) \delta f\left(
X,\theta -1\right) +\frac{\delta \left( \frac{\hat{K}_{X}\left\vert \hat{\Psi%
}\left( X\right) \right\vert ^{2}}{K_{X}\left\vert \Psi \left( X\right)
\right\vert ^{2}}\right) S_{1}\left( X\right) }{\frac{\hat{K}_{X}\left\vert 
\hat{\Psi}\left( X\right) \right\vert ^{2}}{K_{X}\left\vert \Psi \left(
X\right) \right\vert ^{2}}} \\
&\rightarrow &\frac{\delta \left( \frac{\hat{K}_{X}\left\vert \hat{\Psi}%
\left( X\right) \right\vert ^{2}}{K_{X}\left\vert \Psi \left( X\right)
\right\vert ^{2}}\right) S_{1}\left( X\right) }{\frac{\hat{K}_{X}\left\vert 
\hat{\Psi}\left( X\right) \right\vert ^{2}}{K_{X}\left\vert \Psi \left(
X\right) \right\vert ^{2}}}+\frac{h\left( X\right) }{2}\left( \hat{h}\left(
X\right) +\frac{h\left( X\right) }{2}\right) S_{1}\left( X,X\right) \frac{%
\partial f\left( X\right) }{\partial S\left( X,\theta -2\right) }\delta
S\left( X,\theta -2\right)
\end{eqnarray*}%
\begin{eqnarray*}
\delta S\left( X,\theta -1\right) &\simeq &\frac{\hat{h}\left( X\right) }{2}%
\delta f\left( X,\theta -1\right) S\left( X\right) +\frac{\delta \frac{\hat{K%
}_{X}\left\vert \hat{\Psi}\left( X\right) \right\vert ^{2}}{K_{X}\left\vert
\Psi \left( X\right) \right\vert ^{2}}}{\frac{\hat{K}_{X}\left\vert \hat{\Psi%
}\left( X\right) \right\vert ^{2}}{K_{X}\left\vert \Psi \left( X\right)
\right\vert ^{2}}}S\left( X\right) \\
&\simeq &\frac{\hat{h}\left( X\right) }{2}S\left( X\right) \frac{\partial
f\left( X\right) }{\partial S\left( X,\theta -2\right) }\delta S\left(
X,\theta -2\right) +\frac{\delta \frac{\hat{K}_{X}\left\vert \hat{\Psi}%
\left( X\right) \right\vert ^{2}}{K_{X}\left\vert \Psi \left( X\right)
\right\vert ^{2}}}{\frac{\hat{K}_{X}\left\vert \hat{\Psi}\left( X\right)
\right\vert ^{2}}{K_{X}\left\vert \Psi \left( X\right) \right\vert ^{2}}}%
S\left( X\right)
\end{eqnarray*}

We use in first approximation that:%
\begin{eqnarray*}
\delta \frac{1-\left( \hat{S}_{1}\left( X\right) \right) }{1-\left( \hat{S}%
\left( X\right) \right) } &\simeq &\delta \frac{1-\frac{1}{2}\hat{S}\left(
X\right) }{1-\hat{S}\left( X\right) } \\
&\rightarrow &\frac{1}{2\left( 1-\hat{S}\left( X\right) \right) ^{2}}\delta 
\hat{S}\left( X\right) \\
&\simeq &\frac{1}{2\left( 1-\hat{S}\left( X\right) \right) ^{2}}\left( \frac{%
1}{2}\delta \hat{f}\left( X,\theta -1\right) -\frac{\delta \hat{K}%
_{X^{\prime }}\left\vert \hat{\Psi}\left( X^{\prime }\right) \right\vert ^{2}%
}{\hat{K}_{X^{\prime }}\left\vert \hat{\Psi}\left( X^{\prime }\right)
\right\vert ^{2}}\right) \hat{S}\left( X\right)
\end{eqnarray*}%
\begin{equation*}
\frac{\partial _{\hat{f}\left( X\right) }\left( \frac{\hat{K}_{X}\left\vert 
\hat{\Psi}\left( X\right) \right\vert ^{2}}{K_{X}\left\vert \Psi \left(
X\right) \right\vert ^{2}}\right) S\left( X\right) }{\frac{\hat{K}%
_{X}\left\vert \hat{\Psi}\left( X\right) \right\vert ^{2}}{K_{X}\left\vert
\Psi \left( X\right) \right\vert ^{2}}}\simeq -\frac{2S\left( X\right) }{%
\hat{f}\left( X,\theta -1\right) }
\end{equation*}%
\begin{equation*}
\delta \frac{\partial _{\hat{f}\left( X\right) }\left( \frac{\hat{K}%
_{X}\left\vert \hat{\Psi}\left( X\right) \right\vert ^{2}}{K_{X}\left\vert
\Psi \left( X\right) \right\vert ^{2}}\right) S\left( X\right) }{\frac{\hat{K%
}_{X}\left\vert \hat{\Psi}\left( X\right) \right\vert ^{2}}{K_{X}\left\vert
\Psi \left( X\right) \right\vert ^{2}}}\simeq -2\frac{\frac{\hat{h}\left(
X\right) }{2}\delta f\left( X,\theta -1\right) S\left( X\right) -2S\left(
X\right) \frac{\delta \hat{f}\left( X,\theta -1\right) }{\hat{f}\left(
X,\theta -1\right) }-\frac{\frac{\partial f\left( X\right) }{\partial
S\left( X,\theta -2\right) }}{f\left( X\right) +C_{0}}\delta S\left(
X,\theta -2\right) }{\hat{f}\left( X,\theta -1\right) }
\end{equation*}%
\begin{eqnarray}
\delta \alpha &\simeq &\left( \frac{\hat{S}\left( X\right) }{\left( 1-\left( 
\hat{S}\left( X\right) \right) \right) }\frac{1}{2\left( 1+\frac{\Delta \hat{%
f}\left( X^{\prime }\right) +\Delta \hat{r}\left( X^{\prime }\right) }{2}%
\right) }\right. \\
&&\left. -\frac{\hat{S}_{1}\left( X\right) }{\left( 1-\left( \hat{S}%
_{1}\left( X^{\prime }\right) \right) \right) }\frac{1}{1+\Delta \hat{f}%
\left( X\right) }\right) \delta \hat{f}\left( X\right)  \notag
\end{eqnarray}%
In all expressions, due to presence of banks, $\hat{f}\left( X\right) $
stands for:%
\begin{equation*}
\hat{f}\left( X\right) +\frac{\left\langle \hat{S}\left( X^{\prime
},X\right) \right\rangle _{X^{\prime }}\left\langle \hat{f}\right\rangle }{%
1-\left\langle \hat{S}\right\rangle }+\frac{\left( \left\langle \hat{S}%
_{1}^{B}\left( X,X\right) \right\rangle _{X}+\left\langle \hat{S}%
_{2}^{B}\left( X,X\right) \right\rangle _{X}+\left\langle \hat{S}\left(
X^{\prime },X\right) \right\rangle _{X^{\prime }}\frac{\left\langle \hat{S}%
_{1}^{B}\right\rangle +\left\langle \hat{S}_{2}^{B}\right\rangle }{%
1-\left\langle \hat{S}\right\rangle }\right) \frac{\left\langle \bar{K}%
\right\rangle \left\Vert \bar{\Psi}\right\Vert ^{2}}{\left\langle \hat{K}%
\right\rangle \left\Vert \hat{\Psi}\right\Vert ^{2}}\left\langle \bar{f}%
\right\rangle }{1-\left\langle \bar{S}\right\rangle }
\end{equation*}

We obtain:%
\begin{eqnarray}
\delta \alpha _{0} &\simeq &\left( \frac{\hat{S}_{1}\left( X\right) }{\left(
1-\left( \hat{S}_{1}\left( X^{\prime }\right) \right) \right) }-\frac{\hat{S}%
\left( X\right) }{\left( 1-\left( \hat{S}\left( X\right) \right) \right) }%
\right) \delta \hat{f}\left( X\right) \\
&&+\delta \left\{ \frac{1-\left( \hat{S}_{1}\left( X\right) \right) }{%
1-\left( \hat{S}\left( X\right) \right) }S_{1}\left( X,X,\theta -1\right)
S\left( X\right) \right\}  \notag \\
&=&\left( \frac{\hat{S}_{1}\left( X\right) }{\left( 1-\left( \hat{S}%
_{1}\left( X^{\prime }\right) \right) \right) }-\frac{\hat{S}\left( X\right) 
}{\left( 1-\left( \hat{S}\left( X\right) \right) \right) }\right) \delta 
\hat{f}\left( X\right)  \notag \\
&&+\frac{\hat{S}\left( X\right) }{2\left( 1-\hat{S}\left( X\right) \right)
^{2}}\left( \frac{1}{2}\delta \hat{f}\left( X,\theta -1\right) -\frac{\delta 
\hat{K}_{X^{\prime }}\left\vert \hat{\Psi}\left( X^{\prime }\right)
\right\vert ^{2}}{\hat{K}_{X^{\prime }}\left\vert \hat{\Psi}\left( X^{\prime
}\right) \right\vert ^{2}}\right) S_{1}\left( X,X,\theta -1\right) S\left(
X\right)  \notag \\
&&+\frac{1-\left( \hat{S}_{1}\left( X\right) \right) }{1-\left( \hat{S}%
\left( X\right) \right) }\delta \left( S_{1}\left( X,X,\theta -1\right)
S\left( X\right) \right)  \notag
\end{eqnarray}%
\begin{eqnarray*}
&&\delta \left( S_{1}\left( X,X,\theta -1\right) S\left( X\right) \right) \\
&\rightarrow &\left( \frac{h\left( X\right) }{2}\left( \hat{h}\left(
X\right) +\frac{h\left( X\right) }{2}\right) +\frac{\hat{h}\left( X\right) }{%
2}\right) S_{1}\left( X,X\right) S\left( X\right) \frac{\partial f\left(
X\right) }{\partial S\left( X,\theta -2\right) }\delta S\left( X,\theta
-2\right) \\
&&+S\left( X\right) S_{1}\left( X\right) \frac{\delta \left( \frac{\hat{K}%
_{X}\left\vert \hat{\Psi}\left( X\right) \right\vert ^{2}}{K_{X}\left\vert
\Psi \left( X\right) \right\vert ^{2}}\right) }{\frac{\hat{K}_{X}\left\vert 
\hat{\Psi}\left( X\right) \right\vert ^{2}}{K_{X}\left\vert \Psi \left(
X\right) \right\vert ^{2}}}
\end{eqnarray*}%
\begin{eqnarray}
\delta \alpha _{0} &\simeq &\left( \frac{\hat{S}_{1}\left( X\right) }{\left(
1-\left( \hat{S}_{1}\left( X^{\prime }\right) \right) \right) }-\frac{\hat{S}%
\left( X\right) }{\left( 1-\left( \hat{S}\left( X\right) \right) \right) }+%
\frac{1}{2}\frac{\hat{S}\left( X\right) S_{1}\left( X,X\right) S\left(
X\right) }{2\left( 1-\hat{S}\left( X\right) \right) ^{2}}\right) \delta \hat{%
f}\left( X\right) \\
&&-\frac{\hat{S}\left( X\right) S_{1}\left( X,X,\theta -1\right) S\left(
X\right) }{2\left( 1-\hat{S}\left( X\right) \right) ^{2}}\frac{\delta \hat{K}%
_{X^{\prime }}\left\vert \hat{\Psi}\left( X^{\prime }\right) \right\vert ^{2}%
}{\hat{K}_{X^{\prime }}\left\vert \hat{\Psi}\left( X^{\prime }\right)
\right\vert ^{2}}  \notag \\
&&+\frac{1-\left( \hat{S}_{1}\left( X\right) \right) S_{1}\left( X,X\right)
S\left( X\right) }{1-\left( \hat{S}\left( X\right) \right) }\left( \frac{%
h\left( X\right) }{2}\left( \hat{h}\left( X\right) +\frac{h\left( X\right) }{%
2}\right) +\frac{\hat{h}\left( X\right) }{2}\right) \frac{\partial f\left(
X\right) }{\partial S\left( X,\theta -2\right) }\delta S\left( X,\theta
-2\right)  \notag \\
&&+\frac{1-\left( \hat{S}_{1}\left( X\right) \right) S_{1}\left( X,X\right)
S\left( X\right) }{1-\left( \hat{S}\left( X\right) \right) }\frac{\delta
\left( \frac{\hat{K}_{X}\left\vert \hat{\Psi}\left( X\right) \right\vert ^{2}%
}{K_{X}\left\vert \Psi \left( X\right) \right\vert ^{2}}\right) }{\frac{\hat{%
K}_{X}\left\vert \hat{\Psi}\left( X\right) \right\vert ^{2}}{K_{X}\left\vert
\Psi \left( X\right) \right\vert ^{2}}}  \notag
\end{eqnarray}

\begin{eqnarray}
\delta \alpha _{1} &\simeq &-\frac{1}{2}\frac{\hat{S}\left( X\right)
S_{1}\left( X,X\right) S\left( X\right) }{2\left( 1-\hat{S}\left( X\right)
\right) ^{2}}\delta \hat{f}\left( X\right) \\
&&+\frac{\hat{S}\left( X\right) S_{1}\left( X,X,\theta -1\right) S\left(
X\right) }{2\left( 1-\hat{S}\left( X\right) \right) ^{2}}\frac{\delta \hat{K}%
_{X^{\prime }}\left\vert \hat{\Psi}\left( X^{\prime }\right) \right\vert ^{2}%
}{\hat{K}_{X^{\prime }}\left\vert \hat{\Psi}\left( X^{\prime }\right)
\right\vert ^{2}}  \notag \\
&&-\frac{1-\left( \hat{S}_{1}\left( X\right) \right) S_{1}\left( X,X\right)
S\left( X\right) }{1-\left( \hat{S}\left( X\right) \right) }\left( \frac{%
h\left( X\right) }{2}\left( \hat{h}\left( X\right) +\frac{h\left( X\right) }{%
2}\right) +\frac{\hat{h}\left( X\right) }{2}\right) \frac{\partial f\left(
X\right) }{\partial S\left( X,\theta -2\right) }\delta S\left( X,\theta
-2\right)  \notag \\
&&-\frac{1-\left( \hat{S}_{1}\left( X\right) \right) S_{1}\left( X,X\right)
S\left( X\right) }{1-\left( \hat{S}\left( X\right) \right) }\frac{\delta
\left( \frac{\hat{K}_{X}\left\vert \hat{\Psi}\left( X\right) \right\vert ^{2}%
}{K_{X}\left\vert \Psi \left( X\right) \right\vert ^{2}}\right) }{\frac{\hat{%
K}_{X}\left\vert \hat{\Psi}\left( X\right) \right\vert ^{2}}{K_{X}\left\vert
\Psi \left( X\right) \right\vert ^{2}}}  \notag
\end{eqnarray}%
\begin{eqnarray*}
\delta \beta &\simeq &\frac{\hat{h}\left( X\right) }{2}\delta S\left(
X\right) \frac{\partial f\left( X\right) }{\partial S\left( X,\theta
-2\right) }+\frac{\hat{h}\left( X\right) }{2}S\left( X\right) \frac{\partial
^{2}f\left( X\right) }{\partial S^{2}\left( X,\theta -2\right) }\delta
S\left( X,\theta -2\right) \\
&\simeq &\frac{\hat{h}\left( X\right) }{2}S\left( X\right) \left( \frac{\hat{%
h}\left( X\right) }{2}\left( \frac{\partial f\left( X\right) }{\partial
S\left( X,\theta -2\right) }\right) ^{2}+\frac{\partial ^{2}f\left( X\right) 
}{\partial S^{2}\left( X,\theta -2\right) }\right) \delta S\left( X,\theta
-2\right) \\
&&+\frac{\hat{h}\left( X\right) }{2}S\left( X\right) \frac{\partial f\left(
X\right) }{\partial S\left( X,\theta -2\right) }\frac{\delta \frac{\hat{K}%
_{X}\left\vert \hat{\Psi}\left( X\right) \right\vert ^{2}}{K_{X}\left\vert
\Psi \left( X\right) \right\vert ^{2}}}{\frac{\hat{K}_{X}\left\vert \hat{\Psi%
}\left( X\right) \right\vert ^{2}}{K_{X}\left\vert \Psi \left( X\right)
\right\vert ^{2}}}
\end{eqnarray*}%
\begin{equation*}
\delta \beta _{0}\simeq \frac{\hat{h}\left( X\right) }{2}S\left( X\right) 
\frac{\partial f\left( X\right) }{\partial S\left( X,\theta -2\right) }%
\delta S\left( X,\theta -2\right) +\frac{\delta \frac{\hat{K}_{X}\left\vert 
\hat{\Psi}\left( X\right) \right\vert ^{2}}{K_{X}\left\vert \Psi \left(
X\right) \right\vert ^{2}}}{\frac{\hat{K}_{X}\left\vert \hat{\Psi}\left(
X\right) \right\vert ^{2}}{K_{X}\left\vert \Psi \left( X\right) \right\vert
^{2}}}S\left( X\right)
\end{equation*}%
\begin{equation*}
\delta h=\frac{\hat{h}\left( X\right) }{2}S\left( X\right) \frac{\partial
f\left( X\right) }{\partial S\left( X,\theta -2\right) }\delta S\left(
X,\theta -2\right) +\frac{\delta \frac{\hat{K}_{X}\left\vert \hat{\Psi}%
\left( X\right) \right\vert ^{2}}{K_{X}\left\vert \Psi \left( X\right)
\right\vert ^{2}}}{\frac{\hat{K}_{X}\left\vert \hat{\Psi}\left( X\right)
\right\vert ^{2}}{K_{X}\left\vert \Psi \left( X\right) \right\vert ^{2}}}%
S\left( X\right)
\end{equation*}%
\begin{eqnarray*}
\delta &&c\simeq -\frac{\left( 1-\left\langle \hat{S}\left( X^{\prime
}\right) \right\rangle \right) }{1-\left\langle \hat{S}_{1}\left( X^{\prime
}\right) \right\rangle }\frac{\left\langle \hat{h}\left( X^{\prime
},X\right) \right\rangle \left\langle h\left( X\right) \right\rangle }{4}%
\frac{\partial f\left( X\right) }{\partial S\left( X\right) }\delta \hat{f}%
\left( X^{\prime }\right) \\
&&+\frac{\frac{h\left( X\right) }{4}\left( 1+\hat{h}\left( X\right) \right)
S_{1}\left( X\right) }{1+\left( \hat{h}\left( X\right) \left( f\left(
X\right) -\frac{\left\langle \hat{f}\left( X^{\prime }\right) \right\rangle
_{\hat{h}_{1}}+\left\langle \hat{r}\left( X^{\prime }\right) \right\rangle _{%
\hat{h}_{2}}}{2}\right) +\frac{h\left( X\right) }{2}\left( f\left( X\right) -%
\bar{r}\left( X\right) \right) \right) }\left( \frac{\partial f\left(
X\right) }{\partial S\left( X\right) }\right) ^{2}\delta S\left( X,\theta
-2\right) \\
&&+\delta \left( \frac{\hat{h}\left( X\right) }{2}S_{1}\left( X\right)
S\left( X\right) \frac{\partial f\left( X\right) }{\partial S\left( X,\theta
-2\right) }\frac{\partial f\left( X\right) }{\partial S\left( X,\theta
-1\right) }\right)
\end{eqnarray*}%
\begin{eqnarray*}
\delta &&c\simeq -\frac{\left( 1-\left\langle \hat{S}\left( X^{\prime
}\right) \right\rangle \right) }{1-\left\langle \hat{S}_{1}\left( X^{\prime
}\right) \right\rangle }\frac{\left\langle \hat{h}\left( X^{\prime
},X\right) \right\rangle \left\langle h\left( X\right) \right\rangle }{4}%
\frac{\partial f\left( X\right) }{\partial S\left( X\right) }\delta \hat{f}%
\left( X^{\prime }\right) \\
&&+\frac{\frac{h\left( X\right) }{4}\left( 1+\hat{h}\left( X\right) \right)
S_{1}\left( X\right) }{1+\left( \hat{h}\left( X\right) \left( f\left(
X\right) -\frac{\left\langle \hat{f}\left( X^{\prime }\right) \right\rangle
_{\hat{h}_{1}}+\left\langle \hat{r}\left( X^{\prime }\right) \right\rangle _{%
\hat{h}_{2}}}{2}\right) +\frac{h\left( X\right) }{2}\left( f\left( X\right) -%
\bar{r}\left( X\right) \right) \right) }\left( \frac{\partial f\left(
X\right) }{\partial S\left( X\right) }\right) ^{2}\delta S\left( X,\theta
-2\right) \\
&&+\frac{\hat{h}\left( X\right) }{2}S_{1}\left( X\right) S\left( X\right)
\left( \frac{\partial f\left( X\right) }{\partial S\left( X\right) }\right)
^{3} \\
&&\times \left( \left( \frac{h\left( X\right) }{2}\left( \hat{h}\left(
X\right) +\frac{h\left( X\right) }{2}\right) +\frac{\hat{h}\left( X\right) }{%
2}+\frac{\frac{\partial ^{2}f\left( X\right) }{\partial S^{2}\left( X\right) 
}}{\left( \frac{\partial f\left( X\right) }{\partial S\left( X\right) }%
\right) ^{2}}\right) +\frac{\delta \left( \frac{\hat{K}_{X}\left\vert \hat{%
\Psi}\left( X\right) \right\vert ^{2}}{K_{X}\left\vert \Psi \left( X\right)
\right\vert ^{2}}\right) }{\frac{\partial f\left( X\right) }{\partial
S\left( X\right) }\frac{\hat{K}_{X}\left\vert \hat{\Psi}\left( X\right)
\right\vert ^{2}}{K_{X}\left\vert \Psi \left( X\right) \right\vert ^{2}}}%
\right) \delta S\left( X,\theta -2\right)
\end{eqnarray*}%
\begin{eqnarray*}
\delta c_{0} &=&S_{1}\left( X\right) S\left( X\right) \left( \frac{\partial
f\left( X\right) }{\partial S\left( X\right) }\right) ^{2} \\
&&\times \left( \left( \frac{h\left( X\right) }{2}\left( \hat{h}\left(
X\right) +\frac{h\left( X\right) }{2}\right) +\frac{\hat{h}\left( X\right) }{%
2}+\frac{\frac{\partial ^{2}f\left( X\right) }{\partial S^{2}\left( X\right) 
}}{\left( \frac{\partial f\left( X\right) }{\partial S\left( X\right) }%
\right) ^{2}}\right) +\frac{\delta \left( \frac{\hat{K}_{X}\left\vert \hat{%
\Psi}\left( X\right) \right\vert ^{2}}{K_{X}\left\vert \Psi \left( X\right)
\right\vert ^{2}}\right) }{\frac{\partial f\left( X\right) }{\partial
S\left( X\right) }\frac{\hat{K}_{X}\left\vert \hat{\Psi}\left( X\right)
\right\vert ^{2}}{K_{X}\left\vert \Psi \left( X\right) \right\vert ^{2}}}%
\right) \delta S\left( X,\theta -2\right)
\end{eqnarray*}%
\begin{equation*}
\frac{\partial _{\hat{f}\left( X,\theta -1\right) }\hat{K}_{X}\left\vert 
\hat{\Psi}\left( X\right) \right\vert ^{2}}{\hat{K}_{X}\left\vert \hat{\Psi}%
\left( X\right) \right\vert ^{2}}\simeq -\frac{2}{\hat{f}\left( X\right) }
\end{equation*}%
\begin{equation*}
\frac{\partial _{f\left( X\right) }K_{X}\left\vert \Psi \left( X\right)
\right\vert ^{2}}{K_{X}\left\vert \Psi \left( X\right) \right\vert ^{2}}%
\simeq \frac{2}{r}\frac{1}{f\left( X\right) +C_{0}}
\end{equation*}%
\begin{equation*}
\delta \frac{\partial _{\hat{f}\left( X,\theta -1\right) }\hat{K}%
_{X}\left\vert \hat{\Psi}\left( X\right) \right\vert ^{2}}{\hat{K}%
_{X}\left\vert \hat{\Psi}\left( X\right) \right\vert ^{2}}=\frac{2}{\left( 
\hat{f}\left( X\right) \right) ^{2}}\delta \hat{f}\left( X\right)
\end{equation*}%
\begin{eqnarray*}
&&\delta \left( \frac{\partial _{f\left( X\right) }K_{X}\left\vert \Psi
\left( X\right) \right\vert ^{2}}{K_{X}\left\vert \Psi \left( X\right)
\right\vert ^{2}}\frac{\partial f\left( X\right) }{\partial S\left( X,\theta
-2\right) }\right) \\
&\simeq &-\left( 1-\frac{\left( f\left( X\right) +C_{0}\right) \frac{%
\partial ^{2}f\left( X\right) }{\partial S^{2}\left( X\right) }}{\left( 
\frac{\partial f\left( X\right) }{\partial S\left( X\right) }\right) ^{2}}%
\right) \frac{2}{r}\frac{\left( \frac{\partial f\left( X\right) }{\partial
S\left( X\right) }\right) ^{2}}{\left( f\left( X\right) +C_{0}\right) ^{2}}%
\delta S\left( X,\theta -2\right)
\end{eqnarray*}

As before, the derivatives of these coefficients with respect to $\hat{K}%
_{X}\left\vert \hat{\Psi}\left( X\right) \right\vert ^{2}$ and $%
K_{X}\left\vert \Psi \left( X\right) \right\vert ^{2}$ can be neglected.

The matrices can then be obtained:%
\begin{eqnarray*}
\hat{M} &\simeq &\left[ 
\begin{array}{cccc}
\frac{\partial }{\partial \hat{f}\left( X\right) }\alpha & \frac{\hat{h}%
\left( X^{\prime },X\right) }{2}\frac{\partial }{\partial \hat{f}\left(
X\right) }c & \frac{\partial }{\partial \hat{f}\left( X\right) }\left(
\alpha _{0}+c_{0}\right) & \frac{\partial }{\partial \hat{f}\left( X\right) }%
\left( \alpha _{1}-c_{0}\right) \\ 
\frac{\partial }{\partial S\left( X\right) }\alpha & \frac{\hat{h}\left(
X^{\prime },X\right) }{2}\frac{\partial }{\partial S\left( X\right) }c & 
\frac{\partial }{\partial S\left( X\right) }\left( \alpha _{0}+c_{0}\right)
& \frac{\partial }{\partial S\left( X\right) }\left( \alpha _{1}-c_{0}\right)
\\ 
0 & 0 & 0 & 0 \\ 
0 & 0 & 0 & 0%
\end{array}%
\right] \\
M_{S} &=&\left[ 
\begin{array}{cccc}
0 & \frac{\partial }{\partial \hat{f}\left( X\right) }\beta & \frac{\partial 
}{\partial \hat{f}\left( X\right) }\left( \beta _{0}+h\right) & -\frac{%
\partial }{\partial \hat{f}\left( X\right) }\left( \beta _{0}+h\right) \\ 
0 & \frac{\partial }{\partial S\left( X\right) }\beta & \frac{\partial }{%
\partial S\left( X\right) }\left( \beta _{0}+h\right) & -\frac{\partial }{%
\partial S\left( X\right) }\left( \beta _{0}+h\right) \\ 
0 & 0 & 0 & 0 \\ 
0 & 0 & 0 & 0%
\end{array}%
\right] \\
M_{\hat{K}} &=&\left[ 
\begin{array}{cccc}
\frac{\partial _{\hat{f}\left( X,\theta -1\right) }\hat{K}_{X}\left\vert 
\hat{\Psi}\left( X\right) \right\vert ^{2}}{\hat{K}_{X}\left\vert \hat{\Psi}%
\left( X\right) \right\vert ^{2}} & 0 & 0 & 0 \\ 
0 & 0 & 0 & 0 \\ 
0 & 0 & 0 & 0 \\ 
0 & 0 & 0 & 0%
\end{array}%
\right] \\
M_{K} &=&\left[ 
\begin{array}{cccc}
0 & 0 & 0 & 0 \\ 
0 & 
\begin{array}{c}
\partial _{f\left( X\right) }\frac{\partial _{f\left( X\right)
}K_{X}\left\vert \Psi \left( X\right) \right\vert ^{2}}{K_{X}\left\vert \Psi
\left( X\right) \right\vert ^{2}}\left( \frac{\partial f\left( X\right) }{%
\partial S\left( X,\theta -2\right) }\right) ^{2} \\ 
+\frac{\partial _{f\left( X\right) }K_{X}\left\vert \Psi \left( X\right)
\right\vert ^{2}}{K_{X}\left\vert \Psi \left( X\right) \right\vert ^{2}}%
\frac{\partial ^{2}f\left( X\right) }{\partial \left( S\left( X,\theta
-2\right) \right) ^{2}}%
\end{array}
& 0 & 0 \\ 
0 & 0 & 0 & 0 \\ 
0 & 0 & 0 & 0%
\end{array}%
\right]
\end{eqnarray*}

\paragraph*{A5.2.4 Conditions for stability}

The condition for stability becomes:%
\begin{eqnarray*}
&&\left( 1-\left\langle \lambda ^{2}\right\rangle \right) \left( \left(
\delta \hat{S}\left( X\right) \right) ^{2}+\left( \delta S\left( X\right)
\right) ^{2}+\left( \delta \hat{K}_{X}\left\vert \hat{\Psi}\left( X\right)
\right\vert ^{2}\right) ^{2}+\left( \delta K_{X}\left\vert \Psi \left(
X\right) \right\vert ^{2}\right) ^{2}\right) \\
&>&\left\{ \left( \delta \hat{S}\left( X\right) ,\delta S\left( X\right)
\right) \left( \delta \hat{S}\left( X\right) \left( 
\begin{array}{c}
\frac{\partial }{\partial \hat{f}\left( X\right) } \\ 
\frac{\partial }{\partial S\left( X\right) }%
\end{array}%
\right) \alpha +\frac{\hat{h}\left( X^{\prime },X\right) }{2}\delta S\left(
X\right) \left( 
\begin{array}{c}
\frac{\partial }{\partial \hat{f}\left( X\right) } \\ 
\frac{\partial }{\partial S\left( X\right) }%
\end{array}%
\right) c\right. \right. \\
&&+\left. \left. \frac{\delta \frac{\hat{K}_{X}\left\vert \hat{\Psi}\left(
X\right) \right\vert ^{2}}{K_{X}\left\vert \Psi \left( X\right) \right\vert
^{2}}}{\frac{\hat{K}_{X}\left\vert \hat{\Psi}\left( X\right) \right\vert ^{2}%
}{K_{X}\left\vert \Psi \left( X\right) \right\vert ^{2}}}\left( 
\begin{array}{c}
\frac{\partial }{\partial \hat{f}\left( X\right) } \\ 
\frac{\partial }{\partial S\left( X\right) }%
\end{array}%
\right) \left( \alpha _{0}+c_{0}\right) +\frac{\delta K_{X}\left\vert \Psi
\left( X\right) \right\vert ^{2}}{K_{X}\left\vert \Psi \left( X\right)
\right\vert ^{2}}\left( 
\begin{array}{c}
\frac{\partial }{\partial \hat{f}\left( X\right) } \\ 
\frac{\partial }{\partial S\left( X\right) }%
\end{array}%
\right) \left( \alpha _{1}+\alpha _{0}\right) \right) \right\} ^{2} \\
&&+\left\{ \left( \delta \hat{S}\left( X\right) ,\delta S\left( X\right)
\right) \left( \delta S\left( X\right) \left( 
\begin{array}{c}
\frac{\partial }{\partial \hat{f}\left( X\right) } \\ 
\frac{\partial }{\partial S\left( X\right) }%
\end{array}%
\right) \beta +\frac{\delta \frac{\hat{K}_{X}\left\vert \hat{\Psi}\left(
X\right) \right\vert ^{2}}{K_{X}\left\vert \Psi \left( X\right) \right\vert
^{2}}}{\frac{\hat{K}_{X}\left\vert \hat{\Psi}\left( X\right) \right\vert ^{2}%
}{K_{X}\left\vert \Psi \left( X\right) \right\vert ^{2}}}\left( 
\begin{array}{c}
\frac{\partial }{\partial \hat{f}\left( X\right) } \\ 
\frac{\partial }{\partial S\left( X\right) }%
\end{array}%
\right) \left( \beta _{0}+h\right) \right) \right\} ^{2} \\
&&+\left\{ \partial _{\hat{f}\left( X\right) }\left( \frac{\partial _{\hat{f}%
\left( X\right) }\hat{K}_{X}\left\vert \hat{\Psi}\left( X\right) \right\vert
^{2}}{\hat{K}_{X}\left\vert \hat{\Psi}\left( X\right) \right\vert ^{2}}%
\right) \left( \delta \hat{S}\left( X\right) \right) ^{2}\right\} ^{2} \\
&&+\left\{ \left( \partial _{f\left( X\right) }\left( \frac{\partial
_{f\left( X\right) }K_{X}\left\vert \Psi \left( X\right) \right\vert ^{2}}{%
K_{X}\left\vert \Psi \left( X\right) \right\vert ^{2}}\right) \left( \frac{%
\partial f\left( X\right) }{\partial S^{T}\left( X,\theta -2\right) }\right)
^{2}+\frac{\partial _{f\left( X\right) }K_{X}\left\vert \Psi \left( X\right)
\right\vert ^{2}}{K_{X}\left\vert \Psi \left( X\right) \right\vert ^{2}}%
\frac{\partial ^{2}f\left( X\right) }{\partial \left( S^{T}\left( X,\theta
-2\right) \right) ^{2}}\right) \left( \delta S\left( X\right) \right)
^{2}\right\} ^{2}
\end{eqnarray*}

The dominant contributions involving the fluctuations in $\hat{K}%
_{X}\left\vert \hat{\Psi}\left( X\right) \right\vert ^{2}$ \ and $%
K_{X}\left\vert \Psi \left( X\right) \right\vert ^{2}$ are:

\begin{eqnarray*}
&&\left\{ \left( \delta \hat{S}\left( X\right) ,\delta S\left( X\right)
\right) \left( \delta S\left( X\right) \left( 
\begin{array}{c}
\frac{\partial }{\partial \hat{f}\left( X\right) } \\ 
\frac{\partial }{\partial S\left( X\right) }%
\end{array}%
\right) \beta +\frac{\delta \frac{\hat{K}_{X}\left\vert \hat{\Psi}\left(
X\right) \right\vert ^{2}}{K_{X}\left\vert \Psi \left( X\right) \right\vert
^{2}}}{\frac{\hat{K}_{X}\left\vert \hat{\Psi}\left( X\right) \right\vert ^{2}%
}{K_{X}\left\vert \Psi \left( X\right) \right\vert ^{2}}}\left( 
\begin{array}{c}
\frac{\partial }{\partial \hat{f}\left( X\right) } \\ 
\frac{\partial }{\partial S\left( X\right) }%
\end{array}%
\right) \left( \beta _{0}+h\right) \right) \right\} ^{2} \\
&\simeq &\left( \frac{\hat{h}\left( X\right) }{2}S\left( X\right) \right)
^{2}\left\{ \left( \frac{\hat{h}\left( X\right) }{2}\left( \frac{\partial
f\left( X\right) }{\partial S\left( X,\theta -2\right) }\right) ^{2}+\frac{%
\partial ^{2}f\left( X\right) }{\partial S^{2}\left( X,\theta -2\right) }%
\right) \delta S\left( X\right) \right. \\
&&\left. +\frac{\partial f\left( X\right) }{\partial S\left( X,\theta
-2\right) }\frac{\delta \frac{\hat{K}_{X}\left\vert \hat{\Psi}\left(
X\right) \right\vert ^{2}}{K_{X}\left\vert \Psi \left( X\right) \right\vert
^{2}}}{\frac{\hat{K}_{X}\left\vert \hat{\Psi}\left( X\right) \right\vert ^{2}%
}{K_{X}\left\vert \Psi \left( X\right) \right\vert ^{2}}}\right\} ^{2}
\end{eqnarray*}%
\begin{eqnarray*}
&&\left\{ \left( \partial _{f\left( X\right) }\left( \frac{\partial
_{f\left( X\right) }K_{X}\left\vert \Psi \left( X\right) \right\vert ^{2}}{%
K_{X}\left\vert \Psi \left( X\right) \right\vert ^{2}}\right) \left( \frac{%
\partial f\left( X\right) }{\partial S^{T}\left( X,\theta -2\right) }\right)
^{2}+\frac{\partial _{f\left( X\right) }K_{X}\left\vert \Psi \left( X\right)
\right\vert ^{2}}{K_{X}\left\vert \Psi \left( X\right) \right\vert ^{2}}%
\frac{\partial ^{2}f\left( X\right) }{\partial \left( S^{T}\left( X,\theta
-2\right) \right) ^{2}}\right) \left( \delta S\left( X\right) \right)
^{2}\right\} ^{2} \\
&\simeq &\left\{ \left( 1-\frac{\left( f\left( X\right) +C_{0}\right) \frac{%
\partial ^{2}f\left( X\right) }{\partial S^{2}\left( X\right) }}{\left( 
\frac{\partial f\left( X\right) }{\partial S\left( X\right) }\right) ^{2}}%
\right) \frac{2}{r}\frac{\left( \frac{\partial f\left( X\right) }{\partial
S\left( X\right) }\right) ^{2}}{\left( f\left( X\right) +C_{0}\right) ^{2}}%
\right\} ^{2}\left( \left( \delta S\left( X\right) \right) ^{2}\right) ^{2}
\end{eqnarray*}

This leads to the following condition:%
\begin{eqnarray*}
&&\left( 1-\left\langle \lambda ^{2}\right\rangle \right) \left( \left(
\delta \hat{S}\left( X\right) \right) ^{2}+\left( \delta S\left( X\right)
\right) ^{2}+\left( \delta \hat{K}_{X}\left\vert \hat{\Psi}\left( X\right)
\right\vert ^{2}\right) ^{2}+\left( \delta K_{X}\left\vert \Psi \left(
X\right) \right\vert ^{2}\right) ^{2}\right) \\
&>&A\left( \left( \delta S\left( X\right) \right) ^{2}\right) ^{2}+\left(
B_{1}\delta S\left( X\right) +B_{2}\frac{\delta \frac{\hat{K}_{X}\left\vert 
\hat{\Psi}\left( X\right) \right\vert ^{2}}{K_{X}\left\vert \Psi \left(
X\right) \right\vert ^{2}}}{\frac{\hat{K}_{X}\left\vert \hat{\Psi}\left(
X\right) \right\vert ^{2}}{K_{X}\left\vert \Psi \left( X\right) \right\vert
^{2}}}\right) ^{2}
\end{eqnarray*}%
\begin{equation*}
A=\left( 1-\frac{\left( f\left( X\right) +C_{0}\right) \frac{\partial
^{2}f\left( X\right) }{\partial S^{2}\left( X\right) }}{\left( \frac{%
\partial f\left( X\right) }{\partial S\left( X\right) }\right) ^{2}}\right) 
\frac{2}{r}\frac{\left( \frac{\partial f\left( X\right) }{\partial S\left(
X\right) }\right) ^{2}}{\left( f\left( X\right) +C_{0}\right) ^{2}}
\end{equation*}%
\begin{equation*}
B_{1}=\frac{\hat{h}\left( X\right) }{2}S\left( X\right) \left( \frac{\hat{h}%
\left( X\right) }{2}\left( \frac{\partial f\left( X\right) }{\partial
S\left( X,\theta -2\right) }\right) ^{2}+\frac{\partial ^{2}f\left( X\right) 
}{\partial S^{2}\left( X,\theta -2\right) }\right)
\end{equation*}%
\begin{equation*}
B_{2}=\frac{\hat{h}\left( X\right) }{2}S\left( X\right) \frac{\partial
f\left( X\right) }{\partial S\left( X,\theta -2\right) }
\end{equation*}

\end{document}